\documentclass[smallextended]{svjour3}
\usepackage[total={6.5in,9in},top=1in,headsep=0.2in,headheight=1in]{geometry}
\usepackage{url}
\usepackage{graphicx}
\usepackage{amssymb}
\usepackage{natbib}
\usepackage{aas_macros}
\usepackage{lscape} 
\usepackage{makecell}
\usepackage{multirow}
\usepackage{multicol}
\usepackage[separate-uncertainty = true,multi-part-units=single]{siunitx} 
\usepackage{caption}
\DeclareSIUnit\px{px} 
\DeclareSIUnit{\electron}{\text{e}^{\text{-}}} 
\usepackage{hhline}
\usepackage{xcolor}
\usepackage{tcolorbox}
\usepackage{colortbl}
\usepackage{chngcntr}
\usepackage{ulem}
\counterwithin{figure}{section}
\counterwithin{table}{section}

\usepackage{enumitem}

\usepackage{authblk}

\newcounter{institute}

\usepackage{titlesec}

\titleformat{\paragraph}
{\normalfont\normalsize\bfseries}{\theparagraph}{1em}{}
\titlespacing*{\paragraph}
{0pt}{3.25ex plus 1ex minus .2ex}{1.5ex plus .2ex}

\usepackage{color}
\usepackage[hidelinks]{hyperref}
\hypersetup{
     colorlinks   = true,
     citecolor    = blue,
}
\usepackage{tocloft}

\title{\Large ET White Paper: To Find the First Earth 2.0}

\author{ 
{\bf \emph{{\large Edited by :}}}\newline
\newline
Jian Ge, Hui Zhang, Hongping Deng, Kevin Willis, \& Beverly Ge \newline
\newline
{\bf \emph{{\large Contributing Authors:}}} \newline
\newline
Jian Ge$^1$, Hui Zhang$^1$,  Weicheng Zang$^2$, Hongping Deng$^1$, Shude Mao$^{2,17}$, Ji-Wei Xie$^3$, Hui-Gen Liu$^3$, Ji-Lin Zhou$^3$, Kevin Willis$^{20}$, Chelsea Huang$^{26}$
% ...
and the ET consortium
}

\authorrunning{Ge et al.}
\institute{}
\date{\today}

\begin{document}
\maketitle

\clearpage
\vspace{1cm}
\textbf{\emph{\large Institutes:}}
\setcounter{institute}{1}
\begin{enumerate}
{\it 
\item Shanghai Astronomical Observatory, Chinese Academy of Sciences, Shanghai, China\\
jge@shao.ac.cn, zhangh@shao.ac.cn
\item Tsinghua University, Beijing, China
\item Nanjing University, Nanjing, China
\item Sun Yat-Sen University, Zhuhai, China
\item Tsung-Dao Lee Institute, Shanghai Jiao Tong University, Shanghai, China
\item Max Planck Institute for Solar System Research, Göttingen, Germany
\item Max-Planck-Institute f\"ur Astronomie, Heidelberg, Germany 
\item Zhejiang University, Hangzhou, China
\item University of Science and Technology of China, Hefei, China
\item Taiyuan University of Technology, Taiyuan, China
\item University of Maryland, College Park, MD, USA 
\item Boston University, Boston, MA, USA
\item Xinjiang Astronomical Observatory, Chinese Academy of Sciences, Xinjiang, China
\item Huazhong University of Science and Technology, Wuhan, China
\item Purple Mountain Observatories,  Chinese Academy of Sciences, Nanjing, China
\item Guizhou University, Guiyang, China
\item National Astronomical Observatories, Chinese Academy of Sciences, Beijing, China
\item Yunnan University, Kunming, China
\item Tel Aviv University, Tel Aviv, Israel
\item Science Talent Training Center, Gainesville, USA
\item Shanghai Institute of Technical Physics, Chinese Academy of Sciences, Shanghai, China
\item Xi'An Institute of Optics and Precision Mechanics, Chinese Academy of Sciences, Xi'an, China
\item Innovation Academy for Microsatellites, Chinese Academy of Sciences, Shanghai, China
\item Department of Astronomy of the University of Geneva, Versoix, Switzerland
\item University of Southern Queensland, QSL, Australia
\item Research School of Astronomy \& Astrophysics, Australian National University, ACT, Australia
\item Research School of Computer Science, Australian National University, ACT, Australia
\item Beijing Normal University, Beijing, China
\item University of Cambridge, Cambridge, UK
\item Dezhou University, Dezhou,  China
\item Jimei University, Xiamen, China
\item Ohio State University, Columbus, OH, USA
\item Korea Astronomy and Space Science Institute, Daejon, Republic of Korea
\item Kavli Institute for Astronomy and Astrophysics, Peking University, Beijing, China
\item Center for Astrophysics $|$  Harvard \& Smithsonian, Cambridge, MA, USA
\item Weizmann Institute of Science, Rehovot, Israel
\item Columbia University, New York, NY, USA
\item American Museum of Natural History, Central Park West, Manhattan, NY, USA
\item IRAP, Universit\'e de Toulouse, CNRS, CNES, UPS, Toulouse, France
\item Shanxi University, Taiyuan 030006, China
\item NASA Ames Research Center, Moffett Field, CA, USA
\item University of Nevada, Las Vegas, NV 89118, USA
\item Xinjiang Technical Institute of Physics and Chemistry, CAS, Urumqi, China
\item School of Physics, Peking University, Beijing, China
\item University of California, Davis, USA
\item Beijing Planetarium, Beijing Academy of Science and Technology, Beijing, China
\item Xi'an Jiaotong University, Xi'an, China
\item Shaoguan University, 512005 Shaoguan, China
\item Sydney Institute for Astronomy (SIfA), School of Physics, University of Sydney, NSW 2006, Australia
\item Institut f{\"u}r Physik, Karl-Franzens Universit{\"a}t Graz, Graz, Austria
\item Instituto de Astrof\'{\i}sica e Ci\^{e}ncias do Espa\c{c}o, Universidade do Porto, Porto, Portugal
\item Departamento de F\'{\i}sica e Astronomia, Faculdade de Ci\^{e}ncias da Universidade do Porto, Porto, Portugal
\item School of Physics and Astronomy, University of Birmingham, Birmingham, UK
\item Stellar Astrophysics Centre (SAC), Department of Physics and Astronomy, Aarhus University, Aarhus, Denmark
\item Kavli Institute for Theoretical Physics, University of California, Santa Barbara, CA, USA
\item AIM, CEA, CNRS, Universit\'e Paris-Saclay, Universit\'e de Paris, Sorbonne Paris Cit\'e, France
\item Institut f\"{u}r Astrophysik, Georg-August-Universit\"{a}t G\"{o}ttingen, G\"{o}ttingen, Germany
\item Center for Space Science, NYUAD Institute, New York University Abu Dhabi, Abu Dhabi, UAE
\item Landessternwarte K{\"o}nigstuhl (LSW), Heidelberg University, K{\"o}nigstuhl 12, 69117 Heidelberg, Germany
\item Heidelberg Institute for Theoretical Studies (HITS) gGmbH, Heidelberg, Germany
\item Institute for Astronomy, University of Hawai`i, 2680 Woodlawn Drive, Honolulu, HI 96822, USA
\item INAF - Osservatorio Astrofisico di Torino, via Osservatorio 20, 10025 Pino Torinese (TO), Italy
\item Yunnan Observatories, Chinese Academy of Sciences, Kunming, China
\item Key Laboratory for the Structure and Evolution of Celestial Objects, Chinese Academy of Sciences, Kunming, China
\item Center for Astronomical Mega-Science, Chinese Academy of Sciences, Beijing, China
\item University of Chinese Academy of Sciences, Beijing 100049,  China
\item Instituto de Astrofísica de Canarias (IAC), 38205, La Laguna, Tenerife, Spain
\item Universidad de La Laguna (ULL), Departamento de Astrofísica, 38206, La Laguna, Tenerife, Spain
\item Instituto de Astrofísica de Andalucía (IAA-CSIC), 18008, Granada, Spain
\item Dipartimento di Fisica e Astronomia Augusto Righi, Università degli Studi di Bologna, Bologna, Italy
\item INAF-Osservatorio di Astrofisica e Scienza dello Spazio di Bologna, Bologna, Italy
\item LESIA, Observatoire de Paris, PSL Research University, CNRS, Sorbonne Universit\'e, Universit\'e Paris Diderot, 92195, Meudon, France
\item Department of Astronomy, Yale University, 52 Hillhouse Ave., New Haven, CT 06511, USA
\item School of Physics, University of New South Wales, NSW 2052, Australia
\item Institute of Astronomy, KU Leuven, Celestijnenlaan 200D, B-3001 Leuven, Belgium
\item Centre for Astrophysics, University of Southern Queensland, Toowoomba, QLD 4350 Australia
\item Royal Observatory of Belgium, Ringlaan 3, B-1180 Brussel, Belgium
\item Konkoly Observatory, Research Centre for Astronomy and Earth Sciences, E\"otv\"os Lor\'and Research Network (ELKH), Konkoly Thege Mikl\'os \'ut 15-17, H-1121 Budapest, Hungary
\item  ELTE E\"otv\"os Lor\'and University, Institute of Physics, 1117, P\'azm\'any P\'eter s\'et\'any 1/A, Budapest, Hungary
}

\end{enumerate}
\clearpage

\hfill \break
{\bf \emph{\large Acknowledgment of Reviewers}}\newline
\hfill \break
The science and technologies of the ET project have been reviewed by many science and technical experts. These experts provided candid and critical comments, which helped the ET team pinpoint the scientific objectives and improve the systematic design of ET. The ET team is very grateful for the invaluable contributions from all reviewers listed below (in alphabetical order by last name): \\
\hfill \break
Prof. Qi An, {\it University of Science and Technology of China};	\\
Prof. Zhiming Cai, {\it Innovation Academy for Microsatellites, CAS (Chinese Academy of Sciences)};\\
Prof. Jing Chang, {\it National Astronomical Observatories, CAS;}\\
Prof. Xiangqun Cui, {\it Nanjing Institute of Astronomical Optics \& Technology, CAS;}\\
Prof. Mingde Ding, {\it Nanjing University;}\\
Prof. Jiangpei Dou, {\it Nanjing Institute of Astronomical Optics \& Technology, CAS;}\\
Prof. Cheng Fang, {\it Nanjing University;}\\
Prof. Quanlin Fan, {\it Technology and Engineering Center for Space Utilization, CAS;}\\
Prof. Xuewu Fan, {\it Xi’An Institute of Optics and Precision Mechanics, CAS;}\\
Prof. Huixing Gong, {\it Shanghai Institute of Technical Physics, CAS;}\\
Prof. Yidong Gu, {\it Technology and Engineering Center for Space Utilization, CAS;}\\
Prof. Baozhu Guo, {\it China Aerospace Science and Technology Corporation;}\\
Prof. Luis C. Ho, {\it Peking University;}\\
Prof. Xiaoxia Huang, {\it Beijing Institute of Remote Sensing Information;}\\
Prof. Haiying Hu, {\it Innovation Academy for Microsatellites, CAS;}\\
Prof. Aiming Jiang, {\it National Astronomical Observatories, CAS;}\\
Prof. Guang Jin, {\it Changchun Institute of Optics, Fine Mechanics and Physics, CAS;}\\
Prof. Jing Li, {\it Technology and Engineering Center for Space Utilization, CAS;}\\
Prof. Douglas. C. Lin, {\it University of Santa Cruz / Tsinghua University;}\\
Prof. Huawang Li,  {\it Innovation Academy for Microsatellites, CAS;}\\
Prof. Xinfeng Li, {\it Technology and Engineering Center for Space Utilization, CAS;}\\
Prof. Jie Lv, {\it Technology and Engineering Center for Space Utilization, CAS;}\\
Prof. Jianwei Pan, {\it University of Science and Technology of China;}\\	
Prof. Yuntian Pei, {\it Shanghai Institute of Technical Physics, CAS;}\\
Prof. Zhiqiang Shen, {\it Shanghai Astronomical Observatory, CASs;}\\
Prof. Jianzhong Shi, {\it Aerospace Systems Division;}\\
Prof. Shengcai Shi, {\it Purple Mountain Observatories, CAS;}\\
Prof. Huixian Sun, {\it Technology and Engineering Center for Space Utilization, CAS;}\\
Prof. Chi Wang, {\it Technology and Engineering Center for Space Utilization, CAS;}\\
Prof. Jianyu Wang, {\it Shanghai Institute of Technical Physics, CAS;}\\
Prof. Desheng Wen, {\it Xi’An Institute of Optics and Precision Mechanics, CAS;}\\
Prof. Ji Wu, {\it Technology and Engineering Center for Space Utilization, CAS;}\\
Prof. Xiangping Wu, {\it National Astronomical Observatories/Shanghai Astronomical Observatory, CAS;}\\
Prof. Yanqing Wu, {\it Toronto University;}\\
Prof. Weiming Xiong, {\it Technology and Engineering Center for Space Utilization, CAS;}\\
Prof. Genqing Yang, {\it Shanghai Institute of Microsystem and Information Technology, CAS;}\\
Prof. Jianfeng Yang,{\it	Xi’An Institute of Optics and Precision Mechanics, CAS;}\\
Prof. Wengang Yang, {\it	Xi’An Institute of Optics and Precision Mechanics, CAS;}\\
Prof. Shuhua Ye, {\it Shanghai Astronomical Observatory, CAS;}\\
Prof. Jinpei Yu,  {\it Innovation Academy for Microsatellites, CAS;}\\
Prof. Shuangnan Zhang, {\it	Institute of High Energy Physics, CAS; }\\
Prof. Yonghe Zhang,  {\it	Innovation Academy for Microsatellites, CAS;}\\
Prof. Yongwei Zhang, {\it	China Spacesat Co. Ltd.;}\\
Prof. Jianhua Zheng, {\it	Technology and Engineering Center for Space Utilization, CAS;}\\
Prof. Zi Zhu, {\it Nanjing University.}\\

\clearpage
\hfill \break

\hypersetup{linkcolor=}
\tableofcontents
\hypersetup{linkcolor=red}




\hypersetup{
     citecolor    = blue,
     linkcolor    = red
}

%

%
\clearpage
\section{Executive Summary}

The Earth 2.0 (ET) mission is a space telescope being developed in China to detect thousands of terrestrial exoplanets over a wide range of orbital periods and in interstellar space including Earth 2.0s, which are habitable Earth-like planets (0.8-1.25 Earth radius) orbiting solar-type stars. The long-baseline precision photometry enabled by the first space-based ultra-high precision CMOS photometer combined with a microlensing telescope allows the determination of the occurrence rate of Earth 2.0s, for the first time, as well as cold and free-floating low-mass planets. \\

\noindent{\bf Science Goals} 
\begin{itemize}
 \item[1.] The ET mission will explore the diversity of Earth-sized planet populations with different orbital periods including close-in sub-Earths, terrestrial-like planets in habitable zones, cold planets, and free-floating planets, and will accurately determine the occurrence rates of these small/low-mass planets. ET will also study correlations between Earth-sized planets with stellar properties (e.g., mass, multiplicity) and Galactic environments (e.g., thin disks, thick disks, bulge, and halo). These studies will address three key questions: 1) How common are habitable Earth-like planets orbiting around solar-type stars? 2) How do Earth-like planets form and evolve? 3) What is the mass function and likely origin of free-floating low-mass planets?
 \item[2.] The ET mission will significantly expand known planet populations, especially for cold planets with orbital periods up to 4 years. Observations of diverse multiple systems consisting of super-Earths, sub-Neptunes, and gas giants will shed further light on the coevolution of planetary worlds. In addition, there will be a considerable gain in the observation of rare populations such as ultra-short period planets, circumbinary planets, and metal-rich terrestrial planets which will allow for the development of coherent formation theories. The detection of more exotic objects such as exomoons, exorings, and exocomets is attainable with ET allowing for our Solar System to be put into context via censuses for extrasolar systems.  
 \item[3.] ET survey data will also facilitate studies in fields such as asteroseismology, Galactic archeology, time-domain sciences, Solar System objects, and black holes in binaries. Asteroseismology of bright host stars with planet candidates will provide accurate measurements of stellar parameters such as mass, radius, and age. Asteroseismology of different types of stars at different evolution stages will provide a comprehensive understanding of stellar interiors and evolution. ET will enable the rigorous archaeology of the Milky Way by providing state-of-the-art age dating for a substantial set of the oldest stars in our galaxy. ET will also serendipitously observe tens of thousands of Solar System objects over its lifetime, providing uninterrupted monitoring of bodies from the inner Solar System region to the trans-Neptunian region. This unique dataset will enhance our understanding of the dynamical and physical state of Solar System bodies.To get an idea of the breadth of ancillary science possible with a space telescope such as ET, see \cite{Kepler2020nkm..book.....H}.
 
\end{itemize}





\noindent{\bf Payload} 

ET consists of seven \SI{30}{\cm} telescopes, of which six are transit telescopes and one is a microlensing telescope. Each telescope will be equipped with a mosaic of four 9K$\times$9K CMOS detectors with a pixel size of \SI{10}{\um}. The average detector readout noise is 4 e$^-$/pix and the readout time is \SI{1.5}{\s}. An electronic rolling shutter is used for exposure control to avoid image smear.

Each transit telescope consists of eight transmissive lenses with a field of view (FOV) of 500 deg$^2$. The spectral range is 450-900 nm. The telescope works at a focal ratio of 1.57 which offers a pixel scale of 4.38 arcsec/pixel. The planned individual exposure time is \SI{10}{\s}. The image diameter for 90\% of the encircled energy is within 3.8 pixels for the entire FOV. The telescope’s overall transmission exceeds 76\%. The telescope operates at a temperature of about \SI{-30}{\degreeCelsius}  with a stability of $\pm$\SI{0.3}{\degreeCelsius} to minimize image drifts and size changes while the detector is passively cooled to \SI{-40}{\degreeCelsius} with its temperature stability maintained within $\pm$\SI{0.1}{\degreeCelsius}. The focus can be precisely adjusted by the temperature control of the telescope. A sunshield is used to block the Sun from hitting the telescopes and provide a stable thermal environment for all science payloads. With an additional hood on top of each telescope and optical baffles, the telescope’s scattered light level is maintained at less than 3 e$^-$/pix/s. The telescope hood is also used to radiatively cool the CMOS detectors.

The microlensing telescope is based on a Schmidt-Mann catadioptric system, providing a 4 deg$^2$ FOV with a spectral range of 700-\SI{900}{\nm}. Its focal ratio is f/17.2 with a plate scale of 0.4 arcsec/pix. It will be operated at the diffraction limit with FWHM less than 0.85 arcsec. The focus can be precisely adjusted by the temperature control of the telescope. The planned individual exposure time is \SI{10}{\minute}. 

ET will carry out a four-year high precision and uninterrupted photometric monitoring of about 1.2 million FGKM dwarfs (Gaia magnitude G$<$16) in the direction that encompasses the original {\it Kepler} field to obtain light curves of these stars for detecting planetary transits. Planetary radii and orbital periods can be determined from the planetary transits. Light curves of relatively bright stars (G$<$14) will also be used in asteroseismology analyses to derive accurate stellar parameters such as masses, radii, and ages. Most bright stars which are found to have planet candidates can be followed up with ground-based telescopes to determine the masses and densities of these planet candidates while some highly valuable targets will be observed with ground-based and space-based telescopes to measure atmospheric compositions. 

ET will also conduct a four-year high precision photometric monitoring of over 30 million stars ($I \leq 20.6$ mag) in the direction of the Galactic bulge to detect planetary microlensing events. The same field will be simultaneously monitored by three ground-based KMTNet telescopes. The combined data will be used to measure the masses of hundreds of cold planets including free-floating planets.\\

\noindent{\bf Mission Design} 

The ET space telescope craft will be comprised of a spacecraft platform and scientific payloads. The spacecraft is designed to provide an extremely stable platform for high-precision photometry measurements. ET will operate at the L2 halo orbit to ensure a good external heat flow and light environment, with transit telescopes always pointing toward the {\it Kepler} field. Every season, the spacecraft will be rotated by 90 degrees around the optical axis of the transit telescopes to keep the payload in the shade of the sunshield while the Sun powers the solar array of the satellite. The microlensing telescope will observe the Galactic bulge from March 21 to September 21 every year. The satellite will be built with an in-orbit lifetime of at least 4 years. The ET spacecraft will adopt inertial orientation three-axis stabilization, with a telescope pointing accuracy of better than 1.5 arcsec. The telescope pointing will be measured by star trackers, optical gyroscopes, and fine guidance sensors while its stability will be controlled and maintained by reaction wheels and thrusters. This will keep the telescope’s high-frequency jitters (up to 10 Hz) within 0.15 arcsec. By carefully controlling the temperatures of the scientific payloads and their mounting optical bench while thermally isolating the payloads from the spacecraft platform, the telescope’s long-term thermal drift will be controlled to within 0.4 arcsec. X-band will be used for the telemetry, tracking, and command of the spacecraft, as well as science data downloads. A daily scientific data volume of roughly 169 Gb will be downloaded at a rate of 20 Mbps via phased array antennas. The entire spacecraft, including payloads, will weigh $\sim$3.2 tons and the long-term power consumption is $\sim$1500 watts. The ET mission will enter its preliminary design phase after the project’s down-selection this June. The team aims to launch the mission from the Xichang launch site by a CZ-3B rocket at the end of 2026.  \\

\noindent{\bf ET Performance}

ET photometry simulations show that the current design can reach a photometric precision of 34 ppm for a G=13.4 solar-type star with a 6.5-hour integration of the target. This simulation includes photon and instrument noise and has produced consistent results with {\it Kepler} on-sky measurements when {\it Kepler}’s instrument parameters are used. ET survey simulations show that the ET transit survey will be able to detect $\sim 29,000$ new planets, including $\sim 4,900$ Earth-sized planets and 10-20 Earth 2.0s assuming an Earth 2.0 occurrence rate of 10\%. The ET microlensing survey will be able to detect $\sim 400$ bound planets and 600 free-floating planets; $\sim 300$ of these planets will have mass measurements. 

The ET mission consortium consists of more than 300 scientists and engineers from over 40 institutions in China and abroad. The ET technical team has built 
many hardware and software components for space missions and will be building many of the components for ET. These include but are not limited to DAMPE, QUESS, TAIJI-1, SVOM, ASO-S, and EP.

\newpage
\section{Scientific Goals}
\subsection{\bf Planetary Science}
\subsubsection{Earth 2.0s and Terrestrial Planets }\label{sec:Terrestrial} 
{\bf Authors:} \\
\newline
Jian Ge$^1$, Hui Zhang$^1$, Hongping Deng$^1$,  Xinyu Yao$^1$, Jiapeng Zhu$^1$, Tong Fang$^1$, Ji-Wei Xie$^2$, Ji-Lin Zhou$^2$, Hui-Gen Liu$^2$, Steve B. Howell$^3$ \& Jun Yan$^4$ \\
{1.\it Shanghai Astronomical Observatory, Chinese Academy of Sciences, Shanghai 200030, China} \\
{2.\it Department of Astronomy, Nanjing University, Nanjing 210030, China} \\
{3.\it NASA Ames Research Center, Moffett Field, CA 94035, US}\\
{4.\it Department of Atmospheric and Oceanic Sciences, School of Physics, Peking University, Beijing, China}\\

\paragraph{Booming of exoplanets and the elusive Earth 2.0}\label{sec:earth2_background} 
“Are we alone in the universe?” This fundamental question is as old as humankind itself. Humanity has wondered whether there is life elsewhere in the universe and, if so, whether it exists on a second Earth. The discovery of the first extrasolar planet (also called an exoplanet) around a Sun-like star, 51 Pegasi~\citep{Mayor1995},  challenged our Solar System's uniqueness and opened up a new field of astronomy, the study of extrasolar planets, which has expanded exponentially in recent years. Mayor and Queloz were awarded the Nobel Prize in Physics in 2019 for this major achievement. The subsequent hunt for exoplanets has dramatically changed our understanding of extrasolar planetary worlds and Earth-like planets.
\begin{figure}[htbp]
	\centering
	{\includegraphics[scale=.5]{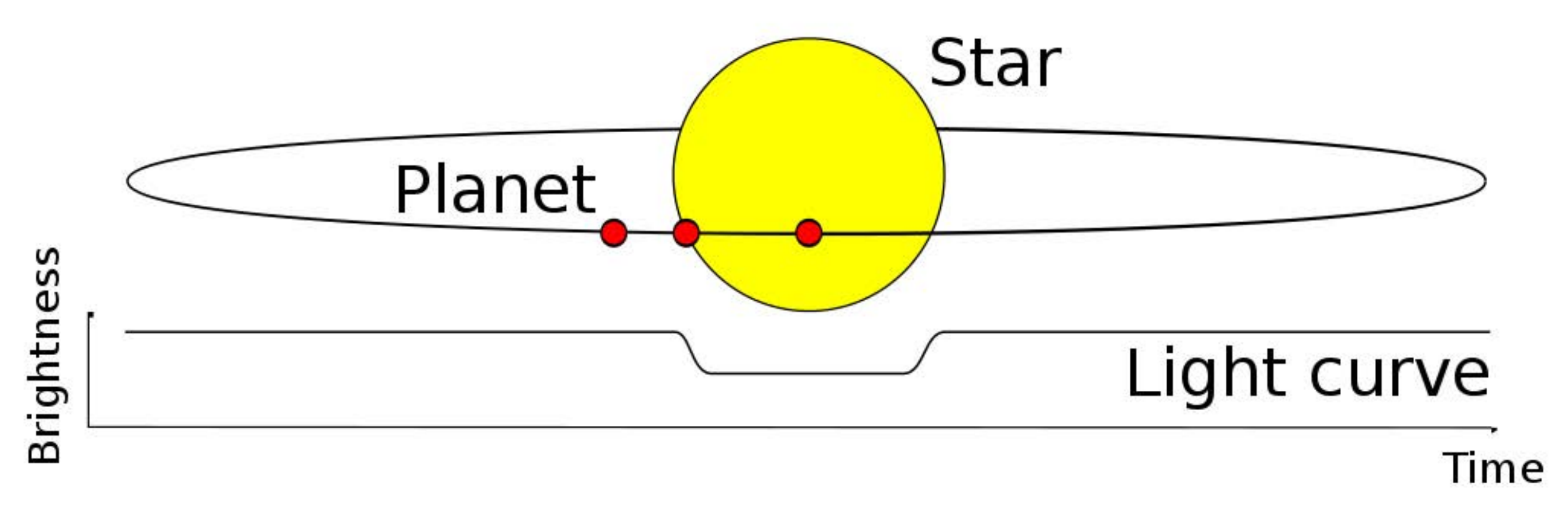}}
	{ \caption{\small The transit method------precision measurements of the dimming of a star's light curve caused by a planet passing in front of the star are used to detect exoplanets (\href{https://commons.wikimedia.org/wiki/File:Planetary_transit.svg}{credit Wikimedia}). }
		\label{transit_method}}. 
\end{figure}

\begin{figure}[htbp]
	\centering
	{\includegraphics[scale=.3]{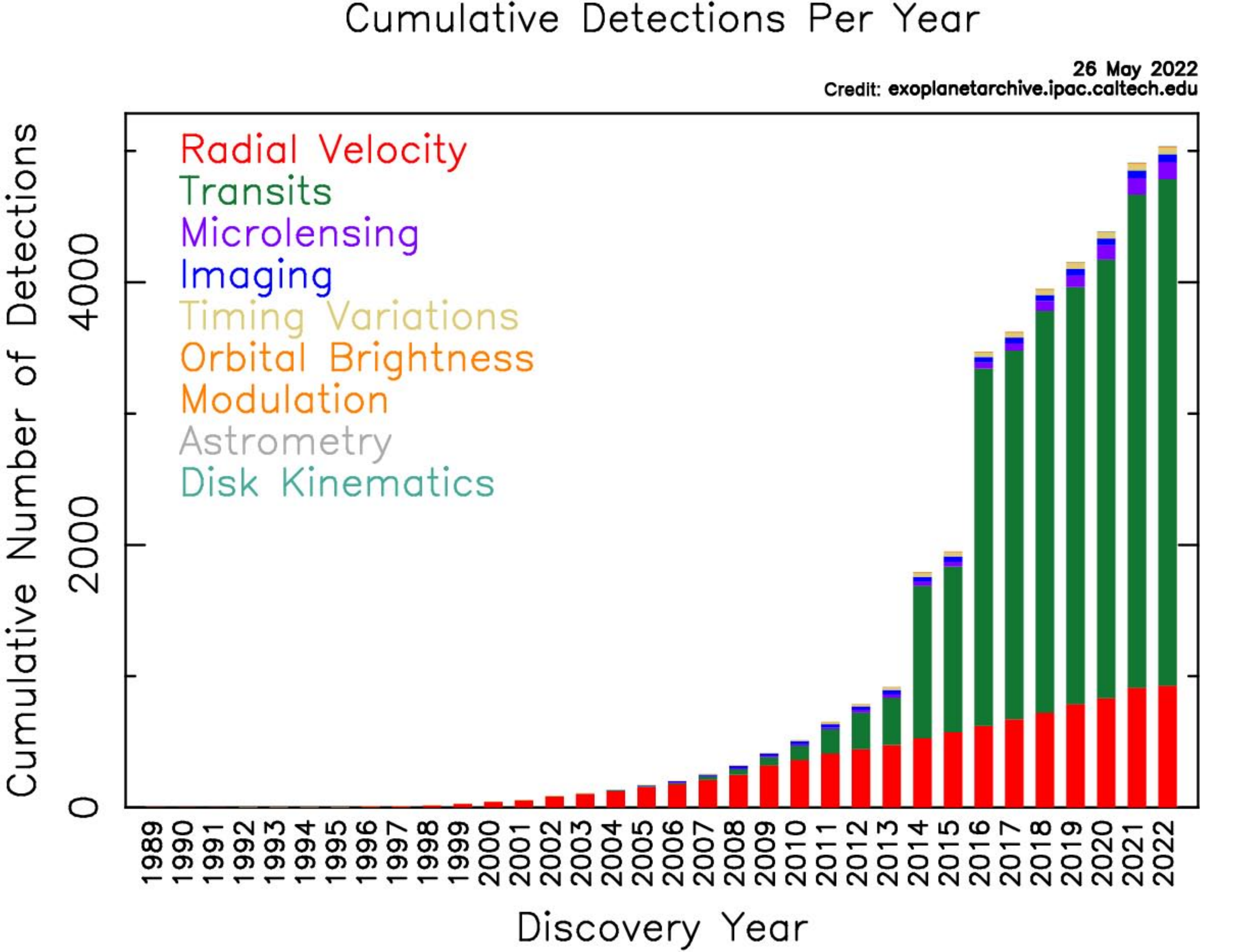}}
	{ \caption{\small The number of confirmed exoplanets with various detection methods over the years. Due to the discovery of a large number of transiting exoplanets by {\it Kepler} in 2014 and 2016, the number of known exoplanets doubled. This also resulted in the transit method becoming the primary method for detecting exoplanets, which is still the most promising way to search for Earth 2.0s (credit exoplanetarchive.ipac.caltech.edu). }
		\label{exo_dischist_cumulative}}. 
\end{figure}

During the early phase of planet-hunting, the radial velocity (RV) method used to discover 51 Peg b was the most efficient and powerful tool for finding massive Jupiter-like planets (giant planets). Continuing improvements in RV precision have allowed the detection of super-Earth mass planets in recent years (e.g., \citealp{Mayor2009}).  
 By the 2010s, hundreds of giant planets had been detected. Through statistical measures, these discoveries helped unveil the following characteristics of giant planets: 1) Jupiter-like and Saturn-like giant planets appear to be very common (occurrence rate about 10\%), especially around solar-type stars with high metallicity; 2) there are other kinds of Jupiter mass planets with unknown origin, such as hot Jupiters and warm Jupiters, in addition to cold Jupiter-like giant planets with long-period orbits; 3) most giant planets likely form via the core accretion processes \citep[one of the two leading theories of planet formation][]{Ida2004,Ida2005}.
 
\begin{figure}[!htbp]
	\centering
	{\includegraphics[width=0.8\textwidth]{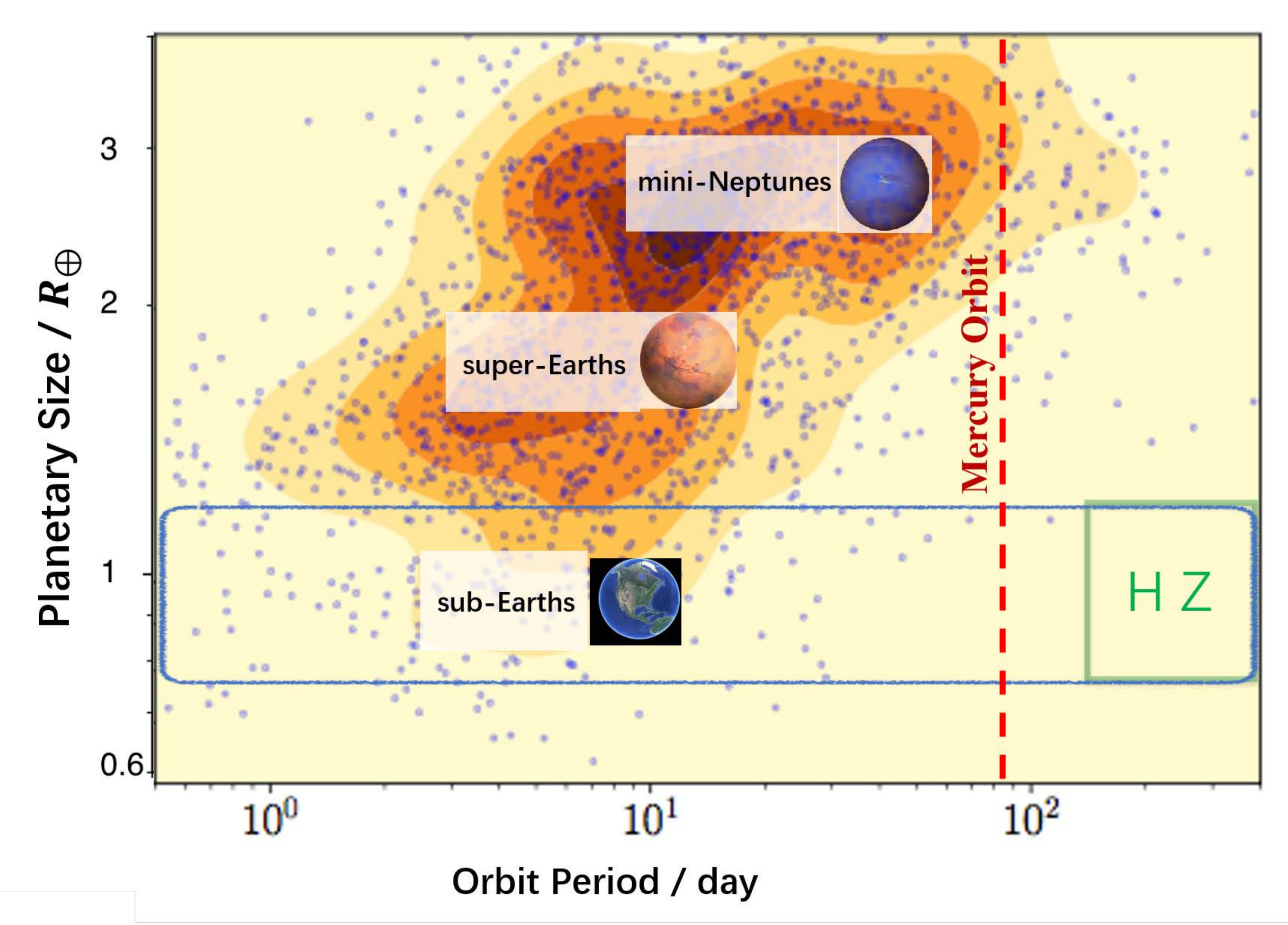}}
	{ \caption{\small The distribution of the radii and orbital periods of the small planets discovered by the {\it Kepler} mission. Most {\it Kepler} planets are the so-called ``super-Earths" and ``sub-Neptunes", which orbit their host stars within the orbit of Mercury and are 1 to 4 Earth radii in size. A few sub-Earths (roughly Earth-size) lie close to their parent stars, and none of them are close to Earth 2.0 (long-period Earth-sized planet in the habitable zone of a solar-type star, as denoted by the green box).}
		\label{sub_super_earth}}
\end{figure}

The dominance of the RV method for detecting planets has declined since the launch of NASA's {\it Kepler} satellite in 2009. With its 0.95-m space telescope, {\it Kepler} achieved an ultra-high photometric precision of around 30 ppm \citep[parts per million,][]{Gilliland2011ApJS} using the transit method (Figure \ref{transit_method}) and quickly identified over 4000 new exoplanets and planet candidates (see Figure \ref{exo_dischist_cumulative}, e.g., \citealp{Borucki2010,Howard2012}). This led to the discovery of two new types of planets that were difficult to detect from ground-based transit observations: planets with sizes between that of Earth and Neptune, also known as super-Earths ($1.25R_\oplus < R_p \leq 2R_\oplus$), and sub-Neptunes ($2R_\oplus < R_p \leq 4R_\oplus$) (see Figure \ref{sub_super_earth}). Many of these planets would orbit within the orbit of Mercury in our Solar System (0.4 AU, where 1 AU is the average distance between the Sun and Earth). These planets are more common than giant planets - about one-third of the stars in the Milky Way host such planets \citep{Zhu2018} - but they are absent in our Solar System. This observation has challenged the traditional planet formation theory and inspired entirely new theoretical models. Observations and theories suggest that most of these planets are born early with atmospheres rich in hydrogen and helium (e.g., \citealp{WuLithwick2013,Owen2017,Fulton2017}) inherited from the gaseous protoplanetary disks, extended flattened structures around young stars, and are thus considered the Generation I planets. Although super-Earths are ubiquitous in the universe, their formation processes and surface environments appear very different from those of our Earth.  Unlike super-Earths, Earth-sized planets likely form from collisional debris after gaseous disks have been dissipated and are likely Generation II planets in the planetary system, thus resembling Earth in origin \citep[see, e.g.,][]{Qian2021,Morbidelli2012,Fang2020,Liu2022}. Therefore, it is speculated that super-Earths are unlikely to harbor life.

The {\it Kepler} space mission was designed to detect Earth 2.0s --- Earth-sized terrestrial planets ($0.8R_\oplus < R_p \leq 1.25R_\oplus$) in the habitable zone of a solar-type star (F8V-K2V, Figure \ref{sub_super_earth}), and measure their occurrence rate \citep{Borucki2009}. However, the mission failed to detect them, although {\it Kepler} successfully detected over 2000 new planets and identified two major planet populations of close-in super-Earths and sub-Neptunes. This was primarily due to: 1) the failure of the two reaction wheels in the fourth year of its operation \citep[see][]{K22014PASP..126..398H}; 2) the unexpectedly high stellar noise of solar-type stars \citep{Gilliland2011ApJS}; and 3) the high readout noise of detectors (on average \SI{86}{\electron}, \citealp{Gilliland2011ApJS}). Specifically, the stellar noise levels for solar-type stars is about 50\% higher than that of our Sun. The high readout noise of the {\it Kepler} CCD detectors led to high photometric errors for solar-type stars fainter than the 12th magnitude exceeding 34 ppm; smaller errors are necessary for detecting Earth 2.0s. As a result, the {\it Kepler} field only contains a small quantity of quiet solar-type stars having a stellar noise of less than 17 ppm in their fields for detecting Earth 2.0s. Higher stellar noises could have been compensated for by monitoring solar-type stars over a longer observation period ($\sim$8 years) than the originally planned mission lifetime (3.5 years) to increase signal-to-noise (SN) ratios until they were high enough for detecting Earth 2.0s. However, the failure of the reaction wheels further impeded {\it Kepler}’s ability to achieve this potential S/N gain. At the conclusion of its primary mission in 2014, {\it Kepler} did not detect any Earth 2.0s as planned.

\begin{figure}[!htbp]  
	\centering
	\includegraphics[width=0.8\textwidth]{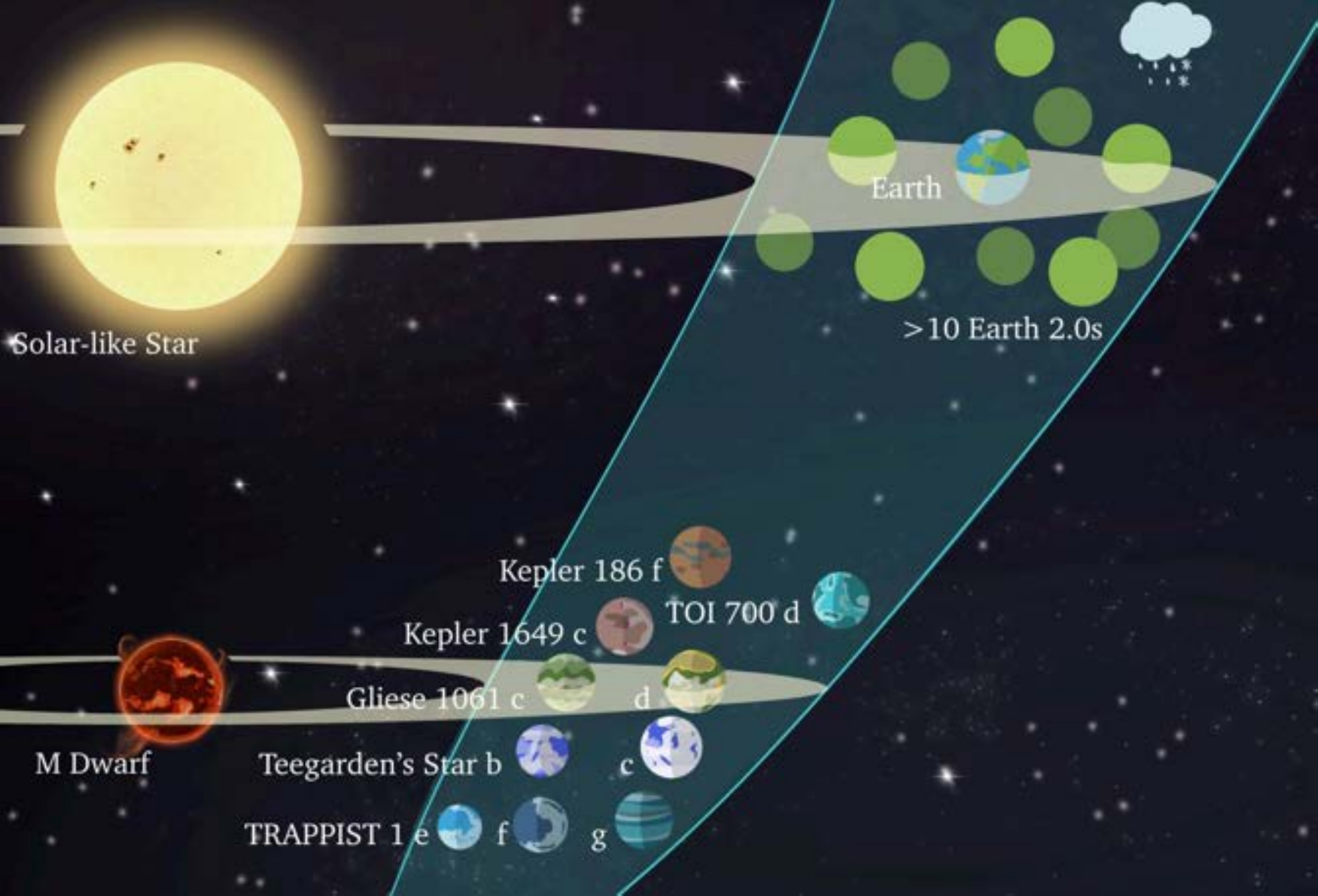}
	 \caption{\small Illustration of all detected Earth-sized ($0.8R_\oplus < R_p \leq 1.25R_\oplus$) planets in habitable zones (blue shaded region) and the expected 2.0 detections (green spheres) by the ET transit telescope. The comparatively active M dwarfs, hosting all current Earth-sized planets, may spoil the habitability of these planets. 	\label{fig:earthsize}}
\end{figure}

Ten Earth-sized planets have been detected in the `habitable zone' around M dwarf stars, i.e., TRAPPIST 1 \citep{Gillon2017seven}, Teegarden's star \citep{Zechmeister2019carmenes}, Gliese 1061 \citep{Dreizler2020reddots}, Kepler 1649 \citep{Vanderburg2020habitable}, TOI 700 \citep{Gilbert2020first}, and Kepler 186 \citep{Quintana2014earth}, as illustrated in Figure \ref{fig:earthsize}. A habitable zone around a main-sequence star is defined as the orbital region at which liquid water could persist on the planet’s surface for a long time. This definition is based on the fact that liquid water is necessary for all types of life on Earth. The inner edge of the habitable zone is determined based on runaway greenhouse or moist greenhouse, and the outer edge of the habitable zone is defined based on CO$_2$ condensation or maximum CO$_2$ greenhouse effect \citep{Kasting1988,Kasting1993,Pierrehumbert2010}. In this regard, solar-like stars have wider habitable zones than M dwarfs, as shown in Figure \ref{fig:earthsize}. However, the Earth-sized planets in the habitable zone of M dwarfs may not be habitable \citep[see, e.g.,][]{Stevenson2019failure,tarter2007reappraisal}. First, the habitable zone of M dwarfs lies close to the star resulting in a strong tidal force on planets which may tidally lock the planet, forming peculiar atmosphere circulations \citep{Yang2013HZ,Yang2014HZ}. Second, M dwarfs are more variable than solar-type stars with intense flares, releasing more X ray and ultraviolet radiation that may deprive the planet of its atmosphere. Last but not least, in the extended formation stage, young M dwarfs may have already evaporated all water in the habitable zones. 


Habitable Earth-like planets around solar-type stars, i.e., Earth 2.0s, are likely the most favorable places to search for extraterrestrial life due to their potentially having physical, chemical, and potentially biological environments similar to Earth. Therefore, it is necessary to identify Earth 2.0s first before extraterrestrial life can be possibly detected. Most current space missions for exoplanets do not cover this key area. For instance, NASA’s Transiting Exoplanet Survey Satellite ({\it TESS}) mission, which is currently in operation in space, has been very successful in detecting thousands of short-period planet candidates including Earth-mass planets (e.g., \citealp{Kunimoto2022}; \citealp{Guerrero2021} and references therein). However, {\it TESS} is highly unlikely to detect any Earth 2.0s as this mission was not designed to monitor the same sky region for multiple years in order to detect Earth 2.0s with a long orbital period. Additionally, due to the {\it TESS} telescope size, it does not reach sufficient photometric precision to detect an Earth orbiting a G star. Similarly, the ESA’s CHaracterising ExOPLanet Satellite (CHEOPS) transit space mission is dedicated to studying bright, nearby stars that are already known to host exoplanets, not to search for planet candidates \citep{Benz2021,Lendl2020}. The {\it Plato} 2.0 (PLAnetary Transits and Oscillation of stars) mission focuses on transit planets around bright stars (4-11 mag) and bulk planet characterization with mass determination from follow-up RV studies \citep{Rauer2014}. Although {\it Plato} has some chance of detecting Earth 2.0s, its short staring time even in the long-pointing mode (2-3 years) limits its ability to robustly detect Earth 2.0s.


This research has been prioritized in multiple programs by different funding agencies. For instance, the US National Academy of Sciences recently announced the study of Habitable Worlds as one of the three highest priority areas for future astronomy in the Decadal Survey on Astronomy and Astrophysics \citep{national2021pathways}. In 2018, the committee on exoplanet science strategy formed by the National Aeronautics and Space Administration (NASA) announced that searching for habitable planets and understanding the formation and evolution of planetary systems along with the diversity of planetary system architectures, planetary compositions, and planetary environments were the two main goals of NASA’s exploration in the next 20 years \citep{ExoplanetScienceStrategy2018}. In addition, the next generation of flagship space telescopes of NASA and the European Space Agency (ESA) (e.g., the James Webb Space Telescope (JWST), the Nancy Grace Roman Space Telescope (ROMAN), the Large UV/Optical/IR Surveyor (LUVOIR), ARIEL, etc. and the next generation ground-based 30-meter class telescopes (e.g., the Thirty Meter Telescope (TMT), the European Extremely Large Telescope (E-ELT), the Giant Magellan Telescope (GMT), etc.) have asserted that the study of the atmospheric compositions, the internal structures of terrestrial planets, and the exploration of these planets' habitability are among their key scientific goals. 

\paragraph{Survey for Earth 2.0s and a census for terrestrial planets }\label{sec:earth2_goals} 


Although Earth 2.0s have not yet been detected, we are confident that they do exist since terrestrial-size planets are already known to exist across a wide range of orbital periods around solar-type stars from close-in orbits all the way out to extremely cold orbits beyond the snowline (where water turns into ice), and even in interstellar space as free-floating terrestrial mass planets. In fact, {\it Kepler} has identified over 300 Earth-sized planets with short-period orbits (most of them shorter than 30 days) around some quiet bright stars. These Earth-sized planets (also called ``sub-Earths") and specifically the ones that orbit in their host stars' habitable zones are the Earth 2.0s that we are looking for. 


On the other hand, our current knowledge about exo-terrestrial-planets is limited to {\it Kepler}'s close-in samples. We know very little about: 1) the occurrence rate of long-period (from several to hundreds of years) terrestrial planets, 2) how terrestrial planets are formed and how mobile these small planets are in their host systems, and 3) the population of free-floating Earths and how they get exiled. To fully understand their formation and evolution, a census for exo-terrestrial planets is crucial. Their small size and low mass pose significant challenges to transit and RV searches, especially so for those with long-period orbits. Microlensing is currently the only method for studying low-mass, long-period and free-floating planets. However, current microlensing planetary studies are limited by small number statistics and a lack of mass measurements  (see section \ref{sec:MicroLensing}). The current largest statistical sample of microlensing planets contains only 23 planets, with a minimum mass of about $10 M_{\oplus}$, assuming a host mass of $0.5 M_{\odot}$ \citep{Suzuki2016}. In addition, for most microlensing planetary events, ground-based light-curve analyses only yield the planet-to-host mass ratio leaving the actual masses of the host star and the planet unknown. Thus, determining the mass function for long period and free-floating Earths from a large statistical sample offers a unique chance to reveal their origin. Microlensing appears to be the most favorable approach for accomplishing this task at the present time.

\begin{figure}[!bp]
   \centering
   \includegraphics[width=0.85\textwidth,trim={0 3cm 0 1cm}, clip]{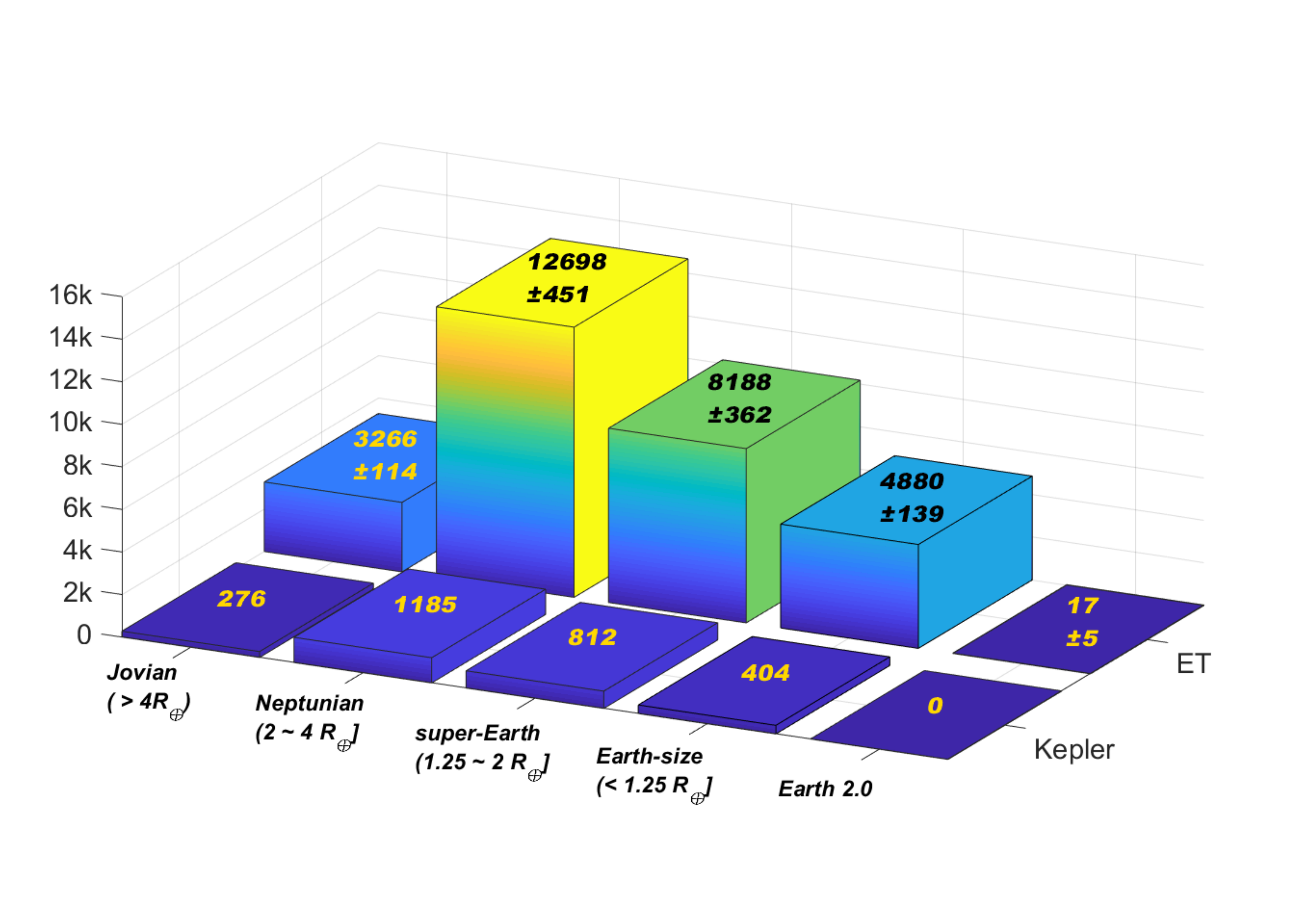} 
   \caption{Predicted ET planets in comparison with {\it Kepler}'s discoveries (see section \ref{sec:yield}). For sub- and super-Earth, ET will increase the sample size by a factor of $\sim$10 in the near future (before 2030s). 
   }\label{fig:ETnumber}
\end{figure}

Today, the exoplanet field is approaching a critical point: the entire astronomical community urgently needs to find Earth 2.0s and collect a large sample of terrestrial planets in various orbits. Fortunately, recent technological advances in space planet detection can readily support the achievement of these scientific goals. We need a new space mission to take full advantage of mature space exoplanet detection technologies to finally explore the domain of terrestrial planets including Earth 2.0s.  The Earth 2.0 (ET) space mission being developed in China is specifically designed to study this key science area. The key scientific goal of this mission is to detect Earth 2.0s for the first time, as well as many hot, warm, and cold sub-Earths (terrestrial planets) and even free-floating Earths, by utilizing the matured transit and microlensing methods together on a single spacecraft platform. 

High precision photometry data from space missions, e.g., {\it Kepler} and {\it TESS}, suggest that a photometry precision of $\sim 34$ ppm is capable of detecting Earth 2.0s given enough target stars and monitoring time. Our proposed ET mission has an optimized space photometer design providing a potential survey capability about 10 times that of {\it Kepler}. This greatly increases ET’s overall capability to detect exoplanets, including the elusive Earth 2.0s. Follow-up observations of Earth 2.0 planet candidates will help to measure their masses, densities, and atmospheric compositions while also allowing the study of the internal structure and the habitability of these planets (see section \ref{sec:targe-follow}). ET's microlensing telescope will add critical long-period planets to the exo-terrestrial planets population. ET's large sample of terrestrial planets, including Earth 2.0s, will be the ``Rosetta Stone" for understanding exo-Earth's origin. In addition, ET will also detect about 30,000 new planets of various types (Figure \ref{fig:ETnumber}), granting insight into different planetary systems and helping to answer the question of whether our Solar System is unique in the Universe.


\subsubsection{Mass Functions for Free-Floating Planets and Long-Period Planets}\label{sec:MicroLensing} 
{\bf Authors:} \\
\newline
Weicheng Zang$^1$, Shude Mao$^{1,2}$, Andrew Gould$^{3,4}$, Chung-Uk Lee$^5$, Subo Dong$^6$, Jennifer Yee$^7$, Wei Zhu$^1$, Yossi Shvartzvald$^8$, Hongjing Yang$^1$, Renkun Kuang$^1$, Jiyuan Zhang$^1$ \\
{1.\it Department of Astronomy, Tsinghua University, Beijing 100084, China}\\
{2.\it National Astronomical Observatories, Chinese Academy of Sciences, Beijing 100101, China}\\
{3.\it Max-Planck-Institute for Astronomy, K\"onigstuhl 17, 69117 Heidelberg, Germany} \\
{4.\it Department of Astronomy, Ohio State University, 140 W. 18th Ave., Columbus, OH 43210, USA} \\
{5.\it  Korea Astronomy and Space Science Institute, Daejon 34055, Republic of Korea}  \\
{6.\it  Kavli Institute for Astronomy and Astrophysics, Peking University, Yi He Yuan Road 5, Hai Dian District, Beijing 100871, China}  \\
{7.\it  Center for Astrophysics $|$ Harvard \& Smithsonian, 60 Garden St.,Cambridge, MA 02138, USA}  \\
{8.\it  Department of Particle Physics and Astrophysics, Weizmann Institute of Science, Rehovot 76100, Israel}  \\

\paragraph{Importance of Microlensing Planets}

The snow line defines the distance from a star beyond which the disk temperature drops below \SI{160}{\kelvin} and water turns to ice in the protoplanetary disk \citep[e.g.,][]{Hayashi1981,Min2011}. The core-accretion models for planet formation \citep[e.g.,][]{Lissauer1987,Mordasini2009,Ida2013} predict that planets beyond the snow line should be abundant and that most gas giants and ice giants reside here. The gravitational microlensing technique is most sensitive to planets around the Einstein ring radius \citep{Shude1991,Andy1992,Bennett1996}
\begin{equation}
    \theta_{\rm E} = \sqrt{\kappa M_{\rm L} \pi_{\rm rel}};\qquad \pi_{\rm rel} = \frac{\rm AU}{D_{\rm L}} - \frac{\rm AU}{D_{\rm S}};\qquad \kappa = \frac{4G}{c^2 \rm AU} = 8.144 \frac{\rm mas}{M_\odot},
\end{equation}
where $M_{\rm L}$ is the lens mass, $D_{\rm L}$ is the lens distance, and $D_{\rm S}$ is the source distance. For typical Galactic microlensing events, the Einstein ring radius is 2 to 3 times the snow line radius, so the gravitational microlensing technique can provide important diagnostics for the formation and evolution of long-period planets beyond the snow line. Microlensing has found that planets near and beyond the snowline are abundant, with an occurrence frequency in the range of $\sim 50\%$ to 100\% \citep[with planetary masses down to several Earth-masses,][]{mufun,Cassan2012,Wise,Suzuki2016}. \citet{Suzuki2016} studied a sample of 23 planets detected by the MOA-II survey (2007--2012) and found that microlensing planets more frequently have smaller planet to host mass ratios, $q$, until the trend reverses at $q \sim 57q_{\oplus}$, where $q_{\oplus} = 3 \times 10^{-6}$ is Earth/Sun mass ratio. This turn over was not expected in the standard core accretion theory \citep[e.g.,][]{Ida2004} and would have important implications for the formation of cold planets \citep{Suzuki2018,Deng2021formation}.

Another unique capability of the gravitational microlensing technique is detecting low-mass free-floating planets (FFPs) that are unbound to any star \citep[e.g.,][]{OB161928}. Many ejection mechanisms, including planet-planet dynamical interactions \citep{Rasio1996,Forgan2017}, ejections from multi-star systems \citep{Kaib2013}, and stellar flybys \citep{Malmberg2011} predict the existence of FFPs, but their abundance and masses vary from model to model \citep[e.g.,][]{MaFFP,Barclay2017}. Recently, \citet{Gould2022} studied the free-floating planet candidates and bound planets detected by the Korea Microlensing Telescope Network \citep[KMTNet,][]{KMT2016} and suggested that FFPs are substantially more frequent than bound planets and thus the ejection mechanisms partly or even largely shape the power-law distribution of bound planets. Therefore, the abundance and mass function of FFPs would give major insights into the formation and evolution of planetary systems \citep{CMST}. 

In addition, the gravitational microlensing technique is complementary to other detection methods in the host properties. Because microlensing does not rely on light emission from the gravitational lenses, it can detect planets around all types of objects \citep[including white dwarfs,][]{MB10477_AO} at various Galactocentric distances. A statistical study of microlensing planets can reveal how different galactic environments \citep[e.g., bulge versus disk,][]{Novati2015,Matthewbulge,Zhu2017spitzer,Naoki2021} or different stellar characteristics \citep[e.g., stellar mass,][]{Naoki2021} relate with planetary frequency. For example, microlensing has demonstrated that giant planets around M dwarfs are common \citep[e.g.,][]{OB050071D,OB050071AO,OB171140}, while the core accretion theory predicts that massive planets around M dwarfs should be rare due to slow core growth \citep{Laughlin2004,Ida2005,Morales2019giant,Liu2020pebble}. 


\paragraph{Current Challenges in Microlensing Planetary Studies}

The main difficulties in utilizing the gravitational microlensing technique to detect bound and free-floating planets are the low probability of microlensing events and the short duration of unpredictable planetary signals. Thus, continuous, wide-area, high-cadence \citep[e.g., a cadence of $\sim 4~{\rm hr}^{-1}$ for Earth-mass planets,][]{Henderson2014} observations toward the Galactic bulge are required. The real implementation of this observation strategy was realized by the KMTNet survey across three continents in 2015. However, current microlensing planetary studies are limited by small number statistics and an absence of mass measurements.

The largest published sample of bound microlensing planets to date contains only 23 planets \citep{Suzuki2016}, with only two low mass-ratio ($q < 33q_{\oplus}$) planets. The ongoing systematic KMTNet planetary anomaly search \citep{OB191053,KB190253,OB180383,KB191042,OB181126} is expected to establish a statistical sample containing more than 100 bound planets, but its current detection rate of cold Earths ($0.5M_{\oplus} \leq M_p \leq 2M_{\oplus}$, corresponding to $q_{\oplus} \lesssim q \lesssim 4q_{\oplus}$) is only $\lesssim 1/$yr \citep{OB191053,KB200414,OB190960}, accompanied by hundreds of times more false positives \citep{OB191053,KB191042}. 


The current microlensing studies of FFPs are also limited by small number statistics. \citet{Sumi2011} analyzed the Einstein timescale ($t_{\rm E}$) distribution of 474 microlensing events observed by the MOA-II survey and found 10 microlensing events with $t_{\rm E} = $ 0.5--2 days, suggesting that the frequency of Jupiter-mass FFPs is about twice the frequency of stellar objects. However, the substantial Jupiter-mass FFPs were later ruled out by \citet{Mroz2017a} using a five-times larger statistical sample with the superior OGLE-IV data. \citet{Mroz2017a} also found six events with $t_{\rm E} < 0.4$ days. Most of these events were poorly sampled, but if they are real microlensing events, it means that low-mass FFPs (from Earth-mass to Neptune-mass) are more frequent than stars. Recently, \citet{Gould2022} probed FFPs by analyzing the $\theta_{\rm E}$ distribution of 2016--2019 KMTNet finite-source/point-lens \citep[FSPL,][]{1994ApJ...421L..75G, Shude1994, Nemiroff1994} events with giant sources. They found a clear gap \citep[named the ``Einstein Desert'',][]{KB172820} in the $\theta_{\rm E}$ distribution at $9~\mu{\rm as} < \theta_{\rm E} < 26~\mu{\rm as}$, and thus confirmed the absence of Jupiter-mass FFPs. Based on four events below this gap, they also estimated that there are substantially more FFPs than known bound planets. The statistics will not improve soon since KMTNet's current detection rate of FFP events is only $\sim 1/$yr, and the substantial false positives limit KMTNet searches to FFP events that occurred on the rare but bright giant sources, although \citet{CMST} showed that G-dwarf and subgiant sources have higher rates on FSPL FFP events. 

For bound and free-floating planets, current microlensing planetary studies only focus on the planet/host mass-ratio function and the $t_{\rm E}$ or $\theta_{\rm E}$ distributions, but no study has been carried out to directly characterize the mass function. The lens mass measurements need two out of three observables that yield mass-distance relations shown in the following equation:  $\theta_{\rm E}$, the microlensing parallax $\pi_{\rm E}$, and the apparent brightness of the lens. Simultaneous measurements of $\theta_{\rm E}$ and $\pi_{\rm E}$ can yield the lens mass by \citep{Gould1992,Gould2000}
\begin{equation}
    M_{\rm L} = \frac{\theta_{\rm E}}{\kappa\pi_{\rm E}}. 
\end{equation}
The measurement of $\theta_{\rm E}$ is mainly through finite-source effects when the source crosses or approaches a caustic, which are frequently detected in planetary events but rare in single-lens events \citep{ZhuPLFS} because the caustic of a single-lens event is only a geometric point. However, for the FFP event, because the angular source radius $\theta_*$ is of the same order as $\theta_{\rm E}$, half of the 12 FFP candidates \citep[][and references therein]{Mroz2017a,KB172820} discovered to date have $\theta_{\rm E}$ measurements. In addition, $\theta_{\rm E}$ can be measured by interferometrically resolving microlensed images \citep{Jan2017,Cassan2022}, but currently this is only feasible for bright ($K \lesssim 11$ mag) events with $\theta_{\rm E} \gtrsim 1$ mas.

The microlensing parallax can be measured by ``orbital microlens parallax'' due to the Earth's orbital acceleration around the Sun, which introduces deviations from rectilinear motion in the lens-source relative motion \citep{Gould1992}. However, this method is generally feasible only for events with long timescales, $t_{\rm E} \gtrsim$ year/$2\pi$ \citep[e.g.,][]{OB171434}. Moreover, systematics in the data, stellar variability of the source star and the microlensing xallarap effect \citep[orbital motion of the source,][]{Griest1992} can all affect the orbital microlens parallax measurements. The most efficient and robust way to measure $\pi_{\rm E}$ is by ``satellite microlens parallax'', which is conducted by observing the same microlensing event from Earth and at least one satellite \citep{1966MNRAS.134..315R,1994ApJ...421L..75G,Gould1995single}. The feasibility of satellite microlens parallax measurements has been demonstrated by the {\it Spitzer} satellite \citep{OB05001} and the two-wheel {\it Kepler} satellite \citep{2017PASP..129j4501Z}. However, there has not been a dedicated satellite microlensing telescope that promises both satellite microlens parallax measurements and a large ($\gtrsim 100$) homogeneous planetary sample from wide-area, high-cadence, and long-duration observations. 

High-resolution (e.g., adaptive optics) imaging can measure the apparent brightness of the lens by resolving the source and lens \citep[e.g.,][]{Alock2001,Kozlowski2007}, and can further constrain $\theta_{\rm E}$ and $\pi_{\rm E}$ by measurements of the lens-source relative proper motion. However, current instruments need to wait $\sim$ 5--30 years for sufficient lens-source separations. In addition, this method is challenging for faint objects such as brown dwarfs and not applicable to dark objects such as FFPs and black holes. 

In summary, measurements of the angular Einstein radius are common, but due to the challenges in parallax measurements and high-resolution imaging, there has not been a study on the mass function for either bound or free-floating microlensing planets.

\paragraph{Expected Yield of the ET + KMTNet Microlensing Survey}

\begin{figure}[htb] 
    \centering
    \includegraphics[width=0.75\columnwidth]{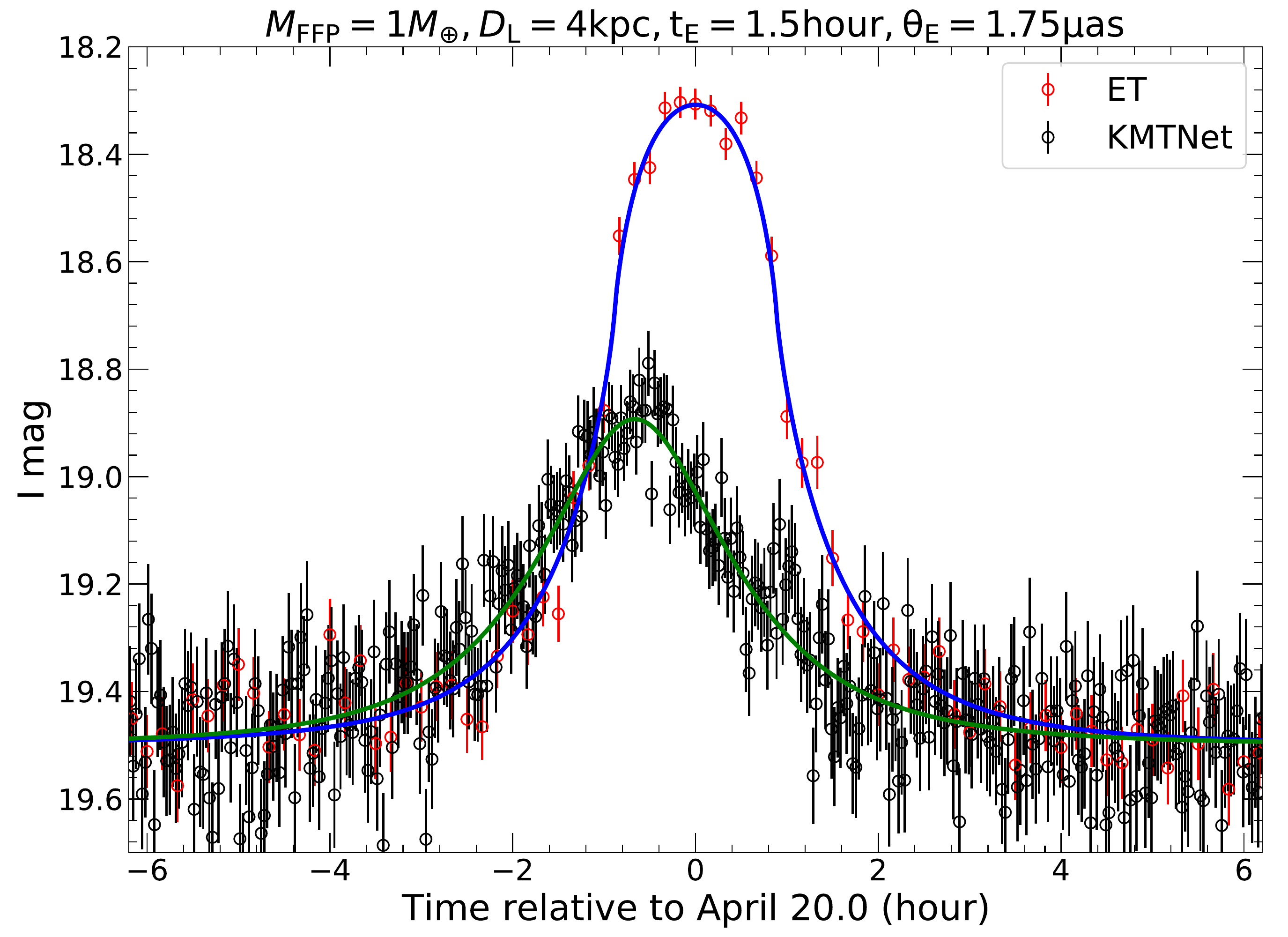}
    \caption{Simulated light curves of a free-floating Earth observed by the ET satellite (red circles) and the ground-based KMTNet telescopes (black circles) with Gaussian noise. The microlensing source is an $I = 19.5$ subgiant in the Galactic bulge with an angular source radius of \SI{1.2}{\mathrm{\mu}as}. The error bars for the ET data are calculated using the telescope parameters given in section \ref{sec:microlensing_yield}. The error bars for the KMTNet data are calculated assuming a seeing of \SI{2}{\arcsecond} and the total background sky brightness of $19.9$ mag per square arcsecond. The continuous KMTNet data have a 1.0-minute exposure and 1.0-minute read-out. For the actual KMTNet observations, some data will be taken in the $V$ band for the source color measurements. About \SI{40}{\percent} of the time during the ET microlensing survey, data will be unavailable due to bad weather, high background, or diurnal cycle \citep[see Figure 4 of][]{CMST}.}
    \label{fig:ET_lc}
\end{figure}

The microlensing telescope on the ET satellite has the same FOV (4 deg$^2$), pixel scale (\SI{0.4}{\arcsecond}), and filter ($I$ band) as KMTNet, and their photometric precisions are comparable. Thanks to the continuous and stable observations from space, the ET microlensing survey will have a higher detection efficiency for low-mass planets than current ground-based surveys. For mass measurements, the only method to determine the mass of FFPs is monitoring the Galactic bulge from two wide-field telescopes separated by $D\sim{\cal O} (0.01 \mathrm{au})$ \citep{CMST}, so the combination of the ET satellite at the Earth-Sun’s L2 point and the ground-based KMTNet survey will be the first to obtain mass measurements for an FFP. Figure \ref{fig:ET_lc} shows simulated light curves for a free-floating Earth observed by ET and KMTNet. In addition, the ET + KMTNet simultaneous observations for caustic crossings can yield the satellite microlens parallax measurements for bound planets.

For an integrated time of two years, the ET + KMTNet Microlensing Survey toward a 4 deg$^2$ bulge field centered on $(\ell, b) \sim (+1.0, -1.8)$ will (see section \ref{sec:microlensing_yield} for details): 
\begin{itemize}
 \item[1.] Measure the masses for about 150 FFPs, including about 50 free-floating Earths ($0.5 M_{\oplus} \leq M_p \leq 2 M_{\oplus}$).

 \item[2.] Measure the masses for about 130 bound planets, including about 10 cold Earths ($0.5 M_{\oplus} \leq M_p \leq 2 M_{\oplus}$).  

 \item[3.] Constrain the index of the power-law mass functions for bound and free-floating microlensing planets to uncertainties within \SI{6}{\percent} and \SI{10}{\percent}, respectively.

 \item[4.] Detect about 20 multiple-planet systems, half of which will have mass measurements. 

 \item[5.] Measure the mass and distance of about 120 host stars for bound planets, including about 60 M dwarfs and 20 brown dwarfs. 
 
\end{itemize}

We note that both the ET satellite and the {\it Nancy Grace Roman Space Telescope} \citep{Spergel2015,MatthewWFIRSTI} are scheduled to be launched to halo orbits at the Earth-Sun's L2 point around 2027, with lifetimes of about four years. About half the field of the ET + KMTNet Microlensing Survey overlaps with that of the {\it Roman} Galactic Exoplanet Survey. The joint observations of the two surveys will enable more satellite microlens parallax and thus mass measurements for more bound and free-floating planets with G-dwarf and subgiant sources. In addition, high-resolution observations from {\it Roman}, the {\it Chinese Space Station Telescope (CSST),} \citealt{CSST_Wei}), the Euclid satellite \citep{Laureijs2011euclid}, and the adaptive optics instruments on ground-based 30m class telescopes will yield more mass measurements for bound planets and identify the putative hosts of FFPs.

In contrast to ET, {\it Roman} observes about five magnitude deeper than current ground-based microlensing surveys so that most {\it Roman} FFP events do not have satellite microlens parallax measurements. With nearly double the survey area and duration of the {\it Roman} Galactic Exoplanet Survey, the ET + KMTNet Microlensing Survey can have 5--10 times (depending on the saturation limit of {\it Roman}) more mass measurements for FFPs.  The ET + KMTNet Microlensing Survey can also better characterize planetary signals thanks to its nearly three times higher cadence and longer sources diameter crossing time than {\it Roman}.

\subsubsection{Super-Earths and sub-Neptunes}\label{sec:super-Earth_sub-Neptune}
{\bf Authors:} \\
\newline
Beibei Liu\\
{\it Zhejiang University, 38 Zheda Road, Hangzhou 310027, China}\\

Our knowledge of planets beyond the Solar System has been greatly improved since the era of the {\it Kepler} space mission. One of the most intriguing findings from {\it Kepler} is that super-Earths are, so far, the most abundant type of planets in our galaxy. Super-Earths have several different definitions in the literature. Sometimes planets with radii slightly larger than Earth  ($1.25~R_{\oplus}{<}R_{\rm p}{<}2~R_{\oplus}$) are regarded as super-Earths, while mini-Neptunes refer to another group of planets with radii of  $2 R_{\oplus}{<}R_{\rm p}{<}4 R_{\oplus}$.  Given such a definition, super-Earths are terrestrial,  rocky planets while mini-Neptunes contain non-negligible hydrogen/helium gaseous envelopes. The other widely-used definition does not distinguish between the above two subgroups, referring to super-Earths as a general type of planets with radii between Earth and Neptune ($1.25~R_{\oplus}{<}R_{\rm p}{<}4~R_{\oplus}$).  Here we adopt the latter convention. Over the last decade, pioneering space missions such as {\it Kepler}, {\it TESS} and CHEOPS have detected thousands of super-Earths. Their physical properties, such as masses, radii, and orbital parameters, are relatively well documented.  We summarize the key findings as follows (also see recent reviews by  \citealt{Liu2020} and \citealt{ZhuDong:2021}). 

\begin{itemize}
  
  \item[--] The occurrence rate of super-Earths, $\eta_{\rm SE}$ is \SI{30}{\percent} around solar-type stars \citep{Zhu2018}.   This $\eta_{\rm SE}$ also varies with the properties of their stellar hosts. For instance, super-Earths are more frequent around low-mass M-dwarfs ($\eta_{\rm SE}\SI{\approx 70}{\percent}$) compared to solar-type or massive stars \citep{Bonfils2013,Dressing2015},  while $\eta_{\rm SE}$ is only modestly dependent on the stellar metallicity \citep{Wang2013,Buchhave2014}.    
  
  \item[--] {\it Kepler} found that ${\sim}$\SI{40}{\percent} of super-Earths are in compact, multi-planet systems \citep{Batalha2013}. The planet multiplicity decreases with increasing stellar mass \citep{Yang2020}. Notably, systems with multiple transiting planets tend to have coplanar and circular orbits while systems with a single planet tend to have much higher eccentric and inclined orbits \citep{Tremaine2012,Xie2016}. 
  
    \item[--]  The period ratios of neighboring planet pairs do not exhibit strong pile-ups at mean motion resonances (MMRs). Most notably, there is an asymmetry around major resonances,  with a deficit just interior to and an excess slightly exterior to the $2$:$1$ and $3$:$2$ MMRs \citep{Fabrycky2014}. 
  
  \item[--] The radii of super-Earths feature a bimodal distribution, with a factor of two drop at $R_{\rm p} {\sim}1.5{-}2 R_{\oplus}$ \citep{Fulton2017}. The above planetary `radius valley' implies a composition transition from rocky planets without hydrogen or helium envelopes to planets with envelopes of a few percent in mass. 
  
    \item[--]  Although direct mass measurements are rare, the masses of super-Earths are inferred to be linearly correlated with the masses of their hosts \citep{Wu2019, Liu2019}. 
 \end{itemize}
 
Although substantial progress has been achieved, several critical questions regarding our understanding of super-Earths remain to be answered.  
The currently observed super-Earths only span a narrow period range up to one hundred days, mainly limited by the operation strategy and lifetime of the previous space missions. As a result, the radial extent of super-Earth orbits and the associated occurrence rates at different radii are highly uncertain, hindering our understanding of 1) whether super-Earths have hot, warm, and cold populations and whether distant populations feature a similar dynamical property as those on close-in orbits,  2) whether all super-Earth populations form the same way or not, and  3) how the stellar environment (e.g., stellar mass and metallicity) affects their formation and evolution?  

As noted above, the period ratios of the observed super-Earth pairs exhibit an asymmetry around major resonances.  Various models have been proposed to explain how these pairs depart from their original resonances. Some of them rely on the stellar tides \citep{Lithwick2012} or stellar magnetic torques \citep{Liu2017} and are thus sensitive to the distance from the central stars. However, other scenarios, e.g., the dynamical instability model resulting from a stochastic process \citep{Izidoro2017},  may barely be affected by the orbital distance. Comparing the period ratios between the super-Earths on both short and wide orbits can naturally disentangle different resonant-breaking mechanisms.  

The previously mentioned radius gap is attributed to envelope mass loss due to stellar photoevaporation \citep{Owen2017} or core-powered heating \citep{Ginzburg2018}, or it may be a direct consequence of late formation through giant impacts \citep{Lopez2018}. 
The stripped core models predict a decreasing trend of the radius's gap with the orbital period in contrast to the late formation models. A sample covering a wider orbital range will help to distinguish whether the planets below the radius's valley once had an early-formed envelope and then stripped from the cores or were originally created rocky. Furthermore, the core-powered mass loss and photoevaporation mechanisms may differ in the dependence on stellar properties (e.g., mass, metallicity, and age, see \citet{Chen2022}). Therefore, a detailed exploration of the stellar parameters for the hosts of super-Earth systems is needed to determine the dominant process shaping the observed planet radius gap.   

A primary task of the ET mission includes the detection and characterization of a high number of super-Earth planets. It will provide the necessary clues to answer these key questions about super-Earths. In particular, ET will be able to detect about 20,000 super-Earths, increasing the sample size by at least a factor of $10$ (see Figure \ref{fig:ETnumber}). In addition, ET will cover the original {\it Kepler} field, revisiting some known {\it Kepler} systems to search for planets with periods up to eight years (four years {\it Kepler} plus four years ET).  Such an enriched planet population with extended orbital periods will greatly improve the constraints on the parameter spaces of the super-Earth population. It will allow follow-up statistical investigations of important correlations of planet formation processes with stellar and planetary system parameters. Hence,  this mission will provide a key breakthrough toward our understanding of the nature of super-Earth systems.   

\subsubsection{Cold Giant Planets}\label{sec:CG} 
{\bf Authors:} \\
\newline
Fabo Feng$^1$, Masataka Aizawa$^1$, Peng Jia$^2$, Shilong Liao$^3$,  Zhaoxiang Qi$^3$ \& Jian Ge$^3$ \\
{1.\it Tsung-Dao Lee Institute, Shanghai Jiao Tong University, 800 Dongchuan Road, Shanghai 200240, People’s Republic of China} \\
{2.\it College of Physics and Optoelectronics, Taiyuan University of Technology, Taiyuan, 030024, China} \\
{3.\it Shanghai Astronomical Observatory, Chinese Academy of Sciences, 80 Nandan Road, Shanghai 200030, P.R.China} \\

While the major scientific goal of ET is to find Earth 2.0s, the
search for Solar System analogs would put Earth 2.0s in the context
of planetary system architectures. In our Solar System, Jupiter and
Saturn are two cold gas giants while Uranus and Neptune are two cold
ice giants. Although more than 5,000 exoplanets\footnote{NASA Exoplanet Archive: \url{https://exoplanets.nasa.gov/}} have been found through various methods, none of them consists
of as many cold giants (CGs) as our Solar System. Here we define CGs as
exoplanets more than \SI{3}{AU} away from their host stars and having masses ranging from 0.1 to $10~\rm{M}_{\rm{Jup}}$. 

It is believed that CGs play an important role in shaping planetary systems and influencing the habitability of planets. For example, CGs may open a
gap in the protoplanetary disc during its formation to prevent outer
materials from migrating inward to form large planets \citep{Bae2019,Drazkowska2019,Li2020CG}. CGs on highly
eccentric and misaligned orbits may significantly increase the orbital
eccentricity and inclination of short-period planets, generate
long-term chaos in the motion of inner planets through Kozai-like
mechanics \citep{lai17} and trigger the migration of inner planets to
form hot Jupiters or hot Neptunes \citep{dong18}. CGs also play a
role in modifying the impact rate of habitable zone planets and shaping the
climate of these planets through the Milankovitch mechanism \citep{horner10}. 

CGs are not only important in shaping planetary structure and habitability
but also are interesting in terms of their formation, evolution,
atmosphere, and magnetic field \citep{Nelson2003,Uribe2011,Haffert2019,Chen2020a,Chen2020b,Li2021CG}. Infrared imaging and spectroscopic
observation of CGs are used to study the properties of their atmosphere such as the
pressure-temperature curve, molecular composition, and cloud coverage
\citep{madhusudhan19}. Radio signals from
hot Jupiters were also tentatively detected by \cite{turner21}. If confirmed,
this would open a new window for us to understand the magnetic field
and internal structure of giant planets. High-frequency radio signals and even
corona from CGs might be observable by FAST and the future SKA \citep{zarka19}. 

The current CGs are mainly detected by the
RV method, which can only estimate the lower limit of the
planetary mass due to the mass-inclination degeneracy. According to
\url{exoplanet.eu}, there are only 9 CGs with dynamical mass and full
orbital information measured. Unlike the RV method, the astrometry
method is able to constrain the mass and full orbit of a planet due to
the projection of planetary orbit onto the 2D sky plane formed by the right
ascension and declination directions. Thus the astrometry method or
its combination with the RV method is crucial for fully constraining CG
masses and orbits. 

As the most precise astrometry mission up to date, Gaia will finish its 10-year
mission in the near future. Since its first data release in 2016, Gaia
has already released three catalogs \citep{Prusti2016,Brown2018,Brown2021}. Due to the long time interval
between Gaia and its precursor, Hipparcos, the astrometry difference
between Gaia and Hipparcos has recently been used to confirm multiple CGs by
multiple groups
\citep{snellen18,kervella19,feng19b,brandt19,feng21}. The
release of 10-year Gaia epoch data would eventually enable the detection of
70,000$\pm$20,000 planets \citep{perryman14}  and about 100 giant planets orbiting the M dwarfs within 30 pc from the Sun \citep{10.1093/mnras/stt1899}. 

However, the 10-year Gaia baseline is still not long enough to cover
Saturn's 30-year orbit, Uranus' 84-year orbit, or Neptune's 165-year
orbit. Fortunately, ET will continuously observe a 500 square degree
area in the vicinity of the {\it Kepler} field for at least 4 years. If
launched around 2027, it would form a nearly 20-year baseline if
combined with Gaia data and a 40-year baseline if combined with
Hipparcos data and archived RV data. Since the baseline is
more important than the precision of astrometry in the detection of CGs,
ET is able to find thousands of CGs with comparable or even slightly worse precision than Gaia if combined with Gaia data. For example, Jupiter, Saturn,
Uranus, and Neptune would, respectively, change the position of the Sun
by about 0.9, 0.5, 0.2, and 0.3 mas over half orbital periods if the Solar System was put 10\,pc away from the observer. Such astrometric signals are detectable with the
current Gaia's precision of 0.04\,mas per field-of-view if the Gaia baseline is extended. 

Thanks to the high cadence and high stability transit observations of a sample
of about 1.2 million FGKM dwarfs, ET can likely achieve the same
astrometric precision as Gaia. Considering a 30\,min integration time and
a long-term drift of \SI{0.4}{\arcsecond}, ET would achieve $\sim$1\,mas
astrometric precision by reducing the jitter using 10,000 reference stars. The monthly precision would be 0.02\,mas if multiple
observations are reliably combined \citep{2000PASPAnderson}. With such precision, ET can detect analogs of Jupiter, Saturn, Uranus,
and Neptune if they are less than 100\,pc away from
the observer. Assuming an occurrence rate of one CG per star, we
expect ET to detect at least 1,000 CGs in combination with Gaia. 

To explore the feasibility of detecting CGs using ET astrometry, we develop a centroiding algorithm to measure the position of a star to 1/1000 pixel precision. 
The accuracy of contemporary centroiding algorithms is closely related to the size of the point spread function, the pixel scale of the camera and the signal-to-noise ratio of the target. The pixel scale of ET's cameras for transit detection is 4.38\,arcsec, and the point spread function has an encircled energy of 90\,\% (EE90) within a 3 to 5 pixel diameter. According to scientific requirements, ET requires around 100\,mas centroiding accuracy for celestial objects with moderate brightness, which is around 0.023 pixel. Since the image is critically sampled, the centroiding accuracy is achievable by contemporary centroiding algorithms, such as the modified moment algorithm or the point spread function fitting algorithm \citep{stone1989comparison}, albeit with some challenges.
 \begin{figure}[!htbp]
\centering
\begin{minipage}[t]{0.45\textwidth}
\centering
\includegraphics[width=\columnwidth]{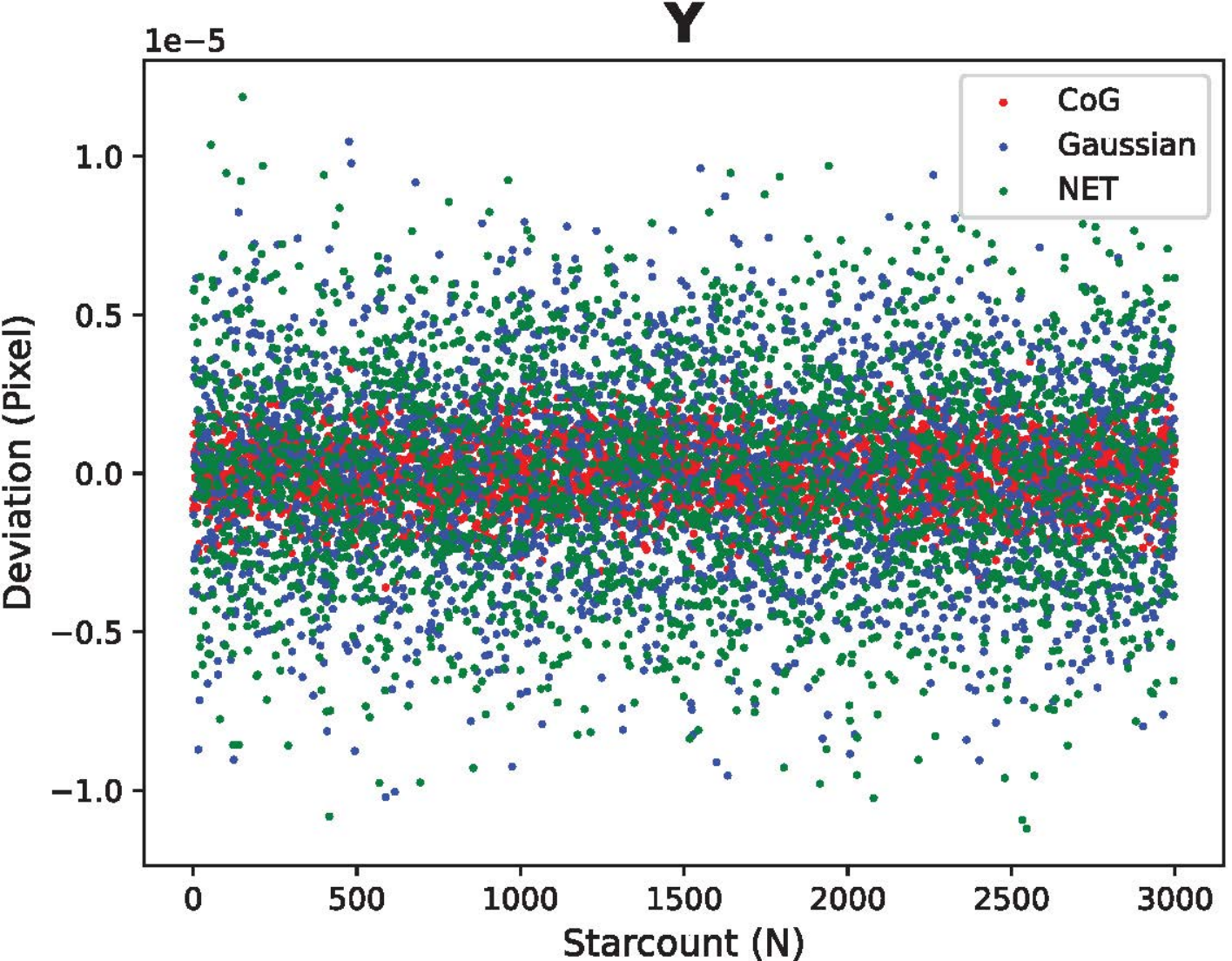}
\end{minipage}
\begin{minipage}[t]{0.45\textwidth}
\centering
\includegraphics[width= \columnwidth]{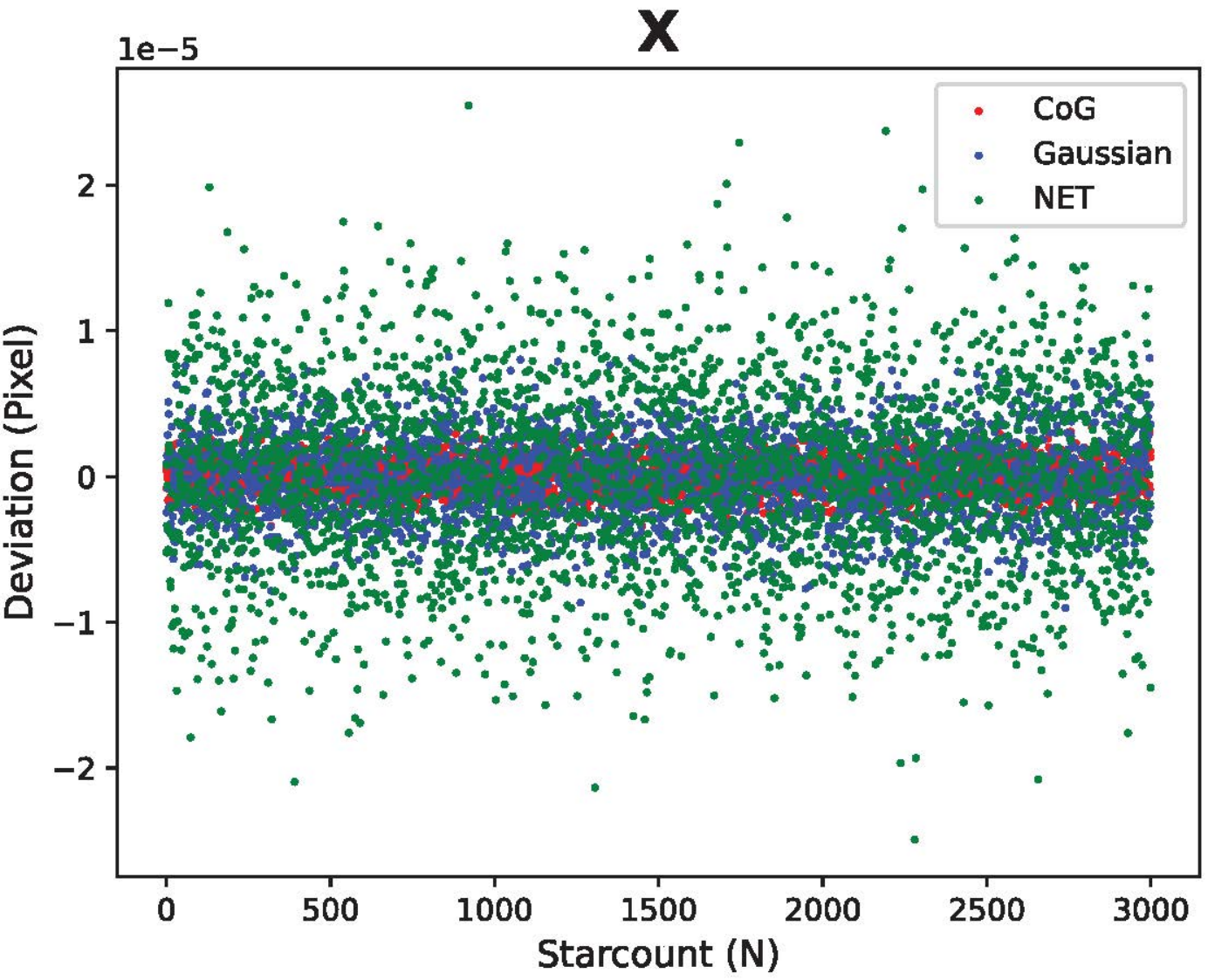}
\end{minipage}
\caption{The astrometry error in the x and y direction (in detector coordinates). We find that both contemporary methods (COG stands for ordinary Centre of Gravity astrometry method, which is a modified moment method; Gaussian represents the PSF fitting astrometry method with a prior Gaussian PSF) and the neural network (NET) could achieve an astrometry accuracy better than $10^{-4}$ pixel. We estimate that we could keep the astrometry error better than $10^{-3}$ pixel, with optimized contemporary centroiding algorithms and newly developed PSF estimation methods and centroiding algorithms.}
    \label{fig:astrometry}
\end{figure}
The modified moment algorithm is an unbiased estimator, which would output stable position measurements, albeit with lower accuracy. The effective point spread function (ePSF) method proposed by \cite{2000PASPAnderson} would firstly build an ePSF model from observation images and then use the ePSF for position measurements. The ePSF method could achieve an astrometry accuracy of around \SI{0.02}{\arcsecond} for isolated celestial objects with moderate brightness. The ePSF method is successfully used for images obtained by the Hubble Space Telescope with carefully reconstructed PSFs from real observation images \citep{2016Anderson}. However, for the under-sampled or critically sampled images obtained by ET, building an ePSF model would be the main challenge for the centroiding algorithm developed for ET.

Some statistical PSF modelling methods based on the principal component analysis \citep{2007PASPJee,2017MNRASJia} are proposed for PSF estimation. These algorithms could provide a reliable PSF model if we could provide enough isolated star images with moderate brightness. The principal component analysis would be one of our selections for PSF modelling. In recent years, the technology of digital twin of optical systems is being rapidly developed with highly reliable optical system simulation algorithms and deep learning algorithms for complex environments or complex components modelling. The digital twin of the wide-field telescopes in ET could estimate the states of the telescopes in orbit with several PSFs extracted from observation images. Then we use these estimated PSFs as prior PSFs for the centroiding algorithm. The digital twin is developed by our groups and would be used as the second PSF modelling method.

The iteration algorithm is a necessary step for PSF-based centroiding algorithms. The iteration algorithm would obtain position estimations from star images by continuously estimating x and y positions to find minimum $\chi^2$. Estimated results could achieve higher accuracy if we use a biased point estimation algorithm, such as deep neural networks \citep{2021Jia}. We are currently developing a Bayesian neural network for the centroiding algorithm. As shown in Figure \ref{fig:astrometry}, we find that both the contemporary method and the neural network could achieve an astrometry accuracy better than $10^{-4}$ pixel with preliminary simulated data. With optimized contemporary centroiding algorithms, newly developed PSF estimation methods, and centroiding algorithms, the accuracy of centroiding algorithm could achieve an accuracy better than $10^{-3}$ pixel. 

In addition to astrometric methods, the ET mission will also significantly increase the number of cold transiting planets with periods longer than 1-2 years. Several tens of such long-period planetary candidates have already been identified by the {\it Kepler} data \citep{Uehara2016,Foreman-Mackey2016,Herman2019,Kawahara2019}. The ET mission will increase the number of these systems by at least a few times due to its wide field of 500 $\mathrm{deg}^{2}$. Moreover, the ET mission will also confirm the candidates of the transiting cold planets observed by {\it Kepler}, most of which showed only single transit or double transits due to their long periods. In doing so, the ET mission uniquely studies the population of cold planets and could validate large gas giant planets orbiting at Jupiter-type distances, thus possibly detecting true Solar System analogs.

\subsubsection{Correlations Between Exoplanets and Their Host Stars}\label{sec:exoplanet_star_correlation} 
{\bf Authors:} \\
\newline
Wei Zhu$^1$ \& Steve B. Howell$^2$\\
{1. \it Tsinghua University, Beijing, China} \\
{2. \it NASA Ames Research Center, Moffett Field, CA 94035, US}\\

The occurrence rate and architecture of planetary systems may be correlated with the properties of their host stars. In particular, numerous studies have been carried out to investigate the dependence on stellar binarity, metallicity, and stellar mass, but many questions remain unresolved. A transit mission like ET will have the potential to resolve these outstanding issues and thus advance our understanding of the formation and evolution of planetary systems across a broad variety of stellar properties.

It has been well established that the presence of giant planets strongly correlates with the metallicity of the host star \citep[e.g.,][]{Santos:2001, Fischer:2005, Johnson:2010}. Specifically, giant planets are more likely to be found around stars with higher metallicities (measured by [Fe/H]). This so-called giant planet--metallicity correlation provides one of the crucial support to the core accretion model \citep[e.g.,][]{IdaLin:2005}. However, more questions remain to be answered by future surveys with more exoplanet detections. Below is an incomplete list of open questions relevant to the science of ET:
\begin{itemize}
    \item How steep is the giant planet--metallicity correlation? The crucial step in giant planet formation is building a core that reaches some critical mass \citep{Pollack:1996}. In the classical core accretion theory, the proto-core grows by colliding with (nearly) equal mass objects in the nearby region, and thus the formation of a core above the critical mass is usually hard. In particular, the giant planet occurrence rate is roughly proportional to the square of the host metallicity \citep{IdaLin:2005}. On the other hand, if the core grows by accreting pebbles that drift inward from the outer region, then the growth of a core up to the critical mass becomes easier, and thus the giant planet occurrence rate may scale approximately linearly with the host metallicity. Together with the spectroscopic survey of LAMOST, ET will provide the largest uniform sample to better study the giant planet--metallicity correlation.
    \item Do eccentric giant planets prefer metal-rich stars? \cite{Buchhave:2018} found that stars with giant planets in nearly circular orbits tend to be less metal-rich than stars with giant planets in eccentric ($e \gtrsim 0.2$) orbits. Their interpretation is that because giant planets obtain eccentric orbits from planet-planet scatterings \citep{Juric:2008, Chatterjee:2008}, eccentric giants may have been formed in systems with many giant planets, which necessitates high metallicities. The dynamical evolution of giant planets probably has important impact on the formation, evolution, and long-term stability of the terrestrial planets in the same system, as has been proposed for the Solar System \citep[e.g.,][]{2005Natur.435..466G, 2011Natur.475..206W}. Therefore, whether or not eccentric giant planets correlate with metal-rich stars will have important implications for the search for Earth 2.0 and planetary systems like our own \citep{Buchhave:2018}. ET's capability to detect both long-period planets (see section \ref{sec:CG}) and terrestrial planets in the habitable zone, assisted with the stellar information from ground-based spectroscopic observations, will be able to settle this issue.
    \item How do small planets depend on the host metallicity? Unlike the well established giant planet--metallicity correlation, it is still under debate whether small (radius below $\sim4\,R_\oplus$) planets correlate with the host metallicity (e.g., \citealp{Wang:2015, Buchhave2014, Petigura:2018}; see a brief review in \citealt{ZhuDong:2021}). The formation of small planets may have a very low metallicity threshold and thus no correlation can be seen at typical metallicities \citep[e.g.,][]{Zhu:2016, Lu:2020}, or the dynamical interactions with the usually co-existing giant planets at wider orbits have erased the imprint of host metallicity on small planets \citep[e.g.,][]{ZhuWu:2018, Zhu:2019}. Alternatively, the formation of small planets may correlate with specific elemental abundances other than the bulk metallicity measured by [Fe/H] \citep[e.g.,][]{Adibekyan:2012, Liu:2016, Teske:2019}. Future discoveries of small planets around a wider range of host metallicities, accompanied by detailed elemental abundances of large numbers of field stars, will be the key.
\end{itemize}

The presence of stellar companions may affect the formation, detection, and statistics of exoplanets (see also section \ref{sec:circumbinary}). Theoretically, a close-in stellar companion can perturb and even break the planet-forming disk when misaligned \citep{Artymowicz:1994,Deng2022non}. As a result, the pathway of planet formation is altered, and the formed systems keep experiencing perturbations from the companion \citep{Holman:1999,childs2021terrestrial,childs2022misalignment}. Through observations, it has been found that {\it Kepler} planet hosts are less likely to have stellar companions within $\sim100\,$AU than random field stars do \citep{Wang:2013, Kraus:2016, Ngo:2016}. \citet{Howell2021AJ....161..164H} and \citet{Lester2021AJ....162...75L} have shown that F,G,K host stars residing in binary star systems have mean orbital separations near 100 AU, larger than that found for field stars \citep[$\sim$50 AU][]{Rag2010}. These results support the idea that the presence of such close stellar companions tends to suppress the formation of {\it Kepler} planets \citep{Kraus:2016}. This suppression effect becomes the strongest for any companions within $\sim10\,$AU \citep{Moe:2021}. On the other hand, dozens of planets have been found to orbit around close binaries in transit surveys like {\it Kepler} and {\it TESS} \citep{Doyle2011, Kostov2020, Howell2021AJ....161..164H,Lester2021AJ....162...75L}. Based on these detections, statistical analysis suggests that planets may be as common around binary stars as single stars \citep{Armstrong:2014}.

In a transit mission like ET, the existence of a stellar companion can also affect the detailed characterization of the transit signal. A stellar companion dilutes the transit signal and leads to shallower transits (i.e., a smaller planet radius), if not properly accounted for. As demonstrated by studies based on the {\it Kepler}, {\it K2}, and {\it TESS} missions, this dilution effect is critical to the overall distribution of planet radii and it can substantially affect the number of Earth-sized planet detections \citep{Furlan:2017,FH2017AJ....154...66F, Bouma:2018, FH2020ApJ...898...47F}, thus posing a challenge to searching for a truly Earth-like planet. \citet{Lester2021AJ....162...75L} has shown that transiting planets of $\le$2 Earth-radii are missed in transit observations
due to ``third-light" contamination diluting their transit signal. ET will mitigate this dilution effect by incorporating results from Gaia, LAMOST, and follow-up observations with high spatial resolution imaging.


\subsubsection{Multi-Planet Exoplanetary System}\label{sec:multi} 
{\bf Authors:} \\
\newline
Hongping Deng$^{1}$ \& Ya-Ping Li$^{1}$ \& Ji-Wei Xie$^{2}$ \\
{\it 1. Shanghai Astronomical Observatory, Chinese Academy of Sciences, Shanghai, China}\\
{\it 2. School of Astronomy and Space Science, Nanjing University, Nanjing 210023, China}\\

\begin{figure}[!htbp]
\centering
\includegraphics[width=0.8\textwidth]{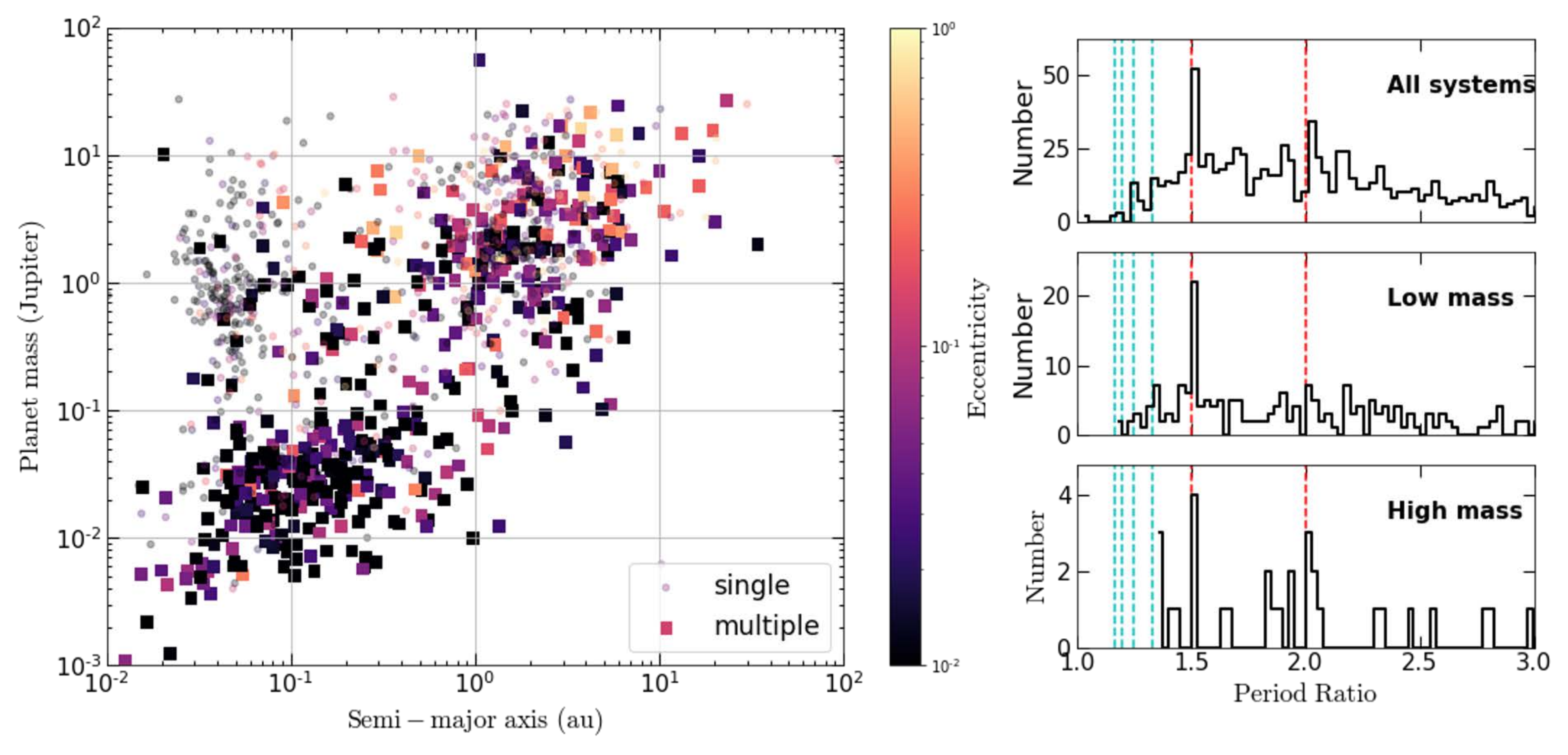}
\caption{Statistical properties of confirmed multi-planet systems with data from the NASA exoplanet archive at \url{https://exoplanetarchive.ipac.caltech.edu/} as of 05/03/2022. Left panel: distribution of exoplanet mass vs. planet semi-major axis for planets with measured eccentricities (color-coded). Here the large squares (small circles) indicate planets in multiple (single) planet systems. Right panels: histogram of orbital period ratios of 1255 adjacent pairs from all the confirmed multi-planet systems. The top panel is for all planet pairs while the middle and bottom panels for low-mass (both masses smaller than $0.1\ M_{\rm J}$) and high-mass (both masses larger than $0.8\ M_{\rm J}$) pairs, respectively. The vertical red dashed lines show the main mean motion resonance (MMR) of 2:1 and 3:2, while some other MMRs are marked with blue dashed lines.}
\label{fig:multipp}
\end{figure}
Roughly 40\% of the {\it Kepler} planets reside in multi-planet systems \citep{Batalha2013} and this also holds true for the entire exoplanet population. To date, there are 5,021 confirmed exoplanets, and 2,087 of them are in multiple planet systems, as shown in the left panel of Figure~\ref{fig:multipp} (planets without eccentricity confirmation are not shown). Hot Jupiters are lonely while super-Earths and sub-Neptunes are frequently found in multi-planet systems.  The system architecture (orbital configuration, mass ordering, etc.) potentially holds crucial information about its evolutionary history involving habitability, which can be deciphered by numerical modeling \citep[see, e.g.,][]{Liu2022}. For example, the Moon, which formed in a giant impact \citep{Canup2001}, is crucial for maintaining a stable spin for Earth and thus its habitability. Some relics of the impactor may still be buried in the deep Earth's mantle, controlling the mantle dynamics \citep{Deng2019, Yuan2021}. In contrast, the hazardous similar-mass Venus may be the witness of planetary encounters \citep{Fang2020, Deng2020} and long-lasting volcanism. ET will find more systems with terrestrial planets complementary to the current population dominated by massive planets, and only by studying the whole system containing an Earth 2.0 can we fairly assess the Earth 2.0's habitability. 

In addition to Earth 2.0s, ET is expected to observe a wide variety of exoplanet systems. Due to this abundance, we can better characterize the eccentricity \citep{Xie2016, Van2019} and mutual-inclination \citep{Zhu2018, He2020} distributions for planets in multi-planet systems. For example, in Figure~\ref{fig:multipp}, low mass multi-planet systems seem to have lower eccentricities. These measures of orbital excitation are crucial for understanding the assembly of multi-planet systems. When combined with the legacy {\it Kepler} data, ET can provide us with long-baseline data allowing the detection of cold planets with periods of several years. Therefore, ET will extend the planet period distribution (see, e.g., Figure~\ref{fig:multipp}) to check if the log-uniform distribution holds for cold super-Earths and sub-Neptunes and consolidate the period distribution for cool Saturns and Jupiters \citep{Petigura2018}. This extension in planet period distribution is crucial in assessing the role of the ice-line ($\sim$1 AU) in planet formation \citep{Drazkowska2017}. 

\begin{figure}[htb!]
   \centering
   \includegraphics[width=0.7\textwidth, angle=0]{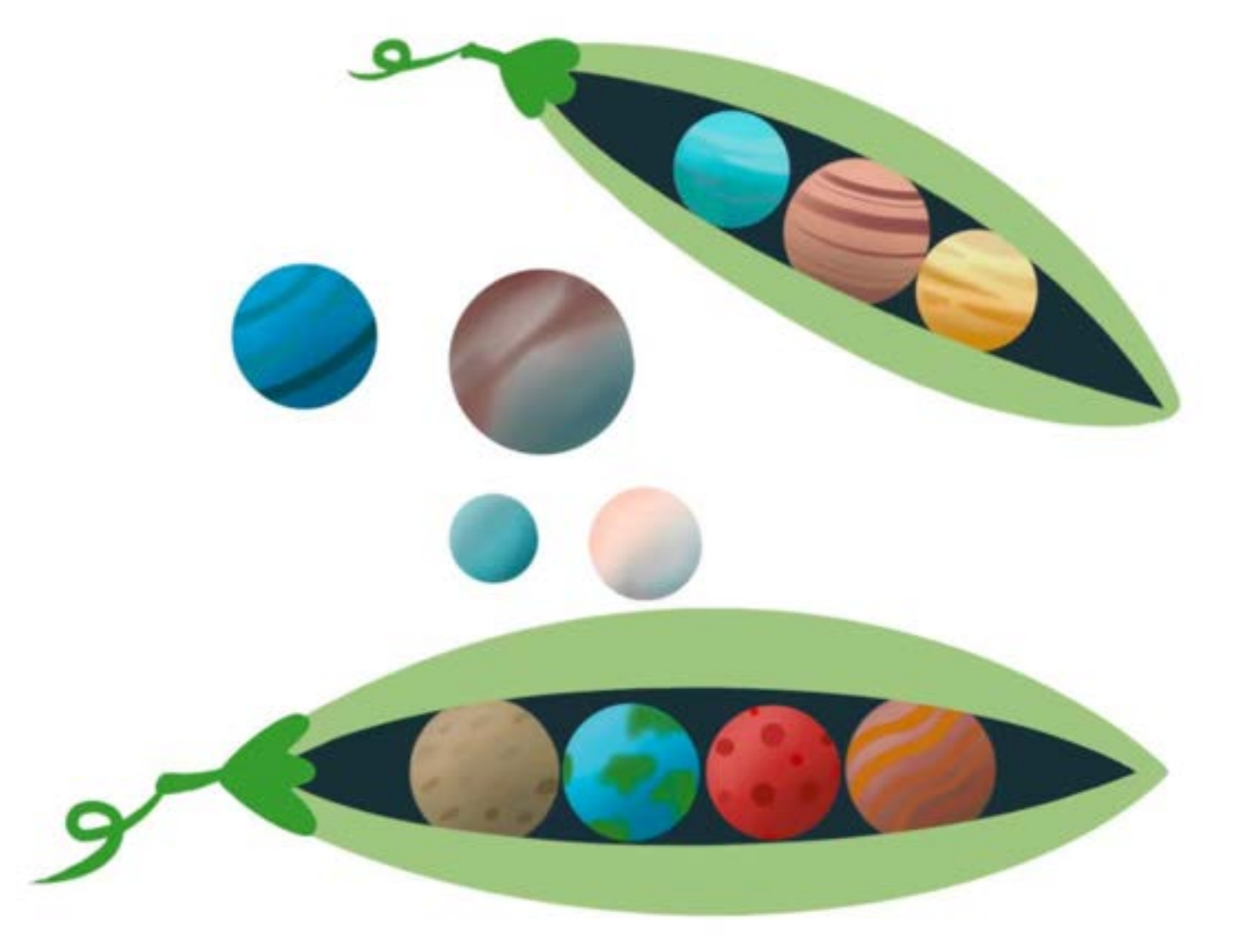}
   \caption{Illustration of the arguable peas-in-a-pod pattern or intra-system uniformity in {\it Kepler} multi-planet systems, courtesy of Dr. Tong Fang.}
   \label{fig:peas}
\end{figure}

With comprehensive data, we can verify a few characteristics of the {\it Kepler} multis \citep{Weiss2022}. First, they show a peas-in-a-pod pattern or intra-system uniformity in both mass \citep{Millholland2017} and size \citep{Weiss2018}. More specifically, planets in a multi-planet system are more similar to each other than to a planet drawn randomly from the distribution of observed planets\citep{Ciardi2013ApJ...763...41C}. This observation has crucial implications for the formation of multi-planet systems \citep{Emsenhuber2021,Mishra2021} (see the illustration in Figure \ref{fig:peas}). However, the claimed pattern is affected by detection noise and the underlying planet radius distribution and thus may be a detection bias \citep{zhu2020patterns, Murchikova2020}. ET, with higher precision (lower noise) and wider field of view than {\it Kepler}, will significantly expand the number of multi-planet systems and lead to a better understanding of intra-system properties.

Another notable feature in the {\it Kepler} multi-planet systems is the apparent paucity of resonance chains. If multiple planets were formed in protoplanetary discs, we might expect them to migrate and be captured into resonance \citep{Mustill2011}. However, the period ratios between adjacent planets show a log-normal distribution except for some peaks and troughs near the first-order resonance \citep{Fabrycky2014}. Specifically, in Figure \ref{fig:multipp}, there are about $\sim$\SI{10}{\percent} of planet pairs in mean motion resonance (MMRs), predominately the 2:1 and 3:2 MMRs with minor offset \citep[see, e.g.,][]{Lithwick2012,Zhang2014,Choksi2020}. Various models are proposed to explain the features of the planet period ratios. The resonance chain instability model, where planets initial in resonance rearranges due to a late-stage instability, can best match the observed period ratio distribution \citep{Izidoro2017}. However, the late-stage instability and how the planet-planet scattering happens in the violent rearrangement remain unknown. New data from ET may help us understand these processes better.

Some systems are likely to harbor giant planets such as on our Solar System. Cold Jupiters are always accompanied by super-Earths \citep{ZhuWu:2018, Bryan2019,Chen2020a}, and they also seem to affect the multiplicity of their host systems \citep{Hansen2017}. Theoretically, giant planets play a vital role in the formation of terrestrial-planet systems by dynamically exciting the planetesimals disk \citep{Raymond2006} or impeding the solid material supply via pebble fluxes \citep{Schlecker2021}. However, we expect a weak impact of wide orbit giant planets on inner terrestrial planet systems. The current Jupiter samples are limited to close-in planets so the observed correlations may be biased. In synergy with Gaia, ET will be able to detect giant planets up to 10 AU (see section \ref{sec:CG}). With such wide-orbit cold Jupiters, we can assess how the strength of Jupiters diminishes with distance and the exact role they played in terrestrial planet formation.


In multi-planet systems, transit-timing-variations (TTVs) can be a powerful tool to determine masses of planets, to derive their orbital properties and to detect non-transiting planets \citep{HM05, Ago05, Lit12, Nes12}. 
By monitoring about 200,000 stars for 4 years, Kepler has found over 260 planets with significant TTVs \citep{Hol2016}, based on which masses and orbital eccentricities have been inferred for over 100 planets\citep{Hadden2017}. 
Considering ET will observe about 7 times more targets with similar precision and time baseline as compared to Kepler, it is expected that ET will detect about 2000 TTV planets and 700 of them will have mass and eccentricity measurements. 
Such a large, homogeneous, and well-characterized planet sample will allow us to establish the mass-radius relationship for planets in the critical transition region where planetary radii are between Earth and Neptune, uncovering their internal composition and revealing the dynamical evidence of their formation and evolution. 
Furthermore, mass measurement is a critical ingredient to understanding planetary habitability. 
By combining ET and Kepler, the TTV time baseline can extend to over 10 years, which allows us to have a census of planet masses in the habitable zone and aids us in answering the question: what’s the frequency of (potentially) habitable planets?

In short, ET can reveal more multi-planet systems, including planets on wide orbits, which are crucial for testing the trends observed for short-period planets and constraining planet formation models.

\subsubsection{Exomoons, Exorings \&  Exocomets}\label{sec:exomoon} 
{\bf Authors:} \\
\newline
Bo Ma$^{1}$, Fabo Feng$^2$, Masataka Aizawa$^2$, Cong Yu$^1$ and Shangfei Liu$^1$ \\
{\it 1.  School of Physics and Astronomy, Sun Yat-Sen University, Zhuhai, 519082, PR China}\\
{\it 2. Tsung-Dao Lee Institute, Shanghai Jiao Tong University, 800 Dongchuan Road, Shanghai 200240, PR China}\\

\begin{figure}[htb!]
   \centering
   \includegraphics[width=0.85\textwidth, angle=0]{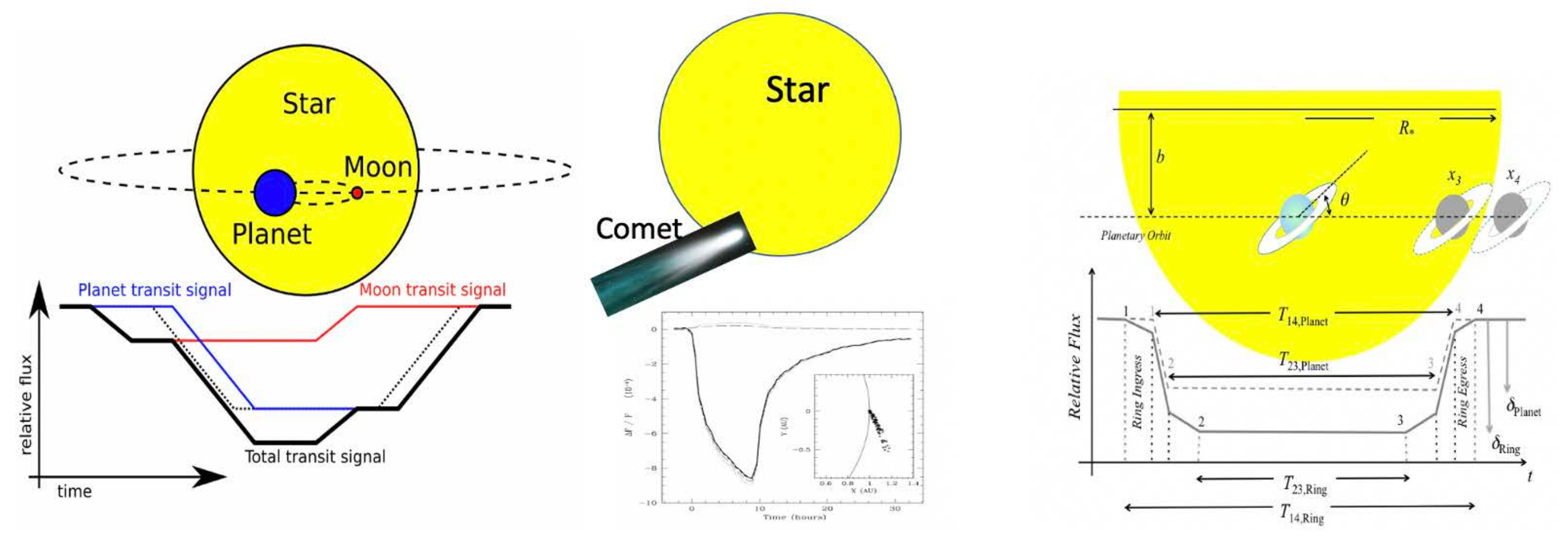}
   \caption{Illustration of a transiting exomoon, exocomet and exoring \citep{Lecavelier1999,Zuluaga2015}.
   \label{fig:exo-objects}}
\end{figure}

To form a comprehensive view of planet formation and evolution, it is important to know the properties of planets and also minor bodies in planetary systems, such as natural satellites, rings, and comets. Due to its highly precise photometry, the ET  mission has the opportunity to detect exotic exo-objects that astronomers have rarely been able to detect in the past, including exomoons, exorings, and exocomets (see Figure \ref{fig:exo-objects}). There are plenty of such objects in our Solar System, and thus it is reasonable to believe that they should be prevalent in exoplanet systems too. In particular, ET will discover an unprecedentedly large number of long-period transiting planets, around which exotic objects are relatively abundant \citep{barnes02,namouni10} and candidates of exomoons and exorings have been identified \citep{Aizawa2017,Teachey2018,Kipping2022}. Moreover, the baseline difference between ET and {\it Kepler} can also be exploited to study transit timing variation (TTV) and transit duration variation (TDV) for cold planets, giving unique data sets for searches of exomoons. Such a study can help reveal a more comprehensive picture of the overall structure of the exoplanet system. Furthermore, the comparison of the properties of minor objects in the Solar System and those of exoplanet systems would constrain the formation mechanisms of exoplanet systems and their habitability. 

There are subtle changes in the transiting light curves of these exotic objects compared with those of an isolated exoplanet. Such deviation from the standard transit light curve can be used to search for exotic objects. Considering that these signals are typically weak and last for a relatively short period of time, we need high temporal resolution (e.g., 1\,min) and high-precision photometric data to detect them: 
\begin{itemize}
\item {\bf Exomoon:} The TTV and TDV methods are frequently used to detect exoplanet systems with satellites \citep{Kipping2009,Rodenbeck2020,Fox2021}. We present a typical transit light curve of a moon-bearing exoplanet in the left panel of Figure~\ref{fig:exo-objects}. The transit light curve sensitively depends on the relative positions of the planet and satellite, leading to TTV and TDV.  By detecting these TTV and TDV signals simultaneously, exomoons can be identified and confirmed. With high cadence and high precision photometric data collected for known transit systems, ET is expected to measure significant TTV and TDV signals and detect several exomoon candidates.
\item {\bf Exocomet:} Due to the asymmetry of cometary tails, \citet{Lecavelier1999} propose that the transit method can be used to detect exocomets. We show a cartoon plot of exocomet and corresponding transit light curve in the center panel of Figure~\ref{fig:exo-objects}. It can be seen from the plot that, compared with typical transiting exoplanets, the transiting light curve of an exocomet has a sharp drop and then a slow recovery. The high cadence and high precision ET data for bright stars collected by ET will enable the detection of rare exocomets from a large sample of nearby systems. 

\item {\bf Exoring:} The existence of planetary rings makes exoplanets no longer spherically symmetric, and thus the cross-section is no longer a circle. Therefore, the transiting light curve is different from the typical light curve of transiting planet without rings during the ingress and egress phases \citep{Barnes2004,Ohta2009,Zuluaga2015,Aizawa2017}. By searching for these subtle differences, it is possible to detect the ring structures around exoplanets, as shown in the right panel of Figure~\ref{fig:exo-objects}. Because the durations of ingress and egress can be as short as a few tens of minutes, such events are likely identified in the short-cadence data collected by ET. 
\end{itemize}

Considering that only one robust exomoon candidate has been identified in a sample of 70 cool giant planets found by {\it Kepler} \citep{Kipping2022}, ET will probably detect at least 10 exomoon candidates in its sample of cool giants,which is 10 times the size of the {\it Kepler} sample due to ET's better strategy in finding wide-orbit planets (see section \ref{sec:yield}). ET will also confirm the exomoon candidates found in {\it Kepler} data by detecting more transits of them with high photometric precision. Since the number of exoring and exocomet candidates detected by {\it Kepler} is comparable with the number of exomoon candidates, we also expect an increase of the sample of exoring and exocomet candidates by ten times. Therefore, ET will probably detect $\gtrsim$10 exomoon candidates, $\gtrsim$10 exocomet candidates, and $\gtrsim$10 exoring candidates. In particular, the ET mission will continuously monitor the systems where candidate exomoons, exorings and exocomets were previously identified \citep{Aizawa2017,Teachey2018,Alam2022,Kipping2022} to confirm and further study these candidates.


\subsubsection{Circumbinary Planets}\label{sec:circumbinary} 
{\bf Authors:} \\
\newline
Mu-Tian Wang$^{1}$, Ya-Ping Li$^2$, Hui-Gen Liu$^1$ \& Ji-Wei Xie$^1$ \\
{1. \it School of Astronomy and Space Science, Nanjing University, Nanjing 210023, China}\\
{2. \it Shanghai Astronomical Observatory, Chinese Academy of Sciences, Shanghai 200030, China}\\

\begin{figure}[htbp]
\centering
\includegraphics[width=0.85\textwidth]{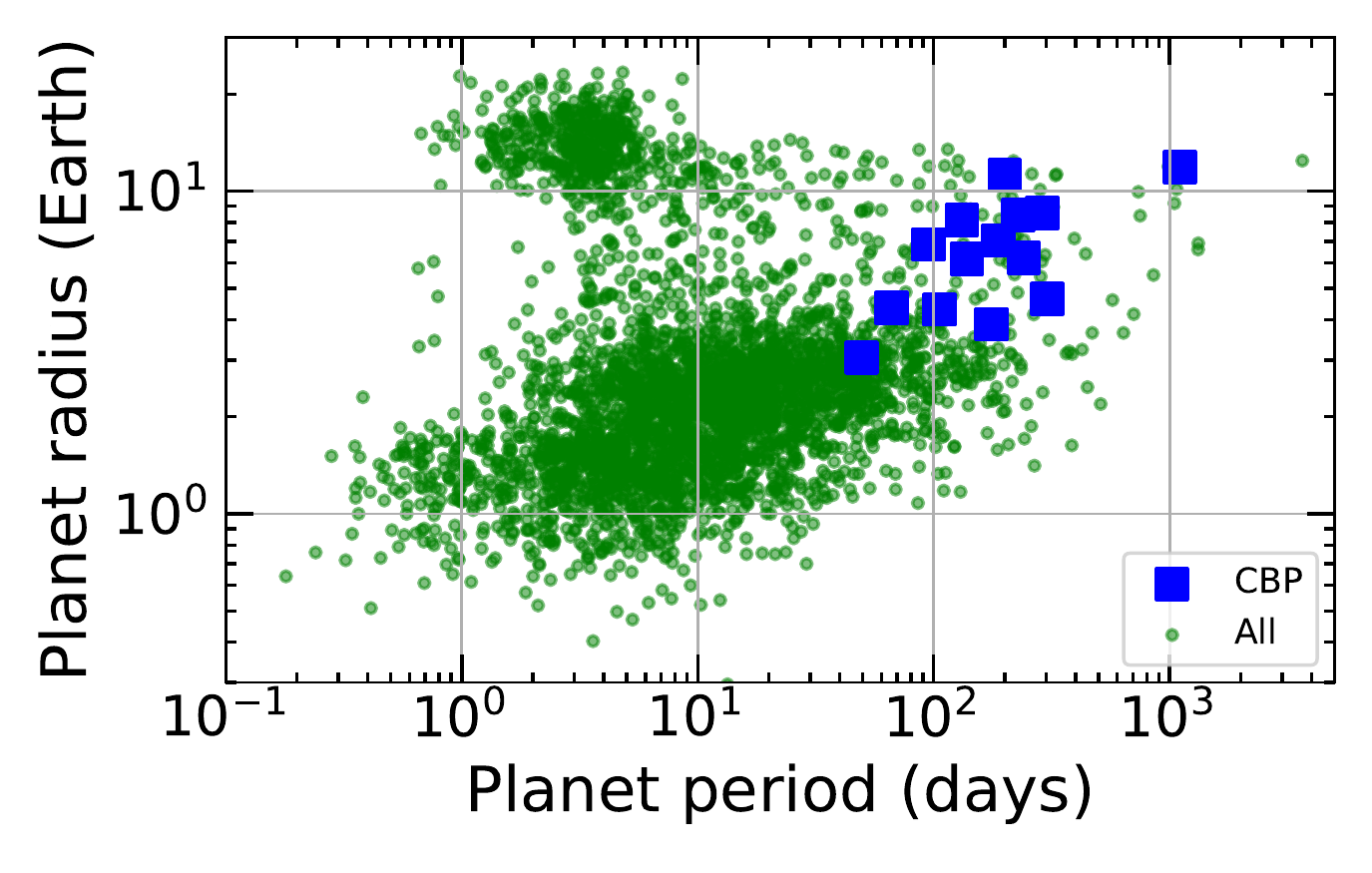}
\caption{Scatter plot of planet periods and radii for all confirmed exoplanets (green dots) and CBPs (blue squares). It is clearly seen that all 14 CBPs are relatively large in size and orbiting the binary with periods longer than 50 days.}
\label{fig:porb_psize}
\end{figure}

Circumbinary planets (hereafter CBPs) refer to planets that are orbiting both stellar objects in binary systems. To date, 14 transiting circumbinary planets have been found in 12 eclipse binary systems by {\it Kepler} and {\it TESS} ({\it Kepler}-16b: \citet{Doyle2011}; {\it Kepler}-34 b and 35 b: \citet{Welsh2012}; {\it Kepler}-38 b: \citet{Orosz2012b}; {\it Kepler}-47 b, c, and d: \citet{Orosz2012a,Orosz2019,Kostov2013}; {\it Kepler}-64 b: \citet{Schwamb2013,Kostov2013}; {\it Kepler}-413 b: \citet{Kostov2014}; {\it Kepler}-453b: \citet{Welsh2015}; {\it Kepler}-1647b: \citet{Kostov2016}; {\it Kepler}-1661 b: \citet{Socia2020}; TOI-1338: \citet{Kostov2020}; TIC 172900988: \citet{Kostov2021}). The mutual inclinations between the CBPs and their host binary are also very small ($\Delta I < 4^\circ$). In addition, all the CBPs are large and likely gas-rich, with radii ranging from 0.3 to 1.1 Jupiter radii (or 3 to 12 Earth radii) and orbital periods longer than 50 days. They stand out among {\it Kepler} planets, which are mainly small on short-period orbits (see Figure~\ref{fig:porb_psize}).  

There is a dearth of small or terrestrial CBPs (Figure  \ref{fig:porb_psize}), which are abundant around single stars \citep[e.g.,][]{Kunimoto2020AJ}. The non-detection of CBPs smaller than 0.3 Jupiter radii is likely an observational bias. Due to the binary motion, CBP transit signals will exhibit considerable TTVs for up to a few days \citep{Welsh2012,Armstrong13,Liu14}. The transit durations could also vary by hours, depending on the binary phase when transits occur \citep{Kostov2014}. Traditional transit searching algorithms, which target periodic signals with, e.g., boxed least squares \citep{Kovacs2002}, would fail to identify CBPs' aperiodic transit signatures. Therefore, it is hard to stack signals of small CBPs to achieve detection significance. Previous CBP detections were all obtained by human inspection, which is not sensitive to small planets. Nevertheless, efforts have been made to develop automatic algorithms specific to CBP transit signals, some of which have approached the detection sensitivity of Earth-like planets in {\it Kepler} data \citep{Windemuth19,Martin_Fabrycky21}. On the other hand, the absence of small CBPs may have theoretical roots: due to the slow migration rate of the low-mass CBPs, they are more likely to be captured in mean motion resonances with central binary and ejected through resonance overlapping. Companion giant planets may also imperil low-mass CBPs by pushing them too close to the binary stars during migration \citep{Sutherland19,Fitzmarice22,Martin22}. With the next-generation photometry precision and the long time baseline ($>$8 years) that ET will provide, small CBPs may be revealed, leading to a more complete radius distribution of CBPs, thereby offering insights into the planet formation pathways in circumbinary environments.

\begin{figure}[!htbp]
    \centering
    \includegraphics[width=0.8\textwidth]{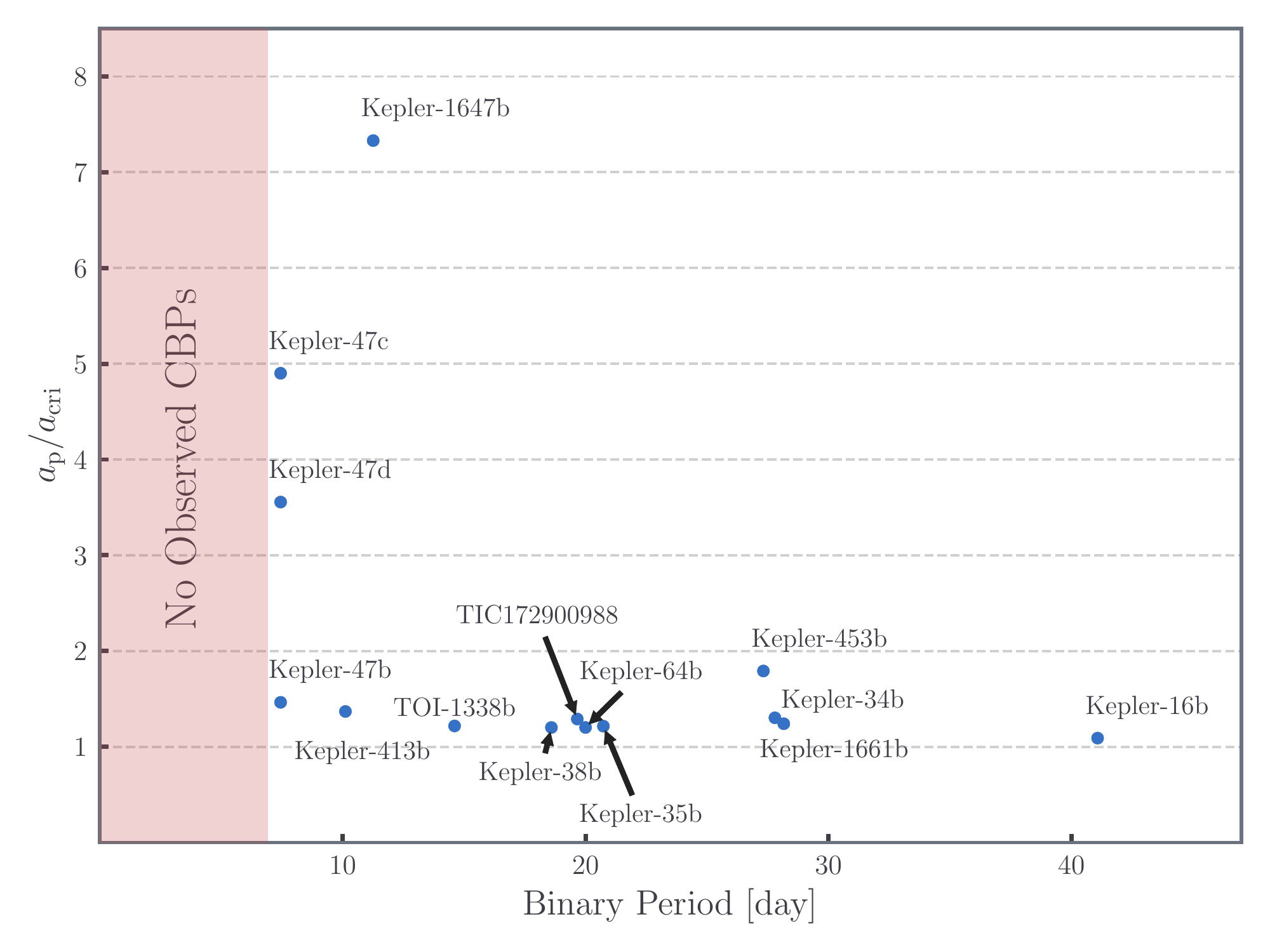}
    \caption{The semi-major axis of circumbinary planets, $a_{\rm p}$ (normalized to the critical separation for stability, $a_{\rm cri}$), against the orbital period of the host binaries. Most of the CBPs are very close to the stability limit of the system ($a_{\rm cri}/a_{\rm p} \approx 1$). No CBPs have been observed around binaries with periods of less than 7 days (red shaded area). This figure is adapted from Figure 1 of \cite{Fleming18}, adding in {\it Kepler}-47 d, {\it Kepler}-1661 b, TOI-1338 b, and TIC 172900988.}
    \label{fig:cbp_period_stability}
\end{figure}

Most of the known transiting CBPs are located near the stability limit, around three times the binary semi-major axis \citep[see Figure \ref{fig:cbp_period_stability},][]{Holman:1999}. If this pile-up feature is real,  it may have crucial implications for circumbinary planets' formation and evolution histories. Due to the turbulent environment in the inner region of the circumbinary disk, most of the CBPs may not have formed in their present location. It is likely that these CBPs formed farther out, migrated, then eventually came to a rest at the edge of the cavity truncated by the binary, which is around the location of the stability limit \citep{KH14,PN13}. However, this orbital pile-up theory is not well established \citep{Quarles18}, although selection bias could not fully account for this pile-up phenomenon \citep{Martin14,Li16}. In general, more CBP observations from ET are needed to build up sufficient evidence of the CBPs' over-densities at their stability limit.

While the CBPs' over-abundance at their stability limit remains inconclusive, there is an apparent absence of CBPs orbiting the shorter-period eclipsing binaries (see Figure \ref{fig:cbp_period_stability}). The shortest known orbital period of all CBP host binaries is around 7 days ({\it Kepler}-47) which is longer than most eclipsing binaries discovered by {\it Kepler} \citep[mostly around 1-2 days,][]{Kirk2016}. A systematic search for CBPs in the entire {\it Kepler} eclipsing binary catalog did not yield any additional discoveries, suggesting coplanar CBPs are indeed less common around shorter-period binaries \citep{Armstrong:2014,Martin14}. These compact binaries are thought to have formed in wider orbits. They evolved to their current location by tidally shrinking their high-eccentricity orbits excited by nearby tertiary stars via the Kozai-Lidov mechanism. CBPs would become unstable during the orbital shrinkage of the binary. Even if they survived, they would likely end up in highly misaligned orbits which are not amenable for transit detection \citep{Munoz15,Martin15}. The evolution of short-period binary orbits during their pre-main-sequence stage may also shape the structure of the CBP system: the orbital expansion of binary orbits caused by stellar-tidal coupling would clear the closest CBP around stability limits \citep{Fleming18}. Therefore, short-period binaries may host a distinct population of CBPs on very long period or misaligned orbits. ET's broad survey area may deliver more insights into the host binary and CBP properties, enriching our knowledge of the star-planet interplay that shaped the CBP populations.

Misaligned CBPs may also exist around binaries with a wide range of periods. The transit patterns of misaligned CBPs may reveal their intrinsic inclination distribution \citep{ChenKipping21}. Misaligned CBPs may form within steadily precessing warped circumstellar disks \citep{Deng2022non}. Even CBPs in the polar configuration can form within disks since such circumbinary disks are theoretically envisaged \citep{Martin2017} and observationally confirmed from gas and debris disks \citep{Kennedy2012,Kennedy2019}. However, all CBPs discovered to date are in orbits almost aligned with the binary orbit. 
In fact, it is difficult to detect misaligned CBPs due to their irregular non-periodic transit patterns  \citep{Martin14,ChenKipping21} and transient transit windows caused by orbital precession \citep[e.g.,][]{Martin14}. Therefore the detection of misaligned CBPs are difficult for the {\it Kepler} mission since the mission only has a lifetime of 4 years. Nevertheless, some omitted {\it Kepler} misaligned CBPs could be found since ET will revisit {\it Kepler} field and increase the observation baseline. The detection of these CBPs may also be possible with eclipse timing variations \citep{ZhangFabrycky2019,Martin_Fabrycky21}. Overall, ET will facilitate the detection of inclined and even polar CBPs and help constrain the inclination distribution and overall occurrence rate of CBPs.

With ET's systematic monitoring of host stars which increases exoplanets of various populations by a factor of $\gtrsim10$ with the longer-duration baseline and broad survey area, we could expect to discover more than 100 CBPs by conservatively assuming the same CBP occurrence rates \citep{Armstrong2014}. With such a diverse population of CBPs, the key questions to be addressed are summarised as follows: 

1) What are the mass, size distribution, and orbital properties of CBPs in general? How are they correlated with the stellar host properties? 

2) Are there some other populations of CBPs that are not detected in the current {\it Kepler} and {\it TESS} surveys, e.g., polar/misaligned CBPs, CBPs with smaller size, and CBPs around short-period binaries? Is the clustering of CBPs around the stability limit intrinsic or due to selection bias? 
All of these will shed light on the formation and evolution of CBPs in circumbinary disks.

\subsubsection{Transiting Planets around White Dwarfs}\label{sec:planet_wd} 
{\bf Authors:} \\
\newline
Ming Yang$^1$, Hui Zhang$^2$ \& Jian Ge$^2$\\
{1.\it School of Astronomy and Space Science, Nanjing University, Nanjing 210023, China}\\
{2.\it Shanghai Astronomical Observatory, Chinese Academy of Sciences, Shanghai, China}\\

Studying the likely fate of our Solar System is one of the fundamental questions to be addressed by astrophysics, which is also linked to the fate of humankind. The Sun was born 4.6 billion years ago and is now in the middle of its main sequence stage. The Sun will eventually  expand to become a red giant in about 5 billion years, likely engulfing the planets within the Earth's orbit. The Sun would then cast off its outer layers and leave behind a compact stellar remnant known as a white dwarf. Key questions related to planets in the white dwarf system remain to be addressed: can the remaining planets in the Solar System survive the late stage of solar evolution? 
How are the remaining planets redistributed in the white dwarf system if they do?  In addition, due to the extremely high surface gravity of a white dwarf (log $g \approx 8$), heavy elements on a white dwarf quickly sink out of the atmosphere within a short time scale \citep[a few days to a few million years; for comparison, it takes several billion years for a white dwarf to cool down,][]{1986ApJS...61..197P}. Therefore, heavy elements detected in white dwarf spectra cannot be primordial and must have been accreted relatively recently. By studying heavy element abundances of white dwarfs, the chemical composition of planetary system objects which were accreted to the white dwarf surface can be effectively studied \citep{2010ApJ...719..803D}. It is important to study planets and planetary objects around white dwarfs as it helps to address these key questions, provide insights in  understanding the long-term evolution of star-planet systems, and address the future destiny of Earth and humanity.


To date, more than 30,000 white dwarfs have been found by SDSS. Analysis of young white dwarfs shows that the occurrence rate of metal pollution is about 25-50\% \citep{2014A&A...566A..34K,2003ApJ...596..477Z,2010ApJ...722..725Z}, indicating the presence of material replenishment among these metal-polluted white dwarfs. In addition, $\sim$2-4\% of single white dwarfs are found associated with dusty disks \citep{2014ApJ...786...77B,2011ApJS..197...38D,2009ApJ...694..805F}, which show infrared excesses. The existence of infrared excesses in these white dwarfs also indicates that the dust is constantly replenished as it is consumed. Combined with the theory of stellar evolution, it is reasonable to conclude that planets or minor bodies exist around some white dwarfs. Materials from the minor bodies continuously feed these white dwarfs, creating dust disks at the early age of these white dwarfs. The accreted materials lead to metal pollution in the atmosphere of these white dwarfs \citep{2019NatAs...3...69C}.

\begin{table}[htbp]
    \renewcommand\arraystretch{1.2}
    \centering
        \captionsetup{justification=centering}
    \captionsetup{justification=centering}
    \caption{Major planets around white dwarfs.} 
    \begin{tabular}{c c c c}
    \hline
    \hline
    Name & Planetary Mass & Planet-Host Separation & Detection Method \\
    \hline
    PSR B1620-26 (AB) b$^1$ & $\sim 2.5 M_J$ & 23 AU & pulsar timing \\
    WD 0806-661 b$^2$ & $\sim 7M_J$ & 2500 AU & direct imaging \\
    WD J0914+1914 b$^3$ & unknown & 0.07 AU & spectral observations \\
    WD 1856+534 b$^4$ & $<11.7 M_J$ & 0.02 AU & transit  \\
    MOA-2010-BLG-477Lb$^5$ & $\sim 1.4 M_J$ & $\sim 2.8$ AU & microlensing \\
    \hline
    \end{tabular}\\
    { $^1$ \citet{1993ApJ...412L..33T,2003Sci...301..193S}; $^2$ \citet{2011ApJ...730L...9L}; $^3$ \citet{2019Natur.576...61G}; $^4$ \citet{2020Natur.585..363V}; $^5$ \citet{MB10477_AO} }
    \label{tab:planet_wd_tab1}
\end{table}

To date, five major planets around white dwarfs have been detected via different methods, as listed in Table \ref{tab:planet_wd_tab1}. Among them, WD 1856+534 b was detected by the transit method \citep{2020Natur.585..363V}. The transit signals caused by a Jupiter-sized planet have a period of 1.4 days. Researchers believe that this planet survived the red giant stage because it was far away from its host star. At the late stage, the planet was scattered towards the host star under dynamic perturbations. This discovery therefore suggests that massive planets can survive the progenitor's red giant evolutionary phase and can be scattered into close orbits around white dwarfs without being tidally disrupted.

Several transiting minor planets have also been detected around confirmed/candidate white dwarfs, as listed in Table \ref{tab:planet_wd_tab2}. These asteroids and tidally disrupted planetesimals have different orbital periods and transit depths. A good example is WD 1145+017, with a distance of 174 pc and a magnitude of 17. Its light curve shows transit signals from both asteroids and tidally-disrupted planetesimals \citep{2015Natur.526..546V,2016MNRAS.458.3904R}. The planetary embryo is being photo-evaporated by its host star and forms a hot dust disk \citep{2016ApJ...816L..22X}.
\begin{table}[htbp]
    \centering
        \captionsetup{justification=centering}
    \caption{Transiting planetary embryos and remnants around white dwarfs.}
    \begin{tabular}{c c c c c c c c c}
    \hline    \hline
    Name & $\alpha$ (deg) & $\delta$ (deg) & $d$ (pc) & $G$ (mag) & $T_{\mathrm{eff}}$ (K) & Tdur & Tdep (\%) \\
    \hline
    WD\,1054$-$226$^1$ & 164.16058 & -22.88092 & 36.17 $\pm$ 0.07 & 16.0 & 7910 $\pm$ 120	&	4$-$15 mins	&	$<$10 \\
WD\,1145+017$^2$ & 177.13994 & 1.48316 & 141.2$^{+2.5}_{-2.5}$ & 17.2 & $15{,}080$ $\pm$ 640	&	$\sim$5 mins	&	$\sim$40 \\
ZTF\,J0139+5245$^3$ & 24.77633 & 52.76027 & 172.9$^{+7.7}_{-7.2}$ & 18.5 & 9420 $\pm$ 580	&	15$-$25 days	&	20$-$45 \\
SDSS\,J0107+2107$^4$ & 16.95550 & 21.12910 & 90.2$^{+3.9}_{-3.5}$ & 19.2 & 7590 $\pm$ 800	&	10 mins	&	10$-$30 \\
ZTF\,J0328$-$1219$^4$ & 52.14013 & $-$12.32930 & 43.3$^{+0.2}_{-0.2}$ & 16.6 & 8550 $\pm$ 160	&	1$-$2 hrs	&	$<$10 \\
ZTF\,J0347$-$1802$^4$ & 56.76399 & $-$18.04825 & 76.4$^{+0.8}_{-0.7}$ & 17.4 & $13{,}370$ $\pm$ 510	&	$\sim$70 days	&	$\sim$25 \\
ZTF\,J0923+4236$^4$ & 140.79749 & 42.60934 & 147.2$^{+2.8}_{-2.7}$ & 17.5 & $13{,}110$ $\pm$ 420 	&	20$-$80 days	&	10$-$35 \\
SBSS\,1232+563$^4$ & 188.63559 & 56.11195 & 173.0$^{+3.7}_{-3.5}$ & 18.1 & $11{,}670$ $\pm$ 600	&	$\sim$1 hr	&	$<$5 \\
    \hline
    \end{tabular} \\
{$^1$\citet{2022MNRAS.511.1647F}; $^2$\citet{2015Natur.526..546V}; $^3$\citet{2020ApJ...897..171V}; $^4$\citet{2021ApJ...912..125G} }
\label{tab:planet_wd_tab2}
\end{table}


These planets are involved in processes such as debris disk dynamics, atmospheric pollution, and tidal interactions with white dwarfs. They provide clues to the different stages of planets around white dwarfs and allow us to better understand the evolution of planetary systems and the future of our Solar System. In addition, dusty disks and metal pollution can reveal the chemical composition, which may retain information about the material of the planet, as well as the original protoplanetary disk. 

Based on Gaia EDR3 data, there are about 360,000 white dwarf candidates of less than 21 mag \citep{2021MNRAS.508.3877G}. About $N_{\rm WD} \approx 7000-8000$ of them are within the ET's FOV. ET has an equivalent diameter of \SI{73.5}{\cm} and can observe all the Gaia's white dwarf candidates in its FOV. The major planets detected around white dwarfs can be as close as \SI{0.02}{AU} to the Roche limit and as far as \SI{2500}{AU}. Therefore, it is possible that most of the giant planets in white dwarf systems may have been dynamically scattered, leading to a nearly uniform distribution along the semi-major axis in logarithmic coordinates. The uniform distribution is calculated to be f(a)= 0.087/a, assuming the scattered planets are between 0.01 and \SI{1000}{AU}. The occurrence rates of Jupiter-size, Neptune-size, and smaller size (super Earth and Earth) planets are $f_{\rm J} \approx $ 10\%, $f_{\rm N} \approx $ 77\%, and $f_{\rm E} \approx $ 200\%, respectively \citep{2018AJ....155..205H}. Considering the occurrence rates and the probability of transit (the ratio of the planet's radius to its orbital semi-major axis), the expected number of transiting major planets detected by ET is 
\begin{eqnarray}
N &=& N_{\rm WD}  \left[ f_{\rm J} * \int f (a)  {R_{\rm J} \over a} {\rm d}a  + f_{\rm N} * \int f (a)  {R_{\rm N} \over a} {\rm d}a + f_{\rm E} * \int f (a)  {R_{\rm E} \over a} {\rm d}a \right],
\end{eqnarray}
where $ R_{\rm J} $, $ R_{\rm N} $ and $R_{\rm E}$ are Jupiter, Neptune and Earth radius, respectively. The integrating range is set to be \SIrange[range-units = single]{0.01}{1}{AU}. ET is expected to detect $N \approx 10$ major planets around white dwarfs in its FOV. Among these detections, 30\% are Earth-size or super-Earth-size planets. Considering that more minor planets or tidally disrupted planetesimals have been detected around white dwarfs than  major planets, it is expected that ET will likely detect more than 10 asteroids and tidally disrupted planetesimals. 

We also applied the above estimation process to the {\it TESS} mission. {\it TESS} can observe white dwarfs of less than 19 mag, resulting in $\sim$2200 TIC targets flagged to be in Gaia's photometric white dwarf catalog. We set the integrating range from \SIrange[range-units = single]{0.01}{0.1}{AU} based on a {\it TESS} one-month section. The calculation shows that two or three transiting planets around white dwarfs are expected to be detected by {\it TESS}, which is consistent with the one real detection by {\it TESS} to date  \citep{2020Natur.585..363V}.

In conclusion, ET is expected to discover several transiting major planets, asteroids, and tidally disrupted planetesimals around white dwarfs. As planets around \SI{1}{AU} have not yet been found around white dwarfs,  it would provide a better comparison with our Solar System if ET can identify one. In addition, the ET + KMTNet Microlensing Survey is expected to provide mass measurements for more than 10 long-period planets orbiting white dwarfs, which can provide additional samples for the statistical study of planet populations around white dwarfs. More details can be seen in  Section \ref{sec:microlensing_yield}.

\subsubsection{The Solar System}\label{sec:ss}
{\bf Author:}\\
\newline
Quanzhi Ye$^{1,2}$ \& Xian Shi$^3$\\
{1.\it Department of Astronomy, University of Maryland, College Park, MD 20742, USA}\\
{2.\it Center for Space Physics, Boston University, 725 Commonwealth Ave, Boston, MA 02215, USA}\\
{3.\it Shanghai Astronomical Observatory, Chinese Academy of Sciences, Shanghai 200030, China} \\

Besides the planets in other planetary systems that it sets to detect, ET will also detect a large number of small bodies in {\it our~own} Solar System that happen to transit its fields. Small Solar System bodies are primitive planetesimals \citep[see, e.g.,][]{Nesvorny2019trans} from the formation stage of the Solar System that encompass all comets and asteroids, including main-belt asteroids, Trojans, Centaurs, comets, various near-Earth objects (NEOs) and trans-Neptunian objects (TNOs). These bodies provide important clues for a number of fundamental questions in planetary science such as the dynamical evolution of the Solar System \citep[e.g.,][]{Liu2022} as well as the accretion of water and organic materials on terrestrial planets, including Earth. As signs of exo-asteroid-belts and exo-Kuiper-belts are being found in other planetary systems \citep{martin2012formation, veras2020linking}, small bodies in the Solar System also provide context for the studies of their exoplanetary counterparts.

Even though ET's observing and data acquisition strategy is not optimal for Solar System science (which involves objects that are mostly found near the ecliptic and move over time), its continuous monitoring of a large part of the sky to a moderate depth from a perspective different from ground-based observatories remains valuable. A number of previous studies \citep[e.g.][]{szabo2015main, szabo2016uninterrupted, ryan2017trojan, molnar2018main,Kalup2021trojans, podlewska2021determination} have successfully analyzed Solar System objects using {\it Kepler}, which used a similar data acquisition strategy as ET. Known Solar System objects can be observed by arranging the masks along the predicted tracks of these objects. The monitoring for exoplanetary transits may also serendipitously discover occultation by known or unknown Solar System objects \citep{ofek2010detectability}. Full images of sufficiently high cadence (a few times a day) as currently planned for the microlensing telescope can be used for these detections and discoveries of Solar System objects. Overall, the microlensing telescope will be the major contributor for Solar System science owing to a field near the ecliptic and its more versatile data acquisition strategy (see section \ref{sec:MicroLensing}).

We use SkyBoT \citep{berthier2006skybot} to search the Solar System objects detectable by ET. We assume two circular fields $10^\circ$ and $1.13^\circ$ in radius centered at 19h22m, $+44^\circ30'$ and 17h55m, $-29^\circ$ as defined for the transit and microlensing telescopes on ET, corresponding to total survey areas of 314~deg$^2$ and 4~deg$^2$. (The modeled survey area for the transit telescopes is smaller than the actual 500~deg$^2$ due to the limit imposed by SkyBoT, but we believe this will not significantly impact the outcome given the limited number of objects visible to these telescopes.) We then simulate the Solar System objects that will transit these fields from October 1, 2026 to September 30, 2030. The seasonal window of the microlensing telescope (only operational from March 21 to September 21 each year) is considered. We adopt the designed detection limit of $I<20.6$ (equivalent to $V<21.3$ for Sun-color objects) for our search, though we note that this is somewhat conservative given that the proposed detection limit is for photometric purposes, and that most Solar System objects are redder than the Sun. This search is by no means exhaustive, since new objects are being found and new surveys are coming online.

\begin{figure}[htbp]
\centering
\includegraphics[width=0.85\textwidth]{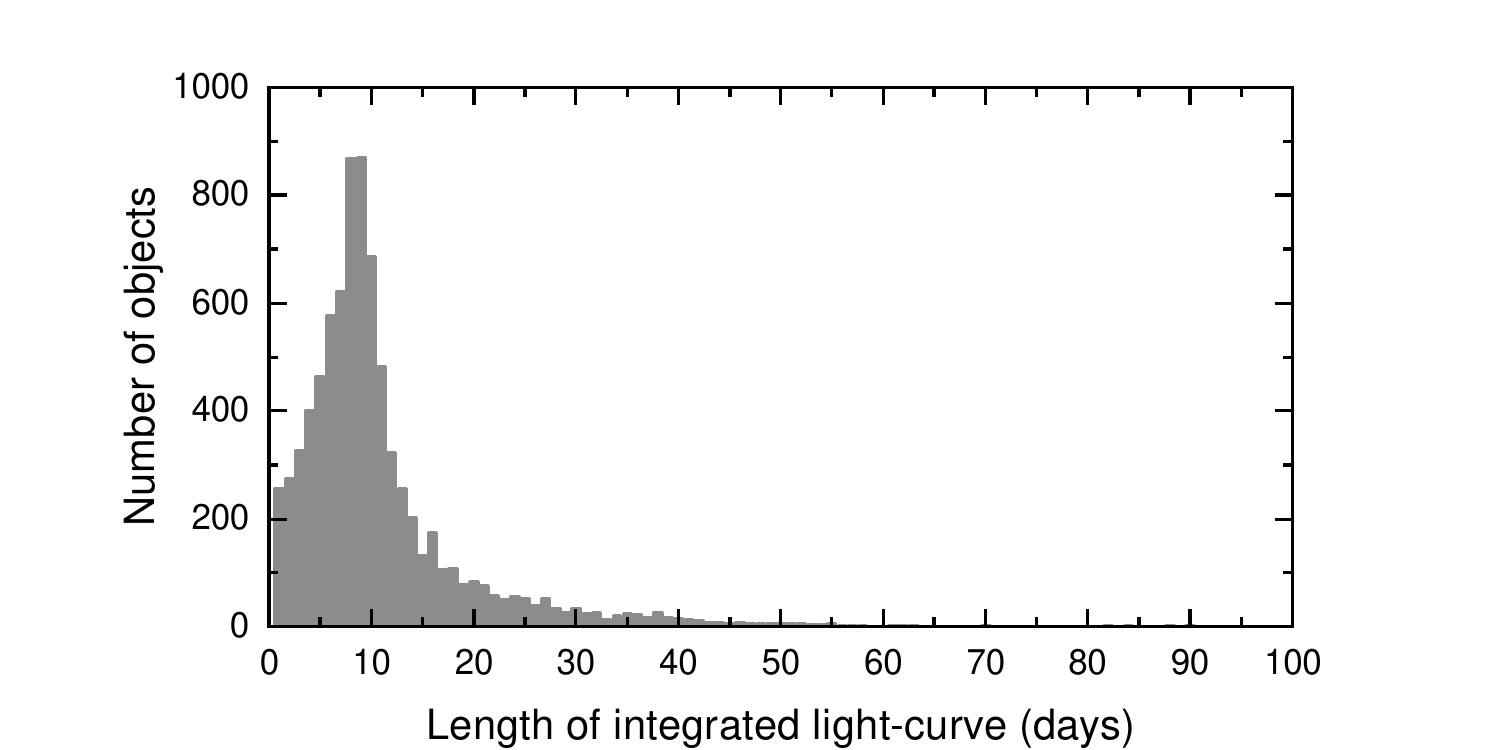}
\caption{The length of light-curve integrated over a 4-yr survey vs. the number of objects in each bin. About 85\% objects accumulate over 100~hr of light-curve.}
\label{fig:TotalDays}
\end{figure}

\begin{figure}[htbp]
\centering
\includegraphics[width=0.85\textwidth]{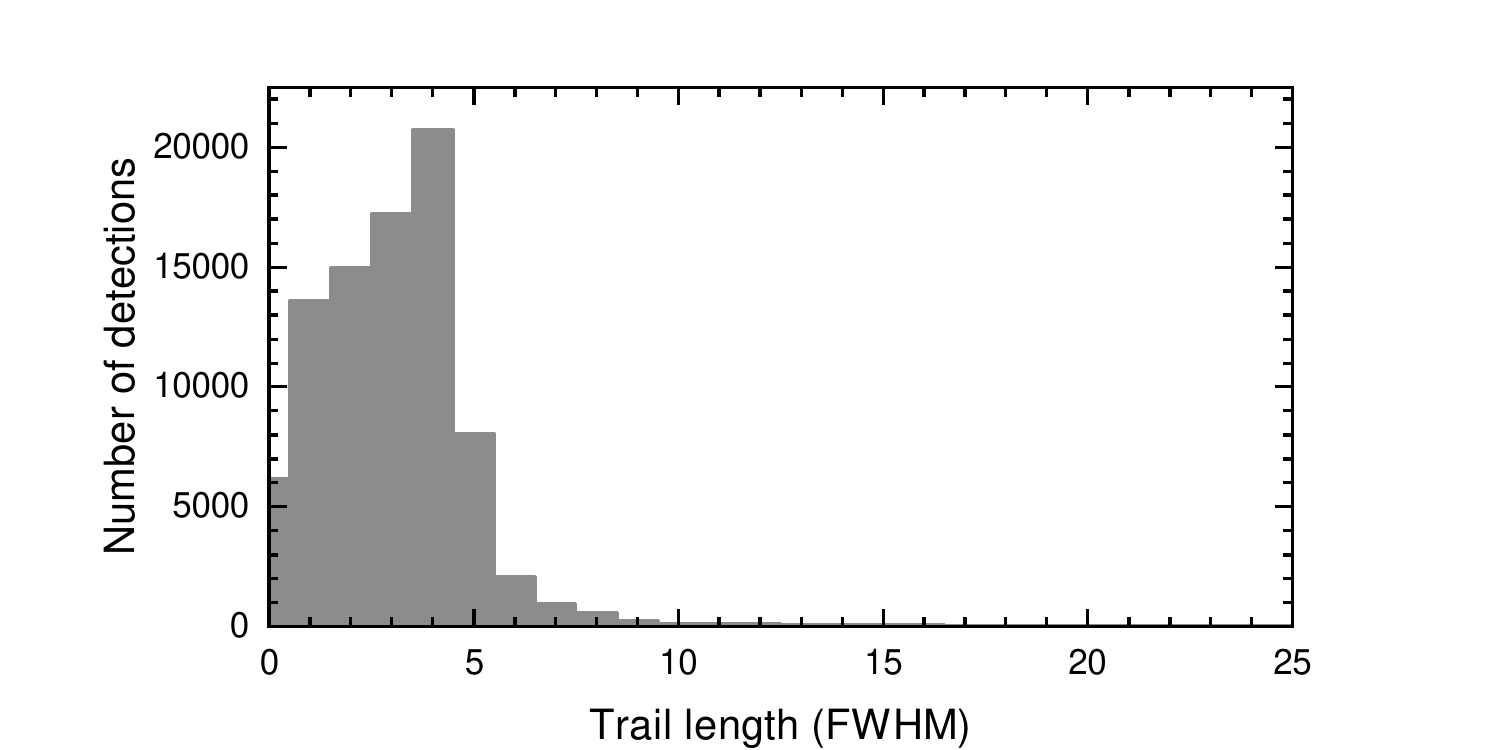}
\caption{Length of trails detected by the microlensing telescope vs. the number of detections in each bin. The trail lengths of detected objects are calculated at 0 hours UT each day. About \SI{98}{\percent} detections have trail lengths below 5 FWHMs.}
\label{fig:Trail}
\end{figure}

We find that ET will be able to detect at least 8,000 unique Solar System objects, \SI{95}{\percent} of which are main-belt asteroids. Among these objects, \SI{85}{\percent} accumulate over 100~hr of light-curve over a 4-yr survey (Figure~\ref{fig:TotalDays}). Since the majority ($\sim$\SI{98}{\percent}) of these objects will be detected by the microlensing telescope, which is planned to have a uniform 10-min exposure, we examine the trailing effect caused by the motion of the objects. We find that \SI{98}{\percent} detections have trail lengths below 5~FWHMs (Figure~\ref{fig:Trail}), a threshold where significant performance impact caused by trailing losses starts to kick in \citep{ye2019toward}. Meanwhile, it is worth noting that the high stellar density in the galactic bulge will likely affect the number of objects that can actually be detected in the actual data.


Here we discuss a few Solar System topics that can be pursued using the data from ET.

\begin{itemize}

\item{\bf Small body discovery.} ET's full images can be searched for unknown small bodies. Typical bodies in the main asteroid belt have a motion rate of 0.5$^\circ$/day, hence small body search is most efficient on groups of images with an interval of a fraction of an hour to a few hours. In this regard, the microlensing telescope (which will return 10 images per day, compared to the transit telescope's 1 image per day) will be most useful for small body searches. The designed capability of retrieving imagettes acquired within the past 3 days is also useful in case a higher cadence is needed (e.g. for fast-moving NEOs). The field of the microlensing telescope is also much closer to the ecliptic where most small bodies appear. ET will have a stable image FWHM nearly comparable to the Legacy Survey of Space and Time (LSST), which will be capable of resolving objects in the crowded galactic bulge field. Finally, the small FWHM will also benefit orbit determination using parallax with Earth-based telescopes (see below).

\item{\bf Improvement of small body orbits through parallax measurements.} The traditional strategy for orbital determination is to measure some orbital ``arc'' of the object of interest, which often involves months to years of observation. The distance between ET and Earth (\SI{0.01}{AU}) provides a sufficiently long baseline such that the parallax of Solar System bodies outward to the Kuiper-belt region (order of several \SI{10}{\arcsecond} or larger) is easily measurable. Effective orbital determination can be made with a few near-simultaneous observations spanning over a course of a day to a month by a telescope at the Sun-Earth's L2 point and a ground-based telescope \citep{rhodes2017scientific, giovinazzi2021enhancing}. This is particularly important for Kuiper-belt objects (KBOs) and distant comets, as their low orbital angular speed typically requires a long baseline on time to determine the orbit using the traditional method.

\item{\bf Monitoring active asteroids in the main asteroid belt.} Active asteroids are dynamical asteroids that episodically exhibit cometary features \citep{jewitt2012active}, possibly due to the sublimation of water ice. High-cadence observation of the onset and evolution of cometary activity provides crucial constraints on the nature of the activity \citep[e.g.][]{hsieh2010superwasp, farnham2019first, ye2019multiple}. ET (more precisely, the microlensing telescope) will provide a series of short (a few days) uninterrupted light-curves of about 8,000 main-belt asteroids, providing early warning of asteroid activity and instantaneous follow-up of any detected events. The occurrence rate of active asteroids is about 1 every 10,000 asteroids \citep{chandler2018safari}, thus ET may detect up to one event during its operational lifetime.

\item{\bf Asteroid rotation.} Light curves of asteroids provide important constraints on their rotation periods, shapes, and mechanical properties. For instance, the 2.2-hr ``spin barrier'' derived from early asteroid light curve surveys provides direct evidence that most km-class asteroids are gravitationally bounded aggregates, commonly known as the ``rubble-piles'' \citep{walsh2018rubble}. The ever-increasing catalog reveals a small number of unique asteroids known as the super-fast rotators which are interpreted as monolithic objects \citep{chang2017asteroid}. As shown in Figure~\ref{fig:TotalDays}, ET will provide robust rotational light curve measurements for most of the 8,000 objects that it will observe. Currently, about 35,000 asteroids have measured light-curves \citep{warner2009asteroid} out of which 6,000 will be observed by ET. ET will provide light-curves for 2,000 previously unmeasured asteroids, representing a $\sim5\%$ increase in the sample. High-accuracy spin rate determined with ET light-curves could also enrich the statistics for understanding small bodies’ YORP effect.

\item{\bf Binary asteroid systems.} Vastly distributed from the near-Earth space to the trans-Neptunian region, binary asteroid systems are a class of small Solar System bodies that consist of a pair of mutually orbiting asteroids. Their diverse dynamical configurations could be the result of different formation scenarios, such as disruption or mass-shedding. ET’s high-precision light-curves would contribute to the identification and quantification of such systems. The retrieved high-fidelity orbital and rotational states will shed light on the nature and history of various binary systems, advancing our understanding of the internal structure of primitive Solar System bodies as well as their varying evolutionary paths.

\end{itemize}

\subsubsection{Planetary Science Conclusions}\label{sec:planetary_conclusion} 
{\bf Authors:}
\newline
Ji-Wei Xie$^1$, Ji-Lin Zhou$^1$, Jian Ge$^2$, Hui Zhang$^2$ \& Hongping Deng$^2$\\
{1.\it School of Astronomy and Space Science, Nanjing University, Nanjing 210023, China}\\
{2.\it Shanghai Astronomical Observatory, Chinese Academy of Sciences, Shanghai 200030, China} \\

ET, as its name implies, will detect tens of temperate terrestrial planets (i.e., Earth 2.0s) around Solar-like stars, putting us on the path toward addressing the fundamental question: are we (Earth) alone? Furthermore, ET will detect tens of thousands of planetary systems of all kinds, including planets with different sizes (from Jupiter-like planets down to Mercury-like planets), planets with different orbital periods (from ultra-short-period lava planets, long-period icy planets, to free-floating planets), and planets around stars with different properties (e.g., mass, metallicity and age) in different stellar and Galactic environments (e.g., binary systems, clusters, the Galactic thin or thick disk, and halo stars). Such an exoplanet census conducted by ET will not only provide crucial observational constraints to deepen our theoretical understanding of planet formation and evolution, but also place our Solar System in context toward answering the fundamental question: how special are we (the Solar System)? In addition, ET will be able to detect exomoons, exorings, exocomets, and so on, thus enriching our knowledge of the diversity of planetary systems. Last but not least, ET will also detect and characterize many small bodies, e.g., asteroids, Trojans, Centaurs, comets, and various near-Earth objects (NEOs) in our Solar System,  which is invaluable to learning the history of the Solar System's formation and shedding light on planet formation in general.

\subsection{\bf Stellar and Milky Way Science}

\subsubsection{Asteroseismology}  



\textbf{Solar-like Oscillations}\\
{\bf Authors:}
\newline
Jie Yu$^1$, Tanda Li$^2$, Chen Jiang$^1$, Gang Li$^3$, Jian-Wen Ou$^4$, Zhao Guo$^5$, Yaguang Li$^6$, Paul Beck$^7$, Timothy R. Bedding$^6$, Shaolan Bi$^2$, Tiago L. Campante$^{8,9}$, William J. Chaplin$^{10,11,12}$, J{\o}rgen Christensen-Dalsgaard$^{11}$, Rafael A. Garc\'{\i}a$^{13}$, Patrick Gaulme$^1$, Laurent Gizon$^{1,14,15}$, Saskia Hekker$^{16, 17}$, Daniel Huber$^{18}$, Shourya Khanna$^{19}$, Yan Li$^{20,21,22,23}$, Savita Mathur$^{24,25}$, Andrea Miglio$^{10,26,27}$, Beno\^it Mosser$^{28}$, J. M. Joel Ong$^{29,18}$, \^Angela R. G. Santos$^{8}$, Dennis Stello$^{6, 11, 30}$, and Maosheng Xiang$^{31,32}$\\
\\
\noindent
{1, \it Max-Planck-Institut f{\"u}r Sonnensystemforschung, Justus-von-Liebig-Weg 3, 37077 G{\"o}ttingen, Germany}\\
{2, \it Beijing Normal University, 100875 Beijing, China}\\
{3, \it IRAP, Université de Toulouse, CNRS, CNES, UPS, 31400 Toulouse, France}\\
{4, \it Shaoguan University, 512005 Shaoguan, China}\\
{5, \it Department of Applied Mathematics and Theoretical Physics, University of Cambridge, Cambridge CB3 0WA, UK }\\
{6, \it Sydney Institute for Astronomy (SIfA), School of Physics, University of Sydney, NSW 2006, Australia}\\
{7, \it Institut f{\"u}r Physik, Karl-Franzens Universit{\"a}t Graz, Universit{\"a}tsplatz 5/II, NAWI Graz, 8010 Graz, Austria}\\
{8, \it Instituto de Astrof\'{\i}sica e Ci\^{e}ncias do Espa\c{c}o, Universidade do Porto,  Rua das Estrelas, 4150-762 Porto, Portugal}\\
{9, \it Departamento de F\'{\i}sica e Astronomia, Faculdade de Ci\^{e}ncias da Universidade do Porto, Rua do Campo Alegre, s/n, 4169-007 Porto, Portugal}\\
{10, \it School of Physics and Astronomy, University of Birmingham, Birmingham, UK}\\
{11, \it Stellar Astrophysics Centre (SAC), Department of Physics and Astronomy, Aarhus University, Aarhus, Denmark}\\
{12, \it Kavli Institute for Theoretical Physics, University of California, Santa Barbara, CA, USA}\\
{13, \it AIM, CEA, CNRS, Universit\'e Paris-Saclay, Universit\'e de Paris, Sorbonne Paris Cit\'e, F-91191 Gif-sur-Yvette, France}\\
{14, \it Institut f\"{u}r Astrophysik, Georg-August-Universit\"{a}t G\"{o}ttingen, Friedrich-Hund-Platz 1, 37077 G\"{o}ttingen, Germany}\\
{15, \it Center for Space Science, NYUAD Institute, New York University Abu Dhabi, PO Box 129188, Abu Dhabi, UAE}\\
{16, \it Landessternwarte K{\"o}nigstuhl (LSW), Heidelberg University, K{\"o}nigstuhl 12, 69117 Heidelberg, Germany}\\
{17, \it Heidelberg Institute for Theoretical Studies (HITS) gGmbH, Schloss-Wolfsbrunnenweg 35, 69118 Heidelberg, Germany}\\
{18, \it Institute for Astronomy, University of Hawai`i, 2680 Woodlawn Drive, Honolulu, HI 96822, USA}\\
{19, \it INAF - Osservatorio Astrofisico di Torino, via Osservatorio 20, 10025 Pino Torinese (TO), Italy}\\
{20, \it Yunnan Observatories, Chinese Academy of Sciences, 396 Yangfangwang, Guandu District, Kunming, 650216, China}\\
{21, \it Key Laboratory for the Structure and Evolution of Celestial Objects, Chinese Academy of Sciences, 396 Yangfangwang, Guandu District, Kunming, 650216, China}\\
{22, \it Center for Astronomical Mega-Science, Chinese Academy of Sciences, 20A Datun Road, Chaoyang District, Beijing, 100012, China}\\
{23, \it University of Chinese Academy of Sciences, Beijing 100049,  China}\\
{24, \it Instituto de Astrofísica de Canarias (IAC), 38205, La Laguna, Tenerife, Spain}\\
{25, \it Universidad de La Laguna (ULL), Departamento de Astrofísica, 38206, La Laguna, Tenerife, Spain}\\
{26, \it Dipartimento di Fisica e Astronomia Augusto Righi, Università degli Studi di Bologna, Via Gobetti 93/2, I-40129 Bologna, Italy}\\
{27, \it INAF-Osservatorio di Astrofisica e Scienza dello Spazio di Bologna, via Gobetti 93/3, I-40129 Bologna, Italy}\\
{28, \it LESIA, Observatoire de Paris, PSL Research University, CNRS, Sorbonne Universit\'e, Universit\'e Paris Diderot, 92195, Meudon, France}\\
{29, \it Department of Astronomy, Yale University, 52 Hillhouse Ave., New Haven, CT 06511, USA}\\
{30, \it School of Physics, University of New South Wales, NSW 2052, Australia}\\
{31, \it Max-Planck-Institut f\"ur Astronomie, K\"onigstuhl 17, D-69117, Heidelberg, Germany} \\
{32, \it National Astronomical Observatories, Chinese Academy of Science, Beijing 100101, China} \\

Asteroseismology is the study of stellar oscillations. Over the past decade, high-quality data collected by ground-based and space telescopes have presented great detail on the stellar oscillations of hundreds of thousands of stars that cover the whole \mbox{H--R} diagram. In particular, solar-like oscillations are those excited by near-surface convection, and occur in FGK dwarfs, subgiants and giants. These asteroseismic observations have subsequently significantly advanced several astrophysical fields, ranging from the physics of stellar interiors, to exoplanetary science, to the formation history of the Galaxy \citep[e.g., see][for reviews]{chaplin2013b, hekker2017a, garcia2019a, aerts2021a}.

The ET space mission aims to lead the dawning of another era of asteroseismology. This is warranted by its dedicated instrumental design. Specifically, ET consists of six 30 cm telescopes for the transit survey, and will continuously point towards an identical field of view of 500 square degrees for 4--8 years. This field covers the \textit{Kepler} field \citep{ borucki2010a}, and overlaps with the TESS northern Continuous Viewing Zone \citep{ricker2015a} and the PLATO Long-Observation Phase North field \citep{nascimbeni2022a}. The telescopes will offer a scale of 4.38 arcsec/pixel. Based on the instrumental requirement of detecting Earth-like exoplanets, ET is expected to offer asteroseismic data with a high photometric precision, comparable with that of \textit{Kepler} (see Figure~\ref{ET and Kepler simulated error budgets}).

We investigated the detectability of solar-like oscillations with ET using the methods applied for the \textit{Kepler}, K2, and TESS missions \citep{chaplin2011a,Chaplin2015,Campante2016a,Schofield2019a}. Our simulation shows that ET can detect solar-like oscillations in $\sim$100,000 main-sequence and subgiant stars, about two orders of magnitude more than all the previous detections combined \citep{Mathur2022}. Furthermore, a major difference in the observing strategy between ET and the PLATO mission is that ET will target a large sample of red giants for 4--8 years, which is not the case for the PLATO core or complementary science program \citep{montalto2021a}. The duration of 4--8 yr for the ET observations will yield Fourier spectra with exquisite frequency resolutions of 4.0--8.0 nHz. This is of great importance, for example, for investigating mixed modes (which have narrow intrinsic linewidths), rotational splittings (particularly for slow rotators), and potential magnetic splittings \citep[e.g.,][]{hekker2017a, bugnet2021a}, all of which are key observational seismic features of luminous red giants \citep[e.g.,][]{yu2020a}. Combining ET and \textit{Kepler} data will increase the signal-to-noise ratio (SNR) of the Fourier spectra, and allow us to study trends and variations over a baseline of two decades, such as magnetic activity cycles. Below, according to the stellar evolutionary stage, we outline some key science cases that ET is likely to provide unique data to address. 

\textit{\textbf{Asteroseismology of Dwarfs and Subgiant Stars}}: (1) Asteroseismology will enhance exoplanet studies by precisely characterising exoplanet hosts \citep[e.g.,][]{Huber2013b, lundkvist2016a, Huber2018a}. We plan to achieve stellar parameters precision of $\sim$2\% in radius, $\sim$5\% in mass, and $\sim$10\% in age. Precise and accurate characterization of exoplanet hosts paves the way for detecting habitable Earth analogs. With sufficiently high SNR, measurements of the relative amplitudes of rotational multiplet components may permit the determination of the projected inclination of the stellar rotational axis \citep{Gizon2003}, at scale \citep[e.g.,][]{Kuszlewicz2019,Gehan2021}. This will be highly complementary to any follow-up studies of the dynamical architectures of associated exoplanetary systems \citep{Chaplin2013, huber2013, Campante2016}, e.g. in conjunction with Rossiter-McLaughlin RV measurements. (2) Metal-poor stars are unfortunately under-represented in the \textit{Kepler} asteroseismic target list \citep{chaplin2020a}. Our pre-selected metal-poor stars will advance Galactic archaeology enabled by asteroseismology. (3) ET will be instrumental for performing seismic analysis of F-type stars. Due to shorter mode lifetimes, radial and quadrupole modes blend, thus complicating asteroseismic modelling of these stars \citep{compton2019a}. ET's high-precision, long-duration observations will provide an opportunity to disentangle the mode blending for mode identification. (4) Another open question that we aim to address with ET data concerns how the transition takes place between stochastic and coherent oscillations near the red edge of the instability strip \citep{huber2011a,balona2020a}. (5) Furthermore, high-precision photometry will extend the detection limit of solar-like oscillations towards K-type dwarfs, which include numerous exoplanet hosts \citep{campante2015a}. (6) Asteroseismology serves as a robust benchmark to calibrate age-rotation-activity relations by measuring precise ages from asteroseismic modelling, and rotational periods either directly from the light curves \citep[e.g.][]{Santos2021a}, or from the frequency splittings of rotational multiplets \citep[e.g.,][]{Hall2021a}.
ET will also provide asteroseismic constraints on the radial and latitudinal differential rotation of many of these stars \citep[e.g.,][]{deheuvels2014a,benomar2018a,bazot_latitudinal_2019}. (7) Finally, ET holds the potential to increase the detections of magnetic activity cycles and investigate the relation between rotation and long activity cycles \citep[e.g.][]{Baliunas1995, Mathur2014,garcia2019a}, given ET's long time baseline (4-8 yr), particularly the combination of the \textit{Kepler} and ET data. Asteroseismology of magnetic cycles will also be possible \citep[e.g.][]{Garcia2010_HD49933,Karoff2018_HD173701,Santos2018_fshifts}, which can provide constraints on the location in depth and latitude of the perturbation related to magnetic activity \citep{Salabert2018_fshifts,Thomas2019_activeLat}.

\textit{\textbf{Asteroseismology of Red Giant Stars}}: Unlike the pressure waves in main-sequence stars, solar-like oscillations in evolved red giants are coupled to an internal mode cavity of buoyancy waves, which permits them to precisely constrain conditions deep in the stellar interior, making red giants particularly inviting targets. A \mbox{long-standing} open question in stellar physics amenable to study in this fashion is that of angular momentum transport \citep{aerts2019b}.  Theoretical predictions of rotation rates differ by about two orders of magnitude from those actually measured from red giants \citep{mosser2012c,aerts2019b}.  Enabled by the high resolution spectra provided by the long temporal baseline of ET, higher-precision asteroseismic constraints on internal and envelope rotational rates of stars along the red giant branch and in the red clump will clarify this tension, and provide clues for its resolution. Precise rotational splitting measurements at the level of individual rotational multiplets may permit going beyond the present two-zone phenomenological model of radial differential rotation, thereby permitting discrimination between various proposed mechanisms for this angular momentum transport. Another interesting puzzle is the unknown mechanism responsible for the amplitude suppression of dipole modes \citep{Garcia2014a,stello2016a}. Are they the result of internal magnetism or unknown damping mechanisms \citep{Fuller2015,mosser2017a}? This remains unclear.  Moreover, \textit{Kepler} data has revealed a large sample of particular oscillating red giants that have high-quality long-coverage \textit{Kepler} data but show complex peaks in the power spectra, mimicking low SNR scenarios (Braun et al. 2022, in prep.). The physics driving this has yet to be revealed. Meanwhile, \citet{bugnet2021a} have predicted detectable asymmetric magnetic splittings. This is crucial for probing magnetic fields deep in the core region. Since these asymmetries are predicted to be most significant for modes most confined to the interior buoyancy cavity, which have the lowest amplitudes and narrowest linewidths, their detection demands power spectra that have high SNR and fine frequency resolution; ET will provide critical asteroseismic observations for this interesting science case.

\textit{\textbf{Asteroseismology of Long Period Variables}}: Long Period Variables (LPVs) are high-luminosity RGB/AGB stars that pulsate with periods $P~\gtrsim~1$ day. The high pulsation amplitudes of LPVs are easily detectable with modern space missions, including ET \citep{yu2020a}. ET will target these stars with the aiming of understanding the following fundamental open questions. Firstly, how do RGB stars lose their mass? Currently, there is no consensus on the amount of mass loss taking place in the RGB phase \citep{miglio2021a, yu2021}. The estimates of integrated RGB mass loss ranges from $\sim$0.10~M$_{\odot}$ in open clusters determined by asteroseismology \citep[e.g.][]{miglio2021a}, to $\sim$0.23~M$_{\odot}$ in globular clusters mainly obtained by horizontal branch morphology approaches \citep[e.g.][]{salaris2016a,tailo++2022-m4}. We recall that integrated mass loss has an important impact on the determination of the ages of red-clump stars. Secondly, the asteroseismic classification of RGB and core-helium-burning (CHeB) stars has been one of the major breakthroughs in stellar physics \citep[e.g.][]{bedding2011a}. However, distinguishing AGB and RGB stars is challenging. This is because mixed modes, which are used to disentangle RGB and CHeB stars, are hardly detectable in LPVs, mainly due to their low intrinsic oscillation amplitudes and small frequency spacing. Pressure modes with high frequency resolution would provide new insights \citep{dreau2021a} and, if possible, would have profound impact on the determination of ages of RGB stars, and on understanding the chemical evolution of such stars. Thirdly, studies have shown that LPVs are valuable distance indicators, and asteroseismology can provide precision distance (15\%) out to 30-50 kpc, based on period-luminosity relations \citep{auge2020a}. The ongoing and future transit surveys, such as OGLE, ASAS-SN, ATLAS, and LSST, can provide light curves with time baselines from a few to $\sim$10 years. However, these light curves are subject to large data gaps (duty cycles $\sim$ 30\%) and low sampling rates (3-5 days cadence), which causes sidelobes, complicating the identification of the pulsation modes used for deriving luminosity with period-luminosity relations. ET has the potential to provide high-cadence long-duration light curves that would guide the synergy of asteroseismology with ground-borne and space-based time-domain photometry, particularly given ET's higher photometry precision towards faint magnitudes compared with \textit{Kepler}.
  
\textit{\textbf{Asteroseismology of Open Clusters and Binaries}}:  The \textit{Kepler} mission observed 4 open clusters, and only two of them have been subject to extensive detailed asteroseismic analysis, exclusively for red giants. In addition,  while \textit{Kepler} observed more than $\sim$3000  binaries, there are only $\sim$100 known red-giant oscillators in binary systems and among them 28 in eclipsing binary systems \citep{beck2022a}. Based on these limitations, we aim to observe more open clusters and binaries, and prioritize their members to span a broad range of evolutionary stage, age, and metallicity. Open clusters and eclipsing binaries are valuable targets for testing complex microscopic and macroscopic stellar physics \citep[e.g., see the review of the \texttt{HAYDN} project,][]{miglio2021a}. Pre-selected open clusters will enable diverse studies, including mass loss, non-standard evolution of under-massive and over-massive stars, determination of helium abundance, internal mixing, asteroseismic analysis of young stars, etc. 
ET will also offer new clues and constraints to understand several open questions for binaries, such as mass transfer, mergers, orbital precession triggered by dynamic tides, orbital decay of heartbeat stars, angular momentum transport between stars and orbits, calibration of seismic scaling relations, stellar activity, among others. 
\\

\noindent{\textbf{Classical pulsating stars}}\\
{\bf Authors:}
\newline
Weikai Zong$^1$, Tao Wu$^{2,3,4,5}$,  Zhao Guo$^6$, Gang Li$^7$, Dominic M. Bowman$^8$, Mariel Lares-Martiz$^{16}$, Simon Murphy$^9$,
Jia-Shu Niu$^{10}$, Ming Yang$^{11}$, Xiao-Yu Ma$^1$, L\'aszl\'o Moln\'ar$^{12,13}$, Jian-Ning, Fu$^1$, Yan Li$^{2,3,4,5}$, Peter De Cat$^{14}$, Timothy R. Bedding$^{15}$, \& Jie Su$^{2,3,4}$ \\
1.{\it Department of Astronomy, Beijing Normal University, Beijing 100875, China}\\
2. {\it Yunnan Observatories, Chinese Academy of Sciences, 396 Yangfangwang, Guandu District, Kunming, 650216, China}\\
3. {\it Key Laboratory for the Structure and Evolution of Celestial Objects, Chinese Academy of Sciences, 396 Yangfangwang, Guandu District, Kunming, 650216, China}\\
4. {\it Center for Astronomical Mega-Science, Chinese Academy of Sciences, 20A Datun Road, Chaoyang District, Beijing, 100012, China}\\
5. {\it University of Chinese Academy of Sciences, Beijing 100049,  China}\\
6. {\it Department of Applied Mathematics and Theoretical Physics, University of Cambridge, Cambridge CB3 0WA, UK}\\
7. {\it IRAP, Université de Toulouse, CNRS, CNES, UPS, 31400 Toulouse, France}\\
8. {\it 
Institute of Astronomy
, KU Leuven, Celestijnenlaan 200D, B-3001 Leuven, Belgium
}\\
9. {\it Centre for Astrophysics, University of Southern Queensland, Toowoomba, QLD 4350 Australia}\\
10. {\it Institute of Theoretical Physics, Shanxi University, Taiyuan 030006, China}\\
11. {\it School of Astronomy and Space Science, Nanjing University, Nanjing 210023, China} \\
12. {\it Konkoly Observatory, Research Centre for Astronomy and Earth Sciences, E\"otv\"os Lor\'and Research Network (ELKH), Konkoly Thege Mikl\'os \'ut 15-17, H-1121 Budapest, Hungary}\\
13. {\it ELTE E\"otv\"os Lor\'and University, Institute of Physics, 1117, P\'azm\'any P\'eter s\'et\'any 1/A, Budapest, Hungary}\\
14. {\it Royal Observatory of Belgium, Ringlaan 3, B-1180 Brussel, Belgium} \\
15. {\it Sydney Institute for Astronomy (SIfA), School of Physics, University of Sydney, NSW 2006, Australia}\\
16. {\it Instituto de Astrofísica de Andalucía (IAA-CSIC), 18008, Granada, Spain} \\

Apart from solar-like oscillations, the majority of pulsating stars have oscillations with mode lifetimes much longer than the oscillation period and the typical length of observed light curves. Thus, a long timeline and high precision photometry would greatly reduce the noise level by $\sqrt{N} $ of $N$ data points and allows the detection of a lot more low-amplitude pulsation modes. We expect to perform asteroseismic investigations of various types of pulsating stars using their rich pulsation frequencies as offered by ET. When we combine {\it Kepler} data with ET data, the timeline enables the characterization of both amplitude and frequency modulations over an unprecedented interval of 20 year with two nearly long and continuous segements of photometry. This would allow for asteroseismic studies of both linear and nonlinear schemes through their frequency contents and dynamics. Note that clear-cut evidence of nonlinear weak interactions is barely detectable from ground-based observations due to its poor precision and low duty cycle \citep{1989A&A...215L..17V}.

Linear asteroseismology focuses on probing the internal structure and chemical profiles, and the strength of different physical processes such as overshooting, diffusion, convection, rotation and magnetism using the pulsation mode frequencies of a star as observables \citep{aerts2021a,2022arXiv220111629K}. The results are expected to have a major impact on the development of the theory of stellar evolution\citep[e.g.][]{Bowman2021a}, the transport of angular momentum and mixing\citep[e.g.][]{aerts2019b}, the improvement of the period-luminosity relationship \citep[e.g.,][]{2019MNRAS.486.4348Z}, and the accuracy of the location of instability strips for stars across the H-R diagram \citep[e.g.,][]{2019MNRAS.485.2380M}.

To date, most results from asteroseismology are mainly obtained only using frequencies and ignoring amplitudes, phases and the relationship between them. The linear nonadiabatic calculations predict the values for stationary eigen-frequencies and show that all modes within the instability range should, in principle, get excited \citep{2014A&A...569A.123B,2003ApJ...597..518F}. However, the mechanisms controlling amplitude saturation are not well known in asteroseismology. For instance, it is still unknown in compact pulsators why the observed number of frequencies is small compared to the theoretical predictions, even after taking geometric cancellation effects into account. More generally, the excitation and number of observed versus expected modes in dwarf stars is not well understood. This is a longstanding question for linear asteroseismology, but can potentially be addressed within a nonlinear framework through 
precisely characterizing of amplitude modulation patterns and timescales. With its long-duration monitoring in conjunction with {\it Kepler}'s high-precision photometry data, ET can potentially help address these questions through nonlinear amplitude equations\citep[e.g.][]{VanHoolst1994b}. Nonlinear asteroseismolgy may play a key role in understanding the dynamics of oscillations along with time and measure the growth rate of oscillation modes through observations for the first time. \\

The linear asteroseismology will mainly focus on the following topics:

\textit{\textbf{O/B type pulsators}}: Among the most massive stars, which have spectral types of O/B, there is a large diversity of variability mechanisms including coherent pulsations and travelling waves \citep{Bowman2020c}. The coherent pulsations of massive stars are driven by the $\kappa$-mechanism operating in the ionization region of the iron group elements \citep{1993MNRAS.265..588D}. Light curves assembled by ET will identify new pulsating stars which are currently rarely observed \citep[e.g.][]{Burssens2019a,Bowman2019b,Handler2019a}.  Precise measurements of pulsation frequencies allow asteroseismic constraints on physical properties of the core, such as convection, overshooting, rotation and diffusion \citep{Moravveji2015b,2018ApJ...867...47W,Szewczuk2018a,2021NatAs...5..715P}. Continuous high precision photometry over a long time baseline is needed to achieve this high frequency resolution. For instance, the current sample size of slowly pulsating B (SPB) stars is about 30, with most being relatively fast rotators\citep{Szewczuk2021a}. Due to the large and asymmetric frequency separation of rotational splittings, it is hard to identify which multiplet each frequency belongs to\citep[e.g.][]{Papics2014},  making it difficult to achieve unambiguous mode identifications. On the other hand, in the case of slow rotators, frequency splittings are hard to resole without long time-base light curves spanning years. This situation will be drastically improved with the combination of {\it Kepler} and ET data, which is expected to provide excellent frequency resolution to study different rotational component for slow rotating SPB stars, and high frequency precision for asteroseismic modelling.

\textit{\textbf{A/F type pulsators}}: These are classical variables with spectral type of A/F, including $\delta$~Scuti, $\gamma$~Dor, roAp, RR Lyrae and Cepheid stars. Mode identification is difficult in these stars as their pressure-mode frequency patterns are quite complicated\citep[see e.g.][]{Paparo2016a}. {\it TESS} has provided a few dozen $\delta$~Scuti stars with regular frequency spacing whereas most are beyond the Nyquist frequency of {\it Kepler} \citep{2020Natur.581..147B}. These can be used to obtain ages for young clusters and associations \citep{2022MNRAS.511.5718M}. On the other hand, the period spacing patterns of asymptotic gravity modes in $\gamma$~Dor stars can be used to measure the interior rotation period \citep{VanReeth2016a,2020MNRAS.491.3586L}. However, such analysis requires continuous photometric data longer than one year, which is beyond the time window for most targets observed by {\it TESS}.  The high cadence provided by ET will cover a much larger frequency range than {\it Kepler} and provide a much larger sample than that produced by {\it TESS} continuous viewing zones (CVZ) observations. With at least 4 years of continuous observations overlapping with the {\bf Kepler} field, ET will allow for the measurements of the interior rotation periods of $\gamma$~Dor stars and the characterization of the physical processes of their convective cores \citep{2020A&A...640A..49O}. ET will also be able to investigate other classical pulsators, such as RR~Lyrae and Cepheid stars, to study their period doubling, Blazkho effect and period-luminosity relationship \citep{Szabo2010}. We hope that the continuous observations will discover more non-radial small-amplitude modes to reveal mode interactions in these stars \citep{Benk2010MNRAS,Plachy2021TESSCepheid,Molnar2022TESSRL}. The large FOV of the ET mission also provide the research possibility of the RR~Lyrae stars and other pulsators in globular clusters \citep{Wallace2019ApJ}.
 
\textit{\textbf{Compact pulsators}}: These faint blue objects, including white dwarfs and hot B subdwarfs (sdB), are only few in number as observed by space missions with year-long durations \citep[see, e.g.,][]{2010MNRAS.409.1470O}. However, they are particularly important since they exist under extreme physical conditions and are the testbed of the terminal phase in stellar and planetary systems \citep[see, e.g.,][]{2011Natur.480..496C}. The discovery of a unique He-dominated pulsating white dwarf from {\it Kepler} was surprising due to its oxygen content and core size \citep{2018Natur.554...73G}. In addition, the rapid oscillations in compact pulsators may exceed the Nyquist frequency of {\it Kepler} SC data. {\it TESS} was able to study a few within its CVZs due to its brightness limit although it has covered a large number of compact \citep[see, e.g.,][]{2020A&A...638A..82B}. ET will extend the sample of {\it Kepler}'s compact pulsators by about 10 times with its faster cadence and wider FOV than {\it Kepler}. Compared to {\it TESS}, ET will excel in both limiting magnitude and observational duration, thus providing better quality data to study a large number of compact pulsators. The seismic cartography of the internal structure of these compact pulsators derived from the ET data is key to tracing stellar evolution processes, such as the mixing process, mass loss rate and angular momentum transport.

\textit{\textbf{Pulsators in binaries}}: Binary systems contain various types of pulsating stars from pre-main-sequence and post-main-sequence to white dwarfs. The dynamical tide within binary star systems is able to induce and excite oscillations or even suppress the magnetic activity \citep{2012MNRAS.420.3126F,2020ApJ...888...95G}. Tides can also tilt the pulsation axis away from the rotational axis and towards the binary companion \citep{2020MNRAS.498.5730F}. In tight contact binaries, there is still insufficient understanding about the relationships among mass transfer, angular momentum transport, and how pulsations are affected by tides\citep{Bowman2019d,Handler2020a}. In addition, several kinds of interesting pulsators such as extreme low-mass white dwarfs \citep{2012ApJ...750L..28H}, blue stragglers, and BLAPs \citep{2017NatAs...1E.166P}, are probably generated through the merger channel of binary evolution. ET will observe a large number of binary systems, which may provide new clues and insights into correlations among tidal forces, mass transfer, angular momentum transport, and oscillation. \\

The nonlinear asteroseismology will mainly focus on the following topics:

\textit{\textbf{Nonlinear mode interactions}}: Nonlinear resonant mode coupling predicts various resonances and modulation patterns of amplitude and frequency using the numeric solutions to amplitude equations when only a few interacting modes are involved \citep{1984ApJ...279..394B}. The observational evidence for such weak modulation is scarce since the timescale of the nonlinear modulation is much longer than what can be obtained from the ground. The advent of {\it Kepler} opened a new window for uncovering amplitude and frequency modulations. In pulsating white dwarfs \citep{2016A&A...585A..22Z}, sdB \citep{2021ApJ...921...37Z}, SPB stars \citep{VanBeeck2021a}, and $\delta$~Scuti stars \citep{2016MNRAS.460.1970B}, various patterns of modulations were detected and characterized, providing new constraints on the nonlinear coefficients and other key parameters. However, the observed modulations are more complex than the simple patterns predicted by the nonlinear theory \citep{2016A&A...594A..46Z}. In addition, quasi-periodic modulations are only noticeable on timescales of years, which means that only one or two cycles were observed by {\it Kepler}. ET will substantially extend the time baseline for such modulations, having a frequency resolution of \SIrange[range-units = single]{0.1}{1}{\nano\hertz} and fully covering at least twice as many cycles as {\it Kepler}. Therefore, ET's observational results will be extremely useful for constraining the nonlinear coupling coefficients, linear growth and damping rates of coherent pulsations in various-type stars.

\textit{\textbf{Nonlinear response of the stellar medium to pulsation}}: 
Apart from resonant mode coupling, oscillation spectra feature patterns of frequency from another physical origin. Another well now source of combination frequencies is the nonlinear response of the stellar medium to pulsation. It is a thoroughly studied nonlinear mechanism known to be operating in white dwarfs stellar surface \citep{Brickhill1992, Brassard1995}. Combination frequencies due to this nonlinear mechanism, have been helpful for mode identification in ZZ Ceti stars \citep{Wu2001, Montgomery2005}.

Combination frequencies from a non-linear response nature are also present in other type of pulsating stars, where an outer convective layer can be expected. This is the case of $\delta$~Scuti stars, where the presence of such non-linearities represent an additional obstacle to their already difficult to interpret power spectra. Under suitable mathematical framework \citep[first  proposed in ][]{Garrido1996}, it was possible to properly identify the non-sinusoidal characteristics of the light curves \citep{Lares-Martiz2020} and to associate specific features to non-linearities of finite amplitude radial pulsations. It can be proved that such characterization is useful for asteroseismic inferences and to determine stellar properties of $\delta$~Scuti stars (Lares-Martiz et al., in prep). Nonlinear interactions between radial modes have also been observed in RR Lyrae stars, and they offer a possible new hypothesis to explain the Blazhko effect \citep{BuchlerKollath2011,Molnar2012RRLyrae}. ET’s ultra-precise observations and its large FoV will contribute to statistically strengthen these heuristic conclusions by providing a larger sample of pulsating stars where to study the non-linearities phenomena.

\textit{\textbf{Phase modulations}}: Pulsation phase modulations are applicable for detecting stellar or planetary companions orbiting pulsating stars \citep{2007Natur.449..189S,2016ApJ...827L..17M,HajduG2021binaryRRL} as well as constraining the changing rate of stellar evolution \citep{2021ApJ...906....7K}. The phase measurement precision determines the mass limit for the former application. However, the phase modulation must be disentangled from the frequency modulation first as phase is an integral of frequency over time. After we have the correct phase information, this technique is suitable for detecting companions in  binary systems with orbital periods on the order of a few hundred days \citep{2018MNRAS.474.4322M}.  This companion detection method through monitoring phase modulations is quite unique and powerful: detected candidates are rarely seen as eclipsing binaries because the eclipsing probability of binaries with such long periods is very low. At the same time, it is hard to reach these period ranges using other detection techniques (i.e., spectroscopic observations). {\it Kepler} has demonstrated that this method is very important for studying binaries containing classical pulsating stars. However, the time span of {\sl Kepler} limits the orbital periods to less than four years. ET can detect binaries with orbital periods from a few hundred to several thousand days after combining with {\sl Kepler} and {\it TESS} data, which is useful to constrain the binary formation and evolution. ET will provide a much larger binary sample and can find new non-transiting multi-planet systems with relatively wide orbits. 

High-precision photometry from ET will also help to disentangle true RR Lyrae binaries from those that experience phase modulation through the Blazhko effect \citep{HajduG2021binaryRRL}. Despite the large number of known RR Lyrae stars, pulsators in binary are very difficult to identify and are still sought after, Dynamical masses for RR Lyrae stars would represent a key reference point for pulsation models. 

With sufficiently long and precise photometric monitoring it is also possible to study stellar evolution through stellar pulsation rate changes. The duration of the {\it Kepler} mission was not long enough for such research. Nevertheless, when combined with ET data and the {\it Kepler} data, it would allow for pulsation rate ($\dot{P}$) measurements with a precision higher than 1 ms per year, considering the 20 years of observation interval. These plusation rates are particularly important for testing the theory of stellar evolution with much improved precision \citep[see e.g.][]{Bowman2021a}.

\subsubsection{Stellar Age and Galactic Archaeology} 
{\bf Authors:}
\newline
Maosheng Xiang$^{1,2}$, Yaqian Wu$^2$, Jinghua Zhang$^2$, Haibo Yuan$^3$, Shaolan Bi$^3$, Ning Gai$^4$, Jian Ge$^5$, Zhishuai Ge$^6$, Zhao Guo$^7$, Yang Huang$^8$, Haining Li$^2$, Tanda Li$^3$, Yuxi (Lucy) Lu$^{9,10}$, Hans-Walter Rix$^{1}$, Jianrong Shi$^2$, Fen Song$^{11}$, Yanke Tang$^4$, Yuan-Sen Ting$^{12,13}$ \& Taozhi Yang$^{14}$, Qing-Zhu Yin$^{15}$, Jie Yu$^{16}$ \\
{1, \it Max-Planck-Institut f\"ur Astronomie, K\"onigstuhl 17, D-69117, Heidelberg, Germany} \\
{2, \it National Astronomical Observatories, Chinese Academy of Science, Beijing 100101, China} \\
{3, \it Department of Astronomy, Beijing Normal University, Beijing 100875, China} \\
{4, \it College of Physics and Electronic information, Dezhou University, Dezhou 253023, China} \\
{5, \it Shanghai Astronomical Observatories, Chinese Academy of Science, Shanghai, China} \\
{6, \it Beijing Planetarium, Beijing Academy of Science and Technology, Beijing 100044, China} \\
{7, \it Department of Applied Mathematics and Theoretical Physics, University of Cambridge, Cambridge CB3 0WA, UK} \\
{8, \it South-Western Institute for Astronomy Research, Yunnan University, Kunming 650500, China} \\
{9, \it Department of Astronomy, Columbia University, 550 West 120th Street, New York, NY 10027, USA} \\
{10, \it American Museum of Natural History, Central Park West, Manhattan, NY 10024, USA} \\
{11, \it College of Science, Jimei University, Xiamen 361021, China} \\
{12, \it Research School of Astronomy \& Astrophysics, Australian National University, Cotter Rd., Weston, ACT 2611, Australia} \\
{13, \it Research School of Computer Science, Australian National University, Acton ACT 2601, Australia} \\
{14, \it Ministry of Education Key Laboratory for Nonequilibrium Synthesis and Modulation of Condensed Matter, School of Physics, Xi’an Jiaotong University, Xi’an 710049, China} \\
{15, \it Department of Earth and Planetary Sciences, University of California, Davis, CA 95616, USA} \\
{16, \it Max-Planck-Institut f{\"u}r Sonnensystemforschung, Justus-von-Liebig-Weg 3, 37077 G{\"o}ttingen, Germany} \\

Galactic archaeology aims to reconstruct the origin, assembly, and evolutionary history of our Galaxy through detailed studies of the chemical fingerprint and orbital distribution of the stellar ``fossils"  that span almost the entire age range of the universe. To achieve this goal, precise and accurate stellar ages are required, which is very challenging. The ET mission, with its large field of view, ultra-high-precision, and long-duration photometry, will open a new window for determining precise stellar ages for large samples of stars through asteroseismology and gyrochronology. A central mission of the ET Galactic archaeology project is to conduct a systematic and complete survey of $\gtrsim$6000 metal-poor stars (${\rm [Fe/H]}<-0.8$) in the ET field, and to obtain asteroseismic ages with a precision of within a few percent.  This sample will provide an unprecedented foundation for unveiling the earliest phases of Galactic formation, where precision and accuracy calibration of stars across a wide range of metallicities matter most. ET will also deliver precision ages, with both asteroseismology and gyrochronology, for more than 100,000 stars that cover the halo, thick and thin disk components of our Galaxy. This will enable archaeological studies of the full Galactic formation history. ET will provide a transformational ``ground-truth" sample of stellar parameters (particularly age) for extensive large-scale stellar surveys.\\

\noindent\textbf{Stellar age as a foundation of Galactic archaeology}\\
Our Galaxy, the Milky Way, is a disk galaxy that has gathered about a hundred billions of stars in the past 14 billion years. The distribution of these stars can be described by a few characteristic structural components, including the halo \citep{Helmi2008, Helmi2020} and bulge (bar) \citep{Barbuy2018}, as well as a number of disk components that range in age, element composition, dynamics, and structure (thick to thin) \citep{Rix2013}. Understanding when and how these characteristic galactic structures assembled and evolved in the past billions of years is a fundamental task of Galactic archaeology \citep[e.g.][]{Freeman2002, Bland-Hawthorn2016}. This requires precise knowledge of the distances, velocities, ages, and chemical abundances for large and complete (or representative) samples of stars \citep[e.g.][]{Rix2013, Xiang2022}.

Galactic archaeology has entered an exciting new era driven by a data revolution of large-scale stellar surveys based on both ground and space telescopes. The ESA's Gaia mission \citep{Prusti2016} has delivered precision astrometric information, including celestial coordinates, parallax, and proper motions, for 1.4 billion stars completed down to $G\simeq20$ \citep{Brown2018, Brown2021}. Meanwhile, the implementation of a number of large spectroscopic surveys, such as the SEGUE \citep[$R\simeq2000$;][]{Yanny2009}, RAVE \citep[$R\simeq7500$;][]{Steinmetz2006}, LAMOST \citep[$R\simeq1800$;][]{Deng2012}, APOGEE \citep[$R\simeq22500$;][]{Majewski2017}, GALAH \citep[$R\simeq24000$;][]{De_Silva2015}, and Gaia-ESO \citep{Gilmore2012}, deliver the stellar line-of-sight velocity, atmospheric parameters (effective temperature, surface gravity, and metallicity), and elemental abundance for millions of stars  \citep[e.g.][]{Lee2011, Luo2015, Casey2017, Xiang2017a, Xiang2019, Ting2019, Buder2021, Abdurrouf2022, WangC2022}. Several spectroscopic surveys that will provide an even larger spectral data set over the full sky are also around the corner, for example, SDSS-V Milky Way Mapper \citep{Kollmeier2017}, 4MOST \citep{deJong2019}, and PFS \citep{Tamura2018}. In addition, multi-band photometric surveys, such as the Skymapper Southern Survey \citep[SMSS;][]{Wolf2018} and the J-PLUS survey \citep{Cenarro2019}, have also shown great potential for obtaining stellar parameters and abundances of massive sets of stars \citep{Yang2022, Xu2022, Huang2022, Fouesneau2022}. It is expected that a number of upcoming multi-color photometric surveys, such as the LSST \citep{2019ApJ...873..111I}, Mephisto \citep{YuanX2020}, and CSST \citep{ZhanH2021}, will further lead to a surge in available data in the coming decade.

Stellar age dating is, however, a far more challenging task, as most stars are in equilibrium with low explicit time-dependence in their properties. Hence, stellar ages cannot be measured directly but must be inferred through models, by comparing measured stellar parameters with theoretical stellar evolution models \citep[e.g.][]{Soderblom2010, Xiang2017b, Mints2017, Sanders2018}. As a consequence, the precision of stellar age dating is dependent on what evolutionary phase the star is in. In other words, different methods are desired for robust age estimation of stars in different evolutionary phases. Currently, stellar ages for large samples of stars with $\lesssim$\SI{10}{\percent} precision are only available for main-sequence turnoff (MSTO) and subgiant stars \citep{Bonaca2020, Xiang2022}. These stars' ages are inferred from stellar isochrones by locating their position in the Hertzsprung-Russell (HR) diagram and utilizing precise stellar atmospheric parameters and luminosity delivered by large surveys. However, MSTO and subgiant stars constitute only a few percent of the stellar population, while most of the low-mass fossil stars are either cool dwarfs or red giants, both of which are difficult to date. 

For red giant stars, asteroseismology provides an effective method of precision age dating \citep[e.g.][]{WuYQ2018, Pinsonneault2018, Silva_Aguirre2018, Bellinger2020}. Asteroseismic analysis of high-precision light curves from the {\it Kepler} mission \citep{Borucki2010} have provided global asteroseismic parameters for 16,000 red giant stars \citep{Yu2018, LiTanda2022} from which precise stellar masses can be inferred. These asteroseismic masses, together with spectroscopic stellar parameters, allow age dating for red giant stars up to a precision of \SI{10}{\percent}. The age determination via asteroseismic modeling can also apply to more massive stars, such as Delta Scuti, roAp stars, Gamma Dor, Slowly Pulsating B-stars, and sdB stars \citep[e.g.][]{Lenz2008, Mombarg2019, Bedding2020}.  

The {\it Kepler} asteroseismic sample for stellar ages is small for Galactic archaeology, as only $\lesssim1000$ stars have age estimates precise to $\sim10$\% \citep{WuYQ2017, Serenelli2017, Silva_Aguirre2017, 2020MNRAS.495.3431L}. Nonetheless, it has served as a crucial benchmark data set for inferring ages of hundreds of thousands of red giant stars from large spectroscopic surveys, such as APOGEE \citep{Martig2016, Ness2016} and LAMOST \citep{Ho2017, WuYQ2019, Huang2020}. This opened a door for Galactic archaeology using red giant stars, which are crucial tracers for probing a large volume of our Galaxy. However,  a serious and in some ways fatal shortcoming of the Kepler sample is that the current asteroseismic sample is restricted to be metal-rich disk stars. The sample size for metal-poor stars that are desirable for archaeology of the ancient Milky Way is limited to less than a hundred \citep{Matsuno2021, Montalban2021}.

The ages of low-mass main-sequence dwarfs (e.g. late G to M dwarfs), which constitute the majority of the Galaxy’s stellar population, can only be poorly constrained with isochrones using their atmospheric parameters as the temperature and luminosity change very slowly on the main-sequence. Nor can the age be inferred from their masses since they are still undergoing the main-sequence evolution process. An effective way to determine the age of a main-sequence dwarf star is gyrochronology: making use of the fact that stars spin down over time due to ``magnetic breaking" \citep{Skumanich1972, Barnes2007, Barnes2010, Barnes2016, Mamajek2008}. The high-precision light curves provided by the {\it Kepler} satellite survey have yielded measurements of the rotation periods to a few percent precision for tens of thousands of dwarf stars \citep{McQuillan2014, Santos2019, Santos2021, Reinhold2020, Gordon2021}, based on which rotation chronology can be used to determine their ages \citep{Meibom2015, Lu2021}. 
Nonetheless, due to the complexity of magnetic fields in stars, gyrochronology relies heavily on empirical calibration. The existing gyrochronology relation is calibrated with sparse sources in a rather restricted parameter space, and there is still a lack of gyrochronology calibration with metallicity. In particular, gyrochronology for metal-poor stars and old stars awaits better constraint and understanding \citep{Garcia2014, Angus2015, van_Saders2016, Metcalfe2019, Amard2020, Hall2021}. Improved calibration of gyrochronology is desired to unlock the potential of the data precision.\\

\noindent \textbf{ET perspective} \\
With a 5 times larger field of view and $\sim0.5$\,mag deeper photometric limiting magnitude than the {\it Kepler} mission, ET offers an opportunity for obtaining precision stellar ages of large samples of stars across the full HR diagram. Asteroseismic analysis with ET light curves can deliver stellar age determination of better than 10 percent for evolved stars, such as subgiant and red giants, while gyrochronology studies with ET photometry will enable precise age determination for main-sequence dwarfs.  

ET plans for Galactic archaeology were made taking into account the needs for high-quality benchmark data set for stellar age determination, ET's unique strengths and limitations, as well as lessons learned from previous and other ongoing missions, such as {\it Kepler} \citep{Borucki2010} and PLATO \citep{Rauer2014, Miglio2017}. Combining these elements, we plan to conduct 
\begin{itemize}
    \item A complete survey of $\gtrsim$6000 metal-poor stars (${\rm [Fe/H]}<-0.8$) down to 15\,mag in the ET field that will yield high-precision, short-cadence light curves that enable age determination precise to a few percent with asteroseismology for both giants and main-sequence (turn-off) stars. This will be the first dedicated asteroseismic survey of a large and complete sample of metal-poor stars. The sample size will be larger than the current metal-poor seismic star sample from the {\it Kepler} mission by more than an order of magnitude. This sample will provide an unprecedented foundation for unveiling the earliest phases of Galactic formation, where the precision and accuracy calibration of stellar ages for stars across a wide range of metallicities matter most.
    \item A systematic, well-defined survey of $>50,000$ relatively metal-rich stars (${\rm [Fe/H]}>-0.8$), for which we will determine their ages precisely with asteroseismology. This will provide a gold standard dataset with which to study the Galactic disk evolution history across the past 10 billion years. Due to the limitation of data downloading ability, we are unable to retrieve the light curves for all stars that suitable for seismic age determination, but will select the targets in an optimal way to ensure the sample covers all the major Galactic components reachable by ET, such as the thin disk, the thick disk, and the metal-rich part of the halo.
    \item A well-defined survey of more than 50,000 main-sequence dwarfs for stellar age determination with gyrochronology. With this sample, we will improve the calibration of gyrochronology in an extensive parameter space, as well as obtain precise ages for a large, representative set of dwarf stars for Galactic archaeology. 
\end{itemize}
 A detailed introduction of the ET Galactic archaeology science goals and the survey targeting strategy will be presented in a dedicated paper, and we note that the numbers presented here may be updated.  \\

\noindent \textbf{ET science cases of Galactic archaeology} \\
An obvious value of the ET sample to the community is that ET will provide key reference data sets for inferring and calibrating stellar ages as well as other stellar parameters from larger surveys, such as large spectroscopic surveys as mentioned above. In particular, the ET metal-poor star sample is expected to be a unique benchmark sample for Galactic archaeology. 

 With state-of-the-art stellar age dating for large samples of stars from ET and its synergy with large spectroscopic surveys, the community can tackle a broad range of key questions from Galactic archaeology, such as
\begin{itemize}
    \item When were the first stars in our Galaxy formed?
    \item When and how were the Galactic structural components, especially the halo and the disk(s), formed? 
    \item What was the chemical enrichment and star-formation history of our Galaxy, particularly at the earliest epoch?
    \item What role has stellar migration played for the Milky Way's disk secular evolution?
\end{itemize}

Below we present a few case studies as examples of the cutting-edge science that ET data can generate.\\

\noindent {\bf The first Galactic stars}\\
It is generally believed that the first stars in the universe were born soon \citep[100-200\,Myr; e.g.][]{Bromm2004} after the Big Bang, which took place about 13.8 billion years ago \citep{Planck2016}. It is still unclear when the very first stars in the Milky Way were formed, though it is a crucial aspect of fully understanding the assembly history of our Galaxy. It is already difficult to determine stellar ages accurately and precisely; to obtain ages for all stars in the Milky Way poses a real challenge. The first challenge arises because stellar age determination is model-dependent, and it is hard to reduce the systematic uncertainty in the age determination with current models to better than a few percent ($\gtrsim0.5$\,Gyr for the old stars) \citep[e.g.][]{Soderblom2010, Dotter2017, Xiang2022}. The second challenge arises because it is almost impossible to obtain high-quality stellar parameters that are needed for the age determination for all stars in our Galaxy. As mentioned above, it is only very recently that age estimates for large samples of stars, mostly in the disk and halo, became available. 

Globular clusters (GCs) are among the oldest population in our Galaxy, and they exhibit ages from about 10\,Gyr to $\sim13.5$\,Gyr, mainly attributed to different origins \citep[e.g.][]{Forbes2010, VandenBerg2013, Massari2019, Valcin2020}. Individual stars of age older than 13\,Gyr have also been reported \citep[e.g.][]{VandenBerg2014, Creevey2015, Schlaufman2018}. In particular, from a sample of a quarter of a million subgiant stars whose ages are determined to 7.5\%  precision on average, \citet{Xiang2022} found clear evidence that the Galactic thick disk at its relatively metal-poor part (${\rm [Fe/H]}\simeq-1$) formed 13\,Gyr ago --- even older than the accreted inner halo stars. Their data further show that on the metal-poor (${\rm [Fe/H]}<-1$) side, a significant fraction of the stars can be even older than 13\,Gyr (see also \cite{Conroy2022}). In summary, the evidence suggests that the first stars in our Galaxy must be older than 13\,Gyr, and they probably belong to a metal-poor tail of the early disk or in-situ halo that formed from the splashes of the early disk due to an early merger event \cite[e.g.][]{Belokurov2020}. Indeed, it is expected that the very first stars in our Galaxy should be metal-poor stars \citep[e.g.][]{Frebel2015, Sestito2019}

We aim to provide asteroseismic ages for $\gtrsim6000$ metal-poor stars in our Galaxy. Their asteroseismic parameters from ET, combining with luminosity from Gaia and atmospheric parameters from spectra, are expected to provide the most accurate age determination ever. Compared with existing asteroseismic age samples such as that from the {\it Kepler} mission, which contain limited numbers of metal-poor stars, ET will undoubtedly enlarge the metal-poor star sample by more than an order of magnitude, shedding important new light on the first stars and thus the early assembly history of our Galaxy. Moreover, high-resolution spectroscopic follow-up observations for this metal-poor star sample will be performed to obtain high-precision full abundance patterns from light through neutron-capture elements, providing key information for uncovering the true origin of the oldest stars in the Milky Way. \\

\noindent {\bf Time-resolved picture of the stellar halo and disk formation}\\
It is believed that the stellar halo was formed via accretion of smaller galaxies that continued in the past 10 billion years \citep[e.g., see a review by][]{Helmi2020}. Most of the halo stars are old and metal-poor \citep[e.g.][]{Frebel2015, Carollo2016, Bonaca2020, Xiang2022}. With data from the Gaia mission and spectroscopic surveys, it was revealed that an ancient single merge event with a galaxy named Gaia-Enceladus (or Gaia-Sausage) \citep{Belokurov2018, Helmi2018} contributed the majority of the halo stars within a few kiloparsec from the Sun \citep{Koppelman2018, Di_Matteo2019, Naidu2020}. Thanks to the availability of high-precision age for large samples of stars, a picture of the early Milky Way formation is becoming clear: although the halo stars are old and metal-poor, they might not be the first-formed major structures of our Galaxy. Instead, the Galactic thick disk occurred already at 13\,Gyr ago, only 800 million years after the Big Bang \citep{Xiang2022}, and it is perhaps the first presented major stellar component of our Galaxy. The merger of the proto-thick disk with Gaia-Enceladus, which leads to the last assembly of the inner halo, has been suggested as happening at around 9--11\,Gyr ago \citep{Gallart2019, Xiang2022}. Besides the assembly of the inner halo, this merger event also leads to some other consequences: first, it caused the formation of a large portion of the thick disk stars \citep{Gallart2019, Xiang2022}; and second, it splashed a significant fraction of the proto-thick disk stars into the halo, causing the so-called in-situ halo population \citep{Bonaca2017, Bonaca2020, Di_Matteo2019, Belokurov2020, Xiang2022}. The formation of the
thick disk was quenched at $\sim$8\,Gyr ago, and the formation of the thin disk started at around a similar epoch \citep[e.g.][]{Haywood2013, Xiang2022}. 

However, the time-resolved picture of our Milky Way formation is reconstructed to only within a few kilo-parsec from the Sun. A main obstacle for painting a more panoramic picture of the Milky Way's formation history is obtaining precise and accurate ages for stars in larger volumes. The current large sample of precision age data is limited to main-sequence turn-off and subgiant stars, for which the ages are estimated using precise luminosity delivered by the Gaia parallax and stellar atmospheric delivered from spectra. ET will deliver precision asteroseismic stellar ages and age calibration for luminous stars covering the full age and metallicity space, which will provide an independent view of the formation of the halo and disk(s). In particular, precision asteroseismic ages for a large sample of red giant stars from ET will serve as a golden benchmark for inferring ages of red giants from spectroscopic surveys with a much fainter magnitude, thus probing a larger volume of our Galaxy. \\

\noindent{\bf Stellar age-metallicity relation (AMR)} \\
Photospheric chemical abundances of a low-mass star provide an archaeological fossil record for the chemistry of its birth environment. The distribution of chemical abundances as a function of age thus allows us to reconstruct the Galactic chemical enrichment history. The stellar AMR has been extensively studied but was mostly restricted to a sparse sample of solar-neighbourhood stars \citep[e.g.][]{Edvardsson1993, Feltzing2001, Haywood2013, Bergemann2014, Nissen2020}. It is only recently that the availability of ages and metallicity from large surveys significantly enlarged the size and volume coverage of the sample, allowed us to draw a more complete stellar age-metallicity map. This leads to clear evidence that the Galactic stars are distributed in a few distinguished, multi-sequence features in the age and metallicity space \citep{Nissen2020, Sahlholdt2022, Xiang2022}. The most prominent features include an old sequence of halo stars on the metal-poor side (${\rm [Fe/H]}\lesssim-1$), an old age-metallicity sequence of disk stars that have a higher metallicity than the halo stars (${\rm [Fe/H]}\gtrsim-1$), and a young sequence of disk stars \citep{Xiang2022}. Metallicity of the halo stars quickly increased from ${\rm [Fe/H]}\lesssim-2.5$ to ${\rm [Fe/H]}\simeq-1$ in the first Gyr of the universe, and the old (thick) disk stars enriched their metallicity by about 30 times, from ${\rm [Fe/H]}\simeq-1$ to ${\rm [Fe/H]}\simeq0.5$, in the first 5\,Gyr of the universe \citep{Xiang2022, Conroy2022}. The age-metallicity distribution of the younger disk stars is, however, more complicated than simply following a single sequence. The distribution exhibits over-densities and unusual subsequences \citep[e.g.][]{Feuillet2018, Feuillet2019, Sahlholdt2022, Xiang2022} which are consequences of both radial migration and merger events \citep[e.g.][]{Lu2022}. Beyond the overall metallicity, the relation between age and individual elemental abundances has been also explored with using data sets of large spectroscopic surveys \citep{Ness2019, Lin2020, Sharma2022}. It was suggested that given the age and metallicity, the abundances of most elements have only a small scatter, of 0.02--0.03\,dex \citep{Ness2019, Sharma2022}. However, whether the conclusions of these studies hold in a larger volume of our Galaxy will depend on more advanced data sets.\\

\noindent {\bf Age-dependent disk stellar metallicity gradient}\\
The Milky Way disk, like the disks of many other spiral galaxies, exhibits a negative radial abundance gradient in its interstellar medium \citep{Shaver1983, Henry1999, Sanchez2014}. This is the result of a radially varied star-formation history of the disk, specifically, the so-called inside-out disk growth \citep[e.g.][]{Matteucci1989, Hou2000, Chiappini2001, Grisoni2018, Toyouchi2018, Molla2019}. The abundance of interstellar medium, however, reflects only a present snapshot of the disk's chemical enrichment history. The stellar metallicity gradient as a function of age provides an approximate constraint of the temporal variation of the abundance gradient, which is particularly true if the stellar migration effect is properly considered \citep[e.g.][]{Minchev2018}. It thus serves as a crucial tool to unravel the star-formation and chemical enrichment history. The stellar metallicity as a function of age has been determined using different data sets \citep{Nordstrom2004, Casagrande2011, Xiang2015, Anders2017, WangC2019, Vickers2021}. A strong temporal evolution of the metallicity gradient is found, as the oldest (thick) disk stars exhibit an essentially flat radial gradient, while the younger (thin) disk stars show negative gradients that vary with both age and height from the disk mid-plane \citep{Xiang2015, WangC2019}. The stellar metallicity gradient in the vertical direction of the disk is also found to vary with age \citep{Xiang2015, Ciuca2018, WangC2019}. While these results provided important constraints and insights to the disk formation and evolution history, the current stellar age sample is still quite limited in the volume coverage. In particular, for old stars, even the largest stellar sample covers only $\sim2$\,kpc around the Sun \citep{Xiang2015, WangC2019, Vickers2021}, and the results thus can be biased.\\      

\noindent {\bf Galactic star formation history}\\
Prior to the era of large stellar samples with well-defined (albeit often complex) selection functions, star formation history was mostly estimated by chemical evolution models \citep[e.g.][]{Chiappini1997, Snaith2014, Snaith2015, Maoz2017, Grisoni2018, Toyouchi2018, Spitoni2019, Lian2020a, Lian2020b,Spitoni2021, Matteucci2021}, using abundance ratios as ``chemical clocks". While in many cases this approach has worked well for the solar neighbourhood by fine-tuning the model parameters, mapping the stellar age distribution may provide a more straightforward measure of the star formation history; this has become reality recently thanks to large surveys \citep[e.g.][]{Xiang2018, Gallart2019, Ruiz-Lara2020}. However, to precisely reveal the Galactic star-formation history, especially for the early epoch, requires very precise age determination and accurate characterization of the selection function. 

Accurate and precise stellar ages from ET will give us an independent view of the chemical enrichment and star formation history of the disk and halo. Furthermore, current studies on the above topics are restricted to small volumes --- within a few ($\lesssim$\,3kpc) kiloparsec from the Sun. Extending our knowledge to a larger part of the Milky Way will be an important step forward. By taking advantage of synergy between ET and large surveys (for instance, LAMOST, APOGEE and other coming spectroscopic surveys), accurate stellar ages for large samples of luminous red giant stars can be delivered to a much deeper magnitude and thus larger volume. \\

\noindent {\bf Stellar migration and disk secular evolution}\\
Stellar migration may significantly alter the position of stars from their birth radii in the Galactic disk \citep[e.g.][]{Sellwood2002, Roskar2008, Minchev2011, Minchev2013}. It has been invoked in solving several problems, such as explaining the observed spread in stellar abundance patterns \citep[e.g.][]{Nordstrom2004, Hayden2015, Spitoni2015, Frankel2018, Frankel2019}, and to explain the [$\alpha$/Fe] bimodality as well as the existence of the thick disk itself \citep[e.g.][]{Schonrich2009, Sharma2021}.
At the solar neighbourhood, stars with super-rich metallicity have been found, and as they have much higher metallicity than the current interstellar medium, they are often referred to as stars that have migrated from the innermost disk \citep[e.g.][]{Grenon1989, Boeche2013, Kordopatis2015, Anders2017, ChenY2019, WangC2019, WuYQ2021}. 
In a more general context, it has been suggested that stellar migrators can be identified from their metallicity --- that is, by comparing their metallicity with what one should expect according to their age and position, given that the interstellar medium forming the (thin) disk stars exhibits a radial metallicity gradient \citep[e.g.][]{Minchev2018, Frankel2018, Frankel2019}. In this way, one can identify stars migrated either from the inner disk or from the outer disk. It has been suggested that all stars born at the outer disk has experienced strong inward migration, possibly triggered by merger events \citep{WuYQ2021}. The stellar kinematic properties as a function of age also provides insights on the stellar migration, as well as the disk heating in general \citep[e.g.][]{Minchev2014, YuJC2018, Ting2019b, Bird2021, WuYQ2021}.   

However, the role stellar migration has played in the disk's secular evolution still awaits better understanding. Models or scenarios without stellar migration can reproduce many major aspects of the observed properties \citep[e.g.][]{Minchev2013, Minchev2014, Xiang2015, Spitoni2015, Matteucci2021}. Recently, it has been also suggested that merger events, such as those with the Sagittarius dwarf galaxy, can bring metal-poor gas to the disk at the solar neighbourhood and trigger the formation of stars with lower metallicity than expected when no merger has happened \citep[e.g.][]{Lu2022}. Attaining a more comprehensive understanding of the impact of radial migration is still underway. Stellar age from ET will facilitate these kinds of studies, as the age distribution is critical in characterizing the migration. \\

\noindent \textbf{Synergy with other Milky Way surveys} \\
ET Galactic archaeology will provide great synergy with other Galactic surveys, such as the LAMOST, APOGEE, Gaia, and upcoming surveys that will target the ET field. These surveys will provide the stellar parameters required for stellar age determination of ET stars. On the other hand, the ET stellar age sample will provide a golden reference for stellar age determination of the extensive large-scale stellar surveys. This is similar to the case of {\it Kepler}, which has greatly promoted the Galactic archaelogy studies via synergy with large spectroscopic surveys, such as APOGEE \citep{Pinsonneault2014, Pinsonneault2018} and LAMOST \citep{DeCat2015, Fu2020}, as these synergies have enabled the inference of stellar ages for hundreds of thousands of red giant star  \citep[e.g.][]{Ness2016, Ho2017, WuYQ2019, Huang2020}.

\subsubsection{Binary and Multiple Stars}
{\bf Authors:}
\newline
Jinzhong Liu$^1$, Yu Zhang$^1$, Liyun Zhang$^2$, Kareem El-Badry$^3$, Jian Ge$^4$ \& Steve B. Howell$^5$  \\
{1. \it Xinjiang Astronomical Observatory, Chinese Academy of Sciences, 150 Science 1-Street, Urumqi, Xinjiang 830011, People's Republic of China} \\
{2. \it College of Physics, Guizhou University, Guiyang 550025, China} \\
{3. \it Center for Astrophysics $|$ Harvard \& Smithsonian, 60 Garden Street, Cambridge, MA 02138, USA} \\
{4. \it Shanghai Astronomical Observatory, Chinese Academy of Sciences, 80 Nandan Road, Shanghai, China} \\
{5. \it NASA Ames Research Center, Moffett Field, CA 94035, US}\\

Binary stars are as common as singles in our universe. Previous studies show that at least \SI{30}{\percent} of all stellar systems are binaries \citep[e.g.,][]{Rag2010,Lee2020}, and binarity is ubiquitous among massive stars \citep{Sana12}. Binary stars are involved in a wide variety of study areas in astrophysics and they are the primary targets for studies of the fundamental properties of stars.

\begin{figure}[htbp]
\centering
\includegraphics[scale=.18,angle=90]{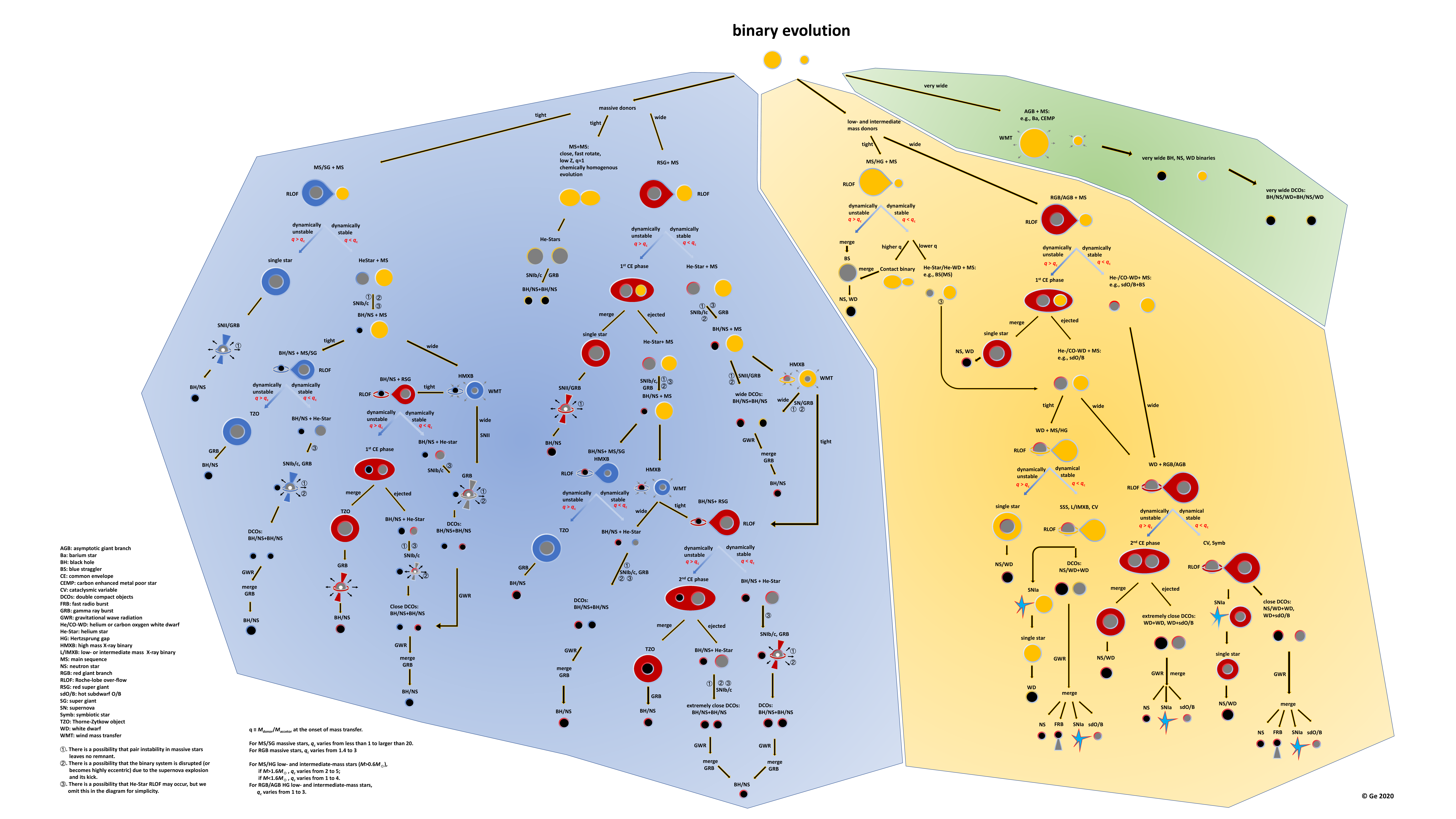}
\caption{\small Binary evolution tree adopted from \cite{Han2020}.
		\label{fig:binary}}. 
\end{figure}

Interaction in a binary system plays an important role in the evolution of binary populations. Physical processes, such as wind accretion, orbital changes, tidal evolution, gravitational wave (GW) radiation, magnetic braking, supernova kicks, Roche lobe overflow, common envelopes, coalescence, and collisions, occur in interacting binaries. Some of the most exotic objects, such as magnetic cataclysmic binaries, hot subdwarfs, symbiotic binaries, binary supernova progenitors, X-ray binaries, gamma ray bursts, millisecond pulsars and Thorne-$\dot{Z}$ytkwo objects, are formed during the binary evolution. Figure \ref{fig:binary} shows the binary evolution processes and the formation of binary-related objects reported in a recent review paper \cite{Han2020}.

ET will be an efficient machine for both the discovery and precision characterization of large samples of binaries through photometric variability. The {\it Kepler} mission precipitated a renaissance in photometric studies of binaries \citep[e.g.][]{Prsa2011, Bloemen2011}. The torch has been carried forward by {\it TESS}, which monitors nearly the whole sky but has relatively short time baselines and performs well only for bright stars. One of the main science objectives of ET in binary studies is to detect faint binary objects such as AM CVn stars, Ultra-compact X-ray binaries, and white dwarfs and provide new insights on their formation and evolution. At the same time, ET with its wide-field ultra-high-precision photometric capabilities is expected to reveal:
\begin{itemize}
 \item[$\bullet$] \textbf{Gravitational wave sources and their progenitors:} 
 Short-period ($\lesssim 1\,\rm hr$) binaries in the Milky Way and nearby galaxies will be the most common gravitational wave sources for upcoming millihertz gravitational wave observatories such as LISA, Taiji, and Tianqin \citep[e.g.][]{Liu2009, Liu2014, Krem2017, Kupfer2018}. Time-domain photometric surveys have already proved very successful in finding and characterizing such binaries \citep{Burdge2020}. Joint modeling of these binaries using optical data from missions such as ET and gravitational waves offers a powerful tool for constraining their properties. For example, joint measurements of orbital period decay from long time baseline photometric surveys and gravitational waves can separate the contribution of gravitational waves and tides, which has thus far been impossible to do \citep[e.g.][]{Burdge2019}. 
 ET will obtain time-series data for binaries in the fixed sky region around the {\it Kepler}'s original field, which is close to the galactic plane, and will likely provide important constraints for GW sources in the galaxy. The ET photometric detection of an electromagnetic counterpart of a compact binary would provide a precise location of a GW source for space-based GW detectors to monitor. Compared to existing space- and ground-based surveys, the exquisite precision of ET will make it possible to detect compact binaries containing only very slightly distorted sources by ellipsoidal variation, such as compact binaries containing two massive white dwarfs.  
 
 ET will also be useful for modeling massive star binaries that are progenitors of binaries containing black holes and neutron stars and are currently being discovered by the LIGO/Virgo collaboration. Detailed modeling of ET light curves for local massive-star binaries will yield population statistics that serve as critical inputs for population synthesis models of these gravitational wave sources.
 \item[$\bullet$] \textbf{Geometry and evolution characteristics:} Binary stars that are close enough ($P_{\rm orb}<10$ yr) transfer mass from one star to another, changing the structures of both stars and their subsequent evolution. Close binary star systems with orbital periods $< 10$ d most likely have circular orbits because of the tidal circularization effect. With the high precision photometry, ET will be able to measure the following important properties: orbital period rate of change, light curve variations and types, and the mechanisms that trigger the evolution of the orbital period. For instance,  light curves with complete phase coverage for eclipsing binary stars can be used to study magnetic activities through measurements of starspots and flares.
 \item[$\bullet$] \textbf{Stellar parameters:} Eclipsing binaries play an important role in measuring basic stellar parameters such as mass and radius. Light curve analysis of a low-mass eclipsing binary is one of the main methods for deriving stellar parameters. Here a low mass star is defined as a star with stellar mass  $\leqslant$ 0.8 M$_\odot$ \citep{Rib2006, Mor2008}. Low mass eclipsing binaries are very important laboratories for measuring stellar physical parameters. For many low mass eclipsing binary stars, we have not yet obtained their orbital parameters or studied their magnetic activities \citep{Zly2014}. The high-precision time-series photometric data released by the ET mission will provide an excellent opportunity for studying these stars and obtaining their physical parameters. We will study low mass eclipsing binary stars using light curves from ET to determine their orbital parameters, and obtain starspot parameters and time evolution. It is also possible to detect flare events and determine their flare parameters. These can help determine the stellar physical parameters and evolution of magnetic activity, and reveal the relationship between the starspots and flares \citep{Pi2019}).
 \item[$\bullet$] \textbf{Statistics:} The statistics of binary stars from ET observations can help the development of binary evolution theory. Large-scale surveys, such as Gaia, {\it Kepler}, and {\it TESS} can provide a large binary sample for studying their stellar properties. The unique contribution from ET is that it will provide a large sample of binaries over a wide orbital period range from hours to hundreds of days. Combined with data from spectral surveys such as LAMOST or CSST,  these missions can measure stellar mass, radius, atmosphere parameters (log g, T$_{eff}$ and metallicity), orbital (or spin) periods, class types and age. In the absence of spectroscopic data, ET's ultra-high-precision light curves can still be used to estimate masses of eclipsing binaries by using a new method developed by \citep{Gaz2009,Lu2020,Li2021}. Furthermore, the derived relations, such as mass-luminosity, period-mass, temperature-period, and mass-orbital angular momentum, can also be studied. 

\item[$\bullet$] \textbf{Multiple Star Systems:} 
While it is well established that about one-half of sun-like stars are binaries, we also know that $\sim$10\% of these are higher order systems, being triple or quadruple in nature. For massive stars, single stars are rare. Multiple star systems, in their way acting as mega-solar systems, can inform us about formation and evolution of stellar systems and thus planetary bodies as well. Resonances between mass ratios, orbital periods, and system inclinations mimic what we see in planetary systems but on a grander scale. The formation of such systems in terms of proto-disk fragmentation or aggregation via gravitational capture are still very much in debate. Using space-based light curves, co-planer multiple stars have been discovered by {\it Kepler}, {\it K2}, and {\it TESS}, each mission providing more than the last due to new software search techniques and better knowledge of what to look for \citep{Kostov2022ApJS..259...66K}. Even an eclipsing sextuple star system has been discovered \citep{Powell2021AJ....161..162P}. Such complex star systems are ideal for providing very precise stellar parameters for the component sizes, masses, temperatures, and orbital inclinations. Such mutual co-planer systems are quite rare, yet ET will eclipse all past missions combined in the discovery of multiple star systems, finding many hundreds of such systems useful for detailed follow-up studies. The detection of such large samples of multiple star systems will provide high statistical value and robust observational evidence for input to theoretical models of multiple star formation and evolution. 
\end{itemize}

\subsubsection{Black Holes with Visible Companions} 
{\bf Authors:}
\newline
Rongfeng Shen$^1$, Pak-Hin Thomas Tam$^1$, Zhecheng Hu$^1$, Yanlv Yang$^1$ \& Kareem El-Badry$^2$\\
{1. \it School of Physics and Astronomy, Sun Yat-Sen University, Zhuhai, China}  \\
{2. \it  Center for Astrophysics $|$ Harvard \& Smithsonian, 60 Garden Street, Cambridge, MA 02138, USA} \\

The number and mass distribution of black holes in the Milky Way, as well as the proportion of black holes in binary stars, is of fundamental importance for a full understanding of the formation process of black holes, supernova explosion mechanisms, and massive stellar evolution. It is estimated that there are hundreds of millions of black holes in the Milky Way \citep{Shapiro83,Timmes96}, which vastly outnumbers known population of black holes (less than 100) that have been detected mainly in interacting binaries~\citep{Corral-Santana16,Tetarenko16}. According to recent population synthesis studies \cite[e.g.,][]{Wiktorowicz19,Olejak20}, the number of black holes in binaries is about 10$^6$, and many of these reside in systems involving an invisible black hole and a main-sequence star in wider orbits than those detected. However, uncertainties in these estimates are enormous. Observational data to constrain the formation models for black hole binaries are lacking.

Due to the difficulty of detecting most black holes, our knowledge of black hole demography is limited, missing especially wide-orbit binaries containing a black hole. On the other hand, mass measurements of black holes come almost exclusively from accreting binary systems selected from radio, X-ray, and gamma-ray surveys which only constitute a small fraction of the overall population. Observations of massive binary systems~\citep{Sana12} show that binary interaction can greatly intervene with the evolution of stars, particularly for mass-transferring systems. This means that the mass of black hole progenitors (and the black holes) in mass-transferring systems is largely modified over its lifetime, while the black hole mass in wide-orbit binaries is less affected by binary interactions, preserving a better picture of the initial masses of massive stars.

The incomplete knowledge of the mass distribution of black holes includes the absence of mass-gap black holes. In theory, precursor stars that can produce 3 to 5 solar mass black holes should be abundant, but observations show that this type of black hole accounts for only a very small portion of known black hole binaries. Knowing how many mass-gap black holes exist in our Milky Way galaxy is an important factor in understanding star evolution and core-collapse supernova mechanism. If more mass-gap black holes can be found, it will help to resolve this problem. 

The observational search for dormant black holes in binaries has been plagued by false positives: such systems are rare, and their imposters are abundant. Of the half-dozen candidates identified in the last few years \citep[e.g.][]{Thompson19, Liu_LB1, Rivinius2020, Lennon2021, Saracino2022, Jayasinghe21}, none are unambiguous, and most have been definitely shown to be astrophysical false-positives containing two luminous stars \citep[e.g.][]{Shenar2020, Bodensteiner2020, El-Badry2022, El-Badry2022unicorn}. It is therefore very important to discover more black holes in detached and wide-orbit systems. ET opens a window on new methods to search for dormant black holes, which will be complementary to existing methods and have different systematic uncertainties.  

\begin{figure}[htbp]
	\centering
	{\includegraphics[scale=.26]{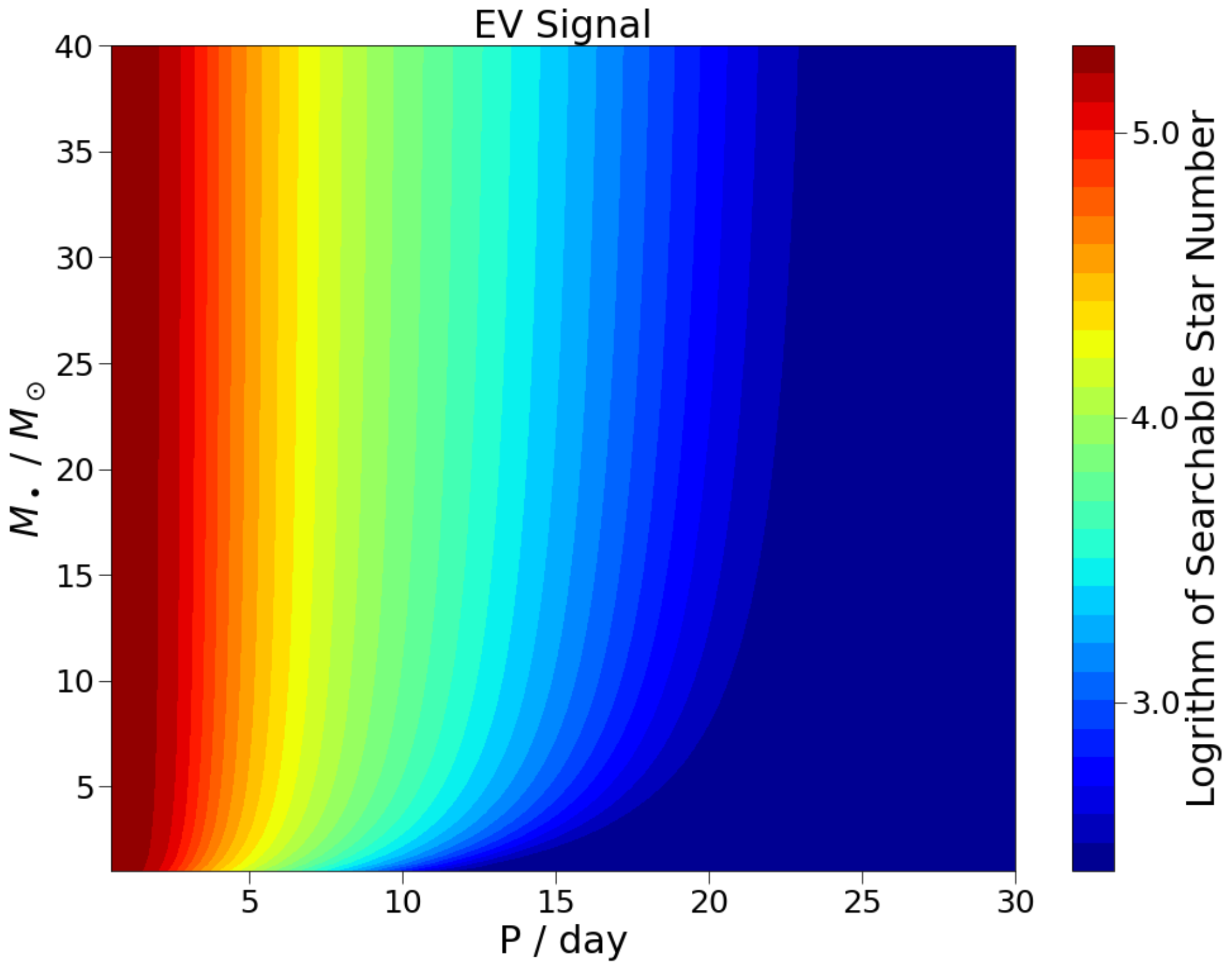}
	\includegraphics[scale=.26]{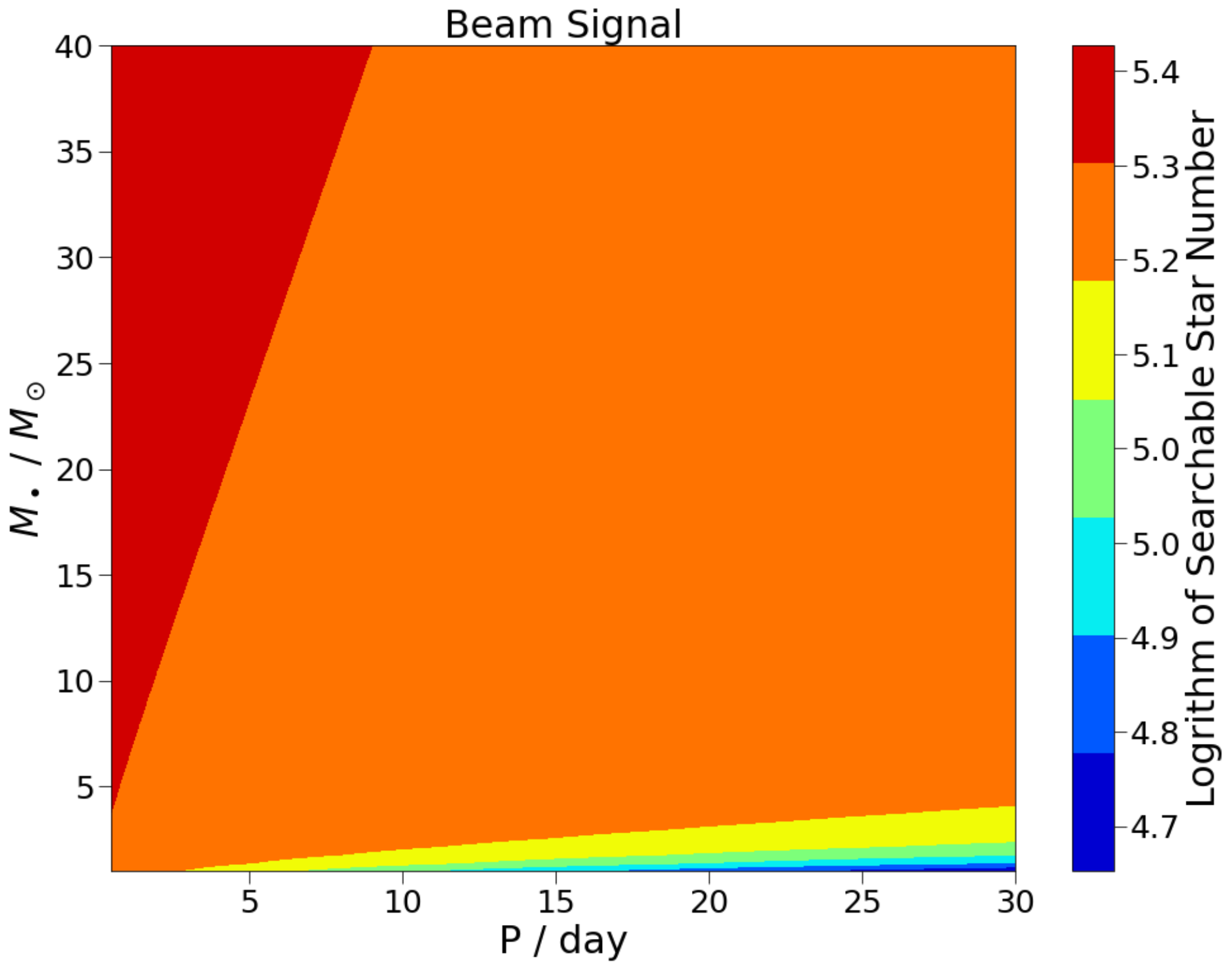}
	\includegraphics[scale=.26]{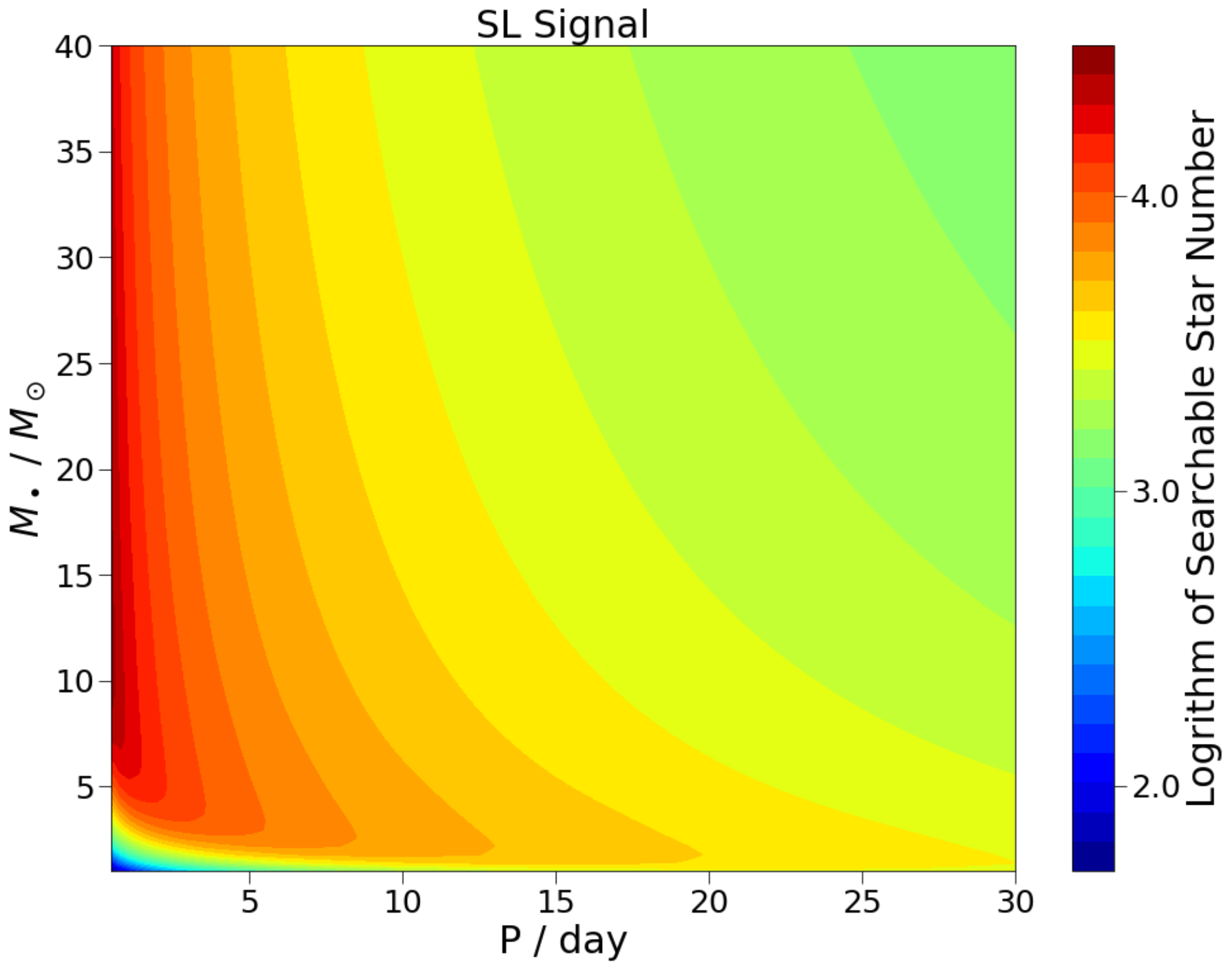}}
	{ \caption{\small The quantity of searchable stars in the {\it Kepler} Input Catalog for the three kinds of signals against black hole mass and orbital period, assuming that every star is accompanied by a black hole. The color bar shows the logarithmic quantity of searchable stars.}	\label{fig:signal_bh}}. 
\end{figure}

 Although black holes in wide binaries are basically quiescent in X-rays, one can rely on the orbital modulation of a companion's brightness to detect them. The three types of signals are ellipsoidal variation (EV), Doppler beaming, and self-lensing (SL). \textit{Ellipsoidal variation signals} are mainly caused by tidal distortions of companion stars due to the gravity of their companion black hole. \textit{Doppler beaming} contains three observable components: Doppler shift, light aberration and time dilation caused by the orbital motion of the companion star. A \textit{self-lensing signal} is seen as a sudden increase in flux when the black hole transits in front of the companion star, due to gravitational lensing. These three types of photometric signals are summarized by \citet{Masuda2019_Prospects}. Figure~\ref{fig:signal_bh} shows the number of searchable stars in the {\it Kepler} Input Catalog (KIC) for the three kinds of signal against black hole mass and orbital period assuming that every star is accompanied by a black hole. Although the orbital inclination angle of any detectable self-lensing signal is restricted to a small range around 90 degrees, i.e., the edge-on case, the signal amplitude remains detectable in wide-orbit and close-orbit systems. All three effects have been detected in the {\it Kepler} light curves of binaries containing white dwarfs and neutron stars \citep[e.g.][]{Bloemen2011, Kawahara2018}; ET will deliver data of similar quality on a much larger scale, paving the way to observe such signals from black holes.
 
\begin{table}[]
\centering
    \captionsetup{justification=centering}
\caption{An estimate of the number of photometric signals from BHs with visible companions in KIC stars by {\it Kepler}.}
\begin{tabular}{lrrr}
\hline
                                        & Flat IMF & STD IMF & Steep IMF \\
Modeled number of BHs in binaries in KIC & 53     & 20    & 7       \\ \hline
Number of detectable EV signals        & 22     & 8     & 3       \\
Number of detectable beaming signals   & 30       & 12    & 4       \\ 
Number of detectable SL signals        & 3      & 1     & 0      \\ \hline

\end{tabular}
    \label{tab:KIC_BHs}
\end{table}

In recent years, space-based high-precision photometric telescopes including {\it Kepler}/\textit{K2}
and {\it TESS} 
have been used to obtain thousands of high quality light curves over a long period of time. These light curves allow the search of black holes from a huge number of stars in a way that could not be done before. \cite{Masuda2019_Prospects} estimated that around 10 to 100 black holes in binaries can be detected in {\it TESS} data. In light-curve searches for black hole companions with {\it TESS}, the largest challenges have been due to the short time baselines: with only a month-long light curve, it is difficult to distinguish the variability originating from beaming and ellipsoidal variability from slowly-evolving variability due to starspots.  Applying similar methodology for the original {\it Kepler} mission yields the same order of magnitude number of detectable signals from black holes. Table~\ref{tab:KIC_BHs} presents a roughly estimated number of photometric signals in KIC, in which the modeled number of BHs in binaries is based on calculations by~\citet{Wiktorowicz19} for different IMFs~\citep{Kroupa2001_variation}. 

The ET mission will have an improved sensitivity on measuring stellar flux variation over its precedents and can potentially detect many undetected black holes in binaries in its wide field of view. A factor of 10 increase in the number of detectable BHs as shown in Table~\ref{tab:KIC_BHs} might be anticipated for ET by simple scaling. The long observing baseline and the supreme photometric accuracy of ET will facilitate efforts to filter out abundant false-positives. 

\subsection{\bf Time-domain Sciences}

Numerous astronomical objects can be classified as transients. Of interest here are those objects for which mass transfer and mass accretion occur, often in the processes of the birth or the death of a star. Over short timescales, fractions of seconds to hours and days, such phenomena provide the observer with a unique laboratory experiment unable to be performed anywhere on Earth. Objects such as cataclysmic variables, novae, gamma ray bursts, and neutron star mergers are examples of transient events ET will be ideally suited to study. Such well sampled time domain light sampling is only available with a space-based telescope and will provide strong constraints for extreme physics modeling processes such as those of relativity and quantum mechanics. We discuss some primary transient sources of interest below.

\subsubsection{High-Energy Transients}
{\bf Authors:}
\newline
Yuan-Chuan Zou$^1$, Jia-Li Wu$^1$, Wei-Hua Lei$^1$, Jun-Jie Wei$^2$, Xue-Feng Wu$^2$, Tian-Rui Sun$^2$, Fa-Yin Wang$^3$, Bin-Bin Zhang$^3$, Dong Xu$^4$, Yuan-Pei Yang$^5$, Tinggui Wang$^6$, Bing Zhang$^{7,8}$ \& Steve B. Howell$^9$ \\
{1.\it Department of Astronomy, School of Physics, Huazhong University of Science and Technology, Wuhan, China} \\
{2.\it Purple Mountain Observatories,  Chinese Academy of Sciences, Nanjing, China} \\
{3.\it School of Astronomy and Space Science, Nanjing University, Nanjing, China} \\
{4.\it National Astronomical Observatories, Chinese Academy of Sciences, Beijing, China}\\
{5.\it South-Western Institute for Astronomy Research, Yunnan University, Kunming, China}\\
{6.\it Department of Astronomy, University of Science and Technology of China, Hefei, China}\\
{7.\it Nevada Center for Astrophysics, University of Nevada, Las Vegas, NV 89118, USA}\\
{8.\it Department of Physics and Astronomy, University of Nevada, Las Vegas, NV 89118, USA}\\
{9.\it NASA Ames Research Center, Moffett Field, CA 94035, USA}\\

High-energy transients usually emit high energy photons and sometimes multi-messenger (e.g. high energy cosmic rays, neutrinos and gravitational waves) signals in a short timescale. Some of them (e.g. gamma-ray bursts, supernovae, kilonovae, and tidal disruption events) are related to the cataclysmic events, such as the births of black holes or neutron stars, or the deaths of massive stars or of compact binaries. Some others (e.g. AGN flares, X-ray binary outbursts and fast radio bursts) invoke repeated catastrophic activities powered by a central engine such as a black hole or a magnetar.
Studying these high-energy transients not only provides insight on the nature of these events themselves, such as particle acceleration, jet production/propagation, and radiation mechanisms, but also provides useful tests or constraints on fundamental physics, such as general relativity and other theories of gravitation, Lorentz invariance violation, photon rest mass, as well as cosmological evolution of physical constants.

\begin{figure}[htbp]
\centering
\includegraphics[width=0.85\textwidth]{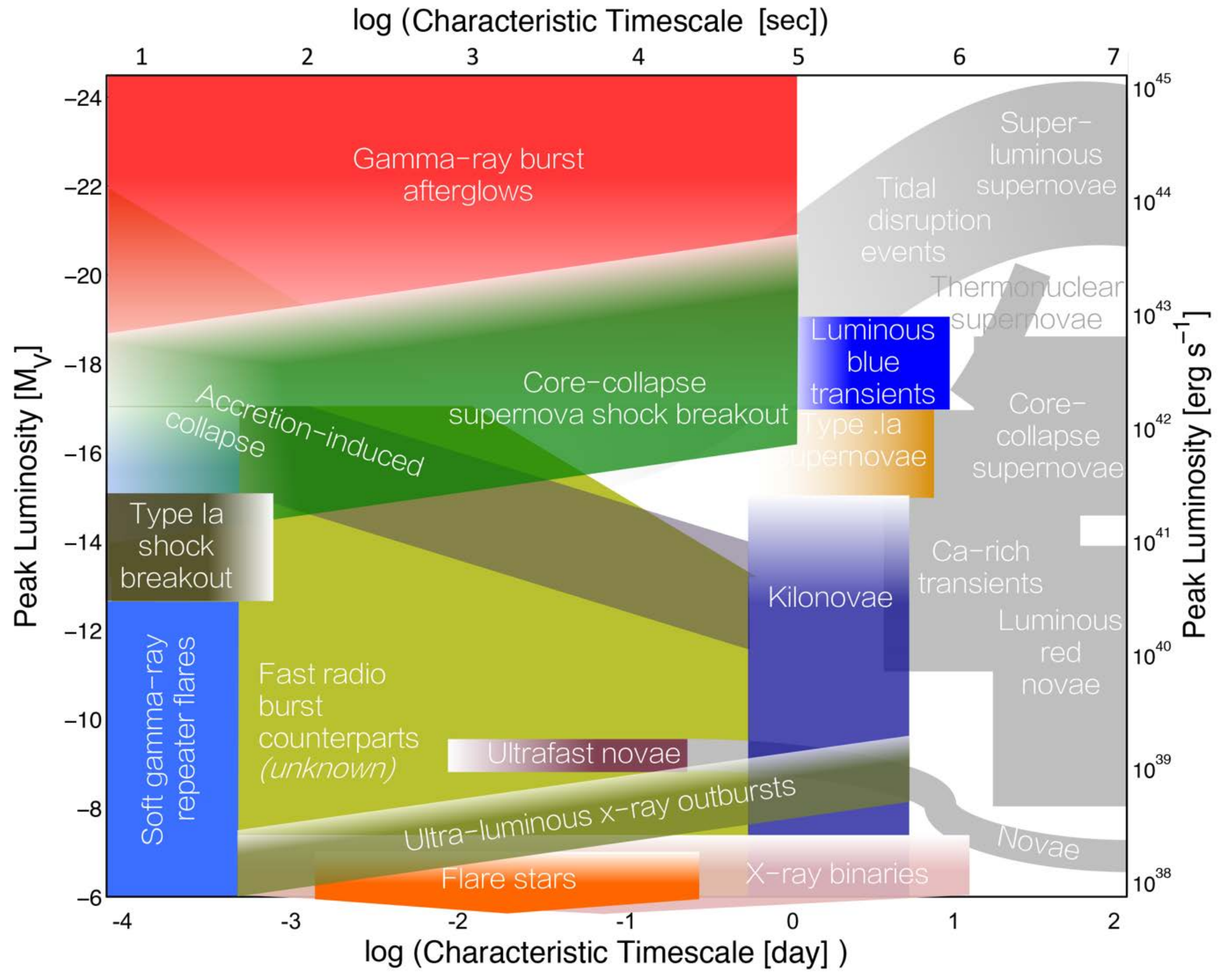}
\caption{\small Characteristic timescales and peak luminosities of transients, adopted from \cite{2021AnABC..93..917B}.
		\label{fig:transients}}. 
\end{figure}

Most high-energy transients are broad-band emitters. Many of them emit optical emission in the ET band. These optical counterparts of high-energy transient sources include, but are not limited to:
\begin{itemize}
    \item prompt optical emission of gamma-ray bursts \citep[e.g., GRB 080319B;][]{2008Natur.455..183R,2009ApJ...691..723B};
    \item optical afterglow emission of both long and short GRBs \citep[e.g., GRB 130427A;][]{2014Sci...343...38V};
    \item orphan optical afterglow of GRBs that is theoretically predicted but has not been observationally confirmed;
    \item engine-powered supernova emission associated or not associated with a detected long GRB;
    \item kilonova emission from a neutron star merger (NS-NS or NS-BH) system associated \citep[e.g., AT 2017gfo/GW170817;][]{2017ApJ...848L..12A} or not associated \citep[e.g., GRB 130603B; ][]{2013Natur.500..547T} with a detected gravitational wave event;
    \item optical emission from tidal disruption events (TDE) by supermassive black holes \citep{2011ApJ...741...73V}; 
    \item optical flaring activities from active galactic nuclei (AGN) at various timescales \citep{2004ApJ...601..692V};
    \item optical variable radiation from X-ray binaries \citep{2011MNRAS.413.1600R};
    \item hitherto undetected optical counterparts of the mysterious fast radio bursts (FRBs) \citep{2007Sci...318..777L}.
\end{itemize}
These optical transients have a wide range of variability timescales and peak luminosities (Figure \ref{fig:transients}), as well as event rate densities and the observed event rates. For example,
GRB optical afterglows last for about tens of days \citep{2013ApJ...774..132W}. However, they decay very quickly. Early observations are crucial to catch these. As another example, there is only one decisively confirmed gravitational wave event accompanied with an electromagnetic counterpart, i.e., GW170817/GRB 170817A/2017gfo \citep{2017PhRvL.119p1101A}. The large directional error boxes of the GW and GRB detectors demand large FOV optical telescopes for follow-up observations. Tidal disruption events happen in the center of galaxies, which last a much longer time up to months. FRBs last for only milliseconds and even down to microseconds, whose optical counterpart may also last for very short time scale \citep{2019ApJ...878...89Y}. This also needs telescopes with a wide FOV for coincident observations. Among the targets in above list, supernovae and fast blue optical transients are likely the dominant contributors in terms of their brightness and detectability. These are discussed separately in the next sub-section. In general, optical band observations can offer unique clues regarding the nature of these high-energy transients. 

Detecting these transients is of great scientific values. For example, the detection or non-detection of optical emission accompanied with the prompt GRB emission can provide unique constraints on the radiation mechanism of GRBs. The detection of more kilonova emission in association with gravitational wave sources will allow much more detailed studies of neutron star merger physics, including r-process nucleosynthesis as well as the possibility of energy injection from the central engine. The early rising phase of the lightcurve of a TDE event carries crucial information about the properties of black holes and the disrupted star. 

An ideal time-domain optical telescope should combine a large FOV, deep limiting magnitude and high cadence. Even though ET is not designed to study transients, its design makes it an ideal transient hunter.
A combination of its wide FOV and large aperture makes ET about 26 times more powerful than {\it Kepler} in detecting transients, allowing it to obtain light curves of detected transients at the densest cadence ever.  

\noindent{The strategies for observing the transients using ET include the following four aspects.}
\begin{itemize}
\item[1.] Follow-up searches if other observatories discover a source in the ET's FOV. From the telescope cache data archives, download all the data in the corresponding direction and process them. The advantage of this strategy is that all the data can be downloaded, which can be used for a deeper search and a more comprehensive analysis. This strategy requires the telescope to store data for two days and trace the source to download the relevant data according to the direction of the triggered source by other telescopes. The required bandwidth is minimal. Only several transients are expected to be observed each day. Each transient may need about 10 days of follow-up observations with images as small as $3\times3$ pixels if the source direction is precisely obtained.

\item[2.] Incremental source selection. When a transient source is discovered in the ET's FOV by other telescopes, if its position is known with high accuracy it is added to the ET's target catalog for further monitoring. A target will be removed from the catalog if it is not detected for a few consecutive days. Assuming that one new source is discovered every day and all the transients are visible for about 30 days on average, one can estimate that only about 30 transient targets are assigned to ET's target catalog at any given time.

\item[3.] Direct source selection. About 1,000 nearby galaxies in the ET's FOV can be assigned to ET's target catalog. Given that transients occur inside galaxies, continuously monitoring these galaxies within the ET's FOV would allow for very early detections of optical variable sources in their host galaxies, such as supernovae, fast blue optical transients, optical counterparts of GRBs, kilonovae, tidal disruption events, and even possible optical counterparts of FRBs. Some Galactic targets within the ET's FOV, such as X-ray binaries, pulsars, and supernova remnants, are also of interest because optical transient emission may also arise from these systems that host a black hole or energetic neutron star engine at the center. 

\item[4.] Integrate daily observations. Sum up all the data for each pixel for one whole day (or each 6 hours for a higher cadence) and download the full-frame data once a day. This ensures that all observations are not wasted. One can use this data for super-deep field searches. Since bright sources occupy too much data storage and since they can be detected with the aforementioned three strategies, they can be ignored for this strategy to allow deep searches for dim sources.
\end{itemize}

With ET's superb observational capabilities including long-term monitoring, large FOV, high cadence, and 24-hour continuous staring observations, one expects to make the following discoveries during the operation time of ET: dozens of cases of very early optical afterglows of GRBs (including a few orphan afterglows) which help with understanding the physics of relativistic jets, their engines, and the interaction with the ambient medium; many optical flaring events from AGNs, which help with the study of the mechanisms of the supermassive black holes, their accretion disks and jets; and likely a small number of optical TDEs, which help with studying the physics of how supermassive BHs tidally disrupt and swallow stars. If lucky enough, ET may pick up an optical counterpart of a GW event, which will provide unprecedented details on the evolution of kilonova emission, unveiling many interesting physics of neutron star mergers; 
a prompt optical counterpart of GRBs analogous with the famous ``naked-eye'' GRB, GRB 080319B, which would provide a rare glance at the physical conditions during the GRB prompt emission phase; a rare type of TDE event such as a white dwarf tidally disrupted by an intermediate-mass black holes, which would provide important evidence of the existence of intermediate-mass black holes and probe the history of BH formation and growth;
and potentially an FRB optical counterpart, which would certainly revolutionize our understanding of these mysterious objects. The estimated numbers of various types of detectable transients are shown in Table \ref{tab-transients}.

\begin{table}[!ht] 
\centering
    \captionsetup{justification=centering}
\caption{Rough estimation of the detection rate of the transients. The estimation are derived based on the observations or the statistics listed in the references.}
    \begin{center}
    \begin{tabular}{|l|l|l|l|l|l|}
    \hline
        Science case name & Magnitude limit & S/N  & Exposure time & No. of targets & Ref.  \\ \hline
        Optical afterglow of GRBs & 25.5 & 3 & 6 hrs & 9/year &  [1] \\ \hline
        Orphan afterglow of GRBs & 20.5 & 5 & 5.5 s & 1/year & [2] \\ \hline
        Optical counterpart of TDEs & 25.5 & 3 & 6 hrs & 0.2/year  &  [3] \\ \hline
        Prompt Optical emission of GRBs & 21 & 3 & 5.5 s & 0.06/year & [4]  \\ \hline
        Optical emission of LIGO GW events & 25.5 & 3 & 6 hrs & 0.01/year & [5]  \\ \hline
        AGN optical counterparts & 25.5 & 3 & 6 hrs & 90 in total  & [6]  \\ \hline
        Optical counterpart of FRBs & 21 & 5 & 5.5 s & Not known & ~ \\ \hline
    \end{tabular}
    \end{center}
    ([1] \citep{2013ApJ...774..132W}, [2] \citep{2007A&A...461..115Z}, [3] \citep{2011ApJ...740L..27L}, [4] \citep{2010ApJ...720.1513K}, [5] \citep{2017PhRvL.119p1101A}, [6] \citep{2018ApJ...857..141S})
    \label{tab-transients}
\end{table}


\subsubsection{Supernovae} 
{\bf Authors:}
\newline
Wen-Xiong Li$^1$, Dan-Feng Xiang$^2$ \& Xiaofeng Wang$^2$\\
{1.\it The School of Physics and Astronomy, Tel Aviv University, Tel Aviv, 69978, Israel}\\
{2.\it Physics Department and Tsinghua Center for Astrophysics, Tsinghua University, Beijing, 100084, China}\\

\paragraph{Progenitor Systems and Explosion Models of Type Ia Supernovae}
Observations of type Ia supernovae (SNe Ia) led to the discovery of an accelerating expansion of the universe \citep{1998AJ....116.1009R,1999ApJ...517..565P}, revolutionizing our view of the structure of the universe. Understanding the nature of the dark energy, which is responsible for the accelerated expansion of the universe, is one of the fundamental questions in astrophysics research and SNe Ia as distance indicators remain an important tool for deciphering this issue.

Further improvement of distance measurements by SNe Ia is limited by systematic uncertainties, including extinction, photometric system differences, and intrinsic uncertainties. To further reduce the intrinsic uncertainties, we need to have a deeper understanding of SNe Ia progenitor systems, but there is still considerable controversy \citep{2012NewAR..56..122W,2013Sci...340..170W,2014ARA&A..52..107M}. Specifically, the companion star of an exploding white dwarf may be a white dwarf star (double degenerate model), main sequence star, helium star, or red giant star (single degenerate model). Different progenitor systems could come from different stellar populations which means the properties of SNe Ia could evolve with redshift.

\begin{figure}[htbp]
\center
\includegraphics[width=0.8\textwidth]{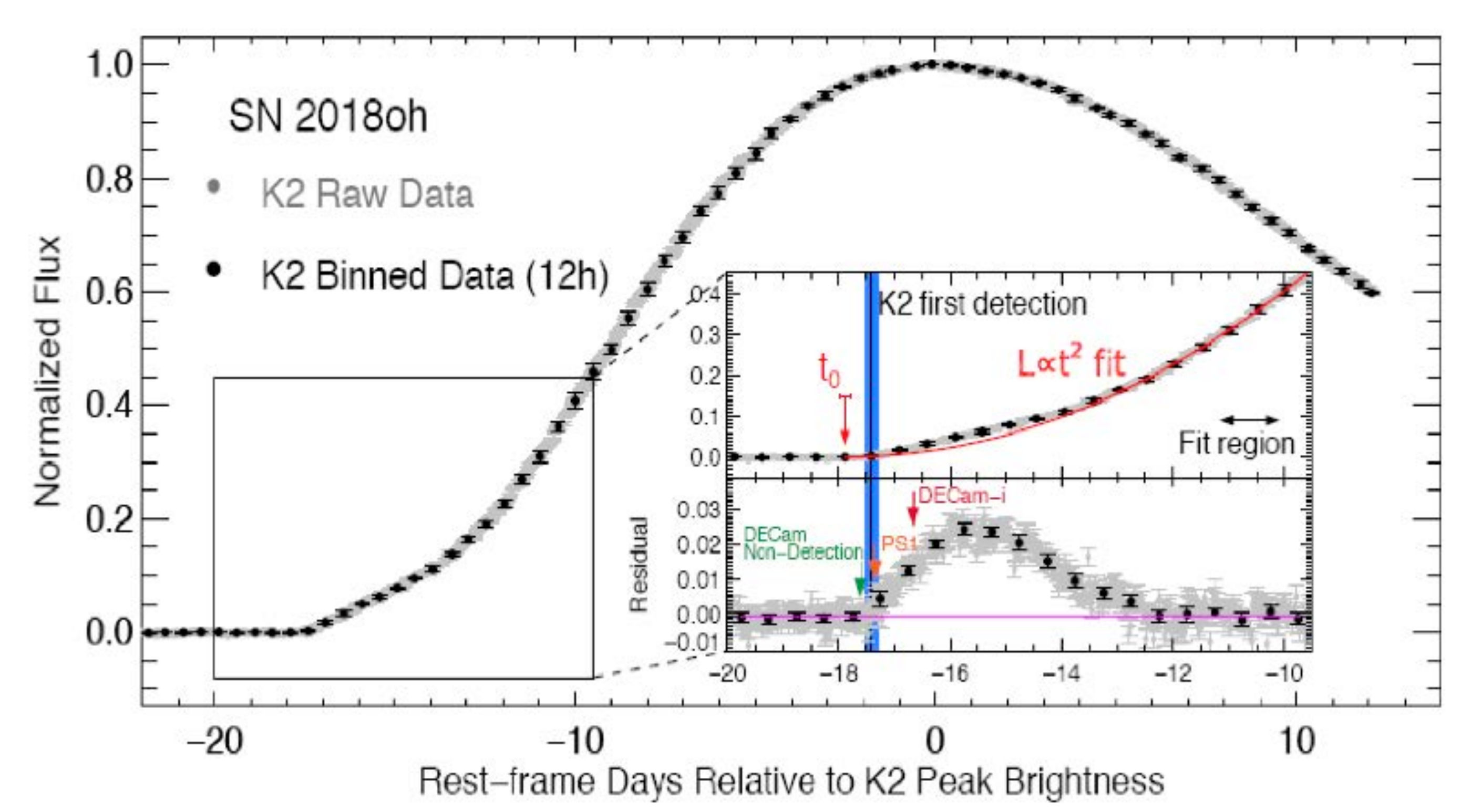}
\vspace{0.2cm}
\caption{The SN 2018oh found in {\it Kepler}'s field has excess flux in the early stage of its light curve \citep{2019ApJ...870...12L,2019ApJ...870L...1D}. The excess flux may be due to supernova ejecta collisions with non-degenerate companion stars, nickel mixing, or detonation of the helium shell. We need early color evolution to distinguish the above models. } \vspace{-0.0cm}
\label{ia}
\end{figure}

In the single degenerate model, the high-speed ejecta produced by a supernova explosion interacts with its companion star or surrounding circumstellar material then emits a shock wave which can generate excess flux within one hour to a few days in the early light curves, depending on the observation angle \citep{2010ApJ...708.1025K}. This excess energy has been observed in some SNe Ia samples and is also considered as evidence in support of single degenerate model (see Figure \ref{ia}) \citep{2015Natur.521..328C,2017ApJ...845L..11H,2018ApJ...852..100M,2019ApJ...870L...1D,2021ApJ...923L...8J}. However, the mixing of synthesized radioactive nickel on the surface of a white dwarf, or the first helium detonation in the sub-Chandrasekhar double-detonation model can also produce this early excess energy \citep{2019ApJ...870...13S,2019ApJ...870...12L,2019ApJ...882...30L,2019ApJ...873...84P}. The temperature evolution in the early color evolution of the explosion can effectively distinguish these models \citep{2019ApJ...873...84P}. {\it TESS} has observed dozens of SNe Ia, placing upper limits on the radii of companion stars \citep{2021ApJ...908...51F}. ET with its high cadence observations and bluer observational wavelength band than {\it TESS} is expected to observe more SNe Ia during its earliest phase to help constrain the radii of companion stars and also different models. 

\paragraph{Progenitors of core-collapse supernovae and evolution theory of massive stars}
Core-collapse supernovae are the products of the late stages of massive stars. Determining the types of pre-explosion stars of supernovae is a key area of study and is of great importance for constraining the evolution theory of massive stars, i.e. the process of core-collapse and explosion of these massive stars. Direct detection of supernova progenitors are rare. At present, only the progenitors of the most frequent type IIP supernovae have been clearly identified to be red supergiants, while those of other types of core-collapse supernovae, especially those with their envelopes stripped, still remain undiscovered (see reviews \citealp{Smartt2009,Smartt2015} and references therein). In addition, determining the relationship between supernovae and their progenitor stars can help solve the problem of using numerical calculations to simulate the explosion process, which will also reduce the huge uncertainty in the evolution theory of massive stars (especially regarding the mass loss mechanism). 

\begin{figure}[htbp]
\centering
    \includegraphics[width=0.4\textwidth]{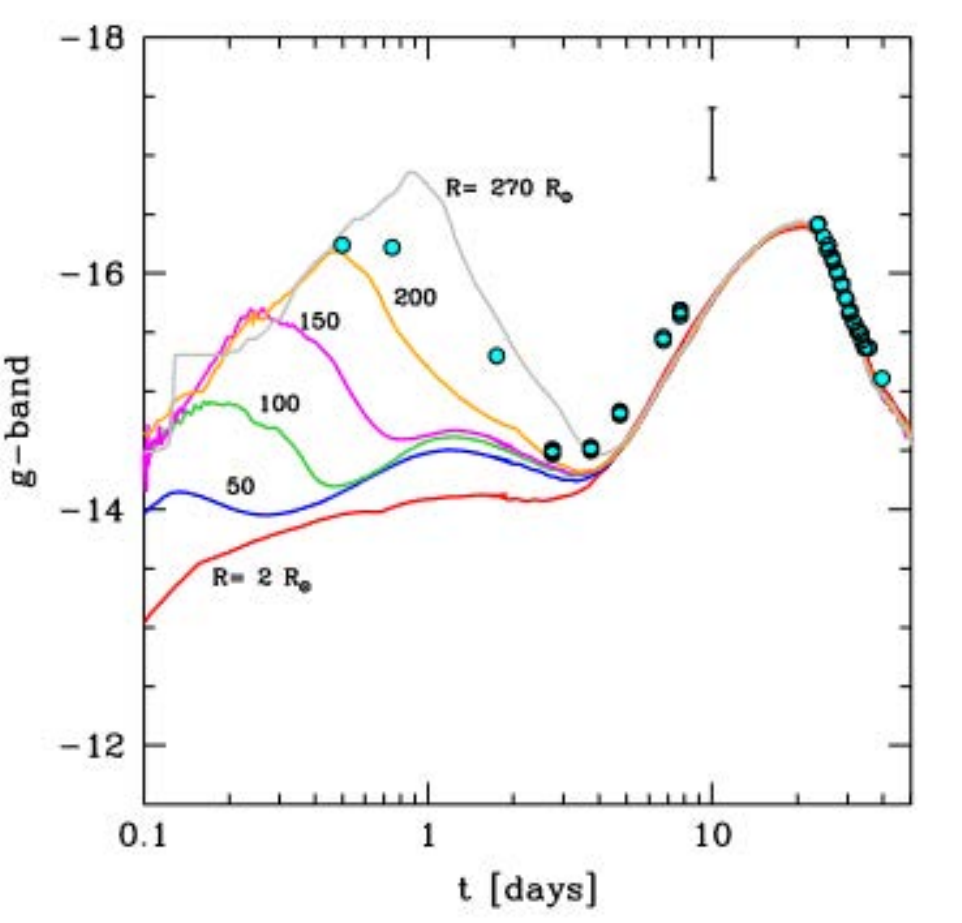}
	\includegraphics[width=0.56\textwidth]{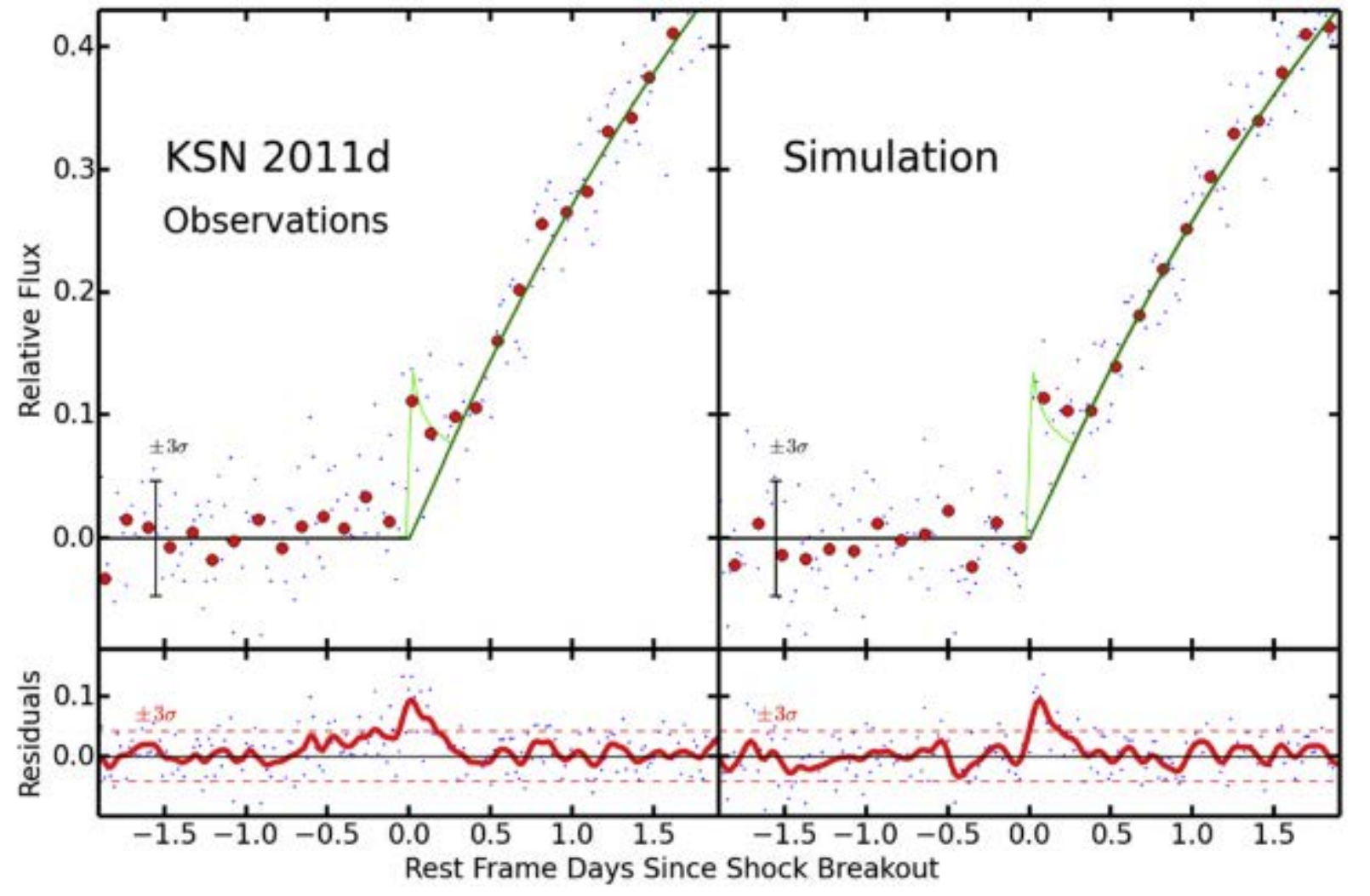}
\caption{{\it Left:} $g$-band light curves of stripped supergiant stars with different stellar radii \citep{Bersten2012}; {\it Right:} {\it Kepler} Space Telescope observations of the possible shock breakout signal of a SN~IIP candidate \citep{Garnavich2016}. \label{fig:ccsn-shock}}
\end{figure}

Another more common method is to use the early-time observational properties of the supernovae. The first light of supernovae is powered by the breakout of the energetic shock which usually lasts for a fraction of an hour, depending on the properties of the progenitor star. The breakout emission carries information of the radius, density profile and composition of the pre-explosion star (Figure \ref{fig:ccsn-shock}) \citep{Nakar+Sari2010,Rabinak+Waxman2011,Bersten2012,Waxman2017}. However, the short duration of the breakout emission makes it extremely difficult to be observed. The fact that it peaks in X-ray band makes the situation even worse in optical observations. By now only one event for stripped supernova in 2016 \citep{Bersten2017} and two {\it Kepler} events for SNe~II in 2011 (Figure \ref{fig:ccsn-shock}) \citep{Garnavich2016} have been discovered. Yet it is much easier to capture the signal right after the breakout during the cooling of the shocked envelope, which lasts days \citep[e.g.][]{Modjaz2009}. To do so, it is crucial to monitor the sky with high cadence.

\begin{figure}[htbp]
\center
\includegraphics[width=0.45\textwidth]{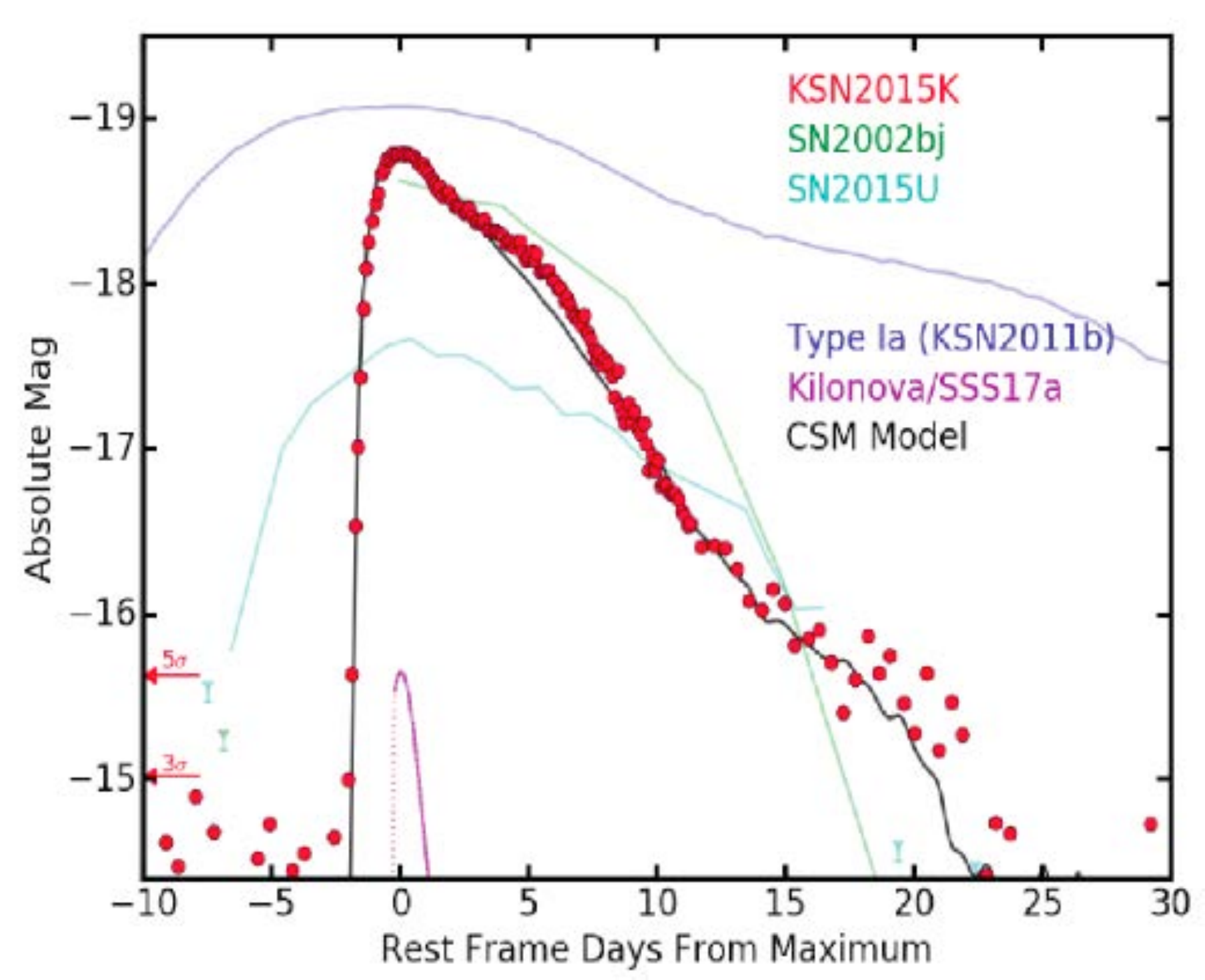}
\includegraphics[width=0.5\textwidth]{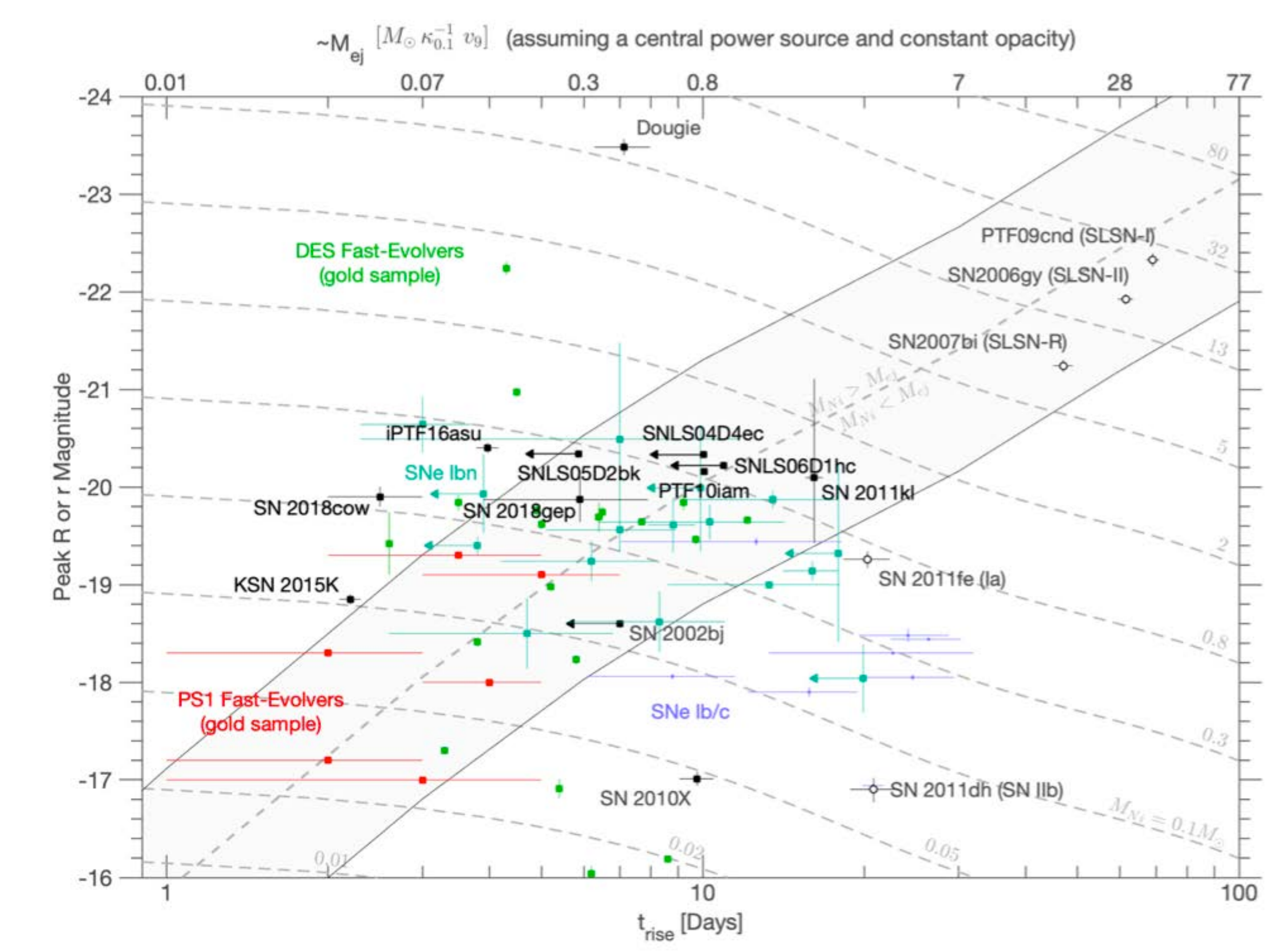}
\vspace{0.2cm}
\caption{{\it Left}: KSN2015K, a rapidly evolving transient source discovered by the {\it Kepler} space telescope \citep{2018NatAs...2..307R}. {\it Right}: Events with a short rise-time (implying a small ejecta mass; top axis), but with a luminous peak (implying a large Ni mass, if powered by Ni decay; dashed horizontal lines), lie in a region of phase space where more Ni than total ejecta mass is required (to the top left of the shaded band). Therefore, these events cannot be powered by the standard central-Ni-decay mechanism responsible for the luminosity of most supernovae. } \vspace{-0.0cm}
\label{fast}
\end{figure}

\paragraph{The Explosion Mechanisms of Rapidly-evolving Transients}
Recent wide-field surveys have found a class of rapidly-evolving transients with an unknown explosion mechanism. The time scale of rising to the maximum light is very short, usually only a few days (note that there are quite a few samples with rising time shorter than two days). Such transient objects evolve very fast and are also known as Fast Blue Optical Transients (FBOTs). FBOTs have peculiar properties and their luminosity evolution present challenges to the known supernova explosion model (see the right panel of Figure \ref{fast}). 

The KSN~2015K discovered by the {\it Kepler} space telescope is one example. Its rise time is only 2.2 days and its peak luminosity is about $-$18.8 mag. The light curve is flat shortly after the explosion and then decays following the power law \citep{2018NatAs...2..307R}. This is very inconsistent with the radioactive decay model (see the left panel of Figure \ref{fast}). Another well-observed FBOT is AT2018cow which defies several proposed scenarios \citep{2018ApJ...864...45M,2019MNRAS.484.1031P,2021NatAs...6..249P}. The probable nature of the central engine includes an accreting black hole, rapidly spinning magnetar, or embedded internal shock \citep{2019ApJ...872...18M}. The research on FBOTs is still in the early stage and viable samples are still relatively sparse.

So far, such transients found in the nearby universe have generally been discovered by high-cadence time-domain surveys, such as the {\it Kepler} and Zwicky Transient Facility (ZTF) surveys. The wide-field high-cadence capabilities of the ET space telescope will hopefully lead to more discoveries of these objects. Timely early alerts after discovery will help for carrying out follow-up spectroscopic observations, further helping us unravel the mystery of these transients.

\section{Payload}
{\bf Authors:}
\newline
Jian Ge$^1$, Dan Zhou$^1$, Congcong Zhang$^1$, Yong Yu$^1$,  Yongshuai Zhang$^1$, Yan Li$^1$, Hui Zhang$^1$, Zhenghong Tang$^1$, Chaoyan Wang$^1$, Yonghe Chen$^2$, Chuanxin Wei$^2$, Yanwu Kang$^2$, Baoyu Yang$^2$, Chao Qi$^2$, Xiaohua Liu$^2$, Quan Zhang$^2$, Yuji Zhu$^2$, Fengtao Wang$^3$, Wei Li$^3$, Pengfei Cheng$^3$, Chao Shen$^3$, Baopeng Li$^3$, Yue Pan$^3$, Sen Yang$^3$, Wei Gao$^3$, Zongxi Song$^3$, Jian Wang$^4$, Hongfei Zhang$^4$, Cheng Chen$^4$, Hui Wang$^4$, Jun Zhang$^4$, Zhiyue Wang$^4$, Feng Zeng$^4$, Zhenhao Zheng$^4$, Jie Zhu$^4$, Yingfan Guo$^4$, Yihao Zhang$^4$, Yudong Li$^5$, Lin Wen$^4$ \& Jie Feng$^5$\\
{1. \it Shanghai Astronomical Observatory, Chinese Academy of Sciences}\\
{2. \it Shanghai Institute of Technical Physics, Chinese Academy of Sciences}\\
{3. \it Xi’An Institute of Optics and Precision Mechanics, Chinese Academy of Sciences}\\
{4. \it University of Science and Technology of China, Chinese Academy of Sciences}\\
{5. \it Xinjiang Technical Institute of Physics and Chemistry, Chinese Academy of Sciences}\\

\subsection{Basic Instrument Overview} 

ET's scientific payload consists of two kinds of instruments. One is the Transit Telescope Array, which will carry out a four-year high precision and continuous photometric monitoring of about 1.2 million FGKM dwarfs (Gaia magnitude G$\leq$16) in the direction that encompasses the original {\it Kepler} field to obtain light curves of stars for detecting planetary transits. The other one is the Microlensing Telescope, which will also conduct a four-year high precision photometric monitoring of over 30 million stars (I$\leq$20.6) toward the Galactic bulge to detect microlensing events produced by planets. The same fields will be simultaneously monitored by three ground-based KMTNet telescopes. The combined data will be used to measure the masses of hundreds of cold planets including free-floating planets.
\begin{figure}[ht]
    \centering
    \includegraphics[width=0.85\textwidth]{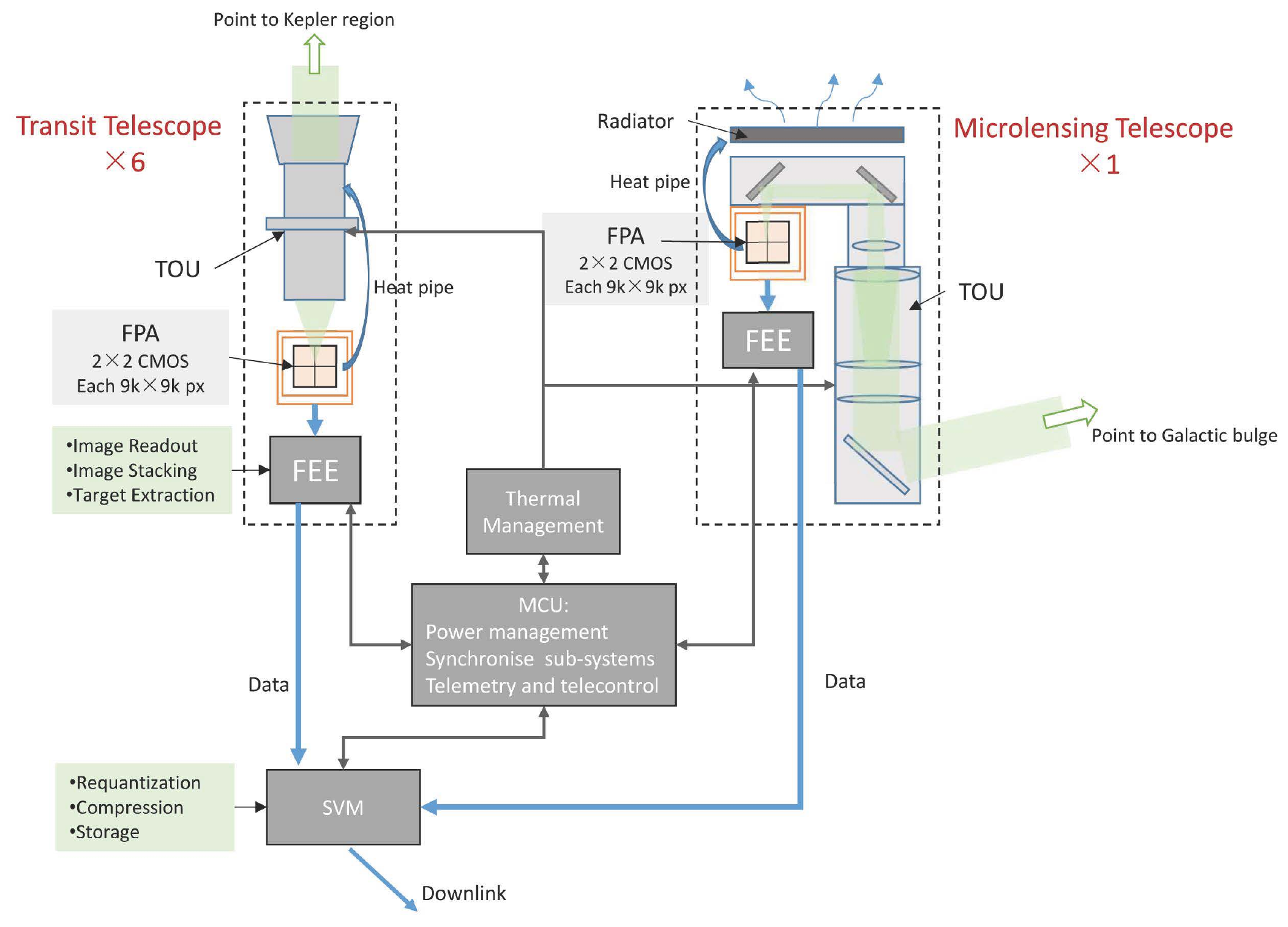}
    \caption{Block diagram of ET's payload, which consists of six transit telescopes and one microlensing telescope. The transit telescopes will be pointed towards the original {\it Kepler}'s field while the microlensing telescope will be pointed towards the Baade's window near the Galactic Center.}
    \label{fig:6.1}
\end{figure}

The Transit Telescope Array consists of six identical telescopes, all of which will observe the same region. Each transit telescope is a dioptric optical system with a field of view (FOV) of 500 $deg^2$. It operates with a focal ratio of 1.57 which offers a pixel scale of 4.38 arcsec/pix. \SI{90}{\percent} encircled energy is within 5×5 pixels for the entire FOV. 

The microlensing telescope is a Schmidt-Mann catadioptric system, providing a 4 $deg^2$ FOV with a spectral range of \SIrange[range-units = single]{700}{900}{\nm}. Its focal ratio is f/17.2 with a plate scale of 0.4 arcsec/pix. It will operate close to the diffraction limit with a PSF FWHM of less than 0.85 arcsecond. 

Each telescope will be equipped with a mosaic of four 9k$\times$9k CMOS detectors with a pixel size of \SI{10}{\um}. An electronic rolling shutter is used for exposure control to avoid image smear. The planned single exposure time is \SI{10}{\s} for transit telescopes and \SI{10}{\minute} for the microlensing telescope. The readout time is \SI{1.5}{\s}.

To reduce the instrument noise, the detectors are passively cooled to \SI{-40}{\degreeCelsius} with their temperature stability maintained within \SI{\pm0.1}{\degreeCelsius}. The telescope optical units (TOUs) will operate at a temperature of about \SI{-30}{\degreeCelsius} with a stability of \SI{\pm0.3}{\degreeCelsius} to minimize image drifts and PSF size changes. A sunshield is used to block sunlight and provide a stable thermal environment for all science payloads.

Data processing includes image stacking, target extraction, data requantization, and data compression, all of which will be carried out onboard to reduce data transmission requirements.

\begin{figure}[htbp]
    \centering
    \includegraphics[width=0.55\textwidth]{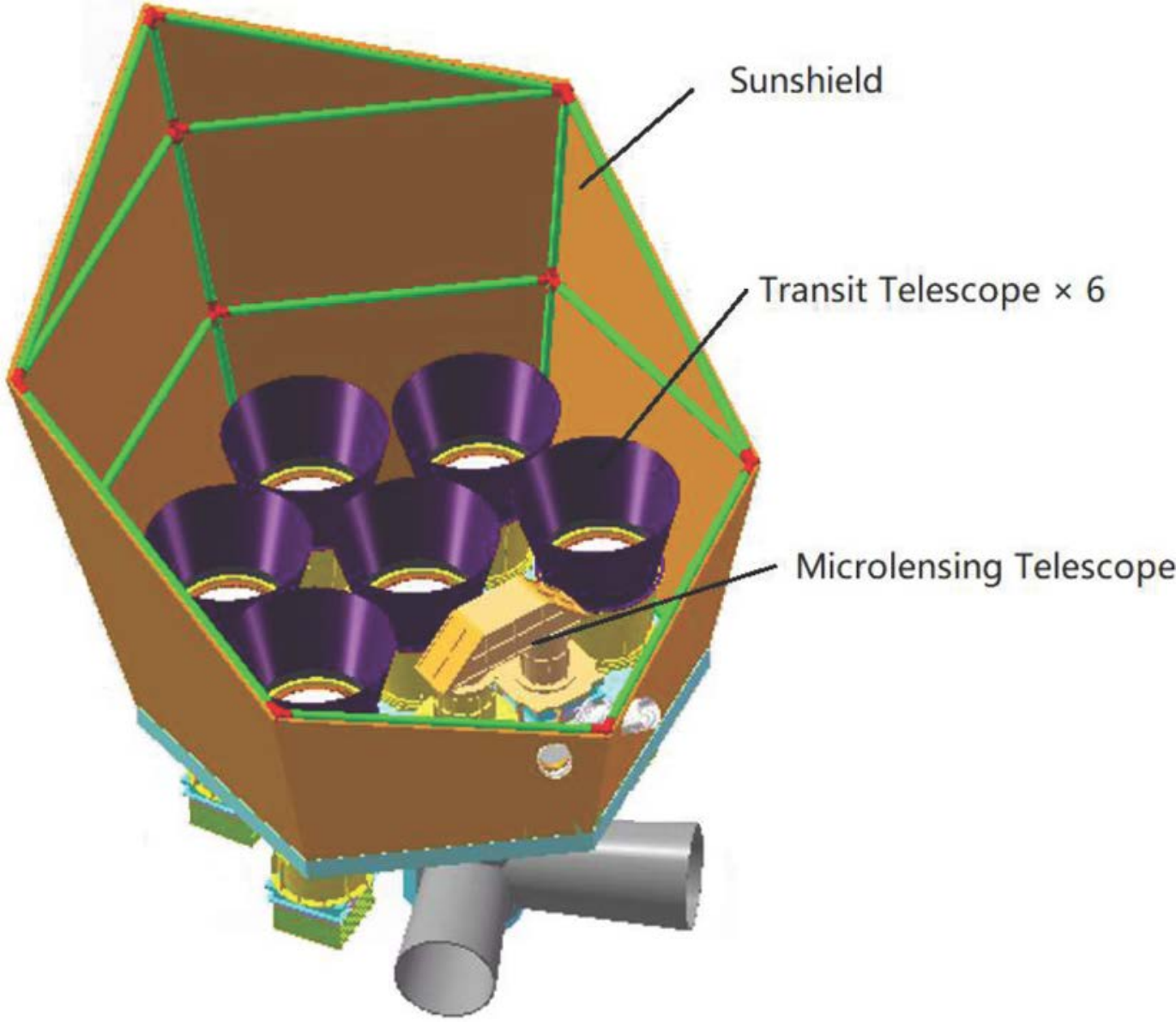}
    \caption{Overall Layout of ET's Scientific Payload. Six transit telescopes and one microlensing telescopes are mounted on an optical bench. The sunshield will block radiation from the Sun, Earth, and Moon. The hood on top of each telescope is used to block light from background stars out of the telescope's FOV and also serves as a radiator to cool detectors at the telescope focal plane. }
    \label{fig:6.2}
\end{figure}

A block diagram and summaries of ET's payload are shown in Figure \ref{fig:6.1} and Table \ref{tab:Summaries_of_ET_payload} respectively while Figure \ref{fig:6.2} shows the overall layout of ET's payload.

\begin{table}[htbp]
    \centering
        \captionsetup{justification=centering}
    \caption{\centering Summaries of ET's payload}
    \begin{tabular}{|c|c|c|}\hline
         Parameters & Transit Telescope	& Microlensing Telescope \\ \hline
         Number of Telescopes & 6 & 1 \\ \hline
         Observation region  &	\makecell[c]{Six telescopes point to the\\ same direction that\\ encompass the {\it Kepler} field} &  Galactic Baade's window \\ \hline
         Spectral range & \SIrange[range-units = single]{450}{900}{\nm}	& \SIrange[range-units = single]{700}{900}{\nm} \\ \hline
         Optical systems & \makecell[c]{Refractive design with 8\\ lenses and 1 window} &	\makecell[c]{Schmidt-Mann catadioptric\\ design with 6 lenses, 2 fold\\ mirrors and 1 pointing mirror} \\ \hline
         Pupil diameter	& \SI{30}{\cm} & \SI{30}{\cm} \\ \hline
         Field of view & 500 $deg^2$ & 4 $deg^2$ \\ \hline
         Plate scale & 4.38 arcsec/pix & 0.4 arcsec/pix \\ \hline
         Image quality & \makecell[c]{\SI{90}{\percent} of enclosed energy\\ (EE90) within 5×5 px} & \makecell[c]{FWHM of PSF within 0.85 arcsecond} \\ \hline
         Focal Plane Array & 2$\times$2 CMOS & 2$\times$2 CMOS \\ \hline
         Detectors & \makecell[c]{9k$\times$9k CMOS\\ \SI{10}{\um} square pixels} & \makecell[c]{9k×9k CMOS\\ \SI{10}{\um} square pixels} \\ \hline
         Detector temperature & \SI{-40}{\degreeCelsius} & \SI{-40}{\degreeCelsius}  \\ \hline
         Telescope temperature & \SI{\sim-30}{\degreeCelsius} & \SI{\sim-30}{\degreeCelsius} \\ \hline
         Exposure time & \SI{10}{\s} & \SI{10}{\minute} \\ \hline
         Readout time & \SI{1.5}{\s} & \SI{1.5}{\s} \\ \hline
         Power needed & \multicolumn{2}{|c|}{Peak$\colon$ 869W   \  \   \             Average$\colon$ 764W} \\ \hline
         Mass of payload & \multicolumn{2}{|c|}{\SI{1240}{\kg}} \\ \hline
         On-board storage & \multicolumn{2}{|c|}{10Tb} \\ \hline
         Data download & \multicolumn{2}{|c|}{95.3 Gb$\sim$156.9 Gb per day} \\ \hline
    \end{tabular}
    \label{tab:Summaries_of_ET_payload}
\end{table}

\subsection{Telescope Optical Units (TOU) } 

\subsubsection{Transit Telescope}

\begin{figure}[htbp]
    \centering
    \includegraphics[width=0.8\textwidth]{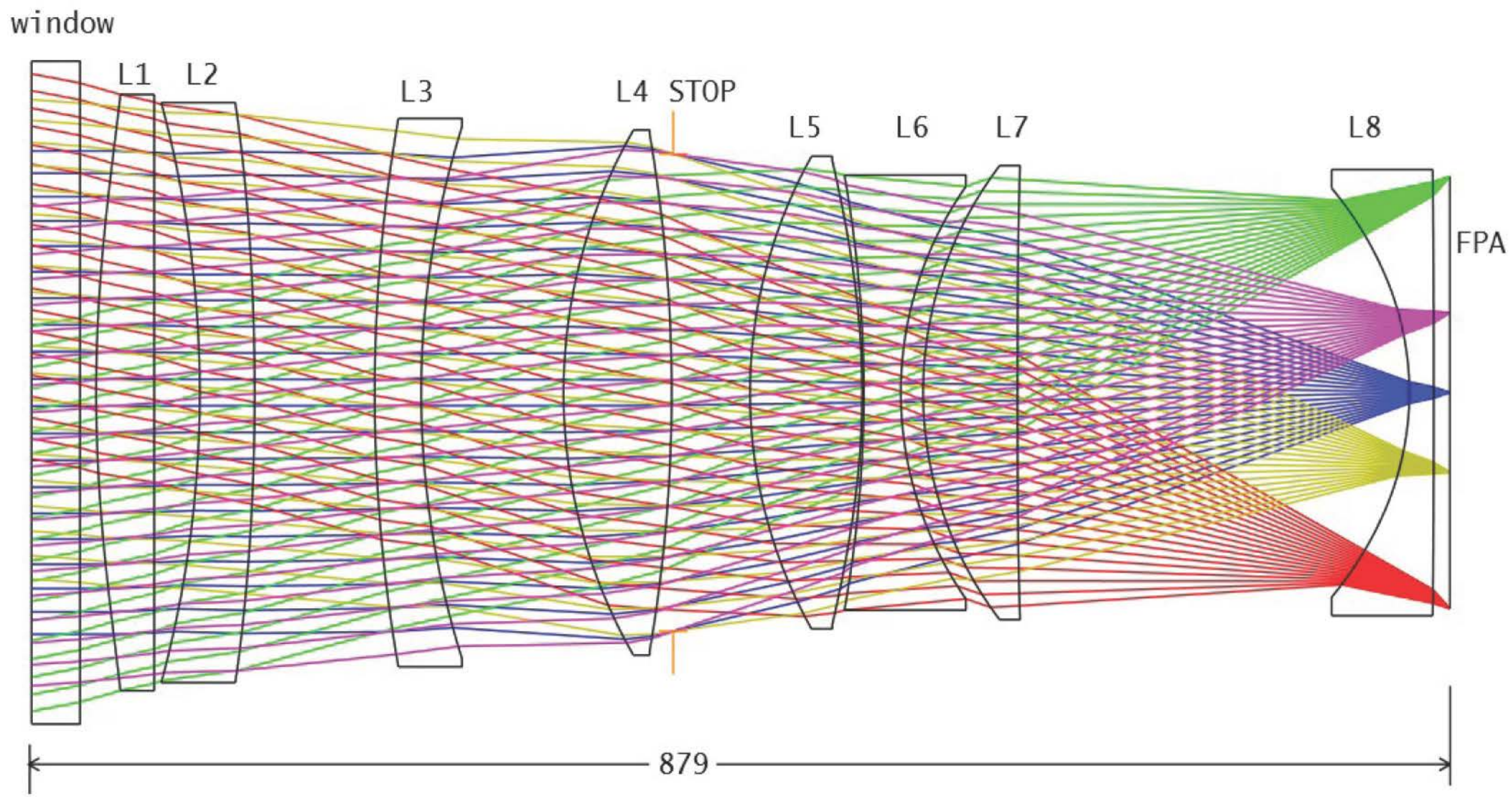}
    \caption{Optical layout of a Transit Telescope. The FOV is 500 deg$^2$. The diameter of EE90 is within 3.8 pixels. }
    \label{fig:6.3}
\end{figure}

\begin{figure}[htbp]
    \centering
    \includegraphics[width=0.65\textwidth]{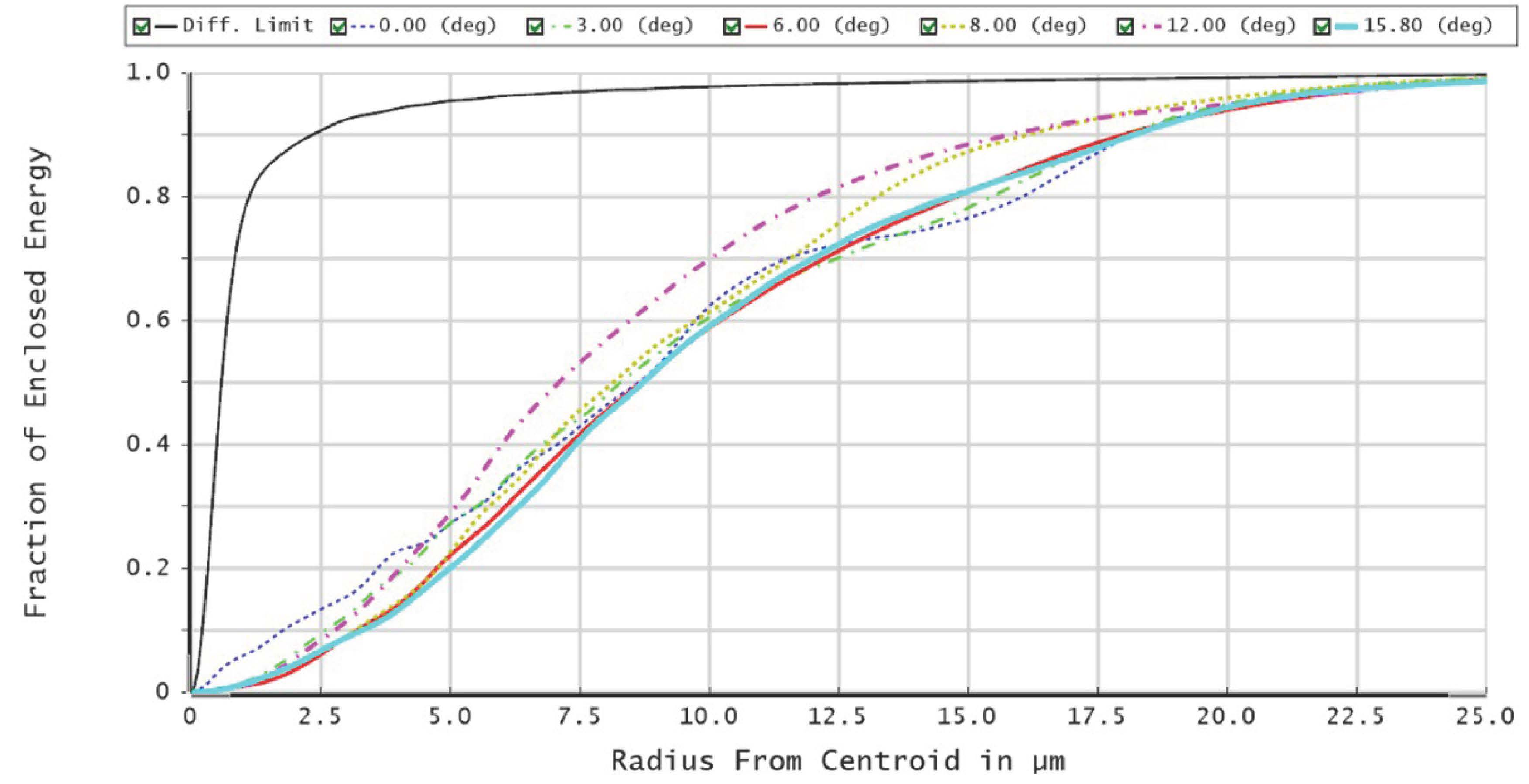}
    \caption{Fraction of enclosed energy for the transit telescope's optical design-- EE90 is within 3.8$\times$3.8 pixels in the entire FOV. }
    \label{fig:6.4}
\end{figure}

ET's payload will be equipped with six \SI{30}{\cm} transit telescopes. Each transit telescope contains eight lenses with a fused silica window in front of the telescope to resist thermal shocks and radiation. The lenses have a maximum diameter of \SI{40}{\cm} (the window) and a minimum diameter of \SI{28}{\cm}. Five even aspherical surfaces are used to improve the image quality. Materials used in the optical system include fused silica, calcium fluoride, N-KZFS11, and LLF1. The overall length from window to image plane is \SI{879}{\cm}. The primary requirements and parameters of the optical system are summarized in Table \ref{tab:Requirements_and_parameters_of_transit_telescope}, while an optical layout of transit telescope is shown in Figure \ref{fig:6.3}. \SI{90}{\percent} enclosed energy (EE90) is within \SI{38}{\um} inside a circular field of view with diameter of \SI{31.6}{\degree}, as shown in Figure \ref{fig:6.4}.  

\begin{table}[htbp]
    \centering
        \captionsetup{justification=centering}
    \caption{\centering Requirements and parameters of transit telescope }
    \begin{tabular}{|c|c|}\hline
       Parameters  & Value \\ \hline
       Spectral range & \SIrange[range-units = single]{450}{900}{\nm} \\ \hline
       Entrance pupil diameter & 30 cm \\ \hline
       Field of view & 500 $deg^2$ \\ \hline 
       Focal ratio & 1.57 \\ \hline
       Plate scale & 4.38 arcsec/pix \\ \hline
       Image quality & EE90$\leq$50 $\mu$m \\ \hline
       Vignetting & \makecell[c]{0 vignetting within central 256$deg^2$  \\ Max 20\% vignetting at four corners of the field} \\ \hline
       Working temperature	& $\sim$ -30$^{\circ}$C  \\ \hline
    \end{tabular}
    \label{tab:Requirements_and_parameters_of_transit_telescope}
\end{table}

\begin{figure}[hb]
    \centering
    \includegraphics[width=0.45\textwidth]{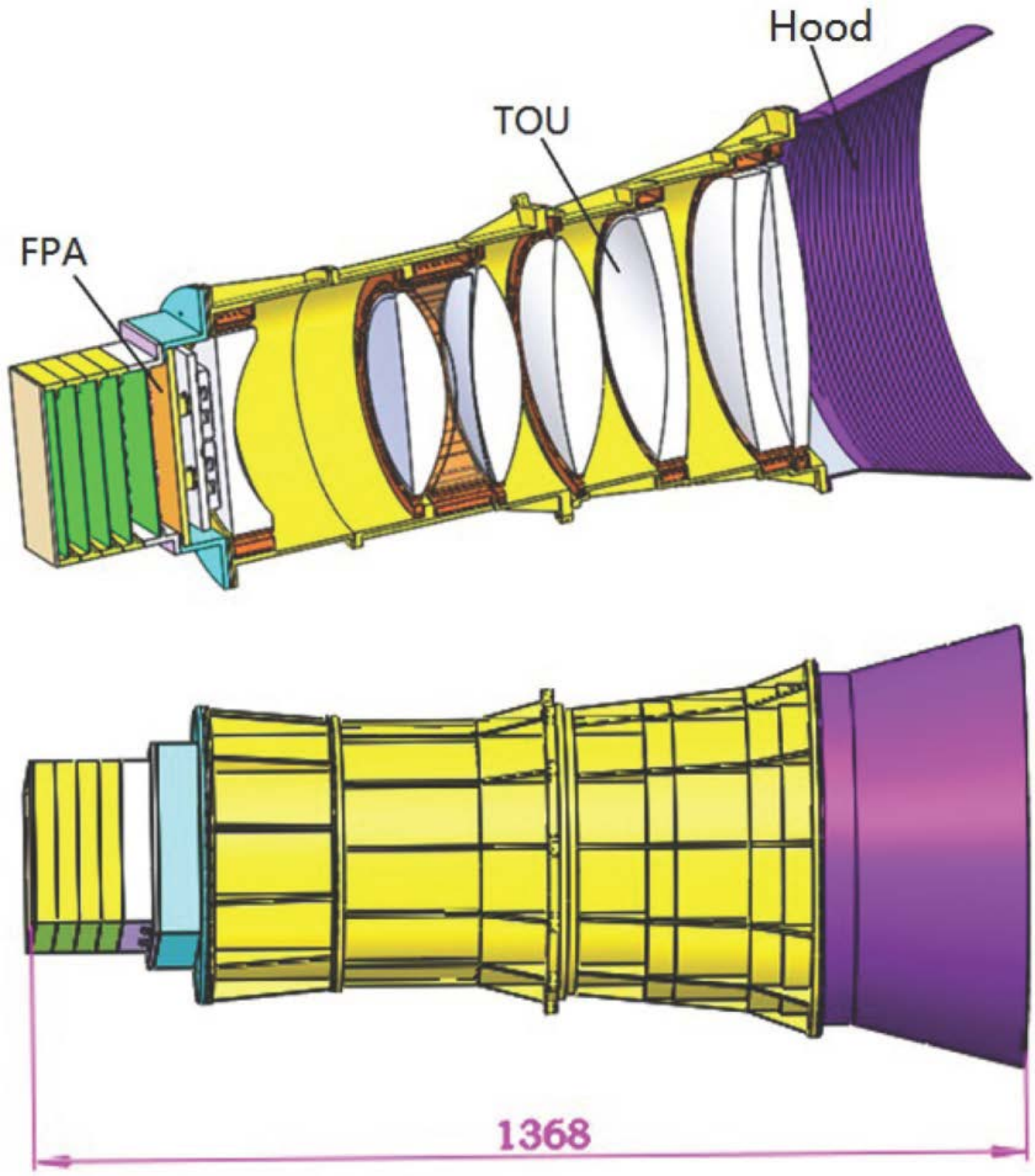}
    \caption{Mechanical structure of a transit telescope. Each lens is mounted in a cell, which is installed in the lens barrel. The telescope will be temperature controlled by adding surface heaters to the outside of the telescope structure. }
    \label{fig:6.5}
\end{figure}
A preliminary error distribution analysis has been carried out. Factors affecting image quality including manufacturing error, alignment error, detector flatness and environmental factors such as gravity, press, and temperature were considered. In addition, optical uniformity, stress, and other material properties were considered. Furthermore, a tolerance analysis based on the Monte Carlo method has been conducted. According to the analysis, the allowed lenses decentration is ±\SI{0.02}{\mm} and the allowed lenses tilt is ±0.3 arcmin. 
A thermal simulation was carried out to evaluate the thermal adaptability of the TOU. The temperature of the TOU should be stable with a variation of less than ±\SI{0.3}{\degreeCelsius} in order to keep the image quality and spot position stable. 

A preliminary design for the mechanical structure of each transit telescope is shown in Figure \ref{fig:6.5}. Each lens is first mounted in a cell which is flexibly supported. Lenses with their cells will then be inserted into the lens barrel. The TOUs will work at about \SI{-30}{\degreeCelsius} in orbit, albeit integrated at room temperature. Therefore, mechanical design and material selection should consider the thermal-elastic stresses resulting from the  coefficient of thermal expansion (CTE) mismatch of different materials. The material of the lens barrel should have low density, high stiffness, high thermal conductivity and moderate CTE. Based on a preliminary analysis of tradeoffs, SiC/Al(55$\%$Vol) will be used for the lens barrel. An aluminium alloy and titanium alloy will be used for lens cells, chosen according to the lens glass.

An additional hood is set on top of each telescope to suppress stray light. The hood is also used as a radiator to cool down the Focal Plane Array (FPA) to \SI{-40}{\degreeCelsius}. At the bottom of the TOU, the camera is mounted and insulated through the flange. The envelope size for each transit telescope is about $\Phi$\SI{680}{\mm}×\SI{1368}{\mm}.

Each transit telescope will be focused by adjusting the operating temperature of the TOU. The TOU nominally operates at about -30$^{\circ}$C and can be adjusted up to \SI{\pm3}{\degreeCelsius}. A change of \SI{1}{\degreeCelsius} produces about \SI{20}{\um} focus shift. After the focus shift is done, the temperature of the TOU will be controlled and remain stable in order to minimize image drifts and size changes during observation.

\subsubsection{Microlensing Telescope}

\begin{figure}[ht]
    \centering
    \includegraphics[width=0.65\textwidth]{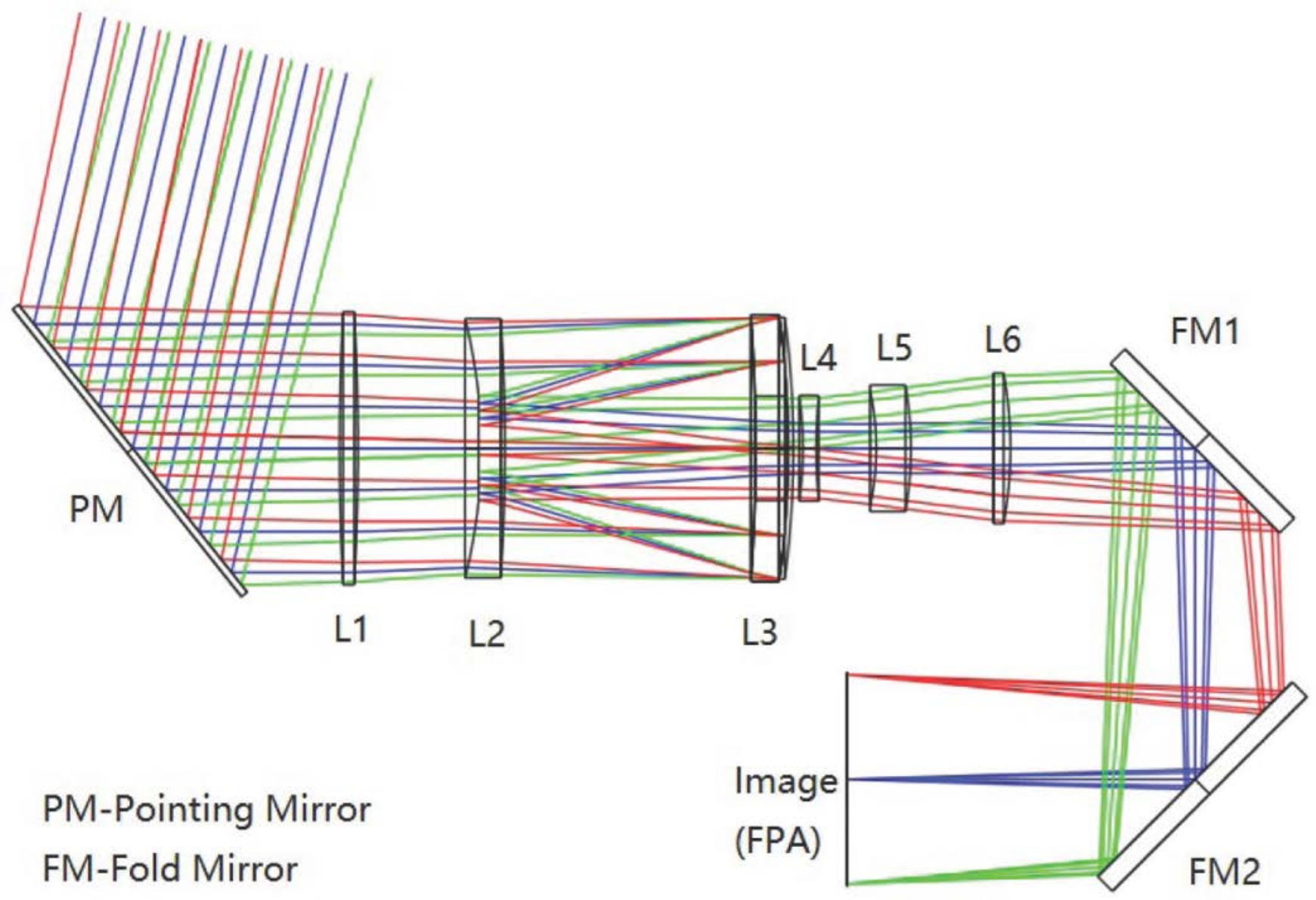}
    \caption{Optical layout of the microlensing telescope. The telescope uses a Schmidt-Mann catadioptric design with an optical performance close to the diffraction limit. }
    \label{fig:6.6}
\end{figure}

Table \ref{tab:Requirements_and_parameters_of_Microlensing_telescope } shows the requirements and parameters of the microlensing telescope. It is a \SI{30}{\cm} telescope with a spectral range of \SIrange[range-units = single]{700}{900}{\nm}, with a plate scale of 0.4 arcsec/pix. It has a focal ratio of 17.2, making it a telephoto system. In order to further compress the size of the instrument, the telescope uses a Schmidt-Mann catadioptric design with six lenses. Two additional fold mirrors are used to reduce the length of the telescope. The telescope will stare at the Baade's window near the Galactic Center to detect microlensing events produced by planets. Since the spacecraft needs to rotate \SI{90}{\degree} around the optical axis of the transit telescopes every quarter to keep the solar array facing the Sun, a pointing mirror is set in front of the microlensing telescope which will rotate as needed to maintain its view of the Galactic bulge. The optical layout of the telescope is shown in Figure \ref{fig:6.6}. Most of the lenses will be composed of fused silica except L4 (PSK50) and L6 (N-KZFS11), while the pointing mirror will use SiC.

\begin{table}[htbp]
    \centering
        \captionsetup{justification=centering}
    \caption{\centering Requirements and parameters of the microlensing telescope}
    \begin{tabular}{|c|c|}\hline
         Parameters & Value \\ \hline
         Spectral range & \SIrange[range-units = single]{700}{900}{\nm} \\ \hline
         Entrance pupil diameter & \SI{30}{\cm} \\ \hline
         Field of view & \SI{4}{deg^2} \\ \hline 
         Focal ratio & 17.2 \\ \hline
         Effective focal length & \SI{5160}{\mm} \\ \hline
         Plate scale & 0.4 arcsec/pix \\ \hline
         Image quality & FWHM of PSF within \SI{0.85}{\arcsecond} \\ \hline
         Working temperature &  \SI{\sim-30}{\degreeCelsius} \\ \hline

    \end{tabular}
    \label{tab:Requirements_and_parameters_of_Microlensing_telescope }
\end{table}

\begin{figure}[htbp]
    \centering
    \includegraphics[width=0.75\textwidth]{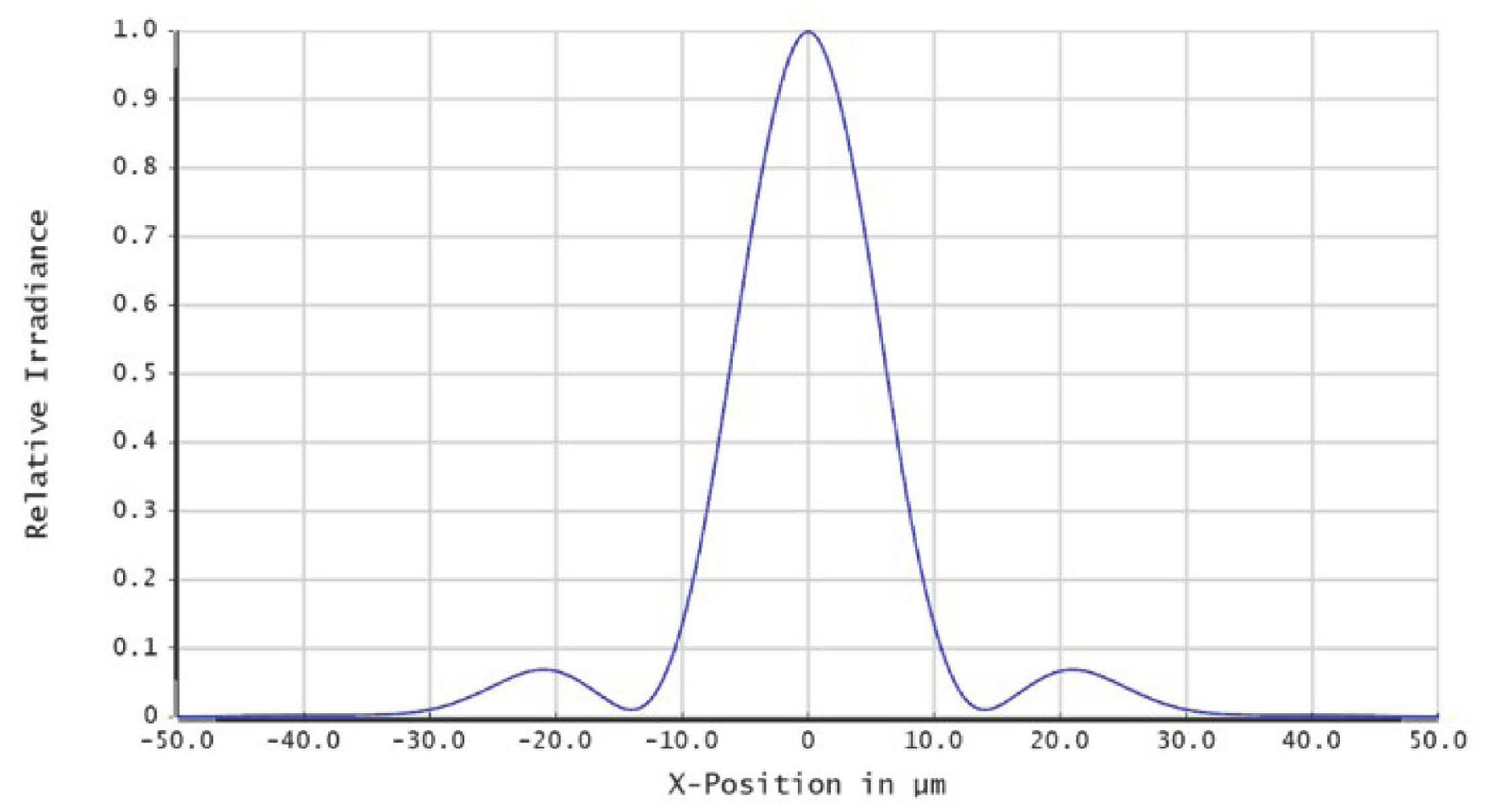}
    \caption{The microlensing telescope's PSF cross section, which is very close to the diffraction limit. The nominal FWHM is about 11 $\mu$m, which is under-sampled with the CMOS' 10 $\mu$m pixels to cover the required FOV of the telescope.}
    \label{fig:6.7(a)}
\end{figure}

\begin{figure}[htbp]
    \centering
    \includegraphics[width=0.65\textwidth]{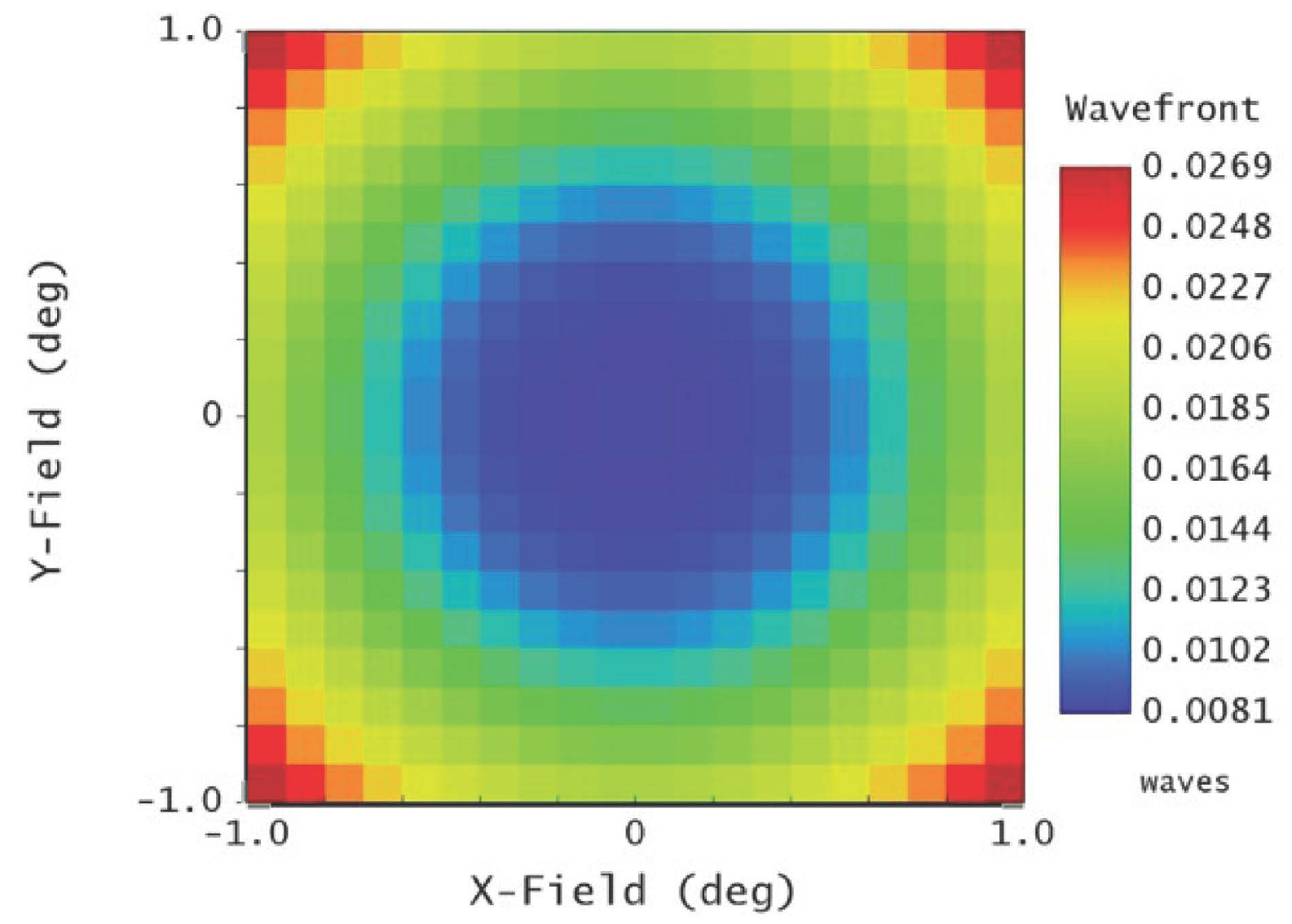}
    \caption{The microlensing telescope's wavefront field map RMS. The RMS wavefront error is less than 0.027$\lambda$@633 nm  in the entire FOV.}
    \label{fig:6.7(b)}
\end{figure}

In contrast to the transit telescopes, the microlensing telescope needs to operate close to the diffraction limit with a PSF FWHM less than \SI{0.85}{\arcsecond}. Therefore, all six lenses are standard spherical surfaces, which can obtain higher surface figure precision. The PSF cross section of the telescope is shown in Figure \ref{fig:6.7(a)}. The nominal FWHM of PSF is \SI{0.44}{\arcsecond} and the wavefront error (WFE) is less than 0.027$\lambda$ @\SI{633}{\nm} RMS in the entire \SI{4}{deg^2} field (see Figure \ref{fig:6.7(b)}).

A preliminary tolerance analysis has been carried out to estimate the image quality. According to the analysis, the allowed lenses decentration is \SI{\pm0.02}{\mm} and the allowed lenses tilt is \SIrange[range-units = repeat]{\pm0.3}{\pm0.5}{\arcminute}. The allowed figure error of lenses is \SI{25}{\nm} RMS, while the figure error of reflective surface on L2 and L3 should be reduced to no more than \SI{10}{\nm} RMS.

\begin{figure}[htbp]
    \centering
    \includegraphics[width=0.5\textwidth]{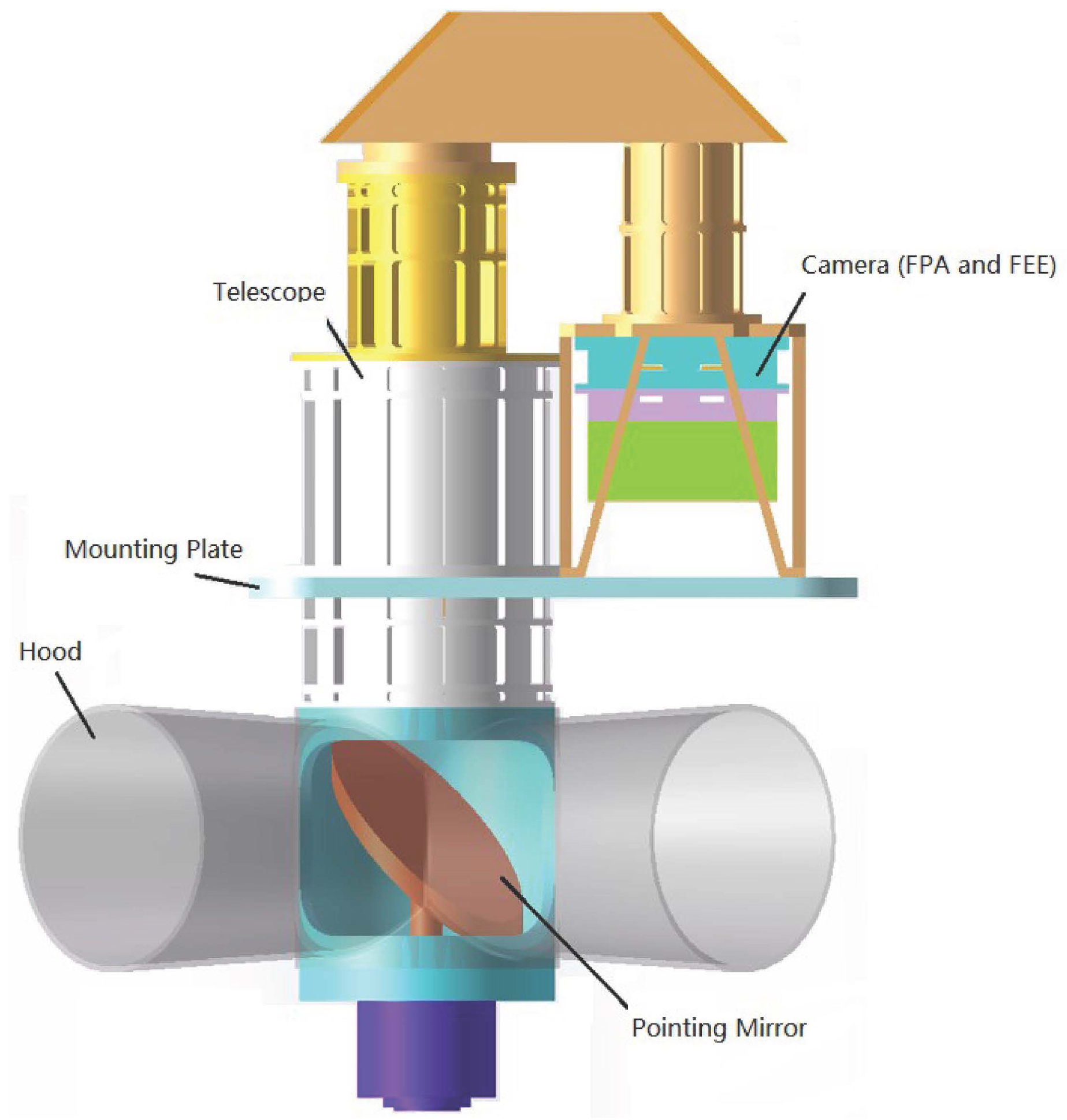}
    \caption{Mechanical structure of the microlensing telescope, which is mounted to the spacecraft optical bench at the middle of the telescope barrel. A pointing mirror can be adjusted by 90 degrees to keep the telescope continuously monitoring microlensing events at the Galactic Baade's window direction for six months each year. }
    \label{fig:6.8}
\end{figure}

A preliminary design for the mechanical structure of the microlensing telescope is shown in Figure \ref{fig:6.8}. This instrument will be fixed to the spacecraft’s optical bench through a mounting structure located in the middle of the telescope. A carbon fiber hood with two orthogonal openings is set in front of the pointing mirror to suppress stray light. The supporting structure is made of SiC/Al(55$\%$Vol) due to its high stiffness and low density. A titanium alloy will be used for lens cells. The envelope size for the microlensing telescope, excluding the hood, is \SI{950}{\mm}$\times$\SI{520}{\mm}$\times$\SI{1680}{\mm}.

\subsection{Focal Plane Array Assembly (FPAA)} 

To meet the desired FOV, the focal plane for each telescope is \SI{182}{\mm}$\times$\SI{182}{\mm}. Attached is a mosaic of four 9k$\times$9k CMOS detectors with a pixel size of \SI{10}{\um}. Detectors are required to have low readout noise, low dark current, high quantum efficiency, and fast readout, all of which are important factors in achieving high precision photometry. Both transit and microlensing telescopes adopt GSENSE1081BSI type CMOS detectors from Gpixel Inc. as the FAAs. The following describes the overall design, composition, functions, and resource requirements of the camera. 

\begin{figure}[htbp]
    \centering
    \includegraphics[width=0.85\textwidth]{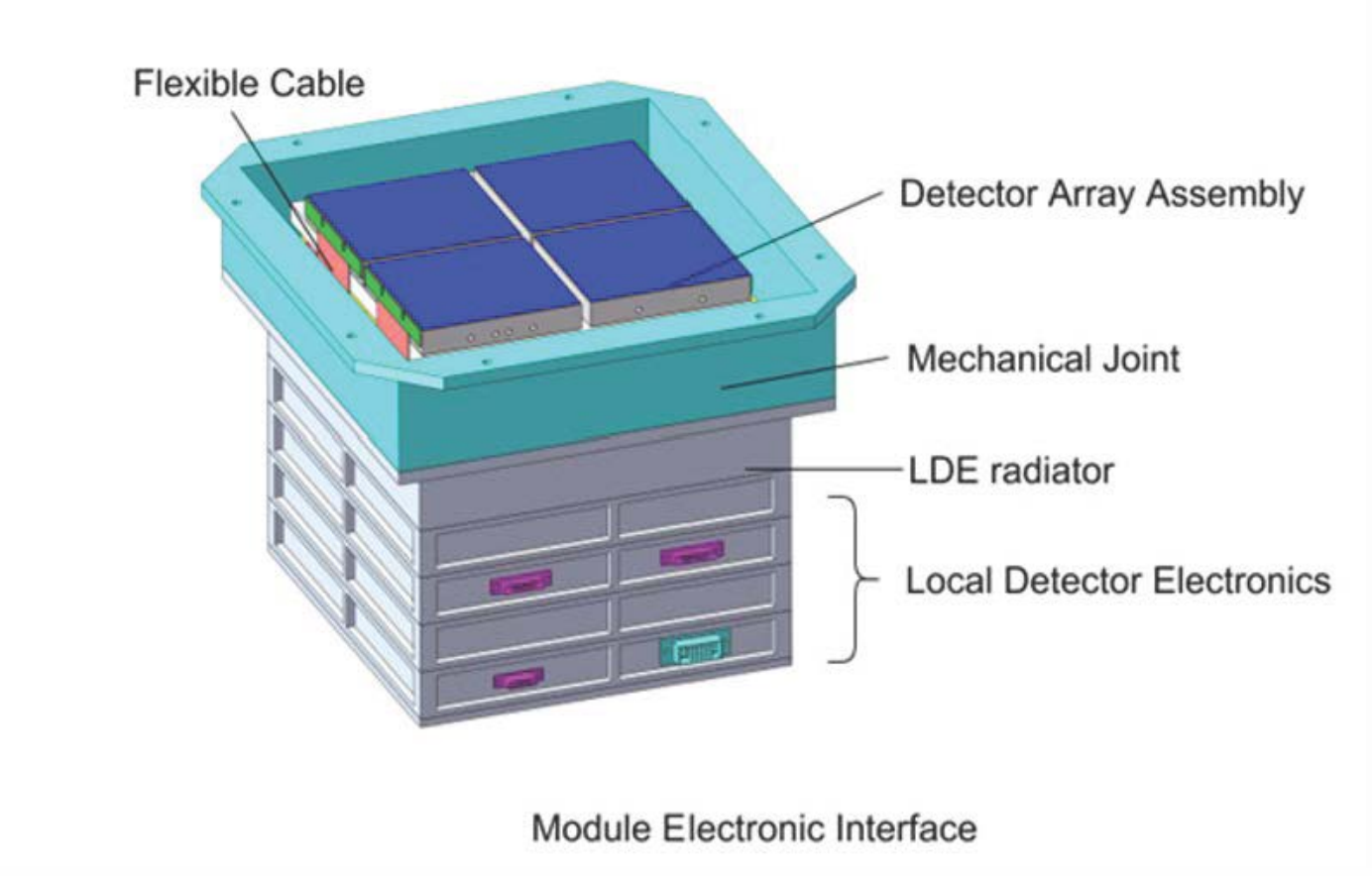}
    \caption{The conceptual design of the Focal Plane Array Assembly, which consists of a mosaic of four 9k$\times$9k CMOS detectors, supporting structure, and control electronics at the back of the detector array.}
    \label{fig:6.9}
\end{figure}

The FPAA consists of the detector array assembly (DAA), local detector electronics (LDE), the LDE power supply, a thermo-electric cooler (TEC), and a mechanical joint which secures the DAA and LDE to the telescope. Figure \ref{fig:6.9} shows the conceptual design of the FPAA.

The LDE receives power from a secondary power supply; generates the timing drive, low-noise bias, and readout of the detector; accurately controls the temperature of the detector; and completes communications with the satellite platform and image output.

The DAA's main mechanical components include four CMOS detectors, a TEC, a cold head and heating resistors. The cold head is connected to the telescope hood through a heat pipe for passive radiation cooling, and a PID control loop is used to achieve precise temperature control of the detector. The four detectors use silicon carbide as the substrate and are spliced to form a focal plane. A flexible cable is connected to each detector. Four of these flexible cables are extended out from two sides of the spliced substrate and connected to the CMOS driver board inside the LDE box. 

The LDE box has a 5-layer metal frames. The LDE radiator is installed at the top layer of the LDE box while four electronic PCB printed boards are installed to the lower layers.

\subsection{Front End Electronics (FEE)} 

\subsubsection{Composition and Main Functions}

The LDE electronics includes a power supply, bias, timing drive, high-speed image data receiving and transmission, data storage and calculation, high-precision temperature control, telemetry, and focal plane module control. 

The DAA electronics consists of four scientific CMOS modules. Each module provides 8,900 × 9,120 format photosensitive pixels. Each module has four output channels with a total of 16 channels for the four modules. The four modules have a total of 3.3 million active pixels, as well as additional masked real pixels. The LDE electronics internally generates all the CMOS clocks and bias voltages, video signal processing, and analog to digital conversions. Figure \ref{fig:6.10} shows a signal flow block diagram of the LDE. The main functions includes the following:

\begin{figure}[htbp]
    \centering
    \includegraphics[width=0.85\textwidth]{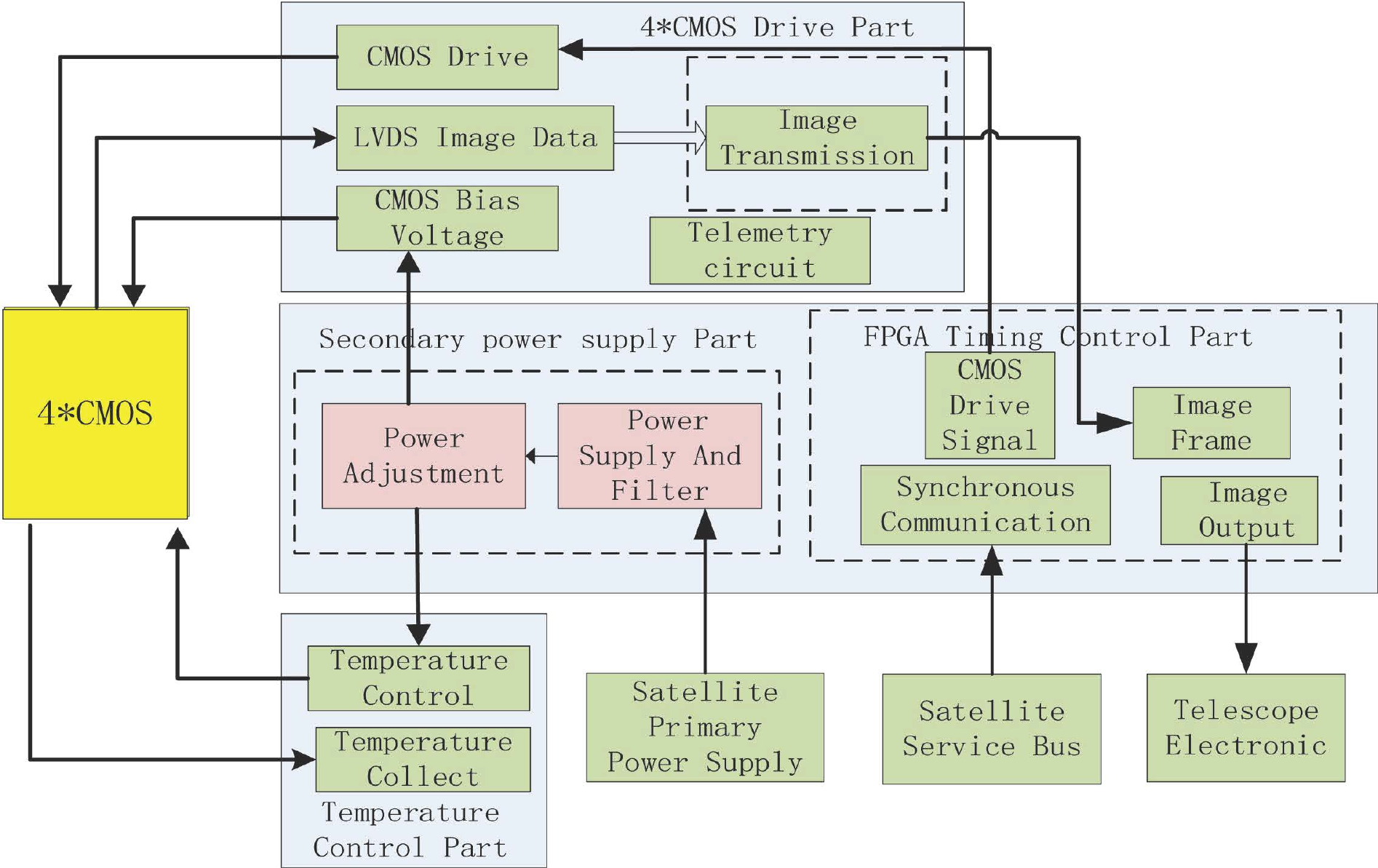}
    \caption{A signal flow diagram of the DAA electronics. FPGA obtains the remote control command of the satellite platform, provides the voltage reference source for the CMOS driver module, and controls the working parameters such as the exposure time of the CMOS module. At the same time FPGA receives LVDS image data from CMOS module, caches and finishes and then transmits them down to telescope electronics.}
    \label{fig:6.10}
\end{figure}

\begin{itemize}
    \item[1.] Providing power supply for each part of the camera focal plane electronic, generating each power supply through DC/DC converter

    \item[2.] Providing the drive signal for CMOS, including the power supply required for CMOS work, various bias power supply, and the CMOS control timing drive signal

    \item[3.] Collecting, combining, and caching the digital image. When the image data are output, working parameters such as time code, gain, and work temperature are packaged

    \item[4.] Receiving telecontrol command and realizing CMOS work parameter setting, work mode switch, and work state control according to the telecontrol command and telecontrol data 

    \item[5.] Completing telemetry data collection, including CMOS work temperature, power supply voltage, power supply current, system work parameters, and outputted telemetry data.
\end{itemize}

\subsubsection{System Software Implementation}

\begin{figure}[htbp]
    \centering
    \includegraphics[width=0.85\textwidth]{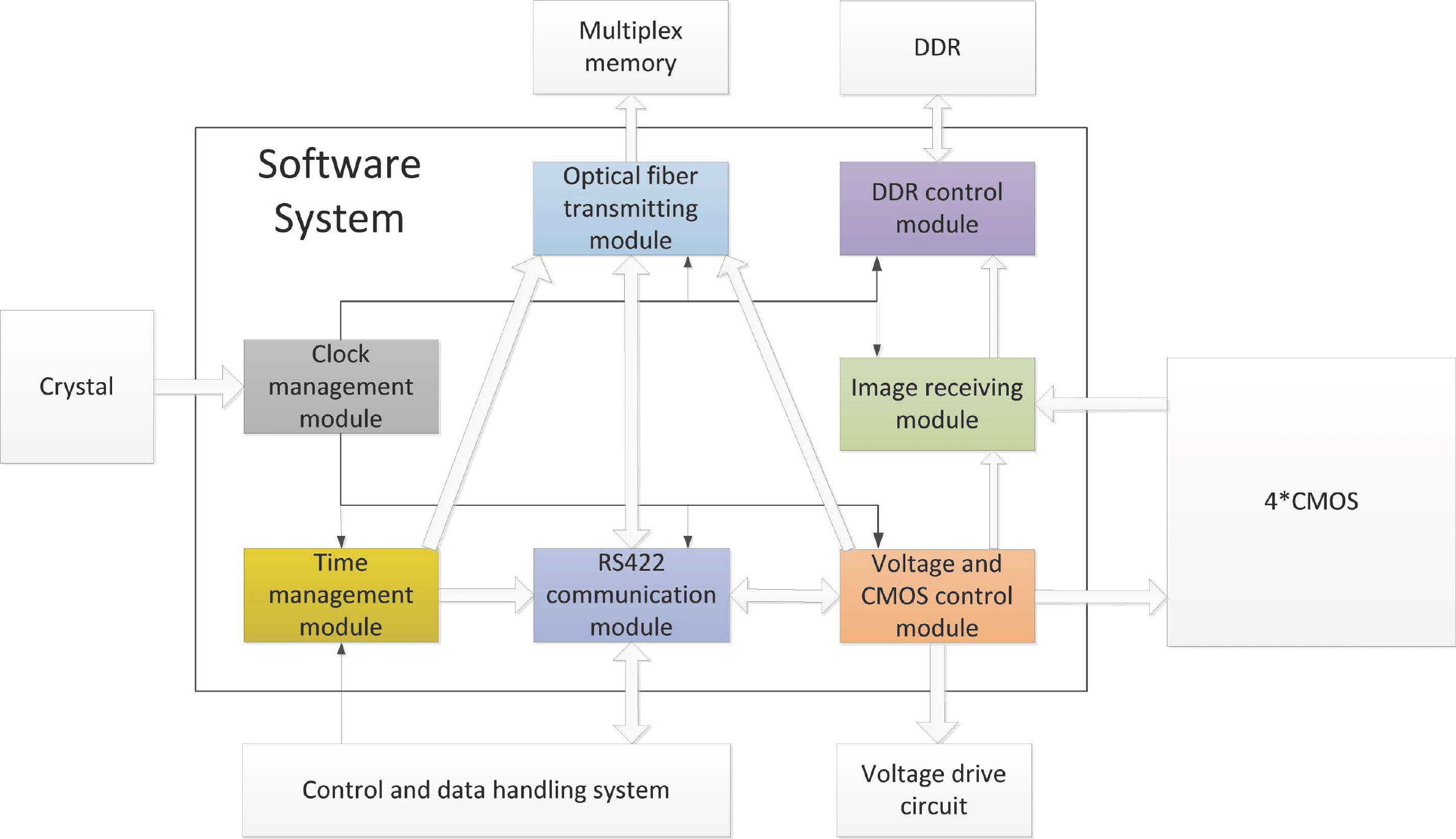}
    \caption{The system software block diagram. The software consists of seven modules to operate the CMOS modules for taking data. }
    \label{fig:6.11}
\end{figure}

Based on the types of system tasks, the operational software is divided into 7 functional modules, as shown in Figure \ref{fig:6.11}. The clock management module generates clock and reset signals for all modules. The time management module is responsible for receiving pulse per second (PPS) signals from the control and data handling system, and managing the system time. A RS422 communication module is responsible for receiving commands and transmitting telemetry information. A voltage and CMOS driver module controls the voltages of the four CMOS detectors, configures their registers, and generates their control timing. The image receiving module is responsible for receiving serial pixel data from four CMOS modules obtained with low-voltage differential signaling (LVDS). A DDR control module is responsible for reading and writing images to DDR memory, and an optical fiber transmitting module is responsible for packaging and transmitting images to a multiplex memory device.

\subsubsection{Camera Resources}

Each of the seven CMOS cameras has a processing circuit envelope of \SI{306}{\mm}$\times$\SI{326}{\mm}$\times$\SI{240}{\mm} and an estimated weight of \SI{14}{\kg}. 


The peak power consumption of the single chip CMOS detector is about \SI{1.59}{\watt}; the power consumption of the single detector driver board is about \SI{1.5}{\watt}; the power consumption of the FPGA board is about \SI{15}{\watt}; the peak power consumption of the TEC control board is about \SI{55}{\watt}; and the average power consumption is about \SI{24}{\watt} (the exact value is related to the temperature of the cooling surface ). The peak power consumption per camera is about \SI{114}{\watt}, and the average power consumption is about \SI{70}{\watt}. Therefore, the total peak and average power consumption of the seven cameras is \SI{798}{\watt} and \SI{490}{\watt}, respectively.

At present, the fastest frame readout frequency of the detector is \SI{0.68}{fps} and the combined size of the image data from the four detectors is \SI{\sim5}Gbit per frame. However, since the data is encoded before being transmitted through the optical fiber, the data will increase to a final size of about \SI{6.25}Gbit (\SI{5}Gbit$\times1.25$) per frame. Since the total read out time for each CMOS camera in 1.5 s, a net data rate for each camera is \SI{\geq 4.17} Gbps (6.25/1.5). Considering invalid data periods such as frame intervals and row intervals, the data rate is required to be $\geq$5.5Gbps. The total data rate of seven cameras is $\geq$38.5 Gbps.

\subsection{Thermal Subsystem}

In order to meet the requirement of high photometry precision, the detectors need to work at \SI{-40\pm0.1}{\degreeCelsius} while the TOUs operate at \SI{-30\pm0.3}{\degreeCelsius}. Heat dissipation from the focal plane assembly on each telescope needs to be at least \SI{6}{\watt}@\SI{-40}{\degreeCelsius}, while heat generated by camera electronics is about \SI{70}{\watt} at room temperature.

ET will operate at the L2 Halo orbit and a sunshield will block radiation from the Sun, Earth, and Moon.
The detectors will be passively cooled to below \SI{-40}{\degreeCelsius}. Heat from the detectors will be conducted to the telescope top hood through two heat pipes where it will be radiated away from the spacecraft. A PID heating loop is used to keep the temperature of the detectors stable at \SI{-40\pm0.1}{\degreeCelsius}.
The heat of camera electronics will be dissipated partly through the external radiation of the camera shell, while the rest will be removed to the front part of the telescope via a loop heat pipe. 

The temperature of the TOUs will be stabilized at \SI{-30\pm0.3}{\degreeCelsius} through active zone heating from the heaters. Each TOU is coated with a multilayer polyimide membrane to reduce heat leakage. Thermal management electronics are used to measure the temperature of each telescope and control power of the heaters, so as to realize the closed-loop control of temperature.  

\begin{figure}[htbp]
    \centering
    \includegraphics[width=0.6\textwidth]{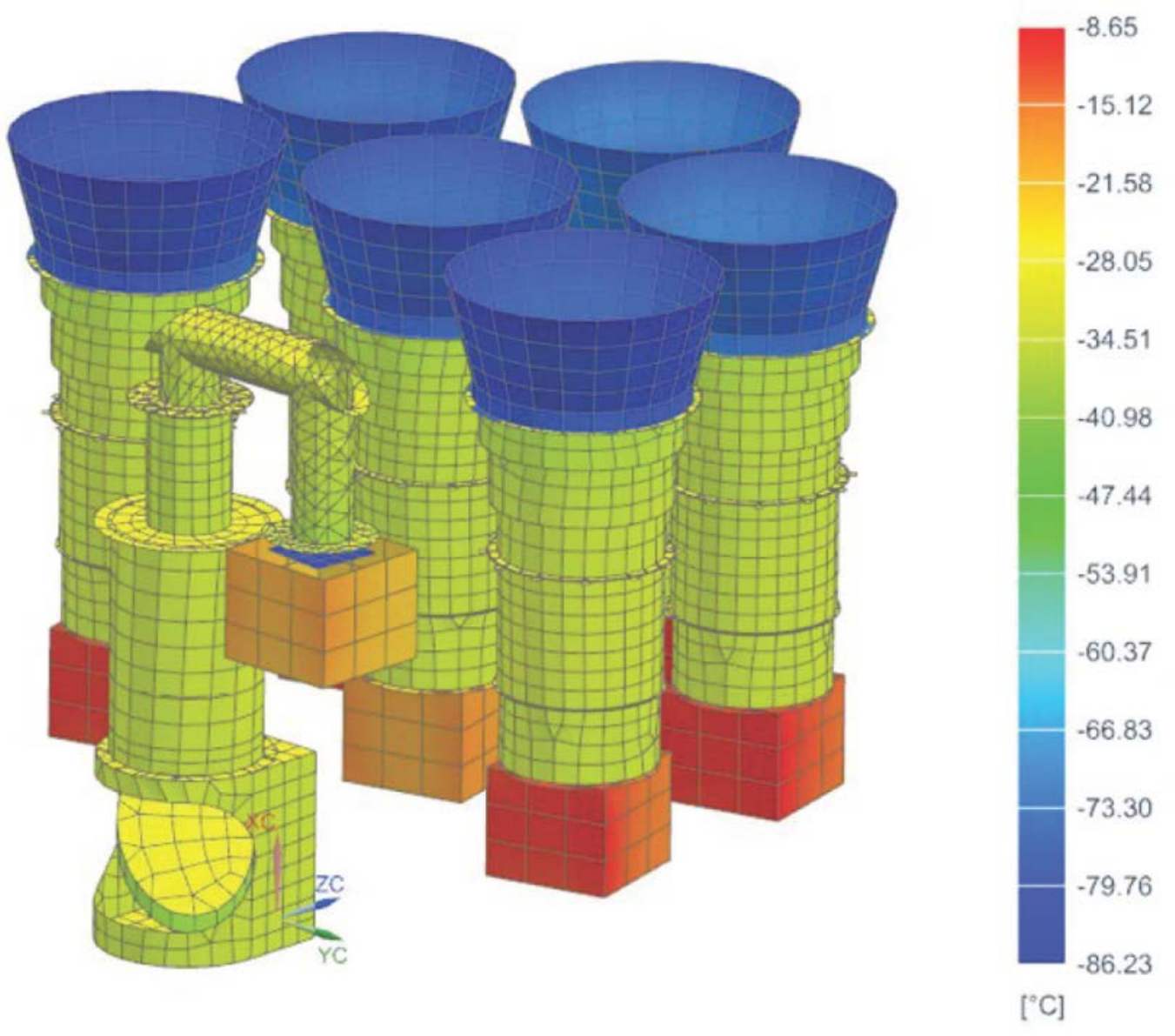}
    \caption{Temperature distribution of the transit and microlensing telescopes with the telescope heater's PID control loop on to keep temperatures constant.}
    \label{fig:6.13}
\end{figure}

Preliminary thermal simulations have been carried out to estimate the temperature distribution of the telescopes; the result is shown in Figure \ref{fig:6.13}. The temperature gradients along the TOUs are controlled to less than \SI{3}{\degreeCelsius}.

\subsection{On-board Data Processing}  

Each telescope uses four 9k$\times$9k CMOS detectors, which provides a total of 18k$\times$18k pixels. Each transit telescope will take exposures at a cadence of \SI{11.5}{\s}, which includes \SI{10}{\s} integration and \SI{1.5}{\s} readout time. Consequently, the six transit telescopes will produce up to \SI{2.81}Tbit of data per day. To reduce data transmission and storage requirements, operations including image stacking, target selection, requantization, and data compression will be preformed on-board. 

The transit telescopes' image data will be stacked at \SI{34.5}{\s} (3 stacking), \SI{10}{\minute} (52 stacking), or \SI{30}{\minute} (156 stacking), according to different scientific goals. Then, sub-images of each star, having an area of $\sim$36 pixels, will be extracted based on a pre-selected subset of 0.8 to 1.2 million target stars. Both image stacking and target extraction are carried out in the FEE. The full frame images (FFIs) and cropped data will then be transmitted to the spacecraft and service module (SVM) for requantization and compression. The size of each FFI is about 1.4Tb while the net size of the imagettes (postage stamps) of pre-selected stars is about 140.4Gb (23.4Gb$\times$6) per day, after requantization and compression. Data processing operations and relative requirements are shown in Table \ref{tab:On-board_data_processing_operations_and_requirements}.

\begin{table}[htbp]
    \centering
        \captionsetup{justification=centering}
    \caption{\centering On-board data processing operations and requirements (for one telescope)}
    \begin{tabular}{|l|l|l|l|l|}\hline
    Operation & Cadence(s) & Algorithm & \makecell[l]{Calculation\\ amount(pixels)} & \makecell[l]{Memory\\ requirements\\ (Gb)} \\ \hline
    \makecell[l]{Image\\ stacking} & 11.5 & \makecell[l]{Matrix\\ addition} & 18k$\times$18k$\times$2 & 10.125 \\ \hline
    \makecell[l]{Target\\ extraction} & \makecell[l]{30$\times$1\\ 30$\times$10\\ 30$\times$30} & \makecell[l]{Matrix\\ extraction} & 18k$\times$18k$\times$2 & 	10.125 \\ \hline
    Requantization & \makecell[l]{30$\times$1\\ 30$\times$10\\ 30$\times$30} & \makecell[l]{Matrix\\ addition\\  Matrix\\ subtraction} & 1200k$\times$36 & $\sim$0.15 \\ \hline
    \makecell[l]{Data\\ compression} & \makecell[l]{30$\times$1\\ 30$\times$10\\ 30$\times$30} &	\makecell[l]{Huffman\\ Coding} & 1200k$\times$36 & $\sim$0.15 \\ \hline
    \end{tabular}
    \label{tab:On-board_data_processing_operations_and_requirements}
\end{table}

The microlensing telescope will take exposures at a cadence of \SI{10}{\minute}, which means it will produce 1.4Tb of data per day after compression. Ten FFIs will be downloaded every day from the microlensing telescope. After real-time processing on the ground, a list of target stars with significant ($\geq5\sigma$) flux increases will be uploaded to the onboard computer. Then FFIs containing these targets which have been stored within the previous five days will be selected. An imagette of size 100$\times$100 pixels around each target will be extracted and downloaded. The daily download data limit is 35Gb, including 10 FFIs and 5-day imagettes of 150 targets.

All saved data will be stored on-board for 2-5 days, depending on different scientific requirements. With a 50$\%$ redundancy backup to ensure data security, the net storage space required on the satellite is about 10Tb. There will be 110 GB to 175.4 GB of data to download per day according to different scientific goals; see details in Table \ref{tab:Data_to_be_downloaded}.

\begin{table}[htbp]
    \centering
        \captionsetup{justification=centering}
    \caption{\centering Data to be downloaded for different scientific goals of ET (each camera)}
    \begin{tabular}{|l|l|l|l|l|l|l|}\hline
    Scientific goal & \makecell[l]{Target\\ quantity} & Star type & \makecell[l]{sample\\ interval} & \makecell[l]{Pixels\\/star} & \makecell[l]{Basic\\ data\\ volume\\ (Gb\/day\\/camera)} & \makecell[l]{Max\\ data\\ volume\\ (Gb\\/day\\/camera)} \\ \hline
    \makecell[l]{Planet\\ Black hole\\ Variable star} & \makecell[l]{800K\\$\sim$1,200K} & \makecell[l]{Main\\ sequence star\\ brighter than\\ mag 16} & 30min & 36 & 5.6 & 8.3 \\ \hline
    \makecell[l]{Time domain\\ astronomy} & 2M & \makecell[l]{Download\\ pixels with\\ varying\\ luminosity\\ only} & 30min & 36 & 0.1 & 0.1 \\ \hline
    \makecell[l]{stellar\\ seismology} & \makecell[l]{10K\\$\sim$20K} & \makecell[l]{Dwarf\\ and subgiants} & 30s & 36 & 4.1 & 8.2 \\ \hline
    \makecell[l]{Stellar\\ archaeology} & \makecell[l]{40K\\$\sim$100K} & Giant stars & 10min & 36 & 0.8 & 2.1 \\ \hline
    \makecell[l]{Stellar\\ archaeology\\ Black hole} & \makecell[l]{400K\\$\sim$900K} & \makecell[l]{Main\\ sequence\\ stars dimmer\\ than mag 16} & 1h & 36 & 1.4 &3.1 \\ \hline
    \makecell[l]{Long period\\ variable star} &	500K & \makecell[l]{Giant stars\\ darker than\\ mag 16} & 6h & 36 & 0.3 & 0.3 \\ \hline
    Calibration data & & & & & 0.02 & 0.02 \\ \hline
    Full images	& & & \makecell[l]{Weekly\\ or daily} & 18k$^2$	& 0.2 & 1.3 \\ \hline
    \makecell[l]{Total of\\ Transit\\ telscope} & & & & & 12.5 & 23.4 \\ \hline
    \makecell[l]{Microlensing\\ telescope} & - & Full image & 10/day & 18k$^2$ & 35 & 35 \\ \hline
    \end{tabular}
    \label{tab:Data_to_be_downloaded}
\end{table}

\section{Mission Design}
{\bf Authors:}
\newline
Wen Chen$^1$, Kun Chen$^1$, Xingbo Han$^1$, Yingquan Yang$^1$, Haoyu Wang$^1$, Xuliang Duan$^1$, Jiangjiang Huang$^1$, Hong Liang$^1$, Yong Yu$^2$, Jian Ge$^2$, \& Zhenghong Tang$^2$\\
{1. \it Innovation Academy for Microsatellites, Chinese Academy of Sciences, Shanghai, China}\\
{2. \it Shanghai Astronomical Observatory, Chinese Academy of Sciences, Shanghai, China}\\

\subsection{Mission Operations} 
\subsubsection{Proposed Mission Profile}
The ET mission requires continuous observation of a fixed region of sky with high precision photometry. The ET spacecraft will operate at the L2 halo orbit to ensure a stable environment for the duration of its $\geq$ 4-year lifetime. ET is an inertial pointing spacecraft with three-axis stabilization, having a telescope pointing accuracy of better than \SI{1.5}{\arcsecond}. The telescope pointing stability will keep high frequency jitters (up to \SI{10}{\hertz}) within \SI{0.15}{\arcsecond} and long-term thermal drift will be controlled to within \SI{0.4}{\arcsecond}. X-band will be used for telemetry, tracking, command, and the science data downlink from the spacecraft. A daily scientific data volume of roughly 169 Gb will be downloaded at a data rate of 20 Mbps via phased array antennas. The entire spacecraft including payloads weighs 3.2 tons and the long-term power consumption is about 1500 Watts. ET is expected to launch from Xichang Satellite Launch Center in a CZ-3B rocket by the end of 2026.

\subsubsection{Mission Orbit}
The ET spacecraft needs stable pointing and continuous observation of a fixed field. The L2 Halo orbit is the selected option for the orbit; it is favorable for overall stability because of its lack of gravity gradients, the time available for observation, and the environmental conditions characterized by low total ionizing doses. The thermal condition over the orbit is stable with small fluctuation, and the capabilities of TT\&C (Telemetry, Tracking and Command) and data transmission from the L2 orbit can meet mission requirements.

Considering the effects of the Sun's transit on spacecraft communication and the need to minimize the fuel required for orbit maintenance, the proposed L2 Halo orbit with amplitude is 110,000 km. 

\begin{figure}[hbp]
\centering
    \includegraphics[width=0.85\textwidth]{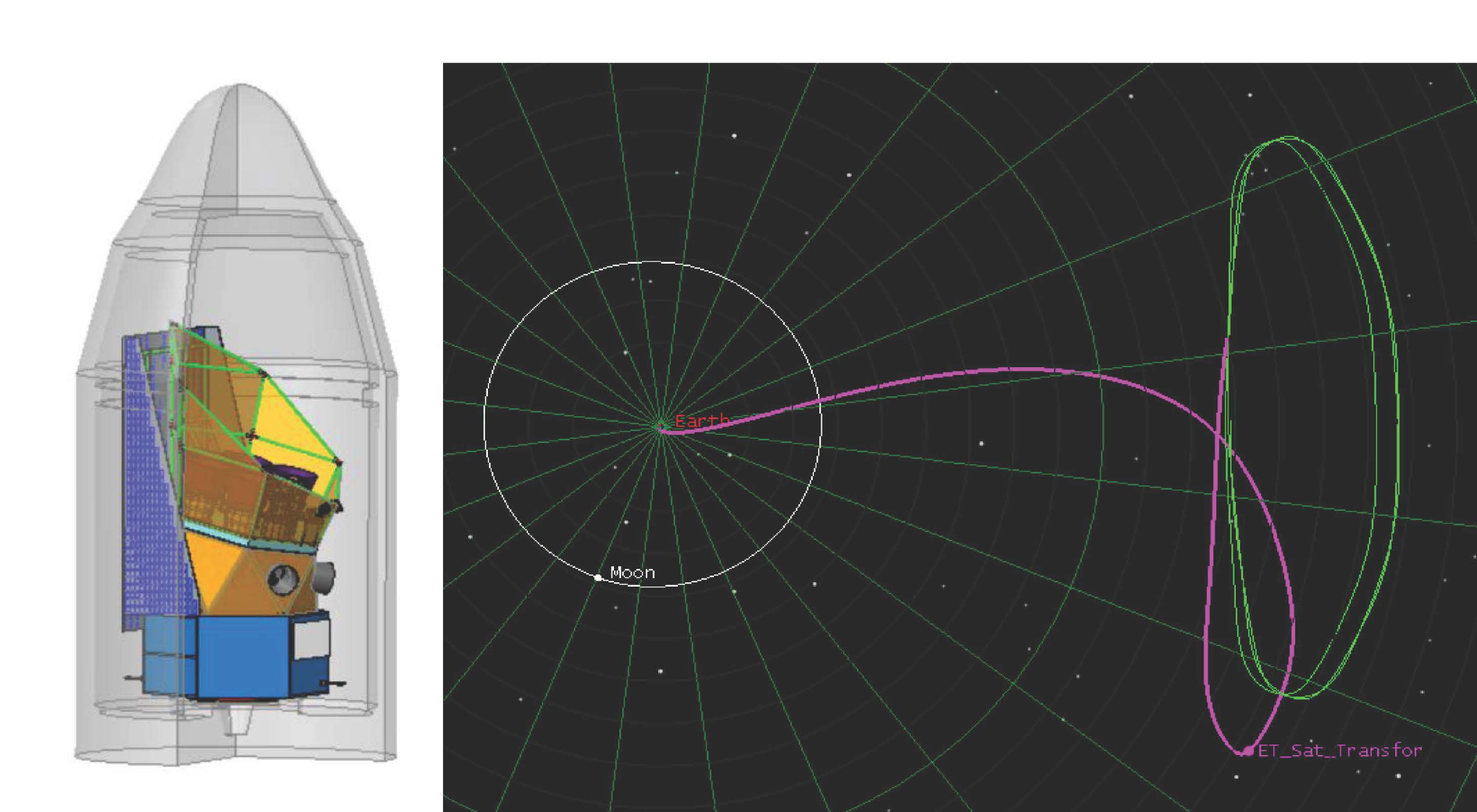}
    \caption{Left: The launch configuration in the 4000F fairing of the CZ-3B rocket. Right: The ET orbit (green curve) and orbital transfer (purple curve) simulation diagram.}
    \label{fig:orbit}
\end{figure}

ET will be sent by CZ-3B directly into the invariant manifold of the L2 orbit. The launcher can provide an envelope size of $\Phi$3650mm (4000F fairing) for the spacecraft, which includes enough space for inspection after closure. The spacecraft will preform orbital corrections and injection to the L2 orbit on its own after being sent to the invariant manifold by the launcher. The launch configuration and orbit simulation are shown in Figure \ref{fig:orbit}.

\subsubsection{Pointing Strategy}
\begin{figure}[htp]
    \centering
    \includegraphics[width=0.6\textwidth]{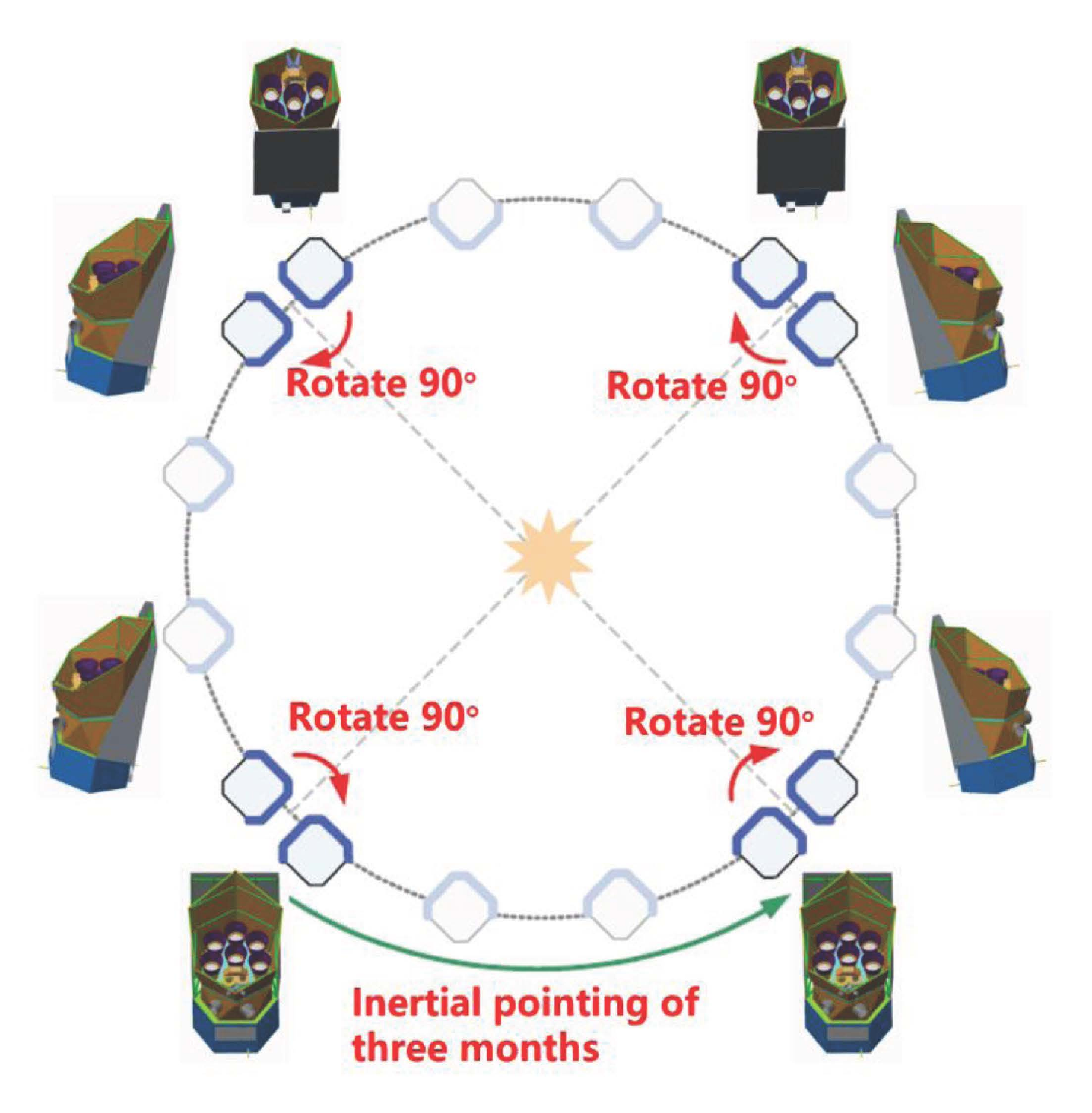}
    \caption{The ET pointing and attitude maneuver strategy. The ET spacecraft will be rotated by \SI{90}{\degree} about every 3 months around the transit telescope's pointing axis to maintain the solar arrays face towards the Sun while the scientific payload away from the Sun to keep its thermal stability. }
    \label{fig:pointing}
\end{figure}

ET's pointing is $RA=19^h22^m$ and $Dec=44^\circ30'$. Since ET observations require lengthy exposures of a fixed sky region, the telescope requires stable pointing for several years. The microlensing telescope will stare in the direction of the Baade's window near the Galactic Center from the spring equinox to autumn equinox. Due to orbital and environmental conditions, ET will be attitude maneuvered by \SI{90}{\degree} about every 3 months about its pointing axis. This strategy will ensure a sufficient energy supply and stable thermal environment for the scientific payload, as well as minimize stray light (see Figure \ref{fig:pointing}).

\subsubsection{Orbit Environment Analysis}
\begin{figure}[htbp]
    \centering
    \includegraphics[width=0.7\textwidth]{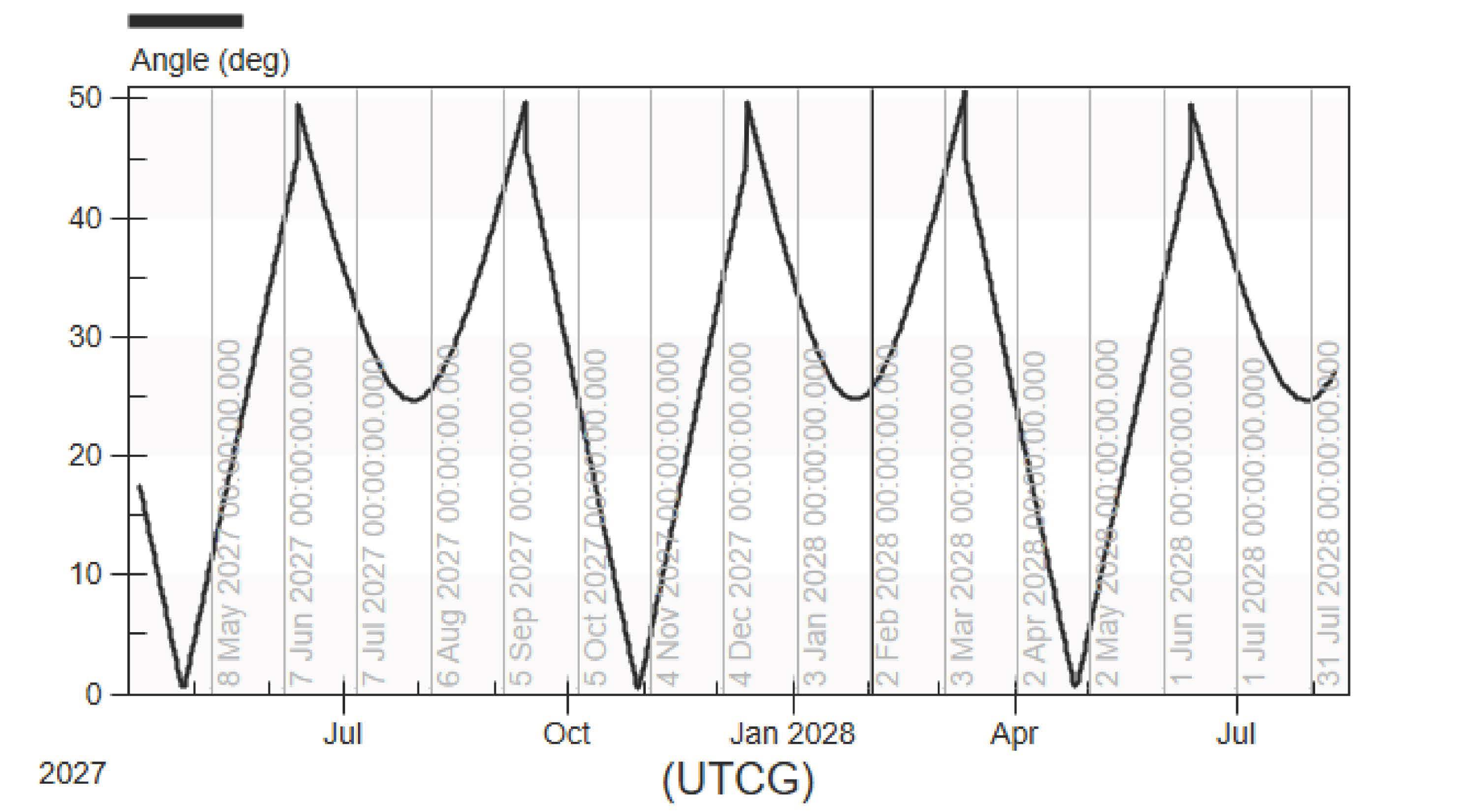}
    \caption{Solar incidence angle changes in one year (i.e., the included angle between the sunlight direction and the normal direction of the spacecraft's sunny side (the +Z axis) )}
    \label{fig:solar_incidence}
\end{figure}

\begin{figure}[htbp]
 \centering
    \includegraphics[width=0.90\textwidth]{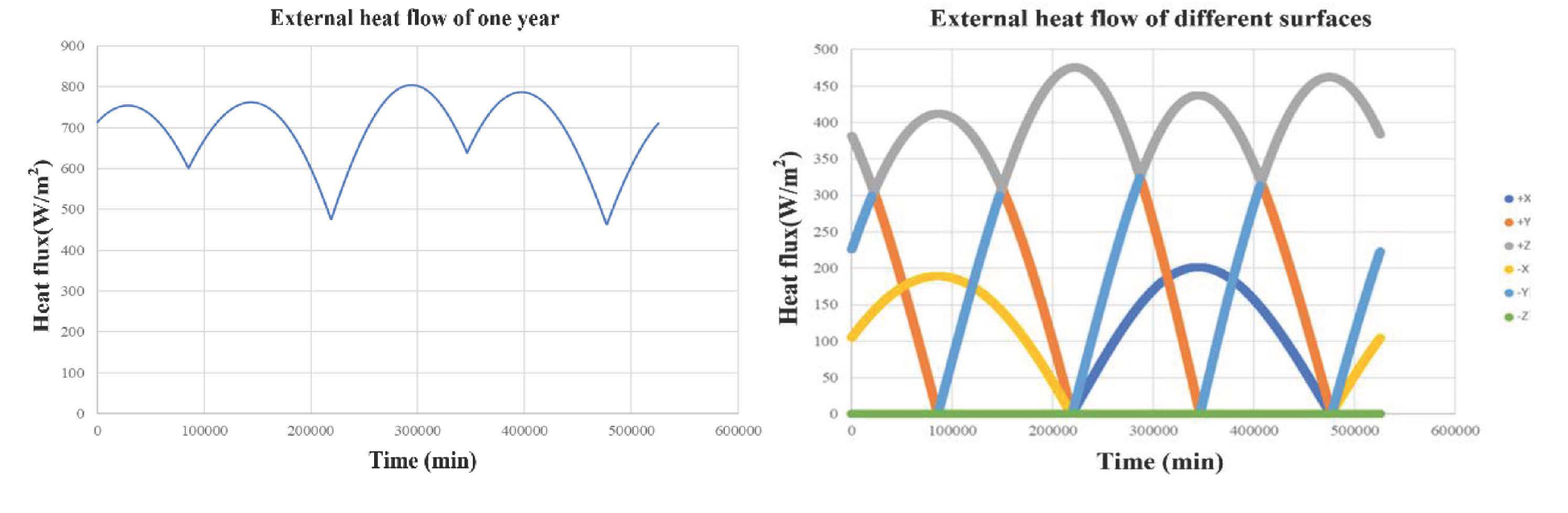}
    \caption{The total external heat flow in one year(left) and the external heat flow on each surface of the spacecraft(right)}
    \label{fig:projection}
\end{figure}

Considering the orbit environment and the pointing strategy during the mission, the ET team conducted an analysis of the Solar incidence angle and external heat flow over one year. The analysis results are shown in Figure \ref{fig:solar_incidence} and Figure \ref{fig:projection}. Heat flow of the spacecraft at the L2 halo orbit is relatively stable throughout the year. The opposite side of the solar arrays of the spacecraft (i.e., the -Z direction) will not receive any direct sunlight over the entire year. This side naturally serves as a radiating surface, and the radiant panel will be installed on this side.

\begin{figure}[htbp]
 \centering
    \includegraphics[width=0.90\textwidth]{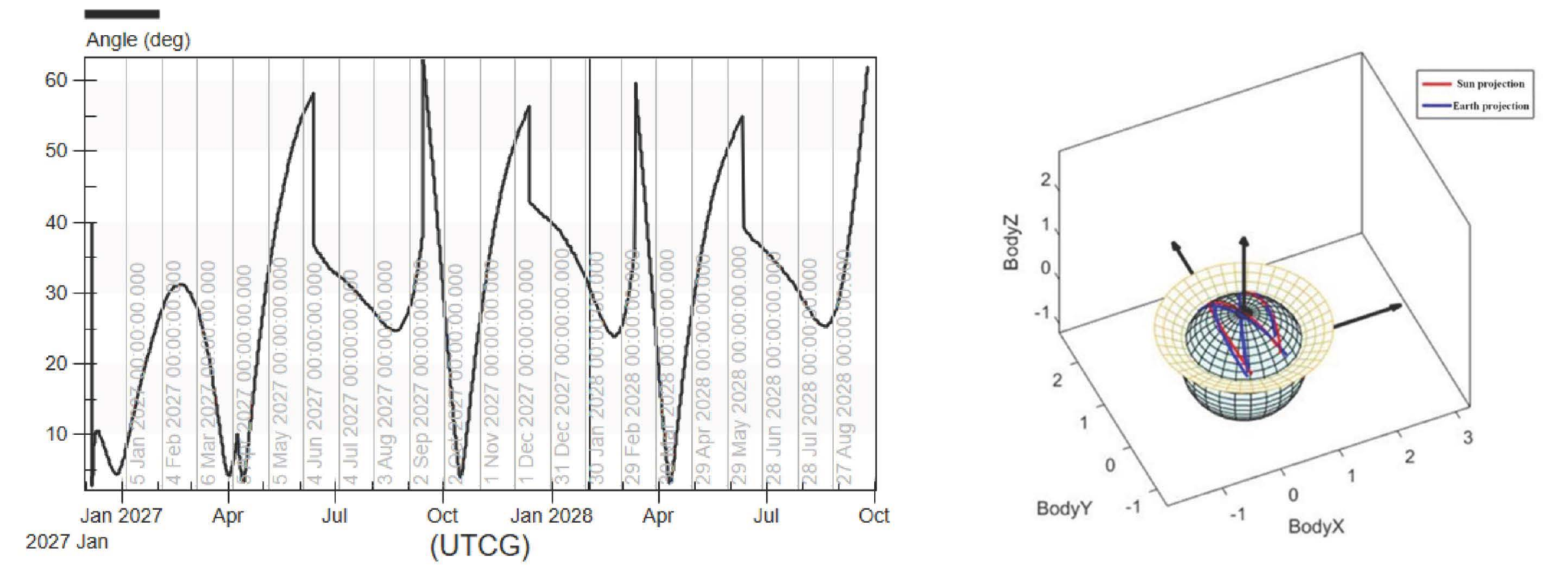}
    \caption{(Left panel) The included angle between the spacecraft +Z axis and the spacecraft-Earth axis between January 2027 and October 2028. The largest angle is smaller than 65 degrees. (Right panel) The spacecraft orientation and the Earth and Sun projection direction. }
    \label{fig:included_angle}
\end{figure}

In addition, the ET team conducted an analysis of the included angle between the spacecraft +Z axis and the spacecraft-Earth axis to study the feasibility of daily data down-linking. The analysis results are shown in Figure \ref{fig:included_angle}. It is clear that data downloading can be conducted on a daily basis by designing a proper antenna scanning range of e.g. 65 degrees.

\subsection{Key Design Challenges}  

\subsubsection{Pointing Stability Requirement}
To detect Earth 2.0s using the transit method, ET must have extremely high photometry precision  (\SI{\sim34}{ppm}/\SI{6.5}{\hour})  with high spacecraft stability. In addition, to detect cold and free-floating terrestrial planets,  ET's microlensing telescope must have high image stability in every \SI{10}{\minute} exposure and thus high spacecraft stability as well.

Based on these requirements for both the transit and microlensing telescopes, the ET spacecraft is required to have inertial pointing with three-axis stabilization. Specifically, the spacecraft's high frequency jitter (up to 10 Hz) must be maintained within \SI{0.15}{\arcsecond} / \SI{5}{\s}, while its long-term thermal drift (i.e., low frequency drift) needs be controlled to within \SI{0.4}{\arcsecond}/\SI{30}{\minute}.

\paragraph{Preliminary Solutions and Strategy}
The challenge of attaining high pointing stability can be addressed through the following key technical developments:
\begin{enumerate}
    \item Micro-vibration suppression: The interference of the reaction wheels
    is the main source of the micro-vibration on the spacecraft. This vibration could be attenuated by a specially designed vibration isolator. One kind of these passively activated wheel vibration isolators (see Figure \ref{fig:wheel}) has been manufactured and has passed an environmental testing, which demonstrated a \SI{90}{\percent} micro-vibration reduction. This vibration isolator will have on-board verification in July of 2022.
    
\begin{figure}[htbp]
 \centering
    \includegraphics[width=0.85\textwidth]{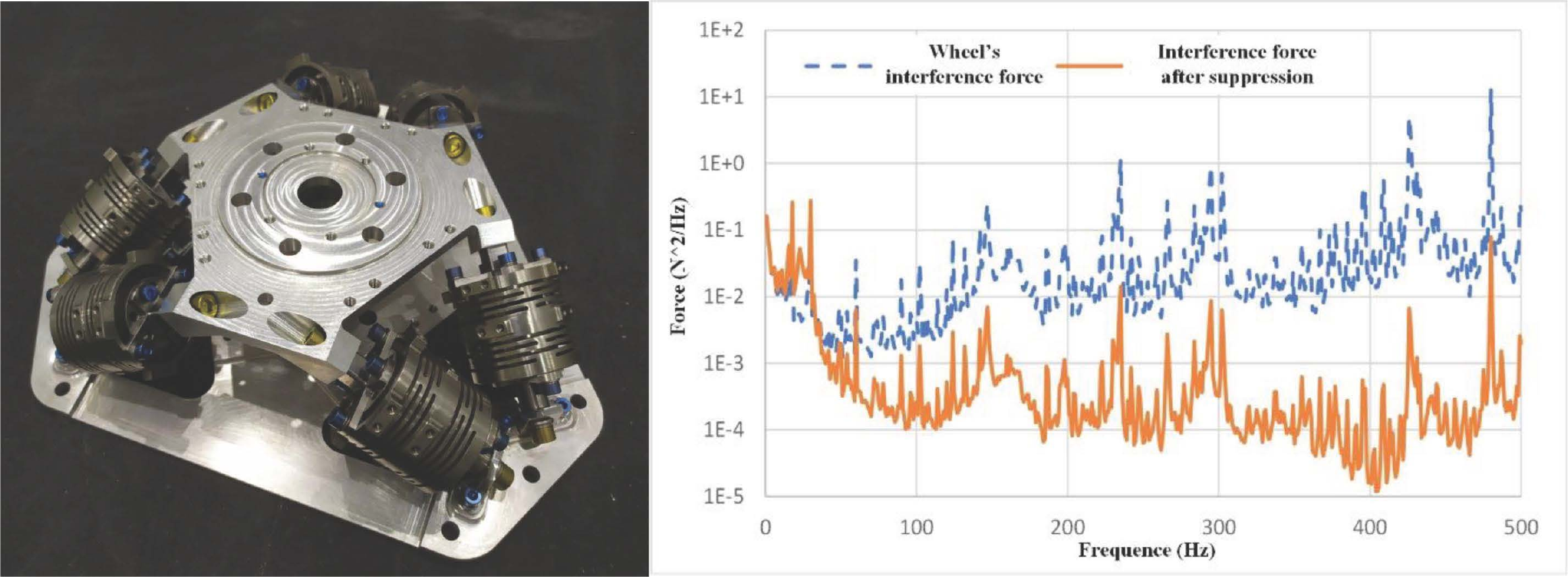}
    \caption{(Left panel) A prototype of a wheel vibration isolator; (Right panel) Micro-vibration reduction test result}
    \label{fig:wheel}
\end{figure}
    
    \item Ultra-high-precision Fine Guidance Sensors (FGSs): At least two sets of FGSs will be used on ET. Each set has four guide star detectors installed around the large-scale main science CMOS detectors (see Figure \ref{fig:detector_layout}). Our simulations show the guiding measurement accuracy could reach \SI{8}{mas} in the translational direction and \SI{31}{mas} for roll measurements.
    
      \begin{figure}[htbp]
        \centering
        \includegraphics[width=0.85\textwidth]{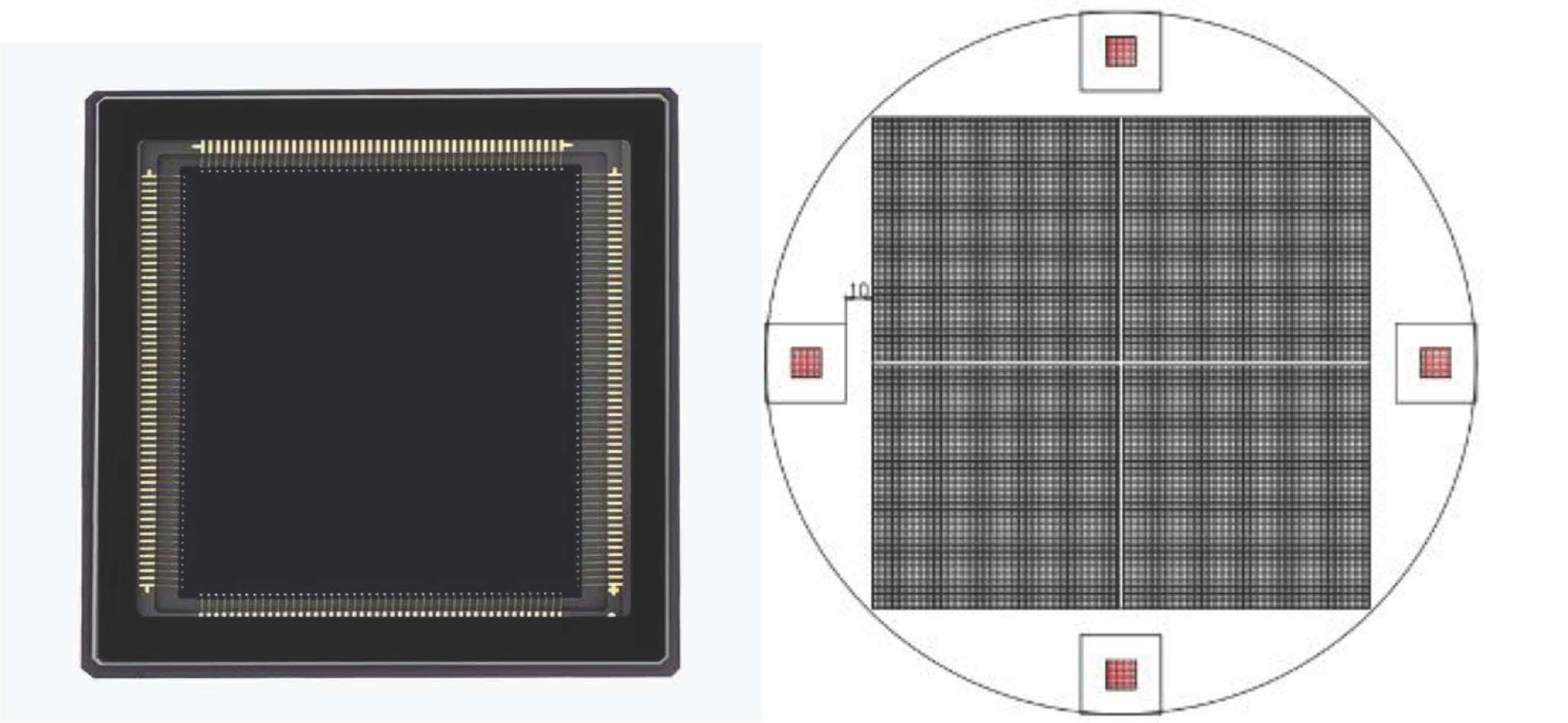}
        \caption{(Left panel) The GSENSE2020BSI CMOS detector to be used as a fine guidance sensor; (Right panel) layout diagram of four guidance detectors around the mosaic of four 9k$\times$9k GSENSE1081BSI CMOS detectors}.
        \label{fig:detector_layout}
    \end{figure}

    \item The high-precision attitude determination algorithm: Data from multi-sensors including FGSs, star trackers, and gyroscopes are fused to improve the attitude measurement precision (see Figure \ref{fig:block}). 
    
    \begin{figure}[htbp]
        \centering
        \includegraphics[width=0.85\textwidth]{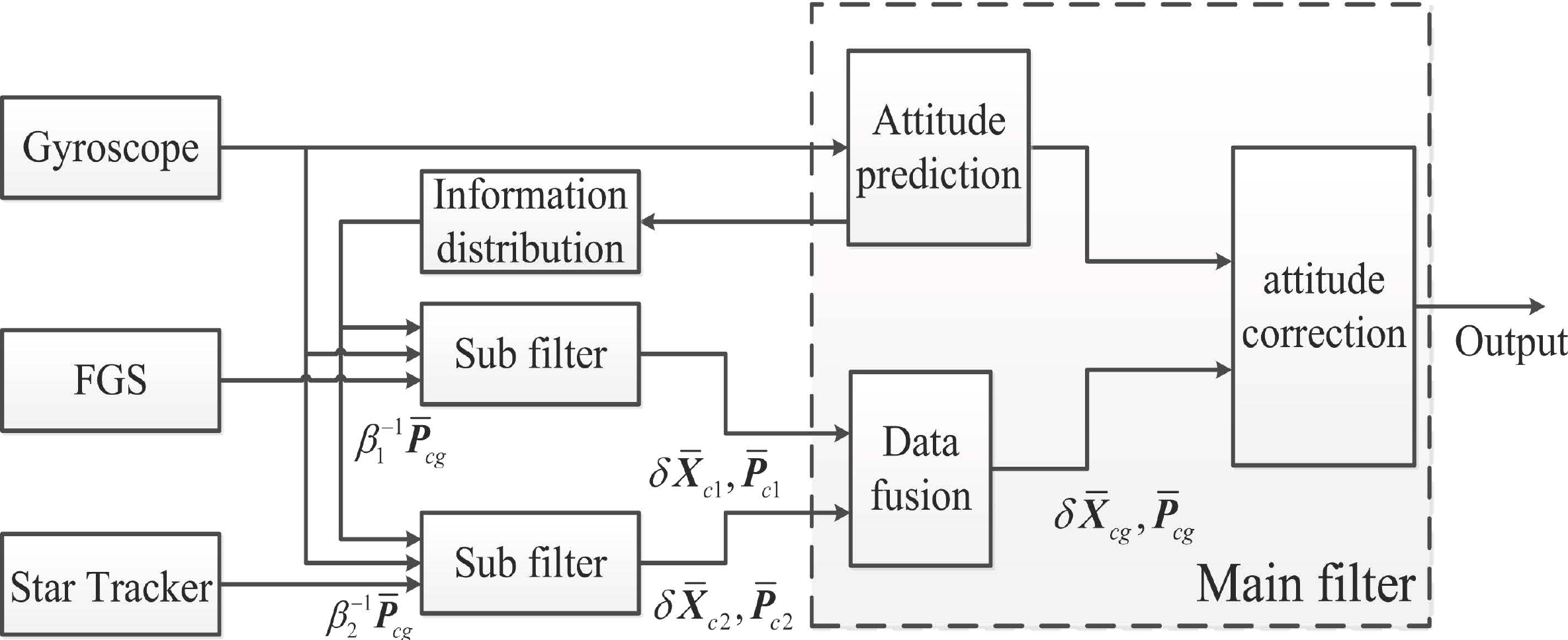}
        \caption{A block diagram for the multi-sensor data fusion algorithm used in ET spacecraft attitude control. Data from FGSs, star-trackers, and gyroscopes are processed for precision attitude measurements. }
        \label{fig:block}
    \end{figure}
    
    \item The high-precision attitude control algorithm: Knowledge from the Hubble satellite, applications of the PID control, structural filter, smoothing filter, and observer can help with achieving better stability control.
\end{enumerate}

\subsection{Spacecraft Design}  

\subsubsection{Spacecraft Overview}
The main technical specifications of the spacecraft design are shown in Table \ref{tab:spacecraft_technical_specification}.

\begin{table}[htbp]
  \centering
      \captionsetup{justification=centering}
  \caption{Spacecraft technical specification} \label{tab:spacecraft_technical_specification}
  \begin{tabular}{|l|l|l|}
  	 \hline
  	 \multicolumn{2}{|l|}{Items} & Technical specification \\ \hline
  	 \multicolumn{2}{|l|}{Launch mass} & $\sim$ 3.2t@CZ-3B \\ \hline
  	 \multicolumn{2}{|l|}{Envelope size} & $\phi2934\times5125$ \\ \hline
     \multirow{4}{*}{\makecell[l]{Thermal \\ control}} & Method & \makecell[l]{Combination of active \\ and passive thermal control} \\ 
     \cline{2-3} & Detectors & -40$\pm0.1$\textcelsius \\ 
     \cline{2-3} & Platform units & -15$\sim$+45\textcelsius \\ 
     \cline{2-3} & Optical components & $\pm$0.3\textcelsius \\ 
     \hline
     \multirow{3}{*}{Power supply} & Solar array & 11.0m$^2$ GaAs cell \\ 
     \cline{2-3} & Battery & 120Ah  lithium ion battery \\
     \cline{2-3} & Bus voltage & 30$\pm$1 V \\
     \hline
     \multirow{4}{*}{\makecell[l]{Attitude and \\ Orbit Control \\ System}} & Type & \makecell[l]{Three-axis stabilization, \\ zero momentum control} \\ 
     \cline{2-3} & Stability  & \makecell[l]{Jitter: 0.15"/30min \\ drift: 0.4"/30min}
 \\
     \cline{2-3} & Thruster & Attitude thruster 12$\times$5N \\
     \cline{2-3} & Propellant & 492.3 kg \\
     \hline
     \multirow{2}{*}{\makecell[l]{Telemetry, Tracking, \\ and Command}} & Telemetry rate & 512bps, 2048bps, 4096bps \\ 
     \cline{2-3} & Telecommand rate & 500bps, 1000bps, 2000bps \\
     \hline
     \multirow{3}{*}{\makecell[l]{Data \\ transmission}} & Working band & X-band \\ 
     \cline{2-3} & Information rate & 20 Mbps \\
     \cline{2-3} & Storage capacity & 10 Tbit \\
     \hline
     \multirow{2}{*}{\makecell[l]{On-board \\ data handling}} & CPU & AT697 \\ 
     \cline{2-3} & Dominant frequency & 80 MHz \\
     \hline
     \multirow{2}{*}{\makecell[l]{Lifetime and \\ reliability}} & Lifetime & $\geq$4 years \\ 
     \cline{2-3} & Reliability & $\geq$0.61 End of life \\
     \hline
  \end{tabular}
\end{table}

\begin{figure}[htbp]
    \centering
    \includegraphics[width=0.6\textwidth]{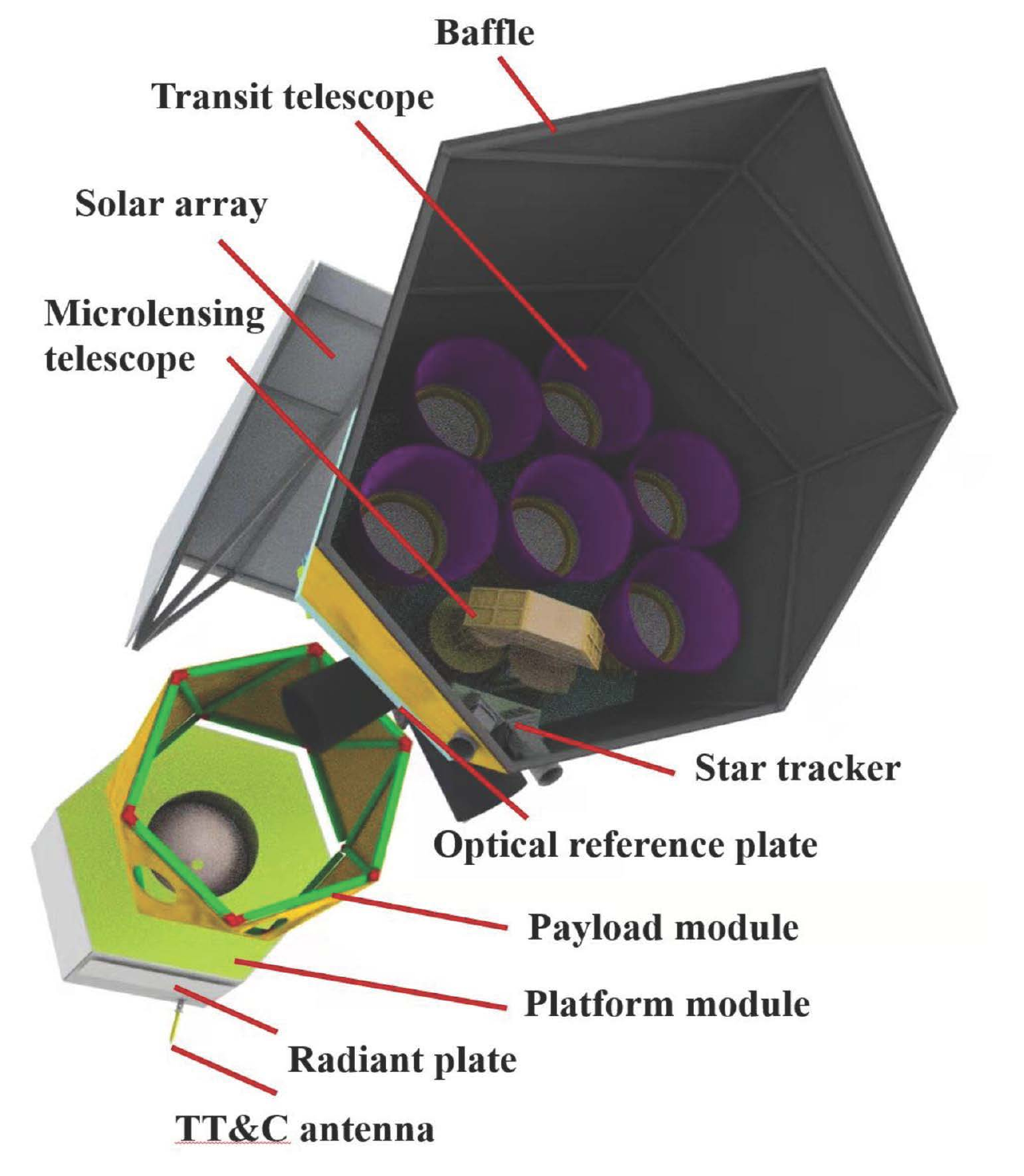}
    \caption{Structure of the ET spacecraft. The spacecraft includes a payload module and a platform module. Six transit and one microlensing telescopes are installed on an reference plate. }
    \label{fig:structure}
\end{figure}

\paragraph{Structure}~{}
The spacecraft structure is designed to meet the payload's stability requirements. The ET spacecraft comprises the payload module and platform module, separately. Seven telescopes are integrated on an optical reference plate which is supported by a carbon fiber bracket, making up the payload module. The platform module uses a hexagonal bearing cylinder structure. This design ensures the minimization of gravity deformation, thermal deformation, and micro vibration. Figure \ref{fig:structure} shows the layout design.

\paragraph{Thermal Control}~{}
The thermal control strategy includes both active and passive ones. During its operation in space, the temperature of the platform units shall be controlled in the range of \SIrange[]{-15}{45}{\degreeCelsius}. For the optical reference plate and payloads, high precision thermal control is required to minimize telescope's image drift and size change for high precision photometry measurements.

\paragraph{AOCS}~{}
The Attitude and Orbit Control System (AOCS) includes a measurement sensor and control actuator. High precision measurements are achieved using the FGSs, gyros, and star trackers. Reaction wheels are used for attitude control while thrusters are mainly used for momentum dump of the spacecraft.

\paragraph{Power Supply}~{}
The spacecraft adopts \SI{30\pm1}{\volt} full adjustment busbar. Power is supplied by $11m^2$ solar arrays with a \SI{120}{Ah} battery capacity.

\paragraph{Communication}~{}
ET adopts an X-band Telemetry, Tracking, and Command (TT\&C), and data transmission integrated communication strategy. A phased array antenna with a \SI{\pm65}{\degree} scanning range is used for data transmission. There are no rotating mechanisms involved to ensure high stability while enabling a large volume of data downloading on a daily basis.
A wide beam omnidirectional antenna is adopted for TT\&C, using a multiple side-tone system capable of sending DOR beacon signals. 

\paragraph{OBDH}~{}
The On-Board Data Handling computer (OBDH) adopts a dual processing system, centralized management strategy, and one is cold standby.

\subsubsection{Spacecraft Budgets}
The ET spacecraft budgets are listed in Table \ref{tab:spacecraft_mass_budgets} and Table \ref{tab:spacecraft_power_consumption_budgets}:

\begin{table}[htbp]
    \centering
        \captionsetup{justification=centering}
    \caption{\centering Spacecraft mass budgets}
    \begin{tabular}{|l|c|c|}\hline
         Subsystems & Mass(kg) & Remark  \\ \hline
         Payloads & 1242 & \\ \hline
         Structure & 432.7 & \\ \hline
         Thermal control & 39 & \\ \hline
         Power supply and harness & 61.8 & \\ \hline
         AOCS & 411.28 & \\ \hline	
         Communication & 65.2 & \\ \hline 	
         OBDH & 27 & \\ \hline
         Dry mass & 2735.9 & with 20\% margin \\ \hline
         Propellant & 492 & with 20\% margin \\ \hline
         Total & 3228 & \\ \hline	
    \end{tabular}
    \label{tab:spacecraft_mass_budgets}
\end{table}

\begin{table}[htbp]
    \centering
    \caption{Spacecraft power consumption budgets}
    \begin{tabular}{|l|c|c|c|c|}\hline
         Subsystem & \makecell[c]{Orbit \\ Injection} & \makecell[c]{Long-term power \\ consumption during \\ mission phase} & \makecell[c]{Peak power \\ consumption} & \makecell[c]{Safe mode} \\ \hline
         Payloads & 0 & 764 & 829 & 0 \\ \hline
         \makecell[l]{Thermal control \\ of platform} & 30 & 30 & 200 & 30 \\ \hline
         AOCS & 75 & 258 & 1161 & 155 \\ \hline
         Power supply & 25 & 25 & 25 & 25 \\ \hline
         OBDH & 45 & 45 & 45 & 45 \\ \hline
         Communication & 35 & 330 & 400 & 25 \\ \hline
         Total & 210 & 1452 & 2660 & 280 \\ \hline
    \end{tabular}
    \label{tab:spacecraft_power_consumption_budgets}
\end{table}

\subsection{Technology Readiness}  

Unlike most mission concepts in the phase A study, the ET mission concept enjoys a high level of technical readiness. The spacecraft structure, thermal control, power supply, and OBDH are already mature deep space technologies. ET's communication system draws lessons from Tianwen-1 technology. 

The key challenge is the high pointing stability of AOCS. To achieve this, we are focusing on these key technologies with the following statuses: (1) The micro-vibration isolator has passed an environmental testing and will attain an on-board verification very soon. (2) The design and simulation analysis of the ultra-high-precision FGS has been finished and is under development by other approved space projects. (3) Numerical simulations have been finished to verify the attitude determination and control algorithms. The Technology Readiness Level has already reached level 5 based on the ISO 16290 standard.

\section{Survey Simulations and Yield Predictions} 
\label{sec:yield}
\subsection{End-to-End Photometry Simulation} \label{sec:photomtry}
{\bf Authors:}
\newline
Kevin Willis$^1$, Jian Ge$^2$ \& Hui Zhang$^2$\\
{1. \it Science Talent Training Center, Florida, USA} \\
{2. \it Shanghai Astronomical Observatory, Chinese Academy of Sciences, Shanghai 200030, China}\\

Development and performance estimation feedback cycles for photometric surveys are largely enabled by simulations of the expected observations. These simulations give valuable insights into the effects caused by individual and combined noise sources that are often impossible to obtain in the physical world. Here we present a photometry simulator built with an emphasis on multi-survey capability and accurate simulations of ET pixel data. 

\subsubsection{Simulator Design and Implementation}
The photometry simulation software was created specifically for ET and has been under development since the genesis of the mission. Written in the Python programming language, this software has maintained a critical role in understanding trade-offs and design limitations. The simulator is capable of modeling point spread function effects such as motion blurring from jitter and thermally induced focus changes. It also simulates telescope pointing drifts, pixel-phase, intra-pixel, and inter-pixel response variations on the detector.

The simulator produces time-series postage-stamp images of simulated stars which need to be processed and analyzed. To enable this, a multitude of modular tools have been built around the simulator to produce an end-to-end photometric data production and analysis pipeline. These tools include optimal aperture creation, pointing drift and jitter generators, light curve detrend and cotrend functions, stellar variability generation and injection, as well as transit injection and recovery tools.

\begin{figure}[!htbp]
  \centering
  \includegraphics[width=1\textwidth]{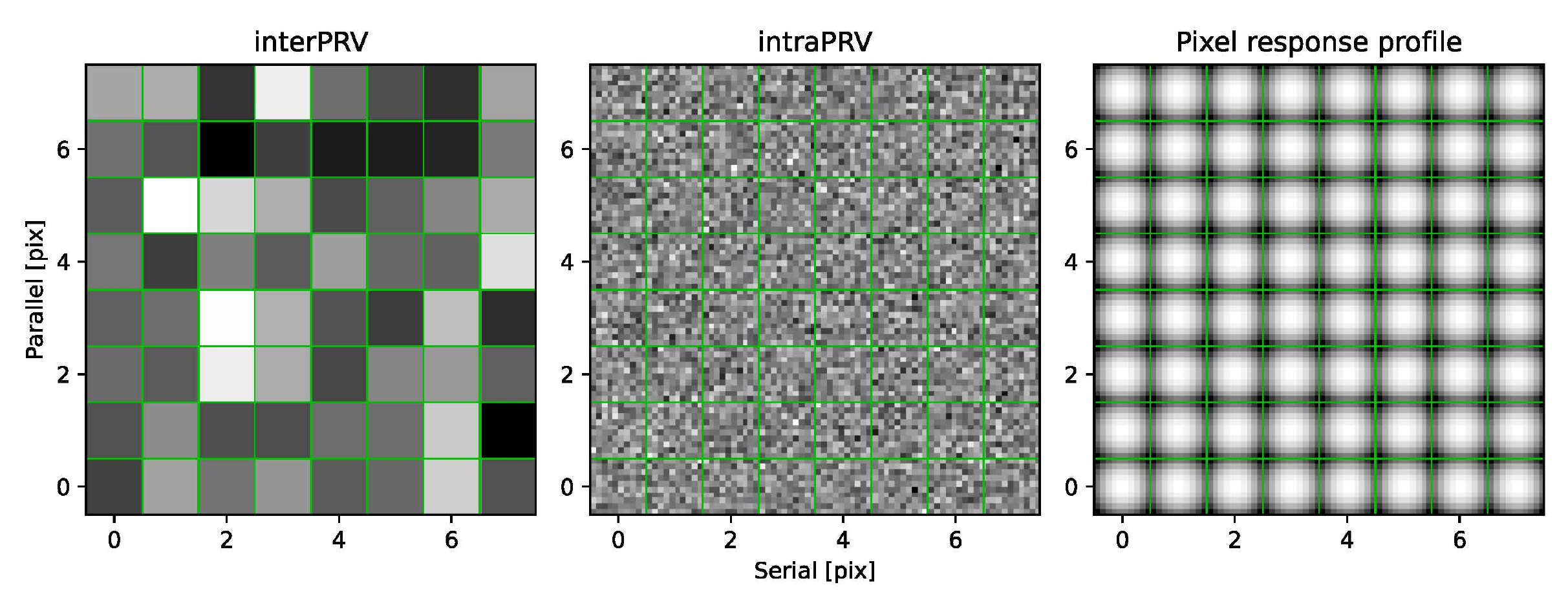}
  \caption{The three quantum efficiency variation layers for a randomly generated detector. Left: inter-pixel response variation. Center: intra-pixel response variations. Right: pixel-phase response profile characterized by diminishing sensitivity to photons as they interact at locations further from the pixel center.}
  \label{Simulated detector quantum efficiency layers}
\end{figure}

The first step in a typical simulation run is to generate a detector by combining three layers that represent major response variations in a CCD or CMOS detector. These three layers are the inter-pixel and intra-pixel response variations as well as what is known as the pixel-phase (see Figure~\ref{Simulated detector quantum efficiency layers}). Inter-pixel variations are simply modeled as net quantum efficiency variations seen on a per-pixel basis. The intra-pixel variation model for ET is a sub-pixel resolution map generated from Gaussian noise with a mean of \SI{100}{\percent} and an RMS of around \SI{1}{\percent}. Lastly, the pixel-phase layer models the non-uniform quantum efficiency spanning each pixel. The model used in ET simulations is derived from measurements of {\it Kepler}’s sub-pixel response, as reported in \citep{Vorobiev:2019}.

\begin{figure}[htbp]
  \centering
  \includegraphics[width=1\textwidth]{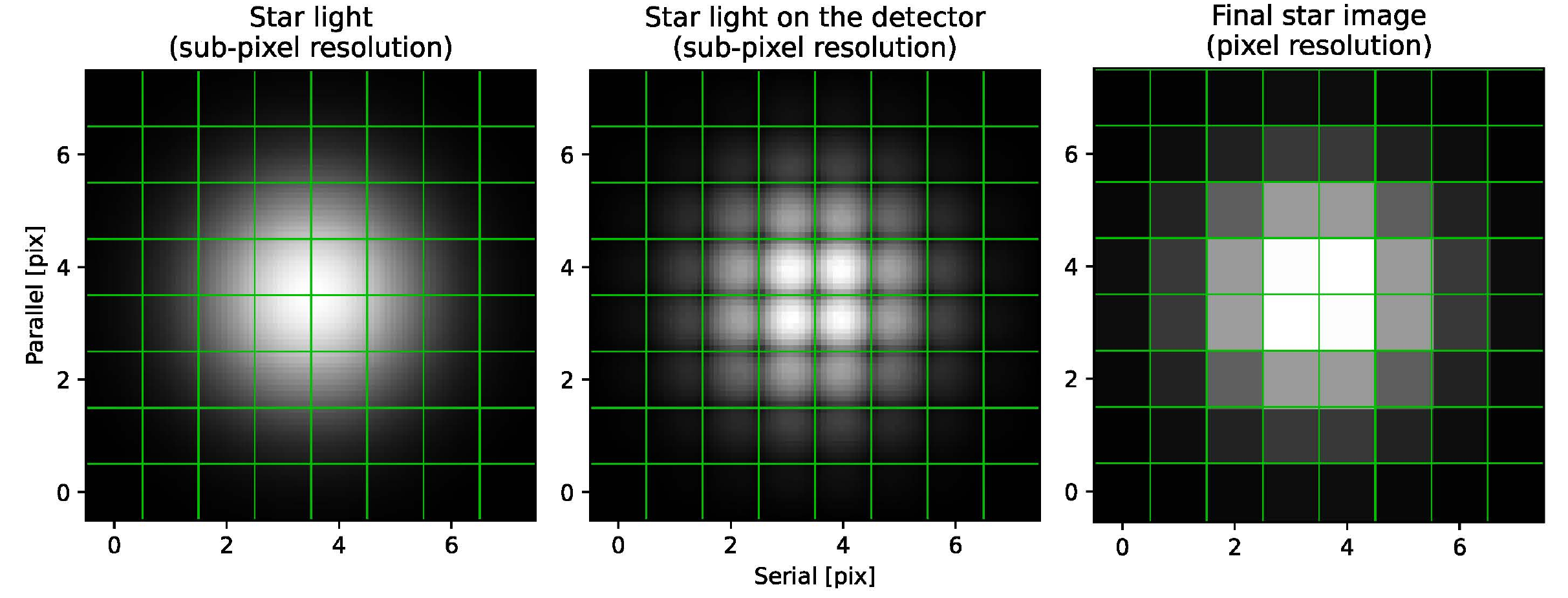}
  \caption{Left: Light from a simulated star, generated by a 2-D Gaussian pixel response function, just before it is combined with the simulated detector. Center: Star light after it interacts with the simulated detector. Right: The final image of the simulated star after the sub-pixels of each pixel have been summed.}
  \label{Stellar image creation stages}
\end{figure}

Synthesizing the final time-series images of a given star is done by first combining the detector frame and star light frames. Then, for each pixel, sub-pixels are summed to produce the final pixel-resolution image (see Figure \ref{Stellar image creation stages}). The star light frames are created at sub-pixel resolution by evaluating the chosen pixel response function (PRF). The simulator has several PRF model options and any given model can be manipulated further depending on the desired jitter, focus changes, or aberration. ET simulations currently use PSFs derived from optical simulations in Zemax while {\it Kepler} simulations use PRF’s supplied by {\it Kepler}. 

The final postage-stamp (typically 8x8 pixels) frames of simulated stars are typically converted to light curves for further analysis. There are several methods to choose from, with the most common being the optimal aperture (OA) method which chooses pixels to include in the aperture mask by maximizing SNR. Other available methods include PSF photometry and basic aperture photometry. Prior to light curve creation, flatfielding can optionally be applied to the images.

Measuring the noise of light curves is an involved process. When combined, noise sources such as inter-pixel response variations and pointing drift cause both low and high-frequency trends which need to be removed to allow for accurate combined differential photometric precision (CDPP) measurements. To do this we employ two standard tools, cotrending and detrending. The cotrending algorithm used in our data reduction routine is based on principal component analysis which is capable of removing the majority of trends common within a set of light curves. After cotrending, light curves are detrended using the LOWESS fitting \citep{Cleveland:1979}. In the final product, nearly all long period trends are removed as well as the high frequency trends common to all light curves in the set.

\begin{table}
\centering
    \captionsetup{justification=centering}
\begin{tabular}{cccc}
\textbf{Parameter}              & \textbf{Unit} & \textbf{ET}    & \textbf{{\it Kepler}} \\ \hline
Read Noise                 & $\mathrm{e^-}$/px     & 4.0   & 86.0            \\
Jitter                     & -             & \SI{0.15}{\arcsecond}/\SI{5}{\s}   & \SI{3}{mas}/\SI{15}{\minute}    \\
Pixel diameter             & arcsec        & 4.38           & 3.98            \\
Intra-PRV                  & \%            & 1              & 1               \\
Inter-PRV                  & \%            & 1              & 1               \\
Background light           & $\mathrm{e^-}$/px/s   & 3    & $\sim$150       \\
PSF breathing              & \% {\it Kepler}'s   & 100            & 100             \\
Pointing drift             & \% {\it Kepler}'s   & 200            & 100             \\
EE90 diameter              & pix           & 4              & $\sim$3         \\
Subpixel count             & ct            & 6x6            & 6x6             \\
Exposure                   & s             & 10             & 6.02            \\
Readout time               & s             & 1.5            & 0.52            \\
Aperture                   & cm            & 73.48          & 95.0            \\
Momentum desaturation rate & day/desat     & 3.0            & 3.0            
\end{tabular}
\caption{Common parameter values used in the ET and {\it Kepler} simulations. The PSF breathing and pointing drifts used in simulation come from per-cadence PSF fitting measurements of stars in {\it Kepler}'s data. Since we expect ET to have $<2$ times larger pointing drifts than {\it Kepler}, we scale the {\it Kepler} drifts by \SI{200}{\percent} for use in simulating ET photometry. Though we expect ET's PSF breathing to be smaller in scale than that of {\it Kepler} due to better temperature stability, we do not downscale the PSF breathing to allow for greater design flexibility where relevant.}
\label{tab:sim_params}
\end{table}

\subsubsection{Testing and Validation of Simulator Output }\label{sec:cdpp}
Building and maintaining confidence in simulation results is crucial. In physical systems where there is complex interplay between multiple noise and signal sources, as well as complex data reduction techniques required, options for verifying simulations of said systems are often limited. Fortunately, there are an abundance of data from prior missions that can be used to gauge the accuracy of the simulator. By measuring individual aspects of carefully selected real data, such as pointing drift and focus changes, and using them as inputs to the simulator we can compare final products such as light curves and CDPPs.

\begin{figure}[htbp]
  \centering
  \includegraphics[width=0.85\textwidth]{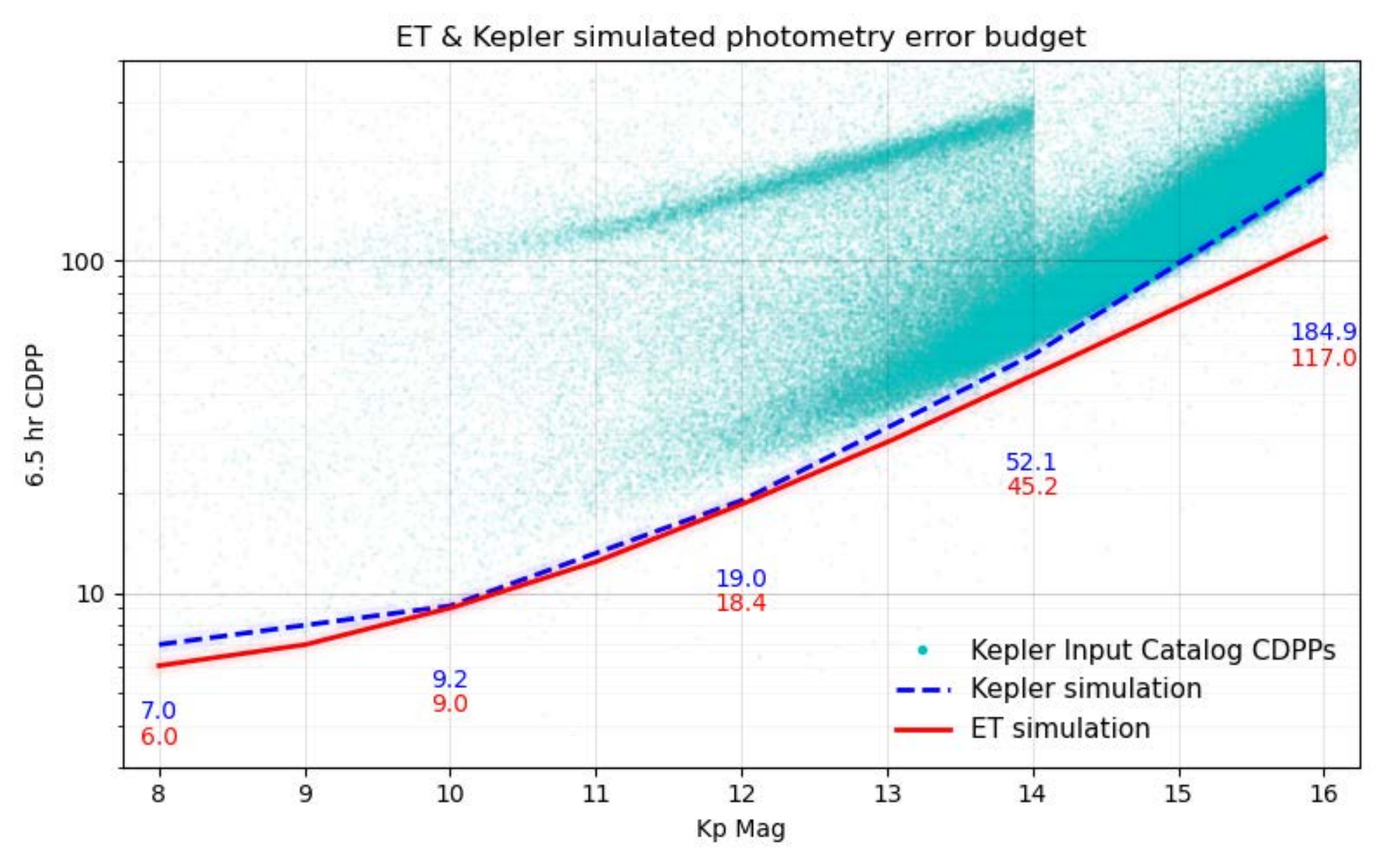}
  \caption{End-to-end simulations of {\it Kepler} and ET photometry provide estimated CDPPs per magnitude. The {\it Kepler} simulation serves to validate the simulator and serves as a valuable benchmark for survey capability comparison.}
  \label{ET and Kepler simulated error budgets}
\end{figure}

We utilized this verification technique with {\it Kepler} data (see Table \ref{tab:sim_params}). To do this, we first carefully selected stars in the {\it Kepler} field which met a set of criteria indicating that they had intrinsically low stellar noise. These criteria included limitations on stellar type, S-index, effective temperature, and more. With a selection of 30 stars randomly scattered throughout the {\it Kepler} field, we preformed robust PSF fitting to each long-cadence frame of each star using high resolution PRF models supplied by {\it Kepler}. This fitting returned time-series data for pointing drift, PSF breathing (a consequence of focus changes caused by the telescope temperature variation), and total photon counts. In addition, {\it Kepler} has supplied approximations for readout noise and background light in each of their CCD channels. Using these as inputs to the simulator, light curves of the selected stars were simulated and their CDPPs measured. These CDPP measurements were within 5 ppm of the measurements of the real light curves. Large scale simulations involving a broad range of magnitudes and thousands of stars also produce results congruent with {\it Kepler}'s survey-wide CDPP measurements, as shown in Figure \ref{ET and Kepler simulated error budgets}.

With this simulator, we conducted photometry simulations with ET and produced measurements shown in Figure \ref{ET and Kepler simulated error budgets}. Both {\it Kepler} and ET produce similar photometry precision for stars brighter than the 13th magnitude. For stars fainter than the 13th mag, ET's photometry precision gets better than {\it Kepler} largely due to having much smaller readout noise than {\it Kepler}. Because {\it Kepler} has a much higher readout noise than ET (an average of 86 e$^-$ vs.  4 e$^-$), readout noise becomes a dominant contributor to CDPPs in Kepler photometry measurements for stars fainter than the 13th mag. ET can achieve the same CDPP for a star about 0.5 mag fainter than {\it Kepler} for stars fainter than 14th mag, allowing ET to observe twice as many targets per solid angle. Additionally, ET will have a larger target catalog (1.2 million stars) allowing it to capture data for more faint targets, further increasing the quantity of observed stars to $> 2$ times as many targets per solid angle.

\begin{table}[hb]
\centering
    \captionsetup{justification=centering}
\begin{tabular}{ccc}
\textbf{Component}       & \textbf{Gilliland} & \textbf{Simulation} \\ \hline
Intrinsic stellar        & 19.5      & 19.5*       \\
Poisson \& readout noise & 16.8      & 16.9       \\
Intrinsic detector       & 10.8      & 10.7       \\
Quarter dependent        & 7.8       & 9.3        \\
Total                    & 29.0      & 30.7       \\
\end{tabular}
\caption{Error budget of real and simulated {\it Kepler} data. The real data was analyzed by \citep{Gilliland2011ApJS}. While the Poisson + readout noise and intrinsic detector categories are very close match to the real data, quarter dependent noise was overestimated. This could be due to the fact that only quarter 6 was simulated, whereas Gilliland used data from quarters 2 through 6. Additionally, it was unclear whether cotrending was applied prior to their measurements. For our simulation we used LOWESS-based detrending without any cotrending. (*) Intrinsic stellar noise was not simulated - the shown value is copied from the Gilliland results.}
\label{tab:sim_gilliland_err_table}
\end{table}

To further validate the simulator and its results, we recreated the analysis shown in Table 4 of \citep{Gilliland2011ApJS}, where noise terms of {\it Kepler}’s Q2 to Q6 light curves where evaluated. In this test, Gilliland et al. measured the noise of stars in the 11.5 – 12.5 Kp magnitude range to give important insights into {\it Kepler}’s error budget. In our simulation, we created a sample of thousands of real stars observed by {\it Kepler} in the same magnitude range, then measured their drifts and PSF scale in every observation of Q6. These measurements were made using a robust PSF fitting method that used {\it Kepler}’s PRF models which approximated the actual PSF seen on the real images. Then these observations were synthetically recreated in our simulator using the {\it Kepler} PRF models, {\it Kepler} detector and telescope characteristics, our measured drifts and PSF scale variations, replicated background stars, and the same aperture mask used in the real data. The simulator was able to simulate multiple versions of these stars with individual noise sources toggled, allowing us to easily determine the noise levels from the various components. Results of the simulation, shown in Table \ref{tab:sim_gilliland_err_table}, indicate that the simulator is able to closely replicate {\it Kepler}’s results for Poisson and readout noises, as well as noises included in the intrinsic detector category. We overestimate in the quarter dependent term causing a higher total noise of 30.7 ppm compared to that of the real data, 29 ppm.


\subsection{Transiting Exoplanet Yield Predictions} 
{\bf Authors:}
\newline
Hui Zhang$^1$, Jian Ge$^1$ \& Chelsea Huang$^1$\\
{1. \it Shanghai Astronomical Observatory, Chinese Academy of Sciences, Shanghai 200030, China}\\
{2. \it University of Southern Queensland, NSW, Australia}\\

\subsubsection{The Input Stellar Population}\label{sec:stellar_population}
\begin{figure}[htbp]
	\centering
	{\includegraphics[width=0.85\textwidth,trim={0 2cm 0 1cm}, clip]{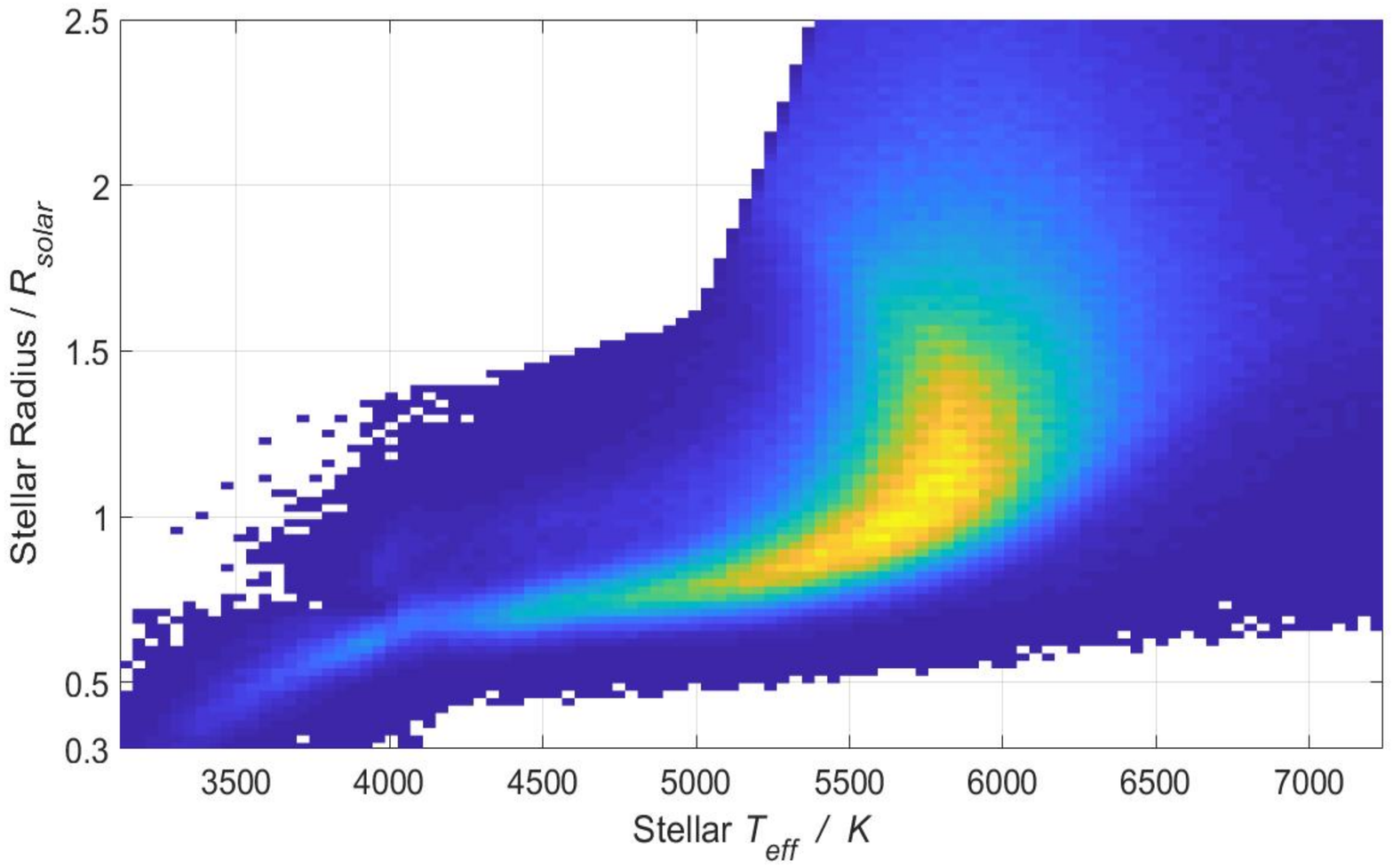}}
	{ \caption{The stellar radius and effective temperature of ET's input stellar population. Most the samples are main-sequence stars ranging from K to F spectral type. The most promising samples conducive to finding Earth 2.0s are those quiet Solar-type stars from K5 to F5. }
		\label{TIC_Rs_Teff}}. 
\end{figure}

To effectively estimate the overall planet yield of the ET survey with a particular focus on Earth 2.0 planets, we first need to build an input stellar population with precisely measured stellar properties such as radius, effective temperature, luminosity, and others. Thanks to the Gaia, {\it Kepler}, and {\it TESS} surveys, a preliminary version of the input stellar population has been created using selections from Gaia DR2, KIC DR25, and TIC V82 catalogs. From TIC V82, all the stellar objects with a magnitude $\leq21$ and a location close to the center of {\it Kepler}'s FOV (less than 16 degrees from $RA=290^{\circ}$, $DEC=45^{\circ}$) are selected to build a base stellar catalog according to the G-band magnitude (very close to the ET band: from \SIrange[]{450}{900}{\nm}) and J2000 coordinates in the Gaia DR2, respectively. Then, we divide this base catalog into two parts: targets and background stars. For the targets, we apply two major criteria to select only main-sequence stars: $R_\ast \in [0.28, 2.5]R_\odot$ and $T_{eff} \in [3210, 10800]$ K, covering the M4 to A0 dwarfs. Moreover, we further limit the magnitude to $\leq 16$ to avoid wasting resources on stars with very low precision. This selection results in a target catalog with over 1.5 million stars. The rest of the stars are categorized as the background sample, which contains more than 18 million stars and is used to calculate the background contamination. Figure \ref{TIC_Rs_Teff} shows the distribution of stellar radius and effective temperature of all stars in the target catalog.

The next step is to locate the stars on ET's focal plane. ET has a wide FOV of 500 square degrees which causes a vignetting effect at the corners. To take this light loss into account, we project all the targets onto ET's focal plane according to the WCS (World Coordinate System) coefficients of ET's detector and calculate their instrument magnitudes according to their pixel coordinates. The largest light-loss fraction at the four vertices is \SI{\leq20}{\percent} and the total light loss integrated over the entire FOV is \SI{\leq8}{\percent}. When measuring the instrument magnitude of a star, its photon and instrument noises are interpolated from the photometric noise profile obtained by a series of photometry simulations (see section~\ref{sec:photomtry}).

\begin{figure}[htbp]
	\centering
	{\includegraphics[width=0.85\textwidth]{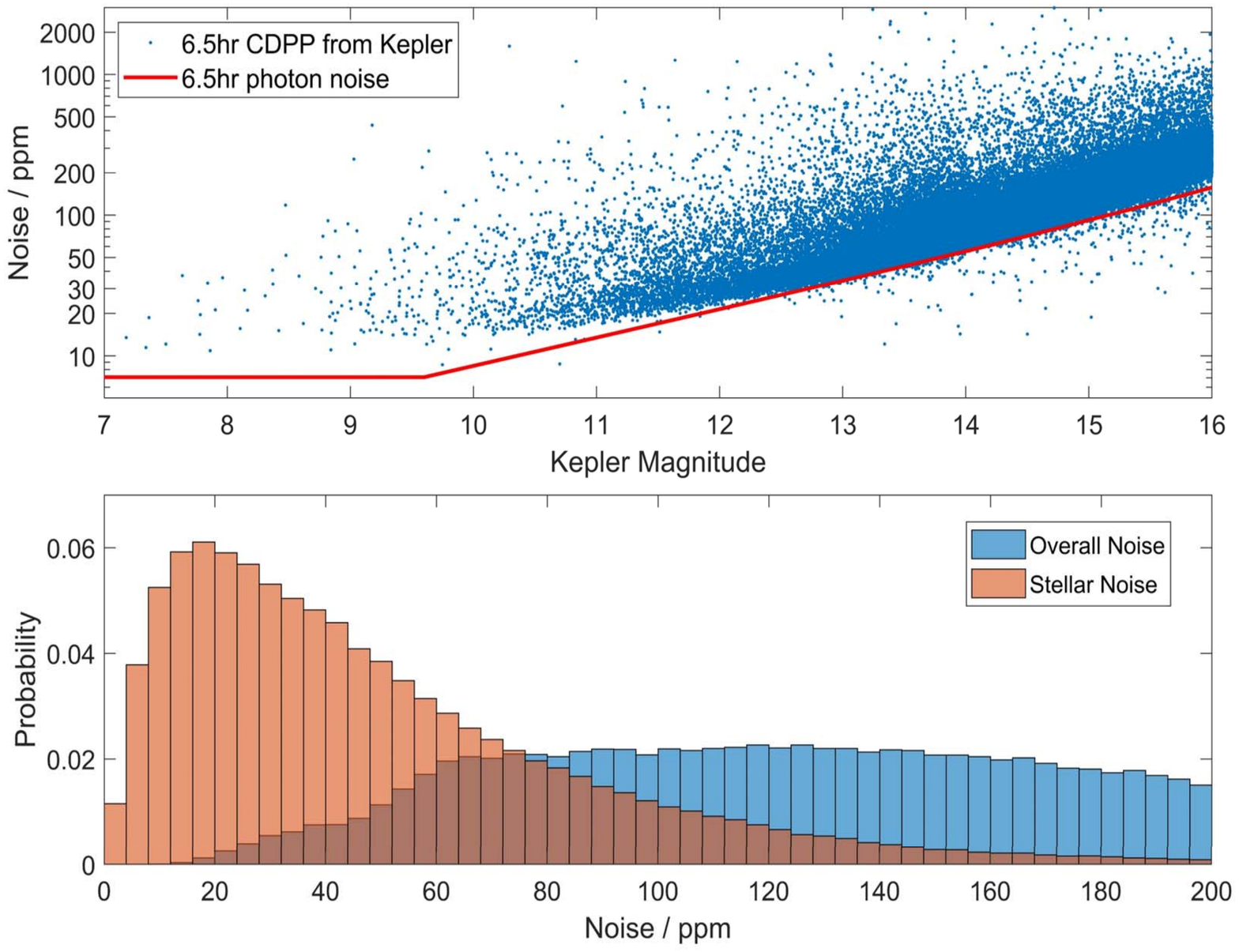}}
	{ \caption{Stellar noise component extracted from {\it Kepler}'s data. Top: The red solid line is the synthetic photon noise curve considering only the brightness of targets, the sky-background and a mean CCD read-out noise of 86$e^-/frame$ \citep{Gilliland2011ApJS}. The blue dots are real {\it Kepler} observations which have been scaled from 6.0-hr to 6.5-hr. Bottom: By subtracting the synthetic photon noise from the real measurement of noise of each target, we get the similar result that the stellar noise peaks around \SI{18}{ppm}.  \citep{Gilliland2011ApJS} }
		\label{stellar_noises}}. 
\end{figure}

We note that this preliminary target catalog is only used to generate the stellar population, providing a statistical prediction of ET's planet yield through a series of Monte Carlo simulations. ET's final high-value target list requires more systematic studies (see section \ref{sec:target_selection}), especially a detailed analysis of stellar activity. At this stage, we only adopt the stellar activity noises from {\it Kepler}'s results. We scale the 6-hour CDPP of {\it Kepler} to 6.5-hour and extract the stellar noise component from it according to \citep{Gilliland2011ApJS}. With all the {\it Kepler} samples $\leq16$th magnitude, we build a Probability Distribution Function (PDF) of the stellar noise, assuming they are independent of stellar magnitudes (see Figure \ref{stellar_noises}). Using this PDF, we randomly draw stellar noise for each target star. This stellar noise is then combined with the vignetting corrected noises caused by photons and instruments (introduced in section \ref{sec:photomtry}) to predict the final photometric precision of ET.

Since ET will cover the same sky field observed by {\it Kepler} for 4 years, some overlapping targets will have an effective observation baseline as long as 8 years. For overlapping targets, we note their 6.5-hour CDPP and duty-cycle parameters reported by {\it Kepler}. If a target star has already been observed by {\it Kepler}, then we assume its final precision will be improved according to:
\begin{equation}
	SNR = \sqrt{SNR_{ET}^2+\frac{T_{et}}{T_{kepler}}SNR_{Kepler}^2},
\end{equation}
where $T_{et}$ and $T_{kepler}$  are the observation baseline of ET and {\it Kepler}, respectively. 

\begin{figure}[htbp]	
	\centering
	{\includegraphics[width=0.95\textwidth,trim={0 3cm 0 1cm}, clip]{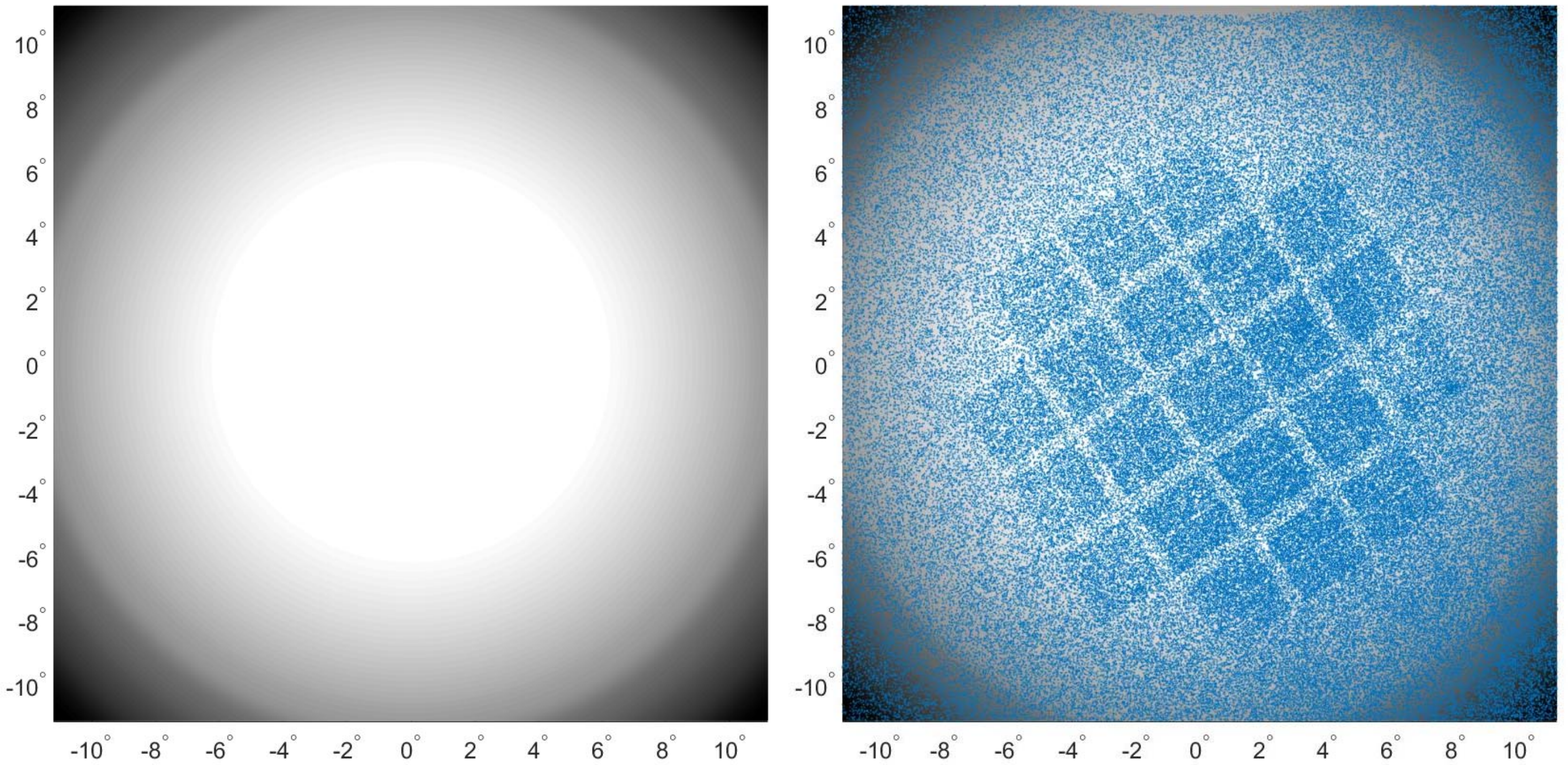}}
	{ \caption{Left: ET's vignetting effect across the 500 square degree FOV. Right: Selected stars on the focal plane. Because of the legacy data available, stars observed by {\it Kepler} have the potential to achieve ultra-high photometric precision and therefore are preferentially selected. Note that this figure shows only 200,000 stars of the 1,200,000 target stars for a clear view of the target distribution. }
		\label{vignetting}}. 
\end{figure}

Because of the bandwidth limitations, we only download data from 1.2 million target stars with the highest photometric precision. With the legacy data, many {\it Kepler} targets are of higher priority for download unless proven to be giants or binaries by the Gaia DR2 and TIC V82 catalogs (see Figure~\ref{vignetting}).

\subsubsection{The Planet Population}
With an adequate stellar population catalog, we can then generate a planet population according to the real planetary occurrence rate function:$f_{p} = O(S_{p}^i,S_{\ast}^j)$, where $S_p^i$ and $S_{\ast}^j$ denote $i$ planetary properties and $j$ stellar properties, respectively. However, there is currently no such accurate function available. In fact, an accurately derived occurrence rate distribution is one of ET's most important scientific goals to be addressed by this survey. Therefore, we adopt the most recent 3-D occurrence rate distribution derived from previous transiting exoplanet surveys, e.g., {\it Kepler} to obtain reasonable predictions for our ET survey. The first two dimensions are the planetary radius ($R_p$) and orbital period ($T_p$), while the third dimension is the stellar type, which can be approximately derived from stellar radius and effective temperature in the input stellar population. For planets around F, G, and K stars, we adopt the occurrence rates from \cite{Kunimoto2020AJ} and for planets around M dwarfs, we follow the result from \cite{Dressing2015}. 

\begin{figure}[htbp]
	\centering
	\includegraphics[width=0.85\textwidth,trim={0 2cm 0 1cm}, clip]{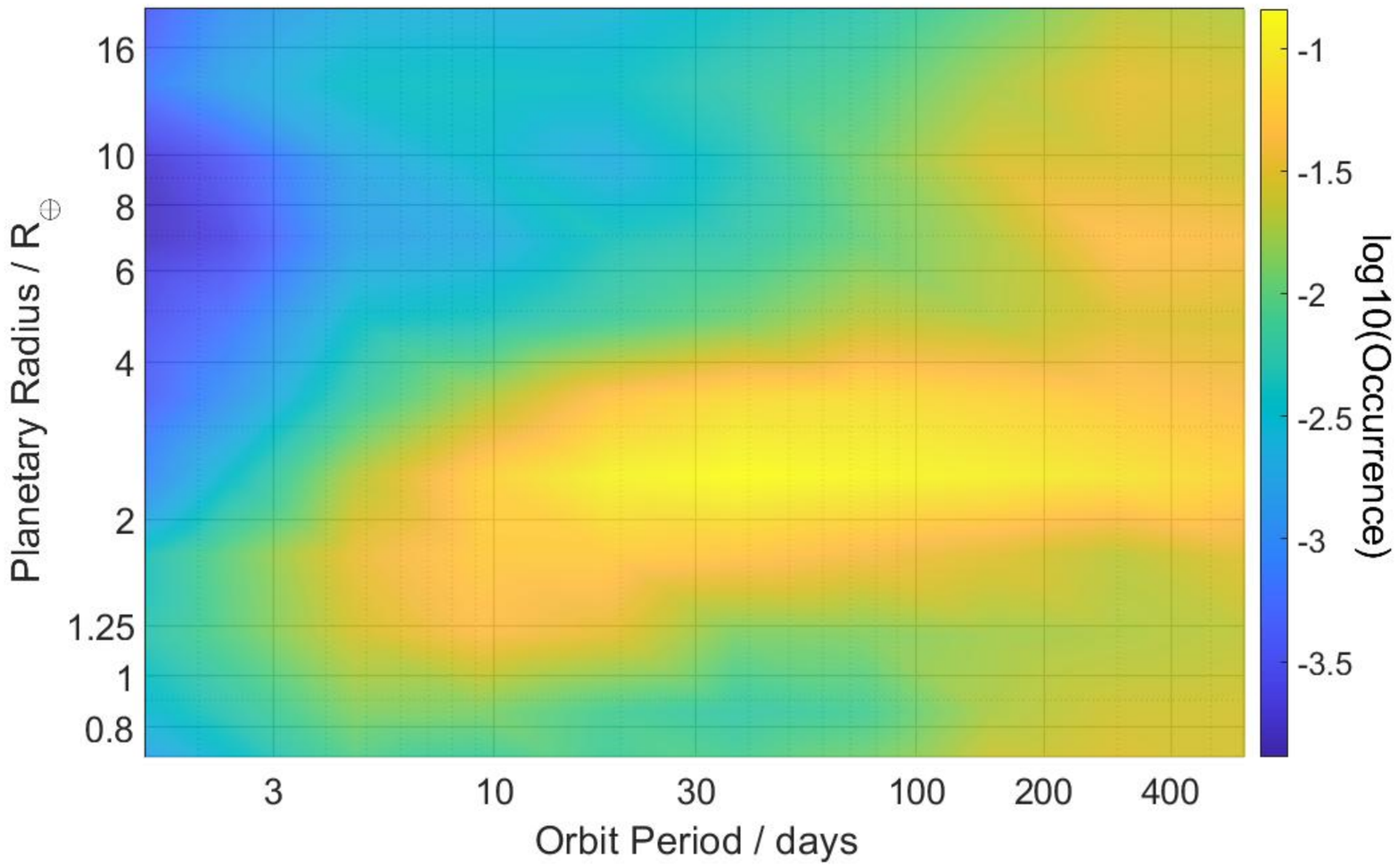}
	\caption{The median occurrence rate distribution of all AFGKM stars in 50 runs. The overall occurrence rate of Earth 2.0s, $\eta_\oplus$ has an upper limit of \SI{\sim 10}{\percent}.\label{OR_planet}}
\end{figure}

Since we use the real stellar properties for each star, the input stellar population is fixed during the entire simulation. The planetary population is randomly generated at the beginning of each run. We set the planetary radius range: $\rm{R}_p\in[0.5, 20] \rm{R}_\oplus$, the orbital period range: $T_p\in[0.5, 730]$ d, and orbital eccentricity range: $e\in[0, 0.5]$. The planet quantity in each system is set to a random integer between [0, 6], given that each system has 3 planets on average \citep{Zhu2018}. When a system has more than one planet, we set $\cos{i}$, $i$ being the first planet's inclination, uniformly distributed between $[0, 1]$ and the inclinations of other planets as within \SI{\pm5}{\degree} of the first planet. Within each $\log{R_p}-\log{T_p}$ grid, we sample the occurrence rate using two half-normal distributions (one for the upper limit and the other for the lower limit, see \cite{Kunimoto2022}) with the median value and $1\sigma$ upper/lower limits provided by \cite{Kunimoto2020AJ}. For grids with only an upper limit, we assume the lower limit is zero and the mean value is $3/4$ of the upper limit, while for those grids outside the boundaries, we extrapolate the upper limit with the value in the closest grid. This implies occurrence rates that are uniformly distributed on a logarithmic scale when the planet period is longer than 400 days for all $R_p$. The occurrence rate of Earth 2.0, $\eta_\oplus$ when $R_p\in[0.8, 1.25] R_\oplus$ and $T_p\sim[250, 400]$ d, the most important value in our occurrence rate distribution, has an upper limit of around \SI{10}{\percent} in our simulations. A typical simulation contains 1.2 million real stars which host about 3.6 million injected planets in total. We usually perform 20 runs, each with different planetary populations, while for some special cases we perform 50 runs to get more reliable uncertainties. Figure~\ref{OR_planet} shows an overall occurrence rate distribution of all AFGKM stars median-combined over 50 runs. 

\subsubsection{Transit Signal Detection}
During each run, we first generate artificial light curves for each star according to the predicted photometric precision (see section~\ref{sec:stellar_population}) and observation time series, e.g. length of baseline, accumulative exposure time, sampling cadence, and duty cycle. In most cases, we set the observation baseline to be 4 years which begins on January 1st, 2026. The cadence and accumulative exposure time are both set to 30 minutes and we get the 0.5-hour precision by scaling down the 6.5-hour value. We assume the same average duty cycle of 0.82  for all ET stars as that for {\it Kepler} . If a target has been observed by {\it Kepler} already, we combine its artificial light curve with {\it Kepler}'s real light curve data. Then, for each injected planet, we derive its transit duration, $T_{dur}$, according to the formula by \cite{Ford2008ApJ}. If $T_{dur} \le 0$, we label it as a "non-transiting planet". Otherwise, we fold the light curve according to its orbital period and calculate how many individual transit windows are covered, $N_{tr}$ and the total number of data points fall in the folded transit window, $N_{dot}$. The signal-to-noise ratio of a folded transit is derived by:
\begin{equation}
\text{SNR}_{tr} = \sqrt{N_{dot}} \delta/\sigma_{0.5h},
\end{equation} 
where $\delta=(R_p/R_\ast)^2$ is the transit depth and $\sigma_{0.5h}$ is the 0.5-hour CDPP.

The SNR loss caused by the combination of data from six telescopes potentially is an issue that requires attention and care. The {\it TESS} project provides critical insight into this issue. Each of the four telescopes of the {\it TESS} satellite covers a single FOV, with small overlapping regions between two neighbouring fields. Stars within this region will suffer different systematic noises from two different telescopes. After cotrending, detrending, and data combination, the final SNR does not always increase by a factor of $\sqrt{2}$, as the residual of different systematic noises causes SNR loss. Moreover, we find that this loss depends on target brightness, where brighter targets suffer larger SNR loss. We fit the loss function with the stellar magnitude from two {\it TESS} telescopes and scale it to mimic the six telescopes of ET. For a 10$^{th}$ magnitude star covered by six telescopes of ET, the maximum SNR loss is around \SI{15}{\percent} and it drops to \SI{\leq5}{\percent} for stars fainter than 14$^{th}$ magnitude. 

\begin{figure}[!htbp] 
	
	\centering
	{\includegraphics[width=0.85\textwidth,trim={0 2cm 0 1cm}, clip]{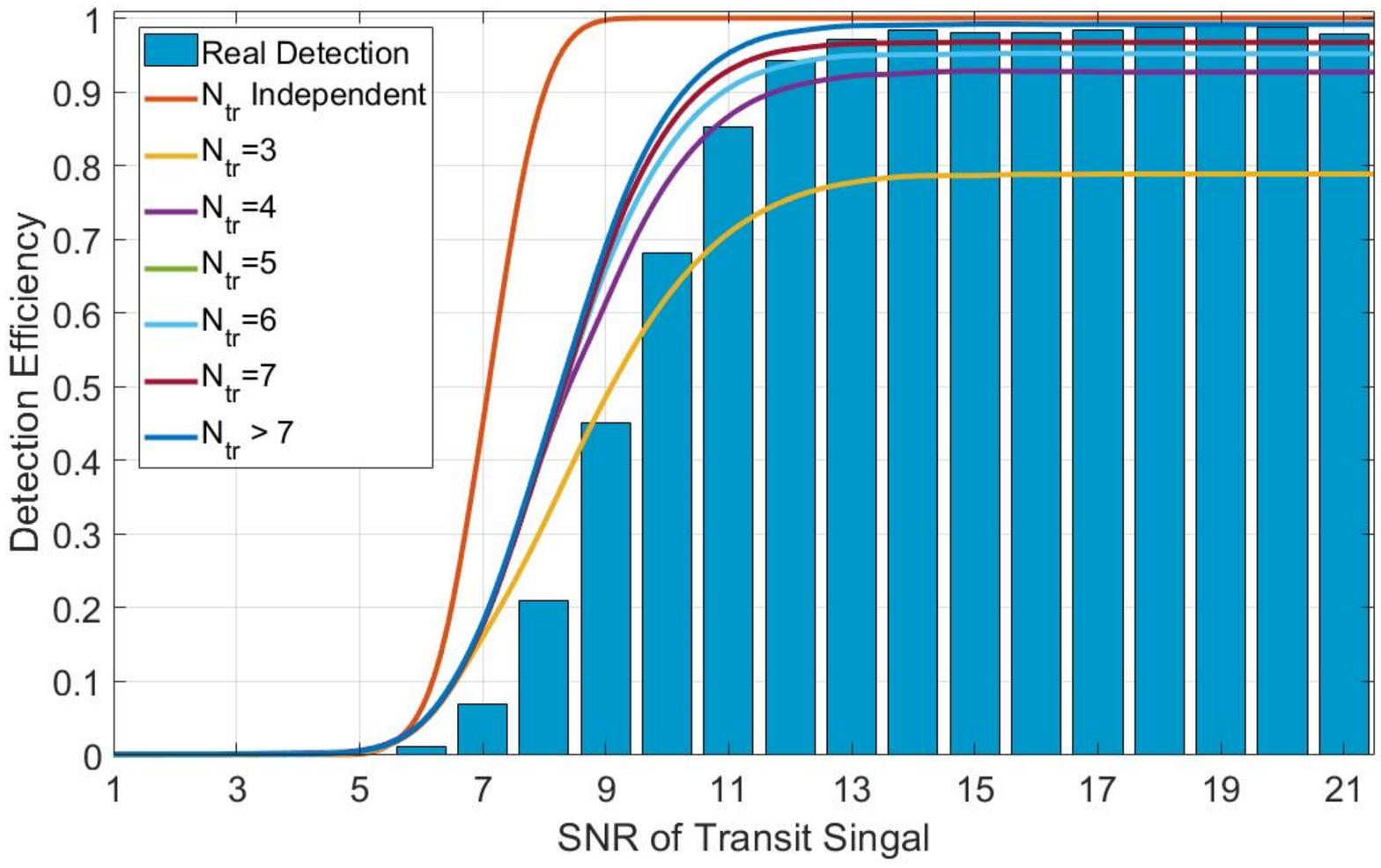}}
	{ \caption{Detection efficiency profile of different transit numbers. Instead of the hypothetical perfect performance for pure white noise (the red solid line which guarantees a \SI{50}{\percent} efficiency at SNR$=7.1$), we adopt different detection profiles depending on how many transit windows would be covered within the observation baseline of ET, assuming an average duty cycle of 82\%. These window-number dependent profiles are from \cite{Christiansen2020AJ} and the histogram is the detection efficiency results from our simulations.}
		\label{DE_profile}}. 
\end{figure}

\begin{figure}[!htbp]
	
	\centering
	{\includegraphics[width=0.85\textwidth,trim={0 2cm 0 1cm}, clip]{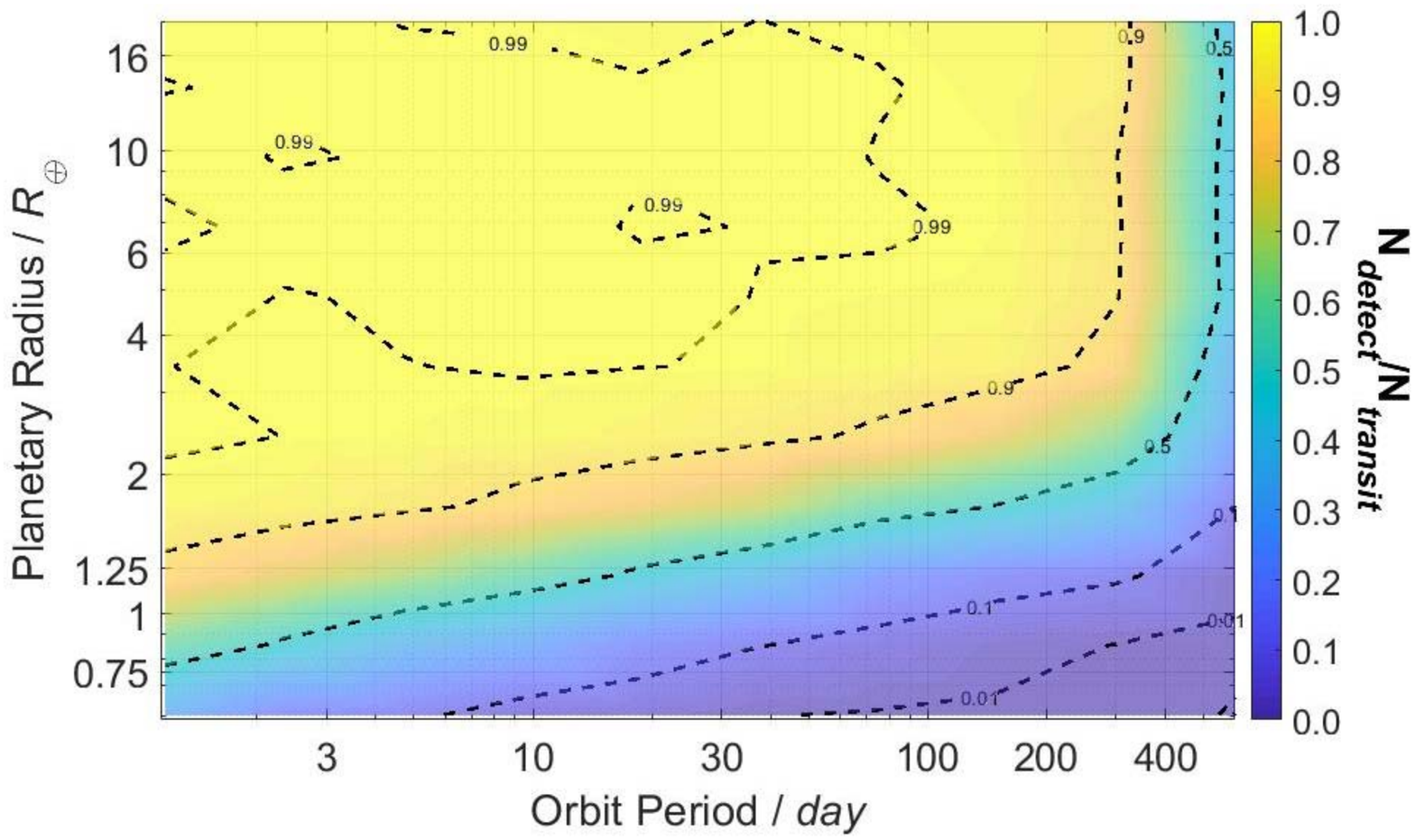}}
	{ \caption{The survey completeness as a function of the planet radius and orbital period. The color denotes the ratio of number of detected planets to number of transiting planets. It is mainly affected by the the photometric precision, the stellar radius, magnitude, noise, and the pipeline efficiency. If an Earth 2.0 transits its host star and that star is in our target list, ET has a probability of $\sim0.01$ to detect it.}
		\label{Ndet_Ntra}}. 
\end{figure}
\begin{figure}[htbp]
	
	\centering
	{\includegraphics[width=0.85\textwidth,trim={0 2cm 0 1cm}, clip]{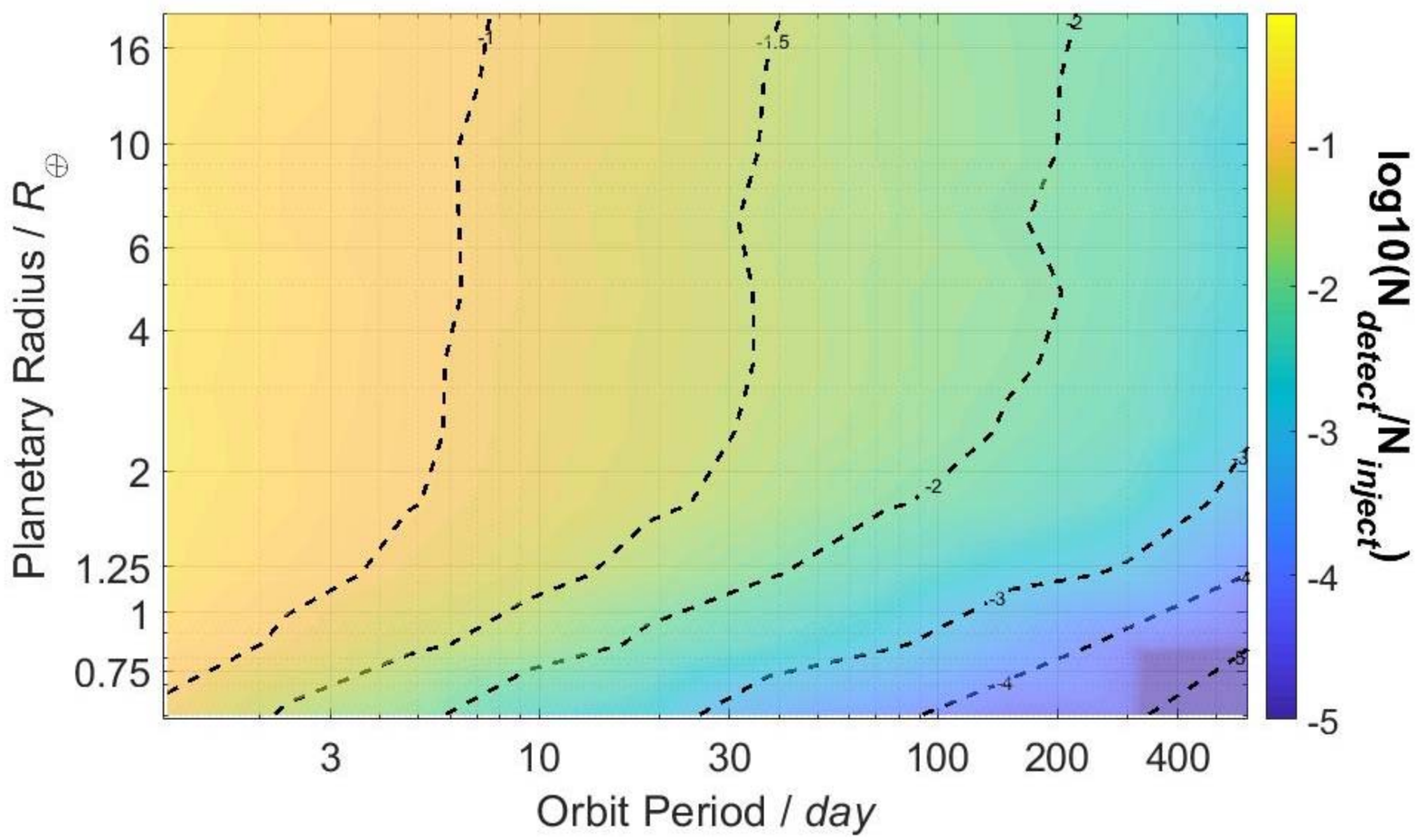}}
	{ \caption{The detection efficiency as a function of the planet radius and orbital period. The color denotes the logarithm of the ratio between the number of revealed planets and the number of all injected planets. Besides the factors affecting survey completeness, it is also affected by factors such as the geometry probability of transit and the observation strategy. The odds of finding an Earth 2.0 is around $10^{-4}$ for ET.}
		\label{log10Ndet_Ninj}}. 
\end{figure}

To avoid overestimating the efficiency of our preliminary pipeline \citep{zhang2019a,zhang2019b}, we adopt a series of detection efficiency profiles that depend on both SNR$_{tr}$ and N$_{tr}$. Figure~\ref{DE_profile} shows the relation between detection efficiency and SNR$_{tr}$ of all recovered transiting planets in our simulations. Instead of the hypothetical value of 7.1 adopted by the {\it Kepler}'s team \citep{Jenkins2010APJ}, we need SNR$_{tr}\geq9.5$ to achieve an efficiency of over \SI{50}{\percent}. The overall detection efficiency is also strongly dependent on the size, orbital period, and orbit inclination of the target planet. Figure~\ref{Ndet_Ntra} shows ET's capability to recover a transiting planet given its planetary radius and period. It is not surprising that the odds of ET finding a transiting Earth 2.0 around different kinds of stars is around 0.01, given that a major part of the planet-host stars are either too big, too faint, or too noisy ($R_\ast \in [0.28, 2.5]R_\odot$, $T_{eff} \in [3210, 10800]$ K, and the magnitude $\leq 16$. See Section \ref{sec:stellar_population} for detailed stellar population.). Figure \ref{log10Ndet_Ninj} shows the probability of finding a planet regardless of whether it transits its host star. The odds drop to $\leq 10^{-4}$ for ET for finding an Earth 2.0 when also considering the geometry probability of transit and the observation strategy. Given a $\eta_\oplus\sim$\SI{10}{\percent}, this means ET has to survey more than 100,000 Solar-type stars to find a single Earth 2.0. If we plan to find ten Earth 2.0s allowing us to constrain $\eta_\oplus$ precisely, then a target list of over 1 million Solar-type stars down to $16^{th}$ magnitude is minimally required.

\subsubsection{Sanity Check with {\it Kepler}'s Results}

%

\begin{table}[htbp]
	\centering
	    \captionsetup{justification=centering}
	\caption{Comparison between simulated {\it Kepler} planet yield and the real discoveries.} \label{tab:sim_kepler_table}
\setlength\tabcolsep{1.5pt}
\begin{tabular}{|l|c|c|c|c|c|c|c|c|c|c|c|c|c|c|c|}
		\hline
		\multirow{2}{*}{ }
		&\multicolumn{3}{|c|}{\makecell[c]{Jupiter-size\\$>4 R_\oplus$}}
		&\multicolumn{3}{|c|}{\makecell[c]{Neptune-size\\$2-4 R_\oplus$}}
		&\multicolumn{3}{|c|}{\makecell[c]{Super-Earth\\$1.25-2 R_\oplus$}}
		&\multicolumn{3}{|c|}{\makecell[c]{Sub-Earth\\$<1.25 R_\oplus$}}
		&\multicolumn{3}{|c|}{\makecell[c]{Sub-Earth in\\Habitable Zone}}\\
		\cline{2-16}
		&{M}&{KGF}&{Giant}
		&{M}&{KGF}&{Giant}
		&{M}&{KGF}&{Giant}
		&{M}&{KGF}&{Giant}
		&{M}&{KGF}&{Giant}\\
		\hline
		\multirow{2}{*}{\makecell[l]{{\it Kepler}\\Observation}}
		&\multicolumn{3}{|c|}{276}&\multicolumn{3}{|c|}{1185}&\multicolumn{3}{|c|}{812}&\multicolumn{3}{|c|}{404}&\multicolumn{3}{|c|}{2}\\
		\cline{2-16}
		& 2 & 263 & 11 & 10 & 1171 & 4 & 32 & 779 & 1 & 22 & 381 & 1 & 2 & 0 & 0 \\
		\hline
		\multirow{2}{*}{\makecell[l]{KIC dr25\\Simulation}}
		&\multicolumn{3}{|c|}{173$\pm$34}&\multicolumn{3}{|c|}{1103$\pm$36}&\multicolumn{3}{|c|}{808$\pm$43}&\multicolumn{3}{|c|}{379$\pm$39}&\multicolumn{3}{|c|}{2$\pm$1}\\
		\cline{2-16}
		& 1 & 144 & 28 & 7 & 1075 & 21 & 41 & 766 & 1 & 21 & 358 & 0 & 1 & 1 & 0 \\
		\hline
	
	\end{tabular}
\end{table}

\begin{figure}[htbp]
	
	\centering
	{\includegraphics[width=0.85\textwidth,trim={0 2cm 0 1cm}, clip]{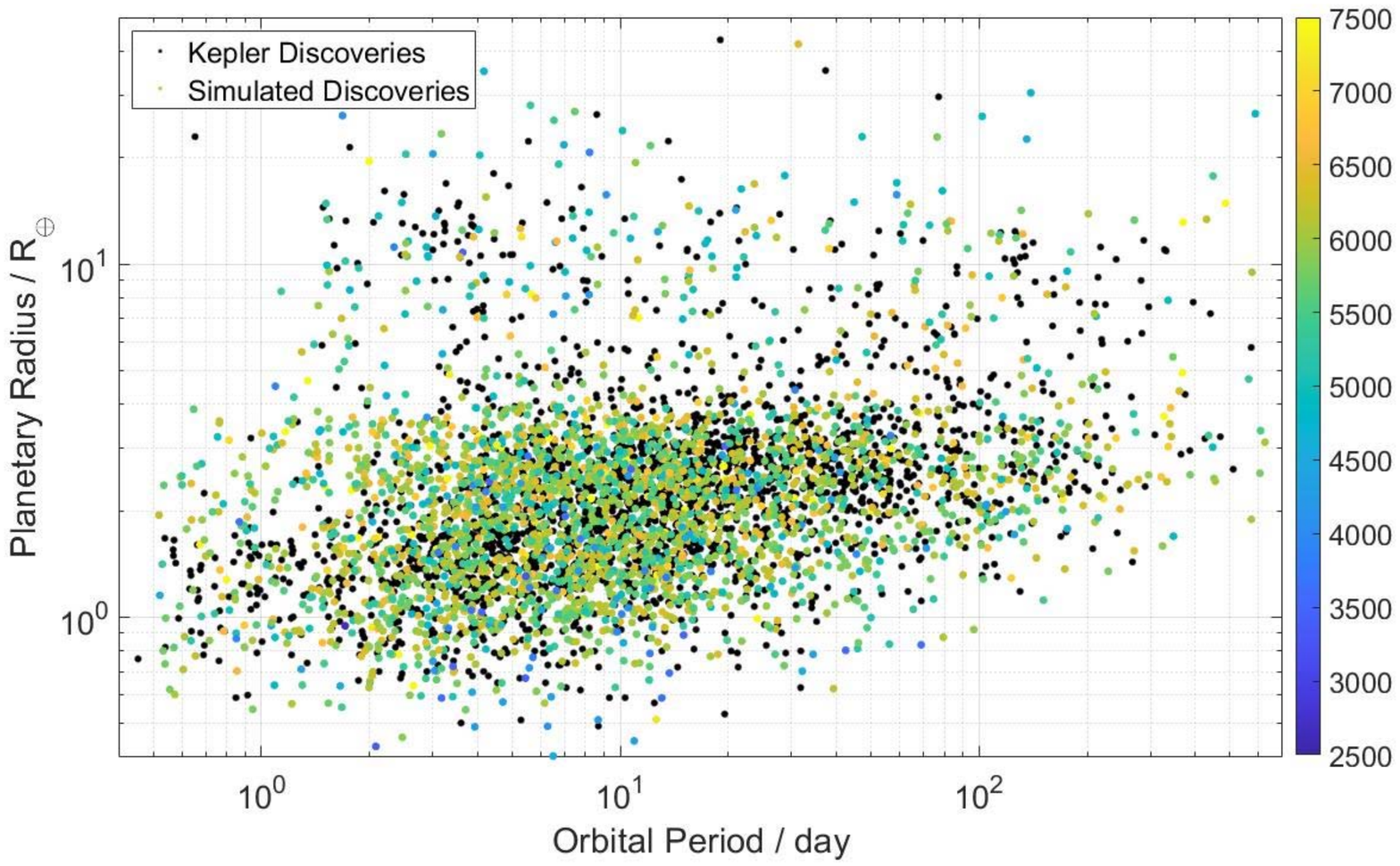}}
	{ \caption{The $\rm{R}_p-\rm{T}_p$ distribution of simulated {\it Kepler} planets. Our simulation has reasonably reproduced the distribution of {\it Kepler} planets including the major populations of mini-Neptunes, super-Earths, hot Jupiters, and warm Jupiters.}
		\label{Sanity_check_Kepler}}. 
\end{figure}

To ensure the reliability of our simulations, we first simulate the planet yield of {\it Kepler} with its latest input catalog (KIC dr25) and compare our results with the published ones (see Table~\ref{tab:sim_kepler_table}). We then divide their planet yield into 4 populations (Jupiter-size, Neptune-size, super-Earth, and sub-Earth) according to planetary radius and count numbers of the planets around M dwarfs, FGK dwarfs, and giant stars in each population. Our simulation results are consistent with {\it Kepler}'s in both the absolute number and relative ratio between planet populations. We also count the number of Earth 2.0 candidates, i.e., Earth-size planets in the habitable zone of a K/G/F dwarf star. Our simulation predicts that {\it Kepler} could marginally detect 1 potential Earth 2.0, but the uncertainty is very large of $\pm2$, which appears to be consistent with {\it Kepler}'s null detection to date. Figure~\ref{Sanity_check_Kepler} shows the $\rm{R}_p-\rm{T}_p$ distribution of our simulated {\it Kepler} planets. The coverage in the parameter space and the trend between the planetary radius and orbital period are also in a good agreement with {\it Kepler}'s detections. The only minor difference is that our simulation predicts fewer giant planets than what were actually discovered by the {\it Kepler} survey.

\subsubsection{The Planet Yield of ET}
To optimize the planet yield, especially the discovery of Earth 2.0s, we explored a huge parameter space with six major dimensions: 
\begin{itemize}
    \item[1.] The number of total telescopes: 4, 5, 6, 7. It is extremely difficult to achieve both a large aperture size and a wide FOV on a single-aperture telescope. Here we focus on designs that combine several small-aperture, wide-field telescopes together to achieve an FOV $\geq256 \rm{deg}^2$ and an effective aperture \SI{>70}{\cm}. 
    
    \item[2.] Aperture size of a single telescope: \SI{12}{\cm}, \SI{20}{\cm}, \SI{30}{\cm}, \SI{35}{\cm}. Larger apertures provide more photons and more  faint stars with high photometry precision, but at the cost of more weight.
    
    \item[3.] Field of view of a single telescope: \SI{256}{deg^2}, \SI{300}{deg^2}, \SI{400}{deg^2}, \SI{500}{deg^2}. Wider FOV provides more bright stars; however this requires a smaller aperture, larger pixel-scale, and potentially increased scattered light level.
    
    \item[4.] Configuration of FOVs: Overlapping FOV for all the telescopes includes more faint stars vs. each telescope covering a unique FOV to achieve the maximal sky coverage. Dividing telescopes into 2 or 3 groups where each group stares at different FOV can reach a balance between the sky coverage and magnitude limit.
    
    \item[5.] Readout noise and readout time of detectors: this will significantly affect the overall photometric precision for faint stars.
    
    \item[6.]stars: we will download as much data  from the satellite as possible. However, the downlink bandwidth is very limited from the Sun-Earth L2 point. We have to determine a feasible quantity of stars for data downloading to maximize the likelihood of detecting Earth 2.0s.
\end{itemize}

\begin{figure}[htbp]
	\centering
	{\includegraphics[width=0.85\textwidth,trim={0 1cm 0 1cm}, clip]{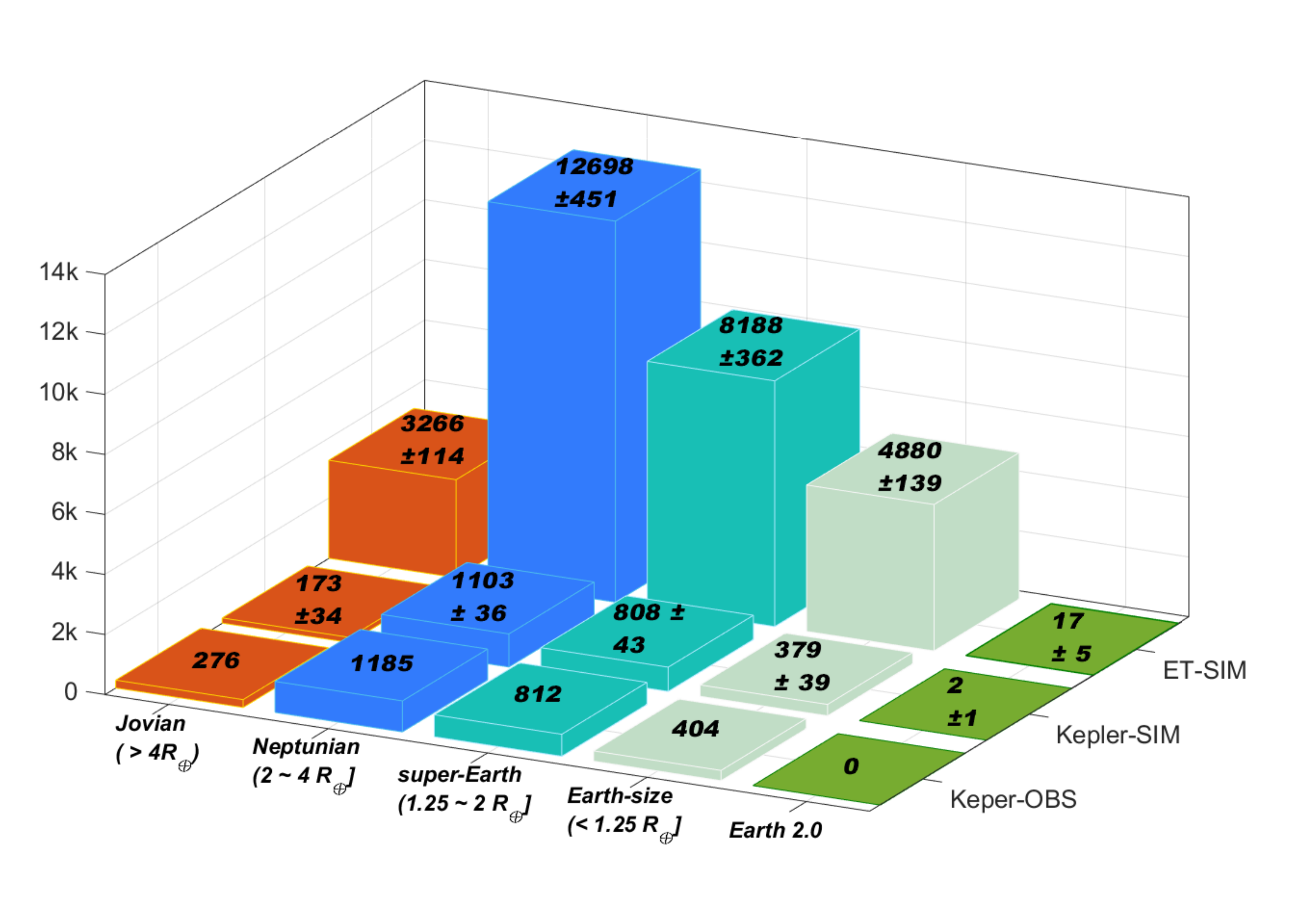}} 
	{ \caption{Predictions of ET's planet yield. We simulated the output of ET's 4-year observation (ET-SIM) of 1.2 million quiet stars. We also show the planets observed by {\it Kepler} ({\it Kepler}-OBS) as a reference. With {\it Kepler} parameters, our simulated {\it Kepler} yield ({\it Kepler}-SIM) matches the actual detection rates well, validating our simulation pipeline.}
	\label{Yield_ET_VS_Kepler}}. 
\end{figure}

\begin{figure}[ht]
	
	\centering
	{\includegraphics[width=0.75\textwidth]{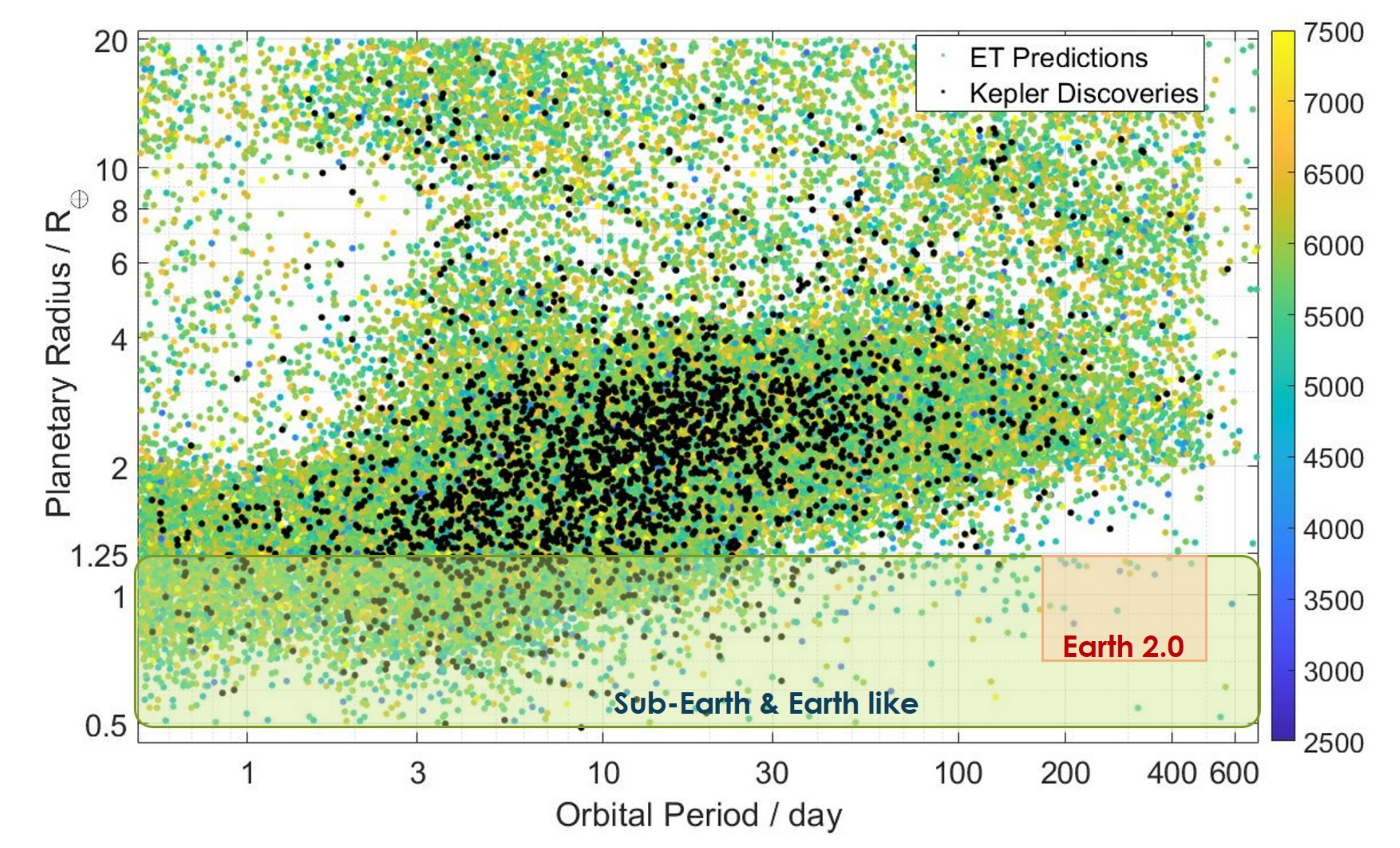}} 
	{ \caption{Comparison between {\it Kepler}'s discoveries and ET's planet yield prediction. The area in the green shaded box is where sub-Earths are located and the light red shaded area is where we should find Earth 2.0s. For clear illustration, we only show the results of 200,000 target stars.}
		\label{ET_Yield_with_Teff}}. 
\end{figure}

\begin{figure}[htbp]
	\centering
	{\includegraphics[width=0.8\textwidth,trim={0 2cm 0 2cm}, clip]{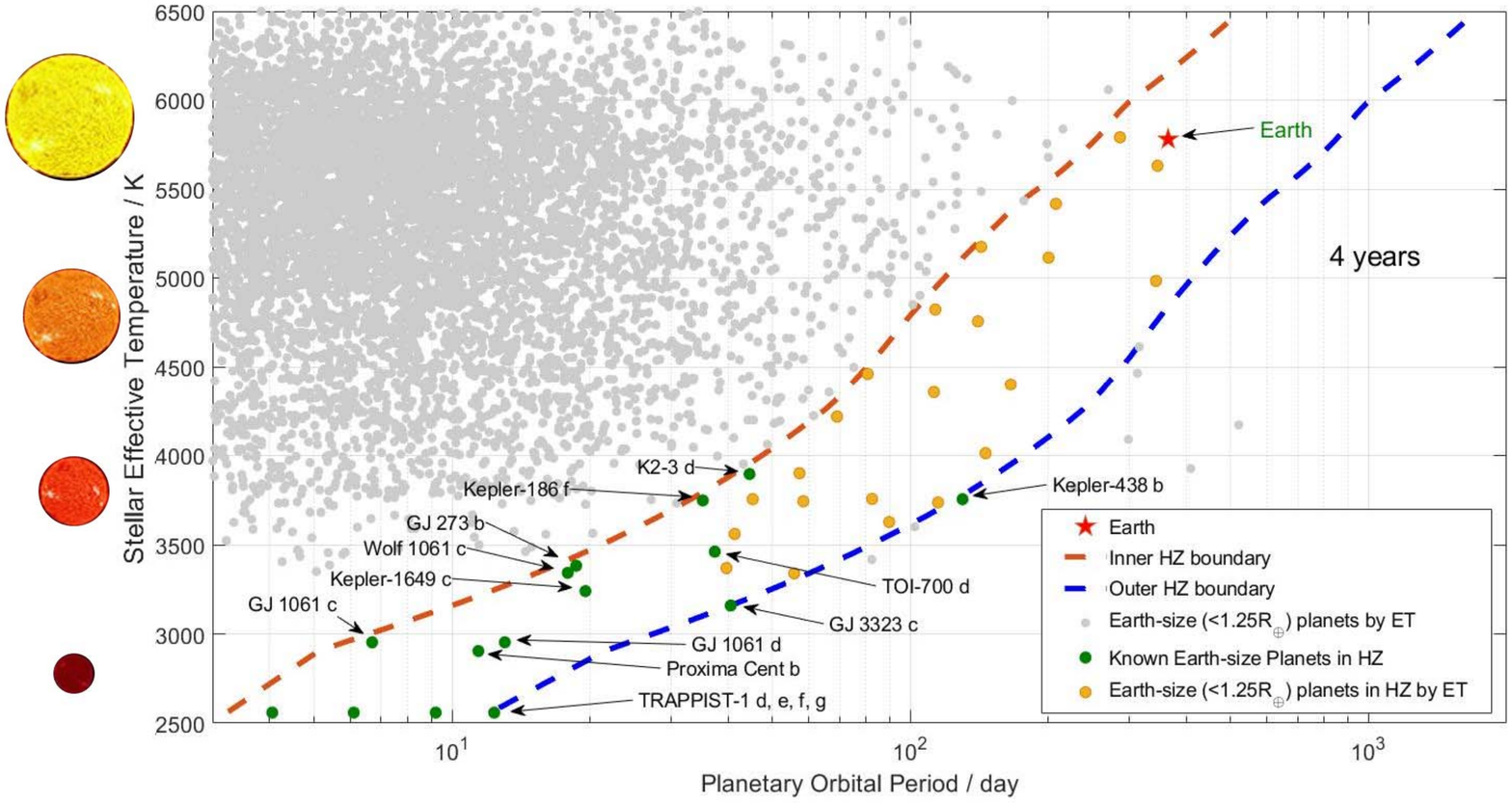}}
	{ \caption{Simulated distribution of Earth-size planets to be detected by ET. ET is expected to find roughly 10 to 20 Earth 2.0s in the habitable zone of F/G/K dwarfs within its 4-year operation.}
		\label{HZ_ET_4yr}}. 
\end{figure}

After a huge series of simulations covering hundreds of combinations of these six major parameters and other minor ones, we decide to adopt an optimal configuration, consisting of six \SI{30}{\cm} telescopes having a \SI{500}{deg^2} FOV. All six telescopes will stack together to monitor an region centered in {\it Kepler}'s field, continuously for a minimum of 4 years. With this configuration, we will obtain light curves of 1.2 million pre-selected relatively quiet main-sequence stars. Our simulation predicts that more than 40,000 new planets of various sizes will be discovered and increase the total number of known planets by approximately ten times (Figure~\ref{Yield_ET_VS_Kepler}). These new planets will cover a wider area in the planetary parameter space than {\it Kepler} and expand the parameter boundary of planets to smaller sizes and longer periods (see Figure~\ref{ET_Yield_with_Teff}). Most importantly, ET is expected to  detect $\sim$5,000 new Earth-size terrestrial-like rocky planets, including about 10 to 20 Eath 2.0s orbiting within the habitable zone of Solar-type stars (see Figure~\ref{HZ_ET_4yr}).

\newpage
\subsection{Microlensing Exoplanet Yield Estimates}\label{sec:microlensing_yield}
{\bf Authors:} 
\newline
Weicheng Zang$^1$, Shude Mao$^{1,2}$, Andrew Gould$^{3,4}$, Chung-Uk Lee$^5$, Subo Dong$^6$, Jennifer Yee$^7$, Wei Zhu$^1$, Yossi Shvartzvald$^8$, Hongjing Yang$^1$, Renkun Kuang$^1$, Jiyuan Zhang$^1$ \\
{1.\it Department of Astronomy, Tsinghua University, Beijing 100084, China}\\
{2.\it National Astronomical Observatories, Chinese Academy of Sciences, Beijing 100101, China}\\
{3.\it Max-Planck-Institute for Astronomy, K\"onigstuhl 17, 69117 Heidelberg, Germany} \\
{4.\it Department of Astronomy, Ohio State University, 140 W. 18th Ave., Columbus, OH 43210, USA} \\
{5.\it  Korea Astronomy and Space Science Institute, Daejon 34055, Republic of Korea}  \\
{6.\it  Kavli Institute for Astronomy and Astrophysics, Peking University, Yi He Yuan Road 5, Hai Dian District, Beijing 100871, China}  \\
{7.\it  Center for Astrophysics $|$ Harvard \& Smithsonian, 60 Garden St.,Cambridge, MA 02138, USA}  \\
{8.\it  Department of Particle Physics and Astrophysics, Weizmann Institute of Science, Rehovot 76100, Israel}  \\

The basic parameters of the ET microlensing telescope are shown in Table \ref{tab:Summaries_of_ET_payload}. The ET microlensing telescope will observe the Galactic bulge from March 21 to September 21, every year. Our estimates are based on an exposure time of \SI{10}{\minute}, readout time of \SI{1.5}{\s}, readout noise of 3.6e$^-$/pixel, and a dark current of 0.02e$^-$/pixel/s. Because the overall throughput and the quantum efficiency for the ET microlensing and the KMTNet telescopes are basically the same, we adopt the $I$-band zero-point of the ET microlensing telescope of 26.80. The current designed image FWHM is \SI{0.82}{\arcsecond} including consideration of jitter and pointing shift. We adopt an extinction of $\rm{A}_I = 2.0$ and a stellar number density of 2.5 times the density of {\it HST} CMD (\citet{HSTCMD}) by comparing the number of red giants in our field and Holtzman et al's field \citep{Nataf2013}. We simulate the ET images using the {\it GalSim} package \citep{GalSim} and find that the background distribution due to ambient stars is $I_{\rm{bkg}} = 18.8^{+0.5}_{-0.9}$ mag/PSF (the median value with the \SI{68}{\percent} uncertainty range).

\begin{figure}[htb] 
    \centering
    \includegraphics[width=0.65\columnwidth]{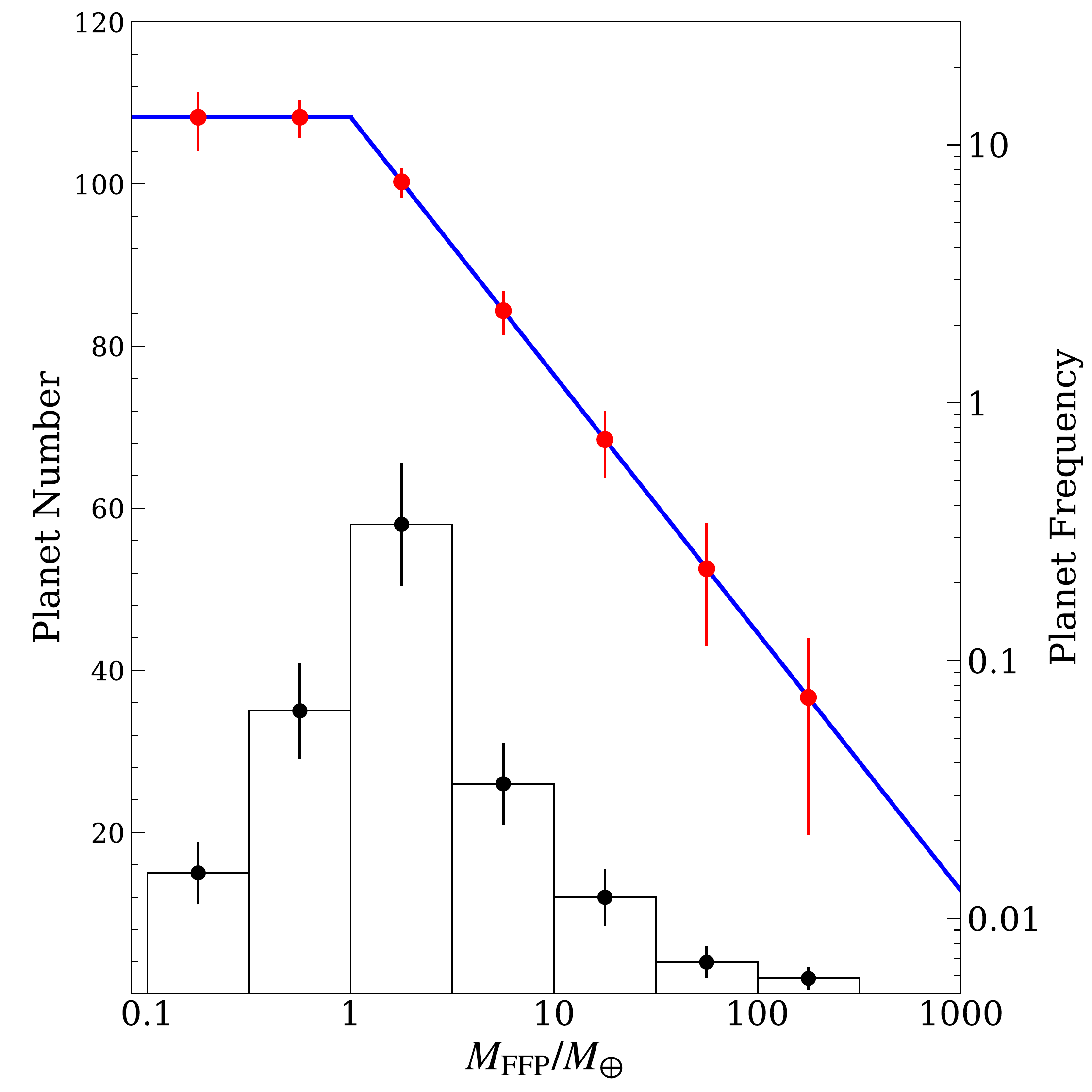}
    \caption{Estimated distribution of FFPs with mass measurements detected by the ET + KMTNet Microlensing Survey. The black histograms show the number of the detected planets in each mass bin, the blue line represents the broken power-law mass function used for the estimate, and the red dots display the estimated constraints in each mass bin.}
    \label{fig:FFPs}
\end{figure}

For FFPs, we follow the procedure of \citet{CMST}, but with two main differences. First, for the simulated ET data, we adopt $\chi^2_{\rm PSPL} - \chi^2_{\rm FSPL} > 16$ as the detection threshold of finite-source effects, where PSPL represents point-source/point-lens. Second, we adopt a broken power-law mass function for FFPs:
\begin{equation}
\frac{dN_{\rm FFP}}{d\log M_{\rm FFP}} = \left\{\begin{array}{ll}
\frac{12.8}{{\rm dex}\ \times\ {\rm star}} \times \left(\frac{M_{\rm FFP}}{M_{\oplus}}\right)^{-1} & {\rm if}\ M_{\rm FFP}\geq 1.0M_{\oplus},\\
\frac{12.8}{{\rm dex}\ \times\ {\rm star}}  & {\rm if}\ M_{\rm FFP} < 1.0M_{\oplus}.
\end{array}\right.
\label{equ:FFPmass}
\end{equation}
This mass function above \SI{1}{M_{\oplus}} is consistent with the mass function derived by the KMTNet FSPL events \citep{Gould2022}. Because current studies had no inference at $M_{\rm FFP}$\SI{<1}{M_{\oplus}}, we adopt a flat power-law function at the low end. For measurements of satellite microlens parallax, Figure 4 of \citet{CMST} displays the fraction of time that the KMTNet telescopes can observe the Galactic bulge. We exclude the bulge-lens events near June 20$^{th}\pm$1 month due to the insufficient projected separation between Earth and the ET satellite. Figure \ref{fig:FFPs} shows the resulting distribution of the FFPs with mass measurements. Considering that the fractions of satellite microlens parallax and $\theta_{\rm E}$ measurements are about \SI{40}{\percent} and \SI{60}{\percent}, respectively, the ET + KMTNet Microlensing Survey will detect about 600 FFP events, of which about 150 will have mass measurements. 

For bound planets, we estimate the yield by a comparison to the simulation of the {\it Roman} Galactic Exoplanet Survey \citep{MatthewWFIRSTI}. Because the two surveys have similar field placements, the event rates are basically the same. We consider three factors that are related to the yield of bound planets, including a linear factor, survey duration (ET with 730 days and {\it Roman} with 432 days), a nearly linear factor, number of microlensing sources, and observational cadences. \citet{MatthewWFIRSTI} showed that sources for most {\it Roman} planetary events have a single-epoch SNR $\geq 5$, corresponding to $\sim$170 million stars. For the ET microlensing survey, $I \approx 20.9$ sources have SNR $\sim 5$, corresponding to 35 million stars. The combined cadence of the ET + KMTNet survey is $\Gamma \geq 12~{\rm hr}^{-1}$, which can detect \SI{\sim 60}{\percent} more planets relative to {\it Roman}'s cadence of $\rm{\Gamma} = 4~{\rm hr}^{-1}$ \citep[see Figure 16 of][]{MatthewWFIRSTI}. Thus, the ultimate ratio for bound planets is ET + KMTNet:{\it Roman} $\approx 5:9$. 

\begin{figure}[htb] 
    \centering
    \includegraphics[width=0.65\columnwidth]{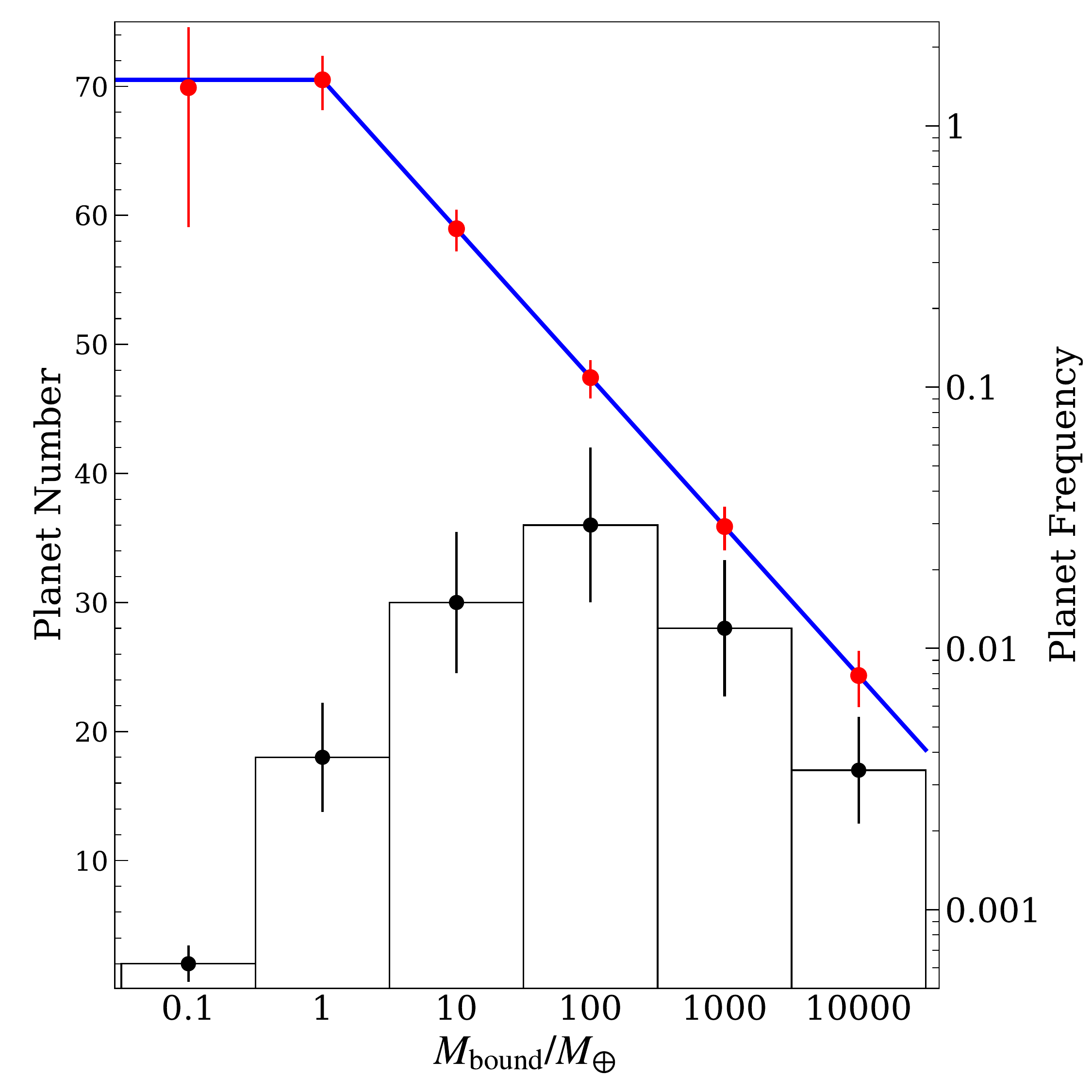}
    \caption{Estimated distribution of bound planets with mass measurements detected by the ET + KMTNet Microlensing Survey. The symbols are similar to those in Figure \ref{fig:FFPs}.}
    \label{fig:bound}
\end{figure}

We use a broken power-law mass function for bound planets:
\begin{equation}
\frac{d^2N_{\rm bound}}{d\log M_{\rm bound}d\log{s}} = \left\{\begin{array}{ll}
\frac{1.5}{{\rm dex}^{-2}\ \times\ {\rm star}} \times \left(\frac{M_{\rm bound}}{M_{\oplus}}\right)^{-0.57} & {\rm if}\ M_{\rm bound}\geq 1.0M_{\oplus},\\
\frac{1.5}{{\rm dex}^{-2}\ \times\ {\rm star}}  & {\rm if}\ M_{\rm bound} < 1.0M_{\oplus},
\end{array}\right.
\label{equ:boundmass}
\end{equation}
where $s$ is the planet-host projected separation in units of $\rm{\theta}_{\rm E}$. This mass function is adapted from the KMTNet mass-ratio function (Zang et al. in prep) with the assumption of a mean host mass of \SI{0.40}{M_{\odot}}. Because current ground-based surveys have no secure detections on $M_{\rm bound}$\SI{<1}{M_{\oplus}}, we adopt a flat power-law function at the low end. We also estimate the fraction of satellite microlens parallax measurements by Figure 4 of \citet{CMST}. The only exception is that we assume only one caustic crossing for all $M$\SI{<10}{M_{\oplus}} events due to their small caustics. In addition, we adopt the simulation results of \cite{Zhu2014ApJ} and assume that \SI{50}{\percent} of planetary events have $\rm{\theta}_{\rm E}$ measurements and \SI{5}{\percent} of planetary events show multiple-planet signatures. Figure \ref{fig:bound} displays the resulting distribution of the bound planets with mass measurements. The ET + KMTNet Microlensing Survey will detect about 430 bound planets, about 130 of which will have mass measurements. 

Here we have estimated the satellite microlens parallax measurements by only considering the KMTNet telescopes, but more bound and free-floating planets can have mass measurements. Other ground-based resources, such as the MOA-II survey, can fill the gap between the KMTNet sites and provide observations when the KMTNet sites are affected by poor weather. Moreover, \citet{Yee2013} showed that high-magnification events, which are sensitive to bound planets \citep{Griest1998}, can yield satellite microlens parallax even if the ground-based telescopes cannot capture the planetary caustic crossings. 

Recently, \citet{MB10477_AO} reported a Jovian analogue orbiting a white dwarf that was detected by a combination of ground-based light curve analysis and Keck adaptive optics imaging, but they did not measure the precise masses of the planetary system due to the poor constraint on microlensing parallax. If we assume that \SI{10}{\percent} of microlensing events are caused by white dwarfs and the planetary occurrence rate is the same for all types of stars, the ET + KMTNet Microlensing Survey, combined with high-resolution imaging, can yield mass measurements for more than 10 planets orbiting white dwarfs.

\section{Target Selection and Follow-up Observations}
\label{sec:targe-follow}
\subsection{ET Target Selection and Input Catalogue}\label{sec:target_selection}  
{\bf Authors:}
\newline
Ji-Wei Xie$^1$, Maosheng Xiang$^2$, Jie Yu$^3$, Haibo Yuan$^4$, Jinghua Zhang$^5$, Hui Zhang$^6$, Jian Ge$^6$, Ruisheng Zhang$^1$ \\
{1.\it School of Astronomy and Space Science, Nanjing University, Nanjing 210023, China}\\
{2.\it Max-Planck-Institut f\"ur Astronomie, K\"onigstuhl 17, D-69117, Heidelberg, Germany }\\
{3.\it Max Planck Institute for Solar System Research, Justus-von-Liebig-Weg 3, 37077 Göttingen, Germany} \\
{4.\it Department of Astronomy, Beijing Normal University, Beijing 100875, China} \\
{5.\it  National Astronomical Observatories, Chinese Academy of Science, Beijing 100101, China }  \\
{6.\it  Shanghai Astronomical Observatories, Chinese Academy of Science, Shanghai 200030, China }  \\

\subsubsection{Overview}
\begin{figure}[htbp]
	\centering
	{\includegraphics[scale=.5]{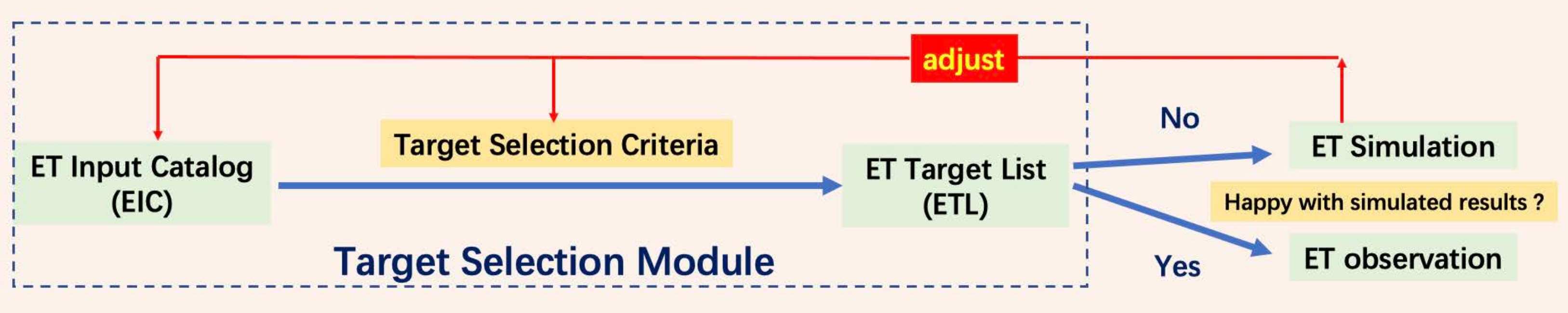}}
	{ \caption{Overview of the ET target selection process.}
		\label{fig:TS_overview}}. 
\end{figure}
Due to the limitations of satellite storage and communication bandwidth, ET cannot downlink all sources in its FOV. Therefore, target selection is an important part of the ET project, whose main goal is to maximize the scientific output of both the core and auxiliary scientific objectives by optimizing the source selection for observation. Figure \ref{fig:TS_overview} shows an overview of ET's target selection process. The first step is to build the input catalog, which contains a comprehensive parameter table of all potential scientific target sources in ET's field of view, including stellar temperature, mass, radius, metal abundance, age, activity index and etc. Next, by applying certain target selection criteria to the catalog, an initial target list is produced and an observational simulation based on the target list will be performed. Then, feedback from the evaluation of simulation results estimating the expected yield will be evaluated by ET's target selection team to make guided adjustments, such as changing the selection criteria. The final ET target list will be produced after, potentially, multiple iterations of these adjustments. 

\begin{figure}[htbp]
	\centering
	{\includegraphics[scale=.6]{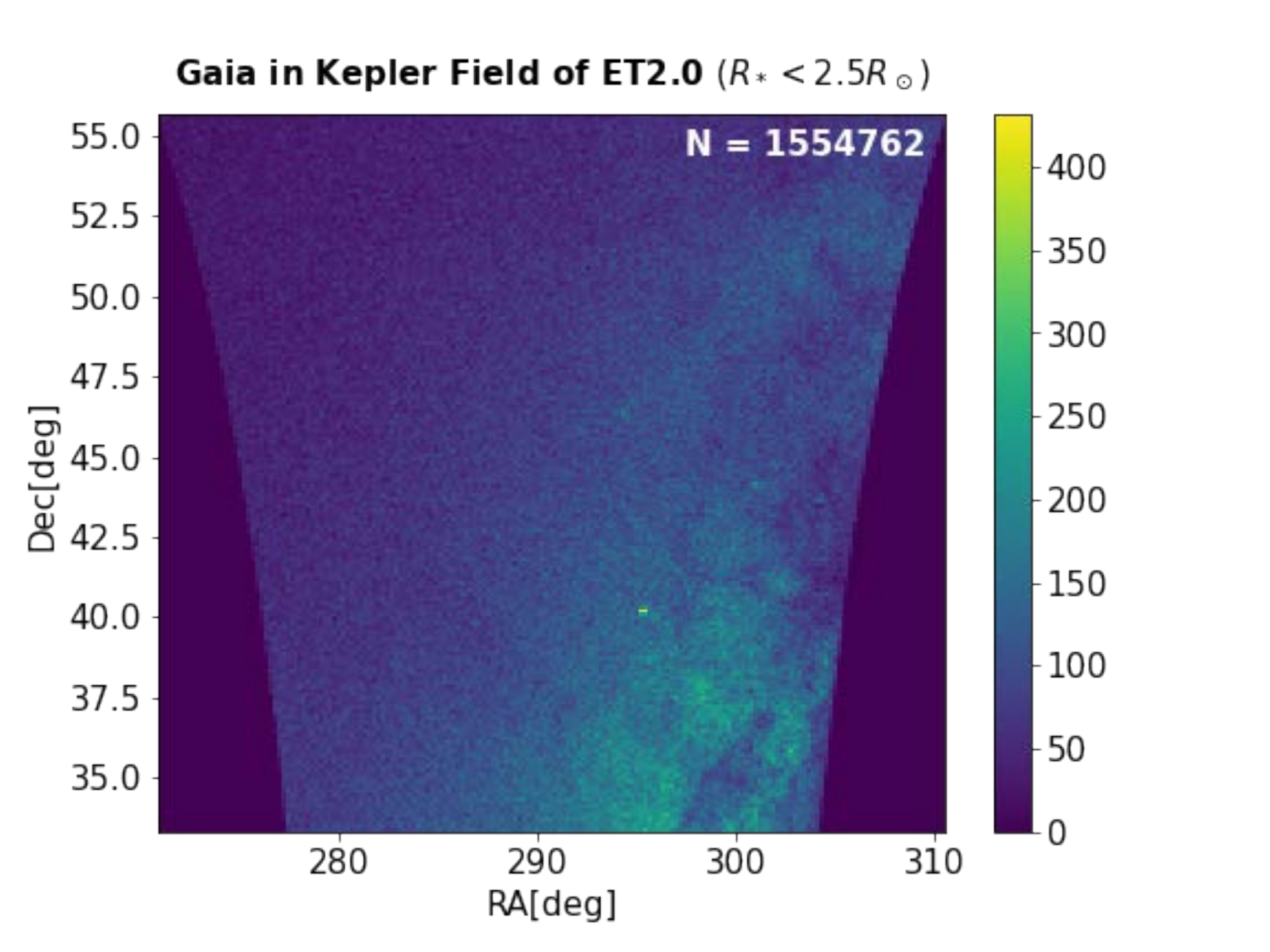}}
	{ \caption{ET's field of view in equatorial coordinates.There are more than 1.5 million dwarf stars ($R_\ast \leq 2.5 R_\odot$) within the 500deg$^2$ FOV. ET plans to select 1.2 million targets according to their stellar activity level. }
		\label{fig:TS_Ra_Dec}}. 
\end{figure}

\subsubsection{Building ET's Input Catalog}
To create ET's initial input catalog we apply some basic selection criteria on the Gaia catalog (currently Gaia eDR3).
\begin{itemize}
    \item[1.] The first selection criterion is the FOV cut. Currently, ET's FOV is set as a square region centered at the center of {\it Kepler} FOV ($\rm{RA}=$\SI{290.67}{\degree}, $\rm{DEC=}$\SI{+44.5}{\degree}) with each side having \SI{22.36}{\degree} and a total area of about \SI{500}{deg^2}.
    
    \item[2.] The second criterion is a brightness cutoff of magnitude G$>16$.
    
    \item[3.] The third criterion is the stellar radius cut. To find exoplanets as small as Earth, we focus on dwarf stars and set the stellar radius criterion to $\leq 2.5\rm{R}_\odot$.
\end{itemize}
After the above three basic criteria, ET's input catalog contains more than 1.5 million stars.
Figure \ref{fig:TS_Ra_Dec} shows the Equatorial coordinate distribution of all the stars in the catalog.

\begin{figure}[htbp]
	\centering
	{\includegraphics[scale=.8]{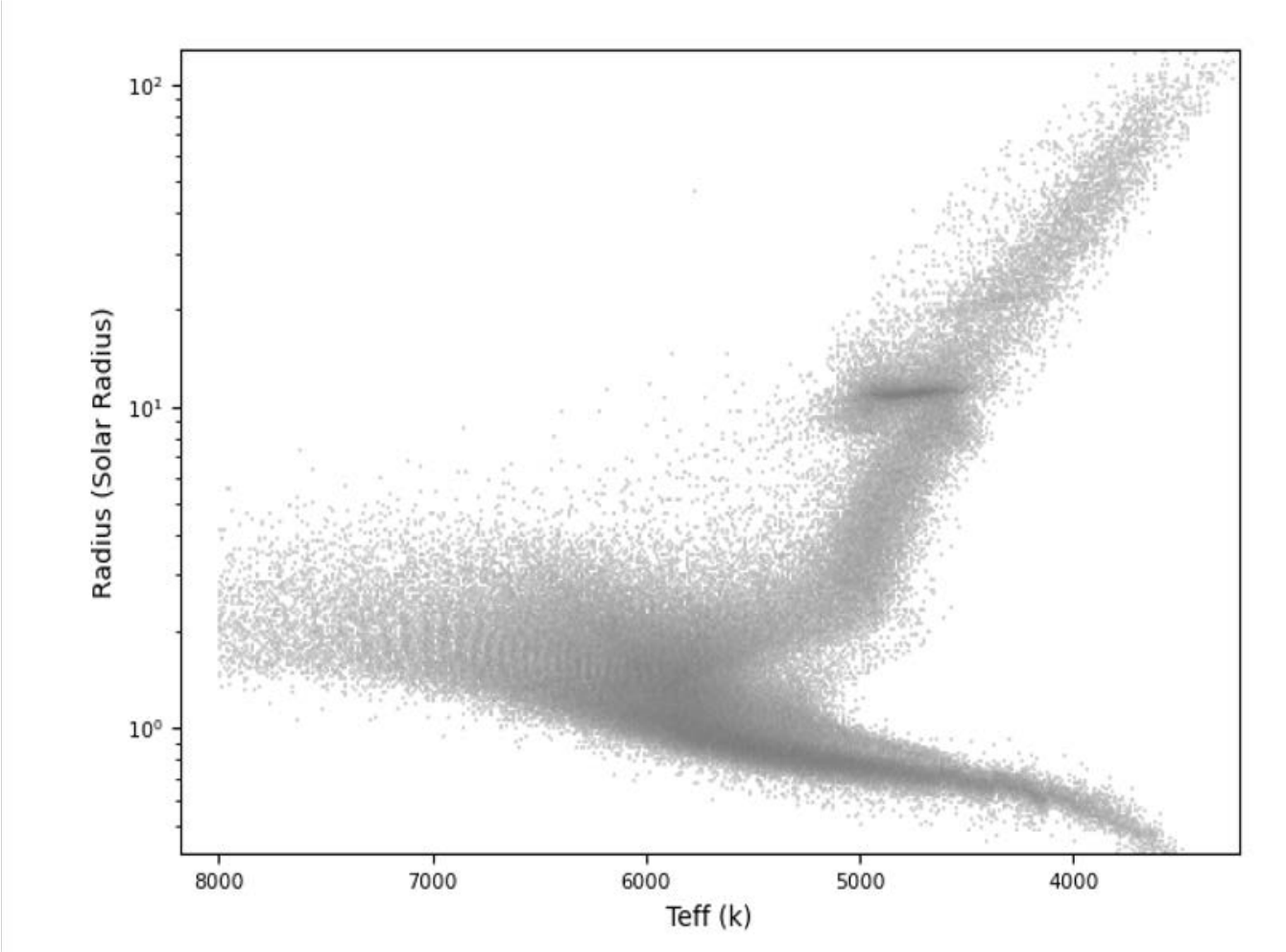}}
	{ \caption{An H-R diagram of stars in ET's input catalog.}
		\label{fig:TS_HR}}. 
\end{figure}

Then, we add more stellar parameters to the basic ET input catalog.
Currently, we have considered the following parameters.
\begin{itemize}
    \item Spectroscopic properties from LAMOST, which provides stellar temperature, surface gravity, and metallicity with typical uncertainties of 50 K, 0.1 dex and 0.1 dex respectively \citep{Luo2015}.
    \item Galactic kinematic properties (e.g., Galactic velocity, thin/thick disk membership) \citep{chen20}.
    \item Stellar properties (e.g., stellar radius with typical uncertainty of 5\%) from SED fitting \citep{Yu2018}.
    \item Magnetic activity indexes (e.g., S-index) \citep{ZhangFabrycky2019}.
    \item Flags indicating binary stars \citep{Xiang2019}.
    \item $[\rm{Fe}/\rm{H}]$ (with typical uncertainty of 0.1-0.2 dex based on Gaia photometry (Yuan et al. in prep.)
\end{itemize}

\begin{figure}[htbp]
	\centering
	{\includegraphics[scale=.4]{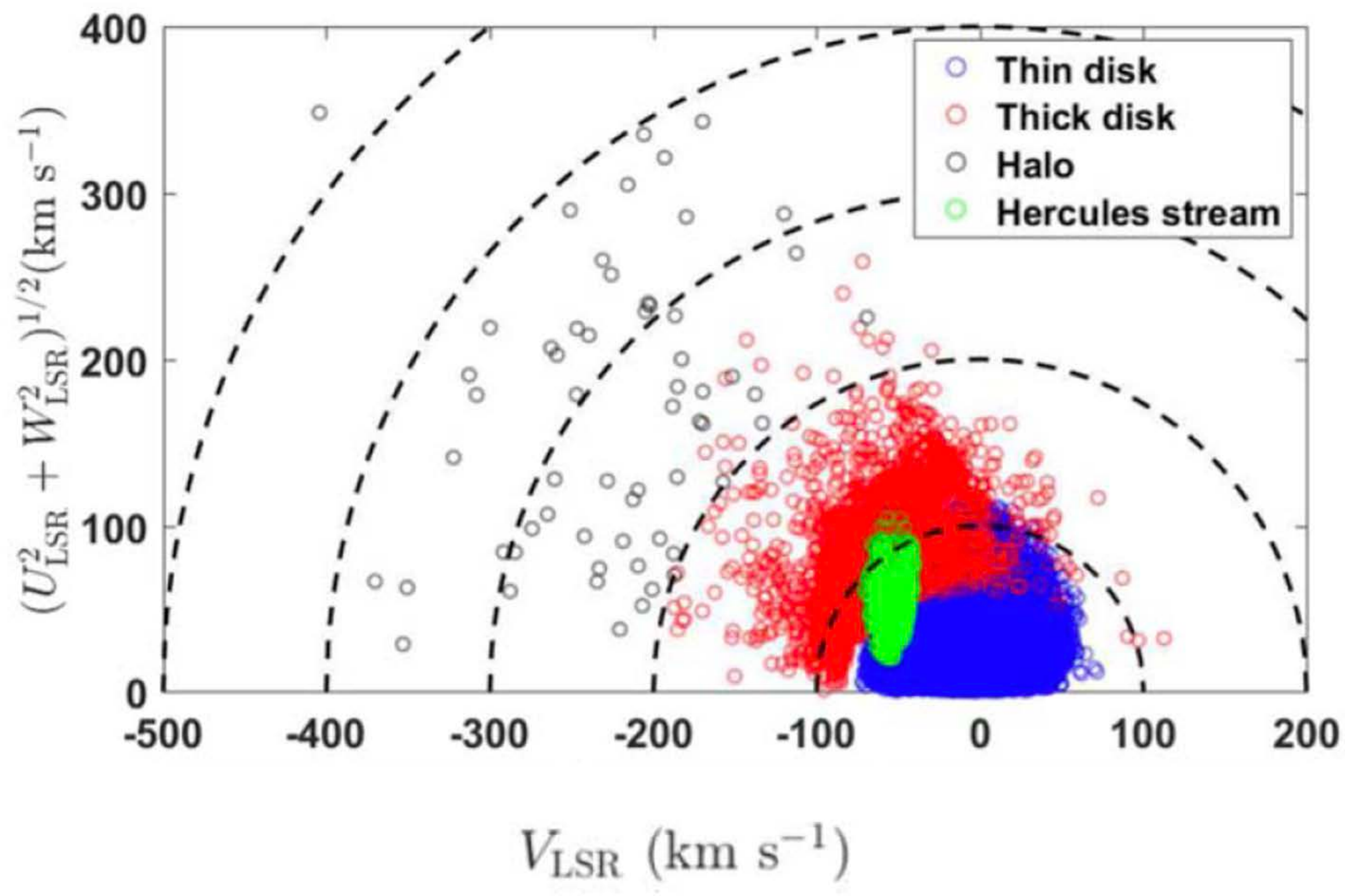}}
	{ \caption{The Toomre diagram of stars for different Galactic components in ET's field of view as characterized by LAMOST and Gaia. Dotted lines show constant values of the total Galactic velocity $V_{\rm tot} = (U_{\rm LSR}^2 + V_{\rm LSR}^2 + W_{\rm LSR}^2)^{1/2}$ in steps of \SI{100}{\km\per\s}.}
	\label{fig:TS_UVW}}
\end{figure}

Figures \ref{fig:TS_HR} and \ref{fig:TS_UVW} show the distributions of some of these added parameters.
The final ET input catalog will be used to select ET targets and characterize planetary systems in the future.

\subsubsection{Building ET's Target List}
ET's target list is generated by applying various selection criteria to the ET input catalog.
Applying appropriate selection criteria is key to the whole target selection process.
Before setting the target selection criteria, we first review how the {\it Kepler} targets were selected \citep{Batalha2010, Wolniewicz2021}, which led to the following main points:
\begin{itemize}
    \item Most {\it Kepler} targets are brighter than 16$^{th}$ magnitude.
    
    \item {\it Kepler} observed nearly all stars brighter than 14$^{th}$ magnitude.
    
    \item For stars between 14$^{th}$ and 16$^{th}$ magnitude, {\it Kepler} tried to observe as many main-sequence stars as possible, but there was sub-giant and giant star contamination; the main sequence stars fraction decreased to \SI{75}{\percent} at 15$^{th}$ magnitude and \SI{55}{\percent} at 16$^{th}$ magnitude.

    \item {\it Kepler} did not apply a filter to preferentially select quiet stars and the majority of its targets are more active than the Sun, which reduced their chances of discovering Earth sized planets.
\end{itemize}

\begin{figure}[htbp]
	\centering
	{\includegraphics[width=0.95\textwidth]{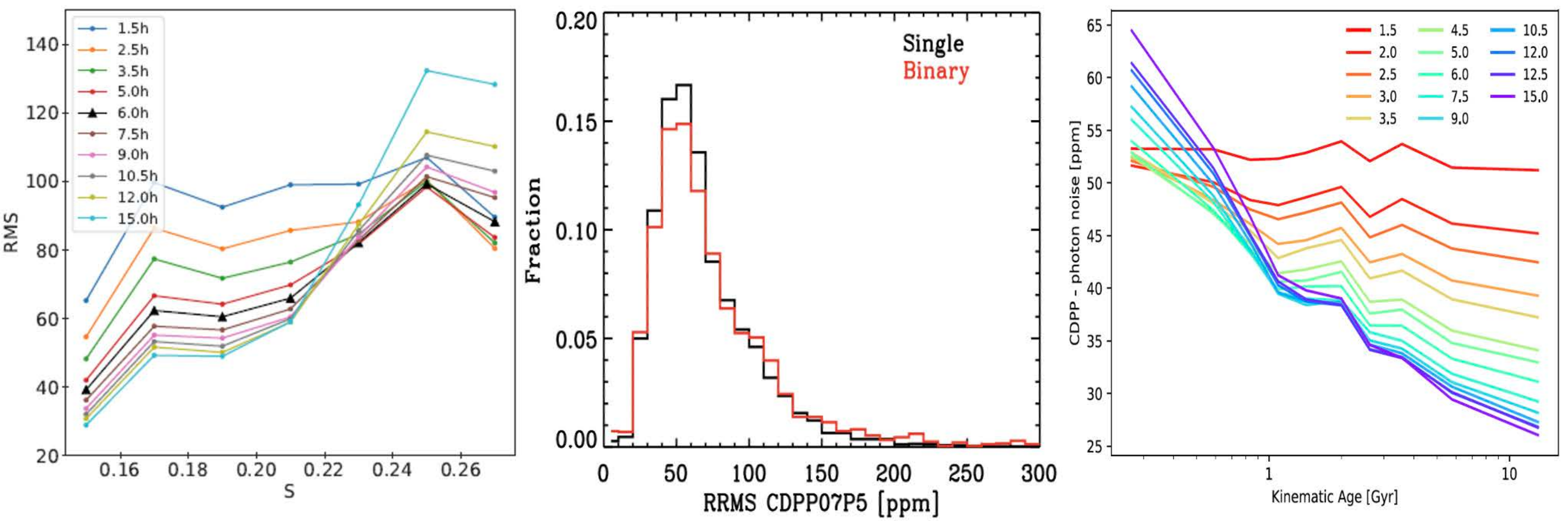}}
	{ \caption{The stellar activity noise, parameterized with the CDPP from {\it Kepler}, as a function of S-index (left) and kinematic age (right), and the CDPP distributions of single and binary systems (middle).}
		\label{fig:TS_CDPP}}. 
\end{figure}

Due to the above lessons from {\it Kepler}, we have worked on the following two aspects to improve the target selection.
\begin{itemize}
    \item \emph{Stellar radius size.}
    For a given transiting planet, a smaller stellar radius leads to a larger transit signal. We have derived radii for most stars in the ET field by performing SED fitting using 26 bandpasses of photometry from eight databases, including GAIA, SDSS, PanSTARRS, and 2MASS. The typical stellar radius uncertainty is \SI{\sim 5}{\percent}, which allows us to accurately select dwarf stars with small radii for planetary transit searching.
    
    \item \emph{Stellar activity noise.}
    Quieter stars with lower stellar activity noises lead to larger transit SNRs, thus enhancing the transit detection efficiency.
    We explored how stellar activity noise, parameterized with the CDPP values in {\it Kepler} photometry data,  depends on some stellar properties by using the data from LAMOST, Gaia, {\it Kepler}, and others (Figure~\ref{fig:TS_CDPP}). 
    We found that the stellar magnetic activity measured with the $S$-index could be a good tracer of the stellar activity noise. 
    Furthermore, older stars (\SI{>2}{Gyr}) are found to be significantly quieter with lower activity noises than younger stars.
    In addition, stellar binaries seems to have little influence on stellar activity noise.
   
\end{itemize}

\subsubsection{Prospect of Future Work}
The ET target selection is still a work in progress.
The ET input catalog will be further improved by incorporating more data from various stellar surveys, such as LAMOST, SDSS-V, Mephisto, and others, to have a more comprehensive characterization of stars in ET's FOV. 
Additionally, the target selection criteria need to be further explored. 
Understanding stellar activity noise is important and although it is impractical to directly measure the stellar activity noise of all stars in ET's field, it may be possible to derive an empirical stellar noise as a function of some general observables or basic parameters, including mass, radius, temperature, and kinematic properties.  
Lastly, it is necessary to find a strategy that balances the core scientific objectives with other scientific objectives.
For example, although giant stars are usually removed from transit search, they are valuable targets for asteroseismology research.

\subsection{Ground and Space-based Follow-up Programs} 
{\bf Authors:}
\newline
Sharon Xuesong Wang$^1$, Tianjun Gan$^1$, Wei Wang$^2$, Yujuan Liu$^2$, Yinan Zhao$^3$, Jian Ge$^4$, Fabo Feng$^5$, Steve B. Howell$^6$ \& Hui-Gen Liu$^7$\\
{1. \it Tsinghua University, Beijing, China}\\
{2. \it National Astronomical Observatories, Chinese Academy of Sciences, Beijing, China}\\
{3. \it Department of Astronomy of the University of Geneva, 51 chemin de Pegasi, 1290 Versoix, Switzerland}\\
{4. \it Shanghai Astronomical Observatories, Chinese Academy of Science, Shanghai 200030, China }\\
{5. \it Tsung-Dao Lee Institute, Shanghai Jiao Tong University, Shanghai, China}\\
{6. \it NASA Ames Research Center, Moffett Field, CA 94035, US}\\
{7. \it School of Astronomy and Space Science, Nanjing University, Nanjing, China}\\

Similar to {\it Kepler} or {\it TESS}, ground-based follow-up observations are critical for validating or confirming ET's exoplanet discoveries and characterizing the host stars and the planetary systems of ET's discoveries. In addition, ET will find targets that are suitable for follow-up atmospheric observations, including cold Jupiters that occupy a relatively empty parameter space in planetary atmospheric characterization. In this subsection, we briefly summarize the follow-up observations that are important to the ET mission.

\subsubsection{Ground-based Photometry} 
Ground-based telescopes have smaller image pixel scales than ET with multi-band photometric capabilities. Thanks to these two features, we can rule out two types of false positive signals with ground-based photometry: (1) Eclipsing binaries near the target source that can be resolved with seeing-limited photometry. Based on previous studies with {\it TESS}, we estimate that around \SIrange[range-units = single]{1}{2}{\percent} of ET triggers fall under this scenario. (2) Blended eclipsing binaries that are too close to the target source to resolve. We can rule out this type of false positive by conducting multi-band photometric follow-up observations of transit signals that are deep enough to be detected using ground-based facilities ($\gtrsim$ \SI{2}{mmag}). Smaller transit signals would have to rely on other types of follow-up observations to confirm or be validated via statistical analyses \citep[e.g.,][]{Giacalone2021}.

Furthermore, given the pixel size of ET ($\sim$4.4"/pixel), some of the target stars will have considerable light contamination from nearby stars. Seeing-limited photometry with ground-based telescopes can help measure the contamination ratio and thus refine the radius estimate of ET-discovered planets. Finally, ground-based photometry could fill in or extend the observational baselines for some of ET's targets, which would refine or keep track of the transit ephemerides \citep[e.g.,][]{Wittenmyer2022}. Additional transit detections on a longer baseline could even reveal or refine the detection of TTVs, which can be important for mass measurement and the detection and characterization of additional planets in the system \citep[see section 2.16, e.g.,][]{Dawson2019}. Global networks of telescopes, such as those operated by the Las Cumbres Observatory \citep{Brown2013}, would be an efficient way to conduct such follow-up observations.


\subsubsection{High-Resolution Imaging} 
Adaptive optics or speckle imaging facilities, such as Robo-AO \citep{Baranec2014} or the Gemini Speckle Imager \citep{Scott2021}, would take on a similar role with the ground-based photometry in the ET follow-up programs.
For the {\it Kepler}, {\it K2}, and {\it TESS} exoplanet missions, \citet{Howell2021FrASS...8...10H} has run a dedicated high-resolution follow-up validation and characterization program leading to global findings for exoplanet host stars \citep[e.g.,][]{Howell2021AJ....161..164H, Lester2021AJ....162...75L}. The high-resolution images provided by these facilities resolve the neighbor or background stars around the target star down to a small angular separation well beyond the seeing limit (\SI{<0.5}{\arcsecond}), reaching inner working angles with speckle observations of the diffraction limit \citep[e.g.,][]{Horch2012AJ....144..165H, Howell2021AJ....161..164H}. The wide-field imaging survey by the upcoming {\it Chinese Space Station Telescope} ({\it CSST}; \citealt{ZhanH2021}) will also provide high spatial resolution (\SI{<0.15}{\arcsecond}) images for the ET targets. High-resolution images often come in two or more bands, providing a color to determine the nature of the blended object \citep[e.g.,][]{clark2022,Lester2021AJ....162...75L}. This is important for ruling out false positives, refining planetary radius measurements, and fully characterizing host stars as well as planetary systems. 

In addition, high-resolution imaging could work in tandem with long-term RV observations and {\it Gaia} astrometric measurements to map out the orbits of stellar or substellar companions in the ET-discovered systems\citep{Colton2021AJ....161...21C,Lester2021AJ....161..164H}. Besides high-resolution AO or speckle imaging, direct imaging systems such as VLT/SPHERE \citep{beuzit2019} or the Gemini Planet Imager \citep{macintosh2014} could be used to follow up on any young planets on wide orbits discovered by ET and measure their spectra. Small transiting planets on moderate to wide orbits around nearby bright stars are ideal targets for space telescopes with coronographs such as {\it CSST} and {\it Roman} \citep{kasdin2020}.

\subsubsection{Reconnaissance Spectroscopy} 
Fundamental stellar parameters such as mass, radius, effective temperature, and surface gravity are critical for deriving planetary properties and estimating the stellar habitable zone \citep{kopparapu2013}. While photometric measurements from stellar catalogs can provide the most basic characterization of a star, spectroscopic observations are more reliable and powerful. In addition to providing basic stellar parameters, spectroscopic line properties, such as those describing the line core flux of H$\alpha$ and Ca H\&K lines, can serve as indicators for the level of stellar magnetic activities, which can have important consequences for habitability \citep[e.g.,][]{airapetian2017}. This is especially important for ET since the primary science goal is to find Earth-like planets in habitable zones.

In addition to characterizing the planet host stars, reconnaissance spectroscopy serves the purpose of selecting targets for follow-up observations that require significantly more telescope resources, such as the RV observations. Reconnaissance spectra with moderate spectral resolution ($\rm{R}\sim 50,000$) can determine the evolutionary stage, binarity in some cases, and stellar activity level of the host star, so that timely decisions can be made on whether or not to pursue follow-up observations. Large spectroscopic surveys such as LAMOST \citep{Cui2012} and SDSS APOGEE V \citep{Majewski2017} would have covered a considerable fraction of ET target stars, but faint or distant stars are often incomplete in such surveys and require additional follow-up efforts. While higher resolution spectra would be ideal for characterization or reconnaissance purposes, they are often not absolutely necessary at the early stage of the follow-up efforts, since they are a natural product of precise RV observations.

\subsubsection{Precise RVs} 
Precise RVs is a powerful tool for confirming transit planet discoveries, as shown by the work done with {\it Kepler} and {\it TESS} targets \citep[e.g.,][]{marcy2014,teske2021}. Precise RVs would provide mass measurements for transiting planets around relatively bright targets discovered by ET and masses are important as they provide a bulk density estimate of the planet and break the degeneracy between surface gravity and mean molecular weight in atmospheric characterization \citep{batalha2019}. Besides confirming the nature of the transit signals and providing the mass measurements, precise RVs are important for characterizing the planetary systems, such as measuring the eccentricity and finding additional non-transiting planets.

\begin{figure}[htbp]
	\centering
	{\includegraphics[scale=.4]{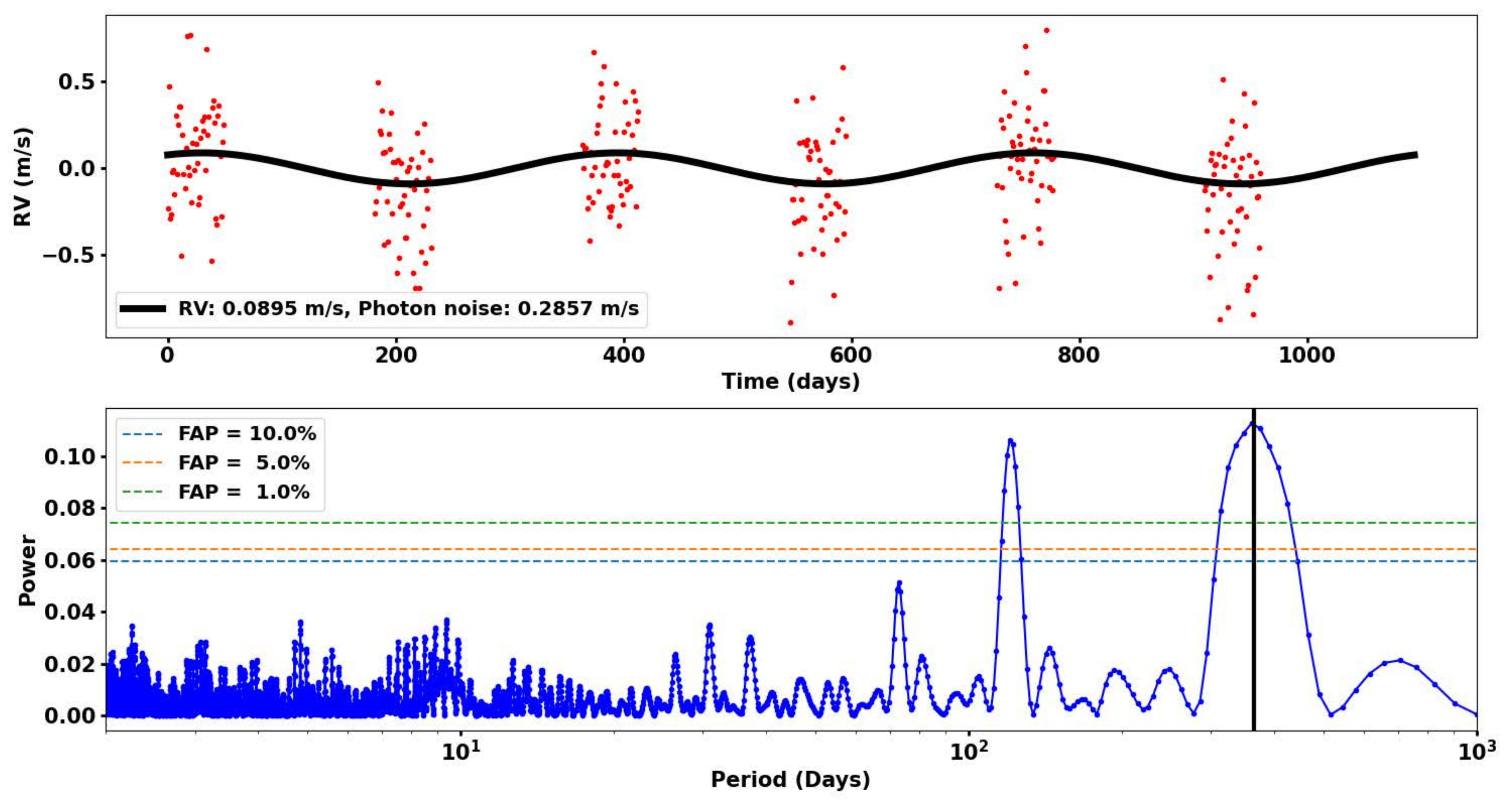}}
	{ \caption{The ability to detect an Earth analog around $V\sim15$ mag one-solar-mass star with an ESPRESSO type instrument on 30-meter class telescopes (TMT). Top: The simulated RV measurements during the 3 year observing window. Each observation is color coded with red and the noise free RV signals is label with a solid black line. The target is observed in each peak and valley phase of the RV curve. 50 days is needed for each phase. For each night, 180 minutes exposure is required to reach a SNR of $\sim 210$. Bottom: The periodogram of simulated RVs. The period of the simulated planet signal is labeled with black line. With 300 nights observation, the signal can reach a 3-$\sigma$ detection.}
		\label{fig:Espresso_limit}}. 
\end{figure}

Similar to {\it Kepler}, not all ET planet candidates are suitable for RV follow-up due to the brightness limit of RV spectroscopy. Typical optical RV spectrographs on 10-meter class telescopes can reach down to \SI{\sim 0.3}{\meter\per\s} precision with a 1-hour exposure of stars at 8 to 9$^{th}$ magnitude and with Solar-like oscillations \citep{chaplin19}, such as ESPRESSO on VLT \citep{pepe10}, Maroon-X on Gemini-North \citep{seifahrt20}, and the upcoming HRS on GTC\footnote{GTC-HRS: \url{http://www.gtc.iac.es/instruments/instrumentat-on.php\#gtchrs}} and the Keck Planet Finder \citep{gibson16}. Most of the ET discovered planets would be challenging for precise RV work due to the faintness of their host stars, but the important ones, such as Earth-like planets around Sun-like stars, are likely to require RV follow-up observations to fully rule out false positive scenarios \citep{fressin13}. This is certainly a resource-intensive effort, but it is feasible within the capability of the current and future 10~cm/s-level instruments. For example, for a $V\sim15$ mag one-solar-mass star hosting a transiting Earth analog (with one Earth mass on a one-year orbit) with an ESPRESSO type instrument on 30-meter class telescopes (TMT), our simulation shows that an observing cadence of one per night with a SNR of 210 over 300 nights would reach a 3-$\sigma$ detection on the RV semiamplitude of this transiting Earth 2.0, as shown in Figure~\ref{fig:Espresso_limit}.

\subsubsection{Planetary Atmosphere Characterization} 
Exo-atmosphere characterization is crucial in understanding the climate and habitability of planets detected by ET. To characterize the atmosphere of transiting planets, especially the Earth-like planets around Solar-type stars, obtaining transmission spectra is a good way to detect absorption of different components.

ET is expected to find planets with long orbital periods around nearby, bright stars, which are great targets for atmospheric characterization. Neither {\it Kepler} or {\it TESS} were sensitive to such planets given their FOV or observational baseline, so ET will fill this relatively empty parameter space and enable studies on the atmosphere of cold or lukewarm planets. In particular, any planets within the habitable zone around nearby stars would be valuable targets for atmospheric characterization.

\begin{figure}[htbp]
	\centering
	{\includegraphics[scale=.6]{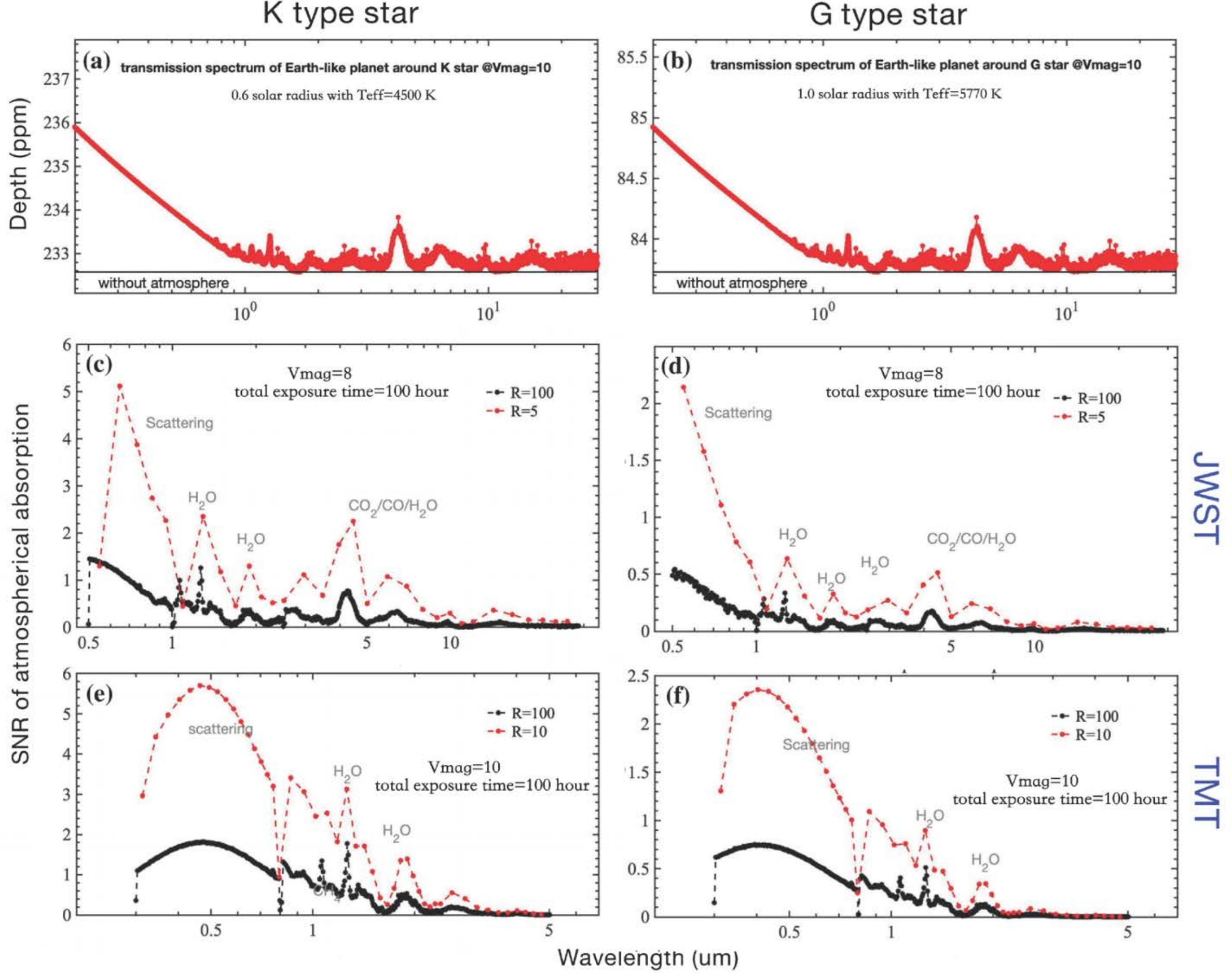}}
	{ \caption{The ability of {\it JWST} and TMT to detect transmission spectrum of an Earth-like planet around G and K type stars. The planet is set as 1 Earth radius and has the same atmosphere as Earth. Panel (a) and (b) are the transiting depth of the planet around G and K type stars, respectively. Note the G star is set exactly as the Sun, while the K star is set as 0.6 solar radius with an effective temperature of \SI{4500}{\kelvin}. Panel (c) and (d) are simulated SNRs of atmospheric absorption via {\it JWST}, assuming the magnitude (V band) of stars are both 8, and with a total exposure time of 100 hours. Panel (e) and (f) are simulated SNRs of atmospheric absorption via TMT, assuming the magnitude (V band) of stars are both 10, and with a total exposure time of 100 hours. Both {\it JWST} and TMT have the ability to characterize the atmosphere of Earth-like planets found by ET, if the host stars are bright enough.}
		\label{fig:atm_SNR_SIM}}. 
\end{figure}

Space missions such as {\it JWST} or {\it ARIEL} would be able to follow up any ET discoveries up to a $J$ magnitude of $\sim$14 and $K$ magnitude of $\sim$12, respectively, down to one Earth radius around red dwarfs \citep[e.g.] []{2022arXiv220203309B, 2019AJ....157..242E}. Nevertheless, bright targets that may be discovered by ET should naturally be more valuable targets for characterization studies, in terms of both required on-target time consumption, and the expected measurement accuracy and thus the yielded physical and chemical constraints. In particular, spectrographs on 30-meter class telescopes (TMT) would play an important role in finding molecular absorption in any nearby, transiting Earth-like planets found by ET  \citep[e.g. ][]{2019MNRAS.484.4855H,2019BAAS...51g.134M}. Figure~\ref{fig:atm_SNR_SIM} show preliminary simulated SNR of transmission spectrum via {\it JWST} and TMT. To follow up the atmosphere of Earth-like planet around G or K type star, both {\it JWST} and TMT have ability to detect the atmospheric absorption if the target star is bright enough.

\section{Summary}  

The ET mission is a wide-field and ultra-high-precision photometric survey mission being developed in China. This mission is designed to measure, for the first time, the occurrence rate and the orbital distributions of Earth-sized planets. ET consists of seven \SI{30}{\cm} telescopes to be launched to the Earth-Sun’s L2 point. Six of these are transit telescopes with a FOV of 500 square degrees. Staring in the direction that encompasses the original {\it Kepler} field for four continuous years, this monitoring will yield tens of thousands of transiting planets, including the elusive Earth twins orbiting solar-type stars. The seventh is a \SI{30}{\cm} microlensing telescope that will monitor an area of 4 square degrees toward the galactic bulge. Combined with simultaneous ground-based KMTNet observations, it will measure masses of hundreds of long-period and free-floating planets. Together, the transit and the microlensing telescopes will revolutionize our understanding of terrestrial planets across a large swath of orbital distances and free space. In addition, the survey data will also facilitate studies in the fields of asteroseismology, Galactic archaeology, time-domain sciences, and black holes in binaries.

\bibliography{ref}{}

\begin{thebibliography}{}
\expandafter\ifx\csname natexlab\endcsname\relax\def\natexlab#1{#1}\fi
\providecommand{\url}[1]{\href{#1}{#1}}

\bibitem[{{Abbott} {et~al.}(2017{\natexlab{a}}){Abbott}, {Abbott}, {Abbott},
  {Acernese}, {Ackley}, {Adams}, {Adams}, {Addesso}, {Adhikari}, {Adya},
  {Affeldt}, {Afrough}, \& etc.}]{2017ApJ...848L..12A}
{Abbott}, B.~P., {Abbott}, R., {Abbott}, T.~D., {et~al.} 2017{\natexlab{a}},
  \apjl, 848, L12

\bibitem[{{Abbott} {et~al.}(2017{\natexlab{b}}){Abbott}, {Abbott}, {Abbott},
  {Acernese}, {Ackley}, {Adams}, {Adams}, {Addesso}, {Adhikari}, {Adya}, \&
  et~al.}]{2017PhRvL.119p1101A}
---. 2017{\natexlab{b}}, \prl, 119, 161101

\bibitem[{{Abdurro'uf} {et~al.}(2022){Abdurro'uf}, {Accetta}, {Aerts}, {Silva
  Aguirre}, {Ahumada}, {Ajgaonkar}, {Filiz Ak}, {Alam}, {Allende Prieto},
  {Almeida}, \& et~al.}]{Abdurrouf2022}
{Abdurro'uf}, {Accetta}, K., {Aerts}, C., {et~al.} 2022, \apjs, 259, 35

\bibitem[{{Adibekyan} {et~al.}(2012){Adibekyan}, {Sousa}, {Santos}, {Delgado
  Mena}, {Gonz{\'a}lez Hern{\'a}ndez}, {Israelian}, {Mayor}, \&
  {Khachatryan}}]{Adibekyan:2012}
{Adibekyan}, V.~Z., {Sousa}, S.~G., {Santos}, N.~C., {et~al.} 2012, \aap, 545,
  A32

\bibitem[{{Aerts}(2021)}]{aerts2021a}
{Aerts}, C. 2021, Reviews of Modern Physics, 93, 015001

\bibitem[{{Aerts} {et~al.}(2019){Aerts}, {Mathis}, \& {Rogers}}]{aerts2019b}
{Aerts}, C., {Mathis}, S., \& {Rogers}, T.~M. 2019, \araa, 57, 35

\bibitem[{{Agol} {et~al.}(2005){Agol}, {Steffen}, {Sari}, \&
  {Clarkson}}]{Ago05}
{Agol}, E., {Steffen}, J., {Sari}, R., \& {Clarkson}, W. 2005, \mnras, 359, 567

\bibitem[{{Airapetian} {et~al.}(2017){Airapetian}, {Glocer}, {Khazanov},
  {Loyd}, {France}, {Sojka}, {Danchi}, \& {Liemohn}}]{airapetian2017}
{Airapetian}, V.~S., {Glocer}, A., {Khazanov}, G.~V., {et~al.} 2017, \apjl,
  836, L3

\bibitem[{{Aizawa} {et~al.}(2017){Aizawa}, {Uehara}, {Masuda}, {Kawahara}, \&
  {Suto}}]{Aizawa2017}
{Aizawa}, M., {Uehara}, S., {Masuda}, K., {Kawahara}, H., \& {Suto}, Y. 2017,
  \aj, 153, 193

\bibitem[{{Alam} {et~al.}(2022){Alam}, {Kirk}, {Dressing}, {L{\'o}pez-Morales},
  {Ohno}, {Gao}, {Akinsanmi}, {Santerne}, {Grouffal}, {Adibekyan}, {Barros},
  {Buchhave}, {Crossfield}, {Dai}, {Deleuil}, {Giacalone}, {Lillo-Box},
  {Marley}, {Mayo}, {Mortier}, {Santos}, {Sousa}, {Turtelboom}, {Wheatley}, \&
  {Vanderburg}}]{Alam2022}
{Alam}, M.~K., {Kirk}, J., {Dressing}, C.~D., {et~al.} 2022, \apjl, 927, L5

\bibitem[{{Alcock} {et~al.}(2001){Alcock}, {Allsman}, {Alves}, {Axelrod},
  {Becker}, {Bennett}, {Cook}, {Drake}, {Freeman}, {Geha}, {Griest}, {Keller},
  {Lehner}, {Marshall}, {Minniti}, {Nelson}, {Peterson}, {Popowski}, {Pratt},
  {Quinn}, {Stubbs}, {Sutherland}, {Tomaney}, {Vandehei}, \&
  {Welch}}]{Alock2001}
{Alcock}, C., {Allsman}, R.~A., {Alves}, D.~R., {et~al.} 2001, \nat, 414, 617

\bibitem[{{Amard} {et~al.}(2020){Amard}, {Roquette}, \& {Matt}}]{Amard2020}
{Amard}, L., {Roquette}, J., \& {Matt}, S.~P. 2020, \mnras, 499, 3481

\bibitem[{{Anders} {et~al.}(2017){Anders}, {Chiappini}, {Minchev}, {Miglio},
  {Montalb{\'a}n}, {Mosser}, {Rodrigues}, {Santiago}, {Baudin}, {Beers}, {da
  Costa}, {Garc{\'\i}a}, {Garc{\'\i}a-Hern{\'a}ndez}, {Holtzman}, {Maia},
  {Majewski}, {Mathur}, {Noels-Grotsch}, {Pan}, {Schneider}, {Schultheis},
  {Steinmetz}, {Valentini}, \& {Zamora}}]{Anders2017}
{Anders}, F., {Chiappini}, C., {Minchev}, I., {et~al.} 2017, \aap, 600, A70

\bibitem[{{Anderson}(2016)}]{2016Anderson}
{Anderson}, J. 2016, {Empirical Models for the WFC3/IR PSF},  Space Telescope
  WFC Instrument Science Report

\bibitem[{{Anderson} \& {King}(2000)}]{2000PASPAnderson}
{Anderson}, J., \& {King}, I.~R. 2000, \pasp, 112, 1360

\bibitem[{{Angus} {et~al.}(2015){Angus}, {Aigrain}, {Foreman-Mackey}, \&
  {McQuillan}}]{Angus2015}
{Angus}, R., {Aigrain}, S., {Foreman-Mackey}, D., \& {McQuillan}, A. 2015,
  \mnras, 450, 1787

\bibitem[{{Armstrong} {et~al.}(2013){Armstrong}, {Martin}, {Brown}, {Faedi},
  {G{\'o}mez Maqueo Chew}, {Mardling}, {Pollacco}, {Triaud}, \&
  {Udry}}]{Armstrong13}
{Armstrong}, D., {Martin}, D.~V., {Brown}, G., {et~al.} 2013, \mnras, 434, 3047

\bibitem[{{Armstrong} {et~al.}(2014{\natexlab{a}}){Armstrong}, {Osborn},
  {Brown}, {Faedi}, {G{\'o}mez Maqueo Chew}, {Martin}, {Pollacco}, \&
  {Udry}}]{Armstrong:2014}
{Armstrong}, D.~J., {Osborn}, H.~P., {Brown}, D.~J.~A., {et~al.}
  2014{\natexlab{a}}, \mnras, 444, 1873

\bibitem[{{Armstrong} {et~al.}(2014{\natexlab{b}}){Armstrong}, {Osborn},
  {Brown}, {Faedi}, {G{\'o}mez Maqueo Chew}, {Martin}, {Pollacco}, \&
  {Udry}}]{Armstrong2014}
---. 2014{\natexlab{b}}, \mnras, 444, 1873

\bibitem[{{Artymowicz} \& {Lubow}(1994)}]{Artymowicz:1994}
{Artymowicz}, P., \& {Lubow}, S.~H. 1994, \apj, 421, 651

\bibitem[{{Auge} {et~al.}(2020){Auge}, {Huber}, {Heinze}, {Shappee}, {Tonry},
  {Chakrabarti}, {Sand erson}, {Denneau}, {Flewelling}, {Holoien}, {Kochanek},
  {Pignata}, {Sickafoose}, {Stalder}, {Stanek}, {Stello}, \&
  {Thompson}}]{auge2020a}
{Auge}, C., {Huber}, D., {Heinze}, A., {et~al.} 2020, \aj, 160, 18

\bibitem[{{Bae} {et~al.}(2019){Bae}, {Zhu}, {Baruteau}, {Benisty}, {Dullemond},
  {Facchini}, {Isella}, {Keppler}, {P{\'e}rez}, \& {Teague}}]{Bae2019}
{Bae}, J., {Zhu}, Z., {Baruteau}, C., {et~al.} 2019, \apjl, 884, L41

\bibitem[{{Baliunas} {et~al.}(1995){Baliunas}, {Donahue}, {Soon}, {Horne},
  {Frazer}, {Woodard-Eklund}, {Bradford}, {Rao}, {Wilson}, {Zhang}, {Bennett},
  {Briggs}, {Carroll}, {Duncan}, {Figueroa}, {Lanning}, {Misch}, {Mueller},
  {Noyes}, {Poppe}, {Porter}, {Robinson}, {Russell}, {Shelton}, {Soyumer},
  {Vaughan}, \& {Whitney}}]{Baliunas1995}
{Baliunas}, S.~L., {Donahue}, R.~A., {Soon}, W.~H., {et~al.} 1995, \apj, 438,
  269

\bibitem[{{Balona}(2020)}]{balona2020a}
{Balona}, L.~A. 2020, Frontiers in Astronomy and Space Sciences, 7, 85

\bibitem[{{Baranec} {et~al.}(2014){Baranec}, {Riddle}, {Law}, {Ramaprakash},
  {Tendulkar}, {Hogstrom}, {Bui}, {Burse}, {Chordia}, {Das}, {Dekany},
  {Kulkarni}, \& {Punnadi}}]{Baranec2014}
{Baranec}, C., {Riddle}, R., {Law}, N.~M., {et~al.} 2014, \apjl, 790, L8

\bibitem[{{Barber} {et~al.}(2014){Barber}, {Kilic}, {Brown}, \&
  {Gianninas}}]{2014ApJ...786...77B}
{Barber}, S.~D., {Kilic}, M., {Brown}, W.~R., \& {Gianninas}, A. 2014, \apj,
  786, 77

\bibitem[{{Barbuy} {et~al.}(2018){Barbuy}, {Chiappini}, \&
  {Gerhard}}]{Barbuy2018}
{Barbuy}, B., {Chiappini}, C., \& {Gerhard}, O. 2018, \araa, 56, 223

\bibitem[{{Barclay} {et~al.}(2017){Barclay}, {Quintana}, {Raymond}, \&
  {Penny}}]{Barclay2017}
{Barclay}, T., {Quintana}, E.~V., {Raymond}, S.~N., \& {Penny}, M.~T. 2017,
  \apj, 841, 86

\bibitem[{{Barnes} \& {Fortney}(2004)}]{Barnes2004}
{Barnes}, J.~W., \& {Fortney}, J.~J. 2004, \apj, 616, 1193

\bibitem[{{Barnes} \& {O'Brien}(2002)}]{barnes02}
{Barnes}, J.~W., \& {O'Brien}, D.~P. 2002, \apj, 575, 1087

\bibitem[{{Barnes}(2007)}]{Barnes2007}
{Barnes}, S.~A. 2007, \apj, 669, 1167

\bibitem[{{Barnes}(2010)}]{Barnes2010}
---. 2010, \apj, 722, 222

\bibitem[{{Barnes} {et~al.}(2016){Barnes}, {Weingrill}, {Fritzewski},
  {Strassmeier}, \& {Platais}}]{Barnes2016}
{Barnes}, S.~A., {Weingrill}, J., {Fritzewski}, D., {Strassmeier}, K.~G., \&
  {Platais}, I. 2016, \apj, 823, 16

\bibitem[{{Batalha} {et~al.}(2019){Batalha}, {Lewis}, {Fortney}, {Batalha},
  {Kempton}, {Lewis}, \& {Line}}]{batalha2019}
{Batalha}, N.~E., {Lewis}, T., {Fortney}, J.~J., {et~al.} 2019, \apjl, 885, L25

\bibitem[{{Batalha} {et~al.}(2010){Batalha}, {Borucki}, {Koch}, {Bryson},
  {Haas}, {Brown}, {Caldwell}, {Hall}, {Gilliland}, {Latham}, {Meibom}, \&
  {Monet}}]{Batalha2010}
{Batalha}, N.~M., {Borucki}, W.~J., {Koch}, D.~G., {et~al.} 2010, \apjl, 713,
  L109

\bibitem[{Batalha {et~al.}(2013)Batalha, Rowe, Bryson, Barclay, Burke,
  Caldwell, Christiansen, Mullally, Thompson, Brown, {et~al.}}]{Batalha2013}
Batalha, N.~M., Rowe, J.~F., Bryson, S.~T., {et~al.} 2013, The Astrophysical
  Journal Supplement Series, 204, 24

\bibitem[{Bazot {et~al.}(2019)Bazot, Benomar, Christensen-Dalsgaard, Gizon,
  Hanasoge, Nielsen, Petit, \& Sreenivasan}]{bazot_latitudinal_2019}
Bazot, M., Benomar, O., Christensen-Dalsgaard, J., {et~al.} 2019, A\&A, 623,
  A125, bibtex: Bazot2019.
\newblock \url{http://adsabs.harvard.edu/abs/2019A%26A...623A.125B}

\bibitem[{{Beck} {et~al.}(2022){Beck}, {Mathur}, {Hambleton}, {Garc{\'\i}a},
  {Steinwender}, {Eisner}, {do Nascimento}, {Gaulme}, \& {Mathis}}]{beck2022a}
{Beck}, P.~G., {Mathur}, S., {Hambleton}, K., {et~al.} 2022, arXiv e-prints,
  arXiv:2202.02373

\bibitem[{{Bedding} {et~al.}(2011){Bedding}, {Mosser}, {Huber},
  {Montalb{\'a}n}, {Beck}, {Christensen-Dalsgaard}, {Elsworth},
  {Garc{\'{\i}}a}, {Miglio}, {Stello}, {White}, {De Ridder}, {Hekker}, {Aerts},
  {Barban}, {Belkacem}, {Broomhall}, {Brown}, {Buzasi}, {Carrier}, {Chaplin},
  {di Mauro}, {Dupret}, {Frandsen}, {Gilliland}, {Goupil}, {Jenkins},
  {Kallinger}, {Kawaler}, {Kjeldsen}, {Mathur}, {Noels}, {Silva Aguirre}, \&
  {Ventura}}]{bedding2011a}
{Bedding}, T.~R., {Mosser}, B., {Huber}, D., {et~al.} 2011, \nat, 471, 608

\bibitem[{{Bedding} {et~al.}(2020{\natexlab{a}}){Bedding}, {Murphy}, {Hey},
  {Huber}, {Li}, {Smalley}, {Stello}, {White}, {Ball}, {Chaplin}, {Colman},
  {Fuller}, {Gaidos}, {Harbeck}, {Hermes}, {Holdsworth}, {Li}, {Li}, {Mann},
  {Reese}, {Sekaran}, {Yu}, {Antoci}, {Bergmann}, {Brown}, {Howard}, {Ireland},
  {Isaacson}, {Jenkins}, {Kjeldsen}, {McCully}, {Rabus}, {Rains}, {Ricker},
  {Tinney}, \& {Vanderspek}}]{2020Natur.581..147B}
{Bedding}, T.~R., {Murphy}, S.~J., {Hey}, D.~R., {et~al.} 2020{\natexlab{a}},
  \nat, 581, 147

\bibitem[{{Bedding} {et~al.}(2020{\natexlab{b}}){Bedding}, {Murphy}, {Hey},
  {Huber}, {Li}, {Smalley}, {Stello}, {White}, {Ball}, {Chaplin}, {Colman},
  {Fuller}, {Gaidos}, {Harbeck}, {Hermes}, {Holdsworth}, {Li}, {Li}, {Mann},
  {Reese}, {Sekaran}, {Yu}, {Antoci}, {Bergmann}, {Brown}, {Howard}, {Ireland},
  {Isaacson}, {Jenkins}, {Kjeldsen}, {McCully}, {Rabus}, {Rains}, {Ricker},
  {Tinney}, \& {Vanderspek}}]{Bedding2020}
---. 2020{\natexlab{b}}, \nat, 581, 147

\bibitem[{{Bellinger}(2020)}]{Bellinger2020}
{Bellinger}, E.~P. 2020, \mnras, 492, L50

\bibitem[{{Belokurov} {et~al.}(2018){Belokurov}, {Erkal}, {Evans}, {Koposov},
  \& {Deason}}]{Belokurov2018}
{Belokurov}, V., {Erkal}, D., {Evans}, N.~W., {Koposov}, S.~E., \& {Deason},
  A.~J. 2018, \mnras, 478, 611

\bibitem[{{Belokurov} {et~al.}(2020){Belokurov}, {Sanders}, {Fattahi}, {Smith},
  {Deason}, {Evans}, \& {Grand}}]{Belokurov2020}
{Belokurov}, V., {Sanders}, J.~L., {Fattahi}, A., {et~al.} 2020, \mnras, 494,
  3880

\bibitem[{{Benk{\H{o}}} {et~al.}(2010){Benk{\H{o}}}, {Kolenberg}, {Szab{\'o}},
  {Kurtz}, {Bryson}, {Bregman}, {Still}, {Smolec}, {Nuspl}, {Nemec},
  {Moskalik}, {Kopacki}, {Koll{\'a}th}, {Guggenberger}, {di Criscienzo},
  {Christensen-Dalsgaard}, {Kjeldsen}, {Borucki}, {Koch}, {Jenkins}, \& {van
  Cleve}}]{Benk2010MNRAS}
{Benk{\H{o}}}, J.~M., {Kolenberg}, K., {Szab{\'o}}, R., {et~al.} 2010, \mnras,
  409, 1585

\bibitem[{{Bennett} \& {Rhie}(1996)}]{Bennett1996}
{Bennett}, D.~P., \& {Rhie}, S.~H. 1996, \apj, 472, 660

\bibitem[{{Bennett} {et~al.}(2020){Bennett}, {Bhattacharya}, {Beaulieu},
  {Blackman}, {Vand orou}, {Terry}, {Cole}, {Henderson}, {Koshimoto}, {Lu},
  {Baptiste Marquette}, {Ranc}, \& {Udalski}}]{OB050071AO}
{Bennett}, D.~P., {Bhattacharya}, A., {Beaulieu}, J.-P., {et~al.} 2020, \aj,
  159, 68

\bibitem[{{Benomar} {et~al.}(2018){Benomar}, {Bazot}, {Nielsen}, {Gizon},
  {Sekii}, {Takata}, {Hotta}, {Hanasoge}, {Sreenivasan}, \&
  {Christensen-Dalsgaard}}]{benomar2018a}
{Benomar}, O., {Bazot}, M., {Nielsen}, M.~B., {et~al.} 2018, Science, 361, 1231

\bibitem[{{Benz} {et~al.}(2021){Benz}, {Broeg}, {Fortier}, {Rando}, {Beck},
  {Beck}, {Queloz}, {Ehrenreich}, {Maxted}, {Isaak}, {Billot}, {Alibert},
  {Alonso}, {Ant{\'o}nio}, {Asquier}, {Bandy}, {B{\'a}rczy}, {Barrado},
  {Barros}, {Baumjohann}, {Bekkelien}, {Bergomi}, {Biondi}, {Bonfils},
  {Borsato}, {Brandeker}, {Busch}, {Cabrera}, {Cessa}, {Charnoz}, {Chazelas},
  {Collier Cameron}, {Corral Van Damme}, {Cortes}, {Davies}, {Deleuil},
  {Deline}, {Delrez}, {Demangeon}, {Demory}, {Erikson}, {Farinato}, {Fossati},
  {Fridlund}, {Futyan}, {Gandolfi}, {Garcia Munoz}, {Gillon}, {Guterman},
  {Gutierrez}, {Hasiba}, {Heng}, {Hernandez}, {Hoyer}, {Kiss}, {Kovacs},
  {Kuntzer}, {Laskar}, {Lecavelier des Etangs}, {Lendl}, {L{\'o}pez}, {Lora},
  {Lovis}, {L{\"u}ftinger}, {Magrin}, {Malvasio}, {Marafatto}, {Michaelis}, {de
  Miguel}, {Modrego}, {Munari}, {Nascimbeni}, {Olofsson}, {Ottacher},
  {Ottensamer}, {Pagano}, {Palacios}, {Pall{\'e}}, {Peter}, {Piazza}, {Piotto},
  {Pizarro}, {Pollaco}, {Ragazzoni}, {Ratti}, {Rauer}, {Ribas}, {Rieder},
  {Rohlfs}, {Safa}, {Salatti}, {Santos}, {Scandariato}, {S{\'e}gransan},
  {Simon}, {Smith}, {Sordet}, {Sousa}, {Steller}, {Szab{\'o}}, {Szoke},
  {Thomas}, {Tschentscher}, {Udry}, {Van Grootel}, {Viotto}, {Walter},
  {Walton}, {Wildi}, \& {Wolter}}]{Benz2021}
{Benz}, W., {Broeg}, C., {Fortier}, A., {et~al.} 2021, Experimental Astronomy,
  51, 109

\bibitem[{{Bergemann} {et~al.}(2014){Bergemann}, {Ruchti}, {Serenelli},
  {Feltzing}, {Alves-Brito}, {Asplund}, {Bensby}, {Gruyters}, {Heiter},
  {Hourihane}, {Korn}, {Lind}, {Marino}, {Jofre}, {Nordlander}, {Ryde},
  {Worley}, {Gilmore}, {Randich}, {Ferguson}, {Jeffries}, {Micela},
  {Negueruela}, {Prusti}, {Rix}, {Vallenari}, {Alfaro}, {Allende Prieto},
  {Bragaglia}, {Koposov}, {Lanzafame}, {Pancino}, {Recio-Blanco}, {Smiljanic},
  {Walton}, {Costado}, {Franciosini}, {Hill}, {Lardo}, {de Laverny}, {Magrini},
  {Maiorca}, {Masseron}, {Morbidelli}, {Sacco}, {Kordopatis}, \&
  {Tautvai{\v{s}}ien{\.{e}}}}]{Bergemann2014}
{Bergemann}, M., {Ruchti}, G.~R., {Serenelli}, A., {et~al.} 2014, \aap, 565,
  A89

\bibitem[{{Bersten} {et~al.}(2012){Bersten}, {Benvenuto}, {Nomoto}, {Ergon},
  {Folatelli}, {Sollerman}, {Benetti}, {Botticella}, {Fraser}, {Kotak},
  {Maeda}, {Ochner}, \& {Tomasella}}]{Bersten2012}
{Bersten}, M.~C., {Benvenuto}, O.~G., {Nomoto}, K., {et~al.} 2012, \apj, 757,
  31

\bibitem[{{Bersten} {et~al.}(2018){Bersten}, {Folatelli}, {Garc{\'\i}a}, {van
  Dyk}, {Benvenuto}, {Orellana}, {Buso}, {S{\'a}nchez}, {Tanaka}, {Maeda},
  {Filippenko}, {Zheng}, {Brink}, {Cenko}, {de Jaeger}, {Kumar}, {Moriya},
  {Nomoto}, {Perley}, {Shivvers}, \& {Smith}}]{Bersten2017}
{Bersten}, M.~C., {Folatelli}, G., {Garc{\'\i}a}, F., {et~al.} 2018, \nat, 554,
  497

\bibitem[{Berthier {et~al.}(2006)Berthier, Vachier, Thuillot, Fernique,
  Ochsenbein, Genova, Lainey, \& Arlot}]{berthier2006skybot}
Berthier, J., Vachier, F., Thuillot, W., {et~al.} 2006, in Astronomical Data
  Analysis Software and Systems XV, Vol. 351, 367

\bibitem[{{Beuzit} {et~al.}(2019){Beuzit}, {Vigan}, {Mouillet}, {Dohlen},
  {Gratton}, {Boccaletti}, {Sauvage}, {Schmid}, {Langlois}, {Petit},
  {Baruffolo}, {Feldt}, {Milli}, {Wahhaj}, {Abe}, {Anselmi}, {Antichi},
  {Barette}, {Baudrand}, {Baudoz}, {Bazzon}, {Bernardi}, {Blanchard}, {Brast},
  {Bruno}, {Buey}, {Carbillet}, {Carle}, {Cascone}, {Chapron}, {Charton},
  {Chauvin}, {Claudi}, {Costille}, {De Caprio}, {de Boer}, {Delboulb{\'e}},
  {Desidera}, {Dominik}, {Downing}, {Dupuis}, {Fabron}, {Fantinel}, {Farisato},
  {Feautrier}, {Fedrigo}, {Fusco}, {Gigan}, {Ginski}, {Girard}, {Giro},
  {Gisler}, {Gluck}, {Gry}, {Henning}, {Hubin}, {Hugot}, {Incorvaia}, {Jaquet},
  {Kasper}, {Lagadec}, {Lagrange}, {Le Coroller}, {Le Mignant}, {Le Ruyet},
  {Lessio}, {Lizon}, {Llored}, {Lundin}, {Madec}, {Magnard}, {Marteaud},
  {Martinez}, {Maurel}, {M{\'e}nard}, {Mesa}, {M{\"o}ller-Nilsson}, {Moulin},
  {Moutou}, {Orign{\'e}}, {Parisot}, {Pavlov}, {Perret}, {Pragt}, {Puget},
  {Rabou}, {Ramos}, {Reess}, {Rigal}, {Rochat}, {Roelfsema}, {Rousset}, {Roux},
  {Saisse}, {Salasnich}, {Santambrogio}, {Scuderi}, {Segransan}, {Sevin},
  {Siebenmorgen}, {Soenke}, {Stadler}, {Suarez}, {Tiph{\`e}ne}, {Turatto},
  {Udry}, {Vakili}, {Waters}, {Weber}, {Wildi}, {Zins}, \&
  {Zurlo}}]{beuzit2019}
{Beuzit}, J.~L., {Vigan}, A., {Mouillet}, D., {et~al.} 2019, \aap, 631, A155

\bibitem[{{Bird} {et~al.}(2021){Bird}, {Loebman}, {Weinberg}, {Brooks},
  {Quinn}, \& {Christensen}}]{Bird2021}
{Bird}, J.~C., {Loebman}, S.~R., {Weinberg}, D.~H., {et~al.} 2021, \mnras, 503,
  1815

\bibitem[{{Birkmann} {et~al.}(2022){Birkmann}, {Ferruit}, {Giardino},
  {Nielsen}, {Garc{\'\i}a Mu{\~n}oz}, {Kendrew}, {Rauscher}, {Beck}, {Keyes},
  {Valenti}, {Jakobsen}, {Dorner}, {Alves de Oliveira}, {Arribas}, {B{\"o}ker},
  {Bunker}, {Charlot}, {de Marchi}, {Kumari}, {L{\'o}pez-Caniego},
  {L{\"u}tzgendorf}, {Maiolino}, {Manjavacas}, {Marston}, {Moseley}, {Prizkal},
  {Proffitt}, {Rawle}, {Rix}, {te Plate}, {Sabbi}, {Sirianni}, {Willott}, \&
  {Zeidler}}]{2022arXiv220203309B}
{Birkmann}, S.~M., {Ferruit}, P., {Giardino}, G., {et~al.} 2022, arXiv
  e-prints, arXiv:2202.03309

\bibitem[{{Blackman} {et~al.}(2021){Blackman}, {Beaulieu}, {Bennett},
  {Danielski}, {Alard}, {Cole}, {Vandorou}, {Ranc}, {Terry}, {Bhattacharya},
  {Bond}, {Bachelet}, {Veras}, {Koshimoto}, {Batista}, \&
  {Marquette}}]{MB10477_AO}
{Blackman}, J.~W., {Beaulieu}, J.~P., {Bennett}, D.~P., {et~al.} 2021, \nat,
  598, 272

\bibitem[{{Bland-Hawthorn} \& {Gerhard}(2016)}]{Bland-Hawthorn2016}
{Bland-Hawthorn}, J., \& {Gerhard}, O. 2016, \araa, 54, 529

\bibitem[{{Bloemen} {et~al.}(2014){Bloemen}, {Hu}, {Aerts}, {Dupret},
  {{\O}stensen}, {Degroote}, {M{\"u}ller-Ringat}, \&
  {Rauch}}]{2014A&A...569A.123B}
{Bloemen}, S., {Hu}, H., {Aerts}, C., {et~al.} 2014, \aap, 569, A123

\bibitem[{{Bloemen} {et~al.}(2011){Bloemen}, {Marsh}, {{\O}stensen},
  {Charpinet}, {Fontaine}, {Degroote}, {Heber}, {Kawaler}, {Aerts}, {Green},
  {Telting}, {Brassard}, {G{\"a}nsicke}, {Handler}, {Kurtz}, {Silvotti}, {Van
  Grootel}, {Lindberg}, {Pursimo}, {Wilson}, {Gilliland}, {Kjeldsen},
  {Christensen-Dalsgaard}, {Borucki}, {Koch}, {Jenkins}, \&
  {Klaus}}]{Bloemen2011}
{Bloemen}, S., {Marsh}, T.~R., {{\O}stensen}, R.~H., {et~al.} 2011, \mnras,
  410, 1787

\bibitem[{{Bloom} {et~al.}(2009){Bloom}, {Perley}, {Li}, {Butler}, {Miller},
  {Kocevski}, {Kann}, {Foley}, {Chen}, {Filippenko}, {Starr}, {Macomber},
  {Prochaska}, {Chornock}, {Poznanski}, {Klose}, {Skrutskie}, {Lopez}, {Hall},
  {Glazebrook}, \& {Blake}}]{2009ApJ...691..723B}
{Bloom}, J.~S., {Perley}, D.~A., {Li}, W., {et~al.} 2009, \apj, 691, 723

\bibitem[{{Bodensteiner} {et~al.}(2020){Bodensteiner}, {Shenar}, {Mahy},
  {Fabry}, {Marchant}, {Abdul-Masih}, {Banyard}, {Bowman}, {Dsilva}, {Frost},
  {Hawcroft}, {Reggiani}, \& {Sana}}]{Bodensteiner2020}
{Bodensteiner}, J., {Shenar}, T., {Mahy}, L., {et~al.} 2020, \aap, 641, A43

\bibitem[{{Boeche} {et~al.}(2013){Boeche}, {Chiappini}, {Minchev}, {Williams},
  {Steinmetz}, {Sharma}, {Kordopatis}, {Bland-Hawthorn}, {Bienaym{\'e}},
  {Gibson}, {Gilmore}, {Grebel}, {Helmi}, {Munari}, {Navarro}, {Parker},
  {Reid}, {Seabroke}, {Siebert}, {Siviero}, {Watson}, {Wyse}, \&
  {Zwitter}}]{Boeche2013}
{Boeche}, C., {Chiappini}, C., {Minchev}, I., {et~al.} 2013, \aap, 553, A19

\bibitem[{{Bogn{\'a}r} {et~al.}(2020){Bogn{\'a}r}, {Kawaler}, {Bell},
  {Schrandt}, {Baran}, {Bradley}, {Hermes}, {Charpinet}, {Handler}, {Mullally},
  {Murphy}, {Raddi}, {S{\'o}dor}, {Tremblay}, {Uzundag}, \&
  {Zong}}]{2020A&A...638A..82B}
{Bogn{\'a}r}, Z., {Kawaler}, S.~D., {Bell}, K.~J., {et~al.} 2020, \aap, 638,
  A82

\bibitem[{{Bonaca} {et~al.}(2017){Bonaca}, {Conroy}, {Wetzel}, {Hopkins}, \&
  {Kere{\v{s}}}}]{Bonaca2017}
{Bonaca}, A., {Conroy}, C., {Wetzel}, A., {Hopkins}, P.~F., \& {Kere{\v{s}}},
  D. 2017, \apj, 845, 101

\bibitem[{{Bonaca} {et~al.}(2020){Bonaca}, {Conroy}, {Cargile}, {Naidu},
  {Johnson}, {Zaritsky}, {Ting}, {Caldwell}, {Han}, \& {van
  Dokkum}}]{Bonaca2020}
{Bonaca}, A., {Conroy}, C., {Cargile}, P.~A., {et~al.} 2020, \apjl, 897, L18

\bibitem[{{Bonfils} {et~al.}(2013){Bonfils}, {Delfosse}, {Udry}, {Forveille},
  {Mayor}, {Perrier}, {Bouchy}, {Gillon}, {Lovis}, {Pepe}, {Queloz}, {Santos},
  {S{\'e}gransan}, \& {Bertaux}}]{Bonfils2013}
{Bonfils}, X., {Delfosse}, X., {Udry}, S., {et~al.} 2013, \aap, 549, A109

\bibitem[{{Borucki} {et~al.}(2010{\natexlab{a}}){Borucki}, {Koch}, \& {Kepler
  Science Team}}]{Borucki2009}
{Borucki}, W.~J., {Koch}, D., \& {Kepler Science Team}. 2010{\natexlab{a}}, in
  AAS/Division for Planetary Sciences Meeting Abstracts, Vol.~42, AAS/Division
  for Planetary Sciences Meeting Abstracts \#42, 47.03

\bibitem[{{Borucki} {et~al.}(2010{\natexlab{b}}){Borucki}, {Koch}, {Basri},
  {Batalha}, {Brown}, {Caldwell}, {Caldwell}, {Christensen-Dalsgaard},
  {Cochran}, {DeVore}, {Dunham}, {Dupree}, {Gautier}, {Geary}, {Gilliland},
  {Gould}, {Howell}, {Jenkins}, {Kondo}, {Latham}, {Marcy}, {Meibom},
  {Kjeldsen}, {Lissauer}, {Monet}, {Morrison}, {Sasselov}, {Tarter}, {Boss},
  {Brownlee}, {Owen}, {Buzasi}, {Charbonneau}, {Doyle}, {Fortney}, {Ford},
  {Holman}, {Seager}, {Steffen}, {Welsh}, {Rowe}, {Anderson}, {Buchhave},
  {Ciardi}, {Walkowicz}, {Sherry}, {Horch}, {Isaacson}, {Everett}, {Fischer},
  {Torres}, {Johnson}, {Endl}, {MacQueen}, {Bryson}, {Dotson}, {Haas},
  {Kolodziejczak}, {Van Cleve}, {Chandrasekaran}, {Twicken}, {Quintana},
  {Clarke}, {Allen}, {Li}, {Wu}, {Tenenbaum}, {Verner}, {Bruhweiler}, {Barnes},
  \& {Prsa}}]{Borucki2010}
{Borucki}, W.~J., {Koch}, D., {Basri}, G., {et~al.} 2010{\natexlab{b}},
  Science, 327, 977

\bibitem[{{Borucki} {et~al.}(2010{\natexlab{c}}){Borucki}, {Koch}, {Basri},
  {Batalha}, {Brown}, {Caldwell}, {Caldwell}, {Christensen-Dalsgaard},
  {Cochran}, {DeVore}, {Dunham}, {Dupree}, {Gautier}, {Geary}, {Gilliland},
  {Gould}, {Howell}, {Jenkins}, {Kondo}, {Latham}, {Marcy}, {Meibom},
  {Kjeldsen}, {Lissauer}, {Monet}, {Morrison}, {Sasselov}, {Tarter}, {Boss},
  {Brownlee}, {Owen}, {Buzasi}, {Charbonneau}, {Doyle}, {Fortney}, {Ford},
  {Holman}, {Seager}, {Steffen}, {Welsh}, {Rowe}, {Anderson}, {Buchhave},
  {Ciardi}, {Walkowicz}, {Sherry}, {Horch}, {Isaacson}, {Everett}, {Fischer},
  {Torres}, {Johnson}, {Endl}, {MacQueen}, {Bryson}, {Dotson}, {Haas},
  {Kolodziejczak}, {Van Cleve}, {Chandrasekaran}, {Twicken}, {Quintana},
  {Clarke}, {Allen}, {Li}, {Wu}, {Tenenbaum}, {Verner}, {Bruhweiler}, {Barnes},
  \& {Prsa}}]{borucki2010a}
---. 2010{\natexlab{c}}, Science, 327, 977

\bibitem[{{Bouma} {et~al.}(2018){Bouma}, {Masuda}, \& {Winn}}]{Bouma:2018}
{Bouma}, L.~G., {Masuda}, K., \& {Winn}, J.~N. 2018, \aj, 155, 244

\bibitem[{{Bowman}(2020)}]{Bowman2020c}
{Bowman}, D.~M. 2020, Frontiers in Astronomy and Space Sciences, 7, 70

\bibitem[{{Bowman} {et~al.}(2021){Bowman}, {Hermans},
  {Daszy{\'n}ska-Daszkiewicz}, {Holdsworth}, {Tkachenko}, {Murphy}, {Smalley},
  \& {Kurtz}}]{Bowman2021a}
{Bowman}, D.~M., {Hermans}, J., {Daszy{\'n}ska-Daszkiewicz}, J., {et~al.} 2021,
  \mnras, 504, 4039

\bibitem[{{Bowman} {et~al.}(2019{\natexlab{a}}){Bowman}, {Johnston},
  {Tkachenko}, {Mkrtichian}, {Gunsriwiwat}, \& {Aerts}}]{Bowman2019d}
{Bowman}, D.~M., {Johnston}, C., {Tkachenko}, A., {et~al.} 2019{\natexlab{a}},
  \apjl, 883, L26

\bibitem[{{Bowman} {et~al.}(2016){Bowman}, {Kurtz}, {Breger}, {Murphy}, \&
  {Holdsworth}}]{2016MNRAS.460.1970B}
{Bowman}, D.~M., {Kurtz}, D.~W., {Breger}, M., {Murphy}, S.~J., \&
  {Holdsworth}, D.~L. 2016, \mnras, 460, 1970

\bibitem[{{Bowman} {et~al.}(2019{\natexlab{b}}){Bowman}, {Burssens},
  {Pedersen}, {Johnston}, {Aerts}, {Buysschaert}, {Michielsen}, {Tkachenko},
  {Rogers}, {Edelmann}, {Ratnasingam}, {Sim{\'o}n-D{\'\i}az}, {Castro},
  {Moravveji}, {Pope}, {White}, \& {De Cat}}]{Bowman2019b}
{Bowman}, D.~M., {Burssens}, S., {Pedersen}, M.~G., {et~al.}
  2019{\natexlab{b}}, Nature Astronomy, 3, 760

\bibitem[{{Brandt} {et~al.}(2019){Brandt}, {Dupuy}, \& {Bowler}}]{brandt19}
{Brandt}, T.~D., {Dupuy}, T.~J., \& {Bowler}, B.~P. 2019, \aj, 158, 140

\bibitem[{Brassard {et~al.}(1995)Brassard, Fontaine, \&
  Wesemael}]{Brassard1995}
Brassard, P., Fontaine, G., \& Wesemael, F. 1995, The Astrophysical Journal
  Supplement Series, 96, 545.
\newblock \url{https://ui.adsabs.harvard.edu/abs/1995ApJS...96..545B}

\bibitem[{Brickhill(1992)}]{Brickhill1992}
Brickhill, A.~J. 1992, Monthly Notices of the Royal Astronomical Society, 259,
  519

\bibitem[{{Bromm} \& {Larson}(2004)}]{Bromm2004}
{Bromm}, V., \& {Larson}, R.~B. 2004, \araa, 42, 79

\bibitem[{{Brown} {et~al.}(2013){Brown}, {Baliber}, {Bianco}, {Bowman},
  {Burleson}, {Conway}, {Crellin}, {Depagne}, {De Vera}, {Dilday}, {Dragomir},
  {Dubberley}, {Eastman}, {Elphick}, {Falarski}, {Foale}, {Ford}, {Fulton},
  {Garza}, {Gomez}, {Graham}, {Greene}, {Haldeman}, {Hawkins}, {Haworth},
  {Haynes}, {Hidas}, {Hjelstrom}, {Howell}, {Hygelund}, {Lister}, {Lobdill},
  {Martinez}, {Mullins}, {Norbury}, {Parrent}, {Paulson}, {Petry}, {Pickles},
  {Posner}, {Rosing}, {Ross}, {Sand}, {Saunders}, {Shobbrook}, {Shporer},
  {Street}, {Thomas}, {Tsapras}, {Tufts}, {Valenti}, {Vander Horst}, {Walker},
  {White}, \& {Willis}}]{Brown2013}
{Brown}, T.~M., {Baliber}, N., {Bianco}, F.~B., {et~al.} 2013, \pasp, 125, 1031

\bibitem[{Bryan {et~al.}(2019)Bryan, Knutson, Lee, Fulton, Batygin, Ngo, \&
  Meshkat}]{Bryan2019}
Bryan, M.~L., Knutson, H.~A., Lee, E.~J., {et~al.} 2019, The Astronomical
  Journal, 157, 52

\bibitem[{{Buchhave} {et~al.}(2018){Buchhave}, {Bitsch}, {Johansen}, {Latham},
  {Bizzarro}, {Bieryla}, \& {Kipping}}]{Buchhave:2018}
{Buchhave}, L.~A., {Bitsch}, B., {Johansen}, A., {et~al.} 2018, \apj, 856, 37

\bibitem[{{Buchhave} {et~al.}(2014){Buchhave}, {Bizzarro}, {Latham},
  {Sasselov}, {Cochran}, {Endl}, {Isaacson}, {Juncher}, \&
  {Marcy}}]{Buchhave2014}
{Buchhave}, L.~A., {Bizzarro}, M., {Latham}, D.~W., {et~al.} 2014, \nat, 509,
  593

\bibitem[{{Buchler} \& {Goupil}(1984)}]{1984ApJ...279..394B}
{Buchler}, J.~R., \& {Goupil}, M.~J. 1984, \apj, 279, 394

\bibitem[{{Buchler} \& {Koll{\'a}th}(2011)}]{BuchlerKollath2011}
{Buchler}, J.~R., \& {Koll{\'a}th}, Z. 2011, \apj, 731, 24

\bibitem[{{Buckley} {et~al.}(2021){Buckley}, {McBride}, {Barres De Almeida},
  {Shustov}, {Pozanenko}, {Lutovinov}, {Omar}, {Murthy}, {Safonova}, {Liu}, \&
  {Soria}}]{2021AnABC..93..917B}
{Buckley}, D., {McBride}, V., {Barres De Almeida}, U., {et~al.} 2021, An. Acad.
  Bras. Ci{\^e}nc, 93, e20200917

\bibitem[{{Buder} {et~al.}(2021){Buder}, {Sharma}, {Kos}, {Amarsi},
  {Nordlander}, {Lind}, {Martell}, {Asplund}, {Bland-Hawthorn}, {Casey}, {de
  Silva}, {D'Orazi}, {Freeman}, {Hayden}, {Lewis}, {Lin}, {Schlesinger},
  {Simpson}, {Stello}, {Zucker}, {Zwitter}, {Beeson}, {Buck}, {Casagrande},
  {Clark}, {{\v{C}}otar}, {da Costa}, {de Grijs}, {Feuillet}, {Horner},
  {Kafle}, {Khanna}, {Kobayashi}, {Liu}, {Montet}, {Nandakumar}, {Nataf},
  {Ness}, {Spina}, {Tepper-Garc{\'\i}a}, {Ting}, {Traven},
  {Vogrin{\v{c}}i{\v{c}}}, {Wittenmyer}, {Wyse}, {{\v{Z}}erjal}, \& {GALAH
  Collaboration}}]{Buder2021}
{Buder}, S., {Sharma}, S., {Kos}, J., {et~al.} 2021, \mnras, 506, 150

\bibitem[{{Bugnet} {et~al.}(2021){Bugnet}, {Prat}, {Mathis}, {Astoul},
  {Augustson}, {Garc{\'\i}a}, {Mathur}, {Amard}, \& {Neiner}}]{bugnet2021a}
{Bugnet}, L., {Prat}, V., {Mathis}, S., {et~al.} 2021, \aap, 650, A53

\bibitem[{{Burdge} {et~al.}(2019){Burdge}, {Coughlin}, {Fuller}, {Kupfer},
  {Bellm}, {Bildsten}, {Graham}, {Kaplan}, {Roestel}, {Dekany}, {Duev},
  {Feeney}, {Giomi}, {Helou}, {Kaye}, {Laher}, {Mahabal}, {Masci}, {Riddle},
  {Shupe}, {Soumagnac}, {Smith}, {Szkody}, {Walters}, {Kulkarni}, \&
  {Prince}}]{Burdge2019}
{Burdge}, K.~B., {Coughlin}, M.~W., {Fuller}, J., {et~al.} 2019, \nat, 571, 528

\bibitem[{{Burdge} {et~al.}(2020){Burdge}, {Prince}, {Fuller}, {Kaplan},
  {Marsh}, {Tremblay}, {Zhuang}, {Bellm}, {Caiazzo}, {Coughlin}, {Dhillon},
  {Gaensicke}, {Rodr{\'\i}guez-Gil}, {Graham}, {Hermes}, {Kupfer},
  {Littlefair}, {Mr{\'o}z}, {Phinney}, {van Roestel}, {Yao}, {Dekany}, {Drake},
  {Duev}, {Hale}, {Feeney}, {Helou}, {Kaye}, {Mahabal}, {Masci}, {Riddle},
  {Smith}, {Soumagnac}, \& {Kulkarni}}]{Burdge2020}
{Burdge}, K.~B., {Prince}, T.~A., {Fuller}, J., {et~al.} 2020, \apj, 905, 32

\bibitem[{{Burssens} {et~al.}(2019){Burssens}, {Bowman}, {Aerts}, {Pedersen},
  {Moravveji}, \& {Buysschaert}}]{Burssens2019a}
{Burssens}, S., {Bowman}, D.~M., {Aerts}, C., {et~al.} 2019, \mnras, 489, 1304

\bibitem[{{Calchi Novati} {et~al.}(2015){Calchi Novati}, {Gould}, {Udalski},
  {Menzies}, {Bond}, {Shvartzvald}, {Street}, {Hundertmark}, {Beichman}, {Yee},
  {Carey}, {Poleski}, {Skowron}, {Koz{\l}owski}, {Mr{\'o}z}, {Pietrukowicz},
  {Pietrzy{\'n}ski}, {Szyma{\'n}ski}, {Soszy{\'n}ski}, {Ulaczyk},
  {Wyrzykowski}, {OGLE Collaboration}, {Albrow}, {Beaulieu}, {Caldwell},
  {Cassan}, {Coutures}, {Danielski}, {Dominis Prester}, {Donatowicz}, {Lon{\v
  c}ari{\'c}}, {McDougall}, {Morales}, {Ranc}, {Zhu}, {PLANET Collaboration},
  {Abe}, {Barry}, {Bennett}, {Bhattacharya}, {Fukunaga}, {Inayama},
  {Koshimoto}, {Namba}, {Sumi}, {Suzuki}, {Tristram}, {Wakiyama}, {Yonehara},
  {MOA Collaboration}, {Maoz}, {Kaspi}, {Friedmann}, {Wise Group}, {Bachelet},
  {Figuera Jaimes}, {Bramich}, {Tsapras}, {Horne}, {Snodgrass}, {Wambsganss},
  {Steele}, {Kains}, {RoboNet Collaboration}, {Bozza}, {Dominik},
  {J{\o}rgensen}, {Alsubai}, {Ciceri}, {D'Ago}, {Haugb{\o}lle}, {Hessman},
  {Hinse}, {Juncher}, {Korhonen}, {Mancini}, {Popovas}, {Rabus}, {Rahvar},
  {Scarpetta}, {Schmidt}, {Skottfelt}, {Southworth}, {Starkey}, {Surdej},
  {Wertz}, {Zarucki}, {MiNDSTEp Consortium}, {Gaudi}, {Pogge}, {DePoy}, \&
  {{$\mu$}FUN Collaboration}}]{Novati2015}
{Calchi Novati}, S., {Gould}, A., {Udalski}, A., {et~al.} 2015, \apj, 804, 20

\bibitem[{{Calchi Novati} {et~al.}(2018){Calchi Novati}, {Skowron}, {Jung},
  {Beichman}, {Bryden}, {Carey}, {Gaudi}, {Henderson}, {Shvartzvald}, {Yee},
  {Zhu}, {Spitzer Team}, {Udalski}, {Szyma{\'n}ski}, {Mr{\'o}z}, {Poleski},
  {Soszy{\'n}ski}, {Koz{\l}owski}, {Pietrukowicz}, {Ulaczyk}, {Pawlak},
  {Rybicki}, {Iwanek}, {OGLE Collaboration}, {Albrow}, {Chung}, {Gould}, {Han},
  {Hwang}, {Ryu}, {Shin}, {Zang}, {Cha}, {Kim}, {Kim}, {Kim}, {Lee}, {Lee},
  {Lee}, {Park}, {Pogge}, \& {KMTNet Collaboration}}]{OB171140}
{Calchi Novati}, S., {Skowron}, J., {Jung}, Y.~K., {et~al.} 2018, \aj, 155, 261

\bibitem[{{Campante} {et~al.}(2015){Campante}, {Barclay}, {Swift}, {Huber},
  {Adibekyan}, {Cochran}, {Burke}, {Isaacson}, {Quintana}, {Davies}, {Silva
  Aguirre}, {Ragozzine}, {Riddle}, {Baranec}, {Basu}, {Chaplin},
  {Christensen-Dalsgaard}, {Metcalfe}, {Bedding}, {Handberg}, {Stello},
  {Brewer}, {Hekker}, {Karoff}, {Kolbl}, {Law}, {Lundkvist}, {Miglio}, {Rowe},
  {Santos}, {Van Laerhoven}, {Arentoft}, {Elsworth}, {Fischer}, {Kawaler},
  {Kjeldsen}, {Lund}, {Marcy}, {Sousa}, {Sozzetti}, \& {White}}]{campante2015a}
{Campante}, T.~L., {Barclay}, T., {Swift}, J.~J., {et~al.} 2015, \apj, 799, 170

\bibitem[{{Campante} {et~al.}(2016{\natexlab{a}}){Campante}, {Schofield},
  {Kuszlewicz}, {Bouma}, {Chaplin}, {Huber}, {Christensen-Dalsgaard},
  {Kjeldsen}, {Bossini}, {North}, {Appourchaux}, {Latham}, {Pepper}, {Ricker},
  {Stassun}, {Vanderspek}, \& {Winn}}]{Campante2016a}
{Campante}, T.~L., {Schofield}, M., {Kuszlewicz}, J.~S., {et~al.}
  2016{\natexlab{a}}, \apj, 830, 138

\bibitem[{{Campante} {et~al.}(2016{\natexlab{b}}){Campante}, {Lund},
  {Kuszlewicz}, {Davies}, {Chaplin}, {Albrecht}, {Winn}, {Bedding}, {Benomar},
  {Bossini}, {Handberg}, {Santos}, {Van Eylen}, {Basu},
  {Christensen-Dalsgaard}, {Elsworth}, {Hekker}, {Hirano}, {Huber}, {Karoff},
  {Kjeldsen}, {Lundkvist}, {North}, {Silva Aguirre}, {Stello}, \&
  {White}}]{Campante2016}
{Campante}, T.~L., {Lund}, M.~N., {Kuszlewicz}, J.~S., {et~al.}
  2016{\natexlab{b}}, \apj, 819, 85

\bibitem[{Canup \& Asphaug(2001)}]{Canup2001}
Canup, R.~M., \& Asphaug, E. 2001, Nature, 412, 708

\bibitem[{{Cao} {et~al.}(2015){Cao}, {Kulkarni}, {Howell}, {Gal-Yam},
  {Kasliwal}, {Valenti}, {Johansson}, {Amanullah}, {Goobar}, {Sollerman},
  {Taddia}, {Horesh}, {Sagiv}, {Cenko}, {Nugent}, {Arcavi}, {Surace},
  {Wo{\'z}niak}, {Moody}, {Rebbapragada}, {Bue}, \&
  {Gehrels}}]{2015Natur.521..328C}
{Cao}, Y., {Kulkarni}, S.~R., {Howell}, D.~A., {et~al.} 2015, \nat, 521, 328

\bibitem[{{Carollo} {et~al.}(2016){Carollo}, {Beers}, {Placco}, {Santucci},
  {Denissenkov}, {Tissera}, {Lentner}, {Rossi}, {Lee}, \&
  {Tumlinson}}]{Carollo2016}
{Carollo}, D., {Beers}, T.~C., {Placco}, V.~M., {et~al.} 2016, Nature Physics,
  12, 1170

\bibitem[{{Casagrande} {et~al.}(2011){Casagrande}, {Sch{\"o}nrich}, {Asplund},
  {Cassisi}, {Ram{\'\i}rez}, {Mel{\'e}ndez}, {Bensby}, \&
  {Feltzing}}]{Casagrande2011}
{Casagrande}, L., {Sch{\"o}nrich}, R., {Asplund}, M., {et~al.} 2011, \aap, 530,
  A138

\bibitem[{{Casey} {et~al.}(2017){Casey}, {Hawkins}, {Hogg}, {Ness}, {Rix},
  {Kordopatis}, {Kunder}, {Steinmetz}, {Koposov}, {Enke}, {Sanders}, {Gilmore},
  {Zwitter}, {Freeman}, {Casagrande}, {Matijevi{\v{c}}}, {Seabroke},
  {Bienaym{\'e}}, {Bland-Hawthorn}, {Gibson}, {Grebel}, {Helmi}, {Munari},
  {Navarro}, {Reid}, {Siebert}, \& {Wyse}}]{Casey2017}
{Casey}, A.~R., {Hawkins}, K., {Hogg}, D.~W., {et~al.} 2017, \apj, 840, 59

\bibitem[{{Cassan} {et~al.}(2012){Cassan}, {Kubas}, {Beaulieu}, {Dominik},
  {Horne}, {Greenhill}, {Wambsganss}, {Menzies}, {Williams}, {J{\o}rgensen},
  {Udalski}, {Bennett}, {Albrow}, {Batista}, {Brillant}, {Caldwell}, {Cole},
  {Coutures}, {Cook}, {Dieters}, {Dominis Prester}, {Donatowicz}, {Fouqu{\'e}},
  {Hill}, {Kains}, {Kane}, {Marquette}, {Martin}, {Pollard}, {Sahu}, {Vinter},
  {Warren}, {Watson}, {Zub}, {Sumi}, {Szyma{\'n}ski}, {Kubiak}, {Poleski},
  {Soszynski}, {Ulaczyk}, {Pietrzy{\'n}ski}, \& {Wyrzykowski}}]{Cassan2012}
{Cassan}, A., {Kubas}, D., {Beaulieu}, J.~P., {et~al.} 2012, \nat, 481, 167

\bibitem[{{Cassan} {et~al.}(2022){Cassan}, {Ranc}, {Absil}, {Wyrzykowski},
  {Rybicki}, {Bachelet}, {Le Bouquin}, {Hundertmark}, {Street}, {Surdej},
  {Tsapras}, {Wambsganss}, \& {Wertz}}]{Cassan2022}
{Cassan}, A., {Ranc}, C., {Absil}, O., {et~al.} 2022, Nature Astronomy, 6, 121

\bibitem[{{Cenarro} {et~al.}(2019){Cenarro}, {Moles},
  {Crist{\'o}bal-Hornillos}, {Mar{\'\i}n-Franch}, {Ederoclite}, {Varela},
  {L{\'o}pez-Sanjuan}, {Hern{\'a}ndez-Monteagudo}, {Angulo}, {V{\'a}zquez
  Rami{\'o}}, {Viironen}, {Bonoli}, {Orsi}, {Hurier}, {San Roman}, {Greisel},
  {Vilella-Rojo}, {D{\'\i}az-Garc{\'\i}a}, {Logro{\~n}o-Garc{\'\i}a},
  {Gurung-L{\'o}pez}, {Spinoso}, {Izquierdo-Villalba}, {Aguerri}, {Allende
  Prieto}, {Bonatto}, {Carvano}, {Chies-Santos}, {Daflon}, {Dupke},
  {Falc{\'o}n-Barroso}, {Gon{\c{c}}alves}, {Jim{\'e}nez-Teja}, {Molino},
  {Placco}, {Solano}, {Whitten}, {Abril}, {Ant{\'o}n}, {Bello}, {Bielsa de
  Toledo}, {Castillo-Ram{\'\i}rez}, {Chueca}, {Civera},
  {D{\'\i}az-Mart{\'\i}n}, {Dom{\'\i}nguez-Mart{\'\i}nez},
  {Garzar{\'a}n-Calderaro}, {Hern{\'a}ndez-Fuertes}, {Iglesias-Marzoa},
  {I{\~n}iguez}, {Jim{\'e}nez Ruiz}, {Kruuse}, {Lamadrid}, {Lasso-Cabrera},
  {L{\'o}pez-Alegre}, {L{\'o}pez-Sainz}, {Ma{\'\i}cas}, {Moreno-Signes},
  {Muniesa}, {Rodr{\'\i}guez-Llano}, {Rueda-Teruel}, {Rueda-Teruel},
  {Soriano-Lagu{\'\i}a}, {Tilve}, {Valdivielso}, {Yanes-D{\'\i}az}, {Alcaniz},
  {Mendes de Oliveira}, {Sodr{\'e}}, {Coelho}, {Lopes de Oliveira}, {Tamm},
  {Xavier}, {Abramo}, {Akras}, {Alfaro}, {Alvarez-Candal}, {Ascaso}, {Beasley},
  {Beers}, {Borges Fernandes}, {Bruzual}, {Buzzo}, {Carrasco}, {Cepa},
  {Cortesi}, {Costa-Duarte}, {De Pr{\'a}}, {Favole}, {Galarza}, {Galbany},
  {Garcia}, {Gonz{\'a}lez Delgado}, {Gonz{\'a}lez-Serrano},
  {Guti{\'e}rrez-Soto}, {Hernandez-Jimenez}, {Kanaan}, {Kuncarayakti},
  {Landim}, {Laur}, {Licandro}, {Lima Neto}, {Lyman}, {Ma{\'\i}z
  Apell{\'a}niz}, {Miralda-Escud{\'e}}, {Morate}, {Nogueira-Cavalcante},
  {Novais}, {Oncins}, {Oteo}, {Overzier}, {Pereira}, {Rebassa-Mansergas},
  {Reis}, {Roig}, {Sako}, {Salvador-Rusi{\~n}ol}, {Sampedro},
  {S{\'a}nchez-Bl{\'a}zquez}, {Santos}, {Schmidtobreick}, {Siffert}, {Telles},
  \& {Vilchez}}]{Cenarro2019}
{Cenarro}, A.~J., {Moles}, M., {Crist{\'o}bal-Hornillos}, D., {et~al.} 2019,
  \aap, 622, A176

\bibitem[{Chandler {et~al.}(2018)Chandler, Curtis, Mommert, Sheppard, \&
  Trujillo}]{chandler2018safari}
Chandler, C.~O., Curtis, A.~M., Mommert, M., Sheppard, S.~S., \& Trujillo,
  C.~A. 2018, Publications of the Astronomical Society of the Pacific, 130,
  114502

\bibitem[{Chang {et~al.}(2017)Chang, Lin, Ip, Prince, Kulkarni, Levitan, Laher,
  \& Surace}]{chang2017asteroid}
Chang, C.-K., Lin, H.-W., Ip, W.-H., {et~al.} 2017, Geoscience Letters, 4, 1

\bibitem[{{Chaplin} {et~al.}(2019){Chaplin}, {Cegla}, {Watson}, {Davies}, \&
  {Ball}}]{chaplin19}
{Chaplin}, W.~J., {Cegla}, H.~M., {Watson}, C.~A., {Davies}, G.~R., \& {Ball},
  W.~H. 2019, \aj, 157, 163

\bibitem[{{Chaplin} \& {Miglio}(2013)}]{chaplin2013b}
{Chaplin}, W.~J., \& {Miglio}, A. 2013, \araa, 51, 353

\bibitem[{{Chaplin} {et~al.}(2011){Chaplin}, {Kjeldsen}, {Bedding},
  {Christensen-Dalsgaard}, {Gilliland}, {Kawaler}, {Appourchaux}, {Elsworth},
  {Garc{\'\i}a}, {Houdek}, {Karoff}, {Metcalfe}, {Molenda-{\.Z}akowicz},
  {Monteiro}, {Thompson}, {Verner}, {Batalha}, {Borucki}, {Brown}, {Bryson},
  {Christiansen}, {Clarke}, {Jenkins}, {Klaus}, {Koch}, {An}, {Ballot}, {Basu},
  {Benomar}, {Bonanno}, {Broomhall}, {Campante}, {Corsaro}, {Creevey}, {Esch},
  {Gai}, {Gaulme}, {Hale}, {Handberg}, {Hekker}, {Huber}, {Mathur}, {Mosser},
  {New}, {Pinsonneault}, {Pricopi}, {Quirion}, {R{\'e}gulo}, {Roxburgh},
  {Salabert}, {Stello}, \& {Suran}}]{chaplin2011a}
{Chaplin}, W.~J., {Kjeldsen}, H., {Bedding}, T.~R., {et~al.} 2011, \apj, 732,
  54

\bibitem[{{Chaplin} {et~al.}(2013){Chaplin}, {Sanchis-Ojeda}, {Campante},
  {Handberg}, {Stello}, {Winn}, {Basu}, {Christensen-Dalsgaard}, {Davies},
  {Metcalfe}, {Buchhave}, {Fischer}, {Bedding}, {Cochran}, {Elsworth},
  {Gilliland}, {Hekker}, {Huber}, {Isaacson}, {Karoff}, {Kawaler}, {Kjeldsen},
  {Latham}, {Lund}, {Lundkvist}, {Marcy}, {Miglio}, {Barclay}, \&
  {Lissauer}}]{Chaplin2013}
{Chaplin}, W.~J., {Sanchis-Ojeda}, R., {Campante}, T.~L., {et~al.} 2013, \apj,
  766, 101

\bibitem[{{Chaplin} {et~al.}(2015){Chaplin}, {Lund}, {Handberg}, {Basu},
  {Buchhave}, {Campante}, {Davies}, {Huber}, {Latham}, {Latham}, {Serenelli},
  {Antia}, {Appourchaux}, {Ball}, {Benomar}, {Casagrande},
  {Christensen-Dalsgaard}, {Coelho}, {Creevey}, {Elsworth}, {Garc{\'\i}a},
  {Gaulme}, {Hekker}, {Kallinger}, {Karoff}, {Kawaler}, {Kjeldsen},
  {Lundkvist}, {Marcadon}, {Mathur}, {Miglio}, {Mosser}, {R{\'e}gulo},
  {Roxburgh}, {Silva Aguirre}, {Stello}, {Verma}, {White}, {Bedding},
  {Barclay}, {Buzasi}, {Dehuevels}, {Gizon}, {Houdek}, {Howell}, {Salabert}, \&
  {Soderblom}}]{Chaplin2015}
{Chaplin}, W.~J., {Lund}, M.~N., {Handberg}, R., {et~al.} 2015, \pasp, 127,
  1038

\bibitem[{{Chaplin} {et~al.}(2020){Chaplin}, {Serenelli}, {Miglio}, {Morel},
  {Mackereth}, {Vincenzo}, {Kjeldsen}, {Basu}, {Ball}, {Stokholm}, {Verma},
  {Mosumgaard}, {Silva Aguirre}, {Mazumdar}, {Ranadive}, {Antia}, {Lebreton},
  {Ong}, {Appourchaux}, {Bedding}, {Christensen-Dalsgaard}, {Creevey},
  {Garc{\'\i}a}, {Handberg}, {Huber}, {Kawaler}, {Lund}, {Metcalfe}, {Stassun},
  {Bazot}, {Beck}, {Bell}, {Bergemann}, {Buzasi}, {Benomar}, {Bossini},
  {Bugnet}, {Campante}, {Orhan}, {Corsaro}, {Gonz{\'a}lez-Cuesta}, {Davies},
  {Di Mauro}, {Egeland}, {Elsworth}, {Gaulme}, {Ghasemi}, {Guo}, {Hall},
  {Hasanzadeh}, {Hekker}, {Howe}, {Jenkins}, {Jim{\'e}nez}, {Kiefer},
  {Kuszlewicz}, {Kallinger}, {Latham}, {Lundkvist}, {Mathur}, {Montalb{\'a}n},
  {Mosser}, {Bed{\'o}n}, {Nielsen}, {{\"O}rtel}, {Rendle}, {Ricker},
  {Rodrigues}, {Roxburgh}, {Safari}, {Schofield}, {Seager}, {Smalley},
  {Stello}, {Szab{\'o}}, {Tayar}, {Theme{\ss}l}, {Thomas}, {Vanderspek}, {van
  Rossem}, {Vrard}, {Weiss}, {White}, {Winn}, \& {Y{\i}ld{\i}z}}]{chaplin2020a}
{Chaplin}, W.~J., {Serenelli}, A.~M., {Miglio}, A., {et~al.} 2020, Nature
  Astronomy, 4, 382

\bibitem[{{Charpinet} {et~al.}(2011){Charpinet}, {Fontaine}, {Brassard},
  {Green}, {Van Grootel}, {Randall}, {Silvotti}, {Baran}, {{\O}stensen},
  {Kawaler}, \& {Telting}}]{2011Natur.480..496C}
{Charpinet}, S., {Fontaine}, G., {Brassard}, P., {et~al.} 2011, \nat, 480, 496

\bibitem[{{Chatterjee} {et~al.}(2008){Chatterjee}, {Ford}, {Matsumura}, \&
  {Rasio}}]{Chatterjee:2008}
{Chatterjee}, S., {Ford}, E.~B., {Matsumura}, S., \& {Rasio}, F.~A. 2008, \apj,
  686, 580

\bibitem[{{Chen} {et~al.}(2020{\natexlab{a}}){Chen}, {Lubow}, \&
  {Martin}}]{chen20}
{Chen}, C., {Lubow}, S.~H., \& {Martin}, R.~G. 2020{\natexlab{a}}, \mnras, 494,
  4645

\bibitem[{{Chen} {et~al.}(2019{\natexlab{a}}){Chen}, {Zhou}, {Xie}, {Yang},
  {Zhang}, {Liu}, {Liang}, {Yu}, \& {Yang}}]{2019NatAs...3...69C}
{Chen}, D.-C., {Zhou}, J.-L., {Xie}, J.-W., {et~al.} 2019{\natexlab{a}}, Nature
  Astronomy, 3, 69

\bibitem[{{Chen} {et~al.}(2022){Chen}, {Xie}, {Zhou}, {Yang}, {Dong}, {Zhu},
  {Zheng}, {Liu}, {Zong}, \& {Luo}}]{Chen2022}
{Chen}, D.-C., {Xie}, J.-W., {Zhou}, J.-L., {et~al.} 2022, \aj, 163, 249

\bibitem[{{Chen} {et~al.}(2019{\natexlab{b}}){Chen}, {Zhao}, {Zhao}, {Liang},
  {Wu}, {Jia}, {Tian}, \& {Liu}}]{ChenY2019}
{Chen}, Y.~Q., {Zhao}, G., {Zhao}, J.~K., {et~al.} 2019{\natexlab{b}}, \aj,
  158, 249

\bibitem[{{Chen} {et~al.}(2020{\natexlab{b}}){Chen}, {Li}, {Li}, \&
  {Lin}}]{Chen2020a}
{Chen}, Y.-X., {Li}, Y.-P., {Li}, H., \& {Lin}, D. N.~C. 2020{\natexlab{b}},
  \apj, 896, 135

\bibitem[{{Chen} {et~al.}(2020{\natexlab{c}}){Chen}, {Zhang}, {Li}, {Li}, \&
  {Lin}}]{Chen2020b}
{Chen}, Y.-X., {Zhang}, X., {Li}, Y.-P., {Li}, H., \& {Lin}, D. N.~C.
  2020{\natexlab{c}}, \apj, 900, 44

\bibitem[{{Chen} \& {Kipping}(2021)}]{ChenKipping21}
{Chen}, Z., \& {Kipping}, D. 2021, arXiv e-prints, arXiv:2112.00966

\bibitem[{{Chiappini} {et~al.}(1997){Chiappini}, {Matteucci}, \&
  {Gratton}}]{Chiappini1997}
{Chiappini}, C., {Matteucci}, F., \& {Gratton}, R. 1997, \apj, 477, 765

\bibitem[{{Chiappini} {et~al.}(2001){Chiappini}, {Matteucci}, \&
  {Romano}}]{Chiappini2001}
{Chiappini}, C., {Matteucci}, F., \& {Romano}, D. 2001, \apj, 554, 1044

\bibitem[{Childs \& Martin(2021)}]{childs2021terrestrial}
Childs, A.~C., \& Martin, R.~G. 2021, Monthly Notices of the Royal Astronomical
  Society, 507, 3461

\bibitem[{Childs \& Martin(2022)}]{childs2022misalignment}
---. 2022, The Astrophysical Journal Letters, 927, L7

\bibitem[{{Choksi} \& {Chiang}(2020)}]{Choksi2020}
{Choksi}, N., \& {Chiang}, E. 2020, \mnras, 495, 4192

\bibitem[{{Christiansen} {et~al.}(2020){Christiansen}, {Clarke}, {Burke},
  {Jenkins}, {Bryson}, {Coughlin}, {Mullally}, {Twicken}, {Batalha},
  {Catanzarite}, {Uddin}, {Zamudio}, {Smith}, {Henze}, \&
  {Campbell}}]{Christiansen2020AJ}
{Christiansen}, J.~L., {Clarke}, B.~D., {Burke}, C.~J., {et~al.} 2020, \aj,
  160, 159

\bibitem[{{Ciardi} {et~al.}(2013){Ciardi}, {Fabrycky}, {Ford}, {Gautier},
  {Howell}, {Lissauer}, {Ragozzine}, \& {Rowe}}]{Ciardi2013ApJ...763...41C}
{Ciardi}, D.~R., {Fabrycky}, D.~C., {Ford}, E.~B., {et~al.} 2013, \apj, 763, 41

\bibitem[{{Ciuc{\v{a}}} {et~al.}(2018){Ciuc{\v{a}}}, {Kawata}, {Lin},
  {Casagrande}, {Seabroke}, \& {Cropper}}]{Ciuca2018}
{Ciuc{\v{a}}}, I., {Kawata}, D., {Lin}, J., {et~al.} 2018, \mnras, 475, 1203

\bibitem[{{Clark} {et~al.}(2022){Clark}, {van Belle}, {Ciardi}, {Lund},
  {Howell}, {Everett}, {Beichman}, \& {Winters}}]{clark2022}
{Clark}, C.~A., {van Belle}, G.~T., {Ciardi}, D.~R., {et~al.} 2022, arXiv
  e-prints, arXiv:2203.12795

\bibitem[{Cleveland(1979)}]{Cleveland:1979}
Cleveland, W.~S. 1979, Journal of the American Statistical Association, 74,
  829.
\newblock
  \url{https://www.tandfonline.com/doi/abs/10.1080/01621459.1979.10481038}

\bibitem[{{Colton} {et~al.}(2021){Colton}, {Horch}, {Everett}, {Howell},
  {Davidson}, {Baptista}, \& {Casetti-Dinescu}}]{Colton2021AJ....161...21C}
{Colton}, N.~M., {Horch}, E.~P., {Everett}, M.~E., {et~al.} 2021, \aj, 161, 21

\bibitem[{{Compton} {et~al.}(2019){Compton}, {Bedding}, \&
  {Stello}}]{compton2019a}
{Compton}, D.~L., {Bedding}, T.~R., \& {Stello}, D. 2019, \mnras, 485, 560

\bibitem[{{Conroy} {et~al.}(2022){Conroy}, {Weinberg}, {Naidu}, {Buck},
  {Johnson}, {Cargile}, {Bonaca}, {Caldwell}, {Chandra}, {Han}, {Johnson},
  {Speagle}, {Ting}, {Woody}, \& {Zaritsky}}]{Conroy2022}
{Conroy}, C., {Weinberg}, D.~H., {Naidu}, R.~P., {et~al.} 2022, arXiv e-prints,
  arXiv:2204.02989

\bibitem[{{Corral-Santana} {et~al.}(2016){Corral-Santana}, {Casares},
  {Mu{\~n}oz-Darias}, {Bauer}, {Mart{\'\i}nez-Pais}, \&
  {Russell}}]{Corral-Santana16}
{Corral-Santana}, J.~M., {Casares}, J., {Mu{\~n}oz-Darias}, T., {et~al.} 2016,
  \aap, 587, A61

\bibitem[{{Creevey} {et~al.}(2015){Creevey}, {Th{\'e}venin}, {Berio}, {Heiter},
  {von Braun}, {Mourard}, {Bigot}, {Boyajian}, {Kervella}, {Morel}, {Pichon},
  {Chiavassa}, {Nardetto}, {Perraut}, {Meilland}, {Mc Alister}, {ten
  Brummelaar}, {Farrington}, {Sturmann}, {Sturmann}, \& {Turner}}]{Creevey2015}
{Creevey}, O.~L., {Th{\'e}venin}, F., {Berio}, P., {et~al.} 2015, \aap, 575,
  A26

\bibitem[{{Cui} {et~al.}(2012){Cui}, {Zhao}, {Chu}, {Li}, {Li}, {Zhang}, {Su},
  {Yao}, {Wang}, {Xing}, {Li}, {Zhu}, {Wang}, {Gu}, {Luo}, {Xu}, {Zhang},
  {Liu}, {Zhang}, {Yang}, {Cao}, {Chen}, {Chen}, {Chen}, {Chen}, {Chu}, {Feng},
  {Gong}, {Hou}, {Hu}, {Hu}, {Hu}, {Jia}, {Jiang}, {Jiang}, {Jiang}, {Jin},
  {Li}, {Li}, {Li}, {Liu}, {Liu}, {Lu}, {Mao}, {Men}, {Qi}, {Qi}, {Shi},
  {Tang}, {Tao}, {Wang}, {Wang}, {Wang}, {Wang}, {Wang}, {Wang}, {Wang},
  {Wang}, {Wang}, {Wang}, {Wang}, {Wang}, {Xu}, {Xu}, {Yang}, {Yu}, {Yuan},
  {Yuan}, {Zhai}, {Zhang}, {Zhang}, {Zhang}, {Zhao}, {Zhou}, {Zhou}, {Zhu}, \&
  {Zou}}]{Cui2012}
{Cui}, X.-Q., {Zhao}, Y.-H., {Chu}, Y.-Q., {et~al.} 2012, Research in Astronomy
  and Astrophysics, 12, 1197

\bibitem[{{Dawson} {et~al.}(2019){Dawson}, {Huang}, {Lissauer}, {Collins},
  {Sha}, {Armstrong}, {Conti}, {Collins}, {Evans}, {Gan}, {Horne}, {Ireland},
  {Murgas}, {Myers}, {Relles}, {Sefako}, {Shporer}, {Stockdale},
  {{\v{Z}}erjal}, {Zhou}, {Ricker}, {Vanderspek}, {Latham}, {Seager}, {Winn},
  {Jenkins}, {Bouma}, {Caldwell}, {Daylan}, {Doty}, {Dynes}, {Esquerdo},
  {Rose}, {Smith}, \& {Yu}}]{Dawson2019}
{Dawson}, R.~I., {Huang}, C.~X., {Lissauer}, J.~J., {et~al.} 2019, \aj, 158, 65

\bibitem[{{De Cat} {et~al.}(2015){De Cat}, {Fu}, {Ren}, {Yang}, {Shi}, {Luo},
  {Yang}, {Wang}, {Zhang}, {Shi}, {Zhang}, {Dong}, {Catanzaro}, {Corbally},
  {Frasca}, {Gray}, {Molenda-{\.Z}akowicz}, {Uytterhoeven}, {Briquet},
  {Bruntt}, {Frandsen}, {Kiss}, {Kurtz}, {Marconi}, {Niemczura}, {{\O}stensen},
  {Ripepi}, {Smalley}, {Southworth}, {Szab{\'o}}, {Telting}, {Karoff}, {Silva
  Aguirre}, {Wu}, {Hou}, {Jin}, \& {Zhou}}]{DeCat2015}
{De Cat}, P., {Fu}, J.~N., {Ren}, A.~B., {et~al.} 2015, \apjs, 220, 19

\bibitem[{{de Jong} {et~al.}(2019){de Jong}, {Agertz}, {Berbel}, {Aird},
  {Alexander}, {Amarsi}, {Anders}, {Andrae}, {Ansarinejad}, {Ansorge}, \&
  et~al.}]{deJong2019}
{de Jong}, R.~S., {Agertz}, O., {Berbel}, A.~A., {et~al.} 2019, The Messenger,
  175, 3

\bibitem[{{De Silva} {et~al.}(2015){De Silva}, {Freeman}, {Bland-Hawthorn},
  {Martell}, {de Boer}, {Asplund}, {Keller}, {Sharma}, {Zucker}, {Zwitter},
  {Anguiano}, {Bacigalupo}, {Bayliss}, {Beavis}, {Bergemann}, {Campbell},
  {Cannon}, {Carollo}, {Casagrande}, {Casey}, {Da Costa}, {D'Orazi}, {Dotter},
  {Duong}, {Heger}, {Ireland}, {Kafle}, {Kos}, {Lattanzio}, {Lewis}, {Lin},
  {Lind}, {Munari}, {Nataf}, {O'Toole}, {Parker}, {Reid}, {Schlesinger},
  {Sheinis}, {Simpson}, {Stello}, {Ting}, {Traven}, {Watson}, {Wittenmyer},
  {Yong}, \& {{\v{Z}}erjal}}]{De_Silva2015}
{De Silva}, G.~M., {Freeman}, K.~C., {Bland-Hawthorn}, J., {et~al.} 2015,
  \mnras, 449, 2604

\bibitem[{{Debes} {et~al.}(2011){Debes}, {Hoard}, {Wachter}, {Leisawitz}, \&
  {Cohen}}]{2011ApJS..197...38D}
{Debes}, J.~H., {Hoard}, D.~W., {Wachter}, S., {Leisawitz}, D.~T., \& {Cohen},
  M. 2011, \apjs, 197, 38

\bibitem[{{Deheuvels} {et~al.}(2014){Deheuvels}, {Do{\u{g}}an}, {Goupil},
  {Appourchaux}, {Benomar}, {Bruntt}, {Campante}, {Casagrande}, {Ceillier},
  {Davies}, {De Cat}, {Fu}, {Garc{\'\i}a}, {Lobel}, {Mosser}, {Reese},
  {Regulo}, {Schou}, {Stahn}, {Thygesen}, {Yang}, {Chaplin},
  {Christensen-Dalsgaard}, {Eggenberger}, {Gizon}, {Mathis},
  {Molenda-{\.Z}akowicz}, \& {Pinsonneault}}]{deheuvels2014a}
{Deheuvels}, S., {Do{\u{g}}an}, G., {Goupil}, M.~J., {et~al.} 2014, \aap, 564,
  A27

\bibitem[{Deng(2019)}]{Deng2020}
Deng, H. 2019, The Astrophysical Journal Letters, 888, L1

\bibitem[{Deng {et~al.}(2019)Deng, Ballmer, Reinhardt, Meier, Mayer, Stadel, \&
  Benitez}]{Deng2019}
Deng, H., Ballmer, M.~D., Reinhardt, C., {et~al.} 2019, The Astrophysical
  Journal, 887, 211

\bibitem[{Deng {et~al.}(2021)Deng, Mayer, \& Helled}]{Deng2021formation}
Deng, H., Mayer, L., \& Helled, R. 2021, Nature Astronomy, 5, 440

\bibitem[{Deng \& Ogilvie(2022)}]{Deng2022non}
Deng, H., \& Ogilvie, G.~I. 2022, Monthly Notices of the Royal Astronomical
  Society, 512, 6078

\bibitem[{{Deng} {et~al.}(2012){Deng}, {Newberg}, {Liu}, {Carlin}, {Beers},
  {Chen}, {Chen}, {Christlieb}, {Grillmair}, {Guhathakurta}, {Han}, {Hou},
  {Lee}, {L{\'e}pine}, {Li}, {Liu}, {Pan}, {Sellwood}, {Wang}, {Wang}, {Yang},
  {Yanny}, {Zhang}, {Zhang}, {Zheng}, \& {Zhu}}]{Deng2012}
{Deng}, L.-C., {Newberg}, H.~J., {Liu}, C., {et~al.} 2012, Research in
  Astronomy and Astrophysics, 12, 735

\bibitem[{{Di Matteo} {et~al.}(2019){Di Matteo}, {Haywood}, {Lehnert}, {Katz},
  {Khoperskov}, {Snaith}, {G{\'o}mez}, \& {Robichon}}]{Di_Matteo2019}
{Di Matteo}, P., {Haywood}, M., {Lehnert}, M.~D., {et~al.} 2019, \aap, 632, A4

\bibitem[{{Dimitriadis} {et~al.}(2019){Dimitriadis}, {Foley}, {Rest}, {Kasen},
  {Piro}, {Polin}, {Jones}, {Villar}, {Narayan}, {Coulter}, {Kilpatrick},
  {Pan}, {Rojas-Bravo}, {Fox}, {Jha}, {Nugent}, {Riess}, {Scolnic}, {Drout},
  {K2 Mission Team}, {Barentsen}, {Dotson}, {Gully-Santiago}, {Hedges}, {Cody},
  {Barclay}, {Howell}, {KEGS}, {Garnavich}, {Tucker}, {Shaya}, {Mushotzky},
  {Olling}, {Margheim}, {Zenteno}, {Kepler spacecraft Team}, {Coughlin}, {Van
  Cleve}, {Cardoso}, {Larson}, {McCalmont-Everton}, {Peterson}, {Ross},
  {Reedy}, {Osborne}, {McGinn}, {Kohnert}, {Migliorini}, {Wheaton}, {Spencer},
  {Labonde}, {Castillo}, {Beerman}, {Steward}, {Hanley}, {Larsen},
  {Gangopadhyay}, {Kloetzel}, {Weschler}, {Nystrom}, {Moffatt}, {Redick},
  {Griest}, {Packard}, {Muszynski}, {Kampmeier}, {Bjella}, {Flynn},
  {Elsaesser}, {Pan-STARRS}, {Chambers}, {Flewelling}, {Huber}, {Magnier},
  {Waters}, {Schultz}, {Bulger}, {Lowe}, {Willman}, {Smartt}, {Smith}, {DECam},
  {Points}, {Strampelli}, {ASAS-SN}, {Brimacombe}, {Chen}, {Mu{\~n}oz},
  {Mutel}, {Shields}, {Vallely}, {Villanueva}, {PTSS/TNTS}, {Li}, {Wang},
  {Zhang}, {Lin}, {Mo}, {Zhao}, {Sai}, {Zhang}, {Zhang}, {Zhang}, {Wang},
  {Zhang}, {Baron}, {DerKacy}, {Li}, {Chen}, {Xiang}, {Rui}, {Wang}, {Huang},
  {Li}, {Cumbres Observatory}, {Hosseinzadeh}, {Howell}, {Arcavi}, {Hiramatsu},
  {Burke}, {Valenti}, {ATLAS}, {Tonry}, {Denneau}, {Heinze}, {Weiland},
  {Stalder}, {Konkoly}, {Vink{\'o}}, {S{\'a}rneczky}, {P{\'a}l}, {B{\'o}di},
  {Bogn{\'a}r}, {Cs{\'a}k}, {Cseh}, {Cs{\"o}rnyei}, {Hanyecz}, {Ign{\'a}cz},
  {Kalup}, {K{\"o}nyves-T{\'o}th}, {Kriskovics}, {Ordasi}, {Rajmon},
  {S{\'o}dor}, {Szab{\'o}}, {Szak{\'a}ts}, {Zsidi}, {ePESSTO}, {Williams},
  {Nordin}, {Cartier}, {Frohmaier}, {Galbany}, {Guti{\'e}rrez}, {Hook},
  {Inserra}, {Smith}, {Arizona}, {Sand}, {Andrews}, {Smith}, \&
  {Bilinski}}]{2019ApJ...870L...1D}
{Dimitriadis}, G., {Foley}, R.~J., {Rest}, A., {et~al.} 2019, \apjl, 870, L1

\bibitem[{{Dong} {et~al.}(2018){Dong}, {Xie}, {Zhou}, {Zheng}, \&
  {Luo}}]{dong18}
{Dong}, S., {Xie}, J.-W., {Zhou}, J.-L., {Zheng}, Z., \& {Luo}, A. 2018,
  Proceedings of the National Academy of Science, 115, 266

\bibitem[{{Dong} {et~al.}(2007){Dong}, {Udalski}, {Gould}, {Reach}, {Christie},
  {Boden}, {Bennett}, {Fazio}, {Griest}, {Szyma{\'n}ski}, {Kubiak},
  {Soszy{\'n}ski}, {Pietrzy{\'n}ski}, {Szewczyk}, {Wyrzykowski}, {Ulaczyk},
  {Wieckowski}, {Paczy{\'n}ski}, {DePoy}, {Pogge}, {Preston}, {Thompson}, \&
  {Patten}}]{OB05001}
{Dong}, S., {Udalski}, A., {Gould}, A., {et~al.} 2007, \apj, 664, 862

\bibitem[{{Dong} {et~al.}(2009){Dong}, {Gould}, {Udalski}, {Anderson},
  {Christie}, {Gaudi}, {OGLE Collaboration}, {Jaroszy{\'n}ski}, {Kubiak},
  {Szyma{\'n}ski}, {Pietrzy{\'n}ski}, {Soszy{\'n}ski}, {Szewczyk}, {Ulaczyk},
  {Wyrzykowski}, {{$\mu$}FUN Collaboration}, {DePoy}, {Fox}, {Gal-Yam}, {Han},
  {L{\'e}pine}, {McCormick}, {Ofek}, {Park}, {Pogge}, {MOA Collaboration},
  {Abe}, {Bennett}, {Bond}, {Britton}, {Gilmore}, {Hearnshaw}, {Itow},
  {Kamiya}, {Kilmartin}, {Korpela}, {Masuda}, {Matsubara}, {Motomura},
  {Muraki}, {Nakamura}, {Ohnishi}, {Okada}, {Rattenbury}, {Saito}, {Sako},
  {Sasaki}, {Sullivan}, {Sumi}, {Tristram}, {Yanagisawa}, {Yock}, {Yoshoika},
  {PLANET/RoboNet Collaborations}, {Albrow}, {Beaulieu}, {Brillant}, {Calitz},
  {Cassan}, {Cook}, {Coutures}, {Dieters}, {Dominis Prester}, {Donatowicz},
  {Fouqu{\'e}}, {Greenhill}, {Hill}, {Hoffman}, {Horne}, {J{\o}rgensen},
  {Kane}, {Kubas}, {Marquette}, {Martin}, {Meintjes}, {Menzies}, {Pollard},
  {Sahu}, {Vinter}, {Wambsganss}, {Williams}, {Bode}, {Bramich}, {Burgdorf},
  {Snodgrass}, {Steele}, {Doublier}, \& {Foellmi}}]{OB050071D}
{Dong}, S., {Gould}, A., {Udalski}, A., {et~al.} 2009, \apj, 695, 970

\bibitem[{{Dong} {et~al.}(2019){Dong}, {M{\'e}rand}, {Delplancke-Str{\"o}bele},
  {Gould}, {Chen}, {Post}, {Kochanek}, {Stanek}, {Christie}, {Mutel},
  {Natusch}, {Holoien}, {Prieto}, {Shappee}, \& {Thompson}}]{Jan2017}
{Dong}, S., {M{\'e}rand}, A., {Delplancke-Str{\"o}bele}, F., {et~al.} 2019,
  \apj, 871, 70

\bibitem[{{Dotter} {et~al.}(2017){Dotter}, {Conroy}, {Cargile}, \&
  {Asplund}}]{Dotter2017}
{Dotter}, A., {Conroy}, C., {Cargile}, P., \& {Asplund}, M. 2017, \apj, 840, 99

\bibitem[{{Doyle} {et~al.}(2011){Doyle}, {Carter}, {Fabrycky}, {Slawson},
  {Howell}, {Winn}, {Orosz}, {P{\v{r}}sa}, {Welsh}, {Quinn}, {Latham},
  {Torres}, {Buchhave}, {Marcy}, {Fortney}, {Shporer}, {Ford}, {Lissauer},
  {Ragozzine}, {Rucker}, {Batalha}, {Jenkins}, {Borucki}, {Koch}, {Middour},
  {Hall}, {McCauliff}, {Fanelli}, {Quintana}, {Holman}, {Caldwell}, {Still},
  {Stefanik}, {Brown}, {Esquerdo}, {Tang}, {Furesz}, {Geary}, {Berlind},
  {Calkins}, {Short}, {Steffen}, {Sasselov}, {Dunham}, {Cochran}, {Boss},
  {Haas}, {Buzasi}, \& {Fischer}}]{Doyle2011}
{Doyle}, L.~R., {Carter}, J.~A., {Fabrycky}, D.~C., {et~al.} 2011, Science,
  333, 1602

\bibitem[{Dr{\c a}{\.z}kowska \& Alibert(2017)}]{Drazkowska2017}
Dr{\c a}{\.z}kowska, J., \& Alibert, Y. 2017, A92

\bibitem[{{Dr{\c a}{\.z}kowska} {et~al.}(2019){Dr{\c a}{\.z}kowska}, {Li},
  {Birnstiel}, {Stammler}, \& {Li}}]{Drazkowska2019}
{Dr{\c a}{\.z}kowska}, J., {Li}, S., {Birnstiel}, T., {Stammler}, S.~M., \&
  {Li}, H. 2019, \apj, 885, 91

\bibitem[{{Dr{\'e}au} {et~al.}(2021){Dr{\'e}au}, {Mosser}, {Lebreton}, {Gehan},
  \& {Kallinger}}]{dreau2021a}
{Dr{\'e}au}, G., {Mosser}, B., {Lebreton}, Y., {Gehan}, C., \& {Kallinger}, T.
  2021, \aap, 650, A115

\bibitem[{Dreizler {et~al.}(2020)Dreizler, Jeffers, Rodr{\'\i}guez,
  Zechmeister, Barnes, Haswell, Coleman, Lalitha, Hidalgo~Soto, Strachan,
  {et~al.}}]{Dreizler2020reddots}
Dreizler, S., Jeffers, S., Rodr{\'\i}guez, E., {et~al.} 2020, Monthly Notices
  of the Royal Astronomical Society, 493, 536

\bibitem[{{Dressing} \& {Charbonneau}(2015)}]{Dressing2015}
{Dressing}, C.~D., \& {Charbonneau}, D. 2015, \apj, 807, 45

\bibitem[{{Dufour} {et~al.}(2010){Dufour}, {Kilic}, {Fontaine}, {Bergeron},
  {Lachapelle}, {Kleinman}, \& {Leggett}}]{2010ApJ...719..803D}
{Dufour}, P., {Kilic}, M., {Fontaine}, G., {et~al.} 2010, \apj, 719, 803

\bibitem[{{Dziembowski} {et~al.}(1993){Dziembowski}, {Moskalik}, \&
  {Pamyatnykh}}]{1993MNRAS.265..588D}
{Dziembowski}, W.~A., {Moskalik}, P., \& {Pamyatnykh}, A.~A. 1993, \mnras, 265,
  588

\bibitem[{{Edvardsson} {et~al.}(1993){Edvardsson}, {Andersen}, {Gustafsson},
  {Lambert}, {Nissen}, \& {Tomkin}}]{Edvardsson1993}
{Edvardsson}, B., {Andersen}, J., {Gustafsson}, B., {et~al.} 1993, \aap, 275,
  101

\bibitem[{{Edwards} {et~al.}(2019){Edwards}, {Mugnai}, {Tinetti}, {Pascale}, \&
  {Sarkar}}]{2019AJ....157..242E}
{Edwards}, B., {Mugnai}, L., {Tinetti}, G., {Pascale}, E., \& {Sarkar}, S.
  2019, \aj, 157, 242

\bibitem[{{El-Badry} \& {Burdge}(2022)}]{El-Badry2022}
{El-Badry}, K., \& {Burdge}, K.~B. 2022, \mnras, 511, 24

\bibitem[{{El-Badry} {et~al.}(2022){El-Badry}, {Seeburger}, {Jayasinghe},
  {Rix}, {Almada}, {Conroy}, {Price-Whelan}, \& {Burdge}}]{El-Badry2022unicorn}
{El-Badry}, K., {Seeburger}, R., {Jayasinghe}, T., {et~al.} 2022, \mnras, 512,
  5620

\bibitem[{{Emsenhuber} {et~al.}(2021){Emsenhuber}, {Mordasini}, {Burn},
  {Alibert}, {Benz}, \& {Asphaug}}]{Emsenhuber2021}
{Emsenhuber}, A., {Mordasini}, C., {Burn}, R., {et~al.} 2021, \aap, 656, A70

\bibitem[{Fabrycky {et~al.}(2014)Fabrycky, Lissauer, Ragozzine, Rowe, Steffen,
  Agol, Barclay, Batalha, Borucki, Ciardi, {et~al.}}]{Fabrycky2014}
Fabrycky, D.~C., Lissauer, J.~J., Ragozzine, D., {et~al.} 2014, The
  Astrophysical Journal, 790, 146

\bibitem[{Fang \& Deng(2020)}]{Fang2020}
Fang, T., \& Deng, H. 2020, Monthly Notices of the Royal Astronomical Society,
  496, 3781

\bibitem[{{Farihi} {et~al.}(2009){Farihi}, {Jura}, \&
  {Zuckerman}}]{2009ApJ...694..805F}
{Farihi}, J., {Jura}, M., \& {Zuckerman}, B. 2009, \apj, 694, 805

\bibitem[{{Farihi} {et~al.}(2022){Farihi}, {Hermes}, {Marsh}, {Mustill},
  {Wyatt}, {Guidry}, {Wilson}, {Redfield}, {Izquierdo}, {Toloza},
  {G{\"a}nsicke}, {Aungwerojwit}, {Kaewmanee}, {Dhillon}, \&
  {Swan}}]{2022MNRAS.511.1647F}
{Farihi}, J., {Hermes}, J.~J., {Marsh}, T.~R., {et~al.} 2022, \mnras, 511, 1647

\bibitem[{Farnham {et~al.}(2019)Farnham, Kelley, Knight, \&
  Feaga}]{farnham2019first}
Farnham, T.~L., Kelley, M.~S., Knight, M.~M., \& Feaga, L.~M. 2019, The
  Astrophysical Journal Letters, 886, L24

\bibitem[{{Fausnaugh} {et~al.}(2021){Fausnaugh}, {Vallely}, {Kochanek},
  {Shappee}, {Stanek}, {Tucker}, {Ricker}, {Vanderspek}, {Latham}, {Seager},
  {Winn}, {Jenkins}, {Berta-Thompson}, {Daylan}, {Doty}, {F{\H{u}}r{\'e}sz},
  {Levine}, {Morris}, {P{\'a}l}, {Sha}, {Ting}, \&
  {Wohler}}]{2021ApJ...908...51F}
{Fausnaugh}, M.~M., {Vallely}, P.~J., {Kochanek}, C.~S., {et~al.} 2021, \apj,
  908, 51

\bibitem[{{Feltzing} {et~al.}(2001){Feltzing}, {Holmberg}, \&
  {Hurley}}]{Feltzing2001}
{Feltzing}, S., {Holmberg}, J., \& {Hurley}, J.~R. 2001, \aap, 377, 911

\bibitem[{{Feng} {et~al.}(2019){Feng}, {Anglada-Escud{\'e}}, {Tuomi}, {Jones},
  {Chanam{\'e}}, {Butler}, \& {Janson}}]{feng19b}
{Feng}, F., {Anglada-Escud{\'e}}, G., {Tuomi}, M., {et~al.} 2019, \mnras, 490,
  5002

\bibitem[{{Feng} {et~al.}(2021){Feng}, {Butler}, {Jones}, {Phillips}, {Vogt},
  {Oppenheimer}, {Holden}, {Burt}, \& {Boss}}]{feng21}
{Feng}, F., {Butler}, R.~P., {Jones}, H. R.~A., {et~al.} 2021, \mnras, 507,
  2856

\bibitem[{{Feuillet} {et~al.}(2019){Feuillet}, {Frankel}, {Lind}, {Frinchaboy},
  {Garc{\'\i}a-Hern{\'a}ndez}, {Lane}, {Nitschelm}, \&
  {Roman-Lopes}}]{Feuillet2019}
{Feuillet}, D.~K., {Frankel}, N., {Lind}, K., {et~al.} 2019, \mnras, 489, 1742

\bibitem[{{Feuillet} {et~al.}(2018){Feuillet}, {Bovy}, {Holtzman}, {Weinberg},
  {Garc{\'\i}a-Hern{\'a}ndez}, {Hearty}, {Majewski}, {Roman-Lopes}, {Rybizki},
  \& {Zamora}}]{Feuillet2018}
{Feuillet}, D.~K., {Bovy}, J., {Holtzman}, J., {et~al.} 2018, \mnras, 477, 2326

\bibitem[{{Fischer} \& {Valenti}(2005)}]{Fischer:2005}
{Fischer}, D.~A., \& {Valenti}, J. 2005, \apj, 622, 1102

\bibitem[{{Fitzmaurice} {et~al.}(2022){Fitzmaurice}, {Martin}, \&
  {Fabrycky}}]{Fitzmarice22}
{Fitzmaurice}, E., {Martin}, D.~V., \& {Fabrycky}, D.~C. 2022, arXiv e-prints,
  arXiv:2202.11719

\bibitem[{{Fleming} {et~al.}(2018){Fleming}, {Barnes}, {Graham}, {Luger}, \&
  {Quinn}}]{Fleming18}
{Fleming}, D.~P., {Barnes}, R., {Graham}, D.~E., {Luger}, R., \& {Quinn}, T.~R.
  2018, \apj, 858, 86

\bibitem[{{Fontaine} {et~al.}(2003){Fontaine}, {Brassard}, {Charpinet},
  {Green}, {Chayer}, {Bill{\`e}res}, \& {Randall}}]{2003ApJ...597..518F}
{Fontaine}, G., {Brassard}, P., {Charpinet}, S., {et~al.} 2003, \apj, 597, 518

\bibitem[{{Forbes} \& {Bridges}(2010)}]{Forbes2010}
{Forbes}, D.~A., \& {Bridges}, T. 2010, \mnras, 404, 1203

\bibitem[{{Ford} {et~al.}(2008){Ford}, {Quinn}, \& {Veras}}]{Ford2008ApJ}
{Ford}, E.~B., {Quinn}, S.~N., \& {Veras}, D. 2008, \apj, 678, 1407

\bibitem[{{Foreman-Mackey} {et~al.}(2016){Foreman-Mackey}, {Morton}, {Hogg},
  {Agol}, \& {Sch{\"o}lkopf}}]{Foreman-Mackey2016}
{Foreman-Mackey}, D., {Morton}, T.~D., {Hogg}, D.~W., {Agol}, E., \&
  {Sch{\"o}lkopf}, B. 2016, \aj, 152, 206

\bibitem[{Forgan {et~al.}(2017)Forgan, Hall, Meru, \& Rice}]{Forgan2017}
Forgan, D.~H., Hall, C., Meru, F., \& Rice, W. K.~M. 2017, Monthly Notices of
  the Royal Astronomical Society, 474, 5036.
\newblock \url{https://doi.org/10.1093/mnras/stx2870}

\bibitem[{{Fouesneau} {et~al.}(2022){Fouesneau}, {Andrae}, {Dharmawardena},
  {Rybizki}, {Bailer-Jones}, \& {Demleitner}}]{Fouesneau2022}
{Fouesneau}, M., {Andrae}, R., {Dharmawardena}, T., {et~al.} 2022, arXiv
  e-prints, arXiv:2201.03252

\bibitem[{{Fox} \& {Wiegert}(2021)}]{Fox2021}
{Fox}, C., \& {Wiegert}, P. 2021, \mnras, 501, 2378

\bibitem[{{Frankel} {et~al.}(2018){Frankel}, {Rix}, {Ting}, {Ness}, \&
  {Hogg}}]{Frankel2018}
{Frankel}, N., {Rix}, H.-W., {Ting}, Y.-S., {Ness}, M., \& {Hogg}, D.~W. 2018,
  \apj, 865, 96

\bibitem[{{Frankel} {et~al.}(2019){Frankel}, {Sanders}, {Rix}, {Ting}, \&
  {Ness}}]{Frankel2019}
{Frankel}, N., {Sanders}, J., {Rix}, H.-W., {Ting}, Y.-S., \& {Ness}, M. 2019,
  \apj, 884, 99

\bibitem[{{Frebel} \& {Norris}(2015)}]{Frebel2015}
{Frebel}, A., \& {Norris}, J.~E. 2015, \araa, 53, 631

\bibitem[{{Freeman} \& {Bland-Hawthorn}(2002)}]{Freeman2002}
{Freeman}, K., \& {Bland-Hawthorn}, J. 2002, \araa, 40, 487

\bibitem[{{Fressin} {et~al.}(2013){Fressin}, {Torres}, {Charbonneau}, {Bryson},
  {Christiansen}, {Dressing}, {Jenkins}, {Walkowicz}, \& {Batalha}}]{fressin13}
{Fressin}, F., {Torres}, G., {Charbonneau}, D., {et~al.} 2013, \apj, 766, 81

\bibitem[{{Fu} {et~al.}(2020){Fu}, {Cat}, {Zong}, {Frasca}, {Gray}, {Ren},
  {Molenda-{\.Z}akowicz}, {Corbally}, {Catanzaro}, {Shi}, {Luo}, \&
  {Zhang}}]{Fu2020}
{Fu}, J.-N., {Cat}, P.~D., {Zong}, W., {et~al.} 2020, Research in Astronomy and
  Astrophysics, 20, 167

\bibitem[{{Fuller} {et~al.}(2015){Fuller}, {Cantiello}, {Stello}, {Garcia}, \&
  {Bildsten}}]{Fuller2015}
{Fuller}, J., {Cantiello}, M., {Stello}, D., {Garcia}, R.~A., \& {Bildsten}, L.
  2015, Science, 350, 423

\bibitem[{{Fuller} {et~al.}(2020){Fuller}, {Kurtz}, {Handler}, \&
  {Rappaport}}]{2020MNRAS.498.5730F}
{Fuller}, J., {Kurtz}, D.~W., {Handler}, G., \& {Rappaport}, S. 2020, \mnras,
  498, 5730

\bibitem[{{Fuller} \& {Lai}(2012)}]{2012MNRAS.420.3126F}
{Fuller}, J., \& {Lai}, D. 2012, \mnras, 420, 3126

\bibitem[{{Fulton} {et~al.}(2017){Fulton}, {Petigura}, {Howard}, {Isaacson},
  {Marcy}, {Cargile}, {Hebb}, {Weiss}, {Johnson}, {Morton}, {Sinukoff},
  {Crossfield}, \& {Hirsch}}]{Fulton2017}
{Fulton}, B.~J., {Petigura}, E.~A., {Howard}, A.~W., {et~al.} 2017, \aj, 154,
  109

\bibitem[{{Furlan} \& {Howell}(2017)}]{FH2017AJ....154...66F}
{Furlan}, E., \& {Howell}, S.~B. 2017, \aj, 154, 66

\bibitem[{{Furlan} \& {Howell}(2020)}]{FH2020ApJ...898...47F}
---. 2020, \apj, 898, 47

\bibitem[{{Furlan} {et~al.}(2017){Furlan}, {Ciardi}, {Everett}, {Saylors},
  {Teske}, {Horch}, {Howell}, {van Belle}, {Hirsch}, {Gautier}, {Adams},
  {Barrado}, {Cartier}, {Dressing}, {Dupree}, {Gilliland}, {Lillo-Box},
  {Lucas}, \& {Wang}}]{Furlan:2017}
{Furlan}, E., {Ciardi}, D.~R., {Everett}, M.~E., {et~al.} 2017, \aj, 153, 71

\bibitem[{{Gaia Collaboration} {et~al.}(2016){Gaia Collaboration}, {Prusti},
  {de Bruijne}, {Brown}, {Vallenari}, {Babusiaux}, {Bailer-Jones}, {Bastian},
  {Biermann}, {Evans}, \& et~al.}]{Prusti2016}
{Gaia Collaboration}, {Prusti}, T., {de Bruijne}, J.~H.~J., {et~al.} 2016,
  \aap, 595, A1

\bibitem[{{Gaia Collaboration} {et~al.}(2018){Gaia Collaboration}, {Brown},
  {Vallenari}, {Prusti}, {de Bruijne}, {Babusiaux}, {Bailer-Jones}, {Biermann},
  {Evans}, {Eyer}, \& et~al.}]{Brown2018}
{Gaia Collaboration}, {Brown}, A.~G.~A., {Vallenari}, A., {et~al.} 2018, \aap,
  616, A1

\bibitem[{{Gaia Collaboration} {et~al.}(2021){Gaia Collaboration}, {Brown},
  {Vallenari}, {Prusti}, {de Bruijne}, {Babusiaux}, {Biermann}, {Creevey},
  {Evans}, {Eyer}, \& et~al.}]{Brown2021}
---. 2021, \aap, 649, A1

\bibitem[{{Gallart} {et~al.}(2019){Gallart}, {Bernard}, {Brook}, {Ruiz-Lara},
  {Cassisi}, {Hill}, \& {Monelli}}]{Gallart2019}
{Gallart}, C., {Bernard}, E.~J., {Brook}, C.~B., {et~al.} 2019, Nature
  Astronomy, 3, 932

\bibitem[{{G{\"a}nsicke} {et~al.}(2019){G{\"a}nsicke}, {Schreiber}, {Toloza},
  {Gentile Fusillo}, {Koester}, \& {Manser}}]{2019Natur.576...61G}
{G{\"a}nsicke}, B.~T., {Schreiber}, M.~R., {Toloza}, O., {et~al.} 2019, \nat,
  576, 61

\bibitem[{García {et~al.}(2010)García, Mathur, Salabert, Ballot, Regulo,
  Metcalfe, \& Baglin}]{Garcia2010_HD49933}
García, R.~A., Mathur, S., Salabert, D., {et~al.} 2010, Science, 329, 1032.
\newblock \url{http://adsabs.harvard.edu/abs/2010Sci...329.1032G}

\bibitem[{{Garc{\'\i}a} \& {Ballot}(2019)}]{garcia2019a}
{Garc{\'\i}a}, R.~A., \& {Ballot}, J. 2019, Living Reviews in Solar Physics,
  16, 4

\bibitem[{{Garc{\'\i}a} {et~al.}(2014{\natexlab{a}}){Garc{\'\i}a}, {P{\'e}rez
  Hern{\'a}ndez}, {Benomar}, {Silva Aguirre}, {Ballot}, {Davies},
  {Do{\u{g}}an}, {Stello}, {Christensen-Dalsgaard}, {Houdek}, {Ligni{\`e}res},
  {Mathur}, {Takata}, {Ceillier}, {Chaplin}, {Mathis}, {Mosser}, {Ouazzani},
  {Pinsonneault}, {Reese}, {R{\'e}gulo}, {Salabert}, {Thompson}, {van Saders},
  {Neiner}, \& {De Ridder}}]{Garcia2014a}
{Garc{\'\i}a}, R.~A., {P{\'e}rez Hern{\'a}ndez}, F., {Benomar}, O., {et~al.}
  2014{\natexlab{a}}, \aap, 563, A84

\bibitem[{{Garc{\'\i}a} {et~al.}(2014{\natexlab{b}}){Garc{\'\i}a}, {Ceillier},
  {Salabert}, {Mathur}, {van Saders}, {Pinsonneault}, {Ballot}, {Beck},
  {Bloemen}, {Campante}, {Davies}, {do Nascimento}, {Mathis}, {Metcalfe},
  {Nielsen}, {Su{\'a}rez}, {Chaplin}, {Jim{\'e}nez}, \& {Karoff}}]{Garcia2014}
{Garc{\'\i}a}, R.~A., {Ceillier}, T., {Salabert}, D., {et~al.}
  2014{\natexlab{b}}, \aap, 572, A34

\bibitem[{{Garnavich} {et~al.}(2016){Garnavich}, {Tucker}, {Rest}, {Shaya},
  {Olling}, {Kasen}, \& {Villar}}]{Garnavich2016}
{Garnavich}, P.~M., {Tucker}, B.~E., {Rest}, A., {et~al.} 2016, \apj, 820, 23

\bibitem[{Garrido \& Rodriguez(1996)}]{Garrido1996}
Garrido, R., \& Rodriguez, E. 1996, Monthly Notices of the Royal Astronomical
  Society, 281, 696

\bibitem[{{Gazeas}(2009)}]{Gaz2009}
{Gazeas}, K.~D. 2009, Communications in Asteroseismology, 159, 129

\bibitem[{{Gehan} {et~al.}(2021){Gehan}, {Mosser}, {Michel}, \&
  {Cunha}}]{Gehan2021}
{Gehan}, C., {Mosser}, B., {Michel}, E., \& {Cunha}, M.~S. 2021, \aap, 645,
  A124

\bibitem[{{Gentile Fusillo} {et~al.}(2021){Gentile Fusillo}, {Tremblay},
  {Cukanovaite}, {Vorontseva}, {Lallement}, {Hollands}, {G{\"a}nsicke},
  {Burdge}, {McCleery}, \& {Jordan}}]{2021MNRAS.508.3877G}
{Gentile Fusillo}, N.~P., {Tremblay}, P.~E., {Cukanovaite}, E., {et~al.} 2021,
  \mnras, 508, 3877

\bibitem[{{Giacalone} {et~al.}(2021){Giacalone}, {Dressing}, {Jensen},
  {Collins}, {Ricker}, {Vanderspek}, {Seager}, {Winn}, {Jenkins}, {Barclay},
  {Barkaoui}, {Cadieux}, {Charbonneau}, {Collins}, {Conti}, {Doyon}, {Evans},
  {Ghachoui}, {Gillon}, {Guerrero}, {Hart}, {Jehin}, {Kielkopf}, {McLean},
  {Murgas}, {Palle}, {Parviainen}, {Pozuelos}, {Relles}, {Shporer}, {Socia},
  {Stockdale}, {Tan}, {Torres}, {Twicken}, {Waalkes}, \&
  {Waite}}]{Giacalone2021}
{Giacalone}, S., {Dressing}, C.~D., {Jensen}, E. L.~N., {et~al.} 2021, \aj,
  161, 24

\bibitem[{{Giammichele} {et~al.}(2018){Giammichele}, {Charpinet}, {Fontaine},
  {Brassard}, {Green}, {Van Grootel}, {Bergeron}, {Zong}, \&
  {Dupret}}]{2018Natur.554...73G}
{Giammichele}, N., {Charpinet}, S., {Fontaine}, G., {et~al.} 2018, \nat, 554,
  73

\bibitem[{{Gibson} {et~al.}(2016){Gibson}, {Howard}, {Marcy}, {Edelstein},
  {Wishnow}, \& {Poppett}}]{gibson16}
{Gibson}, S.~R., {Howard}, A.~W., {Marcy}, G.~W., {et~al.} 2016, in Society of
  Photo-Optical Instrumentation Engineers (SPIE) Conference Series, Vol. 9908,
  Ground-based and Airborne Instrumentation for Astronomy VI, ed. C.~J.
  {Evans}, L.~{Simard}, \& H.~{Takami}, 990870

\bibitem[{Gilbert {et~al.}(2020)Gilbert, Barclay, Schlieder, Quintana, Hord,
  Kostov, Lopez, Rowe, Hoffman, Walkowicz, {et~al.}}]{Gilbert2020first}
Gilbert, E.~A., Barclay, T., Schlieder, J.~E., {et~al.} 2020, The Astronomical
  Journal, 160, 116

\bibitem[{{Gilliland} {et~al.}(2011){Gilliland}, {Chaplin}, {Dunham},
  {Argabright}, {Borucki}, {Basri}, {Bryson}, {Buzasi}, {Caldwell}, {Elsworth},
  {Jenkins}, {Koch}, {Kolodziejczak}, {Miglio}, {van Cleve}, {Walkowicz}, \&
  {Welsh}}]{Gilliland2011ApJS}
{Gilliland}, R.~L., {Chaplin}, W.~J., {Dunham}, E.~W., {et~al.} 2011, \apjs,
  197, 6

\bibitem[{Gillon {et~al.}(2017)Gillon, Triaud, Demory, Jehin, Agol, Deck,
  Lederer, De~Wit, Burdanov, Ingalls, {et~al.}}]{Gillon2017seven}
Gillon, M., Triaud, A.~H., Demory, B.-O., {et~al.} 2017, Nature, 542, 456

\bibitem[{{Gilmore} {et~al.}(2012){Gilmore}, {Randich}, {Asplund}, {Binney},
  {Bonifacio}, {Drew}, {Feltzing}, {Ferguson}, {Jeffries}, {Micela}, \&
  et~al.}]{Gilmore2012}
{Gilmore}, G., {Randich}, S., {Asplund}, M., {et~al.} 2012, The Messenger, 147,
  25

\bibitem[{{Ginzburg} {et~al.}(2018){Ginzburg}, {Schlichting}, \&
  {Sari}}]{Ginzburg2018}
{Ginzburg}, S., {Schlichting}, H.~E., \& {Sari}, R. 2018, \mnras, 476, 759

\bibitem[{Giovinazzi {et~al.}(2021)Giovinazzi, Blake, \&
  Bernardinelli}]{giovinazzi2021enhancing}
Giovinazzi, M.~R., Blake, C.~H., \& Bernardinelli, P.~H. 2021, Publications of
  the Astronomical Society of the Pacific, 133, 114401

\bibitem[{{Gizon} \& {Solanki}(2003)}]{Gizon2003}
{Gizon}, L., \& {Solanki}, S.~K. 2003, \apj, 589, 1009

\bibitem[{{Gomes} {et~al.}(2005){Gomes}, {Levison}, {Tsiganis}, \&
  {Morbidelli}}]{2005Natur.435..466G}
{Gomes}, R., {Levison}, H.~F., {Tsiganis}, K., \& {Morbidelli}, A. 2005, \nat,
  435, 466

\bibitem[{{Gordon} {et~al.}(2021){Gordon}, {Davenport}, {Angus},
  {Foreman-Mackey}, {Agol}, {Covey}, {Ag{\"u}eros}, \& {Kipping}}]{Gordon2021}
{Gordon}, T.~A., {Davenport}, J. R.~A., {Angus}, R., {et~al.} 2021, \apj, 913,
  70

\bibitem[{{Gould}(1992)}]{Gould1992}
{Gould}, A. 1992, \apj, 392, 442

\bibitem[{{Gould}(1994)}]{1994ApJ...421L..75G}
---. 1994, \apjl, 421, L75

\bibitem[{{Gould}(1995)}]{Gould1995single}
---. 1995, \apjl, 441, L21

\bibitem[{{Gould}(2000)}]{Gould2000}
---. 2000, \apj, 542, 785

\bibitem[{{Gould} \& {Loeb}(1992)}]{Andy1992}
{Gould}, A., \& {Loeb}, A. 1992, \apj, 396, 104

\bibitem[{{Gould} {et~al.}(2021){Gould}, {Zang}, {Mao}, \& {Dong}}]{CMST}
{Gould}, A., {Zang}, W.-C., {Mao}, S., \& {Dong}, S.-B. 2021, Research in
  Astronomy and Astrophysics, 21, 133

\bibitem[{{Gould} {et~al.}(2010){Gould}, {Dong}, {Gaudi}, {Udalski}, {Bond},
  {Greenhill}, {Street}, {Dominik}, {Sumi}, {Szyma{\'n}ski}, {Han}, {Allen},
  {Bolt}, {Bos}, {Christie}, {DePoy}, {Drummond}, {Eastman}, {Gal-Yam},
  {Higgins}, {Janczak}, {Kaspi}, {Koz{\l}owski}, {Lee}, {Mallia}, {Maury},
  {Maoz}, {McCormick}, {Monard}, {Moorhouse}, {Morgan}, {Natusch}, {Ofek},
  {Park}, {Pogge}, {Polishook}, {Santallo}, {Shporer}, {Spector}, {Thornley},
  {Yee}, {{$\mu$}FUN Collaboration}, {Kubiak}, {Pietrzy{\'n}ski},
  {Soszy{\'n}ski}, {Szewczyk}, {Wyrzykowski}, {Ulaczyk}, {Poleski}, {OGLE
  Collaboration}, {Abe}, {Bennett}, {Botzler}, {Douchin}, {Freeman}, {Fukui},
  {Furusawa}, {Hearnshaw}, {Hosaka}, {Itow}, {Kamiya}, {Kilmartin}, {Korpela},
  {Lin}, {Ling}, {Makita}, {Masuda}, {Matsubara}, {Miyake}, {Muraki}, {Nagaya},
  {Nishimoto}, {Ohnishi}, {Okumura}, {Perrott}, {Philpott}, {Rattenbury},
  {Saito}, {Sako}, {Sullivan}, {Sweatman}, {Tristram}, {von Seggern}, {Yock},
  {MOA Collaboration}, {Albrow}, {Batista}, {Beaulieu}, {Brillant}, {Caldwell},
  {Calitz}, {Cassan}, {Cole}, {Cook}, {Coutures}, {Dieters}, {Dominis Prester},
  {Donatowicz}, {Fouqu{\'e}}, {Hill}, {Hoffman}, {Jablonski}, {Kane}, {Kains},
  {Kubas}, {Marquette}, {Martin}, {Martioli}, {Meintjes}, {Menzies},
  {Pedretti}, {Pollard}, {Sahu}, {Vinter}, {Wambsganss}, {Watson}, {Williams},
  {Zub}, {PLANET Collaboration}, {Allan}, {Bode}, {Bramich}, {Burgdorf},
  {Clay}, {Fraser}, {Hawkins}, {Horne}, {Kerins}, {Lister}, {Mottram},
  {Saunders}, {Snodgrass}, {Steele}, {Tsapras}, {RoboNet Collaboration},
  {J{\o}rgensen}, {Anguita}, {Bozza}, {Calchi Novati}, {Harps{\o}e}, {Hinse},
  {Hundertmark}, {Kj{\ae}rgaard}, {Liebig}, {Mancini}, {Masi}, {Mathiasen},
  {Rahvar}, {Ricci}, {Scarpetta}, {Southworth}, {Surdej}, {Th{\"o}ne}, \&
  {MiNDSTEp Consortium}}]{mufun}
{Gould}, A., {Dong}, S., {Gaudi}, B.~S., {et~al.} 2010, \apj, 720, 1073

\bibitem[{{Gould} {et~al.}(2022{\natexlab{a}}){Gould}, {Jung}, {Hwang}, {Dong},
  {Albrow}, {Chung}, {Han}, {Ryu}, {Shin}, {Shvartzvald}, {Yang}, {Yee},
  {Zang}, {Cha}, {Kim}, {Kim}, {Lee}, {Lee}, {Lee}, {Park}, \&
  {Pogge}}]{Gould2022}
{Gould}, A., {Jung}, Y.~K., {Hwang}, K.-H., {et~al.} 2022{\natexlab{a}}, arXiv
  e-prints, arXiv:2204.03269

\bibitem[{{Gould} {et~al.}(2022{\natexlab{b}}){Gould}, {Han}, {Zang}, {Yang},
  {Hwang}, {Udalski}, {Bond}, {Albrow}, {Chung}, {Jung}, {Ryu}, {Shin},
  {Shvartzvald}, {Yee}, {Cha}, {Kim}, {Kim}, {Kim}, {Lee}, {Lee}, {Lee},
  {Park}, {Pogge}, {Mr{\'o}z}, {Szyma{\'n}ski}, {Skowron}, {Poleski},
  {Soszy{\'n}ski}, {Pietrukowicz}, {Koz{\l}owski}, {Ulaczyk}, {Rybicki},
  {Iwanek}, {Wrona}, {Abe}, {Barry}, {Bennett}, {Bhattacharya}, {Fujii},
  {Fukui}, {Hirao}, {Ishitani Silva}, {Kirikawa}, {Kondo}, {Koshimoto},
  {Matsubara}, {Matsumoto}, {Miyazaki}, {Muraki}, {Okamura}, {Olmschenk},
  {Ranc}, {Rattenbury}, {Satoh}, {Sumi}, {Suzuki}, {Toda}, {Tristram},
  {Vandorou}, {Yama}, {Beichman}, {Bryden}, {Calchi Novati}, {Gaudi},
  {Henderson}, {Penny}, {Jacklin}, \& {Stassun}}]{OB181126}
{Gould}, A., {Han}, C., {Zang}, W., {et~al.} 2022{\natexlab{b}}, arXiv
  e-prints, arXiv:2204.04354

\bibitem[{{Grenon}(1989)}]{Grenon1989}
{Grenon}, M. 1989, \apss, 156, 29

\bibitem[{{Griest} \& {Hu}(1992)}]{Griest1992}
{Griest}, K., \& {Hu}, W. 1992, \apj, 397, 362

\bibitem[{{Griest} \& {Safizadeh}(1998)}]{Griest1998}
{Griest}, K., \& {Safizadeh}, N. 1998, \apj, 500, 37

\bibitem[{{Grisoni} {et~al.}(2018){Grisoni}, {Spitoni}, \&
  {Matteucci}}]{Grisoni2018}
{Grisoni}, V., {Spitoni}, E., \& {Matteucci}, F. 2018, \mnras, 481, 2570

\bibitem[{{Guerrero} {et~al.}(2021){Guerrero}, {Seager}, {Huang}, {Vanderburg},
  {Garcia Soto}, {Mireles}, {Hesse}, {Fong}, {Glidden}, {Shporer}, {Latham},
  {Collins}, {Quinn}, {Burt}, {Dragomir}, {Crossfield}, {Vanderspek},
  {Fausnaugh}, {Burke}, {Ricker}, {Daylan}, {Essack}, {G{\"u}nther}, {Osborn},
  {Pepper}, {Rowden}, {Sha}, {Villanueva}, {Yahalomi}, {Yu}, {Ballard},
  {Batalha}, {Berardo}, {Chontos}, {Dittmann}, {Esquerdo}, {Mikal-Evans},
  {Jayaraman}, {Krishnamurthy}, {Louie}, {Mehrle}, {Niraula}, {Rackham},
  {Rodriguez}, {Rowden}, {Sousa-Silva}, {Watanabe}, {Wong}, {Zhan},
  {Zivanovic}, {Christiansen}, {Ciardi}, {Swain}, {Lund}, {Mullally},
  {Fleming}, {Rodriguez}, {Boyd}, {Quintana}, {Barclay}, {Col{\'o}n},
  {Rinehart}, {Schlieder}, {Clampin}, {Jenkins}, {Twicken}, {Caldwell},
  {Coughlin}, {Henze}, {Lissauer}, {Morris}, {Rose}, {Smith}, {Tenenbaum},
  {Ting}, {Wohler}, {Bakos}, {Bean}, {Berta-Thompson}, {Bieryla}, {Bouma},
  {Buchhave}, {Butler}, {Charbonneau}, {Doty}, {Ge}, {Holman}, {Howard},
  {Kaltenegger}, {Kane}, {Kjeldsen}, {Kreidberg}, {Lin}, {Minsky}, {Narita},
  {Paegert}, {P{\'a}l}, {Palle}, {Sasselov}, {Spencer}, {Sozzetti}, {Stassun},
  {Torres}, {Udry}, \& {Winn}}]{Guerrero2021}
{Guerrero}, N.~M., {Seager}, S., {Huang}, C.~X., {et~al.} 2021, \apjs, 254, 39

\bibitem[{{Guidry} {et~al.}(2021){Guidry}, {Vanderbosch}, {Hermes}, {Barlow},
  {Lopez}, {Boudreaux}, {Corcoran}, {Bell}, {Montgomery}, {Heintz},
  {Castanheira}, {Reding}, {Dunlap}, {Winget}, {Winget}, \&
  {Kuehne}}]{2021ApJ...912..125G}
{Guidry}, J.~A., {Vanderbosch}, Z.~P., {Hermes}, J.~J., {et~al.} 2021, \apj,
  912, 125

\bibitem[{{Guo} {et~al.}(2020){Guo}, {Shporer}, {Hambleton}, \&
  {Isaacson}}]{2020ApJ...888...95G}
{Guo}, Z., {Shporer}, A., {Hambleton}, K., \& {Isaacson}, H. 2020, \apj, 888,
  95

\bibitem[{Hadden \& Lithwick(2017)}]{Hadden2017}
Hadden, S., \& Lithwick, Y. 2017, The Astronomical Journal, 154, 5

\bibitem[{{Haffert} {et~al.}(2019){Haffert}, {Bohn}, {de Boer}, {Snellen},
  {Brinchmann}, {Girard}, {Keller}, \& {Bacon}}]{Haffert2019}
{Haffert}, S.~Y., {Bohn}, A.~J., {de Boer}, J., {et~al.} 2019, Nature
  Astronomy, 3, 749

\bibitem[{{Hajdu} {et~al.}(2021){Hajdu}, {Pietrzy{\'n}ski}, {Jurcsik},
  {Catelan}, {Karczmarek}, {Pilecki}, {Soszy{\'n}ski}, {Udalski}, \&
  {Thompson}}]{HajduG2021binaryRRL}
{Hajdu}, G., {Pietrzy{\'n}ski}, G., {Jurcsik}, J., {et~al.} 2021, \apj, 915, 50

\bibitem[{{Hall} {et~al.}(2021{\natexlab{a}}){Hall}, {Davies}, {van Saders},
  {Nielsen}, {Lund}, {Chaplin}, {Garc{\'\i}a}, {Amard}, {Breimann}, {Khan},
  {See}, \& {Tayar}}]{Hall2021a}
{Hall}, O.~J., {Davies}, G.~R., {van Saders}, J., {et~al.} 2021{\natexlab{a}},
  Nature Astronomy, 5, 707

\bibitem[{{Hall} {et~al.}(2021{\natexlab{b}}){Hall}, {Davies}, {van Saders},
  {Nielsen}, {Lund}, {Chaplin}, {Garc{\'\i}a}, {Amard}, {Breimann}, {Khan},
  {See}, \& {Tayar}}]{Hall2021}
---. 2021{\natexlab{b}}, Nature Astronomy, 5, 707

\bibitem[{{Han} {et~al.}(2020){Han}, {Ge}, {Chen}, \& {Chen}}]{Han2020}
{Han}, Z.-W., {Ge}, H.-W., {Chen}, X.-F., \& {Chen}, H.-L. 2020, Research in
  Astronomy and Astrophysics, 20, 161

\bibitem[{{Handler} {et~al.}(2019){Handler}, {Pigulski},
  {Daszy{\'n}ska-Daszkiewicz}, {Irrgang}, {Kilkenny}, {Guo}, {Przybilla},
  {Kahraman Ali{\c c}avu{\c s}}, {Kallinger}, {Pascual-Granado}, {Niemczura},
  {R{\'o}{\.z}a{\'n}ski}, {Chowdhury}, {Buzasi}, {Mirouh}, {Bowman},
  {Johnston}, {Pedersen}, {Sim{\'o}n-D{\'{\i}}az}, {Moravveji}, {Gazeas}, {De
  Cat}, {Vanderspek}, \& {Ricker}}]{Handler2019a}
{Handler}, G., {Pigulski}, A., {Daszy{\'n}ska-Daszkiewicz}, J., {et~al.} 2019,
  \apjl, 873, L4

\bibitem[{{Handler} {et~al.}(2020){Handler}, {Kurtz}, {Rappaport}, {Saio},
  {Fuller}, {Jones}, {Guo}, {Chowdhury}, {Sowicka},
  {Ali{\c{c}}avu{\textcommabelow s}}, {Streamer}, {Murphy}, {Gagliano},
  {Jacobs}, \& {Vanderburg}}]{Handler2020a}
{Handler}, G., {Kurtz}, D.~W., {Rappaport}, S.~A., {et~al.} 2020, Nature
  Astronomy, 4, 684

\bibitem[{Hansen(2017)}]{Hansen2017}
Hansen, B.~M. 2017, Monthly Notices of the Royal Astronomical Society, 467,
  1531

\bibitem[{{Hawker} \& {Parry}(2019)}]{2019MNRAS.484.4855H}
{Hawker}, G.~A., \& {Parry}, I.~R. 2019, \mnras, 484, 4855

\bibitem[{{Hayashi}(1981)}]{Hayashi1981}
{Hayashi}, C. 1981, in Fundamental Problems in the Theory of Stellar Evolution,
  ed. D.~{Sugimoto}, D.~Q. {Lamb}, \& D.~N. {Schramm}, Vol.~93, 113--126

\bibitem[{{Hayden} {et~al.}(2015){Hayden}, {Bovy}, {Holtzman}, {Nidever},
  {Bird}, {Weinberg}, {Andrews}, {Majewski}, {Allende Prieto}, {Anders},
  {Beers}, {Bizyaev}, {Chiappini}, {Cunha}, {Frinchaboy},
  {Garc{\'\i}a-Her{\'n}andez}, {Garc{\'\i}a P{\'e}rez}, {Girardi}, {Harding},
  {Hearty}, {Johnson}, {M{\'e}sz{\'a}ros}, {Minchev}, {O'Connell}, {Pan},
  {Robin}, {Schiavon}, {Schneider}, {Schultheis}, {Shetrone}, {Skrutskie},
  {Steinmetz}, {Smith}, {Wilson}, {Zamora}, \& {Zasowski}}]{Hayden2015}
{Hayden}, M.~R., {Bovy}, J., {Holtzman}, J.~A., {et~al.} 2015, \apj, 808, 132

\bibitem[{{Haywood} {et~al.}(2013){Haywood}, {Di Matteo}, {Lehnert}, {Katz}, \&
  {G{\'o}mez}}]{Haywood2013}
{Haywood}, M., {Di Matteo}, P., {Lehnert}, M.~D., {Katz}, D., \& {G{\'o}mez},
  A. 2013, \aap, 560, A109

\bibitem[{He {et~al.}(2020)He, Ford, Ragozzine, \& Carrera}]{He2020}
He, M.~Y., Ford, E.~B., Ragozzine, D., \& Carrera, D. 2020, The Astronomical
  Journal, 160, 276

\bibitem[{{Hekker} \& {Christensen-Dalsgaard}(2017)}]{hekker2017a}
{Hekker}, S., \& {Christensen-Dalsgaard}, J. 2017, \aapr, 25, 1

\bibitem[{{Helmi}(2008)}]{Helmi2008}
{Helmi}, A. 2008, \aapr, 15, 145

\bibitem[{{Helmi}(2020)}]{Helmi2020}
---. 2020, \araa, 58, 205

\bibitem[{{Helmi} {et~al.}(2018){Helmi}, {Babusiaux}, {Koppelman}, {Massari},
  {Veljanoski}, \& {Brown}}]{Helmi2018}
{Helmi}, A., {Babusiaux}, C., {Koppelman}, H.~H., {et~al.} 2018, \nat, 563, 85

\bibitem[{{Henderson} {et~al.}(2014){Henderson}, {Gaudi}, {Han}, {Skowron},
  {Penny}, {Nataf}, \& {Gould}}]{Henderson2014}
{Henderson}, C.~B., {Gaudi}, B.~S., {Han}, C., {et~al.} 2014, \apj, 794, 52

\bibitem[{{Henry} \& {Worthey}(1999)}]{Henry1999}
{Henry}, R.~B.~C., \& {Worthey}, G. 1999, \pasp, 111, 919

\bibitem[{{Herman} {et~al.}(2019){Herman}, {Zhu}, \& {Wu}}]{Herman2019}
{Herman}, M.~K., {Zhu}, W., \& {Wu}, Y. 2019, \aj, 157, 248

\bibitem[{{Hermes} {et~al.}(2012){Hermes}, {Montgomery}, {Winget}, {Brown},
  {Kilic}, \& {Kenyon}}]{2012ApJ...750L..28H}
{Hermes}, J.~J., {Montgomery}, M.~H., {Winget}, D.~E., {et~al.} 2012, \apjl,
  750, L28

\bibitem[{{Ho} {et~al.}(2017){Ho}, {Rix}, {Ness}, {Hogg}, {Liu}, \&
  {Ting}}]{Ho2017}
{Ho}, A. Y.~Q., {Rix}, H.-W., {Ness}, M.~K., {et~al.} 2017, \apj, 841, 40

\bibitem[{{Holczer} {et~al.}(2016){Holczer}, {Mazeh}, {Nachmani},
  {Jontof-Hutter}, {Ford}, {Fabrycky}, {Ragozzine}, {Kane}, \&
  {Steffen}}]{Hol2016}
{Holczer}, T., {Mazeh}, T., {Nachmani}, G., {et~al.} 2016, \apjs, 225, 9

\bibitem[{{Holman} \& {Murray}(2005)}]{HM05}
{Holman}, M.~J., \& {Murray}, N.~W. 2005, Science, 307, 1288

\bibitem[{{Holman} \& {Wiegert}(1999)}]{Holman:1999}
{Holman}, M.~J., \& {Wiegert}, P.~A. 1999, \aj, 117, 621

\bibitem[{{Holtzman} {et~al.}(1998){Holtzman}, {Watson}, {Baum}, {Grillmair},
  {Groth}, {Light}, {Lynds}, \& {O'Neil}}]{HSTCMD}
{Holtzman}, J.~A., {Watson}, A.~M., {Baum}, W.~A., {et~al.} 1998, \aj, 115,
  1946

\bibitem[{{Horch} {et~al.}(2012){Horch}, {Howell}, {Everett}, \&
  {Ciardi}}]{Horch2012AJ....144..165H}
{Horch}, E.~P., {Howell}, S.~B., {Everett}, M.~E., \& {Ciardi}, D.~R. 2012,
  \aj, 144, 165

\bibitem[{{Horner} \& {Jones}(2010)}]{horner10}
{Horner}, J., \& {Jones}, B.~W. 2010, International Journal of Astrobiology, 9,
  273

\bibitem[{{Hosseinzadeh} {et~al.}(2017){Hosseinzadeh}, {Sand}, {Valenti},
  {Brown}, {Howell}, {McCully}, {Kasen}, {Arcavi}, {Bostroem}, {Tartaglia},
  {Hsiao}, {Davis}, {Shahbandeh}, \& {Stritzinger}}]{2017ApJ...845L..11H}
{Hosseinzadeh}, G., {Sand}, D.~J., {Valenti}, S., {et~al.} 2017, \apjl, 845,
  L11

\bibitem[{{Hou} {et~al.}(2000){Hou}, {Prantzos}, \& {Boissier}}]{Hou2000}
{Hou}, J.~L., {Prantzos}, N., \& {Boissier}, S. 2000, \aap, 362, 921

\bibitem[{{Howard} {et~al.}(2012){Howard}, {Marcy}, {Bryson}, {Jenkins},
  {Rowe}, {Batalha}, {Borucki}, {Koch}, {Dunham}, {Gautier}, {Van Cleve},
  {Cochran}, {Latham}, {Lissauer}, {Torres}, {Brown}, {Gilliland}, {Buchhave},
  {Caldwell}, {Christensen-Dalsgaard}, {Ciardi}, {Fressin}, {Haas}, {Howell},
  {Kjeldsen}, {Seager}, {Rogers}, {Sasselov}, {Steffen}, {Basri},
  {Charbonneau}, {Christiansen}, {Clarke}, {Dupree}, {Fabrycky}, {Fischer},
  {Ford}, {Fortney}, {Tarter}, {Girouard}, {Holman}, {Johnson}, {Klaus},
  {Machalek}, {Moorhead}, {Morehead}, {Ragozzine}, {Tenenbaum}, {Twicken},
  {Quinn}, {Isaacson}, {Shporer}, {Lucas}, {Walkowicz}, {Welsh}, {Boss},
  {Devore}, {Gould}, {Smith}, {Morris}, {Prsa}, {Morton}, {Still}, {Thompson},
  {Mullally}, {Endl}, \& {MacQueen}}]{Howard2012}
{Howard}, A.~W., {Marcy}, G.~W., {Bryson}, S.~T., {et~al.} 2012, \apjs, 201, 15

\bibitem[{{Howell}(2020)}]{Kepler2020nkm..book.....H}
{Howell}, S.~B. 2020, {The NASA Kepler Mission}

\bibitem[{{Howell} {et~al.}(2021{\natexlab{a}}){Howell}, {Matson}, {Ciardi},
  {Everett}, {Livingston}, {Scott}, {Horch}, \&
  {Winn}}]{Howell2021AJ....161..164H}
{Howell}, S.~B., {Matson}, R.~A., {Ciardi}, D.~R., {et~al.} 2021{\natexlab{a}},
  \aj, 161, 164

\bibitem[{{Howell} {et~al.}(2021{\natexlab{b}}){Howell}, {Scott}, {Matson},
  {Everett}, {Furlan}, {Gnilka}, {Ciardi}, \&
  {Lester}}]{Howell2021FrASS...8...10H}
{Howell}, S.~B., {Scott}, N.~J., {Matson}, R.~A., {et~al.} 2021{\natexlab{b}},
  Frontiers in Astronomy and Space Sciences, 8, 10

\bibitem[{{Howell} {et~al.}(2014){Howell}, {Sobeck}, {Haas}, {Still},
  {Barclay}, {Mullally}, {Troeltzsch}, {Aigrain}, {Bryson}, {Caldwell},
  {Chaplin}, {Cochran}, {Huber}, {Marcy}, {Miglio}, {Najita}, {Smith},
  {Twicken}, \& {Fortney}}]{K22014PASP..126..398H}
{Howell}, S.~B., {Sobeck}, C., {Haas}, M., {et~al.} 2014, \pasp, 126, 398

\bibitem[{Hsieh {et~al.}(2010)Hsieh, Fitzsimmons, Joshi, Christian, \&
  Pollacco}]{hsieh2010superwasp}
Hsieh, H.~H., Fitzsimmons, A., Joshi, Y., Christian, D., \& Pollacco, D.~L.
  2010, Monthly Notices of the Royal Astronomical Society, 407, 1784

\bibitem[{{Hsu} {et~al.}(2018){Hsu}, {Ford}, {Ragozzine}, \&
  {Morehead}}]{2018AJ....155..205H}
{Hsu}, D.~C., {Ford}, E.~B., {Ragozzine}, D., \& {Morehead}, R.~C. 2018, \aj,
  155, 205

\bibitem[{{Huang} {et~al.}(2020){Huang}, {Sch{\"o}nrich}, {Zhang}, {Wu},
  {Chen}, {Wang}, {Xiang}, {Wang}, {Yuan}, {Li}, {Sun}, {Li}, \&
  {Liu}}]{Huang2020}
{Huang}, Y., {Sch{\"o}nrich}, R., {Zhang}, H., {et~al.} 2020, \apjs, 249, 29

\bibitem[{{Huang} {et~al.}(2022){Huang}, {Beers}, {Wolf}, {Lee}, {Onken},
  {Yuan}, {Shank}, {Zhang}, {Wang}, {Shi}, \& {Fan}}]{Huang2022}
{Huang}, Y., {Beers}, T.~C., {Wolf}, C., {et~al.} 2022, \apj, 925, 164

\bibitem[{{Huber}(2018)}]{Huber2018a}
{Huber}, D. 2018, in Astrophysics and Space Science Proceedings, Vol.~49,
  Asteroseismology and Exoplanets: Listening to the Stars and Searching for New
  Worlds, ed. T.~L. {Campante}, N.~C. {Santos}, \& M.~J.~P.~F.~G. {Monteiro},
  119

\bibitem[{{Huber} {et~al.}(2011){Huber}, {Bedding}, {Stello}, {Hekker},
  {Mathur}, {Mosser}, {Verner}, {Bonanno}, {Buzasi}, {Campante}, {Elsworth},
  {Hale}, {Kallinger}, {Silva Aguirre}, {Chaplin}, {De Ridder}, {Garc{\'\i}a},
  {Appourchaux}, {Frandsen}, {Houdek}, {Molenda-{\.Z}akowicz}, {Monteiro},
  {Christensen-Dalsgaard}, {Gilliland}, {Kawaler}, {Kjeldsen}, {Broomhall},
  {Corsaro}, {Salabert}, {Sanderfer}, {Seader}, \& {Smith}}]{huber2011a}
{Huber}, D., {Bedding}, T.~R., {Stello}, D., {et~al.} 2011, \apj, 743, 143

\bibitem[{{Huber} {et~al.}(2013{\natexlab{a}}){Huber}, {Chaplin},
  {Christensen-Dalsgaard}, {Gilliland}, {Kjeldsen}, {Buchhave}, {Fischer},
  {Lissauer}, {Rowe}, {Sanchis-Ojeda}, {Basu}, {Handberg}, {Hekker}, {Howard},
  {Isaacson}, {Karoff}, {Latham}, {Lund}, {Lundkvist}, {Marcy}, {Miglio},
  {Silva Aguirre}, {Stello}, {Arentoft}, {Barclay}, {Bedding}, {Burke},
  {Christiansen}, {Elsworth}, {Haas}, {Kawaler}, {Metcalfe}, {Mullally}, \&
  {Thompson}}]{Huber2013b}
{Huber}, D., {Chaplin}, W.~J., {Christensen-Dalsgaard}, J., {et~al.}
  2013{\natexlab{a}}, \apj, 767, 127

\bibitem[{{Huber} {et~al.}(2013{\natexlab{b}}){Huber}, {Carter}, {Barbieri},
  {Miglio}, {Deck}, {Fabrycky}, {Montet}, {Buchhave}, {Chaplin}, {Hekker},
  {Montalb{\'a}n}, {Sanchis-Ojeda}, {Basu}, {Bedding}, {Campante},
  {Christensen-Dalsgaard}, {Elsworth}, {Stello}, {Arentoft}, {Ford},
  {Gilliland}, {Handberg}, {Howard}, {Isaacson}, {Johnson}, {Karoff},
  {Kawaler}, {Kjeldsen}, {Latham}, {Lund}, {Lundkvist}, {Marcy}, {Metcalfe},
  {Silva Aguirre}, \& {Winn}}]{huber2013}
{Huber}, D., {Carter}, J.~A., {Barbieri}, M., {et~al.} 2013{\natexlab{b}},
  Science, 342, 331

\bibitem[{{Hwang} {et~al.}(2022){Hwang}, {Zang}, {Gould}, {Udalski}, {Bond},
  {Yang}, {Mao}, {Mao}, {Albrow}, {Chung}, {Han}, {Kil Jung}, {Ryu}, {Shin},
  {Shvartzvald}, {Yee}, {Cha}, {Kim}, {Kim}, {Kim}, {Lee}, {Lee}, {Lee},
  {Park}, {Pogge}, {Pogge}, {Mr{\'o}z}, {Poleski}, {Skowron}, {Szyma{\'n}ski},
  {Soszy{\'n}ski}, {Pietrukowicz}, {Koz{\l}owski}, {Ulaczyk}, {Rybicki},
  {Iwanek}, {Wrona}, {Gromadzki}, {Gromadzki}, {Abe}, {Barry}, {Bennett},
  {Bhattacharya}, {Fujii}, {Fukui}, {Hirao}, {Itow}, {Kirikawa}, {Kondo},
  {Koshimoto}, {Munford}, {Matsubara}, {Miyazaki}, {Muraki}, {Olmschenk},
  {Ranc}, {Rattenbury}, {Satoh}, {Shoji}, {Ishitani Silva}, {Sumi}, {Suzuki},
  {Tristram}, {Yonehara}, {Yonehara}, {Zhang}, {Zhu}, {Penny}, {Fouqu{\'e}}, \&
  {Fouqu{\'e}}}]{KB190253}
{Hwang}, K.-H., {Zang}, W., {Gould}, A., {et~al.} 2022, \aj, 163, 43

\bibitem[{{Ida} \& {Lin}(2004)}]{Ida2004}
{Ida}, S., \& {Lin}, D.~N.~C. 2004, \apj, 604, 388

\bibitem[{{Ida} \& {Lin}(2005{\natexlab{a}})}]{Ida2005}
---. 2005{\natexlab{a}}, \apj, 626, 1045

\bibitem[{{Ida} \& {Lin}(2005{\natexlab{b}})}]{IdaLin:2005}
---. 2005{\natexlab{b}}, \apj, 626, 1045

\bibitem[{{Ida} {et~al.}(2013){Ida}, {Lin}, \& {Nagasawa}}]{Ida2013}
{Ida}, S., {Lin}, D.~N.~C., \& {Nagasawa}, M. 2013, \apj, 775, 42

\bibitem[{{Ivezi{\'c}} {et~al.}(2019){Ivezi{\'c}}, {Kahn}, {Tyson}, {Abel},
  {Acosta}, {Allsman}, {Alonso}, {AlSayyad}, {Anderson}, {Andrew}, \&
  et~al.}]{2019ApJ...873..111I}
{Ivezi{\'c}}, {\v{Z}}., {Kahn}, S.~M., {Tyson}, J.~A., {et~al.} 2019, \apj,
  873, 111

\bibitem[{Izidoro {et~al.}(2017)Izidoro, Ogihara, Raymond, Morbidelli, Pierens,
  Bitsch, Cossou, \& Hersant}]{Izidoro2017}
Izidoro, A., Ogihara, M., Raymond, S.~N., {et~al.} 2017, Monthly Notices of the
  Royal Astronomical Society, 470, 1750

\bibitem[{{Jayasinghe} {et~al.}(2021){Jayasinghe}, {Stanek}, {Thompson},
  {Kochanek}, {Rowan}, {Vallely}, {Strassmeier}, {Weber}, {Hinkle}, {Hambsch},
  {Martin}, {Prieto}, {Pessi}, {Huber}, {Auchettl}, {Lopez}, {Ilyin},
  {Badenes}, {Howard}, {Isaacson}, \& {Murphy}}]{Jayasinghe21}
{Jayasinghe}, T., {Stanek}, K.~Z., {Thompson}, T.~A., {et~al.} 2021, \mnras,
  504, 2577

\bibitem[{{Jee} {et~al.}(2007){Jee}, {Blakeslee}, {Sirianni}, {Martel},
  {White}, \& {Ford}}]{2007PASPJee}
{Jee}, M.~J., {Blakeslee}, J.~P., {Sirianni}, M., {et~al.} 2007, \pasp, 119,
  1403

\bibitem[{{Jenkins} {et~al.}(2010){Jenkins}, {Caldwell}, {Chandrasekaran},
  {Twicken}, {Bryson}, {Quintana}, {Clarke}, {Li}, {Allen}, {Tenenbaum}, {Wu},
  {Klaus}, {Van Cleve}, {Dotson}, {Haas}, {Gilliland}, {Koch}, \&
  {Borucki}}]{Jenkins2010APJ}
{Jenkins}, J.~M., {Caldwell}, D.~A., {Chandrasekaran}, H., {et~al.} 2010,
  \apjl, 713, L120

\bibitem[{Jewitt(2012)}]{jewitt2012active}
Jewitt, D. 2012, The Astronomical Journal, 143, 66

\bibitem[{{Jia} {et~al.}(2017){Jia}, {Sun}, {Wang}, {Cai}, \&
  {Liu}}]{2017MNRASJia}
{Jia}, P., {Sun}, R., {Wang}, W., {Cai}, D., \& {Liu}, H. 2017, \mnras, 470,
  1950

\bibitem[{{Jia} {et~al.}(2021){Jia}, {Sun}, \& {Liu}}]{2021Jia}
{Jia}, P., {Sun}, Y., \& {Liu}, Q. 2021, arXiv e-prints, arXiv:2106.14349

\bibitem[{{Jiang} {et~al.}(2021){Jiang}, {Maeda}, {Kawabata}, {Doi},
  {Shigeyama}, {Tanaka}, {Tominaga}, {Nomoto}, {Niino}, {Sako}, {Ohsawa},
  {Schramm}, {Yamanaka}, {Kobayashi}, {Takahashi}, {Nakaoka}, {Kawabata},
  {Isogai}, {Aoki}, {Kondo}, {Mori}, {Arimatsu}, {Kasuga}, {Okumura},
  {Urakawa}, {Reichart}, {Taguchi}, {Arima}, {Beniyama}, {Uno}, \&
  {Hamada}}]{2021ApJ...923L...8J}
{Jiang}, J.-a., {Maeda}, K., {Kawabata}, M., {et~al.} 2021, \apjl, 923, L8

\bibitem[{{Johnson} {et~al.}(2010){Johnson}, {Aller}, {Howard}, \&
  {Crepp}}]{Johnson:2010}
{Johnson}, J.~A., {Aller}, K.~M., {Howard}, A.~W., \& {Crepp}, J.~R. 2010,
  \pasp, 122, 905

\bibitem[{{Juri{\'c}} \& {Tremaine}(2008)}]{Juric:2008}
{Juri{\'c}}, M., \& {Tremaine}, S. 2008, \apj, 686, 603

\bibitem[{{Kaib} {et~al.}(2013){Kaib}, {Raymond}, \& {Duncan}}]{Kaib2013}
{Kaib}, N.~A., {Raymond}, S.~N., \& {Duncan}, M. 2013, \nat, 493, 381

\bibitem[{{Kalup} {et~al.}(2021){Kalup}, {Moln{\'a}r}, {Kiss}, {Szab{\'o}},
  {P{\'a}l}, {Szak{\'a}ts}, {S{\'a}rneczky}, {Vink{\'o}}, {Szab{\'o}},
  {Kecskem{\'e}thy}, \& {Kiss}}]{Kalup2021trojans}
{Kalup}, C.~E., {Moln{\'a}r}, L., {Kiss}, C., {et~al.} 2021, \apjs, 254, 7

\bibitem[{{Kann} {et~al.}(2010){Kann}, {Klose}, {Zhang}, {Malesani}, {Nakar},
  {Pozanenko}, {Wilson}, {Butler}, {Jakobsson}, {Schulze}, {Andreev},
  {Antonelli}, {Bikmaev}, {Biryukov}, {B{\"o}ttcher}, {Burenin}, {Castro
  Cer{\'o}n}, {Castro-Tirado}, {Chincarini}, {Cobb}, {Covino}, {D'Avanzo},
  {D'Elia}, {Della Valle}, {de Ugarte Postigo}, {Efimov}, {Ferrero}, {Fugazza},
  {Fynbo}, {G{\r{a}}lfalk}, {Grundahl}, {Gorosabel}, {Gupta}, {Guziy},
  {Hafizov}, {Hjorth}, {Holhjem}, {Ibrahimov}, {Im}, {Israel}, {Je{\'l}inek},
  {Jensen}, {Karimov}, {Khamitov}, {Kizilo{\v{g}}lu}, {Klunko}, {Kub{\'a}nek},
  {Kutyrev}, {Laursen}, {Levan}, {Mannucci}, {Martin}, {Mescheryakov},
  {Mirabal}, {Norris}, {Ovaldsen}, {Paraficz}, {Pavlenko}, {Piranomonte},
  {Rossi}, {Rumyantsev}, {Salinas}, {Sergeev}, {Sharapov}, {Sollerman},
  {Stecklum}, {Stella}, {Tagliaferri}, {Tanvir}, {Telting}, {Testa}, {Updike},
  {Volnova}, {Watson}, {Wiersema}, \& {Xu}}]{2010ApJ...720.1513K}
{Kann}, D.~A., {Klose}, S., {Zhang}, B., {et~al.} 2010, \apj, 720, 1513

\bibitem[{Karoff {et~al.}(2018)Karoff, Metcalfe, Santos, Montet, Isaacson,
  Witzke, Shapiro, Mathur, Davies, Lund, Garcia, Brun, Salabert, Avelino, van
  Saders, Egeland, Cunha, Campante, Chaplin, Krivova, Solanki, Stritzinger, \&
  Knudsen}]{Karoff2018_HD173701}
Karoff, C., Metcalfe, T.~S., Santos, A. R.~G., {et~al.} 2018, ApJ, 852, 46.
\newblock \url{http://adsabs.harvard.edu/abs/2018ApJ...852...46K}

\bibitem[{{Kasdin} {et~al.}(2020){Kasdin}, {Bailey}, {Mennesson}, {Zellem},
  {Ygouf}, {Rhodes}, {Luchik}, {Zhao}, {Riggs}, {Seo}, {Krist}, {Kern}, {Tang},
  {Nemati}, {Groff}, {Zimmerman}, {Macintosh}, {Turnbull}, {Debes}, {Douglas},
  \& {Lupu}}]{kasdin2020}
{Kasdin}, N.~J., {Bailey}, V.~P., {Mennesson}, B., {et~al.} 2020, in Society of
  Photo-Optical Instrumentation Engineers (SPIE) Conference Series, Vol. 11443,
  Society of Photo-Optical Instrumentation Engineers (SPIE) Conference Series,
  114431U

\bibitem[{{Kasen}(2010)}]{2010ApJ...708.1025K}
{Kasen}, D. 2010, \apj, 708, 1025

\bibitem[{{Kasting}(1988)}]{Kasting1988}
{Kasting}, J.~F. 1988, \icarus, 74, 472

\bibitem[{{Kasting} {et~al.}(1993){Kasting}, {Whitmire}, \&
  {Reynolds}}]{Kasting1993}
{Kasting}, J.~F., {Whitmire}, D.~P., \& {Reynolds}, R.~T. 1993, \icarus, 101,
  108

\bibitem[{{Kawahara} \& {Masuda}(2019)}]{Kawahara2019}
{Kawahara}, H., \& {Masuda}, K. 2019, \aj, 157, 218

\bibitem[{{Kawahara} {et~al.}(2018){Kawahara}, {Masuda}, {MacLeod}, {Latham},
  {Bieryla}, \& {Benomar}}]{Kawahara2018}
{Kawahara}, H., {Masuda}, K., {MacLeod}, M., {et~al.} 2018, \aj, 155, 144

\bibitem[{{Kennedy} {et~al.}(2012){Kennedy}, {Wyatt}, {Sibthorpe},
  {Duch{\^e}ne}, {Kalas}, {Matthews}, {Greaves}, {Su}, \&
  {Fitzgerald}}]{Kennedy2012}
{Kennedy}, G.~M., {Wyatt}, M.~C., {Sibthorpe}, B., {et~al.} 2012, \mnras, 421,
  2264

\bibitem[{{Kennedy} {et~al.}(2019){Kennedy}, {Matr{\`a}}, {Facchini}, {Milli},
  {Pani{\'c}}, {Price}, {Wilner}, {Wyatt}, \& {Yelverton}}]{Kennedy2019}
{Kennedy}, G.~M., {Matr{\`a}}, L., {Facchini}, S., {et~al.} 2019, Nature
  Astronomy, 3, 230

\bibitem[{{Kepler} {et~al.}(2021){Kepler}, {Winget}, {Vanderbosch},
  {Castanheira}, {Hermes}, {Bell}, {Mullally}, {Romero}, {Montgomery},
  {DeGennaro}, {Winget}, {Chandler}, {Jeffery}, {Fritzen}, {Williams}, {Chote},
  \& {Zola}}]{2021ApJ...906....7K}
{Kepler}, S.~O., {Winget}, D.~E., {Vanderbosch}, Z.~P., {et~al.} 2021, \apj,
  906, 7

\bibitem[{{Kervella} {et~al.}(2019){Kervella}, {Arenou}, {Mignard}, \&
  {Th{\'e}venin}}]{kervella19}
{Kervella}, P., {Arenou}, F., {Mignard}, F., \& {Th{\'e}venin}, F. 2019, \aap,
  623, A72

\bibitem[{{Kim} {et~al.}(2016){Kim}, {Lee}, {Park}, {Kim}, {Cha}, {Lee}, {Han},
  {Chun}, \& {Yuk}}]{KMT2016}
{Kim}, S.-L., {Lee}, C.-U., {Park}, B.-G., {et~al.} 2016, Journal of Korean
  Astronomical Society, 49, 37

\bibitem[{{Kipping} {et~al.}(2022){Kipping}, {Bryson}, {Burke}, {Christiansen},
  {Hardegree-Ullman}, {Quarles}, {Hansen}, {Szul{\'a}gyi}, \&
  {Teachey}}]{Kipping2022}
{Kipping}, D., {Bryson}, S., {Burke}, C., {et~al.} 2022, Nature Astronomy, 6,
  367

\bibitem[{{Kipping}(2009)}]{Kipping2009}
{Kipping}, D.~M. 2009, \mnras, 392, 181

\bibitem[{{Kirk} {et~al.}(2016){Kirk}, {Conroy}, {Pr{\v{s}}a}, {Abdul-Masih},
  {Kochoska}, {Matijevi{\v{c}}}, {Hambleton}, {Barclay}, {Bloemen}, {Boyajian},
  {Doyle}, {Fulton}, {Hoekstra}, {Jek}, {Kane}, {Kostov}, {Latham}, {Mazeh},
  {Orosz}, {Pepper}, {Quarles}, {Ragozzine}, {Shporer}, {Southworth},
  {Stassun}, {Thompson}, {Welsh}, {Agol}, {Derekas}, {Devor}, {Fischer},
  {Green}, {Gropp}, {Jacobs}, {Johnston}, {LaCourse}, {Saetre}, {Schwengeler},
  {Toczyski}, {Werner}, {Garrett}, {Gore}, {Martinez}, {Spitzer}, {Stevick},
  {Thomadis}, {Vrijmoet}, {Yenawine}, {Batalha}, \& {Borucki}}]{Kirk2016}
{Kirk}, B., {Conroy}, K., {Pr{\v{s}}a}, A., {et~al.} 2016, \aj, 151, 68

\bibitem[{{Kley} \& {Haghighipour}(2014)}]{KH14}
{Kley}, W., \& {Haghighipour}, N. 2014, \aap, 564, A72

\bibitem[{{Koester} {et~al.}(2014){Koester}, {G{\"a}nsicke}, \&
  {Farihi}}]{2014A&A...566A..34K}
{Koester}, D., {G{\"a}nsicke}, B.~T., \& {Farihi}, J. 2014, \aap, 566, A34

\bibitem[{{Kollmeier} {et~al.}(2017){Kollmeier}, {Zasowski}, {Rix}, {Johns},
  {Anderson}, {Drory}, {Johnson}, {Pogge}, {Bird}, {Blanc}, {Brownstein},
  {Crane}, {De Lee}, {Klaene}, {Kreckel}, {MacDonald}, {Merloni}, {Ness},
  {O'Brien}, {Sanchez-Gallego}, {Sayres}, {Shen}, {Thakar}, {Tkachenko},
  {Aerts}, {Blanton}, {Eisenstein}, {Holtzman}, {Maoz}, {Nandra}, {Rockosi},
  {Weinberg}, {Bovy}, {Casey}, {Chaname}, {Clerc}, {Conroy}, {Eracleous},
  {G{\"a}nsicke}, {Hekker}, {Horne}, {Kauffmann}, {McQuinn}, {Pellegrini},
  {Schinnerer}, {Schlafly}, {Schwope}, {Seibert}, {Teske}, \& {van
  Saders}}]{Kollmeier2017}
{Kollmeier}, J.~A., {Zasowski}, G., {Rix}, H.-W., {et~al.} 2017, arXiv
  e-prints, arXiv:1711.03234

\bibitem[{{Kopparapu} {et~al.}(2013){Kopparapu}, {Ramirez}, {Kasting}, {Eymet},
  {Robinson}, {Mahadevan}, {Terrien}, {Domagal-Goldman}, {Meadows}, \&
  {Deshpande}}]{kopparapu2013}
{Kopparapu}, R.~K., {Ramirez}, R., {Kasting}, J.~F., {et~al.} 2013, \apj, 765,
  131

\bibitem[{{Koppelman} {et~al.}(2018){Koppelman}, {Helmi}, \&
  {Veljanoski}}]{Koppelman2018}
{Koppelman}, H., {Helmi}, A., \& {Veljanoski}, J. 2018, \apjl, 860, L11

\bibitem[{{Kordopatis} {et~al.}(2015){Kordopatis}, {Binney}, {Gilmore}, {Wyse},
  {Belokurov}, {McMillan}, {Hatfield}, {Grebel}, {Steinmetz}, {Navarro},
  {Seabroke}, {Minchev}, {Chiappini}, {Bienaym{\'e}}, {Bland-Hawthorn},
  {Freeman}, {Gibson}, {Helmi}, {Munari}, {Parker}, {Reid}, {Siebert},
  {Siviero}, \& {Zwitter}}]{Kordopatis2015}
{Kordopatis}, G., {Binney}, J., {Gilmore}, G., {et~al.} 2015, \mnras, 447, 3526

\bibitem[{{Koshimoto} {et~al.}(2021){Koshimoto}, {Bennett}, {Suzuki}, \&
  {Bond}}]{Naoki2021}
{Koshimoto}, N., {Bennett}, D.~P., {Suzuki}, D., \& {Bond}, I.~A. 2021, \apjl,
  918, L8

\bibitem[{{Kostov} {et~al.}(2013){Kostov}, {McCullough}, {Hinse}, {Tsvetanov},
  {H{\'e}brard}, {D{\'\i}az}, {Deleuil}, \& {Valenti}}]{Kostov2013}
{Kostov}, V.~B., {McCullough}, P.~R., {Hinse}, T.~C., {et~al.} 2013, \apj, 770,
  52

\bibitem[{{Kostov} {et~al.}(2014){Kostov}, {McCullough}, {Carter}, {Deleuil},
  {D{\'\i}az}, {Fabrycky}, {H{\'e}brard}, {Hinse}, {Mazeh}, {Orosz},
  {Tsvetanov}, \& {Welsh}}]{Kostov2014}
{Kostov}, V.~B., {McCullough}, P.~R., {Carter}, J.~A., {et~al.} 2014, \apj,
  784, 14

\bibitem[{{Kostov} {et~al.}(2016){Kostov}, {Orosz}, {Welsh}, {Doyle},
  {Fabrycky}, {Haghighipour}, {Quarles}, {Short}, {Cochran}, {Endl}, {Ford},
  {Gregorio}, {Hinse}, {Isaacson}, {Jenkins}, {Jensen}, {Kane}, {Kull},
  {Latham}, {Lissauer}, {Marcy}, {Mazeh}, {M{\"u}ller}, {Pepper}, {Quinn},
  {Ragozzine}, {Shporer}, {Steffen}, {Torres}, {Windmiller}, \&
  {Borucki}}]{Kostov2016}
{Kostov}, V.~B., {Orosz}, J.~A., {Welsh}, W.~F., {et~al.} 2016, \apj, 827, 86

\bibitem[{{Kostov} {et~al.}(2020){Kostov}, {Orosz}, {Feinstein}, {Welsh},
  {Cukier}, {Haghighipour}, {Quarles}, {Martin}, {Montet}, {Torres}, {Triaud},
  {Barclay}, {Boyd}, {Briceno}, {Cameron}, {Correia}, {Gilbert}, {Gill},
  {Gillon}, {Haqq-Misra}, {Hellier}, {Dressing}, {Fabrycky}, {Furesz},
  {Jenkins}, {Kane}, {Kopparapu}, {Hod{\v{z}}i{\'c}}, {Latham}, {Law},
  {Levine}, {Li}, {Lintott}, {Lissauer}, {Mann}, {Mazeh}, {Mardling}, {Maxted},
  {Eisner}, {Pepe}, {Pepper}, {Pollacco}, {Quinn}, {Quintana}, {Rowe},
  {Ricker}, {Rose}, {Seager}, {Santerne}, {S{\'e}gransan}, {Short}, {Smith},
  {Standing}, {Tokovinin}, {Trifonov}, {Turner}, {Twicken}, {Udry},
  {Vanderspek}, {Winn}, {Wolf}, {Ziegler}, {Ansorge}, {Barnet}, {Bergeron},
  {Huten}, {Pappa}, \& {van der Straeten}}]{Kostov2020}
{Kostov}, V.~B., {Orosz}, J.~A., {Feinstein}, A.~D., {et~al.} 2020, \aj, 159,
  253

\bibitem[{{Kostov} {et~al.}(2021){Kostov}, {Powell}, {Orosz}, {Welsh},
  {Cochran}, {Collins}, {Endl}, {Hellier}, {Latham}, {MacQueen}, {Pepper},
  {Quarles}, {Sairam}, {Torres}, {Wilson}, {Bergeron}, {Boyce}, {Bieryla},
  {Buchheim}, {Ben Christiansen}, {Ciardi}, {Collins}, {Conti}, {Dixon},
  {Guerra}, {Haghighipour}, {Herman}, {Hintz}, {Howard}, {Jensen}, {Kielkopf},
  {Kruse}, {Law}, {Martin}, {Maxted}, {Montet}, {Murgas}, {Nelson},
  {Olmschenk}, {Otero}, {Quimby}, {Richmond}, {Schwarz}, {Shporer}, {Stassun},
  {Stephens}, {Triaud}, {Ulowetz}, {Walter}, {Wiley}, {Wood}, {Yenawine},
  {Agol}, {Barclay}, {Beatty}, {Boisse}, {Caldwell}, {Christiansen},
  {Col{\'o}n}, {Deleuil}, {Doyle}, {Fausnaugh}, {F{\H{u}}r{\'e}sz}, {Gilbert},
  {H{\'e}brard}, {James}, {Jenkins}, {Kane}, {Kidwell}, {Kopparapu}, {Li},
  {Lissauer}, {Lund}, {Majewski}, {Mazeh}, {Quinn}, {Quintana}, {Ricker},
  {Rodriguez}, {Rowe}, {Santerne}, {Schlieder}, {Seager}, {Standing},
  {Stevens}, {Ting}, {Vanderspek}, \& {Winn}}]{Kostov2021}
{Kostov}, V.~B., {Powell}, B.~P., {Orosz}, J.~A., {et~al.} 2021, \aj, 162, 234

\bibitem[{{Kostov} {et~al.}(2022){Kostov}, {Powell}, {Rappaport}, {Borkovits},
  {Gagliano}, {Jacobs}, {Kristiansen}, {LaCourse}, {Omohundro}, {Orosz},
  {Schmitt}, {Schwengeler}, {Terentev}, {Torres}, {Barclay}, {Friedman},
  {Kruse}, {Olmschenk}, {Vanderburg}, \& {Welsh}}]{Kostov2022ApJS..259...66K}
{Kostov}, V.~B., {Powell}, B.~P., {Rappaport}, S.~A., {et~al.} 2022, \apjs,
  259, 66

\bibitem[{{Kov{\'a}cs} {et~al.}(2002){Kov{\'a}cs}, {Zucker}, \&
  {Mazeh}}]{Kovacs2002}
{Kov{\'a}cs}, G., {Zucker}, S., \& {Mazeh}, T. 2002, \aap, 391, 369

\bibitem[{{Koz{\l}owski} {et~al.}(2007){Koz{\l}owski}, {Wo{\'z}niak}, {Mao}, \&
  {Wood}}]{Kozlowski2007}
{Koz{\l}owski}, S., {Wo{\'z}niak}, P.~R., {Mao}, S., \& {Wood}, A. 2007, \apj,
  671, 420

\bibitem[{{Kraus} {et~al.}(2016){Kraus}, {Ireland}, {Huber}, {Mann}, \&
  {Dupuy}}]{Kraus:2016}
{Kraus}, A.~L., {Ireland}, M.~J., {Huber}, D., {Mann}, A.~W., \& {Dupuy}, T.~J.
  2016, \aj, 152, 8

\bibitem[{{Kremer} {et~al.}(2017){Kremer}, {Breivik}, {Larson}, \&
  {Kalogera}}]{Krem2017}
{Kremer}, K., {Breivik}, K., {Larson}, S.~L., \& {Kalogera}, V. 2017, \apj,
  846, 95

\bibitem[{{Kroupa}(2001)}]{Kroupa2001_variation}
{Kroupa}, P. 2001, \mnras, 322, 231

\bibitem[{{Kunimoto} \& {Matthews}(2020)}]{Kunimoto2020AJ}
{Kunimoto}, M., \& {Matthews}, J.~M. 2020, \aj, 159, 248

\bibitem[{{Kunimoto} {et~al.}(2022){Kunimoto}, {Daylan}, {Guerrero}, {Fong},
  {Bryson}, {Ricker}, {Fausnaugh}, {Huang}, {Sha}, {Shporer}, {Vanderburg},
  {Vanderspek}, \& {Yu}}]{Kunimoto2022}
{Kunimoto}, M., {Daylan}, T., {Guerrero}, N., {et~al.} 2022, \apjs, 259, 33

\bibitem[{{Kupfer} {et~al.}(2018){Kupfer}, {Korol}, {Shah}, {Nelemans},
  {Marsh}, {Ramsay}, {Groot}, {Steeghs}, \& {Rossi}}]{Kupfer2018}
{Kupfer}, T., {Korol}, V., {Shah}, S., {et~al.} 2018, \mnras, 480, 302

\bibitem[{{Kurtz}(2022)}]{2022arXiv220111629K}
{Kurtz}, D. 2022, arXiv e-prints, arXiv:2201.11629

\bibitem[{{Kuszlewicz} {et~al.}(2019){Kuszlewicz}, {Chaplin}, {North}, {Farr},
  {Bell}, {Davies}, {Campante}, \& {Hekker}}]{Kuszlewicz2019}
{Kuszlewicz}, J.~S., {Chaplin}, W.~J., {North}, T. S.~H., {et~al.} 2019,
  \mnras, 488, 572

\bibitem[{{Lai} \& {Pu}(2017)}]{lai17}
{Lai}, D., \& {Pu}, B. 2017, \aj, 153, 42

\bibitem[{{Lares-Martiz } {et~al.}(2020){Lares-Martiz }, {Garrido}, \&
  {Pascual-Granado}}]{Lares-Martiz2020}
{Lares-Martiz }, M., {Garrido}, R., \& {Pascual-Granado}, J. 2020, Monthly
  Notices of the Royal Astronomical Society, 498, 1194

\bibitem[{{Laughlin} {et~al.}(2004){Laughlin}, {Bodenheimer}, \&
  {Adams}}]{Laughlin2004}
{Laughlin}, G., {Bodenheimer}, P., \& {Adams}, F.~C. 2004, \apjl, 612, L73

\bibitem[{Laureijs {et~al.}(2011)Laureijs, Amiaux, Arduini, Augueres,
  Brinchmann, Cole, Cropper, Dabin, Duvet, Ealet,
  {et~al.}}]{Laureijs2011euclid}
Laureijs, R., Amiaux, J., Arduini, S., {et~al.} 2011, arXiv preprint
  arXiv:1110.3193

\bibitem[{{Lecavelier Des Etangs} {et~al.}(1999){Lecavelier Des Etangs},
  {Vidal-Madjar}, \& {Ferlet}}]{Lecavelier1999}
{Lecavelier Des Etangs}, A., {Vidal-Madjar}, A., \& {Ferlet}, R. 1999, \aap,
  343, 916

\bibitem[{{Lee} {et~al.}(2020){Lee}, {Offner}, {Hennebelle}, {Andr{\'e}},
  {Zinnecker}, {Ballesteros-Paredes}, {Inutsuka}, \& {Kruijssen}}]{Lee2020}
{Lee}, Y.-N., {Offner}, S. S.~R., {Hennebelle}, P., {et~al.} 2020, \ssr, 216,
  70

\bibitem[{{Lee} {et~al.}(2011){Lee}, {Beers}, {Allende Prieto}, {Lai},
  {Rockosi}, {Morrison}, {Johnson}, {An}, {Sivarani}, \& {Yanny}}]{Lee2011}
{Lee}, Y.~S., {Beers}, T.~C., {Allende Prieto}, C., {et~al.} 2011, \aj, 141, 90

\bibitem[{{Lei} \& {Zhang}(2011)}]{2011ApJ...740L..27L}
{Lei}, W.-H., \& {Zhang}, B. 2011, \apjl, 740, L27

\bibitem[{{Lendl} {et~al.}(2020){Lendl}, {Csizmadia}, {Deline}, {Fossati},
  {Kitzmann}, {Heng}, {Hoyer}, {Salmon}, {Benz}, {Broeg}, {Ehrenreich},
  {Fortier}, {Queloz}, {Bonfanti}, {Brandeker}, {Collier Cameron}, {Delrez},
  {Garcia Mu{\~n}oz}, {Hooton}, {Maxted}, {Morris}, {Van Grootel}, {Wilson},
  {Alibert}, {Alonso}, {Asquier}, {Bandy}, {B{\'a}rczy}, {Barrado}, {Barros},
  {Baumjohann}, {Beck}, {Beck}, {Bekkelien}, {Bergomi}, {Billot}, {Biondi},
  {Bonfils}, {Bourrier}, {Busch}, {Cabrera}, {Cessa}, {Charnoz}, {Chazelas},
  {Corral Van Damme}, {Davies}, {Deleuil}, {Demangeon}, {Demory}, {Erikson},
  {Farinato}, {Fridlund}, {Futyan}, {Gandolfi}, {Gillon}, {Guterman}, {Hasiba},
  {Hernandez}, {Isaak}, {Kiss}, {Kuntzer}, {Lecavelier des Etangs},
  {L{\"u}ftinger}, {Laskar}, {Lovis}, {Magrin}, {Malvasio}, {Marafatto},
  {Michaelis}, {Munari}, {Nascimbeni}, {Olofsson}, {Ottacher}, {Ottensamer},
  {Pagano}, {Pall{\'e}}, {Peter}, {Piazza}, {Piotto}, {Pollacco}, {Ratti},
  {Rauer}, {Ragazzoni}, {Rando}, {Ribas}, {Rieder}, {Rohlfs}, {Safa}, {Santos},
  {Scandariato}, {S{\'e}gransan}, {Simon}, {Singh}, {Smith}, {Sordet}, {Sousa},
  {Steller}, {Szab{\'o}}, {Thomas}, {Tschentscher}, {Udry}, {Viotto}, {Walter},
  {Walton}, {Wildi}, \& {Wolter}}]{Lendl2020}
{Lendl}, M., {Csizmadia}, S., {Deline}, A., {et~al.} 2020, \aap, 643, A94

\bibitem[{{Lennon} {et~al.}(2021){Lennon}, {Dufton}, {Villase{\~n}or}, {Evans},
  {Langer}, {Saxton}, {Monageng}, \& {Toonen}}]{Lennon2021}
{Lennon}, D.~J., {Dufton}, P.~L., {Villase{\~n}or}, J.~I., {et~al.} 2021, arXiv
  e-prints, arXiv:2111.12173

\bibitem[{{Lenz} {et~al.}(2008){Lenz}, {Pamyatnykh}, {Breger}, \&
  {Antoci}}]{Lenz2008}
{Lenz}, P., {Pamyatnykh}, A.~A., {Breger}, M., \& {Antoci}, V. 2008, \aap, 478,
  855

\bibitem[{{Lester} {et~al.}(2021{\natexlab{a}}){Lester}, {Matson}, {Howell},
  {Furlan}, {Gnilka}, {Scott}, {Ciardi}, {Everett}, {Hartman}, \&
  {Hirsch}}]{Lester2021AJ....162...75L}
{Lester}, K.~V., {Matson}, R.~A., {Howell}, S.~B., {et~al.} 2021{\natexlab{a}},
  \aj, 162, 75

\bibitem[{{Lester} {et~al.}(2021{\natexlab{b}}){Lester}, {Matson}, {Howell},
  {Furlan}, {Gnilka}, {Scott}, {Ciardi}, {Everett}, {Hartman}, \&
  {Hirsch}}]{Lester2021AJ....161..164H}
---. 2021{\natexlab{b}}, \aj, 162, 75

\bibitem[{{Li} {et~al.}(2016){Li}, {Holman}, \& {Tao}}]{Li16}
{Li}, G., {Holman}, M.~J., \& {Tao}, M. 2016, \apj, 831, 96

\bibitem[{{Li} {et~al.}(2020{\natexlab{a}}){Li}, {Van Reeth}, {Bedding},
  {Murphy}, {Antoci}, {Ouazzani}, \& {Barbara}}]{2020MNRAS.491.3586L}
{Li}, G., {Van Reeth}, T., {Bedding}, T.~R., {et~al.} 2020{\natexlab{a}},
  \mnras, 491, 3586

\bibitem[{{Li} {et~al.}(2021{\natexlab{a}}){Li}, {Xia}, {Kim}, {Gao}, {Hu},
  {Guo}, {Gao}, {Chen}, \& {Guo}}]{Li2021}
{Li}, K., {Xia}, Q.-Q., {Kim}, C.-H., {et~al.} 2021{\natexlab{a}}, \aj, 162, 13

\bibitem[{{Li} {et~al.}(2020{\natexlab{b}}){Li}, {Bedding},
  {Christensen-Dalsgaard}, {Stello}, {Li}, \& {Keen}}]{2020MNRAS.495.3431L}
{Li}, T., {Bedding}, T.~R., {Christensen-Dalsgaard}, J., {et~al.}
  2020{\natexlab{b}}, \mnras, 495, 3431

\bibitem[{{Li} {et~al.}(2022){Li}, {Li}, {Bi}, {Bedding}, {Davies}, \&
  {Du}}]{LiTanda2022}
{Li}, T., {Li}, Y., {Bi}, S., {et~al.} 2022, \apj, 927, 167

\bibitem[{{Li} {et~al.}(2019{\natexlab{a}}){Li}, {Wang}, {Vink{\'o}}, {Mo},
  {Hosseinzadeh}, {Sand}, {Zhang}, {Lin}, {PTSS/TNTS}, {Zhang}, {Wang},
  {Zhang}, {Chen}, {Xiang}, {Rui}, {Huang}, {Li}, {Zhang}, {Li}, {Baron},
  {Derkacy}, {Zhao}, {Sai}, {Zhang}, {Wang}, {LCO}, {Howell}, {McCully},
  {Arcavi}, {Valenti}, {Hiramatsu}, {Burke}, {KEGS}, {Rest}, {Garnavich},
  {Tucker}, {Narayan}, {Shaya}, {Margheim}, {Zenteno}, {Villar}, {UCSC},
  {Dimitriadis}, {Foley}, {Pan}, {Coulter}, {Fox}, {Jha}, {Jones}, {Kasen},
  {Kilpatrick}, {Piro}, {Riess}, {Rojas-Bravo}, {ASAS-SN}, {Shappee},
  {Holoien}, {Stanek}, {Drout}, {Auchettl}, {Kochanek}, {Brown}, {Bose},
  {Bersier}, {Brimacombe}, {Chen}, {Dong}, {Holmbo}, {Mu{\~n}oz}, {Mutel},
  {Post}, {Prieto}, {Shields}, {Tallon}, {Thompson}, {Vallely}, {Villanueva},
  {Pan-STARRS}, {Smartt}, {Smith}, {Chambers}, {Flewelling}, {Huber},
  {Magnier}, {Waters}, {Schultz}, {Bulger}, {Lowe}, {Willman}, {Konkoly/Texas},
  {S{\'a}rneczky}, {P{\'a}l}, {Wheeler}, {B{\'o}di}, {Bogn{\'a}r}, {Cs{\'a}k},
  {Cseh}, {Cs{\"o}rnyei}, {Hanyecz}, {Ign{\'a}cz}, {Kalup},
  {K{\"o}nyves-T{\'o}th}, {Kriskovics}, {Ordasi}, {Rajmon}, {S{\'o}dor},
  {Szab{\'o}}, {Szak{\'a}ts}, {Zsidi}, {Arizona}, {Milne}, {Andrews}, {Smith},
  {Bilinski}, {Swift}, {Brown}, {ePESSTO}, {Nordin}, {Williams}, {Galbany},
  {Palmerio}, {Hook}, {Inserra}, {Maguire}, {Cartier}, {Razza},
  {Guti{\'e}rrez}, {North Carolina}, {Hermes}, {Reding}, {Kaiser}, {ATLAS},
  {Tonry}, {Heinze}, {Denneau}, {Weiland}, {Stalder}, {K2 Mission Team},
  {Barentsen}, {Dotson}, {Barclay}, {Gully-Santiago}, {Hedges}, {Cody},
  {Howell}, {Kepler Spacecraft Team}, {Coughlin}, {Van Cleve}, {Cardoso},
  {Larson}, {McCalmont-Everton}, {Peterson}, {Ross}, {Reedy}, {Osborne},
  {McGinn}, {Kohnert}, {Migliorini}, {Wheaton}, {Spencer}, {Labonde},
  {Castillo}, {Beerman}, {Steward}, {Hanley}, {Larsen}, {Gangopadhyay},
  {Kloetzel}, {Weschler}, {Nystrom}, {Moffatt}, {Redick}, {Griest}, {Packard},
  {Muszynski}, {Kampmeier}, {Bjella}, {Flynn}, \&
  {Elsaesser}}]{2019ApJ...870...12L}
{Li}, W., {Wang}, X., {Vink{\'o}}, J., {et~al.} 2019{\natexlab{a}}, \apj, 870,
  12

\bibitem[{{Li} {et~al.}(2019{\natexlab{b}}){Li}, {Wang}, {Hu}, {Yang}, {Zhang},
  {Mo}, {Chen}, {Zhang}, {Benetti}, {Cappellaro}, {Elias-Rosa}, {Isern},
  {Morales-Garoffolo}, {Huang}, {Ochner}, {Pastorello}, {Reguitti},
  {Tartaglia}, {Terreran}, {Tomasella}, \& {Wang}}]{2019ApJ...882...30L}
{Li}, W., {Wang}, X., {Hu}, M., {et~al.} 2019{\natexlab{b}}, \apj, 882, 30

\bibitem[{{Li} {et~al.}(2021{\natexlab{b}}){Li}, {Chen}, {Lin}, \&
  {Zhang}}]{Li2021CG}
{Li}, Y.-P., {Chen}, Y.-X., {Lin}, D. N.~C., \& {Zhang}, X. 2021{\natexlab{b}},
  \apj, 906, 52

\bibitem[{{Li} {et~al.}(2020{\natexlab{c}}){Li}, {Li}, {Li}, {Birnstiel},
  {Dr{\k{a}}{\.z}kowska}, \& {Stammler}}]{Li2020CG}
{Li}, Y.-P., {Li}, H., {Li}, S., {et~al.} 2020{\natexlab{c}}, \apjl, 892, L19

\bibitem[{{Lian} {et~al.}(2020{\natexlab{a}}){Lian}, {Thomas}, {Maraston},
  {Zamora}, {Tayar}, {Pan}, {Tissera}, {Fern{\'a}ndez-Trincado}, \&
  {Garcia-Hernandez}}]{Lian2020a}
{Lian}, J., {Thomas}, D., {Maraston}, C., {et~al.} 2020{\natexlab{a}}, \mnras,
  494, 2561

\bibitem[{{Lian} {et~al.}(2020{\natexlab{b}}){Lian}, {Zasowski}, {Hasselquist},
  {Nataf}, {Thomas}, {Moni Bidin}, {Fern{\'a}ndez-Trincado},
  {Garcia-Hernandez}, {Lane}, {Majewski}, {Roman-Lopes}, \&
  {Schultheis}}]{Lian2020b}
{Lian}, J., {Zasowski}, G., {Hasselquist}, S., {et~al.} 2020{\natexlab{b}},
  \mnras, 497, 3557

\bibitem[{{Lin} {et~al.}(2020){Lin}, {Asplund}, {Ting}, {Casagrande}, {Buder},
  {Bland-Hawthorn}, {Casey}, {De Silva}, {D'Orazi}, {Freeman}, {Kos}, {Lind},
  {Martell}, {Sharma}, {Simpson}, {Zwitter}, {Zucker}, {Minchev},
  {{\v{C}}otar}, {Hayden}, {Horner}, {Lewis}, {Nordlander}, {Wyse}, \&
  {{\v{Z}}erjal}}]{Lin2020}
{Lin}, J., {Asplund}, M., {Ting}, Y.-S., {et~al.} 2020, \mnras, 491, 2043

\bibitem[{{Lissauer}(1987)}]{Lissauer1987}
{Lissauer}, J.~J. 1987, \icarus, 69, 249

\bibitem[{{Lithwick} \& {Wu}(2012)}]{Lithwick2012}
{Lithwick}, Y., \& {Wu}, Y. 2012, \apjl, 756, L11

\bibitem[{{Lithwick} {et~al.}(2012){Lithwick}, {Xie}, \& {Wu}}]{Lit12}
{Lithwick}, Y., {Xie}, J., \& {Wu}, Y. 2012, \apj, 761, 122

\bibitem[{{Liu} \& {Ji}(2020)}]{Liu2020}
{Liu}, B., \& {Ji}, J. 2020, Research in Astronomy and Astrophysics, 20, 164

\bibitem[{{Liu} {et~al.}(2019{\natexlab{a}}){Liu}, {Lambrechts}, {Johansen}, \&
  {Liu}}]{Liu2019}
{Liu}, B., {Lambrechts}, M., {Johansen}, A., \& {Liu}, F. 2019{\natexlab{a}},
  \aap, 632, A7

\bibitem[{Liu {et~al.}(2020)Liu, Lambrechts, Johansen, Pascucci, \&
  Henning}]{Liu2020pebble}
Liu, B., Lambrechts, M., Johansen, A., Pascucci, I., \& Henning, T. 2020,
  Astronomy \& Astrophysics, 638, A88

\bibitem[{{Liu} {et~al.}(2017){Liu}, {Ormel}, \& {Lin}}]{Liu2017}
{Liu}, B., {Ormel}, C.~W., \& {Lin}, D. N.~C. 2017, \aap, 601, A15

\bibitem[{Liu {et~al.}(2022)Liu, Raymond, \& Jacobson}]{Liu2022}
Liu, B., Raymond, S.~N., \& Jacobson, S.~A. 2022, Nature, 604, 643.
\newblock \url{https://doi.org/10.1038/s41586-022-04535-1}

\bibitem[{{Liu} {et~al.}(2016){Liu}, {Zhang}, \& {Lin}}]{Liu:2016}
{Liu}, B., {Zhang}, X., \& {Lin}, D. N.~C. 2016, \apj, 823, 162

\bibitem[{{Liu} {et~al.}(2014){Liu}, {Wang}, {Zhang}, \& {Zhou}}]{Liu14}
{Liu}, H.-G., {Wang}, Y., {Zhang}, H., \& {Zhou}, J.-L. 2014, \apj, 790, 141

\bibitem[{{Liu}(2009)}]{Liu2009}
{Liu}, J. 2009, \mnras, 400, 1850

\bibitem[{{Liu} \& {Zhang}(2014)}]{Liu2014}
{Liu}, J., \& {Zhang}, Y. 2014, \pasp, 126, 211

\bibitem[{{Liu} {et~al.}(2019{\natexlab{b}}){Liu}, {Zhang}, {Howard}, {Bai},
  {Lu}, {Soria}, {Justham}, {Li}, {Zheng}, {Wang}, {Belczynski}, {Casares},
  {Zhang}, {Yuan}, {Dong}, {Lei}, {Isaacson}, {Wang}, {Bai}, {Shao}, {Gao},
  {Wang}, {Niu}, {Cui}, {Zheng}, {Mu}, {Zhang}, {Wang}, {Heger}, {Qi}, {Liao},
  {Lattanzi}, {Gu}, {Wang}, {Wu}, {Shao}, {Shen}, {Wang}, {Bregman}, {Di
  Stefano}, {Liu}, {Han}, {Zhang}, {Wang}, {Ren}, {Zhang}, {Zhang}, {Wang},
  {Cabrera-Lavers}, {Corradi}, {Rebolo}, {Zhao}, {Zhao}, {Chu}, \&
  {Cui}}]{Liu_LB1}
{Liu}, J., {Zhang}, H., {Howard}, A.~W., {et~al.} 2019{\natexlab{b}}, \nat,
  575, 618

\bibitem[{{Lopez} \& {Rice}(2018)}]{Lopez2018}
{Lopez}, E.~D., \& {Rice}, K. 2018, \mnras, 479, 5303

\bibitem[{{Lorimer} {et~al.}(2007){Lorimer}, {Bailes}, {McLaughlin},
  {Narkevic}, \& {Crawford}}]{2007Sci...318..777L}
{Lorimer}, D.~R., {Bailes}, M., {McLaughlin}, M.~A., {Narkevic}, D.~J., \&
  {Crawford}, F. 2007, Science, 318, 777

\bibitem[{{Lu} {et~al.}(2020{\natexlab{a}}){Lu}, {Schlaufman}, \&
  {Cheng}}]{Lu:2020}
{Lu}, C.~X., {Schlaufman}, K.~C., \& {Cheng}, S. 2020{\natexlab{a}}, \aj, 160,
  253

\bibitem[{{Lu} {et~al.}(2020{\natexlab{b}}){Lu}, {Liu}, {Jiang}, \&
  {Wang}}]{Lu2020}
{Lu}, L.-N., {Liu}, J.-Z., {Jiang}, D.-K., \& {Wang}, Y.-H. 2020{\natexlab{b}},
  Research in Astronomy and Astrophysics, 20, 196

\bibitem[{{Lu} {et~al.}(2021){Lu}, {Angus}, {Curtis}, {David}, \&
  {Kiman}}]{Lu2021}
{Lu}, Y.~L., {Angus}, R., {Curtis}, J.~L., {David}, T.~J., \& {Kiman}, R. 2021,
  \aj, 161, 189

\bibitem[{{Lu} {et~al.}(2022){Lu}, {Ness}, {Buck}, \& {Carr}}]{Lu2022}
{Lu}, Y.~L., {Ness}, M.~K., {Buck}, T., \& {Carr}, C. 2022, \mnras,
  arXiv:2112.05238

\bibitem[{{Luhman} {et~al.}(2011){Luhman}, {Burgasser}, \&
  {Bochanski}}]{2011ApJ...730L...9L}
{Luhman}, K.~L., {Burgasser}, A.~J., \& {Bochanski}, J.~J. 2011, \apjl, 730, L9

\bibitem[{{Lundkvist} {et~al.}(2016){Lundkvist}, {Kjeldsen}, {Albrecht},
  {Davies}, {Basu}, {Huber}, {Justesen}, {Karoff}, {Silva Aguirre}, {van
  Eylen}, {Vang}, {Arentoft}, {Barclay}, {Bedding}, {Campante}, {Chaplin},
  {Christensen-Dalsgaard}, {Elsworth}, {Gilliland}, {Handberg}, {Hekker},
  {Kawaler}, {Lund}, {Metcalfe}, {Miglio}, {Rowe}, {Stello}, {Tingley}, \&
  {White}}]{lundkvist2016a}
{Lundkvist}, M.~S., {Kjeldsen}, H., {Albrecht}, S., {et~al.} 2016, Nature
  Communications, 7, 11201

\bibitem[{{Luo} {et~al.}(2015){Luo}, {Zhao}, {Zhao}, {Deng}, {Liu}, {Jing},
  {Wang}, {Zhang}, {Shi}, {Cui}, {Chu}, {Li}, {Bai}, {Wu}, {Cai}, {Cao}, {Cao},
  {Carlin}, {Chen}, {Chen}, {Chen}, {Chen}, {Chen}, {Chen}, {Chen},
  {Christlieb}, {Chu}, {Cui}, {Dong}, {Du}, {Fan}, {Feng}, {Fu}, {Gao}, {Gong},
  {Gu}, {Guo}, {Han}, {He}, {Hou}, {Hou}, {Hou}, {Hu}, {Hu}, {Hu}, {Huo},
  {Jia}, {Jiang}, {Jiang}, {Jiang}, {Jin}, {Kong}, {Kong}, {Lei}, {Li}, {Li},
  {Li}, {Li}, {Li}, {Li}, {Li}, {Li}, {Li}, {Li}, {Li}, {Li}, {Liang}, {Lin},
  {Liu}, {Liu}, {Liu}, {Liu}, {Lu}, {Luo}, {Mao}, {Newberg}, {Ni}, {Qi}, {Qi},
  {Shen}, {Shi}, {Song}, {Song}, {Su}, {Su}, {Tang}, {Tao}, {Tian}, {Wang},
  {Wang}, {Wang}, {Wang}, {Wang}, {Wang}, {Wang}, {Wang}, {Wang}, {Wang},
  {Wang}, {Wang}, {Wang}, {Wang}, {Wang}, {Wang}, {Wang}, {Wang}, {Wang},
  {Wang}, {Wei}, {Wei}, {Wu}, {Wu}, {Wu}, {Wu}, {Xing}, {Xu}, {Xu}, {Xu},
  {Yan}, {Yang}, {Yang}, {Yang}, {Yang}, {Yao}, {Yu}, {Yuan}, {Yuan}, {Yuan},
  {Yuan}, {Zhai}, {Zhang}, {Zhang}, {Zhang}, {Zhang}, {Zhang}, {Zhang},
  {Zhang}, {Zhang}, {Zhao}, {Zhou}, {Zhou}, {Zhu}, {Zhu}, {Zou}, \&
  {Zuo}}]{Luo2015}
{Luo}, A.~L., {Zhao}, Y.-H., {Zhao}, G., {et~al.} 2015, Research in Astronomy
  and Astrophysics, 15, 1095

\bibitem[{{Ma} {et~al.}(2016){Ma}, {Mao}, {Ida}, {Zhu}, \& {Lin}}]{MaFFP}
{Ma}, S., {Mao}, S., {Ida}, S., {Zhu}, W., \& {Lin}, D.~N.~C. 2016, \mnras,
  461, L107

\bibitem[{{Macintosh} {et~al.}(2014){Macintosh}, {Graham}, {Ingraham},
  {Konopacky}, {Marois}, {Perrin}, {Poyneer}, {Bauman}, {Barman}, {Burrows},
  {Cardwell}, {Chilcote}, {De Rosa}, {Dillon}, {Doyon}, {Dunn}, {Erikson},
  {Fitzgerald}, {Gavel}, {Goodsell}, {Hartung}, {Hibon}, {Kalas}, {Larkin},
  {Maire}, {Marchis}, {Marley}, {McBride}, {Millar-Blanchaer}, {Morzinski},
  {Norton}, {Oppenheimer}, {Palmer}, {Patience}, {Pueyo}, {Rantakyro},
  {Sadakuni}, {Saddlemyer}, {Savransky}, {Serio}, {Soummer},
  {Sivaramakrishnan}, {Song}, {Thomas}, {Wallace}, {Wiktorowicz}, \&
  {Wolff}}]{macintosh2014}
{Macintosh}, B., {Graham}, J.~R., {Ingraham}, P., {et~al.} 2014, Proceedings of
  the National Academy of Science, 111, 12661

\bibitem[{{Madhusudhan}(2019)}]{madhusudhan19}
{Madhusudhan}, N. 2019, \araa, 57, 617

\bibitem[{{Majewski} {et~al.}(2017){Majewski}, {Schiavon}, {Frinchaboy},
  {Allende Prieto}, {Barkhouser}, {Bizyaev}, {Blank}, {Brunner}, {Burton},
  {Carrera}, {Chojnowski}, {Cunha}, {Epstein}, {Fitzgerald}, {Garc{\'\i}a
  P{\'e}rez}, {Hearty}, {Henderson}, {Holtzman}, {Johnson}, {Lam}, {Lawler},
  {Maseman}, {M{\'e}sz{\'a}ros}, {Nelson}, {Nguyen}, {Nidever}, {Pinsonneault},
  {Shetrone}, {Smee}, {Smith}, {Stolberg}, {Skrutskie}, {Walker}, {Wilson},
  {Zasowski}, {Anders}, {Basu}, {Beland}, {Blanton}, {Bovy}, {Brownstein},
  {Carlberg}, {Chaplin}, {Chiappini}, {Eisenstein}, {Elsworth}, {Feuillet},
  {Fleming}, {Galbraith-Frew}, {Garc{\'\i}a}, {Garc{\'\i}a-Hern{\'a}ndez},
  {Gillespie}, {Girardi}, {Gunn}, {Hasselquist}, {Hayden}, {Hekker}, {Ivans},
  {Kinemuchi}, {Klaene}, {Mahadevan}, {Mathur}, {Mosser}, {Muna}, {Munn},
  {Nichol}, {O'Connell}, {Parejko}, {Robin}, {Rocha-Pinto}, {Schultheis},
  {Serenelli}, {Shane}, {Silva Aguirre}, {Sobeck}, {Thompson}, {Troup},
  {Weinberg}, \& {Zamora}}]{Majewski2017}
{Majewski}, S.~R., {Schiavon}, R.~P., {Frinchaboy}, P.~M., {et~al.} 2017, \aj,
  154, 94

\bibitem[{{Malmberg} {et~al.}(2011){Malmberg}, {Davies}, \&
  {Heggie}}]{Malmberg2011}
{Malmberg}, D., {Davies}, M.~B., \& {Heggie}, D.~C. 2011, \mnras, 411, 859

\bibitem[{{Mamajek} \& {Hillenbrand}(2008)}]{Mamajek2008}
{Mamajek}, E.~E., \& {Hillenbrand}, L.~A. 2008, \apj, 687, 1264

\bibitem[{{Mao} \& {Paczynski}(1991)}]{Shude1991}
{Mao}, S., \& {Paczynski}, B. 1991, \apjl, 374, L37

\bibitem[{{Maoz} \& {Graur}(2017)}]{Maoz2017}
{Maoz}, D., \& {Graur}, O. 2017, \apj, 848, 25

\bibitem[{{Maoz} {et~al.}(2014){Maoz}, {Mannucci}, \&
  {Nelemans}}]{2014ARA&A..52..107M}
{Maoz}, D., {Mannucci}, F., \& {Nelemans}, G. 2014, \araa, 52, 107

\bibitem[{{Marcy} {et~al.}(2014){Marcy}, {Isaacson}, {Howard}, {Rowe},
  {Jenkins}, {Bryson}, {Latham}, {Howell}, {Gautier}, {Batalha}, {Rogers},
  {Ciardi}, {Fischer}, {Gilliland}, {Kjeldsen}, {Christensen-Dalsgaard},
  {Huber}, {Chaplin}, {Basu}, {Buchhave}, {Quinn}, {Borucki}, {Koch}, {Hunter},
  {Caldwell}, {Van Cleve}, {Kolbl}, {Weiss}, {Petigura}, {Seager}, {Morton},
  {Johnson}, {Ballard}, {Burke}, {Cochran}, {Endl}, {MacQueen}, {Everett},
  {Lissauer}, {Ford}, {Torres}, {Fressin}, {Brown}, {Steffen}, {Charbonneau},
  {Basri}, {Sasselov}, {Winn}, {Sanchis-Ojeda}, {Christiansen}, {Adams},
  {Henze}, {Dupree}, {Fabrycky}, {Fortney}, {Tarter}, {Holman}, {Tenenbaum},
  {Shporer}, {Lucas}, {Welsh}, {Orosz}, {Bedding}, {Campante}, {Davies},
  {Elsworth}, {Handberg}, {Hekker}, {Karoff}, {Kawaler}, {Lund}, {Lundkvist},
  {Metcalfe}, {Miglio}, {Silva Aguirre}, {Stello}, {White}, {Boss}, {Devore},
  {Gould}, {Prsa}, {Agol}, {Barclay}, {Coughlin}, {Brugamyer}, {Mullally},
  {Quintana}, {Still}, {Thompson}, {Morrison}, {Twicken}, {D{\'e}sert},
  {Carter}, {Crepp}, {H{\'e}brard}, {Santerne}, {Moutou}, {Sobeck}, {Hudgins},
  {Haas}, {Robertson}, {Lillo-Box}, \& {Barrado}}]{marcy2014}
{Marcy}, G.~W., {Isaacson}, H., {Howard}, A.~W., {et~al.} 2014, \apjs, 210, 20

\bibitem[{{Margutti} {et~al.}(2018){Margutti}, {Chornock}, {Metzger},
  {Coppejans}, {Guidorzi}, {Migliori}, {Milisavljevic}, {Berger}, {Nicholl},
  {Zauderer}, {Lunnan}, {Kamble}, {Drout}, \& {Modjaz}}]{2018ApJ...864...45M}
{Margutti}, R., {Chornock}, R., {Metzger}, B.~D., {et~al.} 2018, \apj, 864, 45

\bibitem[{{Margutti} {et~al.}(2019){Margutti}, {Metzger}, {Chornock}, {Vurm},
  {Roth}, {Grefenstette}, {Savchenko}, {Cartier}, {Steiner}, {Terreran},
  {Margalit}, {Migliori}, {Milisavljevic}, {Alexander}, {Bietenholz},
  {Blanchard}, {Bozzo}, {Brethauer}, {Chilingarian}, {Coppejans}, {Ducci},
  {Ferrigno}, {Fong}, {G{\"o}tz}, {Guidorzi}, {Hajela}, {Hurley}, {Kuulkers},
  {Laurent}, {Mereghetti}, {Nicholl}, {Patnaude}, {Ubertini}, {Banovetz},
  {Bartel}, {Berger}, {Coughlin}, {Eftekhari}, {Frederiks}, {Kozlova},
  {Laskar}, {Svinkin}, {Drout}, {MacFadyen}, \&
  {Paterson}}]{2019ApJ...872...18M}
{Margutti}, R., {Metzger}, B.~D., {Chornock}, R., {et~al.} 2019, \apj, 872, 18

\bibitem[{{Martig} {et~al.}(2016){Martig}, {Fouesneau}, {Rix}, {Ness},
  {M{\'e}sz{\'a}ros}, {Garc{\'\i}a-Hern{\'a}ndez}, {Pinsonneault}, {Serenelli},
  {Silva Aguirre}, \& {Zamora}}]{Martig2016}
{Martig}, M., {Fouesneau}, M., {Rix}, H.-W., {et~al.} 2016, \mnras, 456, 3655

\bibitem[{{Martin} \& {Fabrycky}(2021)}]{Martin_Fabrycky21}
{Martin}, D.~V., \& {Fabrycky}, D.~C. 2021, \aj, 162, 84

\bibitem[{{Martin} \& {Fitzmaurice}(2022)}]{Martin22}
{Martin}, D.~V., \& {Fitzmaurice}, E. 2022, \mnras, 512, 602

\bibitem[{{Martin} {et~al.}(2015){Martin}, {Mazeh}, \& {Fabrycky}}]{Martin15}
{Martin}, D.~V., {Mazeh}, T., \& {Fabrycky}, D.~C. 2015, \mnras, 453, 3554

\bibitem[{{Martin} \& {Triaud}(2014)}]{Martin14}
{Martin}, D.~V., \& {Triaud}, A. H.~M.~J. 2014, \aap, 570, A91

\bibitem[{Martin \& Livio(2012)}]{martin2012formation}
Martin, R.~G., \& Livio, M. 2012, Monthly Notices of the Royal Astronomical
  Society: Letters, 428, L11

\bibitem[{{Martin} \& {Lubow}(2017)}]{Martin2017}
{Martin}, R.~G., \& {Lubow}, S.~H. 2017, \apjl, 835, L28

\bibitem[{{Massari} {et~al.}(2019){Massari}, {Koppelman}, \&
  {Helmi}}]{Massari2019}
{Massari}, D., {Koppelman}, H.~H., \& {Helmi}, A. 2019, \aap, 630, L4

\bibitem[{{Masuda} \& {Hotokezaka}(2019)}]{Masuda2019_Prospects}
{Masuda}, K., \& {Hotokezaka}, K. 2019, \apj, 883, 169

\bibitem[{{Mathur} {et~al.}(2014){Mathur}, {Garc{\'\i}a}, {Ballot}, {Ceillier},
  {Salabert}, {Metcalfe}, {R{\'e}gulo}, {Jim{\'e}nez}, \&
  {Bloemen}}]{Mathur2014}
{Mathur}, S., {Garc{\'\i}a}, R.~A., {Ballot}, J., {et~al.} 2014, \aap, 562,
  A124

\bibitem[{{Mathur} {et~al.}(2022){Mathur}, {Garc{\'\i}a}, {Breton}, {Santos},
  {Mosser}, {Huber}, {Sayeed}, {Bugnet}, \& {Chontos}}]{Mathur2022}
{Mathur}, S., {Garc{\'\i}a}, R.~A., {Breton}, S., {et~al.} 2022, \aap, 657, A31

\bibitem[{{Matsuno} {et~al.}(2021){Matsuno}, {Aoki}, {Casagrande}, {Ishigaki},
  {Shi}, {Takata}, {Xiang}, {Yong}, {Li}, {Suda}, {Xing}, \&
  {Zhao}}]{Matsuno2021}
{Matsuno}, T., {Aoki}, W., {Casagrande}, L., {et~al.} 2021, \apj, 912, 72

\bibitem[{{Matteucci}(2021)}]{Matteucci2021}
{Matteucci}, F. 2021, \aapr, 29, 5

\bibitem[{{Matteucci} \& {Francois}(1989)}]{Matteucci1989}
{Matteucci}, F., \& {Francois}, P. 1989, \mnras, 239, 885

\bibitem[{{Mawet} {et~al.}(2019){Mawet}, {Fitzgerald}, {Konopacky}, {Beichman},
  {Jovanovic}, {Dekany}, {Hover}, {Chisholm}, {Ciardi}, {Artigau}, {Banyal},
  {Beatty}, {Benneke}, {Blake}, {Burgasser}, {Canalizo}, {Chen}, {Do},
  {Doppmann}, {Doyon}, {Dressing}, {Fang}, {Greene}, {Hillenbrand}, {Howard},
  {Kane}, {Kataria}, {Kempton}, {Knutson}, {Kotani}, {Lafreni{\`e}re}, {Liu},
  {Nishiyama}, {Pandey}, {Plavchan}, {Prato}, {Rajaguru}, {Robertson}, {Salyk},
  {Sato}, {Schlawin}, {Sengupta}, {Sivarani}, {Skidmore}, {Tamura}, {Terada},
  {Vasisht}, {Wang}, \& {Zhang}}]{2019BAAS...51g.134M}
{Mawet}, D., {Fitzgerald}, M., {Konopacky}, Q., {et~al.} 2019, in Bulletin of
  the American Astronomical Society, Vol.~51, 134

\bibitem[{{Mayor} \& {Queloz}(1995)}]{Mayor1995}
{Mayor}, M., \& {Queloz}, D. 1995, \nat, 378, 355

\bibitem[{{Mayor} {et~al.}(2009){Mayor}, {Bonfils}, {Forveille}, {Delfosse},
  {Udry}, {Bertaux}, {Beust}, {Bouchy}, {Lovis}, {Pepe}, {Perrier}, {Queloz},
  \& {Santos}}]{Mayor2009}
{Mayor}, M., {Bonfils}, X., {Forveille}, T., {et~al.} 2009, \aap, 507, 487

\bibitem[{{McQuillan} {et~al.}(2014){McQuillan}, {Mazeh}, \&
  {Aigrain}}]{McQuillan2014}
{McQuillan}, A., {Mazeh}, T., \& {Aigrain}, S. 2014, \apjs, 211, 24

\bibitem[{{Meibom} {et~al.}(2015){Meibom}, {Barnes}, {Platais}, {Gilliland},
  {Latham}, \& {Mathieu}}]{Meibom2015}
{Meibom}, S., {Barnes}, S.~A., {Platais}, I., {et~al.} 2015, \nat, 517, 589

\bibitem[{{Metcalfe} \& {Egeland}(2019)}]{Metcalfe2019}
{Metcalfe}, T.~S., \& {Egeland}, R. 2019, \apj, 871, 39

\bibitem[{{Miglio} {et~al.}(2017){Miglio}, {Chiappini}, {Mosser}, {Davies},
  {Freeman}, {Girardi}, {Jofr{\'e}}, {Kawata}, {Rendle}, {Valentini},
  {Casagrande}, {Chaplin}, {Gilmore}, {Hawkins}, {Holl}, {Appourchaux},
  {Belkacem}, {Bossini}, {Brogaard}, {Goupil}, {Montalb{\'a}n}, {Noels},
  {Anders}, {Rodrigues}, {Piotto}, {Pollacco}, {Rauer}, {Allende Prieto},
  {Avelino}, {Babusiaux}, {Barban}, {Barbuy}, {Basu}, {Baudin}, {Benomar},
  {Bienaym{\'e}}, {Binney}, {Bland-Hawthorn}, {Bressan}, {Cacciari},
  {Campante}, {Cassisi}, {Christensen-Dalsgaard}, {Combes}, {Creevey}, {Cunha},
  {Jong}, {Laverny}, {Degl'Innocenti}, {Deheuvels}, {Depagne}, {Ridder}, {Di
  Matteo}, {Di Mauro}, {Dupret}, {Eggenberger}, {Elsworth}, {Famaey},
  {Feltzing}, {Garc{\'\i}a}, {Gerhard}, {Gibson}, {Gizon}, {Haywood},
  {Handberg}, {Heiter}, {Hekker}, {Huber}, {Ibata}, {Katz}, {Kawaler},
  {Kjeldsen}, {Kurtz}, {Lagarde}, {Lebreton}, {Lund}, {Majewski}, {Marigo},
  {Martig}, {Mathur}, {Minchev}, {Morel}, {Ortolani}, {Pinsonneault}, {Plez},
  {Prada Moroni}, {Pricopi}, {Recio-Blanco}, {Reyl{\'e}}, {Robin}, {Roxburgh},
  {Salaris}, {Santiago}, {Schiavon}, {Serenelli}, {Sharma}, {Silva Aguirre},
  {Soubiran}, {Steinmetz}, {Stello}, {Strassmeier}, {Ventura}, {Ventura},
  {Walton}, \& {Worley}}]{Miglio2017}
{Miglio}, A., {Chiappini}, C., {Mosser}, B., {et~al.} 2017, Astronomische
  Nachrichten, 338, 644

\bibitem[{{Miglio} {et~al.}(2021){Miglio}, {Chiappini}, {Mackereth}, {Davies},
  {Brogaard}, {Casagrande}, {Chaplin}, {Girardi}, {Kawata}, {Khan}, {Izzard},
  {Montalb{\'a}n}, {Mosser}, {Vincenzo}, {Bossini}, {Noels}, {Rodrigues},
  {Valentini}, \& {Mandel}}]{miglio2021a}
{Miglio}, A., {Chiappini}, C., {Mackereth}, J.~T., {et~al.} 2021, \aap, 645,
  A85

\bibitem[{{Miller} {et~al.}(2018){Miller}, {Cao}, {Piro}, {Blagorodnova},
  {Bue}, {Cenko}, {Dhawan}, {Ferretti}, {Fox}, {Fremling}, {Goobar}, {Howell},
  {Hosseinzadeh}, {Kasliwal}, {Laher}, {Lunnan}, {Masci}, {McCully}, {Nugent},
  {Sollerman}, {Taddia}, \& {Kulkarni}}]{2018ApJ...852..100M}
{Miller}, A.~A., {Cao}, Y., {Piro}, A.~L., {et~al.} 2018, \apj, 852, 100

\bibitem[{Millholland {et~al.}(2017)Millholland, Wang, \&
  Laughlin}]{Millholland2017}
Millholland, S., Wang, S., \& Laughlin, G. 2017, The Astrophysical Journal
  Letters, 849, L33

\bibitem[{{Min} {et~al.}(2011){Min}, {Dullemond}, {Kama}, \&
  {Dominik}}]{Min2011}
{Min}, M., {Dullemond}, C.~P., {Kama}, M., \& {Dominik}, C. 2011, \icarus, 212,
  416

\bibitem[{{Minchev} {et~al.}(2013){Minchev}, {Chiappini}, \&
  {Martig}}]{Minchev2013}
{Minchev}, I., {Chiappini}, C., \& {Martig}, M. 2013, \aap, 558, A9

\bibitem[{{Minchev} {et~al.}(2011){Minchev}, {Famaey}, {Combes}, {Di Matteo},
  {Mouhcine}, \& {Wozniak}}]{Minchev2011}
{Minchev}, I., {Famaey}, B., {Combes}, F., {et~al.} 2011, \aap, 527, A147

\bibitem[{{Minchev} {et~al.}(2014){Minchev}, {Chiappini}, {Martig},
  {Steinmetz}, {de Jong}, {Boeche}, {Scannapieco}, {Zwitter}, {Wyse}, {Binney},
  {Bland-Hawthorn}, {Bienaym{\'e}}, {Famaey}, {Freeman}, {Gibson}, {Grebel},
  {Gilmore}, {Helmi}, {Kordopatis}, {Lee}, {Munari}, {Navarro}, {Parker},
  {Quillen}, {Reid}, {Siebert}, {Siviero}, {Seabroke}, {Watson}, \&
  {Williams}}]{Minchev2014}
{Minchev}, I., {Chiappini}, C., {Martig}, M., {et~al.} 2014, \apjl, 781, L20

\bibitem[{{Minchev} {et~al.}(2018){Minchev}, {Anders}, {Recio-Blanco},
  {Chiappini}, {de Laverny}, {Queiroz}, {Steinmetz}, {Adibekyan}, {Carrillo},
  {Cescutti}, {Guiglion}, {Hayden}, {de Jong}, {Kordopatis}, {Majewski},
  {Martig}, \& {Santiago}}]{Minchev2018}
{Minchev}, I., {Anders}, F., {Recio-Blanco}, A., {et~al.} 2018, \mnras, 481,
  1645

\bibitem[{{Mints} \& {Hekker}(2017)}]{Mints2017}
{Mints}, A., \& {Hekker}, S. 2017, \aap, 604, A108

\bibitem[{Mishra {et~al.}(2021)Mishra, Alibert, Leleu, Emsenhuber, Mordasini,
  Burn, Udry, \& Benz}]{Mishra2021}
Mishra, L., Alibert, Y., Leleu, A., {et~al.} 2021, Astronomy \& Astrophysics,
  656, A74

\bibitem[{{Modjaz} {et~al.}(2009){Modjaz}, {Li}, {Butler}, {Chornock},
  {Perley}, {Blondin}, {Bloom}, {Filippenko}, {Kirshner}, {Kocevski},
  {Poznanski}, {Hicken}, {Foley}, {Stringfellow}, {Berlind}, {Barrado y
  Navascues}, {Blake}, {Bouy}, {Brown}, {Challis}, {Chen}, {de Vries},
  {Dufour}, {Falco}, {Friedman}, {Ganeshalingam}, {Garnavich}, {Holden},
  {Illingworth}, {Lee}, {Liebert}, {Marion}, {Olivier}, {Prochaska},
  {Silverman}, {Smith}, {Starr}, {Steele}, {Stockton}, {Williams}, \&
  {Wood-Vasey}}]{Modjaz2009}
{Modjaz}, M., {Li}, W., {Butler}, N., {et~al.} 2009, \apj, 702, 226

\bibitem[{{Moe} \& {Kratter}(2021)}]{Moe:2021}
{Moe}, M., \& {Kratter}, K.~M. 2021, \mnras, 507, 3593

\bibitem[{{Moll{\'a}} {et~al.}(2019){Moll{\'a}}, {D{\'\i}az}, {Cavichia},
  {Gibson}, {Maciel}, {Costa}, {Ascasibar}, \& {Few}}]{Molla2019}
{Moll{\'a}}, M., {D{\'\i}az}, {\'A}.~I., {Cavichia}, O., {et~al.} 2019, \mnras,
  482, 3071

\bibitem[{{Moln{\'a}r} {et~al.}(2012){Moln{\'a}r}, {Koll{\'a}th}, {Szab{\'o}},
  {Bryson}, {Kolenberg}, {Mullally}, \& {Thompson}}]{Molnar2012RRLyrae}
{Moln{\'a}r}, L., {Koll{\'a}th}, Z., {Szab{\'o}}, R., {et~al.} 2012, \apjl,
  757, L13

\bibitem[{Moln{\'a}r {et~al.}(2018)Moln{\'a}r, P{\'a}l, S{\'a}rneczky,
  Szab{\'o}, Vink{\'o}, Szab{\'o}, Kiss, Hanyecz, Marton, \&
  Kiss}]{molnar2018main}
Moln{\'a}r, L., P{\'a}l, A., S{\'a}rneczky, K., {et~al.} 2018, The
  Astrophysical Journal Supplement Series, 234, 37

\bibitem[{{Moln{\'a}r} {et~al.}(2022){Moln{\'a}r}, {B{\'o}di}, {P{\'a}l},
  {Bhardwaj}, {Hambsch}, {Benk{\H{o}}}, {Derekas}, {Ebadi}, {Joyce},
  {Hasanzadeh}, {Kolenberg}, {Lund}, {Nemec}, {Netzel}, {Ngeow}, {Pepper},
  {Plachy}, {Prudil}, {Siverd}, {Skarka}, {Smolec}, {S{\'o}dor}, {Sylla},
  {Szab{\'o}}, {Szab{\'o}}, {Kjeldsen}, {Christensen-Dalsgaard}, \&
  {Ricker}}]{Molnar2022TESSRL}
{Moln{\'a}r}, L., {B{\'o}di}, A., {P{\'a}l}, A., {et~al.} 2022, \apjs, 258, 8

\bibitem[{{Mombarg} {et~al.}(2019){Mombarg}, {Van Reeth}, {Pedersen},
  {Molenberghs}, {Bowman}, {Johnston}, {Tkachenko}, \& {Aerts}}]{Mombarg2019}
{Mombarg}, J.~S.~G., {Van Reeth}, T., {Pedersen}, M.~G., {et~al.} 2019, \mnras,
  485, 3248

\bibitem[{{Montalb{\'a}n} {et~al.}(2021){Montalb{\'a}n}, {Mackereth}, {Miglio},
  {Vincenzo}, {Chiappini}, {Buldgen}, {Mosser}, {Noels}, {Scuflaire}, {Vrard},
  {Willett}, {Davies}, {Hall}, {Nielsen}, {Khan}, {Rendle}, {van Rossem},
  {Ferguson}, \& {Chaplin}}]{Montalban2021}
{Montalb{\'a}n}, J., {Mackereth}, J.~T., {Miglio}, A., {et~al.} 2021, Nature
  Astronomy, 5, 640

\bibitem[{{Montalto} {et~al.}(2021){Montalto}, {Piotto}, {Marrese},
  {Nascimbeni}, {Prisinzano}, {Granata}, {Marinoni}, {Desidera}, {Ortolani},
  {Aerts}, {Alei}, {Altavilla}, {Benatti}, {B{\"o}rner}, {Cabrera}, {Claudi},
  {Deleuil}, {Fabrizio}, {Gizon}, {Goupil}, {Heras}, {Magrin}, {Malavolta},
  {Mas-Hesse}, {Pagano}, {Paproth}, {Pertenais}, {Pollacco}, {Ragazzoni},
  {Ramsay}, {Rauer}, \& {Udry}}]{montalto2021a}
{Montalto}, M., {Piotto}, G., {Marrese}, P.~M., {et~al.} 2021, \aap, 653, A98

\bibitem[{{Montgomery}(2005)}]{Montgomery2005}
{Montgomery}, M.~H. 2005, \apj, 633, 1142

\bibitem[{{Morales} {et~al.}(2008){Morales}, {Ribas}, \& {Jordi}}]{Mor2008}
{Morales}, J.~C., {Ribas}, I., \& {Jordi}, C. 2008, \aap, 478, 507

\bibitem[{Morales {et~al.}(2019)Morales, Mustill, Ribas, Davies, Reiners,
  Bauer, Kossakowski, Herrero, Rodr{\'\i}guez, L{\'o}pez-Gonz{\'a}lez,
  {et~al.}}]{Morales2019giant}
Morales, J.~C., Mustill, A., Ribas, I., {et~al.} 2019, Science, 365, 1441

\bibitem[{{Moravveji} {et~al.}(2015){Moravveji}, {Aerts}, {P{\'a}pics},
  {Triana}, \& {Vandoren}}]{Moravveji2015b}
{Moravveji}, E., {Aerts}, C., {P{\'a}pics}, P.~I., {Triana}, S.~A., \&
  {Vandoren}, B. 2015, \aap, 580, A27

\bibitem[{Morbidelli {et~al.}(2012)Morbidelli, Lunine, O'Brien, Raymond, \&
  Walsh}]{Morbidelli2012}
Morbidelli, A., Lunine, J.~I., O'Brien, D.~P., Raymond, S.~N., \& Walsh, K.~J.
  2012, Annual Review of Earth and Planetary Sciences, 40, 251

\bibitem[{{Mordasini} {et~al.}(2009){Mordasini}, {Alibert}, \&
  {Benz}}]{Mordasini2009}
{Mordasini}, C., {Alibert}, Y., \& {Benz}, W. 2009, \aap, 501, 1139

\bibitem[{{Mosser} {et~al.}(2012){Mosser}, {Goupil}, {Belkacem}, {Marques},
  {Beck}, {Bloemen}, {De Ridder}, {Barban}, {Deheuvels}, {Elsworth}, {Hekker},
  {Kallinger}, {Ouazzani}, {Pinsonneault}, {Samadi}, {Stello}, {Garc{\'{\i}}a},
  {Klaus}, {Li}, {Mathur}, \& {Morris}}]{mosser2012c}
{Mosser}, B., {Goupil}, M.~J., {Belkacem}, K., {et~al.} 2012, \aap, 548, A10

\bibitem[{{Mosser} {et~al.}(2017){Mosser}, {Belkacem}, {Pin{\c c}on}, {Takata},
  {Vrard}, {Barban}, {Goupil}, {Kallinger}, \& {Samadi}}]{mosser2017a}
{Mosser}, B., {Belkacem}, K., {Pin{\c c}on}, C., {et~al.} 2017, \aap, 598, A62

\bibitem[{{Mr{\'o}z} {et~al.}(2017){Mr{\'o}z}, {}, {Skowron}, {Poleski},
  {Koz{\l}owski}, {Szyma{\'n}ski}, {Soszy{\'n}ski}, {Wyrzykowski},
  {Pietrukowicz}, {Ulaczyk}, {Skowron}, \& {Pawlak}}]{Mroz2017a}
{Mr{\'o}z}, P., {}, A., {Skowron}, J., {et~al.} 2017, \nat, 548, 183

\bibitem[{{Mr{\'o}z} {et~al.}(2020){Mr{\'o}z}, {Poleski}, {Gould}, {Udalski},
  {Sumi}, {and}, {Szyma{\'n}ski}, {Soszy{\'n}ski}, {Pietrukowicz},
  {Koz{\l}owski}, {Skowron}, {Ulaczyk}, {OGLE Collaboration}, {Albrow},
  {Chung}, {Han}, {Hwang}, {Jung}, {Kim}, {Ryu}, {Shin}, {Shvartzvald}, {Yee},
  {Zang}, {Cha}, {Kim}, {Kim}, {Lee}, {Lee}, {Lee}, {Park}, {Pogge}, \& {KMT
  Collaboration}}]{OB161928}
{Mr{\'o}z}, P., {Poleski}, R., {Gould}, A., {et~al.} 2020, \apjl, 903, L11

\bibitem[{{Munoz} \& {Lai}(2015)}]{Munoz15}
{Munoz}, D.~J., \& {Lai}, D. 2015, in AAS/Division for Extreme Solar Systems
  Abstracts, Vol.~47, AAS/Division for Extreme Solar Systems Abstracts, 403.06

\bibitem[{Murchikova \& Tremaine(2020)}]{Murchikova2020}
Murchikova, L., \& Tremaine, S. 2020, The Astronomical Journal, 160, 160

\bibitem[{{Murphy} {et~al.}(2016){Murphy}, {Bedding}, \&
  {Shibahashi}}]{2016ApJ...827L..17M}
{Murphy}, S.~J., {Bedding}, T.~R., \& {Shibahashi}, H. 2016, \apjl, 827, L17

\bibitem[{{Murphy} {et~al.}(2022){Murphy}, {Bedding}, {White}, {Li}, {Hey},
  {Reese}, \& {Joyce}}]{2022MNRAS.511.5718M}
{Murphy}, S.~J., {Bedding}, T.~R., {White}, T.~R., {et~al.} 2022, \mnras, 511,
  5718

\bibitem[{{Murphy} {et~al.}(2019){Murphy}, {Hey}, {Van Reeth}, \&
  {Bedding}}]{2019MNRAS.485.2380M}
{Murphy}, S.~J., {Hey}, D., {Van Reeth}, T., \& {Bedding}, T.~R. 2019, \mnras,
  485, 2380

\bibitem[{{Murphy} {et~al.}(2018){Murphy}, {Moe}, {Kurtz}, {Bedding},
  {Shibahashi}, \& {Boffin}}]{2018MNRAS.474.4322M}
{Murphy}, S.~J., {Moe}, M., {Kurtz}, D.~W., {et~al.} 2018, \mnras, 474, 4322

\bibitem[{Mustill \& Wyatt(2011)}]{Mustill2011}
Mustill, A.~J., \& Wyatt, M.~C. 2011, Monthly Notices of the Royal Astronomical
  Society, 413, 554

\bibitem[{{Naidu} {et~al.}(2020){Naidu}, {Conroy}, {Bonaca}, {Johnson}, {Ting},
  {Caldwell}, {Zaritsky}, \& {Cargile}}]{Naidu2020}
{Naidu}, R.~P., {Conroy}, C., {Bonaca}, A., {et~al.} 2020, \apj, 901, 48

\bibitem[{{Nakar} \& {Sari}(2010)}]{Nakar+Sari2010}
{Nakar}, E., \& {Sari}, R. 2010, \apj, 725, 904

\bibitem[{{Namouni}(2010)}]{namouni10}
{Namouni}, F. 2010, \apjl, 719, L145

\bibitem[{{Nascimbeni} {et~al.}(2022){Nascimbeni}, {Piotto}, {B{\"o}rner},
  {Montalto}, {Marrese}, {Cabrera}, {Marinoni}, {Aerts}, {Altavilla},
  {Benatti}, {Claudi}, {Deleuil}, {Desidera}, {Fabrizio}, {Gizon}, {Goupil},
  {Granata}, {Heras}, {Magrin}, {Malavolta}, {Mas-Hesse}, {Ortolani}, {Pagano},
  {Pollacco}, {Prisinzano}, {Ragazzoni}, {Ramsay}, {Rauer}, \&
  {Udry}}]{nascimbeni2022a}
{Nascimbeni}, V., {Piotto}, G., {B{\"o}rner}, A., {et~al.} 2022, \aap, 658, A31

\bibitem[{{Nataf} {et~al.}(2013){Nataf}, {Gould}, {Fouqu{\'e}}, {Gonzalez},
  {Johnson}, {Skowron}, {}, {Szyma{\'n}ski}, {Kubiak}, {Pietrzy{\'n}ski},
  {Soszy{\'n}ski}, {Ulaczyk}, {Wyrzykowski}, \& {Poleski}}]{Nataf2013}
{Nataf}, D.~M., {Gould}, A., {Fouqu{\'e}}, P., {et~al.} 2013, \apj, 769, 88

\bibitem[{{National Acad Sciences}(2018)}]{ExoplanetScienceStrategy2018}
{National Acad Sciences}. 2018, {Exoplanet Science Strategy} (The National
  Academies Press), doi:10.17226/25187

\bibitem[{{National Acad Sciences}(2021)}]{national2021pathways}
---. 2021, {Pathways to Discovery in Astronomy and Astrophysics for the 2020s}
  (The National Academies Press), doi:10.17226/26141

\bibitem[{{Nelson} \& {Papaloizou}(2003)}]{Nelson2003}
{Nelson}, R.~P., \& {Papaloizou}, J. C.~B. 2003, \mnras, 339, 993

\bibitem[{{Nemiroff} \& {Wickramasinghe}(1994)}]{Nemiroff1994}
{Nemiroff}, R.~J., \& {Wickramasinghe}, W.~A.~D.~T. 1994, \apjl, 424, L21

\bibitem[{{Ness} {et~al.}(2016){Ness}, {Hogg}, {Rix}, {Martig}, {Pinsonneault},
  \& {Ho}}]{Ness2016}
{Ness}, M., {Hogg}, D.~W., {Rix}, H.~W., {et~al.} 2016, \apj, 823, 114

\bibitem[{{Ness} {et~al.}(2019){Ness}, {Johnston}, {Blancato}, {Rix}, {Beane},
  {Bird}, \& {Hawkins}}]{Ness2019}
{Ness}, M.~K., {Johnston}, K.~V., {Blancato}, K., {et~al.} 2019, \apj, 883, 177

\bibitem[{{Nesvorn{\'y}} {et~al.}(2012){Nesvorn{\'y}}, {Kipping}, {Buchhave},
  {Bakos}, {Hartman}, \& {Schmitt}}]{Nes12}
{Nesvorn{\'y}}, D., {Kipping}, D.~M., {Buchhave}, L.~A., {et~al.} 2012,
  Science, 336, 1133

\bibitem[{Nesvorn{\`y} {et~al.}(2019)Nesvorn{\`y}, Li, Youdin, Simon, \&
  Grundy}]{Nesvorny2019trans}
Nesvorn{\`y}, D., Li, R., Youdin, A.~N., Simon, J.~B., \& Grundy, W.~M. 2019,
  Nature Astronomy, 3, 808

\bibitem[{{Ngo} {et~al.}(2016){Ngo}, {Knutson}, {Hinkley}, {Bryan}, {Crepp},
  {Batygin}, {Crossfield}, {Hansen}, {Howard}, {Johnson}, {Mawet}, {Morton},
  {Muirhead}, \& {Wang}}]{Ngo:2016}
{Ngo}, H., {Knutson}, H.~A., {Hinkley}, S., {et~al.} 2016, \apj, 827, 8

\bibitem[{{Nissen} {et~al.}(2020){Nissen}, {Christensen-Dalsgaard},
  {Mosumgaard}, {Silva Aguirre}, {Spitoni}, \& {Verma}}]{Nissen2020}
{Nissen}, P.~E., {Christensen-Dalsgaard}, J., {Mosumgaard}, J.~R., {et~al.}
  2020, \aap, 640, A81

\bibitem[{{Nordstr{\"o}m} {et~al.}(2004){Nordstr{\"o}m}, {Mayor}, {Andersen},
  {Holmberg}, {Pont}, {J{\o}rgensen}, {Olsen}, {Udry}, \&
  {Mowlavi}}]{Nordstrom2004}
{Nordstr{\"o}m}, B., {Mayor}, M., {Andersen}, J., {et~al.} 2004, \aap, 418, 989

\bibitem[{Ofek \& Nakar(2010)}]{ofek2010detectability}
Ofek, E.~O., \& Nakar, E. 2010, The Astrophysical Journal Letters, 711, L7

\bibitem[{{Ohta} {et~al.}(2009){Ohta}, {Taruya}, \& {Suto}}]{Ohta2009}
{Ohta}, Y., {Taruya}, A., \& {Suto}, Y. 2009, \apj, 690, 1

\bibitem[{{Olejak} {et~al.}(2020){Olejak}, {Belczynski}, {Bulik}, \&
  {Sobolewska}}]{Olejak20}
{Olejak}, A., {Belczynski}, K., {Bulik}, T., \& {Sobolewska}, M. 2020, \aap,
  638, A94

\bibitem[{{Orosz} {et~al.}(2012{\natexlab{a}}){Orosz}, {Welsh}, {Carter},
  {Brugamyer}, {Buchhave}, {Cochran}, {Endl}, {Ford}, {MacQueen}, {Short},
  {Torres}, {Windmiller}, {Agol}, {Barclay}, {Caldwell}, {Clarke}, {Doyle},
  {Fabrycky}, {Geary}, {Haghighipour}, {Holman}, {Ibrahim}, {Jenkins},
  {Kinemuchi}, {Li}, {Lissauer}, {Pr{\v{s}}a}, {Ragozzine}, {Shporer}, {Still},
  \& {Wade}}]{Orosz2012b}
{Orosz}, J.~A., {Welsh}, W.~F., {Carter}, J.~A., {et~al.} 2012{\natexlab{a}},
  \apj, 758, 87

\bibitem[{{Orosz} {et~al.}(2012{\natexlab{b}}){Orosz}, {Welsh}, {Carter},
  {Fabrycky}, {Cochran}, {Endl}, {Ford}, {Haghighipour}, {MacQueen}, {Mazeh},
  {Sanchis-Ojeda}, {Short}, {Torres}, {Agol}, {Buchhave}, {Doyle}, {Isaacson},
  {Lissauer}, {Marcy}, {Shporer}, {Windmiller}, {Barclay}, {Boss}, {Clarke},
  {Fortney}, {Geary}, {Holman}, {Huber}, {Jenkins}, {Kinemuchi}, {Kruse},
  {Ragozzine}, {Sasselov}, {Still}, {Tenenbaum}, {Uddin}, {Winn}, {Koch}, \&
  {Borucki}}]{Orosz2012a}
---. 2012{\natexlab{b}}, Science, 337, 1511

\bibitem[{{Orosz} {et~al.}(2019){Orosz}, {Welsh}, {Haghighipour}, {Quarles},
  {Short}, {Mills}, {Satyal}, {Torres}, {Agol}, {Fabrycky}, {Jontof-Hutter},
  {Windmiller}, {M{\"u}ller}, {Hinse}, {Cochran}, {Endl}, {Ford}, {Mazeh}, \&
  {Lissauer}}]{Orosz2019}
{Orosz}, J.~A., {Welsh}, W.~F., {Haghighipour}, N., {et~al.} 2019, \aj, 157,
  174

\bibitem[{{{\O}stensen} {et~al.}(2010){{\O}stensen}, {Silvotti}, {Charpinet},
  {Oreiro}, {Handler}, {Green}, {Bloemen}, {Heber}, {G{\"a}nsicke}, {Marsh},
  {Kurtz}, {Telting}, {Reed}, {Kawaler}, {Aerts}, {Rodr{\'\i}guez-L{\'o}pez},
  {Vu{\v{c}}kovi{\'c}}, {Ottosen}, {Liimets}, {Quint}, {Van Grootel},
  {Randall}, {Gilliland}, {Kjeldsen}, {Christensen-Dalsgaard}, {Borucki},
  {Koch}, \& {Quintana}}]{2010MNRAS.409.1470O}
{{\O}stensen}, R.~H., {Silvotti}, R., {Charpinet}, S., {et~al.} 2010, \mnras,
  409, 1470

\bibitem[{{Ouazzani} {et~al.}(2020){Ouazzani}, {Ligni{\`e}res}, {Dupret},
  {Salmon}, {Ballot}, {Christophe}, \& {Takata}}]{2020A&A...640A..49O}
{Ouazzani}, R.~M., {Ligni{\`e}res}, F., {Dupret}, M.~A., {et~al.} 2020, \aap,
  640, A49

\bibitem[{{Owen} \& {Wu}(2017)}]{Owen2017}
{Owen}, J.~E., \& {Wu}, Y. 2017, \apj, 847, 29

\bibitem[{{Papar{\'o}} {et~al.}(2016){Papar{\'o}}, {Benk{\H o}}, {Hareter}, \&
  {Guzik}}]{Paparo2016a}
{Papar{\'o}}, M., {Benk{\H o}}, J.~M., {Hareter}, M., \& {Guzik}, J.~A. 2016,
  \apj, 822, 100

\bibitem[{{P{\'a}pics} {et~al.}(2014){P{\'a}pics}, {Moravveji}, {Aerts},
  {Tkachenko}, {Triana}, {Bloemen}, \& {Southworth}}]{Papics2014}
{P{\'a}pics}, P.~I., {Moravveji}, E., {Aerts}, C., {et~al.} 2014, \aap, 570, A8

\bibitem[{{Paquette} {et~al.}(1986){Paquette}, {Pelletier}, {Fontaine}, \&
  {Michaud}}]{1986ApJS...61..197P}
{Paquette}, C., {Pelletier}, C., {Fontaine}, G., \& {Michaud}, G. 1986, \apjs,
  61, 197

\bibitem[{{Pasham} {et~al.}(2021){Pasham}, {Ho}, {Alston}, {Remillard}, {Ng},
  {Gendreau}, {Metzger}, {Altamirano}, {Chakrabarty}, {Fabian}, {Miller},
  {Bult}, {Arzoumanian}, {Steiner}, {Strohmayer}, {Tombesi}, {Homan},
  {Cackett}, \& {Harding}}]{2021NatAs...6..249P}
{Pasham}, D.~R., {Ho}, W. C.~G., {Alston}, W., {et~al.} 2021, Nature Astronomy,
  6, 249

\bibitem[{{Pedersen} {et~al.}(2021){Pedersen}, {Aerts}, {P{\'a}pics},
  {Michielsen}, {Gebruers}, {Rogers}, {Molenberghs}, {Burssens}, {Garcia}, \&
  {Bowman}}]{2021NatAs...5..715P}
{Pedersen}, M.~G., {Aerts}, C., {P{\'a}pics}, P.~I., {et~al.} 2021, Nature
  Astronomy, 5, 715

\bibitem[{{Penny} {et~al.}(2019){Penny}, {Gaudi}, {Kerins}, {Rattenbury},
  {Mao}, {Robin}, \& {Calchi Novati}}]{MatthewWFIRSTI}
{Penny}, M.~T., {Gaudi}, B.~S., {Kerins}, E., {et~al.} 2019, \apjs, 241, 3

\bibitem[{{Penny} {et~al.}(2016){Penny}, {Henderson}, \&
  {Clanton}}]{Matthewbulge}
{Penny}, M.~T., {Henderson}, C.~B., \& {Clanton}, C. 2016, \apj, 830, 150

\bibitem[{{Pepe} {et~al.}(2010){Pepe}, {Cristiani}, {Rebolo Lopez}, {Santos},
  {Amorim}, {Avila}, {Benz}, {Bonifacio}, {Cabral}, {Carvas}, {Cirami},
  {Coelho}, {Comari}, {Coretti}, {De Caprio}, {Dekker}, {Delabre}, {Di
  Marcantonio}, {D'Odorico}, {Fleury}, {Garc{\'{\i}}a}, {Herreros Linares},
  {Hughes}, {Iwert}, {Lima}, {Lizon}, {Lo Curto}, {Lovis}, {Manescau},
  {Martins}, {M{\'e}gevand}, {Moitinho}, {Molaro}, {Monteiro}, {Monteiro},
  {Pasquini}, {Mordasini}, {Queloz}, {Rasilla}, {Rebord{\~a}o}, {Santana
  Tschudi}, {Santin}, {Sosnowska}, {Span{\`o}}, {Tenegi}, {Udry}, {Vanzella},
  {Viel}, {Zapatero Osorio}, \& {Zerbi}}]{pepe10}
{Pepe}, F.~A., {Cristiani}, S., {Rebolo Lopez}, R., {et~al.} 2010, in
  \procspie, Vol. 7735, Ground-based and Airborne Instrumentation for Astronomy
  III, 77350F

\bibitem[{{Perley} {et~al.}(2019){Perley}, {Mazzali}, {Yan}, {Cenko}, {Gezari},
  {Taggart}, {Blagorodnova}, {Fremling}, {Mockler}, {Singh}, {Tominaga},
  {Tanaka}, {Watson}, {Ahumada}, {Anupama}, {Ashall}, {Becerra}, {Bersier},
  {Bhalerao}, {Bloom}, {Butler}, {Copperwheat}, {Coughlin}, {De}, {Drake},
  {Duev}, {Frederick}, {Gonz{\'a}lez}, {Goobar}, {Heida}, {Ho}, {Horst},
  {Hung}, {Itoh}, {Jencson}, {Kasliwal}, {Kawai}, {Khanam}, {Kulkarni},
  {Kumar}, {Kumar}, {Kutyrev}, {Lee}, {Maeda}, {Mahabal}, {Murata}, {Neill},
  {Ngeow}, {Penprase}, {Pian}, {Quimby}, {Ramirez-Ruiz}, {Richer},
  {Rom{\'a}n-Z{\'u}{\~n}iga}, {Sahu}, {Srivastav}, {Socia}, {Sollerman},
  {Tachibana}, {Taddia}, {Tinyanont}, {Troja}, {Ward}, {Wee}, \&
  {Yu}}]{2019MNRAS.484.1031P}
{Perley}, D.~A., {Mazzali}, P.~A., {Yan}, L., {et~al.} 2019, \mnras, 484, 1031

\bibitem[{{Perlmutter} {et~al.}(1999){Perlmutter}, {Aldering}, {Goldhaber},
  {Knop}, {Nugent}, {Castro}, {Deustua}, {Fabbro}, {Goobar}, {Groom}, {Hook},
  {Kim}, {Kim}, {Lee}, {Nunes}, {Pain}, {Pennypacker}, {Quimby}, {Lidman},
  {Ellis}, {Irwin}, {McMahon}, {Ruiz-Lapuente}, {Walton}, {Schaefer}, {Boyle},
  {Filippenko}, {Matheson}, {Fruchter}, {Panagia}, {Newberg}, {Couch}, \&
  {Project}}]{1999ApJ...517..565P}
{Perlmutter}, S., {Aldering}, G., {Goldhaber}, G., {et~al.} 1999, \apj, 517,
  565

\bibitem[{{Perryman} {et~al.}(2014){Perryman}, {Hartman}, {Bakos}, \&
  {Lindegren}}]{perryman14}
{Perryman}, M., {Hartman}, J., {Bakos}, G.~{\'A}., \& {Lindegren}, L. 2014,
  \apj, 797, 14

\bibitem[{{Petigura} {et~al.}(2018){Petigura}, {Marcy}, {Winn}, {Weiss},
  {Fulton}, {Howard}, {Sinukoff}, {Isaacson}, {Morton}, \&
  {Johnson}}]{Petigura:2018}
{Petigura}, E.~A., {Marcy}, G.~W., {Winn}, J.~N., {et~al.} 2018, \aj, 155, 89

\bibitem[{Petigura {et~al.}(2018)Petigura, Marcy, Winn, Weiss, Fulton, Howard,
  Sinukoff, Isaacson, Morton, \& Johnson}]{Petigura2018}
Petigura, E.~A., Marcy, G.~W., Winn, J.~N., {et~al.} 2018, The Astronomical
  Journal, 155, 89

\bibitem[{{Pi} {et~al.}(2019){Pi}, {Zhang}, {Bi}, {Han}, {Lu}, {Yue}, {Long},
  \& {Yan}}]{Pi2019}
{Pi}, Q.-f., {Zhang}, L.-y., {Bi}, S.-l., {et~al.} 2019, \apj, 877, 75

\bibitem[{{Pierens} \& {Nelson}(2013)}]{PN13}
{Pierens}, A., \& {Nelson}, R.~P. 2013, \aap, 556, A134

\bibitem[{{Pierrehumbert}(2010)}]{Pierrehumbert2010}
{Pierrehumbert}, R.~T. 2010, {Principles of Planetary Climate}

\bibitem[{{Pietrukowicz} {et~al.}(2017){Pietrukowicz}, {Dziembowski}, {Latour},
  {Angeloni}, {Poleski}, {di Mille}, {Soszy{\'n}ski}, {Udalski},
  {Szyma{\'n}ski}, {Wyrzykowski}, {Koz{\l}owski}, {Skowron}, {Skowron},
  {Mr{\'o}z}, {Pawlak}, \& {Ulaczyk}}]{2017NatAs...1E.166P}
{Pietrukowicz}, P., {Dziembowski}, W.~A., {Latour}, M., {et~al.} 2017, Nature
  Astronomy, 1, 0166

\bibitem[{{Pinsonneault} {et~al.}(2014){Pinsonneault}, {Elsworth}, {Epstein},
  {Hekker}, {M{\'e}sz{\'a}ros}, {Chaplin}, {Johnson}, {Garc{\'\i}a},
  {Holtzman}, {Mathur}, {Garc{\'\i}a P{\'e}rez}, {Silva Aguirre}, {Girardi},
  {Basu}, {Shetrone}, {Stello}, {Allende Prieto}, {An}, {Beck}, {Beers},
  {Bizyaev}, {Bloemen}, {Bovy}, {Cunha}, {De Ridder}, {Frinchaboy},
  {Garc{\'\i}a-Hern{\'a}ndez}, {Gilliland}, {Harding}, {Hearty}, {Huber},
  {Ivans}, {Kallinger}, {Majewski}, {Metcalfe}, {Miglio}, {Mosser}, {Muna},
  {Nidever}, {Schneider}, {Serenelli}, {Smith}, {Tayar}, {Zamora}, \&
  {Zasowski}}]{Pinsonneault2014}
{Pinsonneault}, M.~H., {Elsworth}, Y., {Epstein}, C., {et~al.} 2014, \apjs,
  215, 19

\bibitem[{{Pinsonneault} {et~al.}(2018){Pinsonneault}, {Elsworth}, {Tayar},
  {Serenelli}, {Stello}, {Zinn}, {Mathur}, {Garc{\'\i}a}, {Johnson}, {Hekker},
  {Huber}, {Kallinger}, {M{\'e}sz{\'a}ros}, {Mosser}, {Stassun}, {Girardi},
  {Rodrigues}, {Silva Aguirre}, {An}, {Basu}, {Chaplin}, {Corsaro}, {Cunha},
  {Garc{\'\i}a-Hern{\'a}ndez}, {Holtzman}, {J{\"o}nsson}, {Shetrone}, {Smith},
  {Sobeck}, {Stringfellow}, {Zamora}, {Beers}, {Fern{\'a}ndez-Trincado},
  {Frinchaboy}, {Hearty}, \& {Nitschelm}}]{Pinsonneault2018}
{Pinsonneault}, M.~H., {Elsworth}, Y.~P., {Tayar}, J., {et~al.} 2018, \apjs,
  239, 32

\bibitem[{{Plachy} {et~al.}(2021){Plachy}, {P{\'a}l}, {B{\'o}di}, {Szab{\'o}},
  {Moln{\'a}r}, {Szabados}, {Benk{\H{o}}}, {Anderson}, {Bellinger}, {Bhardwaj},
  {Ebadi}, {Gazeas}, {Hambsch}, {Hasanzadeh}, {Jurkovic}, {Kalaee}, {Kervella},
  {Kolenberg}, {Miko{\l}ajczyk}, {Nardetto}, {Nemec}, {Netzel}, {Ngeow},
  {Ozuyar}, {Pascual-Granado}, {Pilecki}, {Ripepi}, {Skarka}, {Smolec},
  {S{\'o}dor}, {Szab{\'o}}, {Christensen-Dalsgaard}, {Jenkins}, {Kjeldsen},
  {Ricker}, \& {Vanderspek}}]{Plachy2021TESSCepheid}
{Plachy}, E., {P{\'a}l}, A., {B{\'o}di}, A., {et~al.} 2021, \apjs, 253, 11

\bibitem[{{Planck Collaboration} {et~al.}(2016){Planck Collaboration}, {Ade},
  {Aghanim}, {Arnaud}, {Ashdown}, {Aumont}, {Baccigalupi}, {Banday},
  {Barreiro}, {Bartlett}, \& et~al.}]{Planck2016}
{Planck Collaboration}, {Ade}, P.~A.~R., {Aghanim}, N., {et~al.} 2016, \aap,
  594, A13

\bibitem[{Podlewska-Gaca {et~al.}(2021)Podlewska-Gaca, Poleski, Bartczak,
  McDonald, \& P{\'a}l}]{podlewska2021determination}
Podlewska-Gaca, E., Poleski, R., Bartczak, P., McDonald, I., \& P{\'a}l, A.
  2021, The Astrophysical Journal Supplement Series, 255, 4

\bibitem[{{Polin} {et~al.}(2019){Polin}, {Nugent}, \&
  {Kasen}}]{2019ApJ...873...84P}
{Polin}, A., {Nugent}, P., \& {Kasen}, D. 2019, \apj, 873, 84

\bibitem[{{Pollack} {et~al.}(1996){Pollack}, {Hubickyj}, {Bodenheimer},
  {Lissauer}, {Podolak}, \& {Greenzweig}}]{Pollack:1996}
{Pollack}, J.~B., {Hubickyj}, O., {Bodenheimer}, P., {et~al.} 1996, \icarus,
  124, 62

\bibitem[{{Powell} {et~al.}(2021){Powell}, {Kostov}, {Rappaport}, {Borkovits},
  {Zasche}, {Tokovinin}, {Kruse}, {Latham}, {Montet}, {Jensen}, {Jayaraman},
  {Collins}, {Ma{\v{s}}ek}, {Hellier}, {Evans}, {Tan}, {Schlieder}, {Torres},
  {Smale}, {Friedman}, {Barclay}, {Gagliano}, {Quintana}, {Jacobs}, {Gilbert},
  {Kristiansen}, {Col{\'o}n}, {LaCourse}, {Olmschenk}, {Omohundro},
  {Schnittman}, {Schwengeler}, {Barry}, {Terentev}, {Boyd}, {Schmitt}, {Quinn},
  {Vanderburg}, {Palle}, {Armstrong}, {Ricker}, {Vanderspek}, {Seager}, {Winn},
  {Jenkins}, {Caldwell}, {Wohler}, {Shiao}, {Burke}, {Daylan}, \&
  {Villase{\~n}or}}]{Powell2021AJ....161..162P}
{Powell}, B.~P., {Kostov}, V.~B., {Rappaport}, S.~A., {et~al.} 2021, \aj, 161,
  162

\bibitem[{{Pr{\v{s}}a} {et~al.}(2011){Pr{\v{s}}a}, {Batalha}, {Slawson},
  {Doyle}, {Welsh}, {Orosz}, {Seager}, {Rucker}, {Mjaseth}, {Engle}, {Conroy},
  {Jenkins}, {Caldwell}, {Koch}, \& {Borucki}}]{Prsa2011}
{Pr{\v{s}}a}, A., {Batalha}, N., {Slawson}, R.~W., {et~al.} 2011, \aj, 141, 83

\bibitem[{Qian \& Wu(2021)}]{Qian2021}
Qian, Y., \& Wu, Y. 2021, The Astronomical Journal, 161, 201

\bibitem[{{Quarles} {et~al.}(2018){Quarles}, {Satyal}, {Kostov}, {Kaib}, \&
  {Haghighipour}}]{Quarles18}
{Quarles}, B., {Satyal}, S., {Kostov}, V., {Kaib}, N., \& {Haghighipour}, N.
  2018, \apj, 856, 150

\bibitem[{Quintana {et~al.}(2014)Quintana, Barclay, Raymond, Rowe, Bolmont,
  Caldwell, Howell, Kane, Huber, Crepp, {et~al.}}]{Quintana2014earth}
Quintana, E.~V., Barclay, T., Raymond, S.~N., {et~al.} 2014, Science, 344, 277

\bibitem[{{Rabinak} \& {Waxman}(2011)}]{Rabinak+Waxman2011}
{Rabinak}, I., \& {Waxman}, E. 2011, \apj, 728, 63

\bibitem[{{Racusin} {et~al.}(2008){Racusin}, {Karpov}, {Sokolowski}, {Granot},
  {Wu}, {Pal'Shin}, {Covino}, {van der Horst}, {Oates}, {Schady}, {Smith},
  {Cummings}, {Starling}, {Piotrowski}, {Zhang}, {Evans}, {Holland}, {Malek},
  {Page}, {Vetere}, {Margutti}, {Guidorzi}, {Kamble}, {Curran}, {Beardmore},
  {Kouveliotou}, {Mankiewicz}, {Melandri}, {O'Brien}, {Page}, {Piran},
  {Tanvir}, {Wrochna}, {Aptekar}, {Barthelmy}, {Bartolini}, {Beskin}, {Bondar},
  {Bremer}, {Campana}, {Castro-Tirado}, {Cucchiara}, {Cwiok}, {D'Avanzo},
  {D'Elia}, {Della Valle}, {de Ugarte Postigo}, {Dominik}, {Falcone}, {Fiore},
  {Fox}, {Frederiks}, {Fruchter}, {Fugazza}, {Garrett}, {Gehrels},
  {Golenetskii}, {Gomboc}, {Gorosabel}, {Greco}, {Guarnieri}, {Immler},
  {Jelinek}, {Kasprowicz}, {La Parola}, {Levan}, {Mangano}, {Mazets},
  {Molinari}, {Moretti}, {Nawrocki}, {Oleynik}, {Osborne}, {Pagani}, {Pandey},
  {Paragi}, {Perri}, {Piccioni}, {Ramirez-Ruiz}, {Roming}, {Steele}, {Strom},
  {Testa}, {Tosti}, {Ulanov}, {Wiersema}, {Wijers}, {Winters}, {Zarnecki},
  {Zerbi}, {M{\'e}sz{\'a}ros}, {Chincarini}, \&
  {Burrows}}]{2008Natur.455..183R}
{Racusin}, J.~L., {Karpov}, S.~V., {Sokolowski}, M., {et~al.} 2008, \nat, 455,
  183

\bibitem[{{Raghavan} {et~al.}(2010){Raghavan}, {McAlister}, {Henry}, {Latham},
  {Marcy}, {Mason}, {Gies}, {White}, \& {ten Brummelaar}}]{Rag2010}
{Raghavan}, D., {McAlister}, H.~A., {Henry}, T.~J., {et~al.} 2010, \apjs, 190,
  1

\bibitem[{{Rajoelimanana} {et~al.}(2011){Rajoelimanana}, {Charles}, \&
  {Udalski}}]{2011MNRAS.413.1600R}
{Rajoelimanana}, A.~F., {Charles}, P.~A., \& {Udalski}, A. 2011, \mnras, 413,
  1600

\bibitem[{{Rappaport} {et~al.}(2016){Rappaport}, {Gary}, {Kaye}, {Vanderburg},
  {Croll}, {Benni}, \& {Foote}}]{2016MNRAS.458.3904R}
{Rappaport}, S., {Gary}, B.~L., {Kaye}, T., {et~al.} 2016, \mnras, 458, 3904

\bibitem[{{Rasio} \& {Ford}(1996)}]{Rasio1996}
{Rasio}, F.~A., \& {Ford}, E.~B. 1996, Science, 274, 954

\bibitem[{Rauer {et~al.}(2014)Rauer, Catala, Aerts, Appourchaux, Benz,
  Brandeker, Christensen-Dalsgaard, Deleuil, Gizon, Goupil,
  {et~al.}}]{Rauer2014}
Rauer, H., Catala, C., Aerts, C., {et~al.} 2014, Experimental Astronomy, 38,
  249

\bibitem[{Raymond {et~al.}(2006)Raymond, Quinn, \& Lunine}]{Raymond2006}
Raymond, S.~N., Quinn, T., \& Lunine, J.~I. 2006, Icarus, 183, 265

\bibitem[{{Refsdal}(1966)}]{1966MNRAS.134..315R}
{Refsdal}, S. 1966, \mnras, 134, 315

\bibitem[{{Reinhold} \& {Hekker}(2020)}]{Reinhold2020}
{Reinhold}, T., \& {Hekker}, S. 2020, \aap, 635, A43

\bibitem[{{Rest} {et~al.}(2018){Rest}, {Garnavich}, {Khatami}, {Kasen},
  {Tucker}, {Shaya}, {Olling}, {Mushotzky}, {Zenteno}, {Margheim},
  {Strampelli}, {James}, {Smith}, {F{\"o}rster}, \&
  {Villar}}]{2018NatAs...2..307R}
{Rest}, A., {Garnavich}, P.~M., {Khatami}, D., {et~al.} 2018, Nature Astronomy,
  2, 307

\bibitem[{Rhodes {et~al.}(2017)Rhodes, Nichol, Aubourg, Bean, Boutigny, Bremer,
  Capak, Cardone, Carry, Conselice, {et~al.}}]{rhodes2017scientific}
Rhodes, J., Nichol, R.~C., Aubourg, {\'E}., {et~al.} 2017, The Astrophysical
  Journal Supplement Series, 233, 21

\bibitem[{{Ribas}(2006)}]{Rib2006}
{Ribas}, I. 2006, \apss, 304, 89

\bibitem[{{Ricker} {et~al.}(2015){Ricker}, {Winn}, {Vanderspek}, {Latham},
  {Bakos}, {Bean}, {Berta-Thompson}, {Brown}, {Buchhave}, {Butler}, {Butler},
  {Chaplin}, {Charbonneau}, {Christensen-Dalsgaard}, {Clampin}, {Deming},
  {Doty}, {De Lee}, {Dressing}, {Dunham}, {Endl}, {Fressin}, {Ge}, {Henning},
  {Holman}, {Howard}, {Ida}, {Jenkins}, {Jernigan}, {Johnson}, {Kaltenegger},
  {Kawai}, {Kjeldsen}, {Laughlin}, {Levine}, {Lin}, {Lissauer}, {MacQueen},
  {Marcy}, {McCullough}, {Morton}, {Narita}, {Paegert}, {Palle}, {Pepe},
  {Pepper}, {Quirrenbach}, {Rinehart}, {Sasselov}, {Sato}, {Seager},
  {Sozzetti}, {Stassun}, {Sullivan}, {Szentgyorgyi}, {Torres}, {Udry}, \&
  {Villasenor}}]{ricker2015a}
{Ricker}, G.~R., {Winn}, J.~N., {Vanderspek}, R., {et~al.} 2015, Journal of
  Astronomical Telescopes, Instruments, and Systems, 1, 014003

\bibitem[{{Riess} {et~al.}(1998){Riess}, {Filippenko}, {Challis},
  {Clocchiatti}, {Diercks}, {Garnavich}, {Gilliland}, {Hogan}, {Jha},
  {Kirshner}, {Leibundgut}, {Phillips}, {Reiss}, {Schmidt}, {Schommer},
  {Smith}, {Spyromilio}, {Stubbs}, {Suntzeff}, \&
  {Tonry}}]{1998AJ....116.1009R}
{Riess}, A.~G., {Filippenko}, A.~V., {Challis}, P., {et~al.} 1998, \aj, 116,
  1009

\bibitem[{{Rivinius} {et~al.}(2020){Rivinius}, {Baade}, {Hadrava}, {Heida}, \&
  {Klement}}]{Rivinius2020}
{Rivinius}, T., {Baade}, D., {Hadrava}, P., {Heida}, M., \& {Klement}, R. 2020,
  \aap, 637, L3

\bibitem[{{Rix} \& {Bovy}(2013)}]{Rix2013}
{Rix}, H.-W., \& {Bovy}, J. 2013, \aapr, 21, 61

\bibitem[{{Rodenbeck} {et~al.}(2020){Rodenbeck}, {Heller}, \&
  {Gizon}}]{Rodenbeck2020}
{Rodenbeck}, K., {Heller}, R., \& {Gizon}, L. 2020, \aap, 638, A43

\bibitem[{{Ro{\v{s}}kar} {et~al.}(2008){Ro{\v{s}}kar}, {Debattista}, {Quinn},
  {Stinson}, \& {Wadsley}}]{Roskar2008}
{Ro{\v{s}}kar}, R., {Debattista}, V.~P., {Quinn}, T.~R., {Stinson}, G.~S., \&
  {Wadsley}, J. 2008, \apjl, 684, L79

\bibitem[{{Rowe} {et~al.}(2015){Rowe}, {Jarvis}, {Mandelbaum}, {Bernstein},
  {Bosch}, {Simet}, {Meyers}, {Kacprzak}, {Nakajima}, {Zuntz}, {Miyatake},
  {Dietrich}, {Armstrong}, {Melchior}, \& {Gill}}]{GalSim}
{Rowe}, B.~T.~P., {Jarvis}, M., {Mandelbaum}, R., {et~al.} 2015, Astronomy and
  Computing, 10, 121

\bibitem[{{Ruiz-Lara} {et~al.}(2020){Ruiz-Lara}, {Gallart}, {Bernard}, \&
  {Cassisi}}]{Ruiz-Lara2020}
{Ruiz-Lara}, T., {Gallart}, C., {Bernard}, E.~J., \& {Cassisi}, S. 2020, Nature
  Astronomy, 4, 965

\bibitem[{Ryan {et~al.}(2017)Ryan, Sharkey, \& Woodward}]{ryan2017trojan}
Ryan, E.~L., Sharkey, B.~N., \& Woodward, C.~E. 2017, The Astronomical Journal,
  153, 116

\bibitem[{{Ryu} {et~al.}(2021){Ryu}, {Mr{\'o}z}, {Gould}, {Hwang}, {Kim},
  {Yee}, {Albrow}, {Chung}, {Jung}, {Shin}, {Shvartzvald}, {Zang}, {Cha},
  {Kim}, {Kim}, {Lee}, {Lee}, {Lee}, {Park}, {Han}, {Pogge}, {KMTNet
  Collaboration}, {Udalski}, {Poleski}, {Skowron}, {Szyma{\'n}ski},
  {Soszy{\'n}ski}, {Pietrukowicz}, {Koz{\l}owski}, {Ulaczyk}, {Rybicki},
  {Iwanek}, \& {OGLE Collaboration}}]{KB172820}
{Ryu}, Y.-H., {Mr{\'o}z}, P., {Gould}, A., {et~al.} 2021, \aj, 161, 126

\bibitem[{{Sahlholdt} {et~al.}(2022){Sahlholdt}, {Feltzing}, \&
  {Feuillet}}]{Sahlholdt2022}
{Sahlholdt}, C.~L., {Feltzing}, S., \& {Feuillet}, D.~K. 2022, \mnras, 510,
  4669

\bibitem[{Salabert {et~al.}(2018)Salabert, Régulo, Pérez~Hernández, \&
  García}]{Salabert2018_fshifts}
Salabert, D., Régulo, C., Pérez~Hernández, F., \& García, R.~A. 2018, A\&A,
  611, A84.
\newblock \url{http://adsabs.harvard.edu/abs/2018A%26A...611A..84S}

\bibitem[{{Salaris} {et~al.}(2016){Salaris}, {Cassisi}, \&
  {Pietrinferni}}]{salaris2016a}
{Salaris}, M., {Cassisi}, S., \& {Pietrinferni}, A. 2016, \aap, 590, A64

\bibitem[{{Sana} {et~al.}(2012){Sana}, {de Mink}, {de Koter}, {Langer},
  {Evans}, {Gieles}, {Gosset}, {Izzard}, {Le Bouquin}, \& {Schneider}}]{Sana12}
{Sana}, H., {de Mink}, S.~E., {de Koter}, A., {et~al.} 2012, Science, 337, 444

\bibitem[{{S{\'a}nchez} {et~al.}(2014){S{\'a}nchez}, {Rosales-Ortega},
  {Iglesias-P{\'a}ramo}, {Moll{\'a}}, {Barrera-Ballesteros}, {Marino},
  {P{\'e}rez}, {S{\'a}nchez-Blazquez}, {Gonz{\'a}lez Delgado}, {Cid Fernandes},
  {de Lorenzo-C{\'a}ceres}, {Mendez-Abreu}, {Galbany}, {Falcon-Barroso},
  {Miralles-Caballero}, {Husemann}, {Garc{\'\i}a-Benito}, {Mast}, {Walcher},
  {Gil de Paz}, {Garc{\'\i}a-Lorenzo}, {Jungwiert}, {V{\'\i}lchez},
  {J{\'\i}lkov{\'a}}, {Lyubenova}, {Cortijo-Ferrero}, {D{\'\i}az}, {Wisotzki},
  {M{\'a}rquez}, {Bland-Hawthorn}, {Ellis}, {van de Ven}, {Jahnke},
  {Papaderos}, {Gomes}, {Mendoza}, \& {L{\'o}pez-S{\'a}nchez}}]{Sanchez2014}
{S{\'a}nchez}, S.~F., {Rosales-Ortega}, F.~F., {Iglesias-P{\'a}ramo}, J.,
  {et~al.} 2014, \aap, 563, A49

\bibitem[{{Sanders} \& {Das}(2018)}]{Sanders2018}
{Sanders}, J.~L., \& {Das}, P. 2018, \mnras, 481, 4093

\bibitem[{{Santos} {et~al.}(2021{\natexlab{a}}){Santos}, {Breton}, {Mathur}, \&
  {Garc{\'\i}a}}]{Santos2021a}
{Santos}, A.~R.~G., {Breton}, S.~N., {Mathur}, S., \& {Garc{\'\i}a}, R.~A.
  2021{\natexlab{a}}, \apjs, 255, 17

\bibitem[{{Santos} {et~al.}(2021{\natexlab{b}}){Santos}, {Breton}, {Mathur}, \&
  {Garc{\'\i}a}}]{Santos2021}
---. 2021{\natexlab{b}}, \apjs, 255, 17

\bibitem[{{Santos} {et~al.}(2019){Santos}, {Garc{\'\i}a}, {Mathur}, {Bugnet},
  {van Saders}, {Metcalfe}, {Simonian}, \& {Pinsonneault}}]{Santos2019}
{Santos}, A.~R.~G., {Garc{\'\i}a}, R.~A., {Mathur}, S., {et~al.} 2019, \apjs,
  244, 21

\bibitem[{Santos {et~al.}(2018)Santos, Campante, Chaplin, Cunha, Lund, Kiefer,
  Salabert, Garcia, Davies, Elsworth, \& Howe}]{Santos2018_fshifts}
Santos, A. R.~G., Campante, T.~L., Chaplin, W.~J., {et~al.} 2018, ApJS, 237,
  17.
\newblock \url{http://adsabs.harvard.edu/abs/2018ApJS..237...17S}

\bibitem[{{Santos} {et~al.}(2001){Santos}, {Israelian}, \&
  {Mayor}}]{Santos:2001}
{Santos}, N.~C., {Israelian}, G., \& {Mayor}, M. 2001, \aap, 373, 1019

\bibitem[{{Saracino} {et~al.}(2022){Saracino}, {Kamann}, {Guarcello}, {Usher},
  {Bastian}, {Cabrera-Ziri}, {Gieles}, {Dreizler}, {Da Costa}, {Husser}, \&
  {H{\'e}nault-Brunet}}]{Saracino2022}
{Saracino}, S., {Kamann}, S., {Guarcello}, M.~G., {et~al.} 2022, \mnras, 511,
  2914

\bibitem[{{Schlaufman} {et~al.}(2018){Schlaufman}, {Thompson}, \&
  {Casey}}]{Schlaufman2018}
{Schlaufman}, K.~C., {Thompson}, I.~B., \& {Casey}, A.~R. 2018, \apj, 867, 98

\bibitem[{{Schlecker} {et~al.}(2021){Schlecker}, {Mordasini}, {Emsenhuber},
  {Klahr}, {Henning}, {Burn}, {Alibert}, \& {Benz}}]{Schlecker2021}
{Schlecker}, M., {Mordasini}, C., {Emsenhuber}, A., {et~al.} 2021, \aap, 656,
  A71

\bibitem[{{Schofield} {et~al.}(2019){Schofield}, {Chaplin}, {Huber},
  {Campante}, {Davies}, {Miglio}, {Ball}, {Appourchaux}, {Basu}, {Bedding},
  {Christensen-Dalsgaard}, {Creevey}, {Garc{\'\i}a}, {Handberg}, {Kawaler},
  {Kjeldsen}, {Latham}, {Lund}, {Metcalfe}, {Ricker}, {Serenelli}, {Silva
  Aguirre}, {Stello}, \& {Vanderspek}}]{Schofield2019a}
{Schofield}, M., {Chaplin}, W.~J., {Huber}, D., {et~al.} 2019, \apjs, 241, 12

\bibitem[{{Sch{\"o}nrich} \& {Binney}(2009)}]{Schonrich2009}
{Sch{\"o}nrich}, R., \& {Binney}, J. 2009, \mnras, 396, 203

\bibitem[{{Schwamb} {et~al.}(2013){Schwamb}, {Orosz}, {Carter}, {Welsh},
  {Fischer}, {Torres}, {Howard}, {Crepp}, {Keel}, {Lintott}, {Kaib}, {Terrell},
  {Gagliano}, {Jek}, {Parrish}, {Smith}, {Lynn}, {Simpson}, {Giguere}, \&
  {Schawinski}}]{Schwamb2013}
{Schwamb}, M.~E., {Orosz}, J.~A., {Carter}, J.~A., {et~al.} 2013, \apj, 768,
  127

\bibitem[{{Scott} {et~al.}(2021){Scott}, {Howell}, {Gnilka}, {Stephens},
  {Salinas}, {Matson}, {Furlan}, {Horch}, {Everett}, {Ciardi}, {Mills}, \&
  {Quigley}}]{Scott2021}
{Scott}, N.~J., {Howell}, S.~B., {Gnilka}, C.~L., {et~al.} 2021, Frontiers in
  Astronomy and Space Sciences, 8, 138

\bibitem[{{Seifahrt} {et~al.}(2020){Seifahrt}, {Bean}, {St{\"u}rmer}, {Kasper},
  {Gers}, {Schwab}, {Zechmeister}, {Stef{\'a}nsson}, {Montet}, {Dos Santos},
  {Peck}, {White}, \& {Tapia}}]{seifahrt20}
{Seifahrt}, A., {Bean}, J.~L., {St{\"u}rmer}, J., {et~al.} 2020, in Society of
  Photo-Optical Instrumentation Engineers (SPIE) Conference Series, Vol. 11447,
  Society of Photo-Optical Instrumentation Engineers (SPIE) Conference Series,
  114471F

\bibitem[{{Sellwood} \& {Binney}(2002)}]{Sellwood2002}
{Sellwood}, J.~A., \& {Binney}, J.~J. 2002, \mnras, 336, 785

\bibitem[{{Serenelli} {et~al.}(2017){Serenelli}, {Johnson}, {Huber},
  {Pinsonneault}, {Ball}, {Tayar}, {Silva Aguirre}, {Basu}, {Troup}, {Hekker},
  {Kallinger}, {Stello}, {Davies}, {Lund}, {Mathur}, {Mosser}, {Stassun},
  {Chaplin}, {Elsworth}, {Garc{\'\i}a}, {Handberg}, {Holtzman}, {Hearty},
  {Garc{\'\i}a-Hern{\'a}ndez}, {Gaulme}, \& {Zamora}}]{Serenelli2017}
{Serenelli}, A., {Johnson}, J., {Huber}, D., {et~al.} 2017, \apjs, 233, 23

\bibitem[{{Sestito} {et~al.}(2019){Sestito}, {Longeard}, {Martin},
  {Starkenburg}, {Fouesneau}, {Gonz{\'a}lez Hern{\'a}ndez}, {Arentsen},
  {Ibata}, {Aguado}, {Carlberg}, {Jablonka}, {Navarro}, {Tolstoy}, \&
  {Venn}}]{Sestito2019}
{Sestito}, F., {Longeard}, N., {Martin}, N.~F., {et~al.} 2019, \mnras, 484,
  2166

\bibitem[{{Shapiro} {et~al.}(1983){Shapiro}, {Teukolsky}, \&
  {Lightman}}]{Shapiro83}
{Shapiro}, S.~L., {Teukolsky}, S.~A., \& {Lightman}, A.~P. 1983, Physics Today,
  36, 89

\bibitem[{{Shappee} {et~al.}(2019){Shappee}, {Holoien}, {Drout}, {Auchettl},
  {Stritzinger}, {Kochanek}, {Stanek}, {Shaya}, {Narayan}, {ASAS-SN}, {Brown},
  {Bose}, {Bersier}, {Brimacombe}, {Chen}, {Dong}, {Holmbo}, {Katz},
  {Mu{\~n}oz}, {Mutel}, {Post}, {Prieto}, {Shields}, {Tallon}, {Thompson},
  {Vallely}, {Villanueva}, {ATLAS}, {Denneau}, {Flewelling}, {Heinze}, {Smith},
  {Stalder}, {Tonry}, {Weiland}, {Kepler/K2}, {Barclay}, {Barentsen}, {Cody},
  {Dotson}, {Foerster}, {Garnavich}, {Gully-Santiago}, {Hedges}, {Howell},
  {Kasen}, {Margheim}, {Mushotzky}, {Rest}, {Tucker}, {Villar}, {Zenteno},
  {Kepler Spacecraft Team}, {Beerman}, {Bjella}, {Castillo}, {Coughlin},
  {Elsaesser}, {Flynn}, {Gangopadhyay}, {Griest}, {Hanley}, {Kampmeier},
  {Kloetzel}, {Kohnert}, {Labonde}, {Larsen}, {Larson}, {McCalmont-Everton},
  {McGinn}, {Migliorini}, {Moffatt}, {Muszynski}, {Nystrom}, {Osborne},
  {Packard}, {Peterson}, {Redick}, {Reedy}, {Ross}, {Spencer}, {Steward}, {Van
  Cleve}, {Cardoso}, {Weschler}, {Wheaton}, {Pan-STARRS}, {Bulger}, {Chambers},
  {Flewelling}, {Huber}, {Lowe}, {Magnier}, {Schultz}, {Waters}, {Willman},
  {PTSS/TNTS}, {Baron}, {Chen}, {Derkacy}, {Huang}, {Li}, {Li}, {Li}, {Mo},
  {Rui}, {Sai}, {Wang}, {Wang}, {Wang}, {Xiang}, {Zhang}, {Zhang}, {Zhang},
  {Zhang}, {Zhang}, {Zhao}, {Brown}, {Hermes}, {Nordin}, {Points}, {S{\'o}dor},
  {Strampelli}, \& {Zenteno}}]{2019ApJ...870...13S}
{Shappee}, B.~J., {Holoien}, T.~W.~S., {Drout}, M.~R., {et~al.} 2019, \apj,
  870, 13

\bibitem[{{Sharma} {et~al.}(2021){Sharma}, {Hayden}, \&
  {Bland-Hawthorn}}]{Sharma2021}
{Sharma}, S., {Hayden}, M.~R., \& {Bland-Hawthorn}, J. 2021, \mnras, 507, 5882

\bibitem[{{Sharma} {et~al.}(2022){Sharma}, {Hayden}, {Bland-Hawthorn},
  {Stello}, {Buder}, {Zinn}, {Spina}, {Kallinger}, {Asplund}, {De Silva},
  {D'Orazi}, {Freeman}, {Kos}, {Lewis}, {Lin}, {Lind}, {Martell},
  {Schlesinger}, {Simpson}, {Zucker}, {Zwitter}, {Chen}, {Cotar}, {Kafle},
  {Khanna}, {Tepper-Garcia}, {Wang}, \& {Wittenmyer}}]{Sharma2022}
{Sharma}, S., {Hayden}, M.~R., {Bland-Hawthorn}, J., {et~al.} 2022, \mnras,
  510, 734

\bibitem[{{Shaver} {et~al.}(1983){Shaver}, {McGee}, {Newton}, {Danks}, \&
  {Pottasch}}]{Shaver1983}
{Shaver}, P.~A., {McGee}, R.~X., {Newton}, L.~M., {Danks}, A.~C., \&
  {Pottasch}, S.~R. 1983, \mnras, 204, 53

\bibitem[{{Shenar} {et~al.}(2020){Shenar}, {Bodensteiner}, {Abdul-Masih},
  {Fabry}, {Mahy}, {Marchant}, {Banyard}, {Bowman}, {Dsilva}, {Hawcroft},
  {Reggiani}, \& {Sana}}]{Shenar2020}
{Shenar}, T., {Bodensteiner}, J., {Abdul-Masih}, M., {et~al.} 2020, \aap, 639,
  L6

\bibitem[{{Shvartzvald} {et~al.}(2016){Shvartzvald}, {Maoz}, {Udalski}, {Sumi},
  {Friedmann}, {Kaspi}, {Poleski}, {Szyma{\'n}ski}, {Skowron}, {Koz{\l}owski},
  {Wyrzykowski}, {Mr{\'o}z}, {Pietrukowicz}, {Pietrzy{\'n}ski},
  {Soszy{\'n}ski}, {Ulaczyk}, {Abe}, {Barry}, {Bennett}, {Bhattacharya},
  {Bond}, {Freeman}, {Inayama}, {Itow}, {Koshimoto}, {Ling}, {Masuda}, {Fukui},
  {Matsubara}, {Muraki}, {Ohnishi}, {Rattenbury}, {Saito}, {Sullivan},
  {Suzuki}, {Tristram}, {Wakiyama}, \& {Yonehara}}]{Wise}
{Shvartzvald}, Y., {Maoz}, D., {Udalski}, A., {et~al.} 2016, \mnras, 457, 4089

\bibitem[{{Sigurdsson} {et~al.}(2003){Sigurdsson}, {Richer}, {Hansen},
  {Stairs}, \& {Thorsett}}]{2003Sci...301..193S}
{Sigurdsson}, S., {Richer}, H.~B., {Hansen}, B.~M., {Stairs}, I.~H., \&
  {Thorsett}, S.~E. 2003, Science, 301, 193

\bibitem[{{Silva Aguirre} {et~al.}(2017){Silva Aguirre}, {Lund}, {Antia},
  {Ball}, {Basu}, {Christensen-Dalsgaard}, {Lebreton}, {Reese}, {Verma},
  {Casagrande}, {Justesen}, {Mosumgaard}, {Chaplin}, {Bedding}, {Davies},
  {Handberg}, {Houdek}, {Huber}, {Kjeldsen}, {Latham}, {White}, {Coelho},
  {Miglio}, \& {Rendle}}]{Silva_Aguirre2017}
{Silva Aguirre}, V., {Lund}, M.~N., {Antia}, H.~M., {et~al.} 2017, \apj, 835,
  173

\bibitem[{{Silva Aguirre} {et~al.}(2018){Silva Aguirre}, {Bojsen-Hansen},
  {Slumstrup}, {Casagrande}, {Kawata}, {Ciuc{\v{a}}}, {Handberg}, {Lund},
  {Mosumgaard}, {Huber}, {Johnson}, {Pinsonneault}, {Serenelli}, {Stello},
  {Tayar}, {Bird}, {Cassisi}, {Hon}, {Martig}, {Nissen}, {Rix},
  {Sch{\"o}nrich}, {Sahlholdt}, {Trick}, \& {Yu}}]{Silva_Aguirre2018}
{Silva Aguirre}, V., {Bojsen-Hansen}, M., {Slumstrup}, D., {et~al.} 2018,
  \mnras, 475, 5487

\bibitem[{{Silvotti} {et~al.}(2007){Silvotti}, {Schuh}, {Janulis}, {Solheim},
  {Bernabei}, {{\O}stensen}, {Oswalt}, {Bruni}, {Gualandi}, {Bonanno},
  {Vauclair}, {Reed}, {Chen}, {Leibowitz}, {Paparo}, {Baran}, {Charpinet},
  {Dolez}, {Kawaler}, {Kurtz}, {Moskalik}, {Riddle}, \&
  {Zola}}]{2007Natur.449..189S}
{Silvotti}, R., {Schuh}, S., {Janulis}, R., {et~al.} 2007, \nat, 449, 189

\bibitem[{{Skumanich}(1972)}]{Skumanich1972}
{Skumanich}, A. 1972, \apj, 171, 565

\bibitem[{{Smartt}(2015)}]{Smartt2015}
{Smartt}, S.~J. 2015, \pasa, 32, e016

\bibitem[{{Smartt} {et~al.}(2009){Smartt}, {Eldridge}, {Crockett}, \&
  {Maund}}]{Smartt2009}
{Smartt}, S.~J., {Eldridge}, J.~J., {Crockett}, R.~M., \& {Maund}, J.~R. 2009,
  \mnras, 395, 1409

\bibitem[{{Smith} {et~al.}(2018){Smith}, {Mushotzky}, {Boyd}, {Malkan},
  {Howell}, \& {Gelino}}]{2018ApJ...857..141S}
{Smith}, K.~L., {Mushotzky}, R.~F., {Boyd}, P.~T., {et~al.} 2018, \apj, 857,
  141

\bibitem[{{Snaith} {et~al.}(2015){Snaith}, {Haywood}, {Di Matteo}, {Lehnert},
  {Combes}, {Katz}, \& {G{\'o}mez}}]{Snaith2015}
{Snaith}, O., {Haywood}, M., {Di Matteo}, P., {et~al.} 2015, \aap, 578, A87

\bibitem[{{Snaith} {et~al.}(2014){Snaith}, {Haywood}, {Di Matteo}, {Lehnert},
  {Combes}, {Katz}, \& {G{\'o}mez}}]{Snaith2014}
{Snaith}, O.~N., {Haywood}, M., {Di Matteo}, P., {et~al.} 2014, \apjl, 781, L31

\bibitem[{{Snellen} \& {Brown}(2018)}]{snellen18}
{Snellen}, I.~A.~G., \& {Brown}, A.~G.~A. 2018, Nature Astronomy, 2, 883

\bibitem[{{Socia} {et~al.}(2020){Socia}, {Welsh}, {Orosz}, {Cochran}, {Endl},
  {Quarles}, {Short}, {Torres}, {Windmiller}, \& {Yenawine}}]{Socia2020}
{Socia}, Q.~J., {Welsh}, W.~F., {Orosz}, J.~A., {et~al.} 2020, \aj, 159, 94

\bibitem[{{Soderblom}(2010)}]{Soderblom2010}
{Soderblom}, D.~R. 2010, \araa, 48, 581

\bibitem[{Sozzetti {et~al.}(2013)Sozzetti, Giacobbe, Lattanzi, Micela,
  Morbidelli, \& Tinetti}]{10.1093/mnras/stt1899}
Sozzetti, A., Giacobbe, P., Lattanzi, M.~G., {et~al.} 2013, Monthly Notices of
  the Royal Astronomical Society, 437, 497.
\newblock \url{https://doi.org/10.1093/mnras/stt1899}

\bibitem[{{Spergel} {et~al.}(2015){Spergel}, {Gehrels}, {Baltay}, {Bennett},
  {Breckinridge}, {Donahue}, {Dressler}, {Gaudi}, {Greene}, {Guyon}, {Hirata},
  {Kalirai}, {Kasdin}, {Macintosh}, {Moos}, {Perlmutter}, {Postman},
  {Rauscher}, {Rhodes}, {Wang}, {Weinberg}, {Benford}, {Hudson}, {Jeong},
  {Mellier}, {Traub}, {Yamada}, {Capak}, {Colbert}, {Masters}, {Penny},
  {Savransky}, {Stern}, {Zimmerman}, {Barry}, {Bartusek}, {Carpenter}, {Cheng},
  {Content}, {Dekens}, {Demers}, {Grady}, {Jackson}, {Kuan}, {Kruk}, {Melton},
  {Nemati}, {Parvin}, {Poberezhskiy}, {Peddie}, {Ruffa}, {Wallace}, {Whipple},
  {Wollack}, \& {Zhao}}]{Spergel2015}
{Spergel}, D., {Gehrels}, N., {Baltay}, C., {et~al.} 2015, arXiv e-prints,
  arXiv:1503.03757

\bibitem[{{Spitoni} {et~al.}(2015){Spitoni}, {Romano}, {Matteucci}, \&
  {Ciotti}}]{Spitoni2015}
{Spitoni}, E., {Romano}, D., {Matteucci}, F., \& {Ciotti}, L. 2015, \apj, 802,
  129

\bibitem[{{Spitoni} {et~al.}(2019){Spitoni}, {Silva Aguirre}, {Matteucci},
  {Calura}, \& {Grisoni}}]{Spitoni2019}
{Spitoni}, E., {Silva Aguirre}, V., {Matteucci}, F., {Calura}, F., \&
  {Grisoni}, V. 2019, \aap, 623, A60

\bibitem[{{Spitoni} {et~al.}(2021){Spitoni}, {Verma}, {Silva Aguirre},
  {Vincenzo}, {Matteucci}, {Vai{\v{c}}ekauskait{\.{e}}}, {Palla}, {Grisoni}, \&
  {Calura}}]{Spitoni2021}
{Spitoni}, E., {Verma}, K., {Silva Aguirre}, V., {et~al.} 2021, \aap, 647, A73

\bibitem[{{Steinmetz} {et~al.}(2006){Steinmetz}, {Zwitter}, {Siebert},
  {Watson}, {Freeman}, {Munari}, {Campbell}, {Williams}, {Seabroke}, {Wyse},
  {Parker}, {Bienaym{\'e}}, {Roeser}, {Gibson}, {Gilmore}, {Grebel}, {Helmi},
  {Navarro}, {Burton}, {Cass}, {Dawe}, {Fiegert}, {Hartley}, {Russell},
  {Saunders}, {Enke}, {Bailin}, {Binney}, {Bland-Hawthorn}, {Boeche}, {Dehnen},
  {Eisenstein}, {Evans}, {Fiorucci}, {Fulbright}, {Gerhard}, {Jauregi}, {Kelz},
  {Mijovi{\'c}}, {Minchev}, {Parmentier}, {Pe{\~n}arrubia}, {Quillen}, {Read},
  {Ruchti}, {Scholz}, {Siviero}, {Smith}, {Sordo}, {Veltz}, {Vidrih}, {von
  Berlepsch}, {Boyle}, \& {Schilbach}}]{Steinmetz2006}
{Steinmetz}, M., {Zwitter}, T., {Siebert}, A., {et~al.} 2006, \aj, 132, 1645

\bibitem[{{Stello} {et~al.}(2016){Stello}, {Cantiello}, {Fuller}, {Huber},
  {Garc{\'{\i}}a}, {Bedding}, {Bildsten}, \& {Silva Aguirre}}]{stello2016a}
{Stello}, D., {Cantiello}, M., {Fuller}, J., {et~al.} 2016, \nat, 529, 364

\bibitem[{Stevenson(2019)}]{Stevenson2019failure}
Stevenson, D.~S. 2019, in Red Dwarfs (Springer), 285--312

\bibitem[{Stone(1989)}]{stone1989comparison}
Stone, R.~C. 1989, The Astronomical Journal, 97, 1227

\bibitem[{{Sumi} {et~al.}(2011){Sumi}, {Kamiya}, {Bennett}, {Bond}, {Abe},
  {Botzler}, {Fukui}, {Furusawa}, {Hearnshaw}, {Itow}, {Kilmartin}, {Korpela},
  {Lin}, {Ling}, {Masuda}, {Matsubara}, {Miyake}, {Motomura}, {Muraki},
  {Nagaya}, {Nakamura}, {Ohnishi}, {Okumura}, {Perrott}, {Rattenbury}, {Saito},
  {Sako}, {Sullivan}, {Sweatman}, {Tristram}, {}, {Szyma{\'n}ski}, {Kubiak},
  {Pietrzy{\'n}ski}, {Poleski}, {Soszy{\'n}ski}, {Wyrzykowski}, {Ulaczyk}, \&
  {Microlensing Observations in Astrophysics (MOA) Collaboration}}]{Sumi2011}
{Sumi}, T., {Kamiya}, K., {Bennett}, D.~P., {et~al.} 2011, \nat, 473, 349

\bibitem[{{Sutherland} \& {Kratter}(2019)}]{Sutherland19}
{Sutherland}, A.~P., \& {Kratter}, K.~M. 2019, \mnras, 487, 3288

\bibitem[{{Suzuki} {et~al.}(2016){Suzuki}, {Bennett}, {Sumi}, {Bond}, {Rogers},
  {Abe}, {Asakura}, {Bhattacharya}, {Donachie}, {Freeman}, {Fukui}, {Hirao},
  {Itow}, {Koshimoto}, {Li}, {Ling}, {Masuda}, {Matsubara}, {Muraki},
  {Nagakane}, {Onishi}, {Oyokawa}, {Rattenbury}, {Saito}, {Sharan}, {Shibai},
  {Sullivan}, {Tristram}, {Yonehara}, \& {MOA Collaboration}}]{Suzuki2016}
{Suzuki}, D., {Bennett}, D.~P., {Sumi}, T., {et~al.} 2016, \apj, 833, 145

\bibitem[{{Suzuki} {et~al.}(2018){Suzuki}, {Bennett}, {Ida}, {Mordasini},
  {Bhattacharya}, {Bond}, {Donachie}, {Fukui}, {Hirao}, {Koshimoto},
  {Miyazaki}, {Nagakane}, {Ranc}, {Rattenbury}, {Sumi}, {Alibert}, \&
  {Lin}}]{Suzuki2018}
{Suzuki}, D., {Bennett}, D.~P., {Ida}, S., {et~al.} 2018, \apj, 869, L34

\bibitem[{{Szab{\'o}} {et~al.}(2010){Szab{\'o}}, {Koll{\'a}th}, {Moln{\'a}r},
  {Kolenberg}, {Kurtz}, {Bryson}, {Benk{\H o}}, {Christensen-Dalsgaard},
  {Kjeldsen}, {Borucki}, {Koch}, {Twicken}, {Chadid}, {di Criscienzo}, {Jeon},
  {Moskalik}, {Nemec}, \& {Nuspl}}]{Szabo2010}
{Szab{\'o}}, R., {Koll{\'a}th}, Z., {Moln{\'a}r}, L., {et~al.} 2010, \mnras,
  409, 1244

\bibitem[{Szab{\'o} {et~al.}(2015)Szab{\'o}, S{\'a}rneczky, Szab{\'o}, P{\'a}l,
  Kiss, Cs{\'a}k, Ill{\'e}s, R{\'a}cz, \& Kiss}]{szabo2015main}
Szab{\'o}, R., S{\'a}rneczky, K., Szab{\'o}, G.~M., {et~al.} 2015, The
  Astronomical Journal, 149, 112

\bibitem[{Szab{\'o} {et~al.}(2016)Szab{\'o}, P{\'a}l, S{\'a}rneczky, Szab{\'o},
  Moln{\'a}r, Kiss, Hanyecz, Plachy, \& Kiss}]{szabo2016uninterrupted}
Szab{\'o}, R., P{\'a}l, A., S{\'a}rneczky, K., {et~al.} 2016, Astronomy \&
  Astrophysics, 596, A40

\bibitem[{{Szewczuk} \& {Daszy{\'n}ska-Daszkiewicz}(2018)}]{Szewczuk2018a}
{Szewczuk}, W., \& {Daszy{\'n}ska-Daszkiewicz}, J. 2018, \mnras, 478, 2243

\bibitem[{{Szewczuk} {et~al.}(2021){Szewczuk}, {Walczak}, \&
  {Daszy{\'n}ska-Daszkiewicz}}]{Szewczuk2021a}
{Szewczuk}, W., {Walczak}, P., \& {Daszy{\'n}ska-Daszkiewicz}, J. 2021, \mnras,
  503, 5894

\bibitem[{{Tailo} {et~al.}(2022){Tailo}, {Corsaro}, {Miglio}, {Montalb{\'a}n},
  {Brogaard}, {Milone}, {Stokholm}, {Casali}, \& {Bragaglia}}]{tailo++2022-m4}
{Tailo}, M., {Corsaro}, E., {Miglio}, A., {et~al.} 2022, arXiv e-prints,
  arXiv:2205.06645

\bibitem[{{Tamura} {et~al.}(2018){Tamura}, {Takato}, {Shimono}, {Moritani},
  {Yabe}, {Ishizuka}, {Kamata}, {Ueda}, {Aghazarian}, {Arnouts}, {Barkhouser},
  {Balard}, {Barette}, {Belhadi}, {Burnham}, {Caplar}, {Carr}, {Chabaud},
  {Chang}, {Chen}, {Chou}, {Chu}, {Cohen}, {de Almeida}, {de Oliveira}, {de
  Oliveira}, {Dekany}, {Dohlen}, {dos Santos}, {dos Santos}, {Ellis},
  {Fabricius}, {Ferreira}, {Furusawa}, {Garcia-Carpio}, {Golebiowski}, {Gross},
  {Gunn}, {Hammond}, {Harding}, {Hart}, {Heckman}, {Ho}, {Hope}, {Hover},
  {Hsu}, {Hu}, {Huang}, {Jamal}, {Jaquet}, {Jeschke}, {Jing}, {Kado-Fong},
  {Karr}, {Kimura}, {King}, {Koike}, {Komatsu}, {Le Brun}, {Le F{\`e}vre}, {Le
  Fur}, {Le Mignant}, {Ling}, {Loomis}, {Lupton}, {Madec}, {Mao}, {Marchesini},
  {Marrara}, {Medvedev}, {Mineo}, {Minowa}, {Murayama}, {Murray}, {Ohyama},
  {Onodera}, {Orndorff}, {Pascal}, {Peebles}, {Pernot}, {Pourcelot}, {Reiley},
  {Reinecke}, {Roberts}, {Rosa}, {Rousselle}, {Schmitt}, {Schwochert},
  {Seiffert}, {Siddiqui}, {Smee}, {Sodr{\'e}}, {Steinkraus}, {Strauss},
  {Surace}, {Tait}, {Takada}, {Tamura}, {Tanaka}, {Tanaka}, {Thakar},
  {Verducci}, {Vibert}, {Wang}, {Wang}, {Wen}, {Werner}, {Yamada}, {Yan},
  {Yasuda}, {Yoshida}, \& {Yoshida}}]{Tamura2018}
{Tamura}, N., {Takato}, N., {Shimono}, A., {et~al.} 2018, in Society of
  Photo-Optical Instrumentation Engineers (SPIE) Conference Series, Vol. 10702,
  Ground-based and Airborne Instrumentation for Astronomy VII, ed. C.~J.
  {Evans}, L.~{Simard}, \& H.~{Takami}, 107021C

\bibitem[{{Tanvir} {et~al.}(2013){Tanvir}, {Levan}, {Fruchter}, {Hjorth},
  {Hounsell}, {Wiersema}, \& {Tunnicliffe}}]{2013Natur.500..547T}
{Tanvir}, N.~R., {Levan}, A.~J., {Fruchter}, A.~S., {et~al.} 2013, \nat, 500,
  547

\bibitem[{Tarter {et~al.}(2007)Tarter, Backus, Mancinelli, Aurnou, Backman,
  Basri, Boss, Clarke, Deming, Doyle, {et~al.}}]{tarter2007reappraisal}
Tarter, J.~C., Backus, P.~R., Mancinelli, R.~L., {et~al.} 2007, Astrobiology,
  7, 30

\bibitem[{{Teachey} \& {Kipping}(2018)}]{Teachey2018}
{Teachey}, A., \& {Kipping}, D.~M. 2018, Science Advances, 4, eaav1784

\bibitem[{{Teske} {et~al.}(2021){Teske}, {Wang}, {Wolfgang}, {Gan},
  {Plotnykov}, {Armstrong}, {Butler}, {Cale}, {Crane}, {Howard}, {Jensen},
  {Law}, {Shectman}, {Plavchan}, {Valencia}, {Vanderburg}, {Ricker},
  {Vanderspek}, {Latham}, {Seager}, {Winn}, {Jenkins}, {Adibekyan}, {Barrado},
  {Barros}, {Benkhaldoun}, {Brown}, {Bryant}, {Burt}, {Caldwell},
  {Charbonneau}, {Cloutier}, {Collins}, {Collins}, {Colon}, {Conti},
  {Demangeon}, {Eastman}, {Elmufti}, {Feng}, {Flowers}, {Guerrero},
  {Hojjatpanah}, {Irwin}, {Isopi}, {Lillo-Box}, {Mallia}, {Massey}, {Mori},
  {Mullally}, {Narita}, {Nishiumi}, {Osborn}, {Paegert}, {de Leon}, {Quinn},
  {Reefe}, {Schwarz}, {Shporer}, {Soubkiou}, {Sousa}, {Stockdale}, {Str{\o}m},
  {Tan}, {Tang}, {Tenenbaum}, {Wheatley}, {Wittrock}, {Yahalomi}, \&
  {Zohrabi}}]{teske2021}
{Teske}, J., {Wang}, S.~X., {Wolfgang}, A., {et~al.} 2021, \apjs, 256, 33

\bibitem[{{Teske} {et~al.}(2019){Teske}, {Thorngren}, {Fortney}, {Hinkel}, \&
  {Brewer}}]{Teske:2019}
{Teske}, J.~K., {Thorngren}, D., {Fortney}, J.~J., {Hinkel}, N., \& {Brewer},
  J.~M. 2019, \aj, 158, 239

\bibitem[{{Tetarenko} {et~al.}(2016){Tetarenko}, {Sivakoff}, {Heinke}, \&
  {Gladstone}}]{Tetarenko16}
{Tetarenko}, B.~E., {Sivakoff}, G.~R., {Heinke}, C.~O., \& {Gladstone}, J.~C.
  2016, \apjs, 222, 15

\bibitem[{Thomas {et~al.}(2019)Thomas, Chaplin, Davies, Howe, Santos, Elsworth,
  Miglio, Campante, \& Cunha}]{Thomas2019_activeLat}
Thomas, A. E.~L., Chaplin, W.~J., Davies, G.~R., {et~al.} 2019, MNRAS, 485,
  3857.
\newblock \url{https://academic.oup.com/mnras/article/485/3/3857/5398537}

\bibitem[{{Thompson} {et~al.}(2019){Thompson}, {Kochanek}, {Stanek}, {Badenes},
  {Post}, {Jayasinghe}, {Latham}, {Bieryla}, {Esquerdo}, {Berlind}, {Calkins},
  {Tayar}, {Lindegren}, {Johnson}, {Holoien}, {Auchettl}, \&
  {Covey}}]{Thompson19}
{Thompson}, T.~A., {Kochanek}, C.~S., {Stanek}, K.~Z., {et~al.} 2019, Science,
  366, 637

\bibitem[{{Thorsett} {et~al.}(1993){Thorsett}, {Arzoumanian}, \&
  {Taylor}}]{1993ApJ...412L..33T}
{Thorsett}, S.~E., {Arzoumanian}, Z., \& {Taylor}, J.~H. 1993, \apjl, 412, L33

\bibitem[{{Timmes} {et~al.}(1996){Timmes}, {Woosley}, \& {Weaver}}]{Timmes96}
{Timmes}, F.~X., {Woosley}, S.~E., \& {Weaver}, T.~A. 1996, \apj, 457, 834

\bibitem[{{Ting} {et~al.}(2019){Ting}, {Conroy}, {Rix}, \&
  {Cargile}}]{Ting2019}
{Ting}, Y.-S., {Conroy}, C., {Rix}, H.-W., \& {Cargile}, P. 2019, \apj, 879, 69

\bibitem[{{Ting} \& {Rix}(2019)}]{Ting2019b}
{Ting}, Y.-S., \& {Rix}, H.-W. 2019, \apj, 878, 21

\bibitem[{{Toyouchi} \& {Chiba}(2018)}]{Toyouchi2018}
{Toyouchi}, D., \& {Chiba}, M. 2018, \apj, 855, 104

\bibitem[{{Tremaine} \& {Dong}(2012)}]{Tremaine2012}
{Tremaine}, S., \& {Dong}, S. 2012, \aj, 143, 94

\bibitem[{{Turner} {et~al.}(2021){Turner}, {Zarka}, {Grie{\ss}meier}, {Lazio},
  {Cecconi}, {Emilio Enriquez}, {Girard}, {Jayawardhana}, {Lamy}, {Nichols}, \&
  {de Pater}}]{turner21}
{Turner}, J.~D., {Zarka}, P., {Grie{\ss}meier}, J.-M., {et~al.} 2021, \aap,
  645, A59

\bibitem[{{Udalski} {et~al.}(2018){Udalski}, {Ryu}, {Sajadian}, {Gould},
  {Mr{\'o}z}, {Poleski}, {Szyma{\'n}ski}, {Skowron}, {Soszy{\'n}ski},
  {Koz{\l}owski}, {Pietrukowicz}, {Ulaczyk}, {Pawlak}, {Rybicki}, {Iwanek},
  {Albrow}, {Chung}, {Han}, {Hwang}, {Jung}, {}, {Shvartzvald}, {Yee}, {Zang},
  {Zhu}, {Cha}, {Kim}, {Kim}, {Kim}, {Lee}, {Lee}, {Lee}, {Park}, {Pogge},
  {Bozza}, {Dominik}, {Helling}, {Hundertmark}, {J{\o}rgensen},
  {Longa-Pe{\~n}a}, {Lowry}, {Burgdorf}, {Campbell-White}, {Ciceri}, {Evans},
  {Figuera Jaimes}, {Fujii}, {Haikala}, {Henning}, {Hinse}, {Mancini},
  {Peixinho}, {Rahvar}, {Rabus}, {Skottfelt}, {Snodgrass}, {Southworth}, \&
  {von Essen}}]{OB171434}
{Udalski}, A., {Ryu}, Y.-H., {Sajadian}, S., {et~al.} 2018, \actaa, 68, 1

\bibitem[{{Uehara} {et~al.}(2016){Uehara}, {Kawahara}, {Masuda}, {Yamada}, \&
  {Aizawa}}]{Uehara2016}
{Uehara}, S., {Kawahara}, H., {Masuda}, K., {Yamada}, S., \& {Aizawa}, M. 2016,
  \apj, 822, 2

\bibitem[{{Uribe} {et~al.}(2011){Uribe}, {Klahr}, {Flock}, \&
  {Henning}}]{Uribe2011}
{Uribe}, A.~L., {Klahr}, H., {Flock}, M., \& {Henning}, T. 2011, \apj, 736, 85

\bibitem[{{Valcin} {et~al.}(2020){Valcin}, {Bernal}, {Jimenez}, {Verde}, \&
  {Wandelt}}]{Valcin2020}
{Valcin}, D., {Bernal}, J.~L., {Jimenez}, R., {Verde}, L., \& {Wandelt}, B.~D.
  2020, \jcap, 2020, 002

\bibitem[{{Van Beeck} {et~al.}(2021){Van Beeck}, {Bowman}, {Pedersen}, {Van
  Reeth}, {Van Hoolst}, \& {Aerts}}]{VanBeeck2021a}
{Van Beeck}, J., {Bowman}, D.~M., {Pedersen}, M.~G., {et~al.} 2021, \aap, 655,
  A59

\bibitem[{Van~Eylen {et~al.}(2019)Van~Eylen, Albrecht, Huang, MacDonald,
  Dawson, Cai, Foreman-Mackey, Lundkvist, Aguirre, Snellen, {et~al.}}]{Van2019}
Van~Eylen, V., Albrecht, S., Huang, X., {et~al.} 2019, The Astronomical
  Journal, 157, 61

\bibitem[{{Van Hoolst}(1994)}]{VanHoolst1994b}
{Van Hoolst}, T. 1994, \aap, 292, 471

\bibitem[{{Van Reeth} {et~al.}(2016){Van Reeth}, {Tkachenko}, \&
  {Aerts}}]{VanReeth2016a}
{Van Reeth}, T., {Tkachenko}, A., \& {Aerts}, C. 2016, \aap, 593, A120

\bibitem[{{van Saders} {et~al.}(2016){van Saders}, {Ceillier}, {Metcalfe},
  {Silva Aguirre}, {Pinsonneault}, {Garc{\'\i}a}, {Mathur}, \&
  {Davies}}]{van_Saders2016}
{van Saders}, J.~L., {Ceillier}, T., {Metcalfe}, T.~S., {et~al.} 2016, \nat,
  529, 181

\bibitem[{{van Velzen} {et~al.}(2011){van Velzen}, {Farrar}, {Gezari},
  {Morrell}, {Zaritsky}, {{\"O}stman}, {Smith}, {Gelfand}, \&
  {Drake}}]{2011ApJ...741...73V}
{van Velzen}, S., {Farrar}, G.~R., {Gezari}, S., {et~al.} 2011, \apj, 741, 73

\bibitem[{{Vanden Berk} {et~al.}(2004){Vanden Berk}, {Wilhite}, {Kron},
  {Anderson}, {Brunner}, {Hall}, {Ivezi{\'c}}, {Richards}, {Schneider}, {York},
  {Brinkmann}, {Lamb}, {Nichol}, \& {Schlegel}}]{2004ApJ...601..692V}
{Vanden Berk}, D.~E., {Wilhite}, B.~C., {Kron}, R.~G., {et~al.} 2004, \apj,
  601, 692

\bibitem[{{VandenBerg} {et~al.}(2014){VandenBerg}, {Bond}, {Nelan}, {Nissen},
  {Schaefer}, \& {Harmer}}]{VandenBerg2014}
{VandenBerg}, D.~A., {Bond}, H.~E., {Nelan}, E.~P., {et~al.} 2014, \apj, 792,
  110

\bibitem[{{VandenBerg} {et~al.}(2013){VandenBerg}, {Brogaard}, {Leaman}, \&
  {Casagrande}}]{VandenBerg2013}
{VandenBerg}, D.~A., {Brogaard}, K., {Leaman}, R., \& {Casagrande}, L. 2013,
  \apj, 775, 134

\bibitem[{{Vanderbosch} {et~al.}(2020){Vanderbosch}, {Hermes}, {Dennihy},
  {Dunlap}, {Izquierdo}, {Tremblay}, {Cho}, {G{\"a}nsicke}, {Toloza}, {Bell},
  {Montgomery}, \& {Winget}}]{2020ApJ...897..171V}
{Vanderbosch}, Z., {Hermes}, J.~J., {Dennihy}, E., {et~al.} 2020, \apj, 897,
  171

\bibitem[{{Vanderburg} {et~al.}(2015){Vanderburg}, {Johnson}, {Rappaport},
  {Bieryla}, {Irwin}, {Lewis}, {Kipping}, {Brown}, {Dufour}, {Ciardi}, {Angus},
  {Schaefer}, {Latham}, {Charbonneau}, {Beichman}, {Eastman}, {McCrady},
  {Wittenmyer}, \& {Wright}}]{2015Natur.526..546V}
{Vanderburg}, A., {Johnson}, J.~A., {Rappaport}, S., {et~al.} 2015, \nat, 526,
  546

\bibitem[{Vanderburg {et~al.}(2020)Vanderburg, Rowden, Bryson, Coughlin,
  Batalha, Collins, Latham, Mullally, Col{\'o}n, Henze,
  {et~al.}}]{Vanderburg2020habitable}
Vanderburg, A., Rowden, P., Bryson, S., {et~al.} 2020, The Astrophysical
  Journal Letters, 893, L27

\bibitem[{{Vanderburg} {et~al.}(2020){Vanderburg}, {Rappaport}, {Xu},
  {Crossfield}, {Becker}, {Gary}, {Murgas}, {Blouin}, {Kaye}, {Palle}, {Melis},
  {Morris}, {Kreidberg}, {Gorjian}, {Morley}, {Mann}, {Parviainen}, {Pearce},
  {Newton}, {Carrillo}, {Zuckerman}, {Nelson}, {Zeimann}, {Brown},
  {Tronsgaard}, {Klein}, {Ricker}, {Vanderspek}, {Latham}, {Seager}, {Winn},
  {Jenkins}, {Adams}, {Benneke}, {Berardo}, {Buchhave}, {Caldwell},
  {Christiansen}, {Collins}, {Col{\'o}n}, {Daylan}, {Doty}, {Doyle},
  {Dragomir}, {Dressing}, {Dufour}, {Fukui}, {Glidden}, {Guerrero}, {Guo},
  {Heng}, {Henriksen}, {Huang}, {Kaltenegger}, {Kane}, {Lewis}, {Lissauer},
  {Morales}, {Narita}, {Pepper}, {Rose}, {Smith}, {Stassun}, \&
  {Yu}}]{2020Natur.585..363V}
{Vanderburg}, A., {Rappaport}, S.~A., {Xu}, S., {et~al.} 2020, \nat, 585, 363

\bibitem[{{Vauclair} {et~al.}(1989){Vauclair}, {Goupil}, {Baglin}, {Auvergne},
  \& {Chevreton}}]{1989A&A...215L..17V}
{Vauclair}, G., {Goupil}, M.~J., {Baglin}, A., {Auvergne}, M.~., \&
  {Chevreton}, M. 1989, \aap, 215, L17

\bibitem[{Veras {et~al.}(2020)Veras, Reichert, Flammini~Dotti, Cai, Mustill,
  Shannon, McDonald, Portegies~Zwart, Kouwenhoven, \&
  Spurzem}]{veras2020linking}
Veras, D., Reichert, K., Flammini~Dotti, F., {et~al.} 2020, Monthly Notices of
  the Royal Astronomical Society, 493, 5062

\bibitem[{{Vestrand} {et~al.}(2014){Vestrand}, {Wren}, {Panaitescu}, {Wozniak},
  {Davis}, {Palmer}, {Vianello}, {Omodei}, {Xiong}, {Briggs}, {Elphick},
  {Paciesas}, \& {Rosing}}]{2014Sci...343...38V}
{Vestrand}, W.~T., {Wren}, J.~A., {Panaitescu}, A., {et~al.} 2014, Science,
  343, 38

\bibitem[{{Vickers} {et~al.}(2021){Vickers}, {Shen}, \& {Li}}]{Vickers2021}
{Vickers}, J.~J., {Shen}, J., \& {Li}, Z.-Y. 2021, \apj, 922, 189

\bibitem[{Vorobiev {et~al.}(2019)Vorobiev, Irwin, Ninkov, Donlon, Caldwell, \&
  Mochnacki}]{Vorobiev:2019}
Vorobiev, D., Irwin, A., Ninkov, Z., {et~al.} 2019, Journal of Astronomical
  Telescopes, Instruments, and Systems, 5, 1 .
\newblock \url{https://doi.org/10.1117/1.JATIS.5.4.041507}

\bibitem[{{Wallace} {et~al.}(2019){Wallace}, {Hartman}, {Bakos}, \&
  {Bhatti}}]{Wallace2019ApJ}
{Wallace}, J.~J., {Hartman}, J.~D., {Bakos}, G.~{\'A}., \& {Bhatti}, W. 2019,
  \apjl, 870, L7

\bibitem[{Walsh(2018)}]{walsh2018rubble}
Walsh, K.~J. 2018, Annual Review of Astronomy and Astrophysics, 56, 593

\bibitem[{{Walsh} {et~al.}(2011){Walsh}, {Morbidelli}, {Raymond}, {O'Brien}, \&
  {Mandell}}]{2011Natur.475..206W}
{Walsh}, K.~J., {Morbidelli}, A., {Raymond}, S.~N., {O'Brien}, D.~P., \&
  {Mandell}, A.~M. 2011, \nat, 475, 206

\bibitem[{{Wang} \& {Han}(2012)}]{2012NewAR..56..122W}
{Wang}, B., \& {Han}, Z. 2012, \nar, 56, 122

\bibitem[{{Wang} {et~al.}(2022{\natexlab{a}}){Wang}, {Huang}, {Yuan}, {Zhang},
  {Xiang}, \& {Liu}}]{WangC2022}
{Wang}, C., {Huang}, Y., {Yuan}, H., {et~al.} 2022{\natexlab{a}}, arXiv
  e-prints, arXiv:2201.09442

\bibitem[{{Wang} {et~al.}(2019){Wang}, {Liu}, {Xiang}, {Huang}, {Chen}, {Yuan},
  {Ren}, {Zhang}, \& {Tian}}]{WangC2019}
{Wang}, C., {Liu}, X.~W., {Xiang}, M.~S., {et~al.} 2019, \mnras, 482, 2189

\bibitem[{{Wang} {et~al.}(2022{\natexlab{b}}){Wang}, {Zang}, {Zhu}, {Hwang},
  {Udalski}, {Gould}, {Han}, {Albrow}, {Chung}, {Jung}, {Kim}, {Ryu}, {Shin},
  {Shvartzvald}, {Yee}, {Cha}, {Kim}, {Kim}, {Kim}, {Lee}, {Lee}, {Lee},
  {Park}, {Pogge}, {Poleski}, {Mr{\'o}z}, {Skowron}, {Szyma{\'n}ski},
  {Soszy{\'n}ski}, {Pietrukowicz}, {Koz{\l}owski}, {Ulaczyk}, {Rybicki},
  {Iwanek}, {Wrona}, {Gromadzki}, {Yang}, {Mao}, \& {Zhang}}]{OB180383}
{Wang}, H., {Zang}, W., {Zhu}, W., {et~al.} 2022{\natexlab{b}}, \mnras, 510,
  1778

\bibitem[{{Wang} \& {Fischer}(2013)}]{Wang2013}
{Wang}, J., \& {Fischer}, D.~A. 2013, ArXiv:1310.7830, arXiv:1310.7830

\bibitem[{{Wang} \& {Fischer}(2015)}]{Wang:2015}
---. 2015, \aj, 149, 14

\bibitem[{{Wang} {et~al.}(2013{\natexlab{a}}){Wang}, {Fischer}, {Barclay},
  {Boyajian}, {Crepp}, {Schwamb}, {Lintott}, {Jek}, {Smith}, {Parrish},
  {Schawinski}, {Schmitt}, {Giguere}, {Brewer}, {Lynn}, {Simpson}, {Hoekstra},
  {Jacobs}, {LaCourse}, {Schwengeler}, {Chopin}, \& {Herszkowicz}}]{Wang:2013}
{Wang}, J., {Fischer}, D.~A., {Barclay}, T., {et~al.} 2013{\natexlab{a}}, \apj,
  776, 10

\bibitem[{{Wang} {et~al.}(2013{\natexlab{b}}){Wang}, {Wang}, {Filippenko},
  {Zhang}, \& {Zhao}}]{2013Sci...340..170W}
{Wang}, X., {Wang}, L., {Filippenko}, A.~V., {Zhang}, T., \& {Zhao}, X.
  2013{\natexlab{b}}, Science, 340, 170

\bibitem[{{Wang} {et~al.}(2013{\natexlab{c}}){Wang}, {Liang}, {Li}, {Lu},
  {Wei}, \& {Zhang}}]{2013ApJ...774..132W}
{Wang}, X.-G., {Liang}, E.-W., {Li}, L., {et~al.} 2013{\natexlab{c}}, \apj,
  774, 132

\bibitem[{Warner {et~al.}(2009)Warner, Harris, \& Pravec}]{warner2009asteroid}
Warner, B.~D., Harris, A.~W., \& Pravec, P. 2009, Icarus, 202, 134

\bibitem[{{Waxman} \& {Katz}(2017)}]{Waxman2017}
{Waxman}, E., \& {Katz}, B. 2017, in Handbook of Supernovae, ed. A.~W.
  {Alsabti} \& P.~{Murdin} (Springer), 967

\bibitem[{Weiss {et~al.}(2022)Weiss, Millholland, Petigura, Adams, Batygin,
  Bloch, \& Mordasini}]{Weiss2022}
Weiss, L.~M., Millholland, S.~C., Petigura, E.~A., {et~al.} 2022, arXiv
  preprint arXiv:2203.10076

\bibitem[{Weiss {et~al.}(2018)Weiss, Marcy, Petigura, Fulton, Howard, Winn,
  Isaacson, Morton, Hirsch, Sinukoff, {et~al.}}]{Weiss2018}
Weiss, L.~M., Marcy, G.~W., Petigura, E.~A., {et~al.} 2018, The Astronomical
  Journal, 155, 48

\bibitem[{{Welsh} {et~al.}(2012){Welsh}, {Orosz}, {Carter}, {Fabrycky}, {Ford},
  {Lissauer}, {Pr{\v{s}}a}, {Quinn}, {Ragozzine}, {Short}, {Torres}, {Winn},
  {Doyle}, {Barclay}, {Batalha}, {Bloemen}, {Brugamyer}, {Buchhave},
  {Caldwell}, {Caldwell}, {Christiansen}, {Ciardi}, {Cochran}, {Endl},
  {Fortney}, {Gautier}, {Gilliland}, {Haas}, {Hall}, {Holman}, {Howard},
  {Howell}, {Isaacson}, {Jenkins}, {Klaus}, {Latham}, {Li}, {Marcy}, {Mazeh},
  {Quintana}, {Robertson}, {Shporer}, {Steffen}, {Windmiller}, {Koch}, \&
  {Borucki}}]{Welsh2012}
{Welsh}, W.~F., {Orosz}, J.~A., {Carter}, J.~A., {et~al.} 2012, \nat, 481, 475

\bibitem[{{Welsh} {et~al.}(2015){Welsh}, {Orosz}, {Short}, {Cochran}, {Endl},
  {Brugamyer}, {Haghighipour}, {Buchhave}, {Doyle}, {Fabrycky}, {Hinse},
  {Kane}, {Kostov}, {Mazeh}, {Mills}, {M{\"u}ller}, {Quarles}, {Quinn},
  {Ragozzine}, {Shporer}, {Steffen}, {Tal-Or}, {Torres}, {Windmiller}, \&
  {Borucki}}]{Welsh2015}
{Welsh}, W.~F., {Orosz}, J.~A., {Short}, D.~R., {et~al.} 2015, \apj, 809, 26

\bibitem[{{Wiktorowicz} {et~al.}(2019){Wiktorowicz}, {Wyrzykowski},
  {Chruslinska}, {Klencki}, {Rybicki}, \& {Belczynski}}]{Wiktorowicz19}
{Wiktorowicz}, G., {Wyrzykowski}, {\L}., {Chruslinska}, M., {et~al.} 2019,
  \apj, 885, 1

\bibitem[{{Windemuth} {et~al.}(2019){Windemuth}, {Agol}, {Carter}, {Ford},
  {Haghighipour}, {Orosz}, \& {Welsh}}]{Windemuth19}
{Windemuth}, D., {Agol}, E., {Carter}, J., {et~al.} 2019, \mnras, 490, 1313

\bibitem[{{Witt} \& {Mao}(1994)}]{Shude1994}
{Witt}, H.~J., \& {Mao}, S. 1994, \apj, 430, 505

\bibitem[{{Wittenmyer} {et~al.}(2022){Wittenmyer}, {Clark}, {Trifonov},
  {Addison}, {Wright}, {Stassun}, {Horner}, {Lowson}, {Kielkopf}, {Kane},
  {Plavchan}, {Shporer}, {Zhang}, {Bowler}, {Mengel}, {Okumura}, {Rabus},
  {Johnson}, {Harbeck}, {Tronsgaard}, {Buchhave}, {Collins}, {Collins}, {Gan},
  {Jensen}, {Howell}, {Furlan}, {Gnilka}, {Lester}, {Matson}, {Scott},
  {Ricker}, {Vanderspek}, {Latham}, {Seager}, {Winn}, {Jenkins}, {Rudat},
  {Quintana}, {Rodriguez}, {Caldwell}, {Quinn}, {Essack}, \&
  {Bouma}}]{Wittenmyer2022}
{Wittenmyer}, R.~A., {Clark}, J.~T., {Trifonov}, T., {et~al.} 2022, \aj, 163,
  82

\bibitem[{{Wolf} {et~al.}(2018){Wolf}, {Onken}, {Luvaul}, {Schmidt}, {Bessell},
  {Chang}, {Da Costa}, {Mackey}, {Martin-Jones}, {Murphy}, {Preston}, {Scalzo},
  {Shao}, {Smillie}, {Tisserand}, {White}, \& {Yuan}}]{Wolf2018}
{Wolf}, C., {Onken}, C.~A., {Luvaul}, L.~C., {et~al.} 2018, \pasa, 35, e010

\bibitem[{{Wolniewicz} {et~al.}(2021){Wolniewicz}, {Berger}, \&
  {Huber}}]{Wolniewicz2021}
{Wolniewicz}, L.~M., {Berger}, T.~A., \& {Huber}, D. 2021, \aj, 161, 231

\bibitem[{{Wu} {et~al.}(2018{\natexlab{a}}){Wu}, {Li}, \&
  {Deng}}]{2018ApJ...867...47W}
{Wu}, T., {Li}, Y., \& {Deng}, Z.-m. 2018{\natexlab{a}}, \apj, 867, 47

\bibitem[{Wu(2001)}]{Wu2001}
Wu, Y. 2001, Monthly Notices of the Royal Astronomical Society, 323, 248

\bibitem[{{Wu}(2019)}]{Wu2019}
{Wu}, Y. 2019, \apj, 874, 91

\bibitem[{{Wu} \& {Lithwick}(2013)}]{WuLithwick2013}
{Wu}, Y., \& {Lithwick}, Y. 2013, \apj, 772, 74

\bibitem[{{Wu} {et~al.}(2021){Wu}, {Xiang}, {Chen}, {Zhao}, {Bi}, {Li}, {Li},
  \& {Huang}}]{WuYQ2021}
{Wu}, Y., {Xiang}, M., {Chen}, Y., {et~al.} 2021, \mnras, 501, 4917

\bibitem[{{Wu} {et~al.}(2018{\natexlab{b}}){Wu}, {Xiang}, {Bi}, {Liu}, {Yu},
  {Hon}, {Sharma}, {Li}, {Huang}, {Liu}, {Zhang}, {Li}, {Ge}, {Tian}, {Zhang},
  \& {Zhang}}]{WuYQ2018}
{Wu}, Y., {Xiang}, M., {Bi}, S., {et~al.} 2018{\natexlab{b}}, \mnras, 475, 3633

\bibitem[{{Wu} {et~al.}(2019){Wu}, {Xiang}, {Zhao}, {Bi}, {Liu}, {Shi},
  {Huang}, {Yuan}, {Wang}, {Chen}, {Huo}, {Ren}, {Tian}, {Liu}, {Zhang}, {Li},
  \& {Zhang}}]{WuYQ2019}
{Wu}, Y., {Xiang}, M., {Zhao}, G., {et~al.} 2019, \mnras, 484, 5315

\bibitem[{{Wu} {et~al.}(2017){Wu}, {Xiang}, {Zhang}, {Li}, {Bi}, {Liu}, {Fu},
  {Huang}, {Tian}, {Liu}, {Ge}, {He}, \& {Zhang}}]{WuYQ2017}
{Wu}, Y.-Q., {Xiang}, M.-S., {Zhang}, X.-F., {et~al.} 2017, Research in
  Astronomy and Astrophysics, 17, 5

\bibitem[{{Xiang} \& {Rix}(2022)}]{Xiang2022}
{Xiang}, M., \& {Rix}, H.-W. 2022, Nature, 603, 599–603

\bibitem[{{Xiang} {et~al.}(2017{\natexlab{a}}){Xiang}, {Liu}, {Shi}, {Yuan},
  {Huang}, {Chen}, {Wang}, {Tian}, {Wu}, {Yang}, {Zhang}, {Huo}, \&
  {Ren}}]{Xiang2017b}
{Xiang}, M., {Liu}, X., {Shi}, J., {et~al.} 2017{\natexlab{a}}, \apjs, 232, 2

\bibitem[{{Xiang} {et~al.}(2018){Xiang}, {Shi}, {Liu}, {Yuan}, {Chen}, {Huang},
  {Wang}, {Wu}, {Tian}, {Huo}, {Zhang}, \& {Zhang}}]{Xiang2018}
{Xiang}, M., {Shi}, J., {Liu}, X., {et~al.} 2018, \apjs, 237, 33

\bibitem[{{Xiang} {et~al.}(2019){Xiang}, {Ting}, {Rix}, {Sandford}, {Buder},
  {Lind}, {Liu}, {Shi}, \& {Zhang}}]{Xiang2019}
{Xiang}, M., {Ting}, Y.-S., {Rix}, H.-W., {et~al.} 2019, \apjs, 245, 34

\bibitem[{{Xiang} {et~al.}(2015){Xiang}, {Liu}, {Yuan}, {Huang}, {Wang}, {Ren},
  {Chen}, {Sun}, {Zhang}, {Huo}, \& {Rebassa-Mansergas}}]{Xiang2015}
{Xiang}, M.-S., {Liu}, X.-W., {Yuan}, H.-B., {et~al.} 2015, Research in
  Astronomy and Astrophysics, 15, 1209

\bibitem[{{Xiang} {et~al.}(2017{\natexlab{b}}){Xiang}, {Liu}, {Yuan}, {Huo},
  {Huang}, {Wang}, {Chen}, {Ren}, {Zhang}, {Tian}, {Yang}, {Shi}, {Zhao}, {Li},
  {Zhao}, {Cui}, {Li}, {Hou}, {Zhang}, {Zhang}, {Wang}, {Wu}, {Cao}, {Yan},
  {Yan}, {Luo}, {Zhang}, {Bai}, {Yuan}, {Dong}, {Lei}, \& {Li}}]{Xiang2017a}
{Xiang}, M.~S., {Liu}, X.~W., {Yuan}, H.~B., {et~al.} 2017{\natexlab{b}},
  \mnras, 467, 1890

\bibitem[{Xie {et~al.}(2016)Xie, Dong, Zhu, Huber, Zheng, De~Cat, Fu, Liu, Luo,
  Wu, {et~al.}}]{Xie2016}
Xie, J.-W., Dong, S., Zhu, Z., {et~al.} 2016, Proceedings of the National
  Academy of Sciences, 113, 11431

\bibitem[{{Xu} {et~al.}(2016){Xu}, {Jura}, {Dufour}, \&
  {Zuckerman}}]{2016ApJ...816L..22X}
{Xu}, S., {Jura}, M., {Dufour}, P., \& {Zuckerman}, B. 2016, \apjl, 816, L22

\bibitem[{{Xu} {et~al.}(2022){Xu}, {Yuan}, {Niu}, {Yang}, {Beers}, \&
  {Huang}}]{Xu2022}
{Xu}, S., {Yuan}, H., {Niu}, Z., {et~al.} 2022, \apjs, 258, 44

\bibitem[{{Yan} \& {Zhu}(2022)}]{CSST_Wei}
{Yan}, S., \& {Zhu}, W. 2022, Research in Astronomy and Astrophysics, 22,
  025006

\bibitem[{{Yang} {et~al.}(2014){Yang}, {Bou{\'e}}, {Fabrycky}, \&
  {Abbot}}]{Yang2014HZ}
{Yang}, J., {Bou{\'e}}, G., {Fabrycky}, D.~C., \& {Abbot}, D.~S. 2014, \apjl,
  787, L2

\bibitem[{{Yang} {et~al.}(2013){Yang}, {Cowan}, \& {Abbot}}]{Yang2013HZ}
{Yang}, J., {Cowan}, N.~B., \& {Abbot}, D.~S. 2013, \apjl, 771, L45

\bibitem[{{Yang} {et~al.}(2020){Yang}, {Xie}, \& {Zhou}}]{Yang2020}
{Yang}, J.-Y., {Xie}, J.-W., \& {Zhou}, J.-L. 2020, \aj, 159, 164

\bibitem[{{Yang} {et~al.}(2022){Yang}, {Yuan}, {Xiang}, {Duan}, {Huang}, {Liu},
  {Beers}, {Galarza}, {Daflon}, {Fern{\'a}ndez-Ontiveros}, {Cenarro},
  {Crist{\'o}bal-Hornillos}, {Hern{\'a}ndez-Monteagudo}, {L{\'o}pez-Sanjuan},
  {Mar{\'\i}n-Franch}, {Moles}, {Varela}, {Rami{\'o}}, {Alcaniz}, {Dupke},
  {Ederoclite}, {Sodr{\'e}}, \& {Angulo}}]{Yang2022}
{Yang}, L., {Yuan}, H., {Xiang}, M., {et~al.} 2022, \aap, 659, A181

\bibitem[{{Yang} {et~al.}(2019){Yang}, {Zhang}, \& {Wei}}]{2019ApJ...878...89Y}
{Yang}, Y.-P., {Zhang}, B., \& {Wei}, J.-Y. 2019, \apj, 878, 89

\bibitem[{{Yanny} {et~al.}(2009){Yanny}, {Rockosi}, {Newberg}, {Knapp},
  {Adelman-McCarthy}, {Alcorn}, {Allam}, {Allende Prieto}, {An}, {Anderson},
  {Anderson}, {Bailer-Jones}, {Bastian}, {Beers}, {Bell}, {Belokurov},
  {Bizyaev}, {Blythe}, {Bochanski}, {Boroski}, {Brinchmann}, {Brinkmann},
  {Brewington}, {Carey}, {Cudworth}, {Evans}, {Evans}, {Gates}, {G{\"a}nsicke},
  {Gillespie}, {Gilmore}, {Nebot Gomez-Moran}, {Grebel}, {Greenwell}, {Gunn},
  {Jordan}, {Jordan}, {Harding}, {Harris}, {Hendry}, {Holder}, {Ivans},
  {Ivezi{\v{c}}}, {Jester}, {Johnson}, {Kent}, {Kleinman}, {Kniazev},
  {Krzesinski}, {Kron}, {Kuropatkin}, {Lebedeva}, {Lee}, {French Leger},
  {L{\'e}pine}, {Levine}, {Lin}, {Long}, {Loomis}, {Lupton}, {Malanushenko},
  {Malanushenko}, {Margon}, {Martinez-Delgado}, {McGehee}, {Monet}, {Morrison},
  {Munn}, {Neilsen}, {Nitta}, {Norris}, {Oravetz}, {Owen}, {Padmanabhan},
  {Pan}, {Peterson}, {Pier}, {Platson}, {Re Fiorentin}, {Richards}, {Rix},
  {Schlegel}, {Schneider}, {Schreiber}, {Schwope}, {Sibley}, {Simmons},
  {Snedden}, {Allyn Smith}, {Stark}, {Stauffer}, {Steinmetz}, {Stoughton},
  {SubbaRao}, {Szalay}, {Szkody}, {Thakar}, {Sivarani}, {Tucker}, {Uomoto},
  {Vanden Berk}, {Vidrih}, {Wadadekar}, {Watters}, {Wilhelm}, {Wyse}, {Yarger},
  \& {Zucker}}]{Yanny2009}
{Yanny}, B., {Rockosi}, C., {Newberg}, H.~J., {et~al.} 2009, \aj, 137, 4377

\bibitem[{Ye {et~al.}(2019{\natexlab{a}})Ye, Masci, Lin, Bolin, Chang, Duev,
  Helou, Ip, Kaplan, Kramer, {et~al.}}]{ye2019toward}
Ye, Q., Masci, F.~J., Lin, H.~W., {et~al.} 2019{\natexlab{a}}, Publications of
  the Astronomical Society of the Pacific, 131, 078002

\bibitem[{Ye {et~al.}(2019{\natexlab{b}})Ye, Kelley, Bodewits, Bolin, Jones,
  Lin, Bellm, Dekany, Duev, Groom, {et~al.}}]{ye2019multiple}
Ye, Q., Kelley, M.~S., Bodewits, D., {et~al.} 2019{\natexlab{b}}, The
  Astrophysical Journal Letters, 874, L16

\bibitem[{{Yee}(2013)}]{Yee2013}
{Yee}, J.~C. 2013, \apjl, 770, L31

\bibitem[{{Yee} {et~al.}(2021){Yee}, {Zang}, {Udalski}, {Ryu}, {Green},
  {Hennerley}, {Marmont}, {Sumi}, {Mao}, {Gromadzki}, {Mr{\'o}z}, {Skowron},
  {Poleski}, {Szyma{\'n}ski}, {Soszy{\'n}ski}, {Pietrukowicz}, {Koz{\l}owski},
  {Ulaczyk}, {Rybicki}, {Iwanek}, {Wrona}, {Albrow}, {Chung}, {Gould}, {Han},
  {Hwang}, {Jung}, {Kim}, {Shin}, {Shvartzvald}, {Cha}, {Kim}, {Kim}, {Lee},
  {Lee}, {Lee}, {Park}, {Pogge}, {Bachelet}, {Christie}, {Hundertmark}, {Maoz},
  {McCormick}, {Natusch}, {Penny}, {Street}, {Tsapras}, {Beichman}, {Bryden},
  {Novati}, {Carey}, {Gaudi}, {Henderson}, {Johnson}, {Zhu}, {Bond}, {Abe},
  {Barry}, {Bennett}, {Bhattacharya}, {Donachie}, {Fujii}, {Fukui}, {Hirao},
  {Silva}, {Itow}, {Kirikawa}, {Kondo}, {Koshimoto}, {Alex Li}, {Matsubara},
  {Muraki}, {Miyazaki}, {Olmschenk}, {Ranc}, {Rattenbury}, {Satoh}, {Shoji},
  {Suzuki}, {Tanaka}, {Tristram}, {Yamawaki}, {Yonehara}, \& {MOA
  Collaboration}}]{OB190960}
{Yee}, J.~C., {Zang}, W., {Udalski}, A., {et~al.} 2021, \aj, 162, 180

\bibitem[{{Yu} {et~al.}(2020){Yu}, {Bedding}, {Stello}, {Huber}, {Compton},
  {Gizon}, \& {Hekker}}]{yu2020a}
{Yu}, J., {Bedding}, T.~R., {Stello}, D., {et~al.} 2020, \mnras, 493, 1388

\bibitem[{{Yu} {et~al.}(2021){Yu}, {Hekker}, {Bedding}, {Stello}, {Huber},
  {Gizon}, {Khanna}, \& {Bi}}]{yu2021}
{Yu}, J., {Hekker}, S., {Bedding}, T.~R., {et~al.} 2021, \mnras, 501, 5135

\bibitem[{{Yu} {et~al.}(2018){Yu}, {Huber}, {Bedding}, {Stello}, {Hon},
  {Murphy}, \& {Khanna}}]{Yu2018}
{Yu}, J., {Huber}, D., {Bedding}, T.~R., {et~al.} 2018, \apjs, 236, 42

\bibitem[{{Yu} \& {Liu}(2018)}]{YuJC2018}
{Yu}, J., \& {Liu}, C. 2018, \mnras, 475, 1093

\bibitem[{{Yuan} {et~al.}(2021){Yuan}, {Li}, {Desch}, \& {Ko}}]{Yuan2021}
{Yuan}, Q., {Li}, M.~M., {Desch}, S.~J., \& {Ko}, B. 2021, in 52nd Lunar and
  Planetary Science Conference, Lunar and Planetary Science Conference, 1980

\bibitem[{{Yuan} {et~al.}(2020){Yuan}, {Li}, {Liu}, {Niu}, {Lu}, {Jiang},
  {Wang}, {Li}, {Liang}, {Wang}, {Zhang}, {Wang}, {Li}, {Tian}, {Lu}, {Chen},
  {Huang}, {Liu}, {Yao}, {Cui}, \& {Li}}]{YuanX2020}
{Yuan}, X., {Li}, Z., {Liu}, X., {et~al.} 2020, in Society of Photo-Optical
  Instrumentation Engineers (SPIE) Conference Series, Vol. 11445, Society of
  Photo-Optical Instrumentation Engineers (SPIE) Conference Series, 114457M

\bibitem[{{Zang} {et~al.}(2021{\natexlab{a}}){Zang}, {Hwang}, {Udalski},
  {Wang}, {Zhu}, {Sumi}, {Yee}, {Gould}, {Mao}, {Zhang}, {Albrow}, {Chung},
  {Han}, {Jung}, {Ryu}, {Shin}, {Shvartzvald}, {Cha}, {Kim}, {Kim}, {Kim},
  {Lee}, {Lee}, {Lee}, {Park}, {Pogge}, {Mr{\'o}z}, {Skowron}, {Poleski},
  {Szyma{\'n}ski}, {Soszy{\'n}ski}, {Pietrukowicz}, {Koz{\l}owski}, {Ulaczyk},
  {Rybicki}, {Iwanek}, {Wrona}, {Gromadzki}, {Bond}, {Abe}, {Barry}, {Bennett},
  {Bhattacharya}, {Donachie}, {Fujii}, {Fukui}, {Hirao}, {Itow}, {Kirikawa},
  {Kondo}, {Koshimoto}, {Li}, {Matsubara}, {Muraki}, {Miyazaki}, {Olmschenk},
  {Ranc}, {Rattenbury}, {Satoh}, {Shoji}, {Ishitani Silva}, {Suzuki}, {Tanaka},
  {Tristram}, {Yamawaki}, {Yonehara}, {Beichman}, {Bryden}, {Calchi Novati},
  {Carey}, {Gaudi}, {Henderson}, {Johnson}, \& {Spitzer Team}}]{OB191053}
{Zang}, W., {Hwang}, K.-H., {Udalski}, A., {et~al.} 2021{\natexlab{a}}, \aj,
  162, 163

\bibitem[{{Zang} {et~al.}(2021{\natexlab{b}}){Zang}, {Han}, {Kondo}, {Yee},
  {Lee}, {Gould}, {Mao}, {de Almeida}, {Shvartzvald}, {Zhang}, {Albrow},
  {Chung}, {Hwang}, {Jung}, {Ryu}, {Shin}, {Cha}, {Kim}, {Kim}, {Kim}, {Lee},
  {Lee}, {Park}, {Pogge}, {Drummond}, {Tan}, {Nascimento J{\'u}nior}, {Maoz},
  {Penny}, {Zhu}, {Bond}, {Abe}, {Barry}, {Bennett}, {Bhattacharya},
  {Donachie}, {Fujii}, {Fukui}, {Hirao}, {Itow}, {Kirikawa}, {Koshimoto}, {Alex
  Li}, {Matsubara}, {Muraki}, {Miyazaki}, {Olmschenk}, {Ranc}, {Rattenbury},
  {Satoh}, {Shoji}, {Silva}, {Sumi}, {Suzuki}, {Tanaka}, {Tristram},
  {Yamawaki}, {Yonehara}, {Petric}, {Burdullis}, \& {Fouqu{\'e}}}]{KB200414}
{Zang}, W., {Han}, C., {Kondo}, I., {et~al.} 2021{\natexlab{b}}, Research in
  Astronomy and Astrophysics, 21, 239

\bibitem[{{Zang} {et~al.}(2022){Zang}, {Yang}, {Han}, {Lee}, {Udalski},
  {Gould}, {Mao}, {Zhang}, {Zhu}, {Albrow}, {Chung}, {Hwang}, {Jung}, {Ryu},
  {Shin}, {Shvartzvald}, {Yee}, {Cha}, {Kim}, {Kim}, {Kim}, {Lee}, {Lee},
  {Park}, {Mr{\'o}z}, {Skowron}, {Poleski}, {Szyma{\'n}ski}, {Soszy{\'n}ski},
  {Pietrukowicz}, {Koz{\l}owski}, {Ulaczyk}, {Rybicki}, {Iwanek}, {Wrona}, \&
  {Gromadzki}}]{KB191042}
{Zang}, W., {Yang}, H., {Han}, C., {et~al.} 2022, arXiv e-prints,
  arXiv:2204.02017

\bibitem[{{Zarka} {et~al.}(2019){Zarka}, {Li}, {Grie{\ss}meier}, {Lamy},
  {Girard}, {Hess}, {Lazio}, \& {Hallinan}}]{zarka19}
{Zarka}, P., {Li}, D., {Grie{\ss}meier}, J.-M., {et~al.} 2019, Research in
  Astronomy and Astrophysics, 19, 023

\bibitem[{Zechmeister {et~al.}(2019)Zechmeister, Dreizler, Ribas, Reiners,
  Caballero, Bauer, B{\'e}jar, Gonz{\'a}lez-Cuesta, Herrero, Lalitha,
  {et~al.}}]{Zechmeister2019carmenes}
Zechmeister, M., Dreizler, S., Ribas, I., {et~al.} 2019, Astronomy \&
  Astrophysics, 627, A49

\bibitem[{{Zhan}(2021)}]{ZhanH2021}
{Zhan}, H. 2021, Chinese Science Bulletin, 66, 1290

\bibitem[{{Zhang} {et~al.}(2019{\natexlab{a}}){Zhang}, {Yu}, {Liang}, {Yang},
  {Ashley}, {Cui}, {Du}, {Fu}, {Gong}, {Gu}, {Hu}, {Jiang}, {Liu}, {Lawrence},
  {Liu}, {Li}, {Li}, {Ma}, {Mould}, {Shang}, {Suntzeff}, {Tao}, {Tian},
  {Tinney}, {Uddin}, {Wang}, {Wang}, {Wang}, {Wei}, {Wright}, {Wu},
  {Wittenmyer}, {Xu}, {Yang}, {Yu}, {Yuan}, {Zheng}, {Zhou}, {Zhou}, \&
  {Zhu}}]{zhang2019a}
{Zhang}, H., {Yu}, Z., {Liang}, E., {et~al.} 2019{\natexlab{a}}, \apjs, 240, 16

\bibitem[{{Zhang} {et~al.}(2019{\natexlab{b}}){Zhang}, {Yu}, {Liang}, {Yang},
  {Ashley}, {Cui}, {Du}, {Fu}, {Gong}, {Gu}, {Hu}, {Jiang}, {Liu}, {Lawrence},
  {Liu}, {Li}, {Li}, {Ma}, {Mould}, {Shang}, {Suntzeff}, {Tao}, {Tian},
  {Tinney}, {Uddin}, {Wang}, {Wang}, {Wang}, {Wei}, {Wright}, {Wu},
  {Wittenmyer}, {Xu}, {Yang}, {Yu}, {Yuan}, {Zheng}, {Zhou}, {Zhou}, \&
  {Zhu}}]{zhang2019b}
---. 2019{\natexlab{b}}, \apjs, 240, 17

\bibitem[{{Zhang} {et~al.}(2014{\natexlab{a}}){Zhang}, {Pi}, \&
  {Yang}}]{Zly2014}
{Zhang}, L.-Y., {Pi}, Q.-f., \& {Yang}, Y.-G. 2014{\natexlab{a}}, \mnras, 442,
  2620

\bibitem[{{Zhang} {et~al.}(2014{\natexlab{b}}){Zhang}, {Li}, {Li}, \&
  {Lin}}]{Zhang2014}
{Zhang}, X., {Li}, H., {Li}, S., \& {Lin}, D. N.~C. 2014{\natexlab{b}}, \apjl,
  789, L23

\bibitem[{{Zhang} \& {Fabrycky}(2019)}]{ZhangFabrycky2019}
{Zhang}, Z., \& {Fabrycky}, D.~C. 2019, \apj, 879, 92

\bibitem[{{Zhu}(2019)}]{Zhu:2019}
{Zhu}, W. 2019, \apj, 873, 8

\bibitem[{Zhu(2020)}]{zhu2020patterns}
Zhu, W. 2020, The Astronomical Journal, 159, 188

\bibitem[{{Zhu} \& {Dong}(2021)}]{ZhuDong:2021}
{Zhu}, W., \& {Dong}, S. 2021, \araa, 59, arXiv:2103.02127

\bibitem[{{Zhu} {et~al.}(2014){Zhu}, {Penny}, {Mao}, {Gould}, \&
  {Gendron}}]{Zhu2014ApJ}
{Zhu}, W., {Penny}, M., {Mao}, S., {Gould}, A., \& {Gendron}, R. 2014, \apj,
  788, 73

\bibitem[{{Zhu} {et~al.}(2018){Zhu}, {Petrovich}, {Wu}, {Dong}, \&
  {Xie}}]{Zhu2018}
{Zhu}, W., {Petrovich}, C., {Wu}, Y., {Dong}, S., \& {Xie}, J. 2018, \apj, 860,
  101

\bibitem[{{Zhu} {et~al.}(2016{\natexlab{a}}){Zhu}, {Wang}, \&
  {Huang}}]{Zhu:2016}
{Zhu}, W., {Wang}, J., \& {Huang}, C. 2016{\natexlab{a}}, \apj, 832, 196

\bibitem[{{Zhu} \& {Wu}(2018)}]{ZhuWu:2018}
{Zhu}, W., \& {Wu}, Y. 2018, \aj, 156, 92

\bibitem[{{Zhu} {et~al.}(2016{\natexlab{b}}){Zhu}, {Calchi Novati}, {Gould},
  {Udalski}, {Han}, {Shvartzvald}, {Ranc}, {J{\o}rgensen}, {Poleski}, {Bozza},
  {Beichman}, {Bryden}, {Carey}, {Gaudi}, {Henderson}, {Pogge}, {Porritt},
  {Wibking}, {Yee}, {SPITZER Team}, {Pawlak}, {Szyma{\'n}ski}, {Skowron},
  {Mr{\'o}z}, {Koz{\l}owski}, {Wyrzykowski}, {Pietrukowicz}, {Pietrzy{\'n}ski},
  {Soszy{\'n}ski}, {Ulaczyk}, {OGLE Group}, {Choi}, {Park}, {Jung}, {},
  {Albrow}, {Park}, {Kim}, {Lee}, {Cha}, {Kim}, {Lee}, {KMTNET Group},
  {Friedmann}, {Kaspi}, {Maoz}, {WISE Group}, {Hundertmark}, {Street},
  {Tsapras}, {Bramich}, {Cassan}, {Dominik}, {Bachelet}, {Dong}, {Figuera
  Jaimes}, {Horne}, {Mao}, {Menzies}, {Schmidt}, {Snodgrass}, {Steele},
  {Wambsganss}, {RoboNeT Team}, {Skottfelt}, {Andersen}, {Burgdorf}, {Ciceri},
  {D'Ago}, {Evans}, {Gu}, {Hinse}, {Kerins}, {Korhonen}, {Kuffmeier},
  {Mancini}, {Peixinho}, {Popovas}, {Rabus}, {Rahvar}, {Tronsgaard},
  {Scarpetta}, {Southworth}, {Surdej}, {von Essen}, {Wang}, {Wertz}, \&
  {MiNDSTEP Group}}]{ZhuPLFS}
{Zhu}, W., {Calchi Novati}, S., {Gould}, A., {et~al.} 2016{\natexlab{b}}, \apj,
  825, 60

\bibitem[{{Zhu} {et~al.}(2017{\natexlab{a}}){Zhu}, {Udalski}, {Calchi Novati},
  {Chung}, {Jung}, {Ryu}, {}, {Gould}, {Lee}, {Albrow}, {Yee}, {Han}, {Hwang},
  {Cha}, {Kim}, {Kim}, {Kim}, {Kim}, {Lee}, {Park}, {Pogge}, {KMTNet
  Collaboration}, {Poleski}, {Mr{\'o}z}, {Pietrukowicz}, {Skowron},
  {Szyma{\'n}ski}, {Koz{\l}owski}, {Ulaczyk}, {Pawlak}, {OGLE Collaboration},
  {Beichman}, {Bryden}, {Carey}, {Fausnaugh}, {Gaudi}, {Henderson},
  {Shvartzvald}, {Wibking}, \& {Spitzer Team}}]{Zhu2017spitzer}
{Zhu}, W., {Udalski}, A., {Calchi Novati}, S., {et~al.} 2017{\natexlab{a}},
  \aj, 154, 210

\bibitem[{{Zhu} {et~al.}(2017{\natexlab{b}}){Zhu}, {Huang}, {Udalski},
  {Soares-Furtado}, {Poleski}, {Skowron}, {Mr{\'o}z}, {Szyma{\'n}ski},
  {Soszy{\'n}ski}, {Pietrukowicz}, {Koz{\L}owski}, {Ulaczyk}, \&
  {Pawlak}}]{2017PASP..129j4501Z}
{Zhu}, W., {Huang}, C.~X., {Udalski}, A., {et~al.} 2017{\natexlab{b}}, \pasp,
  129, 104501

\bibitem[{{Ziaali} {et~al.}(2019){Ziaali}, {Bedding}, {Murphy}, {Van Reeth}, \&
  {Hey}}]{2019MNRAS.486.4348Z}
{Ziaali}, E., {Bedding}, T.~R., {Murphy}, S.~J., {Van Reeth}, T., \& {Hey},
  D.~R. 2019, \mnras, 486, 4348

\bibitem[{{Zong} {et~al.}(2016{\natexlab{a}}){Zong}, {Charpinet}, \&
  {Vauclair}}]{2016A&A...594A..46Z}
{Zong}, W., {Charpinet}, S., \& {Vauclair}, G. 2016{\natexlab{a}}, \aap, 594,
  A46

\bibitem[{{Zong} {et~al.}(2021){Zong}, {Charpinet}, \&
  {Vauclair}}]{2021ApJ...921...37Z}
---. 2021, \apj, 921, 37

\bibitem[{{Zong} {et~al.}(2016{\natexlab{b}}){Zong}, {Charpinet}, {Vauclair},
  {Giammichele}, \& {Van Grootel}}]{2016A&A...585A..22Z}
{Zong}, W., {Charpinet}, S., {Vauclair}, G., {Giammichele}, N., \& {Van
  Grootel}, V. 2016{\natexlab{b}}, \aap, 585, A22

\bibitem[{{Zou} {et~al.}(2007){Zou}, {Wu}, \& {Dai}}]{2007A&A...461..115Z}
{Zou}, Y.~C., {Wu}, X.~F., \& {Dai}, Z.~G. 2007, \aap, 461, 115

\bibitem[{{Zuckerman} {et~al.}(2003){Zuckerman}, {Koester}, {Reid}, \&
  {H{\"u}nsch}}]{2003ApJ...596..477Z}
{Zuckerman}, B., {Koester}, D., {Reid}, I.~N., \& {H{\"u}nsch}, M. 2003, \apj,
  596, 477

\bibitem[{{Zuckerman} {et~al.}(2010){Zuckerman}, {Melis}, {Klein}, {Koester},
  \& {Jura}}]{2010ApJ...722..725Z}
{Zuckerman}, B., {Melis}, C., {Klein}, B., {Koester}, D., \& {Jura}, M. 2010,
  \apj, 722, 725

\bibitem[{{Zuluaga} {et~al.}(2015){Zuluaga}, {Kipping}, {Sucerquia}, \&
  {Alvarado}}]{Zuluaga2015}
{Zuluaga}, J.~I., {Kipping}, D.~M., {Sucerquia}, M., \& {Alvarado}, J.~A. 2015,
  \apjl, 803, L14

\end{thebibliography}
\bibliographystyle{aasjournal}

\end{document}